%% file: Tesis.tex
\documentclass[a4paper,12pt,twoside,openright]{book}

\usepackage[T1]{fontenc}
\usepackage[utf8x]{inputenc}
\usepackage[english]{babel}
\usepackage{lmodern}

\usepackage{graphicx}
\usepackage{fancyhdr}
\usepackage{indentfirst}
\usepackage{setspace}
\usepackage{amsfonts}
\usepackage{amsmath}
\usepackage{amssymb}
\usepackage{cite}
\usepackage{booktabs}
\usepackage{tabularx}
\usepackage{caption}
\usepackage{subcaption}
\usepackage{float}
\usepackage{color}
\usepackage{epstopdf}
\usepackage{anysize}
\usepackage{float}
\usepackage{nccmath}
\usepackage{diagbox}
\usepackage{multirow}
\usepackage{rotating}
\usepackage{mathrsfs}
\usepackage{slashed}
\usepackage{eufrak}
\usepackage{capt-of}
\usepackage{lastpage}
\usepackage{appendix}
\usepackage{epigraph}
\usepackage{enumerate}
\usepackage{stackengine}
\usepackage{lipsum}
\usepackage{mathtools}
\usepackage{xspace}
\newcommand\xrowht[2][0]{\addstackgap[.5\dimexpr#2\relax]{\vphantom{#1}}}
\usepackage{bm}
\usepackage{lineno}

\definecolor{darkred}{rgb}{0.6,0,0}
\definecolor{darkpurple}{rgb}{0.5,0,0.5}
\definecolor{darkgreen}{rgb}{0, 0.6, 0}

\usepackage[]{hyperref}
\hypersetup{
  colorlinks=true,
  allcolors=darkpurple,
  urlcolor=darkred,
  citecolor=darkgreen,
  pdfborder={0 0 1},
  linktocpage=false,
  bookmarks=true,
  pdftitle={Tesis}
}
\usepackage{tikz}

\usetikzlibrary{decorations.markings}
\usetikzlibrary{decorations.pathmorphing}

\tikzset{
    vector/.style={decorate, decoration={snake}, draw},
    fermion/.style={draw=black, postaction={decorate},
        decoration={markings,mark=at position .55 with {\arrow[scale=1.,>=stealth]{>}}}},
    fermionbar/.style={draw=black, postaction={decorate},
        decoration={markings,mark=at position .58 with {\arrow[scale=1.,>=stealth]{<}}}},
    fermionline/.style={draw=black},
    gluon/.style={decorate, draw=black,
        decoration={coil,amplitude=4pt, segment length=5pt}},
    scalar/.style={dashed, draw=black, postaction={decorate},
        decoration={markings,mark=at position .55 with {\arrow[scale=1.,>=stealth]{>}}}},
    scalarbar/.style={dashed, draw=black, postaction={decorate},
        decoration={markings,mark=at position .58 with {\arrow[scale=1.,>=stealth]{<}}}},
    scalarline/.style={dashed,draw=black},
    fmassin/.style={draw=black, postaction={decorate},
    decoration={markings, mark=at position .75 with {\arrow[scale=1.,>=stealth]{>}}, mark=at position .30 with {\arrow[scale=1.,>=stealth]{<}}}},
}

\makeatletter
\def \cleardoublepage {\clearpage \if@twoside
\ifodd \c@page
\else
\null\thispagestyle{empty}\clearpage
\fi
\fi}
\makeatother

\newcommand{\fig}[1]{figure~\ref{#1}}
\newcommand{\Fig}[1]{Figure~\ref{#1}}
\newcommand{\figs}[2]{figures~\ref{#1}-\ref{#2}}
\newcommand{\tab}[1]{table~\ref{#1}}
\newcommand{\Tab}[1]{Table~\ref{#1}}
\renewcommand{\eq}[1]{\eqref{#1}}
\renewcommand{\eqs}[2]{\eqref{#1}-\eqref{#2}}
\newcommand{\ch}[1]{Chapter~\ref{#1}}
\newcommand{\sect}[1]{section~\ref{#1}}
\newcommand{\sects}[2]{sections~\ref{#1}~and~\ref{#2}}
\newcommand{\Sect}[1]{Section~\ref{#1}}
\newcommand{\app}[1]{appendix~\ref{#1}}

\newcommand{\nn}{\nonumber}
\def\vev#1{\left\langle #1\right\rangle}
\def\ket#1{\left\vert #1\right\rangle}
\def\bra#1{\left\langle #1\right\vert}
\newcommand{\hc}{\mathrm{h.c.}}
\newcommand{\mueg}{$\mu \rightarrow e \gamma$\xspace}
\newcommand{\taueg}{$\tau \rightarrow e \gamma$\xspace}
\newcommand{\taumug}{$\tau \rightarrow \mu \gamma$\xspace}
\newcommand{\mueee}{$\mu \rightarrow 3 \, e $\xspace}

\begin{document}

%%%%%%%%%%%%%%%%%%%%%%%%%%%%%%%%%%%%%%%%%%%%%%%%%%%%%%%%%%%%
%%%%%%%%%%%%%%%%% Portada interior %%%%%%%%%%%%%%%%%%%%%%%%%
%%%%%%%%%%%%%%%%%%%%%%%%%%%%%%%%%%%%%%%%%%%%%%%%%%%%%%%%%%%%
\setlength{\unitlength}{1cm} %Especificar unidad de trabajo
\thispagestyle{empty}
\begin{center}

\begin{picture}(3.5,2.4)
\put(0,-3){\includegraphics[scale=0.06]{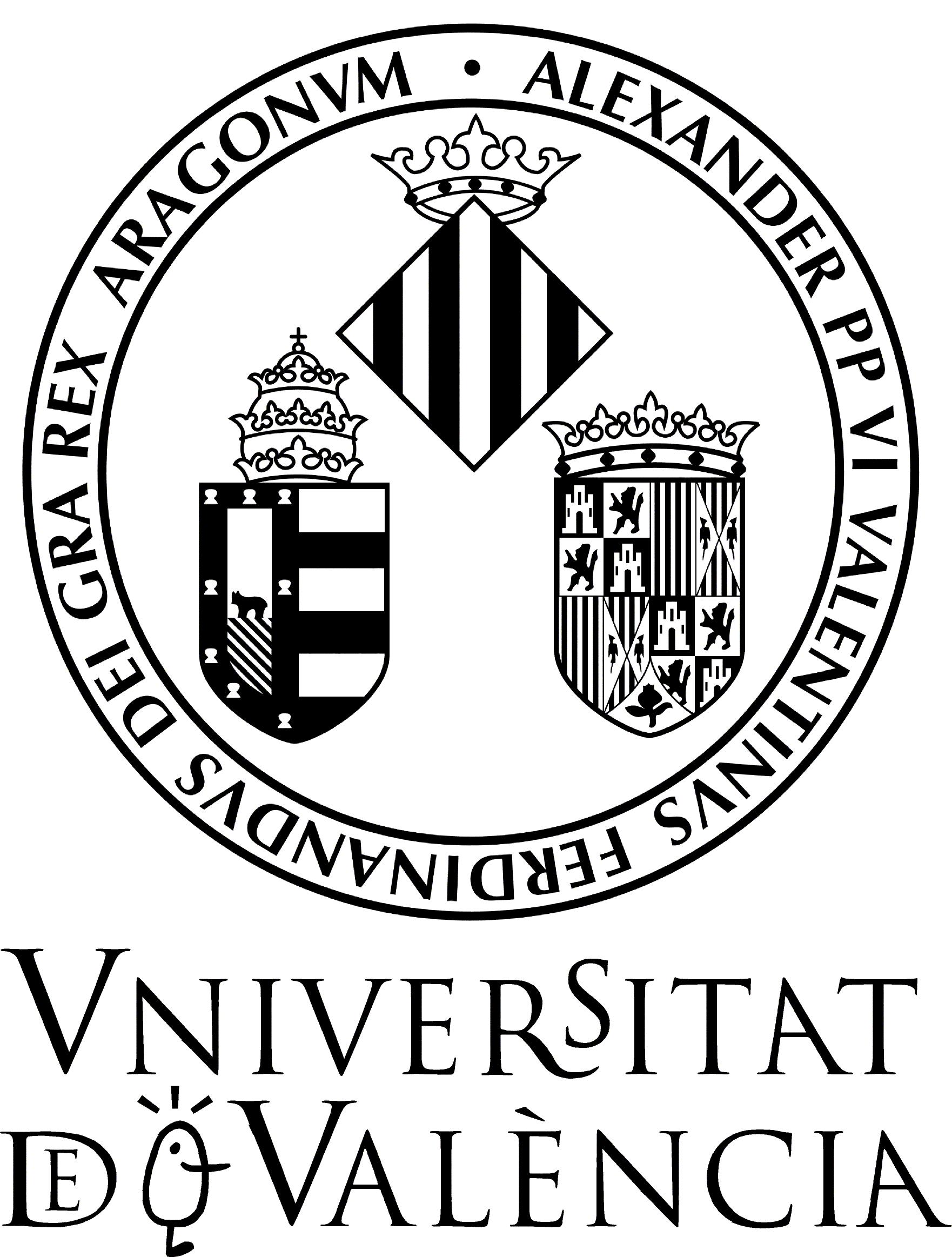}}
\end{picture}
\vspace*{4.55cm}

\textbf{\huge{Radiative neutrino masses:}} \\[1.5ex]
\textbf{\LARGE{A window to new physics}} \\[4ex]
{\Large Tesis Doctoral\\[1ex]
Programa de Doctorat en Física}

\vspace{\fill}

{\LARGE \bf{Ricardo Cepedello Pérez}} \\[4ex]

{\Large Director: Dr. Martin Hirsch}\\[8ex]

{IFIC - CSIC/Universitat de València}\\
{Departament de Física Teòrica}\\[2ex]

{\Large València, enero 2021}

\end{center}
%%%%%%%%%%%%%%%%%%%%%%%%%%%%%%%%%%%%%%%%%%%%%%%%%%%%%%%%%%%%

\newpage
\thispagestyle{empty}

%\cleardoublepage
%\thispagestyle{empty}

%\vfill

%\begin{flushright}
%    \textbf{Ricardo Cepedello Pérez}
%    
%    \textit{Radiative neutrino masses: A window to new physics}
%    
%    Enero, 2021
%    
%    Director: Dr. Martin Hirsch
%    \\[1cm]
%    
%    \textbf{Universitat de València}
%    
%    \textit{Astroparticle and high energy physics group}
%    
%    Institut de Física Corpuscular (CSIC-UV)
%    
%    Departament de Física Teòrica
%    
%\end{flushright}

%%%%%%%%%%%%%%%%%%%% Dedicatoria %%%%%%%%%%%%%%%%%%%%%%%%%%%
\begin{titlepage}
\cleardoublepage
\thispagestyle{empty}
\vspace*{4cm}
\epigraph{\textit{A Laura, a mi familia y, en especial, a mis abuelos.\\A mis amigos, para que vean que al final algo sí fue concluyente.}}{}

\end{titlepage}
%%%%%%%%%%%%%%%%%%%%%%%%%%%%%%%%%%%%%%%%%%%%%%%%%%%%%%%%%%%%

%\singlespacing 
%\frontmatter
%\thispagestyle{empty}
%\vspace*{2cm}

%%%%%%%%%%%%%%%%%% Certifica Tesis %%%%%%%%%%%%%%%%%%%%%%%%%
%\noindent Dr. Martin Hirsch,\\
%\noindent investigador titular del Consejo Superior de Investigaciones Científicas (CSIC),\\[2ex]
%%
%\noindent CERTIFICA:\\[2ex]
%%
%\noindent Que la presente memoria "Radiative neutrino masses: A window to new physics" ha sido realizada bajo su direcci\'on en el Institut de F\'isica Corpuscular, centro mixto del CSIC y de la Universitat de València, por Ricardo Cepedello Pérez y constituye su Tesis para optar al grado de Doctor en F\'isica.\\[2ex]
%%
%\noindent Y para que as\'i conste, en cumplimiento de la legislaci\'on vigente, presenta en el Departament de F\'isica Teòrica de la Universidad de Valencia la referida Tesis Doctoral, y firma el presente certificado.\\[2ex]

%Paterna (Valencia), a 6 de noviembre de 2020.

%\begin{center}
%\includegraphics[width=0.4\textwidth]{firma_Martin}

%\vspace*{-0.6cm}
%Dr. Martin Hirsch
%\end{center}
%%%%%%%%%%%%%%%%%%%%%%%%%%%%%%%%%%%%%%%%%%%%%%%%%%%%%%%%%%%%

%%%%%%%%%%%%%%%%% Acknowledgements %%%%%%%%%%%%%%%%%%%%%%%%%
%\chapter*{Acknowledgements}
%\addcontentsline{toc}{chapter}{Acknowledgements} 

%I would like to thank ...

%%%%%%%%%%%%%%%%%%%%%%%%%%%%%%%%%%%%%%%%%%%%%%%%%%%%%%%%%%%%

%%%%%%%%%%%%%%%%% List of Publications %%%%%%%%%%%%%%%%%%%%%

\chapter*{List of scientific publications}
\addcontentsline{toc}{chapter}{List of scientific publications} 

This thesis is based on the following publications:
\begin{enumerate}
    \item \textit{Loop neutrino masses from $d = 7$ operator}\\
          R. Cepedello, J.C. Helo, M. Hirsch\\
          \textbf{Journal of High-Energy Physics (2017)}, Volume 07, p. 79

    \item \textit{Lepton number violating phenomenology of $d = 7$ neutrino mass models}\\
          R. Cepedello, J.C. Helo, M. Hirsch\\
          \textbf{Journal of High-Energy Physics (2018)}, Volume 01, p. 9

    \item \textit{Systematic classification of three-loop realizations of the Weinberg operator}\\
          R. Cepedello, R.M. Fonseca, M. Hirsch\\
          \textbf{Journal of High-Energy Physics (2018)}, Volume 10, p. 197

    \item \textit{Neutrinoless Double-$\beta$ Decay with Nonstandard Majoron Emission}\\
          R. Cepedello, F.F. Deppisch, L. Gonz\'alez, C. Hati, M. Hirsch\\
          \textbf{Physical Review Letters (2019)}, Volume 122, Issue 18, p. 1801

    \item \textit{Systematic classification of two loop $d = 4$ Dirac neutrino mass models and the Diracness-dark matter stability connection}\\
          S. Centelles-Chuli\`a, R. Cepedello, E. Peinado, R. Srivastava\\
          \textbf{Journal of High-Energy Physics (2019)}, Volume 10, p. 93

    \item \textit{Radiative type-I seesaw neutrino masses}\\
          C. Arbel\'aez, A.E. C\'arcamo, R. Cepedello, M. Hirsch, S. Kovalenko\\
          \textbf{Physical Review D (2019)}, Volume 100, Issue 11, p. 5021
          
    \item \textit{Dark matter stability and Dirac neutrinos using only Standard Model symmetries}\\
          C. Bonilla, S. Centelles-Chuli\`a, R. Cepedello, E. Peinado, R. Srivastava\\
          \textbf{Physical Review D (2020)}, Volume 101, Issue 3, p. 033011

    \item \textit{Scotogenic Dark Symmetry as a residual subgroup of Standard Model Symmetries}\\
          S. Centelles-Chuli\`a, R. Cepedello, E. Peinado, R. Srivastava\\
          \textbf{Chinese Physics C (2020)}, Volume 44, Issue 8, p. 3110

    \item \textit{Minimal 3-loop neutrino mass models and charged lepton flavor violation}\\
          R. Cepedello, M. Hirsch, P. Rocha-Mor\'an, A. Vicente\\
          \textbf{Journal of High-Energy Physics (2020)}, Volume 08, p. 067

\end{enumerate}

Other publications not included in the thesis:
\begin{enumerate}
    \item \textit{Sequentially loop suppressed fermion masses from a single discrete symmetry}\\
          C. Arbel\'aez, A.E. C\'arcamo, R. Cepedello, S. Kovalenko, I. Schmidt\\
          \textbf{Journal of High-Energy Physics (2020)}, Volume 06, p. 43
    
    \item \textit{$(g-2)$ and neutrino masses}\\
          C. Arbel\'aez, R. Cepedello, R.M. Fonseca, M. Hirsch\\
          \textbf{Physical Review D (2020)}, Volume 102, Issue 07, p. 5005\\
\end{enumerate}

%%%%%%%%%%%%%%%%%%%%%%%%%%%%%%%%%%%%%%%%%%%%%%%%%%%%%%%%%%%%

%%%%%%%%%%%%%%%%%%%%%%%% Índice %%%%%%%%%%%%%%%%%%%%%%%%%%%%
{\hypersetup{hidelinks}
\tableofcontents
}

%%%%%%%%%%%%%%%%%%%%%%%%%%%%%%%%%%%%%%%%%%%%%%%%%%%%%%%%%%%%
%%%%%%%%%%%%%%%%%%% Cuerpo principal %%%%%%%%%%%%%%%%%%%%%%%
%%%%%%%%%%%%%%%%%%%%%%%%%%%%%%%%%%%%%%%%%%%%%%%%%%%%%%%%%%%%
\mainmatter

%%%%%%%%%%%%%%%%%%%%%%% Introduccion %%%%%%%%%%%%%%%%%%%%%%%

\fancyhf{}
\fancyhead[LE,RO]{\thepage}
\input{Introduction/Introduction}

%%%%%%%%%%%%%%%%%%%%%%%% Capítulos %%%%%%%%%%%%%%%%%%%%%%%%%

%\input{Chapters/SM_intro/Chapter}

\fancyhf{}
\fancyhead[LE,RO]{\thepage}
\fancyhead[RE]{\slshape\nouppercase{\leftmark}}
\fancyhead[LO]{\slshape\nouppercase{\rightmark}}

\input{Chapters/Neutrino_physics/Chapter_nuphys}

\input{Chapters/Neutrino_masses/Chapter_numass}

\input{Chapters/Dim7_1loop/Chapter_dim7}

\input{Chapters/Dim5_3loop/Chapter_3loop}

\input{Chapters/Dirac_2loop/Chapter_Dirac2loop}

\input{Chapters/Dirac_Majorana_DM/Chapter_DM}

\input{Chapters/Loop_seesaw/Chapter_loopseesaw}

\input{Chapters/Dim7_pheno/Chapter_dim7pheno}

\input{Chapters/cLFV_3loop/Chapter_clfv}

\input{Chapters/0vBB_majoron/Chapter_majoron}

%%%%%%%%%%%%%%%%%%%%%%% Conclusiones %%%%%%%%%%%%%%%%%%%%%%%%

\input{Conclusions/Conclusions}

\clearpage

%%%%%%%%%%%%%%%%%%%%%%%%% Appendices %%%%%%%%%%%%%%%%%%%%%%%%

\begin{appendices}
\noappendicestocpagenum
\addappheadtotoc

\input{Appendices/Appendix_topos}

%\lhead[\thepage]{Appendix \thechapter. \rightmark}
%\rhead[Appendix \thechapter \leftmark]{\thepage}

\input{Appendices/Appendix_loops}
%\lhead[\thepage]{Appendix \thechapter. \rightmark}
%\rhead[Appendix \thechapter \leftmark]{\thepage}

\end{appendices}

%%%%%%%%%%%%%%%%%%%%%%%%%%%%%%%%%%%%%%%%%%%%%%%%%%%%%%%%%%%%
%%%%%%%%%%%%%%%%%%%%%%%%%%%%%%%%%%%%%%%%%%%%%%%%%%%%%%%%%%%%

%%%%%%%%%%%%%%%%%%%%%%%%% Resumen %%%%%%%%%%%%%%%%%%%%%%%%%%

\input{Resumen_Tesis/Resumen_Tesis}

%%%%%%%%%%%%%%%%%%%%%%%%%%%%%%%%%%%%%%%%%%%%%%%%%%%%%%%%%%%%

\fancyhf{}
\fancyhead[LE,RO]{\thepage}

\backmatter

\cleardoublepage
\addcontentsline{toc}{chapter}{\refname}

\phantomsection
\bibliographystyle{utphys}
\bibliography{BibFiles/biblio}

\end{document}

%% file: Introduction/Introduction.tex
\chapter*{Introduction}
\chaptermark{Introduction}
\addcontentsline{toc}{chapter}{Introduction}  

%The quest of high-energy physics to describe and understand the elementary components of matter and their interactions is one of the main parts of fundamental physics. It provides a deep understanding of the physics governing the universe and it may be our best option to shed light on a possible ultimate theory of everything.

Remarkably, our current understanding of matter, contained in the Standard Model (SM) of Fundamental Particles and Interactions \cite{Glashow:1961tr, Weinberg:1967tq, Salam:1968rm, Feynman:1969ej, Bjorken:1969ja, Gross:1973id, Politzer:1973fx}, provides an incredibly accurate picture of subatomic physics. The Standard Model is based on the principle of gauge invariance, which establishes a relation between local symmetries and forces mediated by spin-1 particles, as well as the interactions between the known fermions, once they are assigned to well defined representation of the gauge group. This approach based on gauge symmetries has proved to give a reliable picture of nature. The Standard Model describes a large number of observed phenomena with very precise predictions on particle properties and process rates, which predominantly agree with the experimental data to a great accuracy, representing one of the most successful theories in science of all time.

Despite the success, the Standard Model unfortunately fails to explain all current experimental results. As a consequence, it is generally considered to be an effective realisation of a more complete high-scale theory. The search of this theory is the goal of Beyond the Standard Model (BSM) Physics.

A major inconsistency of the Standard Model with current experimental results is the prediction of massless neutrinos, which has clearly proven to be wrong by the observation of neutrino oscillations. Starting from the observation of a deficit of electron neutrinos from the Sun \cite{Davis:1968cp}, and continuing with the \textit{atmospheric neutrino problem} and the verification by reactor neutrino experiments, the discovery of neutrino oscillations is one of the most important experimental achievements of high-energy physics in recent decades and it establishes beyond doubt the existence of non-zero neutrino masses \cite{Hirata:1990xa, Hirata:1992ku, Fukuda:1998mi, Cleveland:1998nv, Hampel:1998xg, Abdurashitov:1999zd, Fukuda:2001nj, Ahmad:2002jz, Eguchi:2002dm, Altmann:2005ix, Ahn:2006zza, Michael:2006rx, Abe:2008aa, Abe:2011sj, Abe:2011fz, An:2012eh, Ahn:2012nd, Abe:2014ugx}.

Although neutrino flavour oscillation data prove the mixing of the neutrino weak eigenstates $\nu_{e,\mu,\tau}$ as they propagate as mass eigenstates, oscillations cannot give a definite answer to the still unknown absolute neutrino mass scale or the dynamical origin of neutrino masses, including whether neutrinos are Dirac or Majorana fermions. Data from solar and atmospheric oscillations \cite{deSalas:2020pgw}, along with cosmology and other particle physics experiments \cite{Aghanim:2018eyx}, point to mass scales in the range $m_\nu \sim 0.01 - 1$ eV, although the last word should come from experiments directly measuring the absolute neutrino mass, like KATRIN \cite{Aker:2019uuj}.

Neutrinos are generally very elusive and intriguing particles still evading our complete understanding. The smallness of the neutrino masses compared to other fermions in the Standard Model and the fact that they are the only fermions in the SM that can be Majorana, guide the direction to be followed in BSM. How neutrino masses are generated is still unknown, opening a window of possibilities for new physics. The idea that the neutrino mass generating mechanism is connected to other open questions of the Standard Model is very compelling from a theoretical and phenomenological point of view. Starting from the seesaw neutrino mass mechanisms, plenty of models have been built and studied over the last decades. The different seesaw mechanisms can explain neutrino masses with a minimal particle content, but they require small couplings or high mass scales beyond the scope of current and planned experiments. Among the many possibilities in order to lower the new physics mass scale and move towards falsifiable models, an attractive way is to raise the dimensionality of the mass operator or generate masses radiatively via loops.
\\

The thesis is structured as follows. After providing some brief introduction to neutrino physics in \ch{ch:nuphys}, we focus on the generation of neutrino masses. In \ch{ch:numass} we review some of the most important neutrino mass mechanisms and introduce the main concepts and results of systematic classifications of neutrino mass models, essential for the following chapters. Chapters \ref{ch:Dim7_1loop}-\ref{ch:dirac2l} are dedicated to the systematic classification of radiative neutrino mass models. \ch{ch:Dim7_1loop} contains the classification of dimension 7 Majorana neutrino mass models generated at one-loop order, in \ch{ch:3loop} we study systematically the decomposition of the Weinberg operator at three-loop order, while in \ch{ch:dirac2l} we study the generation  of dimension four Dirac neutrino mass models at the two-loop level. We then move to more specific models and mechanisms in Chapters \ref{ch:dm_DiracMajo} and \ref{ch:loop_seesaw}. In the former, we study the connection between the generation of neutrino masses and the stability of dark matter analysing the breaking pattern of lepton number. In \ch{ch:loop_seesaw} we study, from a model-independent point of view, a particular realisation of the type-I seesaw which relies on the radiative generation of neutrino Dirac couplings. Finally, we end with the phenomenological analysis of some of the models discussed in the previous chapters. Chapters \ref{ch:Dim7_pheno} and \ref{ch:clfv} are devoted to the phenomenological analysis of dimension 7 one-loop neutrino mass models and  three-loop neutrino mass models, respectively, while in \ch{ch:0vbb} we discuss a very particular case of neutrinoless double-$\beta$ decay where electrons are emitted with opposite chiralities along with a light scalar (Majoron). Two appendices can be found at the end of the thesis with more details about the computation of the loop integrals, discussed throughout the whole thesis, as well as the list of some of the topologies and diagrams obtained from the classifications of neutrino mass models, not shown in the main text due to their large numbers.

\pagebreak
\fancyhf{}

%% file: Chapters/Neutrino_physics/Chapter_nuphys.tex
\fancyhf{}
\fancyhead[LE,RO]{\thepage}
\fancyhead[RE]{\slshape\nouppercase{\leftmark}}
\fancyhead[LO]{\slshape\nouppercase{\rightmark}}

\chapter{Neutrino physics}
\label{ch:nuphys}
\graphicspath{ {Chapters/Neutrino_physics/Figures/} }

In 1930, Pauli proposed neutrinos to explain the apparent non conservation of energy and spin in $\beta$ decay \cite{Pauli:1930pc}. These elusive particles were then observed in 1953 by Reines and Cowan escaping from a nuclear reactor in Savannah River \cite{Reines:1953pu}.

More than a decade later, the Homestake experiment directed by Davis \cite{Davis:1968cp}, observed a discrepancy between the theoretical solar neutrino flux and the one detected. This led to the hypothesis that neutrinos may be massive particles that mix while propagating \cite{Gribov:1968kq}. The explanation was confirmed in 1998 when Super-Kamiokande reported evidence for neutrino oscillations in the atmospheric neutrino flux \cite{Fukuda:1998mi} and later SNO \cite{Ahmad:2002jz} and KamLAND \cite{Eguchi:2002dm} corroborated solar neutrino oscillations.
\\

The purpose of this chapter is to briefly review the key facts about neutrino physics, paying special attention to neutrino flavour oscillations. We refer the interested reader to \cite{Mohapatra:1998rq, Giunti:2007ry, Valle:2015pba} for details.

%%%%%%%%%%%%%%%%%%%%%%%%%%%%%%%%%%%%%%%%%%%%%%%%%%%%%%%%%%%%%%%
%%%%%%%%%%%%%%%%%%%%%%%%%%%%%%%%%%%%%%%%%%%%%%%%%%%%%%%%%%%%%%%
\section{Neutrino oscillations}
\label{sec:nuphys:osc}

The concept of neutrino oscillations was first proposed in 1957 by Bruno Pontecorvo \cite{Pontecorvo:1957cp, Pontecorvo:1957qd} considering neutrino–antineutrino mixing, in analogy to the oscillation of neutral kaons. In 1962, Maki, Nakagawa and Sakata developed the idea that there is a mismatch between the interaction and propagation eigenstates \cite{Maki:1962mu}. The hypothesis that oscillations occur due to interference between different massive neutrinos was then further developed by Pontecorvo \cite{Pontecorvo:1967fh}.

%%%%%%%%%%%%%%%%%%%%%%%%%%%%%%%%%%%%%%%%%%%%%%%%%%%%%%%%%%%%%%%
\subsection{The leptonic mixing matrix}
\label{subsec:nuphys:numixing}

The oscillation comes from the fact that neutrinos are massive and, in general, their mass and flavour eigenstates do not coincide. Both basis, flavour and mass, are related to each other via a unitary mixing matrix, analogous to the Cabibbo-Kobayashi-Maskawa (CKM) mixing matrix in the quark sector \cite{Cabibbo:1963yz, Kobayashi:1973fv}. Thus, the three neutrinos $\nu_i$ ($i=1,2,3$), which propagate with masses $m_i$, are linear combinations of the three neutrino flavour or weak eigenstates $\nu_\alpha$ ($\alpha = e, \mu, \tau$), or vice-versa. In the basis in which the charged lepton Yukawas are diagonal,
\begin{equation} \label{eq:nuphys:Umix}
    \ket{ \nu_\alpha } = U_{\alpha i}^* \, \ket{ \nu_i } \, ,
\end{equation}
where the summation over repeated indices is understood. $U$ is the lepton mixing matrix or PMNS (after Pontecorvo, Maki, Nakagawa and Sakata) matrix \cite{Tanabashi:2018oca}, a $3 \times 3$ unitary mixing matrix that relates both basis. This matrix is usually parametrised as the product of three different rotations,
\begin{eqnarray} \label{eq:nuphys:Ulep}
    U &=& \begin{pmatrix}
        c_{12} c_{13} & s_{12} c_{13} & s_{13} e^{-i \delta}
        \\
        -s_{12} c_{23} - c_{12} s_{23} s_{13} e^{i \delta} & c_{12} c_{23} - s_{12} s_{23} s_{13} e^{i \delta} & s_{23} c_{13}
        \\
        s_{12} s_{23} - c_{12} c_{23} s_{13} e^{i \delta} & -c_{12} s_{23} - s_{12} c_{23} s_{13} e^{i \delta} & c_{23} c_{13}
    \end{pmatrix}
    \\[8pt] \nn
    && \times \,\,\, {\rm diag}(\, 1, \, e^{-i\, \alpha_{21}/2}, \, e^{-i\, \alpha_{31}/2} \, ) \, ,
\end{eqnarray}
where $s_{ij} = \sin \theta_{ij}$ and $c_{ij} = \cos \theta_{ij}$. $\delta$ is the Dirac CP phase, while $\alpha_{21}$ and $\alpha_{31}$ (Majorana phases) are two additional phases, consequence of the special properties of Majorana fermions \cite{Bilenky:1980cx}. These phases are zero for massive Dirac neutrinos.

%%%%%%%%%%%%%%%%%%%%%%%%%%%%%%%%%%%%%%%%%%%%%%%%%%%%%%%%%%%%%%%
\subsection{Oscillation in vacuum}
\label{subsec:nuphys:osc}

Neutrinos are produced as flavour eigenstates via charged currents (CC) or neutral currents (NC) mediated by $W^{\pm}$ bosons or $Z$ bosons, respectively. However, in vacuum the mass-defined states $\nu_i$ are eigenstates of the free Hamiltonian,
\begin{equation} %\label{eq:}
    \mathcal{H}_0 \ket{\nu_i} = E_{i} \ket{\nu_i} \, ,
\end{equation}
with eigenvalues $E_i^2 = \vec{p}\,^2+m_i^2$. These eigenstates evolve with the usual time-evolution operator $e^{-i\, E_i t}$. This time dependency can be related directly to the weak eigenstates by means of \eq{eq:nuphys:Umix}.
%,
%%
%\begin{eqnarray} \label{eq:nuphys:nu_t}
%    \ket{ \nu_\alpha(t) } &=& U_{\alpha i}^* \, e^{-i\, E_i t} \, \ket{\nu_i}
%    \\ \nn
%                          &=& U_{\alpha i}^* \, U_{\beta i} \, e^{-i\, E_i t} \, \ket{\nu_\beta} \, ,
%\end{eqnarray}
%%
%where in the second step we have introduced the inverse of \eq{eq:nuphys:Umix}. Here, we have defined $\ket{\nu_\beta (0) } \equiv \ket{\nu_\beta}$. 
Therefore, we see that flavour eigenstates at time $t$ are a superposition of flavour eigenstates at time $t=0$, and we can calculate the probability of an initial neutrino of flavour $\alpha$ oscillating to a flavour $\beta$ at a time $t$,
\begin{equation} \label{eq:nuphys:prob0}
    P_{\alpha \beta} (t) = | \vev{\nu_\beta | \nu_\alpha(t)} |^2 = U_{\alpha i}^* U_{\beta i} U_{\beta j}^* U_{\alpha j} \, e^{-i\,(E_i - E_j) t} \, .
\end{equation}
For relativistic neutrinos, and using the equal momentum approximation $E = | \vec{p}_i |$,
\begin{equation} %\label{eq:}
    E_i \approx E + \frac{m_i^2}{2E} \, .
\end{equation}
Considering also that neutrinos travel very close to the speed of light, we can take $t=L$, with $L$ the distance travelled. We then rewrite \eq{eq:nuphys:prob0} as,
\begin{equation} \label{eq:nuphys:prob}
    P_{\alpha \beta} (t) = | \vev{\nu_\beta | \nu_\alpha(t)} |^2 = U_{\alpha i}^* U_{\beta i} U_{\beta j}^* U_{\alpha j} \, e^{-i\,\frac{\Delta m_{ij}^2}{2E} \, L} \, ,
\end{equation}
with $\Delta m_{ij}^2 = m_i^2 - m_j^2$.

%%%%%%%%%%%%%%%%%%%%%%%%%%%%%%%%%%%%%%%%%%%%%%%%%%%%%%%%%%%%%%%
\subsection{Experimental status}
\label{subsec:nuphys:exp}

Neutrino oscillations depend on several parameters that need to be measured. From \eq{eq:nuphys:Ulep} we have three angles and the $\delta$ CP phase, as the oscillation probability \eq{eq:nuphys:prob} does not depend on the diagonal Majorana phases. The probability \eq{eq:nuphys:prob} shows that oscillation experiments are also sensitive to two mass-squared splittings, normally taken to be $\Delta m_{21}^2 = \Delta m_{\rm sol}^2$ and $\Delta m_{31}^2 = \Delta m_{\rm atm}^2$. This last remark leaves some open questions, for instance, the hierarchy of the masses is compatible with: 
\begin{itemize}
    \item Normal Hierarchy (NH) or Ordering (NO): $m_1 < m_2 < m_3$
    \item Inverted Hierarchy (IH) or Ordering (IO): $m_3 < m_1 < m_2$
    %
%    \item Degenerate spectrum: $m_1 \approx m_2 \approx m_3$
    %
\end{itemize}
Oscillation experiments cannot measure the absolute neutrino mass scale, they can only set lower bounds to the sum of the masses depending on the ordering: $\sim 0.06$ eV for NO and $\sim 0.1$ eV for IO, where the best fit values in \tab{tab:nuphys:globalfit} have been used.

\begin{figure}
    \centering
    \includegraphics[width=1\textwidth]{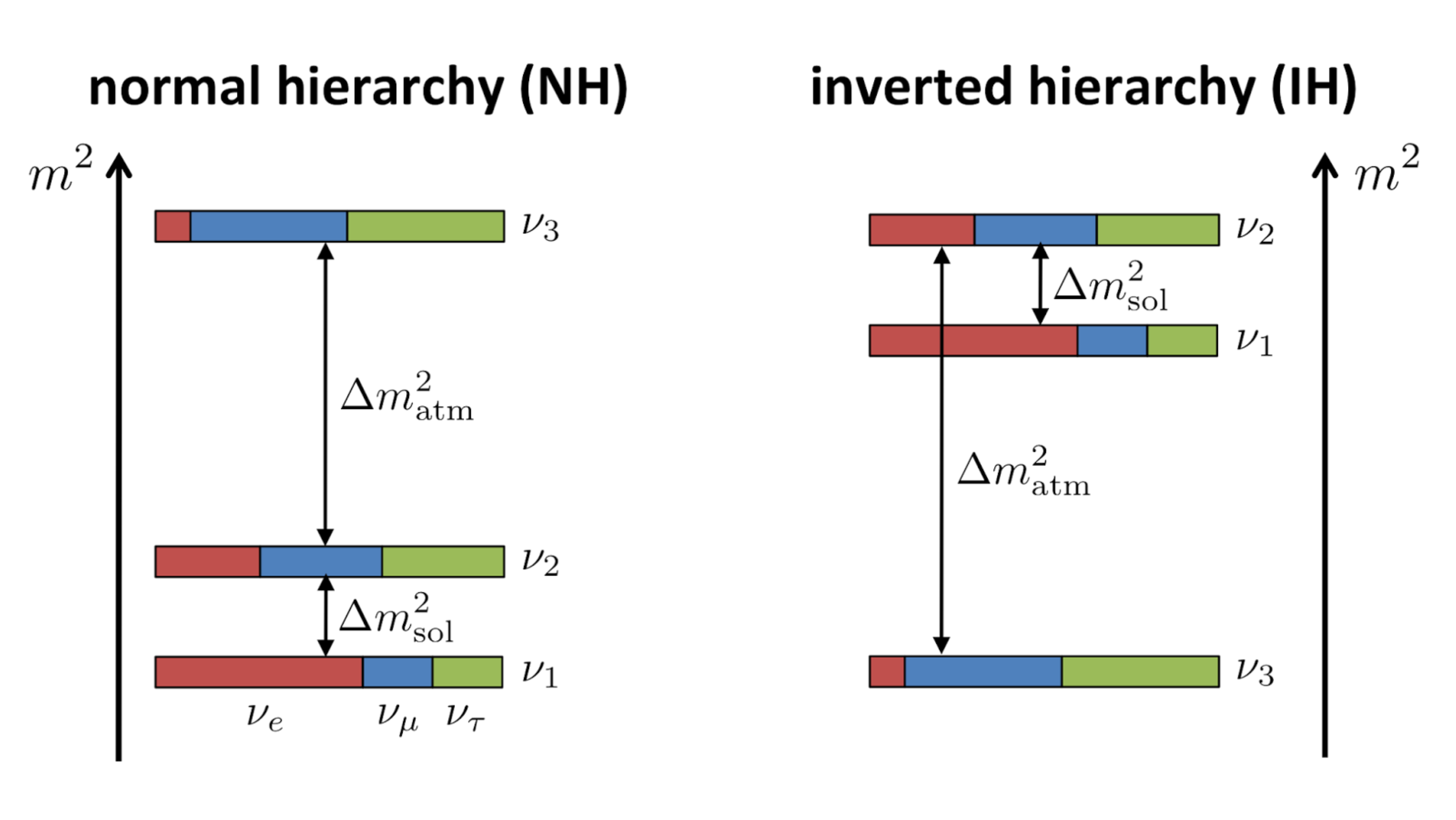}
    \caption{Ordering of the neutrino mass eigenstates. Colours represent the contribution of each flavour: electron, muon and tau are given in red, blue and green, respectively.}
    \label{fig:nuphys:hierarchy}
\end{figure}

\begin{table}
    \centering
    \setlength{\tabcolsep}{20pt}
    \renewcommand{\arraystretch}{1.5}
    \begin{tabular}{| c | c |}
        \hline
        \textbf{Parameter}  &  \textbf{Best fit} $\mathbf{\pm}$ $\mathbf{1\sigma}$
        \\
        \hline
        $\Delta m_{21}^2 \, [10^{-5}\,{\rm eV}^2]$  &  $7.50_{-0.20}^{+0.22}$
        \\
        $|\Delta m_{31}^2| \, [10^{-3}\,{\rm eV}^2]$ (NO)  &  $2.56_{-0.04}^{+0.03}$
        \\
        $|\Delta m_{31}^2| \, [10^{-3}\,{\rm eV}^2]$ (IO)  &  $2.46\, \pm 0.03$
        \\
        $\sin^2 \theta_{12} \, / \, 10^{-1}$  &  $3.18\, \pm 0.16$
        \\
        $\sin^2 \theta_{23} \, / \, 10^{-1}$ (NO)  &  $5.66_{-0.22}^{+0.16}$
        \\
        $\sin^2 \theta_{23} \, / \, 10^{-1}$ (IO)  &  $5.66_{-0.23}^{+0.18}$
        \\
        $\sin^2 \theta_{13} \, / \, 10^{-2}$ (NO)  &  $2.225_{-0.078}^{+0.055}$
        \\
        $\sin^2 \theta_{13} \, / \, 10^{-2}$ (IO)  &  $2.250_{-0.076}^{+0.056}$
        \\
        $\delta \, / \, \pi$ (NO)  &  $1.20_{-0.14}^{+0.23}$
        \\
        $\delta \, / \, \pi$ (IO)  &  $1.54\, \pm 0.13$
        \\
        \hline
    \end{tabular}
    \caption{Neutrino parameters from the global fit \cite{deSalas:2020pgw,Ternes:2020bvy} (see also \cite{Capozzi:2016rtj, Esteban:2018azc} for other global fits).}
    \label{tab:nuphys:globalfit}
\end{table}

Finally, considering the absolute neutrino mass scale, other experiments set constraints to its value. Cosmological observations provide the tightest limit on the sum of the neutrino masses, as neutrinos contribute to the energy density of the universe and the growth of large scale structures. However, this bound depends on assumptions made about the expansion history, as well as on the data included in the analysis, ranging from $\Sigma m_\nu < (0.11 - 0.54)$ eV ($95\%$ C.L.) \cite{Aghanim:2018eyx}. Tritium $\beta$ decay (KATRIN experiment) provides bounds on the so-called effective electron neutrino mass, $m_{\nu_e}^2 = \sum m_i^2 \, |U_{ei}|^2 < 1.1$ eV ($95\%$ C.L.) \cite{Aker:2019uuj}. Moreover, for Majorana neutrinos, neutrinoless double-beta ($0\nu\beta\beta$) decay is proportional to the effective Majorana mass of $\nu_e$, i.e. $m_{\beta\beta} = \left\vert \sum m_{i} \, U_{ei}^2 \right\vert$. Currently the best limit is $m_{\beta\beta} = 100$ meV \cite{KamLAND-Zen:2016pfg}, sensitive to the masses, mixings and also phases.

%%%%%%%%%%%%%%%%%%%%%%%%%%%%%%%%%%%%%%%%%%%%%%%%%%%%%%%%%%%%%%%
%%%%%%%%%%%%%%%%%%%%%%%%%%%%%%%%%%%%%%%%%%%%%%%%%%%%%%%%%%%%%%%
\section{Dirac vs. Majorana}
\label{sec:nuphys:dir_vs_maj}

Neutrinos are Weyl fermions, i.e. two-component spinors. Given that a local quantum field theory must be invariant under the $CPT$ transformation (charge conjugation, parity and time reversal), the theory must contain the $CPT$ conjugate of a neutrino field with spin $\pm 1/2$ and momentum $\vec{p}$, i.e.
\begin{equation} %\label{eq:}
    CPT \, \ket{ \nu(\vec{p},\pm 1/2)} = \ket{ \bar\nu (\vec{p},\mp 1/2)} \, ,
\end{equation}
which describes an antineutrino. Consequently, one cannot distinguish the initial neutrino from its $CPT$ conjugate, unless there is an additional quantum number. Thus, it is said that neutrinos are their own antiparticles, and they are represented by a Majorana spinors.

On the other hand, if an additional symmetry exists, for instance lepton number, it may help distinguishing neutrinos from their $CPT$ conjugates. Then neutrinos are Dirac particles. Note the difference between neutrinos and other Standard Model fermions, as for the latter electric charge differ for particles and antiparticles.
\\

In two-component notation,\footnote{Two-component spinor notation is described in detail in \cite{Dreiner:2008tw}.} the mass term for a Majorana neutrino reads,
\begin{equation} \label{eq:nuphys:majmass}
    m_L \, \bar\nu_L^c \nu_L \, + \, \hc \, .
\end{equation}
Majorana neutrinos have only two internal degrees of freedom and their mass term violates any symmetry, except $(Z_2)^n$.

For Dirac neutrinos, two more degrees of freedom, in the form of another Weyl spinor, are needed. This new right-handed state is normally denoted by $\nu_R$ or $N$. The mass term connects opposite chiralities as,
\begin{equation} %\label{eq:}
    m_D \, \bar\nu_R \nu_L \, + \, \hc \, .
\end{equation}
Indeed, a Dirac neutrino correspond to two Weyl spinors degenerate in mass.

In full generality, considering also the right-handed neutrino $\nu_R$, we can have the following mass terms:
\begin{equation} %\label{eq:}
    m_D \, \bar \nu_R \nu_L \, + \, \frac 12 \, m_L \, \bar\nu_L^c \nu_L \, + \, \frac 12 \, m_R \, \bar\nu_R^c \nu_R \, + \, \hc \, ,
\end{equation}
with several possibilities depending on $m_L$, $m_R$ and $m_D$.
\begin{itemize}
    \item $m_L = m_R = 0$: Dirac neutrinos with mass $m_D$.
    \item $m_D = 0$: $\nu_L$ and $\nu_R$ are both Majorana neutrinos with masses $m_L$ and $m_R$, respectively.
    \item $m_D \neq 0$ and, at least one of $m_L$, $m_R$ is not vanishing: neutrinos are Majorana with masses obtained by diagonalising the corresponding mass matrix (see \sect{sec:numass:majorana}).\\
\end{itemize}

Whether neutrinos are Dirac or Majorana particles is still an open question. Neutrino oscillations do not distinguish between the Dirac and Majorana nature of neutrinos, as the same probabilities are obtained for both options. A possible way to distinguish whether neutrinos are Dirac or Majorana fermions, would be an observation of a lepton number violating process, as we will discuss later.

%%%%%%%%%%%%%%%%%%%%%%%%%%%%%%%%%%%%%%%%%%%%%%%%%%%%%%%%%%%%%%%
%%%%%%%%%%%%%%%%%%%%%%%%%%%%%%%%%%%%%%%%%%%%%%%%%%%%%%%%%%%%%%%
\section{The Weinberg operator}
\label{sec:nuphys:weinberg}

There has been a huge effort in the field to understand how the masses of the neutrinos are generated and why it is so small compared to those of the Standard Model fermions. In the vast majority of the literature, neutrino are considered to be Majorana particles. There are several reasons behind this bias in favour of Majorana neutrinos. For instance, they have a far richer phenomenology than Dirac neutrinos, due mainly to the violation of lepton number. Also, Dirac neutrinos require the addition of $\nu_R$ and, consequently, we have to forbid their mass term $m_R$. This can be done by imposing an extra symmetry or forcing an accidental symmetry to be exact, like lepton number. Even if $m_R$ is absent, we will still have to deal with a severe hierarchy of several orders of magnitude to explain, via the Higgs mechanism, the masses of the charged leptons and the neutrinos. These arguments led people to argue that it does not seem very natural for neutrinos to be Dirac particles. However, there has been a recent increase of interest in Dirac neutrinos, as we will explain in the next chapter.

For now, we consider the Standard Model, where lepton number is an accidental global symmetry. What is meant by accidental symmetry, is that taking the Standard Model as a low-energy effective theory, one can find at least one higher dimensional operator which is invariant under the Standard Model gauge group, but violates the accidental symmetry. For lepton number, one finds that the lowest dimensional operator which violates lepton number appears at dimension $5$ and it is unique. This operator is the Weinberg operator \cite{Weinberg:1979sa},
\begin{equation} \label{eq:nuphys:weinberg}
    \mathcal{O}_W = \frac{c_{\alpha\beta}}{\Lambda} \, \bar{L_\alpha^c} \, L_\beta \, \tilde{H^*} \, \tilde{H^\dagger} \, + \, \hc \, ,
\end{equation}
with the usual notation $\tilde H = i \sigma^2 \, H^*$. Here, $c_{\alpha\beta}$ is some coefficient which is model-dependent, while $\Lambda$ is the scale of new physics.

After electroweak symmetry breaking, the Weinberg operator leads to the Majorana mass matrix \eq{eq:nuphys:majmass}. The neutrino mass scale is then approximately,
\begin{equation} \label{eq:nuphys:mnuWein}
    m_L \sim c_{\alpha \beta} \, \frac{v^2}{\Lambda} \, ,
\end{equation}
with $v$ the vacuum expectation value (VEV) of the Standard Model Higgs. The smallness of the neutrino masses corresponds then to choosing for $\Lambda$ a very large scale, say $\Lambda \sim O(10^{14})$ GeV, in which case neutrino masses are of the (sub-)eV order for $c_{\alpha \beta} \sim O(1)$. However, $c_{\alpha\beta}$ could be naturally much smaller than one, resulting in correspondingly lower values for the energy scale $\Lambda$ at which lepton number is violated. There are two simple ways of realising such a suppression: (i) neutrino masses might be radiatively generated, in which case $c_{\alpha\beta} \propto 1/(16 \pi^2)^n$, where $n$ is the number of loops; (ii) higher d-dimensional operators might be responsible for neutrino mass generation, note that such operators are always of the form ${\cal O}_W \times (H^{\dagger}H)^{\frac{d-5}{2}}$; or (iii) $c_{\alpha\beta}$ could be small either due to some small coupling in the corresponding model or due to some nearly conserved symmetry.\footnote{R-parity violating supersymmetry is an example of the former \cite{Dreiner:1997uz,Hirsch:2000ef}, models such as the inverse \cite{Mohapatra:1986bd} or the linear \cite{Akhmedov:1995ip,Akhmedov:1995vm} seesaw are examples of the latter.}

In the literature, there is a huge variety of models which address the UV completion of Weinberg operator or higher order operators based on $\mathcal{O}_W$. We will discuss the most relevant models in \ch{ch:numass}. Also, several systematic studies of all the realisations of the Weinberg operator considering a fixed number of loops and dimensions have been done \cite{Ma:1998dn, Babu:2001ex, Bonnet:2009ej, Bonnet:2012kz, Farzan:2012ev, Sierra:2014rxa, Cepedello:2017eqf, Cai:2017jrq, Anamiati:2018cuq, Cepedello:2018rfh, Klein:2019iws}. We will briefly review some of them in the next chapter, while in \ch{ch:Dim7_1loop} and \ch{ch:3loop} we will explain in detail the classification of all the realisations of the operator ${\cal O}_W \times (H^{\dagger}H)$ at one-loop level, and the Weinberg operator at three-loop level, respectively. But before, we shall discuss one of the most interesting process in neutrino physics, namely neutrinoless double-$\beta$ decay.

%%%%%%%%%%%%%%%%%%%%%%%%%%%%%%%%%%%%%%%%%%%%%%%%%%%%%%%%%%%%%%%
%%%%%%%%%%%%%%%%%%%%%%%%%%%%%%%%%%%%%%%%%%%%%%%%%%%%%%%%%%%%%%%
\section{Neutrinoless double-$\beta$ decay}
\label{sec:nuphys:0vbb}

Double-$\beta$ decay is a nuclear process in which a nucleus transitions to another one with two more protons and the same mass number, emitting two electrons and, possibly, other light particles. In the Standard Model these light particles are two antineutrinos and the process is called two-neutrino double-$\beta$ ($2\nu\beta\beta$) decay. $2\nu\beta\beta$ conserves lepton number and it can be thought as two simultaneous beta decays. This decay is among the rarest processes ever observed with half-lives around $10^{21}$ years and longer \cite{Barabash:2019nnr}.

Of special interest is the case where only the two electrons are emitted, i.e. the process called neutrinoless double-$\beta$ ($0\nu\beta\beta$) decay,\footnote{For reviews, see for example \cite{GomezCadenas:2011it, Deppisch:2012nb}.}
\begin{equation} %\label{eq:}
    _Z^A {\rm X} \, \to \, _{Z+2}^A {\rm X} \, + \, 2 \, e^- \, .
\end{equation}
The process violates clearly lepton number by two units, since there are only two leptons emitted. Consequently, it univocally requires neutrinos to be Majorana, as will be discussed later. For this reason, there has been a strong experimental effort to observe $0\nu\beta\beta$, though unfortunately, it is predicted to be extremely rare. Current experimental searches set bounds on its half-life around $10^{26}$ years.

\begin{table}[h]
    \centering
    \setlength{\tabcolsep}{15pt}
    \renewcommand{\arraystretch}{1.5}
    \begin{tabular}{ | c | c | c | }
        \hline
        \textbf{Experiment}  &  \textbf{Isotope}  &  $\mathbf{T_{1/2}}$ \textbf{[y]} (limit)
        \\
        \hline
        CUORE \cite{Adams:2019jhp}  &  $^{130}$Te  &  $3.2 \times 10^{25}$
        \\
        EXO-200 \cite{Anton:2019wmi}  &  $^{136}$Xe  &  $3.5 \times 10^{25}$
        \\
        nEXO \cite{Albert:2017hjq}  &  $^{136}$Xe  &  [$\sim 10^{27}-10^{28}$]
        \\
        GERDA \cite{Agostini:2020xta}  &  $^{76}$Ge  &  $1.8 \times 10^{26}$
        \\
        KamLAND-Zen \cite{KamLAND-Zen:2016pfg}  &  $^{136}$Xe  &  $1.1 \times 10^{26}$
        \\
        LEGEND \cite{Abgrall:2017syy}  &  $^{176}$Ge  &  [$\sim 10^{27}-10^{28}$] 
        \\
        NEXT-100 \cite{Martin-Albo:2015rhw}  &  $^{136}$Xe  &  [$6.0 \times 10^{25}$]
        \\
        SNO+ \cite{Andringa:2015tza}  &  $^{130}$Te  &  [$7.0 \times 10^{26}$]
        \\
        SuperNEMO \cite{Arnold:2010tu}  &  $^{150}$Nd / $^{82}$Se / $^{48}$Ca  &  [$\sim 10^{26}$]
        \\
        \hline
    \end{tabular}
    \caption{Overview of some of the main experiments searching for $0\nu\beta\beta$ decay. We give the current and future limits on the half-life of the process. The latter shown in brackets. All the values are given at $90\%$ confidence level. A more complete and detailed list can be found in \cite{Barabash:2019suz}.}
    \label{tab:nuphys:0vbb}
\end{table}

An updated list of current and future major experiments searching for $0\nu\beta\beta$ can be found in \tab{tab:nuphys:0vbb}. The majority of the experiments use $^{76}$Ge, $^{130}$Te, or $^{136}$Xe. One of the most important quantities that characterises each isotope is the available energy in the process, i.e.
\begin{equation} %\label{eq:}
    Q_{\beta\beta} = E_I - E_F - 2 \, m_e \, ,
\end{equation}
with $E_I$ and $E_F$ the energies of the initial and final nuclei. The decay rate $\Gamma_{0\nu\beta\beta}$ is proportional to the fifth power of $Q_{\beta\beta}$, so isotopes with a higher value of $Q_{\beta\beta}$ are better for $0\nu\beta\beta$ decay searches. For example, germanium detectors have the best energy resolution, but a low $Q_{\beta\beta}$ value and high fabrication costs. Meanwhile, $^{130}$Te is quite abundant and has a relatively high $Q_{\beta\beta}$. However, the most used isotope is $^{136}$Xe, because its $Q_{\beta\beta}$ is similar to tellurium and it is a scintillating noble gas.

Currently, the most stringent bounds on $0\nu\beta\beta$ decay come from $^{136}$Xe at KamLAND-Zen experiment \cite{KamLAND-Zen:2016pfg} and $^{76}$Ge at GERDA \cite{Agostini:2020xta}:
\begin{eqnarray} %\label{eq:}
    T_{1/2}^{\rm Ge} \, &>& \, 1.8 \, \times \, 10^{26} \, {\rm y} \,\,\, (90\% \, {\rm CL}) \, ,
    \nn \\ \nn
    T_{1/2}^{\rm Xe} \, &>& \, 1.1 \, \times \, 10^{26} \, {\rm y} \,\,\, (90\% \, {\rm CL}) \, .
\end{eqnarray}

%%%%%%%%%%%%%%%%%%%%%%%%%%%%%%%%%%%%%%%%%%%%%%%%%%%%%%%%%%%%%%%
\subsection{The black-box theorem}
\label{subsec:nuphys:blackbox}

\begin{figure}
    \centering
    \includegraphics[width=0.8\textwidth]{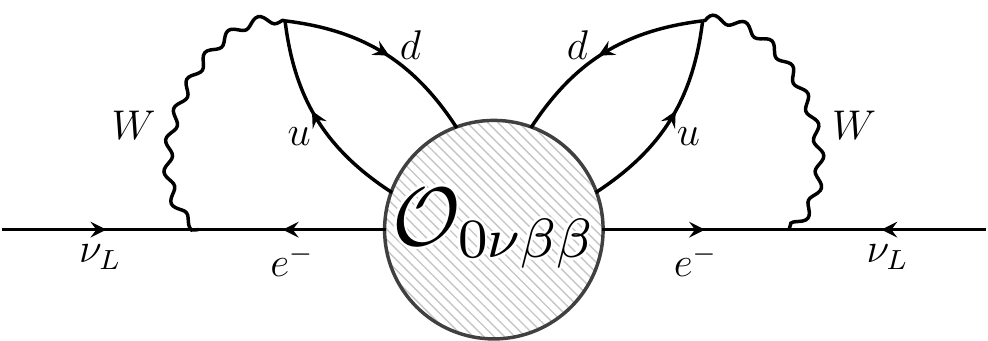}
    \caption{Diagrammatic representation of the black box theorem \cite{Schechter:1981bd}. The existence of the dimension 9 operator responsible for $0\nu\beta\beta$ decay, implies the presence of the Majorana mass operator $\nu_L \nu_L$.}
    \label{fig:nuphys:blackbox}
\end{figure}

We discuss now one of the most important theoretical results in neutrinoless double-$\beta$ decay, the black-box or Schechter-Valle theorem \cite{Schechter:1981bd}. The theorem connects tightly Majorana neutrinos with $0\nu\beta\beta$, i.e. it states the equivalence
\begin{equation*} %\label{eq:}
    \text{Majorana neutrinos} \quad \Longleftrightarrow \quad 0\nu\beta\beta \text{ decay}
\end{equation*}
To the right, the statement that a Majorana mass term $\nu_L \nu_L$ automatically implies the existence of $0\nu\beta\beta$ is straightforward. The corresponding diagram is given in \fig{fig:nuphys:0vbb}. Similarly, the reverse statement indicates that given the effective $0\nu\beta\beta$ decay operator,
\begin{equation} %\label{eq:}
    \mathcal{O}_{0\nu\beta\beta} = \bar e^c \bar e^c d^c d^c \bar u^c \bar u^c \, ,
\end{equation}
the operator $\nu_L \nu_L$ can be generated by attaching only Standard Model vertices. This implication is known as the black-box theorem \cite{Schechter:1981bd}, depicted in \fig{fig:nuphys:blackbox}.

%%%%%%%%%%%%%%%%%%%%%%%%%%%%%%%%%%%%%%%%%%%%%%%%%%%%%%%%%%%%%%%
\subsection{The mass mechanism}
\label{subsec:nuphys:mec}

\begin{figure}
    \centering
    \includegraphics[width=0.5\textwidth]{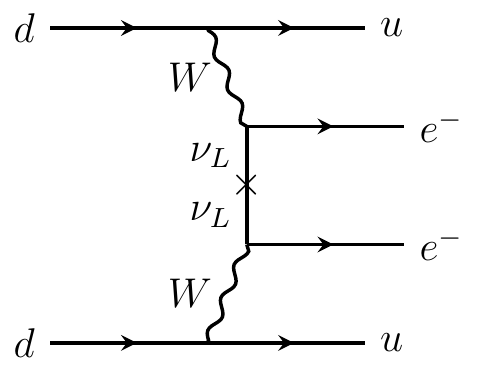}
    \caption{The standard mass mechanism of $0\nu\beta\beta$ decay. Note that when including the mass insertion via the operator $\nu_L \nu_L$ ($\mathcal{O}_W$), represented by a cross, the operator $\mathcal{O}_{0\nu\beta\beta}$ is automatically generated within the Standard Model.}
    \label{fig:nuphys:0vbb}
\end{figure}

In the literature, $0\nu\beta\beta$ decay which is mediated by the Majorana mass of the light neutrinos is often called the mass mechanism, depicted in \fig{fig:nuphys:0vbb}. Neutrino masses are generated by the Weinberg operator \eq{eq:nuphys:weinberg} (mainly, $\nu_L \nu_L$ at low energy), and introduces a chirality flip or mass insertion denoted by a cross in \fig{fig:nuphys:0vbb}.

\begin{figure}[h]
    \centering
    \includegraphics[width=0.95\textwidth]{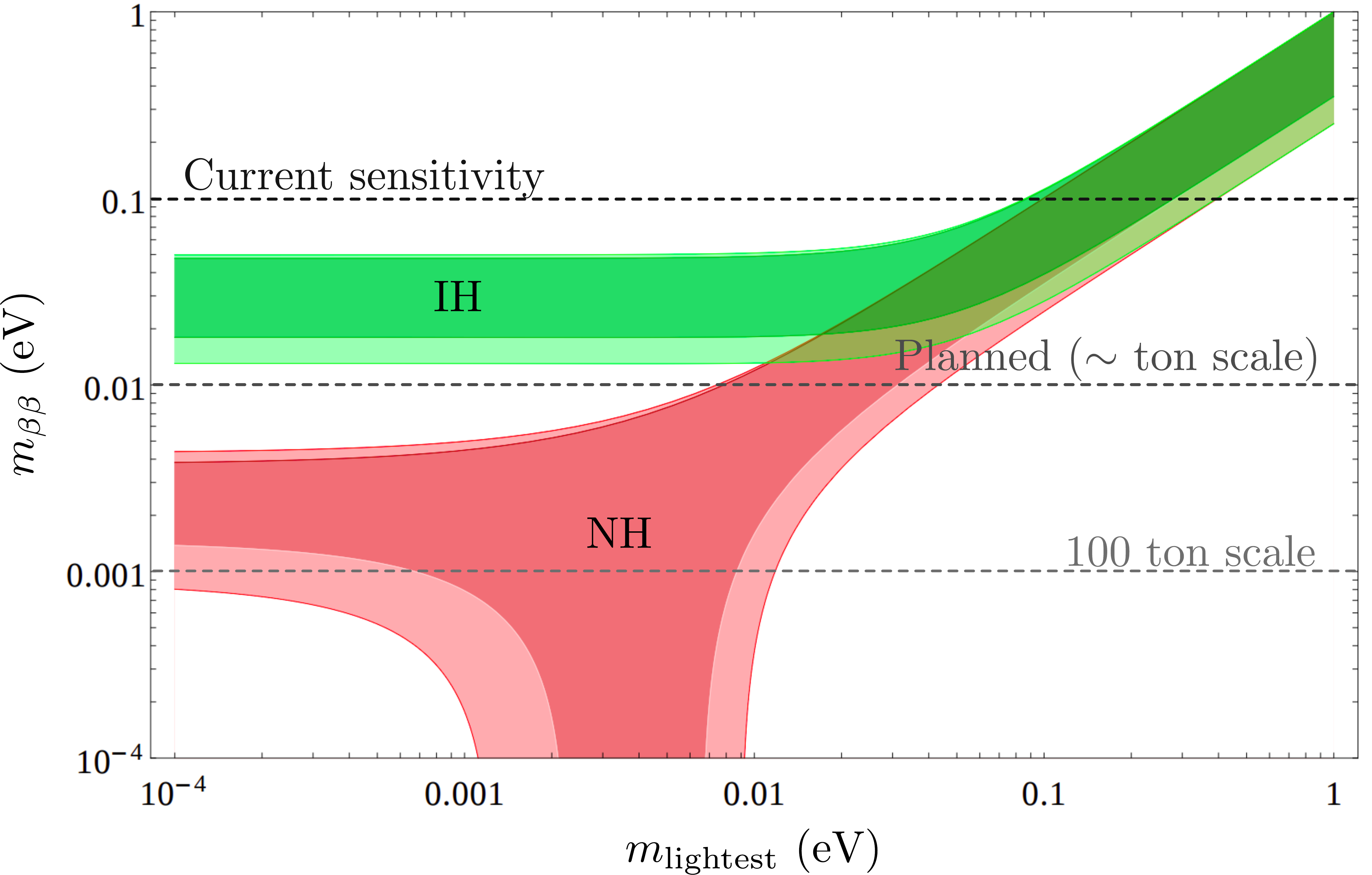}
    \caption{Effective Majorana mass \eq{eq:nuphys:mee} as a function of the lightest neutrino mass for both hierarchies. The lighter shadow corresponds to $3\sigma$ uncertainties in the oscillation parameters. The plot scans over the Majorana phases. We give the most optimistic current and planned searches sensitivities, as well as a rough estimate of the sensitivity of a \textit{hypothetical} future experiment with a mass of 100 tons \cite{Benato:2015via}. Plot adapted from \cite{DellOro:2014ysa}.}
    \label{fig:nuphys:mee}
\end{figure}

How the $0\nu\beta\beta$ decay rate depends on the masses of the neutrinos can be seen by computing the amplitude, or at least a piece of it. We can write the internal neutrino line and the $W$ vertices, rotating to the mass basis by means of the leptonic mixing matrix \eq{eq:nuphys:Ulep}, as
\begin{equation} %\label{eq:}
    \sum_i \bar e \, U_{ei}^2 \, \gamma_\mu \, P_L \, \frac{\slashed{q} + m_{i}}{q^2 + m_{i}^2} \, P_L \, \gamma_\nu \, e^c \approx \sum_i \bar e \, U_{ei}^2 \, \gamma_\mu \, \frac{m_{i}}{q^2} \, \gamma_\nu \, e^c \, ,
\end{equation}
where we have neglected the light neutrino mass $m_i^2$ in the denominator. Thus, the standard mass mechanism is sensitive to what is known as the effective Majorana neutrino mass,
\begin{equation} %\label{eq:}
    m_{ee} \equiv \sum_i U_{ei}^2 m_i \, .
\end{equation}
Indeed, $m_{ee}$ is the first entry of the neutrino mass matrix. Note that as the decay rate is the absolute value squared of the amplitude, results are normally given and denoted in terms of $m_{\beta\beta} \equiv |m_{ee}|$. Finally, considering the parametrisation of the leptonic mixing matrix given in \eq{eq:nuphys:Ulep}, the effective Majorana mass can be written as a function of the oscillation parameters as,
\begin{equation} \label{eq:nuphys:mee}
    m_{ee} = c_{12}^2 c_{13}^2 m_1 + c_{13}^2 s_{12}^2 m_2 e^{-i \alpha_{21}} + s_{13}^2 m_3 e^{-i \alpha _{31}- 2 i \delta } \, .
\end{equation}
The dependence of $m_{ee}$ ($m_{\beta\beta}$) on the lightest neutrino mass is depicted in \fig{fig:nuphys:mee}. The figure is scanned over all possible values of the phases. A cancellation of $m_{\beta\beta}$ appears around $\alpha_{21} = (\alpha_{31} + 2 \delta) = (2n+1) \, \pi$ ($n \in \mathbb{Z}$) only for normal hierarchy for a small range of the lightest neutrino mass. In the unfortunate case where $m_{\beta\beta}$ is very small, neutrinos can be still Majorana but the observation of $0\nu\beta\beta$ would be impossible.

Currently, experiments are starting to test the inverse ordering region. The non-observation of a signal would imply that the hierarchy is not inverted or that neutrinos are Dirac, if the inverted hierarchy is confirmed by other measurements. For inverse hierarchy, the minimal value of the effective neutrino mass corresponds to $\sim 0.02$ eV and a half-life of order $10^{27}$ y. Therefore, inverted hierarchy could be probed by future experiments, see \tab{tab:nuphys:0vbb}.

\pagebreak
\fancyhf{}

%% file: Chapters/Neutrino_masses/Chapter_numass.tex
\fancyhf{}
\fancyhead[LE,RO]{\thepage}
\fancyhead[RE]{\slshape\nouppercase{\leftmark}}
\fancyhead[LO]{\slshape\nouppercase{\rightmark}}

\chapter{Neutrino mass generation}
\label{ch:numass}
\graphicspath{ {Chapters/Neutrino_masses/} }

In this chapter we briefly review some of the most popular models for Majorana and Dirac neutrinos. We start with the usual seesaw models for Majorana neutrinos, then introduce radiative models and other relevant frameworks, such as left-right symmetric models, Majoron models, etc. We take a similar approach for Dirac neutrinos in \sect{sec:numass:dirac}. The last section is devoted to systematic classifications of radiative models. We introduce several important concepts, as well as previous classifications, useful for the following chapters.

%%%%%%%%%%%%%%%%%%%%%%%%%%%%%%%%%%%%%%%%%%%%%%%%%%%%%%%%%%%%%%%
%%%%%%%%%%%%%%%%%%%%%%%%%%%%%%%%%%%%%%%%%%%%%%%%%%%%%%%%%%%%%%%
\section{Majorana neutrinos}
\label{sec:numass:majorana}

As pointed out in the previous chapter, one possible way to explain neutrino masses is by generating the Weinberg operator $\mathcal{O}_W$ \eq{eq:nuphys:weinberg}, or some generalisation of it, as the dimension $d$ Weinberg-like operator $\mathcal{O}^d \equiv \mathcal{O}_W \times (H^\dagger H)^{\frac{d-5}{2}}$. Now the task is to derive one of these operators from a UV complete theory, i.e. to \textit{open up the operator}. The operator can then be realised at tree-level or via loops by adding new particles to the Standard Model. It is important to note that normally an infinite number of such operators will be generated and one should identify the dominant contribution to neutrino masses.

\begin{figure}
    \centering
    \includegraphics[width=0.8\textwidth]{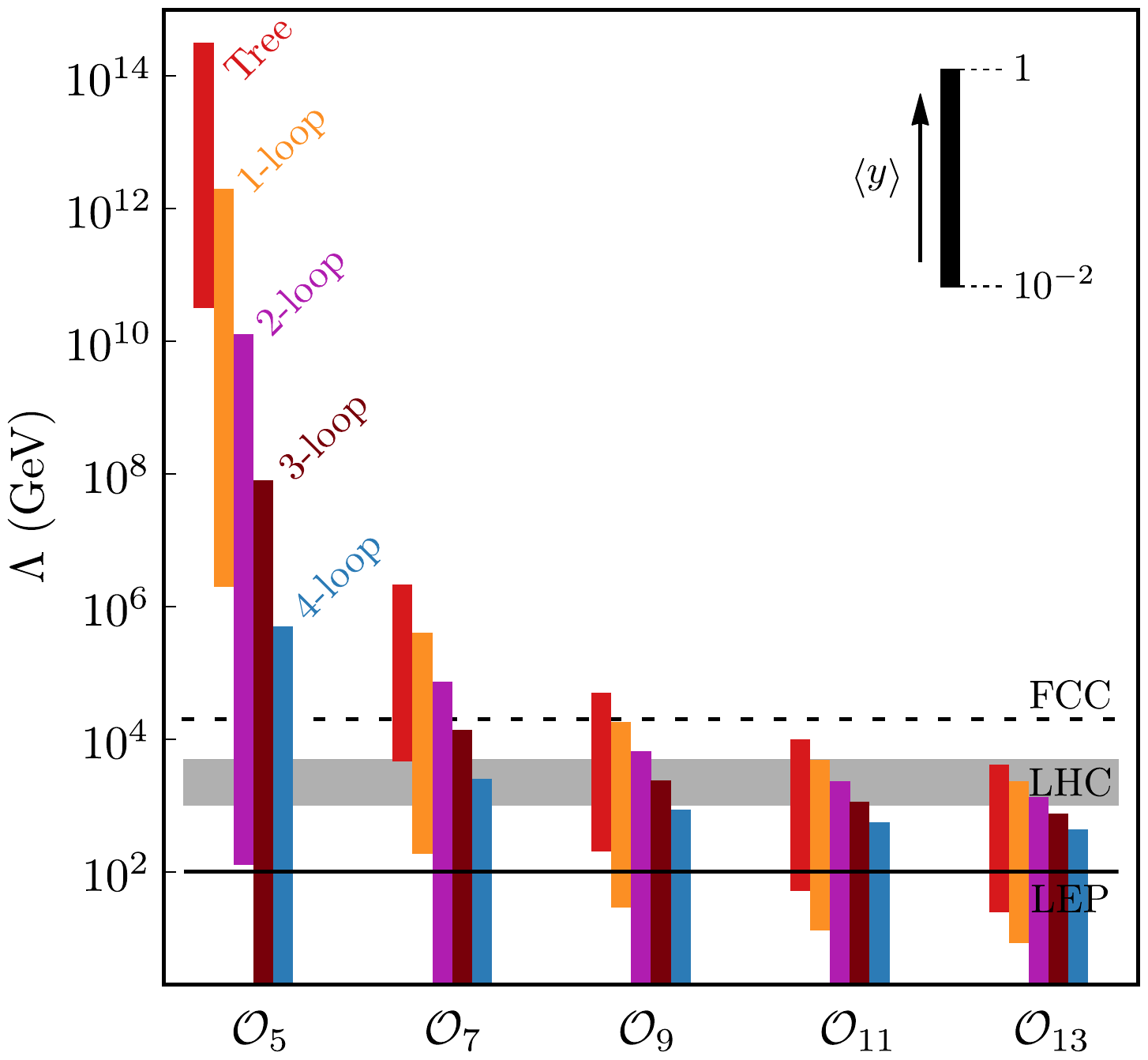}
    \caption{Estimate of the mass scale $\Lambda$ for which a neutrino mass model with dimension $d$ and $n$ loops can explain correctly the observed neutrino masses. The height of the colour bars is due to variation of the average values of the couplings $\vev{y}$. A rough estimate of the bounds set by colliders are given. Figure from \cite{Anamiati:2018cuq}.}
    \label{fig:numass:estimate}
\end{figure}

In full generality, the neutrino mass scale generated by the operator $\mathcal{O}^d$ induced at $n$-loop level is roughly,
\begin{equation} \label{eq:numass:estimate}
    m_\nu \, \sim \, \epsilon \, \frac{v^2}{\Lambda} \, \left( \frac{1}{16 \pi^2} \right)^n \, \left( \frac{v}{\Lambda} \right)^{d-5} \, ,
\end{equation}
with $\epsilon$ some small number that may arise, for instance, due to an additional suppression of lepton number violation, or small dimensionless couplings. From \eq{eq:numass:estimate} we can roughly estimate the typical mass scale $\Lambda$ for a given dimension $d$ and number of loops $n$, considering different values for the couplings (encoded in $\epsilon$). This is done in \fig{fig:numass:estimate} up to four-loops and $d=13$, varying the \textit{average} of the couplings $\vev{y}$ by two orders of magnitude.

Different estimates of the bounds set by colliders are shown in \fig{fig:numass:estimate}. LEP gives a lower limit of roughly $100$ GeV on the mass of electrically charged particle that couple to the Standard Model fermions \cite{Tanabashi:2018oca}. For the LHC, the limit goes from the most conservative estimate considering pair production of charged particles, to the upper edge of the grey band that shows roughly the reach of the LHC for particles produced in s-channel diagrams with or without colour. Finally, the reach of a hypothetical $\sqrt{s} = 100$ TeV collider (FCC), is denoted as a dashed line. Thus, neutrino mass models generated at $d = 9$ or higher, as well as higher loop realisations, should be testable in the near future.

%%%%%%%%%%%%%%%%%%%%%%%%%%%%%%%%%%%%%%%%%%%%%%%%%%%%%%%%%%%%%%%
\subsection{Tree-level models}
\label{subsec:numass:maj_tree}

\begin{figure}
    \centering
    \begin{subfigure}[t]{0.32\textwidth}
        \centering
        \includegraphics[width=1\textwidth]{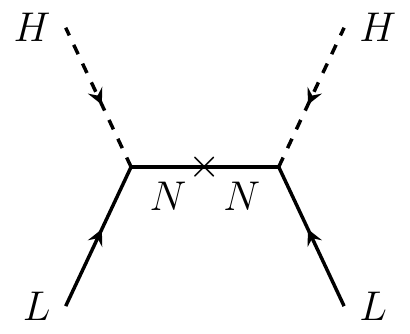}
        \caption{Type-I seesaw}
        \label{fig:numass:seesawI}
    \end{subfigure}
    \hfill
    \begin{subfigure}[t]{0.32\textwidth}
        \centering
        \includegraphics[width=1\textwidth]{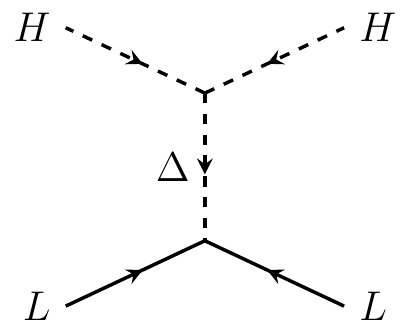}
        \caption{Type-II seesaw}
        \label{fig:numass:seesawII}
    \end{subfigure}
    \hfill
    \begin{subfigure}[t]{0.32\textwidth}
        \centering
        \includegraphics[width=\textwidth]{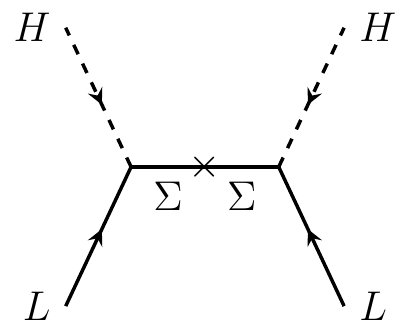}
        \caption{Type-III seesaw}
        \label{fig:numass:seesawIII}
    \end{subfigure}
    \caption{Seesaw neutrino mass mechanisms. This corresponds to the minimal openings of the Weinberg operator at tree-level.}
    \label{fig:numass:seesaws}
\end{figure}

The simplest way to open the Weinberg operator is at tree-level. Starting from the expression \eq{eq:nuphys:weinberg}, by doing Fierz transformations and taking into account that neutrinos are Majorana,
\begin{equation} %\label{eq:}
    \left( \bar{L_\alpha^c} \tilde{H^*} \right) \, \left( \tilde{H^\dagger} L_\beta \right) \, = \, \frac 12 \left( \bar{L_\alpha^c} \boldsymbol{\sigma} L_\beta \right) \, \left( \tilde{H^\dagger} \boldsymbol{\sigma} \tilde{H^*} \right) \, = \, - \left( \bar{L_\alpha^c} \boldsymbol{\sigma} \tilde{H^*} \right) \, \left( \tilde{H^\dagger} \boldsymbol{\sigma} L_\beta \right) \, ,
\end{equation}
where $\boldsymbol{\sigma}$ is the $3$-vector of Pauli matrices. The first corresponds to the coupling of $LH$ to a hyperchargeless fermion singlet, the second equality to a scalar triplet with hypercharge $1$ and the last to a hyperchargeless fermion triplet. In the simplest scenario, these cases correspond to the well-known type-I, -II and -III seesaws \cite{Minkowski:1977sc, Yanagida:1979as, GellMann:1980vs, Mohapatra:1979ia, Schechter:1980gr, Schechter:1981cv, Foot:1988aq}. The diagram for each seesaw is depicted in \fig{fig:numass:seesaws}. Of course, other realisations of the Weinberg operator, or higher dimensional Weinberg-like operators, at tree-level are possible, as we will discuss later.

%%%%%%%%%%%%%%%%
\subsubsection*{Type-I seesaw}

By adding to the Standard Model a gauge singlet (right-handed neutrino) $N$ with a Majorana mass $M_N$ neutrino masses are generated by the diagram in \fig{fig:numass:seesawI}. The number of copies of $N$ is in principle free, though at least two are needed in order to fit neutrino oscillation data.

The relevant Lagrangian terms for the neutrino mass are given by,
\begin{equation} %\label{eq:}
    \mathcal{L}_\nu \, = \, \frac{1}{2} \, ( {\bar \nu}_L^c \; \bar N ) \, \mathcal{M}_\nu \, \begin{pmatrix} \nu_L \\ N^c \end{pmatrix} \, + \, \hc \, ,
\end{equation}
with the neutrino mass matrix,
\begin{equation} \label{eq:numass:mnutypeI}
    \mathcal{M}_\nu \, = \, \begin{pmatrix} 0 & m_D \\ m_D^T & M_N \end{pmatrix} \, .
\end{equation}
Here $m_D = Y_\nu v / \sqrt{2}$ corresponds to the Dirac mass term for neutrinos. For $M_N >> m_D$ one can approximate the light and heavy neutrino mass eigenvalues as,
\begin{eqnarray} \label{eq:numass:typeI}
    m_\nu &\approx& - m_D \cdot M_N^{-1} \cdot m_D^T \, ,
    \\
    m_R &\approx& M_N \, .
\end{eqnarray}
Thus, explaining the smallness of the neutrino masses by the hierarchy between the electroweak scale and the new mass scale $M_N$. Indeed, for couplings order one and considering $m_\nu \sim 0.1$ eV, $M_N$ should be around $10^{14}$ GeV, unfortunately far beyond the experimental reach.

%%%%%%%%%%%%%%%%
\subsubsection*{Type-II seesaw}

In the type-II seesaw the opening of the Weinberg operator is done with a scalar triplet with hypercharge $1$, called $\Delta = ( \Delta^{++}, \,\, \Delta^+, \,\, \Delta^0 )$. This triplet is commonly given in the doublet representation of ${\rm SU(2)_L}$ as,
\begin{equation} %\label{eq:}
    \Delta \, = \, 
    \begin{pmatrix} \Delta^+ / \sqrt{2}  &  \Delta^{++}
                    \\
                    \Delta^0  &  -\Delta^+ / \sqrt{2}
    \end{pmatrix} \, .
\end{equation}
The relevant Lagrangian terms are then,
\begin{equation} \label{eq:numass:lagtypeII}
    \mathcal{L}_\nu \, = \, ( Y_{\Delta} \, L \Delta L \, + \, \mu \, H \Delta H \, + \, \hc ) \, + \, M_\Delta^2 \, \Delta^\dagger \Delta \, ,
\end{equation}
where indices have been omitted for the sake of simplicity. As required for Majorana neutrinos, lepton number is broken in two units by the simultaneous presence of $Y_{\Delta}$ and $\mu$.

The presence of the coupling $\mu$ in \eq{eq:numass:lagtypeII} implies that $\Delta$ gets an induced VEV, $\vev{\Delta} = v_\Delta / \sqrt{2}$ for $\vev{H} \neq 0$. This is straightforward from the tadpole equations, which links both VEVs as,
\begin{equation} %\label{eq:}
    v_\Delta \, = \, \frac{\mu \, v^2}{\sqrt{2} \, M_\Delta^2} \, .
\end{equation}
The smallness of the neutrino masses is then controlled by the smallness of the induced VEV,
\begin{equation} %\label{eq:}
    m_\nu \, = \, Y_\Delta \frac{v_\Delta}{\sqrt{2}} \, ,
\end{equation}
and it is precisely where the seesaw relation is manifest, as the larger the triplet mass is, its VEV gets smaller. The heaviness of the triplet is well motivated by the fact that $v_\Delta$ should be roughly below $1$ GeV due to constraints from the Standard Model $\rho$ parameter, and that $\Delta$ mediates lepton number and lepton flavour violating processes at tree-level. These can be translated into strong bounds for its mass and couplings.

%%%%%%%%%%%%%%%%
\subsubsection*{Type-III seesaw}

This seesaw is analogous to the type-I seesaw, but with a triplet hyperchargeless fermion $\Sigma = ( \Sigma^{+}, \,\, \Sigma^0, \,\, \Sigma^- )$ instead of the right-handed neutrino $N$. The Lagrangian is then,
\begin{equation} %\label{eq:}
    \mathcal{L}_\nu \, = \, \frac{1}{2} \, ( \bar\nu_L^c \; \bar \Sigma^0 ) \, \mathcal{M}_\nu \, \begin{pmatrix} \nu_L \\ \Sigma^{0c} \end{pmatrix} \, + \, \hc \, ,
\end{equation}
with $\mathcal{M}_\nu$ given in \eq{eq:numass:mnutypeI} by substituting $M_N$ by $M_\Sigma$. The structure of this model is identical to the type-I seesaw, although the gauge indices (not explicitly shown) in these two models contract differently, due to the introduction of triplets instead of singlets. The light neutrino masses follow an analogous expression for $M_\Sigma >> m_D$,
\begin{equation} %\label{eq:}
    m_\nu \approx - m_D \cdot M_\Sigma^{-1} \cdot m_D^T \, .
\end{equation}

The main difference between the type-I and -III seesaws is the phenomenology. As $\Sigma$ has charged components with the same Majorana mass that appears in the expression for the light neutrino masses, there are stringent lower bounds on this mass scale from direct searches \cite{Aad:2020fzq}.

%%%%%%%%%%%%%%%%
\subsubsection*{Inverse and linear seesaw}

In general, any number of singlets can be added to any gauge theory. The inverse and linear seesaw models rely on a non-minimal lepton content, adding extra lepton singlets to the type-I seesaw model. In these models, lepton number is broken by a small parameter protected from radiative corrections. This allows to fit neutrino masses with a much lighter right-handed neutrino than in type-I seesaw.

In the basis $( \nu_L, \, N^c, \, S)$, the mass matrix reads,\footnote{Note that in full generality the second element of the diagonal is not necessarily zero, but it is normally set to zero in the literature as it is irrelevant for the generation of neutrino masses at tree-level.}
\begin{equation} %\label{eq:numass}
    \mathcal{M}_\nu \, = \, \begin{pmatrix} 0 & m_D & m_L \\ m_D^T & 0 & M \\ m_L^T & M^T & \mu \end{pmatrix} \, .
\end{equation}
with $M$ and $\mu$ mass matrices corresponding to the singlets, and $m_L$ a small term mixing $\nu$ and $S$. Here, $\mu$ and/or $m_L$ are responsible of the breaking of lepton number. If $m_L = 0$ and $\mu \neq 0$, then the models is called \textit{inverse seesaw}, while in the limit where $\mu = 0$ and $m_L \neq 0$, it is called \textit{linear seesaw}.\\

\begin{figure}
    \centering
    \includegraphics[width=0.5\textwidth]{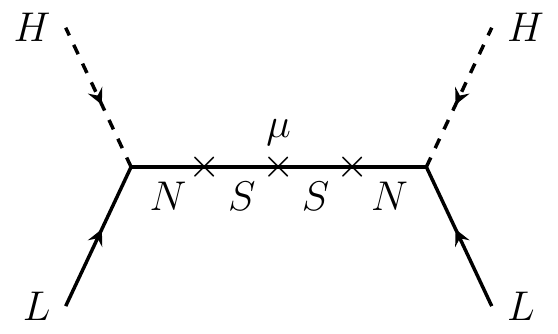}
    \caption{Inverse seesaw neutrino mass mechanism. The smallness of neutrino masses is associated to the smallness of the mass parameter $\mu$.}
    \label{fig:numass:inverse_seesaw}
\end{figure}

The inverse seesaw was initially motivated by string theories \cite{Witten:1985xc,Mohapatra:1986bd} as low-scale tree-level scheme for generating light neutrino masses. Provided the hierarchy $\mu << m_D << M$, the mass matrix can be easily diagonalised. The light neutrino mass matrix can be written as,
\begin{equation} %\label{eq:}
    M_\nu \, = \, m_D \, M^{-1} \, \mu \, (M^T)^{-1} \, m_D^T \, .
\end{equation}
The corresponding neutrino mass diagram is given in \fig{fig:numass:inverse_seesaw}. Contrary to the \textit{standard} seesaws where the smallness of the neutrino masses is related to a large mass, here it is linked to the smallness of $\mu$. In this kind of scenarios, a small $\mu$ is considered to be natural, as in the limit $\mu \to 0$ the symmetry of the theory is enhanced \cite{tHooft:1979rat,GonzalezGarcia:1988rw}.\\

The linear seesaw was originally realised within left-right symmetry models \cite{Akhmedov:1995ip,Akhmedov:1995vm}, although later employed also in supersymmetric SO(10) scenarios \cite{Malinsky:2005bi}. For this seesaw, the light neutrino mass matrix is given by,
\begin{equation} %\label{eq:}
    M_\nu \, = \, m_D^T \, M^{-1} \, m_L \, + \, m_L^T \, (M^T)^{-1} \, m_D \, .
\end{equation}
Here the smallness of neutrino masses can be connected to the large scale $M$, leaving the lepton number violating scale $m_L$ free. This is the most interesting feature of this model, as it allows a low scale breaking of lepton number which may be experimentally accessible, while keeping the neutrino masses small.

%%%%%%%%%%%%%%%%
\subsubsection*{Higher dimensional scenarios}

As mentioned in the introduction of this section, other higher dimensional Weinberg-like operators with dimension $d$, $\mathcal{O}^d \equiv \mathcal{O}_W \times (H^\dagger H)^{\frac{d-5}{2}}$, are also a possible way to explain the smallness of neutrino masses. Here, we show just some examples to support the previous statement.

\begin{figure}
    \centering
    \includegraphics[width=0.5\textwidth]{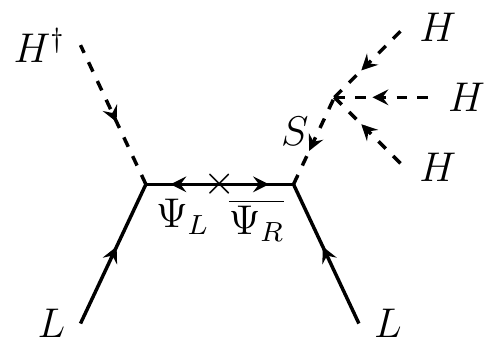}
    \caption{The only genuine tree-level diagram for $d=7$ \cite{Babu:2009aq}.}
    \label{fig:numass:bnt}
\end{figure}

For the simplest possibility, $d=7$, systematically studied in \cite{Bonnet:2009ej}, only one model exists which can account on its own for the neutrino masses. The model, first discussed in \cite{Babu:2009aq} and usually called BNT (Babu-Nasri-Tavartkiladze) for the authors, incorporates to the Standard Model particle content a scalar quadruplet $S$ and a vector-like pair of fermion triplets $\Psi_{L,R}$ with hypercharges $3/2$ and $1$, respectively. The corresponding neutrino mass diagram is shown in \fig{fig:numass:bnt}. The model has a very particular phenomenology due to the highly charged exotic particle it contains, along with the usual lepton number and flavour phenomenology characteristic of Majorana neutrino mass models \cite{Liao:2010rx, Ghosh:2017jbw, Cepedello:2017lyo, Ghosh:2018drw}.

For higher dimension operators ($d=9$ and beyond), there are few examples of models in the literature \cite{Picek:2009is, Liao:2010cc, Kumericki:2012bh, McDonald:2013kca, McDonald:2013hsa, Nomura:2017abu}. For $d=9$, $11$ and $13$ a systematic classification of all possible diagrams and models has been recently done \cite{Anamiati:2018cuq}. In this classification they found a total of $2$, $2$ and $6$ models for $d=9$, $11$ and $13$, respectively. It is worth to mention that among them, there are two models (for $d=9$ and $13$) which can explain neutrino data using only new exotic vector-like fermions. The neutrino mass diagrams for both models are depicted in \fig{fig:numass:d9,13}. As well as before, all these classes of models required large ${\rm SU(2)_L}$ representations and hypercharges and, consequently, they have a very rich phenomenology \cite{Anamiati:2018cuq}.

\begin{figure}[h]
    \centering
    \includegraphics[width=0.6\textwidth]{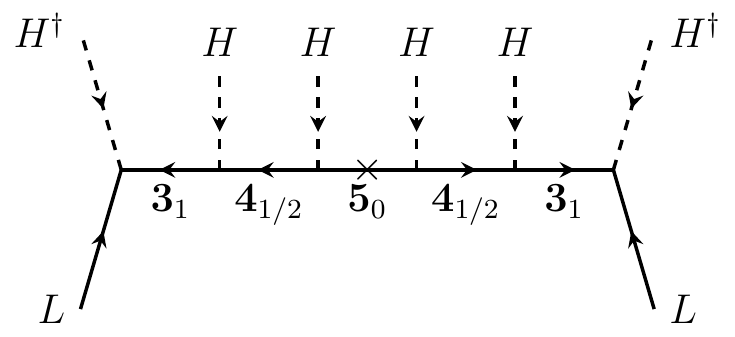}
    \\
    \includegraphics[width=0.9\textwidth]{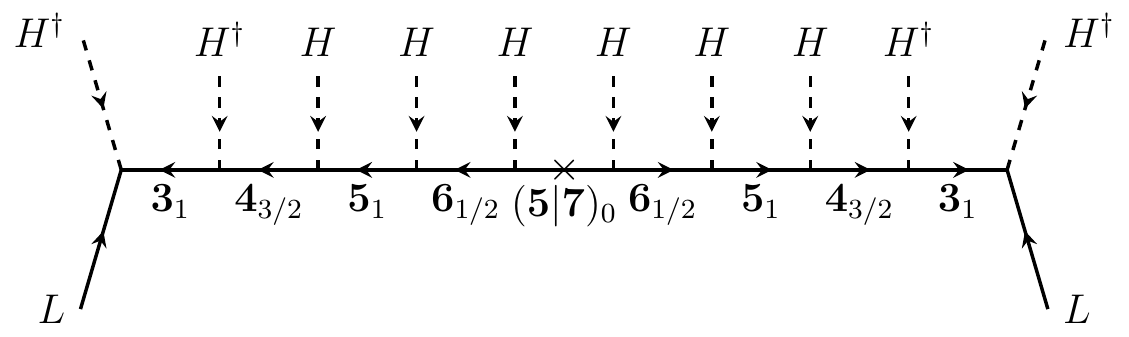}
    \caption{Two examples of genuine tree-level neutrino mass diagrams for $d=9$ and $d=13$ \cite{Anamiati:2018cuq}. Note that only fermions beyond the Standard Model are needed and that for every new fermion, its vector-like partner should be included. All fields are colour blind and the notations indicates the ${\rm SU(2)_L}$ representation and the hypercharge as a subscript.}
    \label{fig:numass:d9,13}
\end{figure}

%%%%%%%%%%%%%%%%%%%%%%%%%%%%%%%%%%%%%%%%%%%%%%%%%%%%%%%%%%%%%%%
\subsection{Radiative models}
\label{subsec:numass:maj_loop}

Small Majorana neutrino masses may be also induced by generating radiatively $\mathcal{O}^d$. As shown in \eq{eq:numass:estimate}, each loop adds a suppression factor of $1/16\pi^2$, lowering the new physics scale $\Lambda$ (see \fig{fig:numass:estimate}).

In the following sections we shall show some examples of one-, two- and three-loop models. We will focus in some of the most studied cases to illustrated how radiative models work.\footnote{For a complete review see \cite{Cai:2017jrq}.} Here, we consider only loop realisations of the Weinberg operator ($d=5$), although radiative high-dimensional examples of models exist in the literature, and we will show some of them in the following chapters.

%%%%%%%%%%%%%%%%
\subsubsection*{One-loop realisations}

\begin{figure}[h]
    \centering
    \begin{subfigure}[t]{0.48\textwidth}
        \centering
        \includegraphics[width=1\textwidth]{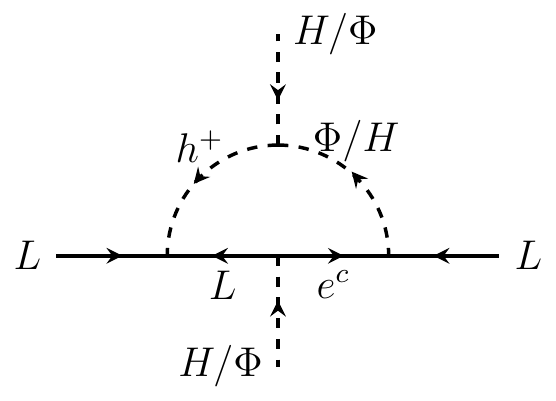}
        \caption{Zee model}
        \label{fig:numass:zee}
    \end{subfigure}
    \hfill
    \begin{subfigure}[t]{0.48\textwidth}
        \centering
        \includegraphics[width=1\textwidth]{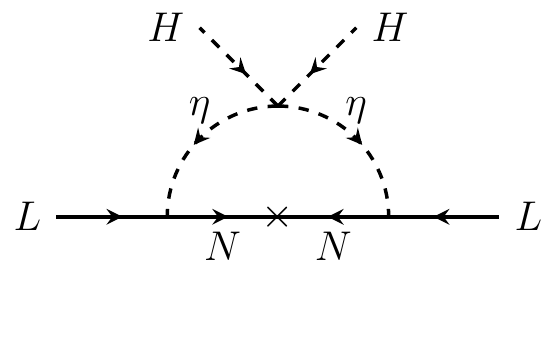}
        \caption{Scotogenic model}
        \label{fig:numass:scotogenic}
    \end{subfigure}
    \caption{One-loop neutrino mass diagrams generated in the Zee model \cite{Zee:1980ai} and the scotogenic model \cite{Ma:1998dn}.}
    \label{fig:numass:ZeeScoto}
\end{figure}

Following the tree-level, the simplest realisation is the Weinberg operator at one-loop. The two most studied neutrino mass models in this category are the Zee model \cite{Zee:1980ai} and the scotogenic model \cite{Ma:1998dn}. The neutrino mass diagrams for both models are depicted in \fig{fig:numass:ZeeScoto}.
\\

The Zee model contains an extra Higgs doublet $\Phi$ and a singly-charged scalar singlet $h^+$. Both Higgses couple to leptons, while the singlet couples to $LL$ with an antisymmetric Yukawa in flavour space. Lepton number is broken in two units by the trilinear coupling,
\begin{equation} %\label{eq:}
    \mu_{\rm Zee} \,  \tilde{H}^\dagger \, \Phi h^{+*} \, + \, \hc \, ,
\end{equation}
which induces a mixing between the two physical charged scalars.

The neutrino mass matrix is approximately given by,
\begin{equation} %\label{eq:}
    {\cal M}_\nu \, \propto \, \mu_{\rm Zee} \, C \, (f_{ik} m_k Y_{kj} - Y_{ki} m_k f_{kj}) \, ,
\end{equation}
with $f$ the antisymmetric Yukawa $f \, \bar{\tilde{L}} L h^+$, $Y$ the matrix of couplings of leptons to $\Phi$, i.e. $Y \, L \bar{e}_R \Phi^\dagger$, and $m_{k}$ the SM charged lepton masses. $C$ contains some information of the mixings.

Note that in the limit where $Y$ is diagonal, the diagonal elements of $m_\nu$ vanish, resulting in neutrino mixing angles which are not compatible with observations. This limit corresponds to the Zee-Wolfenstein model \cite{Wolfenstein:1980sy}. Moreover, it implies that at least some of the non-diagonal entries of $Y$ must be different from zero inducing LFV, like $\mu - e$ conversion in nuclei and \taumug, and especially a significantly large BR($h \to \tau \mu$) \cite{Herrero-Garcia:2017xdu}.
\\

The scotogenic model is by far one of the most popular neutrino mass models mainly because it links dark matter to the radiative generation of neutrino masses. The model adds to the SM particle content three fermions $N_i$ ($i=1,2,3$), pure singlets under the SM gauge group, and a Higgs doublet $\eta = (\eta^+ \, \eta^0)$. The symmetry content is extended with a $Z_2$ under which the SM particles are even, while the new fields are odd. This symmetry forbids the tree-level type-I seesaw (see \fig{fig:numass:seesawI}) and stabilises the lightest particle odd under $Z_2$.\footnote{$Z_2$ is assumed to be preserved after electroweak symmetry breaking \cite{Merle:2015gea}.} The latter, if neutral, will constitute a potentially good dark matter candidate. This can be either the lightest $N$ or the lightest neutral scalar $\eta_R$ or $\eta_I$, the CP-even and CP-odd components of $\eta^0$.

In the scotogenic model, lepton number is violated by the Majorana mass term of the fermion singlets $M_N$ and by the four scalar interaction,
\begin{equation} %\label{eq:}
    \frac{\lambda_5}{2} \, \left[ \left( H^\dagger \eta \right)^2 + \hc \right] \, .
\end{equation}
$\lambda_5$ is responsible of the mass splitting between $\eta_R$ and $\eta_I$, with $\Delta m_{R,I}^2 = \lambda_5 \, v^2$.

Finally, the neutrino mass matrix is then given by,
\begin{eqnarray} %\label{eq:}
    ({\cal M}_\nu)_{ij} &=&  \frac{1}{16\pi^2} \, \sum_{k} (Y_N)_{ik} \, (Y_N)_{jk} \, M_{Nk} 
    \\ \nn && \times \, \left[ \frac{m_R^2}{m_R^2-M_{Nk}^2} \, \ln{\frac{m_R^2}{M_{Nk}^2}} \, - \, \frac{m_I^2}{m_I^2-M_{Nk}^2} \, \ln{\frac{m_I^2}{M_{Nk}^2}} \right] \, ,
\end{eqnarray}
with $Y_N$ the Yukawa matrix associated to the coupling $\eta \bar N L$.

In the limit where $\lambda_5 v^2 << m_{R,I}^2 \equiv m_0^2$ and all the new mass scales are equal, i.e. $m_0 = M_{Nk} = M_k$,
\begin{equation} %\label{eq:}
    ({\cal M}_\nu)_{ij} =  \frac{1}{16\pi^2} \, \lambda_5 v^2 \, \sum_{k} \frac{(Y_N)_{ik} \, (Y_N)_{jk}}{M_k}\, .
\end{equation}
Comparing to the typical mass scale of the seesaw type-I \eq{eq:numass:typeI}, the scotogenic model gets an extra suppression factor of $\lambda_5/16\pi^2$. This means that a new physics scale of order TeV can be achieved with $\lambda_5 \sim 10^{-8}$, which although small can be thought as natural as in the limit $\lambda \to 0$ lepton number is conserved.

%%%%%%%%%%%%%%%%
\subsubsection*{Two-loop realisations}

Radiative Majorana neutrino mass models at two-loop level have been studied over the last years connecting them to dark matter, see for example \cite{Sierra:2016qfa, Simoes:2017kqb}. Here we present Zee-Babu model \cite{Babu:2002uu} as an example, although it does not contain a dark matter candidate, it has been extensively analysed in the literature due to its \textit{minimal} particle content and interesting phenomenology.

\begin{figure}
    \centering
    \includegraphics[width=0.5\textwidth]{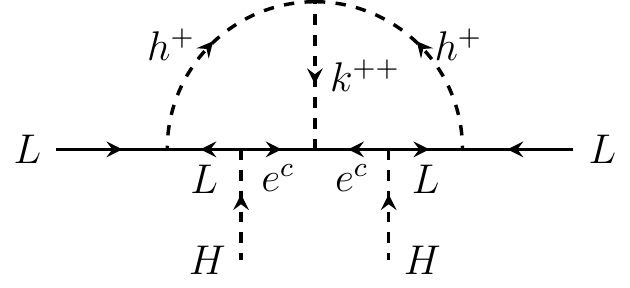}
    \caption{Two-loop diagram that generates neutrino masses in the Zee-Babu model \cite{Babu:2002uu}.}
    \label{fig:numass:Zeebabu}
\end{figure}

Similarly to the Zee model, the Zee-Babu model only contains scalars beyond the Standard Model particle content. In particular two singlets: a singly charged $h^+$ and a doubly charged $k^{++}$. Lepton number is violated by the trilinear term $\mu_{ZB} \, h^+ \, h^+ \, (k^{++})^*$. Note that this term cannot be arbitrarily large, as it participates at loop level in the charged scalar masses, roughly $|\mu_{ZB}| < 4 \, M$, with $M = {\rm max}(m_h, \, m_k)$ \cite{Babu:2002uu}.

The neutrino mass matrix is,
\begin{equation} %\label{eq:}
    (\mathcal{M}_\nu)_{ij} \, = \, \frac{1}{(16\pi^2)^2} \, \frac{16 \mu_{ZB}}{M^2} \, m_\alpha m_\beta \, f_{i\alpha} \, g_{\alpha \beta} \, f_{j\beta} \; \mathcal{I} \, ,
\end{equation}
where $m_{\alpha, \beta}$ are the charged lepton masses, $f$ is the antisymmetric Yukawa, identical to the Zee model, and $g$ the symmetric Yukawa associated to the interaction ${\bar e}_R^c \, e_R \, k^{++}$. $\mathcal{I}$ is the two-loop integral, function of the ratio of masses $m_k^2/m_h^2$ \cite{Herrero-Garcia:2014hfa}. Given that $f$ is an antisymmetric $3 \times 3$ matrix, the rank of the neutrino mass matrix is 2, i.e. at least one neutrino will remain massless. Another interesting feature of this model is due to the presence of the doubly charged singlet. $k^{++}$ mediates the charged lepton flavour violating (CLFV) decay $l_i \to l_j {\bar l}_k l_n$ at tree-level. This provides very stringent limits on the Yukawa $g$.

%%%%%%%%%%%%%%%%
\subsubsection*{Three-loop realisations}

There are few examples of three-loop realisations of the Weinberg operator in the literature. Although compelling models exists, the complexity of three-loop integrals becomes a major drawback, as very few analytical solutions are yet known. Nevertheless, there are famous examples, widely studied, like the KNT (Krauss-Narsi-Trodden) model \cite{Krauss:2002px} or the Cocktail model \cite{Gustafsson:2012vj}. Moreover, the former was the first radiative neutrino mass model with a stable dark matter candidate participating in the loop.

We shall not enter into detail here, but refer to \ch{ch:3loop} and \ch{ch:clfv}.

%%%%%%%%%%%%%%%%%%%%%%%%%%%%%%%%%%%%%%%%%%%%%%%%%%%%%%%%%%%%%%%
\subsection{Other frameworks}
\label{subsec:numass:maj_others}

%%%%%%%%%%%%%%%%
\subsubsection*{Majoron models}

The basic idea behind Majoron models is that the global lepton number symmetry of the Standard Model is spontaneously broken, giving a Majorana mass term for neutrinos. As a consequence, after lepton number is broken, a massless Nambu–Goldstone boson is generated, the Majoron $J$.

The simplest version of the Majoron models includes three right-handed neutrinos $N$ and a complex scalar Higgs singlet $\sigma$ with $2$ units of lepton number. This singlet couples to the right-handed neutrinos through,
\begin{equation} %\label{eq:}
    \frac{1}{2} \, Y_1 \, N^T \sigma N \, + \, \hc \, .
\end{equation}
When $\sigma$ acquires a VEV, $\vev{\sigma} \neq 0$, lepton number is broken and $N$ gets a Majorana mass term, dynamically realising the type-I seesaw neutrino mass matrix \eq{eq:numass:mnutypeI}, with $M_N = Y_1 \vev{\sigma}$. The complex field $\sigma$ can then be written in terms of two real fields,
\begin{equation} %\label{eq:}
    \sigma \, =  \, \frac{1}{\sqrt{2}} \, ( \vev{\sigma} + \rho + i J) \, ,
\end{equation}
where $J$ is the massless Goldstone boson associated to the spontaneous breaking of lepton number\footnote{It has been conjectured that gravity violates global symmetries, then the Majoron will acquire a mass through nonpertubative gravitational effects \cite{Akhmedov:1992hi,Kallosh:1995hi} and it can be considered a source of dark matter \cite{Lattanzi:2007ux,Bazzocchi:2008fh}.} and $\rho$ is a massive boson field.

%%%%%%%%%%%%%%%%
\subsubsection*{Left-Right symmetric models}

One minimal extension of the Standard Model is the so called left-right symmetric models \cite{Mohapatra:1974hk,Mohapatra:1974gc,Pati:1974yy,Senjanovic:1975rk}. Parity is broken in the Standard Model, but it seems attractive to hypothesise that it is restored at some higher energy. The minimal left-right extension of the Standard Model is based on the gauge group,
\begin{equation} %\label{eq:}
    \mathcal{G}_{\rm LR} = {\rm SU(3)_C} \times {\rm SU(2)_L} \times {\rm SU(2)_R} \times {\rm U(1)_{B-L}} \, ,
\end{equation}
where ${\rm B-L}$ is the difference between baryon and lepton number. Now, right-handed fermions are doublets under ${\rm SU(2)_R}$, which automatically introduces a right-handed neutrino, as well as a new gauge boson $W_R$.

Regarding the Higgs sector of left-right symmetric models, the minimal model contains a scalar bi-doublet $\phi$ transforming as $(\mathbf{1},\mathbf{2},\mathbf{2},0)$ under $\mathcal{G}_{\rm LR}$ and, either a pair of triplets $\Delta_L \equiv (\mathbf{1},\mathbf{3},\mathbf{1},-2)$ and $\Delta_R \equiv (\mathbf{1},\mathbf{1},\mathbf{3},-2)$, or a pair of doublets $\chi_L \equiv (\mathbf{1},\mathbf{2},\mathbf{1},-2)$ and $\chi_R \equiv (\mathbf{1},\mathbf{1},\mathbf{2},-2)$.

The left-right symmetry is broken in two steps. First, the right-handed scalar gets a VEV $v_R$ breaking parity and leaving the Standard Model gauge group. This gives a mass to $W_R$ proportional to $v_R$. Afterwards, the bi-doublet acquires a VEV and breaks the Standard Model gauge group as usual to ${\rm SU(3)_C} \times {\rm U(1)_{Q}}$. Note that the left-handed scalar triplet (doublet) acquires too a VEV $v_L$ with $v_R >> v_L$, in agreement with the $\rho$ parameter.

Neutrino masses are then a combination of type-I and type-II seesaws, which gives the mass matrix,
\begin{equation} %\label{eq:}
    \begin{pmatrix} m_L & m_D \\ m_D^T & m_R \end{pmatrix} \, ,
\end{equation}
where $m_R$ ($m_L$) is proportional to $v_R$ ($v_L$), while $m_D$ is a function of the VEVs of the bi-doublet $\phi$. The resulting light neutrino mass matrix is then given by,
\begin{equation} %\label{eq:}
    \mathcal{M}_\nu^{I+II} \, = \, m_L \, - \, m_D (m_R)^{-1} m_D^T \, .
\end{equation}
%

%%%%%%%%%%%%%%%%
\subsubsection*{Grand Unified Theories}

Left-right symmetric models and other extensions of the Standard Model can be understood as a step towards a Grand Unified Theory (GUT), like ${\rm SU(5)}$ or ${\rm SO(10)}$ \cite{Georgi:1974sy,Fritzsch:1974nn}. In general, a GUT is a theory based on a simple Lie group that contains the Standard Model as a subgroup. GUTs always contain exotic fields that couple to quarks and leptons, therefore violating lepton and baryon numbers. Phenomenologically, new processes like proton decay or neutron-antineutron oscillations become possible, which constrain the theory considerably.

Among all the possible Grand Unified Theories, a famous framework are models based on SO(10). These models contain the $\mathcal{G}_{\rm LR}$ as a subgroup and, consequently, a right-handed neutrino and neutrino masses through a seesaw mechanism. However, it is remarkable that the right-handed neutrino is needed to complete the spinorial 16 representation of SO(10), unifying all fermions of each family into a single representation.

For a detailed discussion about neutrino masses in GUTs we refer to \cite{Mohapatra:1998rq}.

%%%%%%%%%%%%%%%%%%%%%%%%%%%%%%%%%%%%%%%%%%%%%%%%%%%%%%%%%%%%%%%
%%%%%%%%%%%%%%%%%%%%%%%%%%%%%%%%%%%%%%%%%%%%%%%%%%%%%%%%%%%%%%%
\section{Dirac neutrinos}
\label{sec:numass:dirac}

Neutrino masses and the existence of dark matter are the two most important pieces of evidence that the Standard Model is not the final theory of nature. Among the open questions about neutrino physics, probably the most important one is the nature of neutrinos, namely if neutrinos are Dirac or Majorana particles. So far we lack any experimental or observational evidence in favour of one or the other, in spite of the big experimental effort in the last decades, as for instance, in neutrinoless double-$\beta$ decay. 

From a theoretical point of view, Majorana neutrinos have garnered much more attention. Several seesaw  and radiative mass generation mechanisms for Majorana neutrinos have been known for a long time. In contrast, Dirac neutrinos have received relatively little attention. However, in the last few years there has been a renewed interest in looking at mass models for Dirac neutrinos. In this direction, several seesaw mechanisms and loop models for Dirac neutrinos have been recently proposed, see for example \cite{Farzan:2012sa, Ma:2014qra, Valle:2016kyz, Bonilla:2016diq, Helo:2018bgb, Reig:2018mdk, Saad:2019bqf}. In this section, we will briefly review some cases to illustrate Dirac neutrino mass models.

%%%%%%%%%%%%%%%%%%%%%%%%%%%%%%%%%%%%%%%%%%%%%%%%%%%%%%%%%%%%%%%
\subsection{Tree-level models}
\label{subsec:numass:dir_tree}

Differently from the Majorana case, to generate Dirac neutrino masses we need to introduce at least two copies of the right-handed neutrino $\nu_R$. We need a symmetry that allows the Dirac mass term $Y_\nu \, L H {\bar \nu}_R$, but forbids a Majorana mass for $\nu_R$, in order to not realise a type-I seesaw. This symmetry can in principle be lepton number or some subgroup of it. The main drawback from this simple approach is that $Y_\nu$ should be ridiculously small, compared to other Yukawas in the Standard Model, to explain neutrino oscillation data.

Recently, several models for Dirac neutrinos have been proposed in analogy to Majorana seesaw scenarios \cite{CentellesChulia:2018gwr}. Forbidding the Dirac mass term $L H {\bar \nu}_R$ by some symmetry, they add some new scalar(s) so that a dimension 5 \textit{Weinberg-like} Dirac operator is allowed, i.e. an operator of the type,
\begin{equation} \label{eq:numass:opDiracSeesaw}
    \frac{1}{\Lambda} \, L \, H \, {\bar \nu}_R \, X \, ,
\end{equation}
where $X$ can either be a scalar singlet $\chi$ or triplet $\Delta$ with no hypercharge. The operators are then \textit{opened} in all possible gauge invariant ways getting models that resemble the three seesaw Majorana models, with an analogous mass suppression mechanisms. Either the smallness of neutrino masses is due to a small induced VEV or to a new heavy mediator with a large vector-like mass.

\begin{figure}
    \centering
    \begin{subfigure}[t]{0.32\textwidth}
        \centering
        \includegraphics[width=1\textwidth]{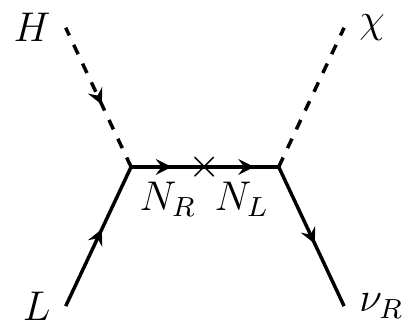}
        \caption{Type-I seesaw}
        \label{fig:numass:diracseesawI}
    \end{subfigure}
    \hfill
    \begin{subfigure}[t]{0.32\textwidth}
        \centering
        \includegraphics[width=1\textwidth]{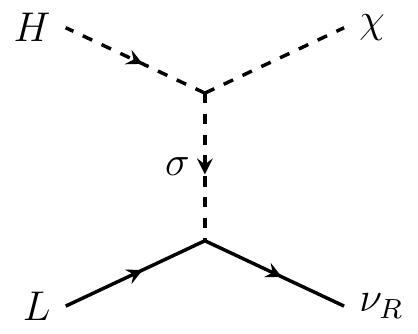}
        \caption{Type-II seesaw}
        \label{fig:numass:diracseesawII}
    \end{subfigure}
    \hfill
    \begin{subfigure}[t]{0.32\textwidth}
        \centering
        \includegraphics[width=\textwidth]{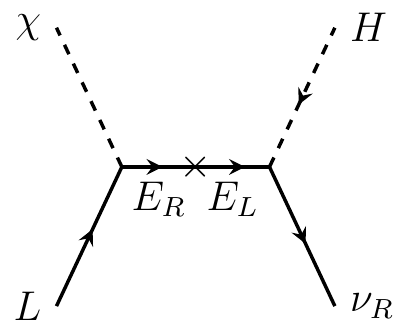}
        \caption{Type-III seesaw}
        \label{fig:numass:diracseesawIII}
    \end{subfigure}
    \caption{Neutrino mass diagrams representing the Dirac analogues to the standard Majorana seesaws. $N_{L,R}$ and $E_{L,R}$ are vector-like fermions with SM quantum numbers $(\mathbf{1},\mathbf{1},0)$ and $(\mathbf{1},\mathbf{2},-1/2)$. $\sigma$ is a doublet with hypercharge $1/2$ with an induced VEV, proportional to $\vev{H}$ and $\vev{\chi}$.}
    \label{fig:numass:diracseesaw1}
\end{figure}

The diagrams for all possible realisations of the operator \eq{eq:numass:opDiracSeesaw} for $X = \chi$ are depicted in \fig{fig:numass:diracseesaw1}. For the triplet version with $\Delta$ we refer to the original publication \cite{CentellesChulia:2018gwr}. Note that one of the main differences between Majorana and Dirac models is that the latter always need some extra symmetry allowing \eq{eq:numass:opDiracSeesaw} but broken to effectively generate the Dirac mass term $L H {\bar \nu}_R$.

%%%%%%%%%%%%%%%%%%%%%%%%%%%%%%%%%%%%%%%%%%%%%%%%%%%%%%%%%%%%%%%
\subsection{Radiative models}
\label{subsec:numass:dir_loop}

Radiative Dirac neutrino models have been proposed with one-loop and two-loops to explain the smallness of neutrino masses, along with a stable dark matter candidate. Similarly to the scotogenic model, a symmetry is needed to ensure the stability of a dark matter candidate participating in the loop. This symmetry is usually taken to be a simple $Z_n$, either enforced by hand or residual from the breaking of some ${\rm U(1)}$ symmetry, like for example lepton number.

\begin{figure}
    \centering
    \includegraphics[width=0.5\textwidth]{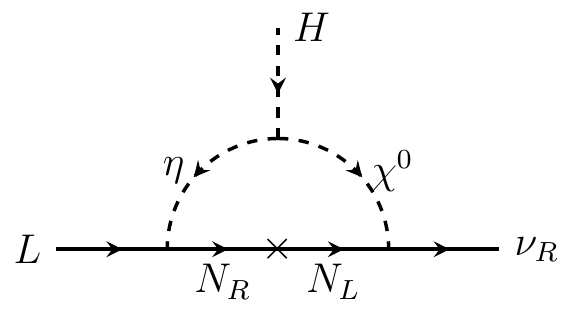}
    \caption{One-loop Dirac neutrino mass diagram. It is analogous to the well-known scotogenic model, in the sense that it contains mainly the same particle content (except $\chi^0$) and that a $Z_2$ symmetry running in the loop stabilises dark matter.}
    \label{fig:numass:dirac1loop}
\end{figure}

In \fig{fig:numass:dirac1loop} we can find an example of a one-loop Dirac mass diagram with a dark matter candidate. The model is thought of as a Dirac analogy of the scotogenic model, proposed by the same author \cite{Farzan:2012sa}. Apart from the field already appearing in \fig{fig:numass:ZeeScoto}, it contains an extra hyperchargeless colourblind singlet $\chi^0$. The model includes two extra $Z_2$ symmetries: one softly broken in the trilinear term $H^\dagger \eta \, \chi^0$ to forbid the tree-level Dirac mass term but allow the loop, and other exact $Z_2$, in the spirit of the scotogenic model, under which only the particles inside the loop are odd. This last $Z_2$ stabilises the dark matter candidate that can be either of the three neutral particles running in the loop, depending on the mass hierarchy.
\\

A further discussion about radiative Dirac neutrino mass models at one- and two-loop level can be found in \ch{ch:dirac2l} and \ch{ch:dm_DiracMajo}.

%%%%%%%%%%%%%%%%%%%%%%%%%%%%%%%%%%%%%%%%%%%%%%%%%%%%%%%%%%%%%%%
%\subsection{Other frameworks}
%\label{subsec:numass:dir_others}

%We finish this section about Dirac neutrino models with a brief comment regarding other mass mechanisms beyond those already explained. 

%%%%%%%%%%%%%%%%%%%%%%%%%%%%%%%%%%%%%%%%%%%%%%%%%%%%%%%%%%%%%%%
%%%%%%%%%%%%%%%%%%%%%%%%%%%%%%%%%%%%%%%%%%%%%%%%%%%%%%%%%%%%%%%
\section{Systematic classifications}
\label{sec:numass:class}

In this section we give an introduction to classifications of neutrino mass models, as some of the chapters of this thesis will be devoted to them. Systematic classifications are a useful tool, especially for model-building. The main idea behind them is to build all possible diagrams that generate a certain neutrino mass operator and classify them according to some common characteristics. This categorisation allows then to extract general conclusions that can be applied to a whole set of diagrams or models. The different classes of diagrams depend on the operator that is generated, the classes can be organised according to whether the corresponding diagram is the dominant contribution to neutrino masses, an estimate of the neutrino mass scale they can fit or even regarding the minimal particle content they need in order to explain neutrino masses.

%%%%%%%%%%%%%%%%%%%%%%%%%%%%%%%%%%%%%%%%%%%%%%%%%%%%%%%%%%%%%%%
\subsection{Genuine topologies, diagrams and models}
\label{subsec:numass:genuine}

As a starting point, we introduce the following nomenclature, that will be used throughout the text. We shall call topologies to be those Feynman diagrams where no property of the fields is considered (in graph theory, these are also known as un-directed multi-graphs). If scalars are differentiated from fermions, we will call them diagrams. If additionally the quantum numbers of the internal particles are specified, we will use the expression model-diagrams (or just model when it is clear from the context what we mean by this word).

Let us discuss now the concept of \textit{genuine} model-diagram, diagram and topology in the context of classifications. Essentially, we want to identify this concept with those model-diagrams (plus their associated diagrams and topologies) for which the leading contribution to neutrino masses arises at a particular loop level, without the need to introduce extra symmetries.

It is important to keep in mind that, in general, loop integrals have a finite and an infinite part. Infinite integrals require a lower order counter term in a consistent renormalisation scheme. Thus, models with infinite $n$-loop amplitudes must necessarily also generate neutrino mass diagrams with fewer loops, and for that reason model-diagrams with an infinite amplitude are not genuine in our sense. However, certain diagrams lead automatically only to finite loop integrals, in which case it might be possible to build genuine model-diagrams from them.

Finiteness of the amplitude of a model-diagram is therefore a necessary condition for it to be genuine. However, it is not sufficient: it is also necessary to ensure that there are no other automatically generated model-diagrams with less loops. Consider neutrino masses generated via the dimension $d=5+2n$ operator
\begin{equation}
    LLHH \left(H^{\dagger}H\right)^{n} \, ,
\end{equation}
through a diagram with $\ell$ loops. It is expected that
\begin{equation}
    M_{\nu} \sim \frac{1}{\left(4\pi\right)^{2\ell}}
    \left(\frac{\left\langle H\right\rangle}{\Lambda}\right)^{1+2n}
    \left\langle H\right\rangle \, .
\end{equation}
As such, for diagrams with a characteristic scale $\Lambda\sim$ TeV, removing a loop ($\ell \to \ell-1$) and increasing the operator dimension by two units ($n \to n+1$) leaves $M_\nu$ with roughly the same value. So in order to have a dominant $\left(d,\ell\right)$ contribution to neutrinos masses, those with $\left(d^\prime,\ell^\prime\right)=\left(d-2i,\ell-j\right)$ and $j>i$ should be absent.\footnote{Note that by closing some pairs of $H$/$H^\dagger$ external lines it will always be possible to find other contributions with $\left(d^\prime,\ell^\prime\right)=\left(d-2i,\ell-i\right)$.} \textit{Genuine} model-diagrams are those associated to these cases; in other words, the combination of fields participating in genuine model-diagrams must not be sufficient, by itself, to generate other more important neutrino mass contributions. For example, a model with a right-handed neutrino, $\nu_R$, will also give a $d=5$ tree-level contribution to the neutrino mass (unless an additional symmetry is used to eliminate some unwanted couplings), which will likely be the more important one. Stated in this way, genuineness is a concept which applies to model-diagrams. However, the list of such cases is infinite, in principle, as there are endless possibilities of assigning quantum numbers to the internal particles. We will therefore be interested in cataloguing those diagrams and topologies for which there exists at least one genuine model-diagram. These topologies and diagrams will be considered genuine themselves.

%%%%%%%%%%%%%%%%%%%%%%%%%%%%%%%%%%%%%%%%%%%%%%%%%%%%%%%%%%%%%%%
\subsection{Realisations of the Weinberg operators}
\label{subsec:numass:maj_class}

We start with the classifications of the Weinberg ($d=5$) and dimension $d>5$ Weinberg-like operators, i.e. $\mathcal{O}_W$ and $\mathcal{O}^d$. Classifications exist up to $d=13$, but only for $d=5$ and $d=7$ at the loop level, up to three-loops and one-loop, respectively.

\subsubsection*{One-loop $d=5$}

\begin{figure}
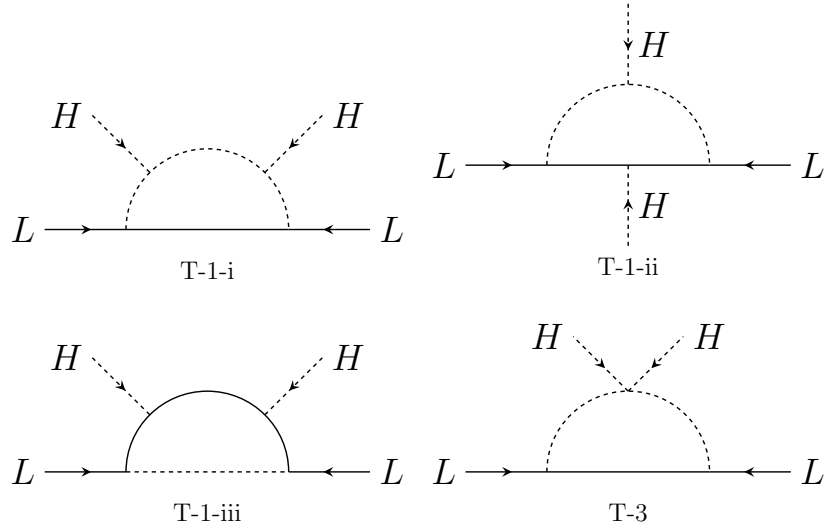

    \centering
    \includegraphics[width=0.4\textwidth]{figures/diagram_dim5_T1-i.pdf}
    \includegraphics[width=0.4\textwidth]{figures/diagram_dim5_T1-ii.pdf}
    \\
    \includegraphics[width=0.4\textwidth]{figures/diagram_dim5_T1-iii.pdf}
    \includegraphics[width=0.4\textwidth]{figures/diagram_dim5_T3.pdf}
    \caption{Genuine $d=5$ one-loop neutrino mass diagrams in the notation of \cite{Bonnet:2012kz}.}
    \label{fig:numass:loopd5}
\end{figure}

A systematic analysis of all one-loop $d=5$ topologies has been given in \cite{Bonnet:2012kz}. In total, 6 topologies where found, but only two of them (denoted as T-1 and T-3) can give genuine models. All other topologies lead either to non-renormalisable models or diagrams with infinite loop integrals (thus representing loop corrections to tree-level quantities) \footnote{Topology T-4 has two divergent and two finite diagrams. In \cite{Fraser:2015mhb} the diagram T-4-2-ii was used to generate a coupling $L \Delta L$ at one-loop. This diagram is classified as divergent in \cite{Bonnet:2012kz}. However, in \cite{Fraser:2015mhb} {\em two} diagrams of this type appear, with the infinity cancelled between diagrams. This can not be justified in terms of symmetry, instead it is due to the fact that lepton number is broken {\em softly} in the model of \cite{Fraser:2015mhb}.} or can be understood as finite one-loop realisations of some particular vertex of one of the tree-level $d=5$ seesaws. Topologies T-1 and T-3 lead to a total of four diagrams shown in \fig{fig:numass:loopd5}. The Zee model \cite{Zee:1980ai} falls within category T-1-ii, the scotogenic model of Ma \cite{Ma:2006km} is an example of T-3.

While at tree-level the size of the representations as well as the hypercharge of the new fields is fixed, at loop level there always exists a ``tower'' of possible models. This is easily understood. Consider, for example, the diagram T-3. The outside leptons couple to a scalar and a fermion. Since $L$ is a ${\rm SU(2)}$ doublet, the representation of the scalar and the fermion can be: ${\bf 1}\otimes {\bf 2}$, ${\bf 2}\otimes{\bf 3}$, ${\bf 3}\otimes {\bf 4}$, etc. Similarly at the four scalar vertex the smallest possibility is ${\bf 2}\otimes {\bf 2}$, but larger representations can be inserted, with the only constraint that the product of the two scalars can build a triplet. In the same way, the hypercharge of the internal particles is fixed only up to an additive constant $x$, that runs in the loop.\footnote{Of course, not all choices of $x$ will lead to phenomenologically acceptable models.}

The minimal possibility to build a model for T-3 is then that the fermion is a $\nu_R$, while the scalar is a doublet $({\bf 1}, {\bf 2}, 1/2)$, i.e. the well-known scotogenic model. This is minimal in the sense that it uses the smallest representations and the smallest value of the hypercharge possible, i.e. $x=0$. However, $({\bf 1}, {\bf 2}, 1/2)$ can not be the Standard Model Higgs, it must be an additional inert doublet.

\subsubsection*{Two-loop $d=5$}

All possible two-loop topologies of the Weinberg operator have been discussed in \cite{Sierra:2014rxa}. From a starting set of 29 topologies, they reduce the number to 6 genuine topologies, out of which 20 renormalisable diagrams can be built. They realised that all the genuine diagrams can be understood as variations of three models: (i) the Cheng-Li-Babu-Zee (CLBZ) \cite{Cheng:1980qt,Zee:1985id,Babu:1988ki}, (ii) the Petcov-Toshev-Babu-Ma (PTBM) \cite{Petcov:1984nz,Babu:1988ig}, and (iii) what they called Rainbow topology (RB).

The different categories of the classification are thought to serve for a systematic study of neutrino mass model signals at several scenarios. For instance, at colliders like LHC, testing the origin of neutrino masses \cite{Cai:2014kra}, or to systematically build models with neutrino masses and dark matter \cite{Sierra:2016qfa, Simoes:2017kqb}, analogous to the one-loop cases.
\\

We have deliberately omitted the discussion about one-loop $d=7$ and three-loop $d=5$ as the following chapters are devoted to their classifications.

%%%%%%%%%%%%%%%%%%%%%%%%%%%%%%%%%%%%%%%%%%%%%%%%%%%%%%%%%%%%%%%
\subsection{The Dirac side of the classifications}
\label{subsec:numass:dirac_class}

Due to the recent interest in Dirac neutrino mass models, several classifications have been done of the dimension 4 operator $L H \bar{\nu}_R$ up to two-loop level \cite{Ma:2016mwh,CentellesChulia:2019xky}, as well as the dimension 5 and 6 Dirac mass operators at tree and one-loop level \cite{Yao:2017vtm,Yao:2018ekp}. Tree-level models are discussed in \sect{subsec:numass:dir_tree}, while \ch{ch:dirac2l} is dedicated to two-loop Dirac neutrino mass models. For one-loop models, discussed in \cite{Ma:2016mwh,Yao:2017vtm,Yao:2018ekp}, they find that the smallness of neutrino masses can be naturally explained via loops analogously to the Majorana case. They argue that lower order contributions to neutrino masses can be forbidden by introducing an abelian or finite non-abelian symmetry and breaking it appropriately. As in previous classifications, they separate infinite diagrams from finite non-genuine and genuine diagrams, as only the latter can be the leading contribution to neutrino masses, and list every possible dark matter candidate that can participate in the genuine one-loop diagrams.

One main difference between Majorana and Dirac neutrino mass models is that the latter always requires some symmetry beyond the Standard Model. As already discussed, this symmetry forbids the operator $L H \bar{\nu}_R$ at tree-level, but once broken it allows an effective realisation of it, either coming from a higher dimension operator or generating it via loops. This implies that by an appropriate choice of the particle content and symmetry transformations, one can ensure the genuineness of every renormalisable amputated Feynman diagram.\footnote{For the dimension 4 operator $L H \bar{\nu}_R$, this reduces to every renormalisable 1 Particle Irreducible (1PI) diagram. See \ch{ch:dirac2l} for details.}

\pagebreak
\fancyhf{}

%% file: Chapters/Dim7_1loop/Chapter_dim7.tex
\fancyhf{}
\fancyhead[LE,RO]{\thepage}
\fancyhead[RE]{\slshape\nouppercase{\leftmark}}
\fancyhead[LO]{\slshape\nouppercase{\rightmark}}

\chapter{Loop neutrino masses from $d = 7$ operator}
\label{ch:Dim7_1loop}
\graphicspath{ {Chapters/Dim7_1loop/} }

In this chapter, based on \cite{Cepedello:2017eqf}, we will focus on dimension $7$ neutrino mass models at one-loop order. Our aim is to give a systematic analysis of such models, constructing first all possible one-loop topologies and then identify those topologies that generate genuine models. We use the usual definition of genuine models, recall \sect{subsec:numass:genuine}. In this case, it refers specifically to models for which the one-loop $d = 7$ contribution to the neutrino mass gives the leading order contribution. This assumption implies, of course, that both the $d = 5$ and the $d = 7$ tree-level, as well as the $d = 5$ one-loop, contributions should be absent. We shall rely on the previous systematic analysis of the Weinberg operator at tree-level and one-loop, as well as the analysis of the $d=7$ operator at tree-level, already introduced in \sect{sec:numass:majorana}.

The classification shown here is different from preceding classifications. In previous works, classifications were done separating genuine models from non-genuine models, and, if applicable, regarding general differences in the loop integrals. For higher dimension ($d>5$) neutrino mass operators, the presence of the $H^\dagger$ makes it possible to do a further classification of the genuine models in terms of particle content. This forces particle contents with a characteristic phenomenology for each class of models, as will be analysed later in the chapter. The phenomenology of these models will be studied in \ch{ch:Dim7_pheno}.
\\

The chapter is organised as follows. In the next section, we give a short introduction, pointing out the assumptions we are considering in the classification. \Sect{sec:dim7loop:class} provides the core of the chapter. It discusses all possible topologies and diagrams, and classifies them into different groups. Finally, we introduce the three most minimal example models that one can construct at $d=7$ one-loop order in \sect{sec:dim7loop:models}.

%%%%%%%%%%%%%%%%%%%%%%%%%%%%%%%%%%%%%%%%%%%%%%%%%%%%%%%%%%%%%%%
%%%%%%%%%%%%%%%%%%%%%%%%%%%%%%%%%%%%%%%%%%%%%%%%%%%%%%%%%%%%%%%
\section{Preliminary considerations}
\label{sec:dim7loop:intro}

Disregarding derivative and gauge operators, the authors of \cite{Babu:2001ex} have written down the complete list of $\Delta L=2$ operators up to dimension $11$. Among them, only one of these operators is important for us here,
\begin{equation}\label{eq:dim7loop:defo7}
    {\cal O}^{d=7} \propto LLHH (H^{\dagger}H) \, ,
\end{equation}
where indices have been omitted for simplicity. All other dimension $7$ operators in the list of \cite{Babu:2001ex} will lead to $d=5$ one-loop neutrino mass models.

As noted in \cite{Bonnet:2009ej}, where they analysed in detail the $d=7$ operator \eq{eq:dim7loop:defo7} at tree-level, this operator will always also generate a higher loop order $d=5$ neutrino mass,
\begin{equation}\label{eq:dim7loop:nlp1}
    \frac{1}{\Lambda^3}LLHH(H^{\dagger}H) \rightarrow \frac{1}{16 \pi^2} \frac{1}{\Lambda}LLHH \, ,
\end{equation}
as explained in \sect{sec:numass:majorana}. One can straightforwardly estimate that this loop contribution will become more important than the tree-level one if $(\Lambda/v) \gtrsim 4 \pi$. This means $\Lambda \lesssim 2$ TeV is required for the $d=7$ contribution to dominate. Since this is unavoidable in the Standard Model, the authors of \cite{Bonnet:2009ej} considered a two Higgs doublet extension of the Standard Model in their discussion of the $d=7$ tree-level neutrino mass. $H^{\dagger}H$ is a singlet under any discrete symmetry. With more than one Higgs it is possible to introduce an additional discrete symmetry, under which the two Higgses transform differently. We will stick, instead, to only the Standard Model Higgs and take \eq{eq:dim7loop:nlp1} as a motivation that any $d=7$ model of neutrino mass must have new particles below $2$ TeV, otherwise it will not give the leading contribution to the neutrino mass matrix.

As mentioned above, both $d=5$ and $d=7$ tree-level contributions should be forbidden, otherwise the $d=7$ one-loop contribution might be just some minor correction to the neutrino mass matrix. Absence of these lower order contributions could be attributed to either: (i) the existence of some symmetry; or (ii) absence of fields which generate neutrino masses at lower order. An example of the former at dimension $5$ is the scotogenic model \cite{Ma:2006km}. In this model, a right-handed neutrino, plus a scalar doublet, are introduced, but due to a $Z_2$ symmetry, there is no tree-level $d=5$ neutrino mass,\footnote{The well-known bonus of the $Z_2$ symmetry is that it allows to ``stabilise'' the lightest $Z_2$ odd particle, thus relating the stability of the dark matter to the radiative generation of neutrino masses.} and neutrinos have mass at one-loop order. The classic example for (ii) is the Zee model \cite{Zee:1980ai}. In the Zee model, none of the particles necessary for a tree-level seesaw exist. Instead, an additional charged singlet scalar, plus an additional doublet scalar, generate neutrino masses radiatively.

Two comments might be in order. First, with the use of non-Abelian discrete symmetries it is possible to construct viable models, genuine in the sense that they can give the leading contribution to the neutrino mass matrix. We will not discuss in detail such models, since the use of discrete symmetries for $d=7$ neutrino masses has been discussed in this context already in a number of references, see for example \cite{Bonnet:2009ej,Chen:2006hn,Kanemura:2010bq,Krauss:2011ur,Krauss:2013gy}.

Second, the example models, which we will discuss later on, all have explicit lepton number violation (LNV). One could construct extensions of these models, in which the LNV is spontaneous. In that case, a massless Goldstone boson would appear, usually called Majoron in the literature. We will not discuss the phenomenology of Majorons here and only note in passing that no new topologies would be generated in such models, with respect to the explicitly LNV models we consider.

For a more compact notation we will also use a notation which gives the ${\rm SU(2)_L}$ representation and hypercharge in the form ${\bf R}_Y$ with a superscript $S$ or $F$, where necessary, i.e. for example ${\bf 5}_1^S$ is a scalar {\bf 5-plet} with $Y=1$. Note that, for some of the fields, particular symbols are common in the literature, such as $\nu_R$, $\Delta$ and $\Sigma$ for the type-I, type-II and type-III seesaw.

%%%%%%%%%%%%%%%%%%%%%%%%%%%%%%%%%%%%%%%%%%%%%%%%%%%%%%%%%%%%%%%
%%%%%%%%%%%%%%%%%%%%%%%%%%%%%%%%%%%%%%%%%%%%%%%%%%%%%%%%%%%%%%%
\section{Classification of $d = 7$ neutrino mass models}
\label{sec:dim7loop:class}

In this section, we will discuss the classification of the different $d=7$ one-loop diagrams. We first construct all possible one-loop topologies with six external legs and then discard in different steps those topologies that cannot lead to genuine models. For the remaining topologies, we order all possible diagrams into different classes, depending on the minimum size of the largest required ${\rm SU(2)_L}$ representation appearing in the corresponding diagram.

We note in passing that we will not discuss colour in detail, because colour assignments can be trivially added: All particles outside loops must be necessarily colour singlets, while pairs of particles in loops can always be assigned colour in combinations ${\bf X}+{\bf\bar X}$, for ${\bf X}={\bf 1}$, ${\bf 3}$, $\cdots$, which then couple to ``outside'' colour-blind particles.

We adopt the following notation: we denote topologies with $T_n$, where $n$ is the number given in the figure with the corresponding graph of the topology. For diagrams, we follow the notation $D_n^{(x)}$, with $n$ the corresponding topology which generates the diagram and $x = i, ii, iii, ...$ to differentiate between different diagrams generated from the same topology. In the figures, we denote diagrams by $n$-$i$.

%%%%%%%%%%%%%%%%%%%%%%%%%%%%%%%%%%%%%%%%%%%%%%%%%%%%%%%%%%%%%%%
\subsection{Topologies}
\label{subsect:dim7loop:topos}

We construct all possible one-loop topologies with six external legs, discarding from the start all self-energy corrections. This construction can be done in different ways, and we used two different procedures to assure that all possible topologies were found. In total there are $48$ possible topologies. The complete list of topologies is given in \app{app:topos}.

We will briefly describe the methods we used to find the topologies. The first procedure consists in taking the five $d=7$ tree-level topologies and to generate loops in all possible combinations, connecting either lines to lines or lines to vertices or vertices to vertices, using only 3-point and 4-point vertices. From this list one has to discard in the end all duplicates.

\begin{figure}
    \centering
    \includegraphics[width=0.3\textwidth]{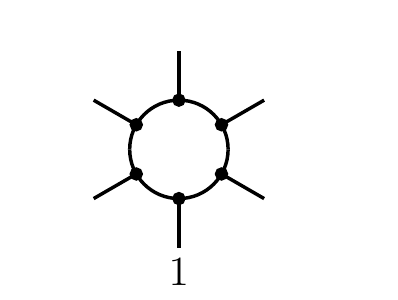}
    \includegraphics[width=0.3\textwidth]{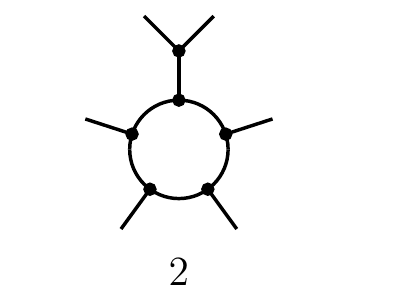}
    \includegraphics[width=0.3\textwidth]{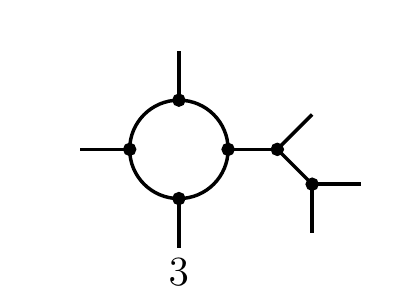}
    \caption{Generating topologies, starting from the simplest topology, containing only 3-point vertices with all 6 particles connected to the loop, $T_1$. Subsequent topologies are found by removing systematically particles attached to the loop and reconnecting them to outside particles, as shown for the examples $T_2$ and $T_3$.}
    \label{fig:dim7loop:T1T2T3}
\end{figure}

\begin{figure}
    \centering 
    \includegraphics[width=0.3\textwidth]{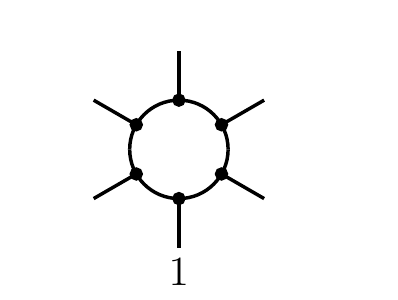}
    \includegraphics[width=0.3\textwidth]{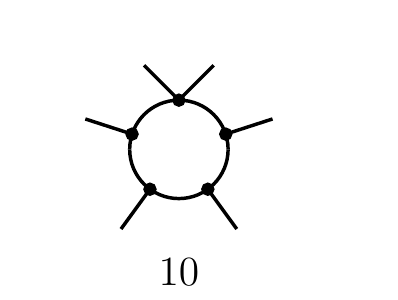}
    \caption{Constructing topologies with 4-point vertices from existing topologies with only 3-point vertices by ``shrinking'' one connecting line. Here shown for the example how $T_1$ generates $T_{10}$.}
    \label{fig:dim7loop:T1T10}
\end{figure}

The second procedure starts from the simplest realisation of a one-loop topology adding six external lines to the loop using only 3-point vertices, as shown in \fig{fig:dim7loop:T1T2T3}. From this topology, all other topologies can be found by systematically removing lines attached to the loop and adding them to an outside particle generating a new 3-point vertex, as shown in the figure for the examples of $T_2$, $T_3$, etc. Once all possible topologies with only 3-point vertices are found, all remaining topologies can be generated from the earlier ones by shrinking one line connecting two 3-point vertices to one new 4-point vertex, see the example \fig{fig:dim7loop:T1T10}. Again, this procedure produces duplicates, which have to be identified and discarded. This procedure provides a systematic construction of all possible topologies in a more intuitive way than the first one described above. It allows to classify topologies according to the number of lines entering the loop, creating subgroups with the same number of 4-leg vertices. This criterion is the one we use for ordering the complete list of topologies. For instance, topologies $T_1$ to $T_9$ are the only ones with just 3-point vertices ordered by decreasing number of legs sticking out of the loop. 

\begin{figure}
    \centering
    \includegraphics[width=0.3\textwidth]{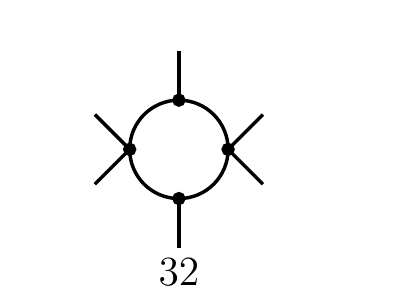}
    \includegraphics[width=0.3\textwidth]{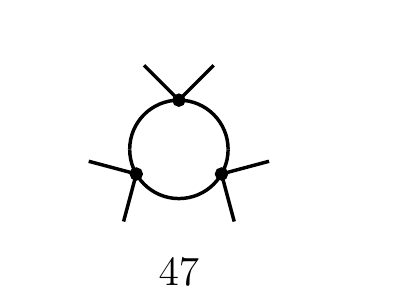}
    \caption{Examples of non-renormalisable topologies. These topologies cannot accommodate two external fermion and scalar lines with only renormalisable vertices.}
    \label{fig:dim7loop:TopologiesNRO}
\end{figure}

We then proceed to order topologies into different groups. We can discard immediately topologies like $T_{32}$ and $T_{47}$, shown in \fig{fig:dim7loop:TopologiesNRO}, because none of them can lead to a renormalisable model. The six topologies given in \fig{fig:app:topos:TopologiesNRO} are excluded because of this argument.

\begin{figure}
    \centering 
    \includegraphics[width=0.3\textwidth]{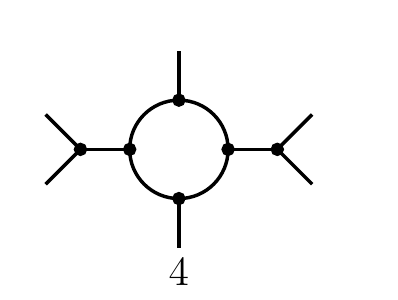}
    \includegraphics[width=0.3\textwidth]{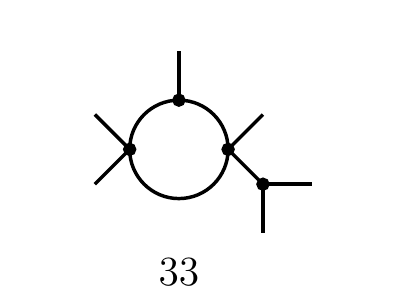}
    \caption{Two example $d=7$ topologies which will always be accompanied by a tree-level seesaw $d=5$ contributions to the neutrino mass matrix. For discussion see the text.}
    \label{fig:dim7loop:T4T33}
\end{figure}

The next step is to generate all possible diagrams and check if any field which generates neutrino masses at lower order is required. Two examples of topologies, which always necessarily will be accompanied by a tree-level $d=5$ seesaw, are shown in \fig{fig:dim7loop:T4T33}. These diagrams can be easily understood. Every topology with at least two 3-leg vertices on two external lines will always generate a vertex $LH{\bar \nu_R}$, $LH{\bar \Sigma}$ or $H\Delta^\dagger H$, and thus a seesaw at tree-level, as in the example $T_4$ in \fig{fig:dim7loop:T4T33} to the left. Topologies which contain one 3-leg vertex with two external lines isolated by a 4-leg vertex will always have a coupling of the type $L\Delta L$, as for example the topology $T_{33}$ in \fig{fig:dim7loop:T4T33} to the right. The 27 topologies, for which all diagrams can be excluded due to this argument, are given in \fig{fig:app:topos:TopologiesSeesaw}. The topologies $T_7$, $T_{22}$, $T_{23}$ and $T_{24}$ in this figure are somewhat particular examples, as they always contain a one-loop realisation of the coupling $H\Delta^\dagger H$ or $LL\Delta$. One might think to bypass the $\Delta$ and introduce a quintuplet instead. However, the coupling of two doublets to a 5-plet is zero at any loop order due to ${\rm SU(2)_L}$.

Before moving on, a brief comment might be in order. Integrals for the diagrams in topologies $T_4$ and $T_{33}$ are finite. One might therefore wonder whether it is possible to forbid one of the ``ingredients'' of the tree-level $d=5$ seesaw, say one particular vertex, via a discrete symmetry, only to generate it at one-loop order. This was discussed at length in the classification of the one-loop realisations of the Weinberg operator in \cite{Bonnet:2012kz}. At the $d=7$ level, however, this will not be possible, since $H^{\dagger} H$ is a singlet under any discrete symmetry.

\begin{figure}
    \centering
    \includegraphics[width=0.5\textwidth]{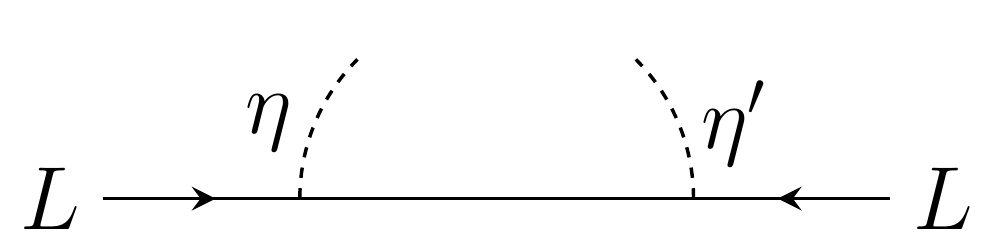}
    \caption{The particular piece of diagram that generates the one-loop $d=5$ diagram T-3, see \fig{fig:numass:loopd5}. If this structure exists in any diagram, the vertex with the two scalars $\eta$, $\eta'$ and two Higgses also always exist. This structure appears in many diagrams of the $d=7$ topologies.}
    \label{fig:dim7loop:piece_dim5-1l}
\end{figure}

Next we turn to identifying topologies which generate diagrams reducible to one-loop $d=5$ models. This is not as straightforward as the tree-level case. In particular, in this class of topologies many diagrams lead to $d=5$ tree-level models, while only the remaining diagrams can lead to $d=5$ one-loop models. However, when the topology is highly symmetric, as for example in $T_1$, one can always find a coupling between two internal fields and an external field which bypasses the $H^\dagger$, giving one of the diagrams of \fig{fig:numass:loopd5}. In addition, any diagram containing the structure given in \fig{fig:dim7loop:piece_dim5-1l} can be reduced to the well-known diagram T-3 in \fig{fig:numass:loopd5}. We list all topologies excluded due to these arguments in \fig{fig:app:topos:Topos_d=5_1-loop}.

Then, there are diagrams which always contain the fields $S = \textbf{4}^S_{3/2}$ and $\Psi = \textbf{3}^F_1$, responsible for generating neutrino mass at tree-level $d=7$ \cite{Babu:2009aq}. Examples are diagrams of the topologies $T_{25}$, $T_{29}$ and $T_{35}$ (\fig{fig:app:topos:Topologiesd=7tree}), which are excluded as genuine ones due to this reason.

\begin{figure}
    \centering
    \includegraphics[width=0.3\textwidth]{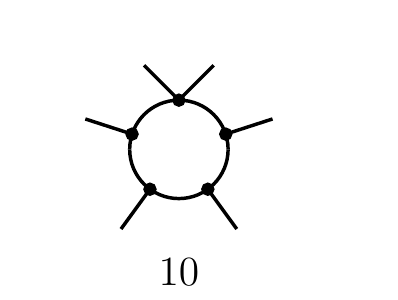}
    \hspace*{1.5cm}
    \includegraphics[width=0.3\textwidth]{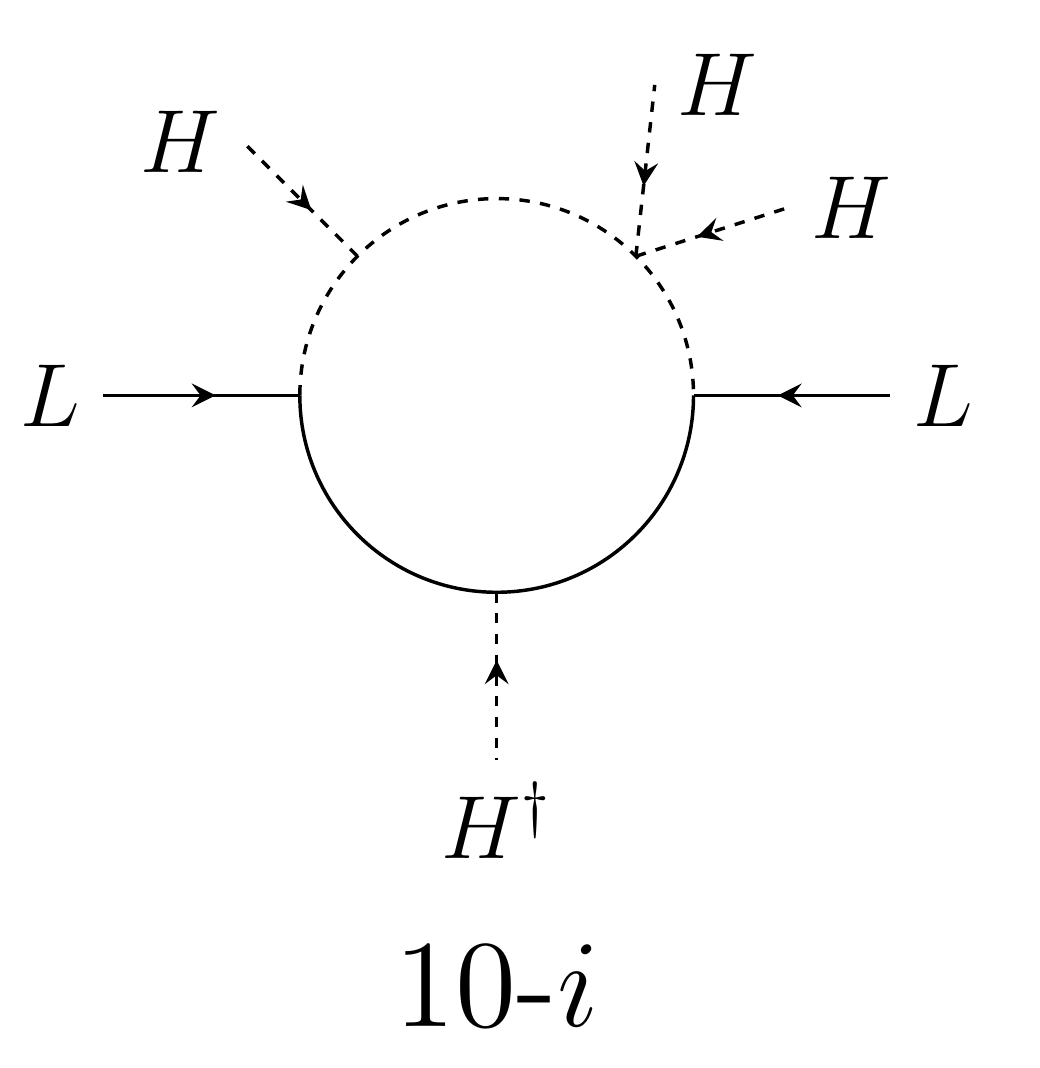}
    \caption{Topology $T_{10}$ and the only diagram derived from this topology, that can give a genuine model.}
    \label{fig:dim7loop:T10}
\end{figure}

Finally, in the remaining 8 topologies, depicted in \fig{fig:app:topos:TopoGenuine_d7}, that are not completely excluded by one of the above arguments, many but not all the diagrams lead to genuine models. For instance, from the 10 different diagrams that one can generate from topology $T_{10}$, only one is not reducible to one-loop $d=5$, shown in \fig{fig:dim7loop:T10}.

In summary, from the initial 48 topologies only 8 have, at least, one genuine one-loop $d=7$ diagram. The excluded topologies are listed in \app{app:topos}. The next step is to classify the surviving topologies in terms of the minimal ${\rm SU(2)_L}$ representation needed to realise a genuine model.

%%%%%%%%%%%%%%%%%%%%%%%%%%%%%%%%%%%%%%%%%%%%%%%%%%%%%%%%%%%%%%%
\subsection{Diagrams: Minimal ${\rm SU(2)_L}$ representations}
\label{subsec:dim7loop:diagrams}

In all $d=7$ diagrams, in order to avoid neutrino masses at lower order, a minimal size for the ${\rm SU(2)_L}$ representations of the model is required. We order the possible models according to the largest representation present in a given model. The ``smallest'' or minimal model that one can construct is then a model in which no representation larger than ${\rm SU(2)_L}$ triplets is needed. The next smallest possibility is models with representations up to quadruplets. Here, one can distinguish three different subgroups:
\begin{enumerate}[i.]
    \item Diagrams in which one quadruplet is needed inside the loop.
    \item Diagrams in which one quadruplet appears outside the loop and internal particles need not be larger than triplets.
    \item Models in which at least two quadruplets are needed.
\end{enumerate}
We will discuss these three possibilities in reverse order and then proceed to briefly discuss the triplet diagram. The complete list of diagrams is given in \app{app:topos}.

\begin{figure}[h]
    \centering
    \includegraphics[width=0.35\textwidth]{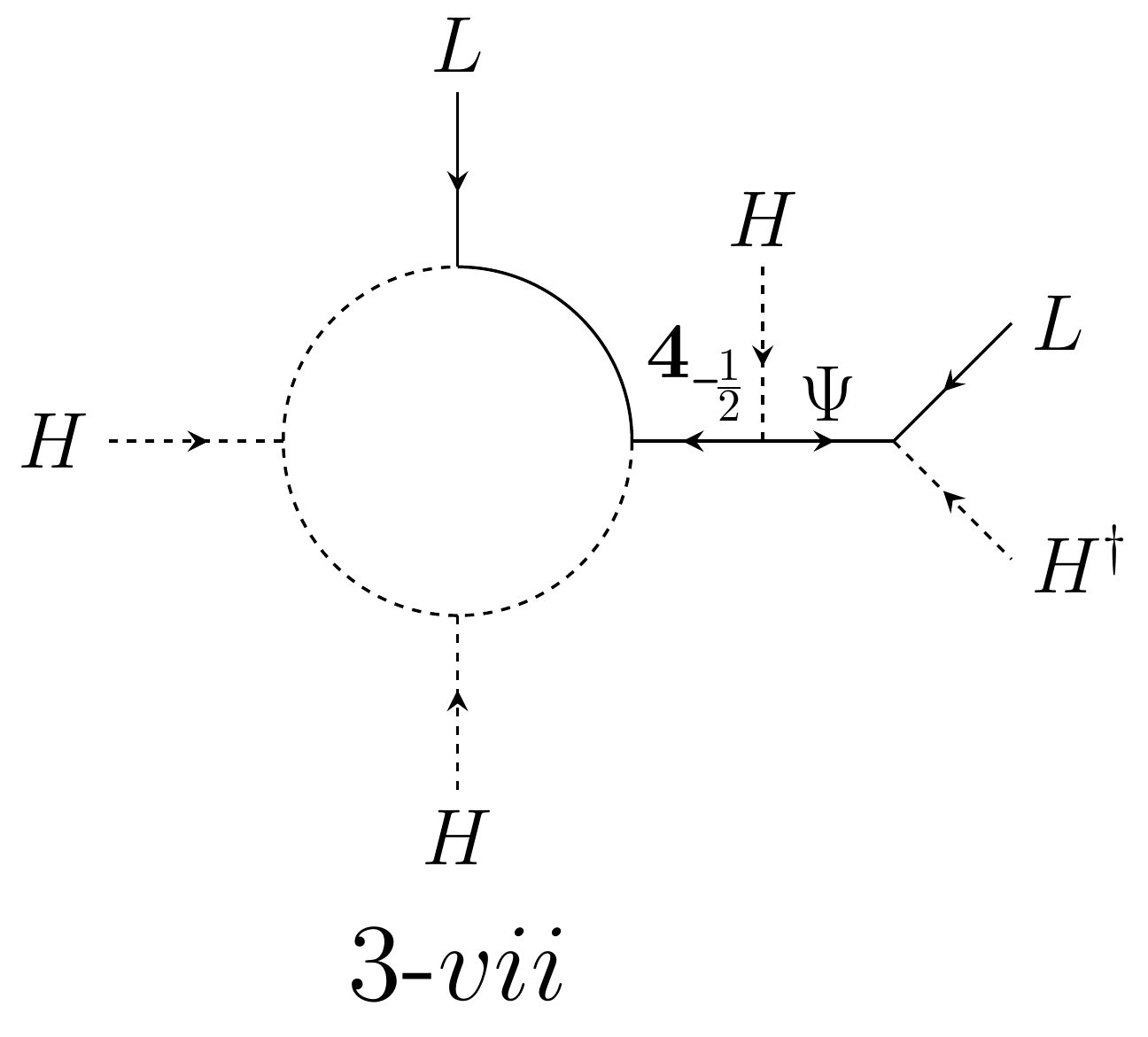}
    \hspace*{1.5cm}
    \includegraphics[width=0.33\textwidth]{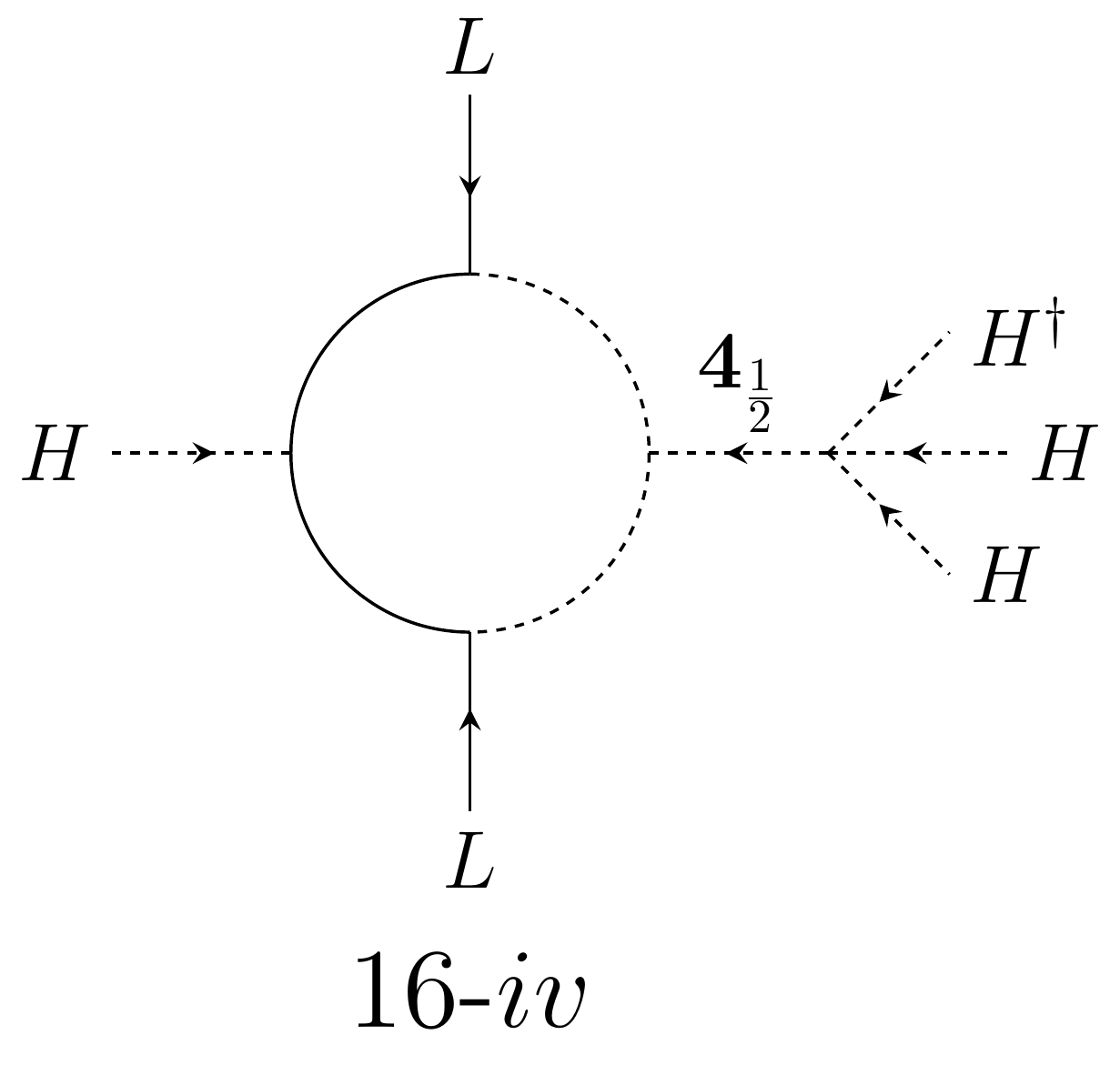}
    \caption{Examples of diagrams which need at least two quadruplets to be genuine, i.e. to not generate a $d=5$ one-loop contribution. Along with the external fermion or scalar quadruplet, a genuine models needs an internal quadruplet to distinguish between a $\textbf{4}^S_{1/2}$ ($\textbf{4}^F_{-1/2}$) and a Higgs ($L$). $\Psi$ denotes the triplet $\textbf{3}^F_1$.}
    \label{fig:dim7loop:Diags4pletsInOut}
\end{figure}

For external fields, finding the minimal representation is straightforward. A recurrent example in most of the diagrams is that of the vertex $HH^\dagger$-scalar. Since $2 \otimes 2 = 3 + 1$, the scalar could be a trivial singlet or the triplet $\phi \equiv {\bf 3}^S_0$. The former case is directly reducible to a $d=5$ diagram, the latter is the one we are interested in. The same principle applies to the diagrams given in \fig{fig:dim7loop:Diags4pletsInOut}, case (iii), for which the largest necessary representation is a quadruplet. In order to avoid lower order contributions, one needs the quadruplet $\textbf{4}^S_{1/2}$ ($\textbf{4}^F_{-1/2}$) outside the loop. Moreover, one should be able to distinguish between these quadruplets and a Higgs or a lepton doublet. For this reason, the external quadruplet must couple, at least, to a singlet and another quadruplet running inside the loop. All the diagrams of this type are depicted in \fig{fig:app:topos:Diags4pletsInOut}, and they always contain two quadruplets, one outside the loop and another inside.

\begin{figure}[h]
    \centering
    \includegraphics[width=0.35\textwidth]{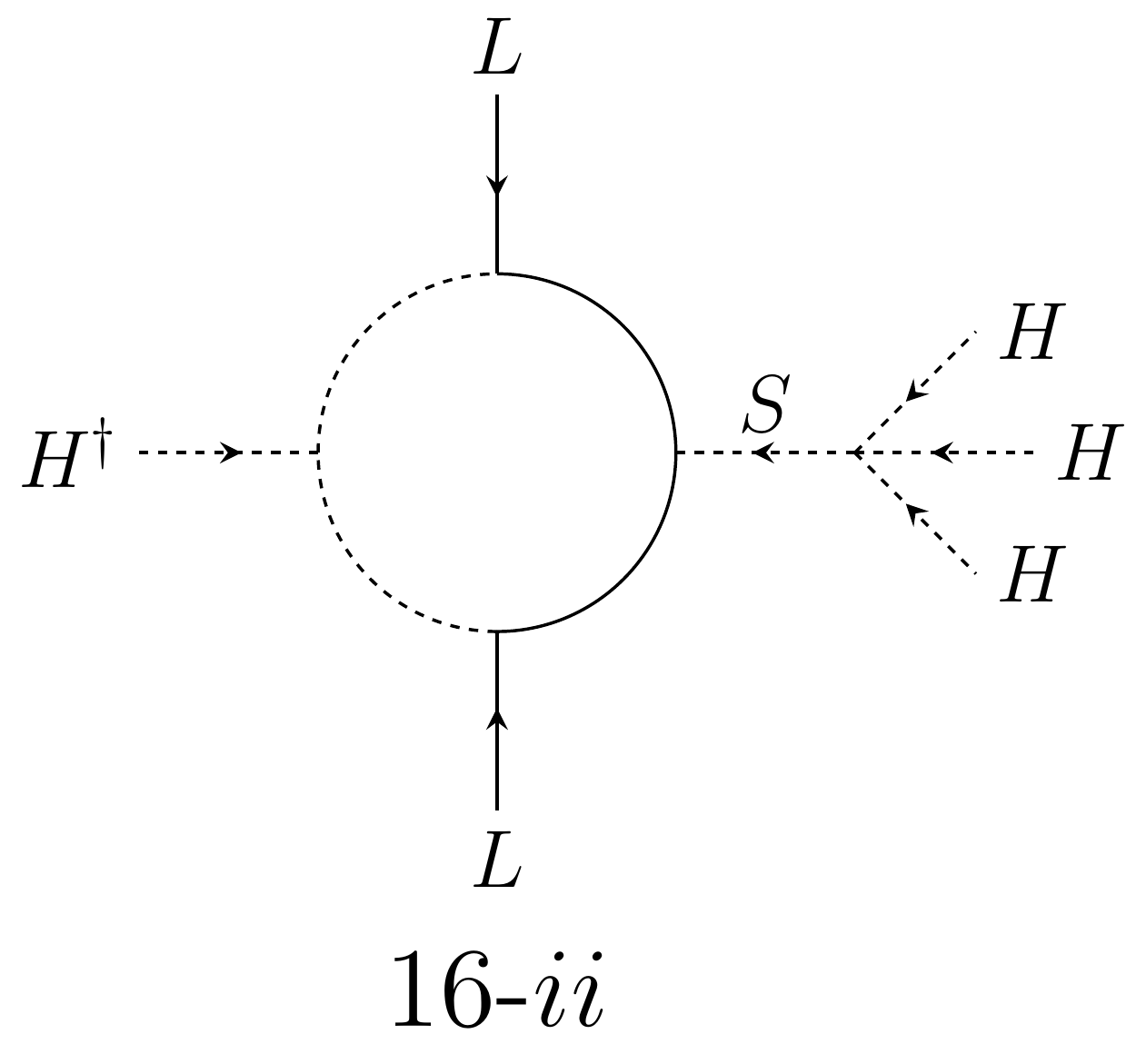}
    \hspace*{1.5cm}
    \includegraphics[width=0.315\textwidth]{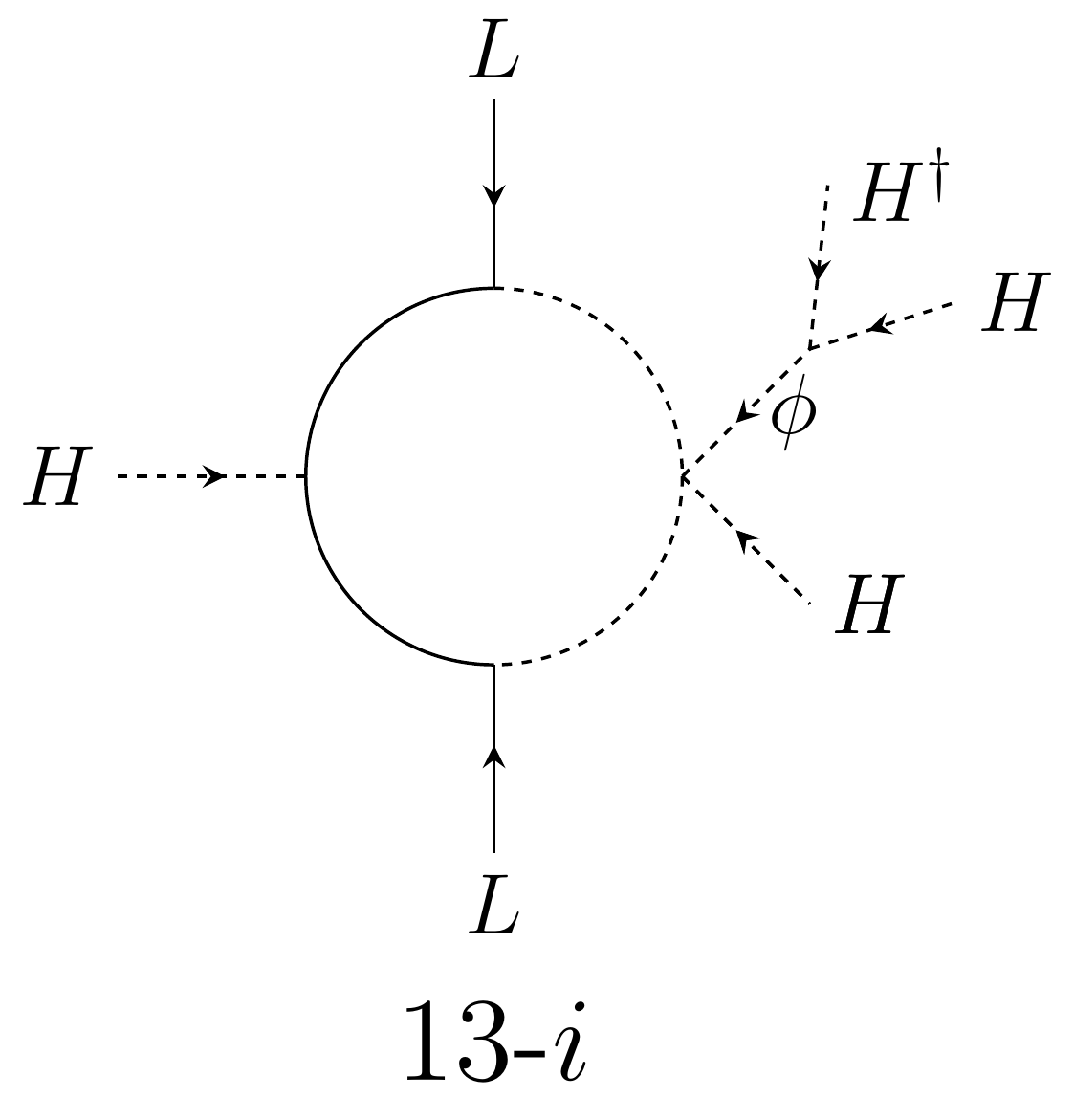}
    \caption{Examples of diagrams which need at least one quadruplets to be genuine. Diagram to the left has the quadruplet ($S=\textbf{4}^S_{3/2}$) outside the loop. Diagram to the right, with $\phi = \textbf{3}^S_0$, needs a quadruplet running in the loop. See text for discussion.}
    \label{fig:dim7loop:Diags4plets}
\end{figure}

As we are dealing with the operator $LLHH(H^\dagger H)$, the maximum hypercharges of an external quadruplet is $3/2$, i.e. the scalar $S$ present in the tree-level $d=7$ neutrino mass diagram (see \sect{subsec:numass:maj_tree}). These diagrams correspond to the class (ii) defined above. For example, the diagrams given in \fig{fig:dim7loop:Diags4plets} to the left. All the diagrams in this class, depicted in \fig{fig:app:topos:Diags4pletsS} contains the scalar $S$ entering the loop, and they belong to the same topology $T_{16}$. Note that the hypercharge $3/2$ of $S$ and the $H^\dagger$ entering the loop, prevents the possibility to reduce these models to $d=5$ one-loop.

\begin{figure}[h]
    \centering
    \includegraphics[width=0.35\textwidth]{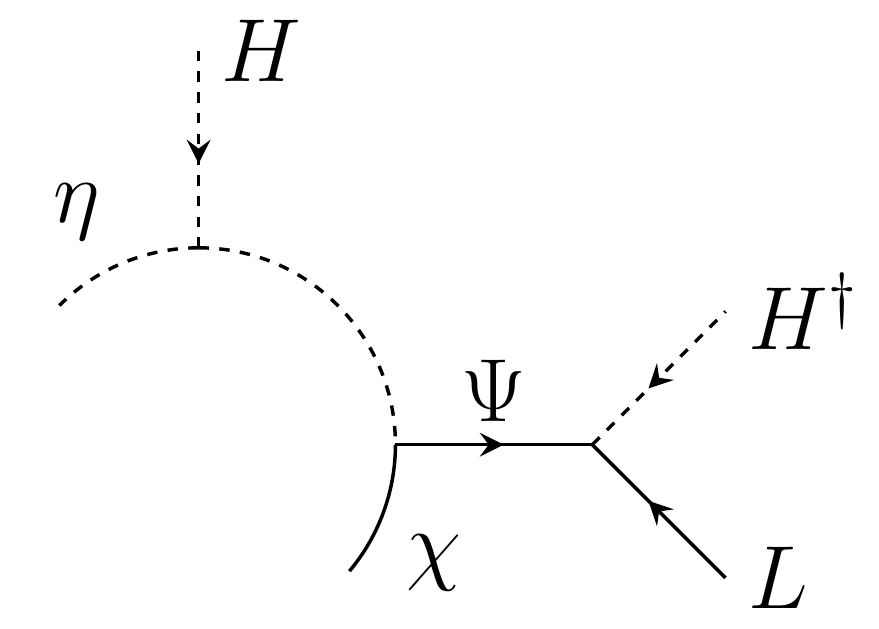}
    \hspace*{1.5cm}
    \includegraphics[width=0.23\textwidth]{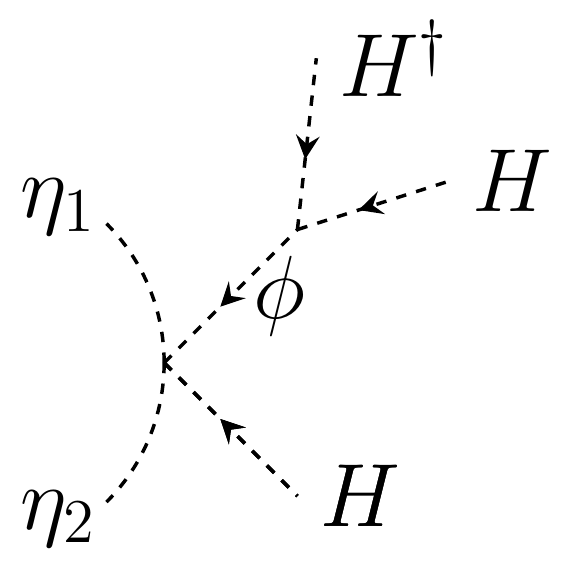}
    \caption{Structures that appear in topologies $T_{12}$ (left) and $T_{13}$ (right) which require at least a quadruplet and a singlet running inside the loop to avoid lower order contributions.}
    \label{fig:dim7loop:piece_4plet}
\end{figure}

The rest of the diagrams do not contain a external quadruplet. In the minimal case, they just need one quadruplet running in the loop. We show an example of this class in \fig{fig:dim7loop:Diags4plets} to the right. The complete list is given in \fig{fig:app:topos:Diags4pletsIn}. Diagrams generated from the topologies $T_2$, $T_{12}$ and $T_{13}$ with a triplet entering the loop have all similar structures to those in \fig{fig:dim7loop:piece_4plet}. The minimum representations for the fields ($\chi$, $\eta$) or ($\eta_1$, $\eta_2$) in \fig{fig:dim7loop:piece_4plet}, are then a singlet and a quadruplet, in order to prevent a coupling of these fields with a lepton doublet $L$ or the Higgs $H$, respectively.

The remaining diagram $D_{10}^{(i)}$ of \fig{fig:app:topos:Diags4pletsIn} is a rather singular case of this group (i), with only one internal quadruplet. Given the isolated $H^\dagger$ and the upper asymmetric structure with three Higgses, the representation between the Higgs vertices needs to be (at least) a quadruplet, otherwise a $d=5$ one-loop contribution is possible.

\begin{figure}
    \centering 
    \includegraphics{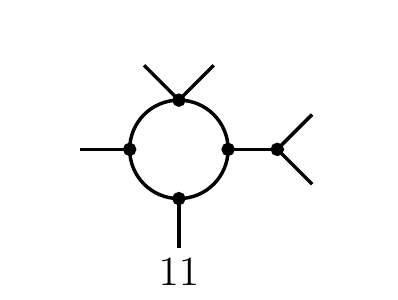}
    \qquad \qquad
    \includegraphics[scale=0.35]{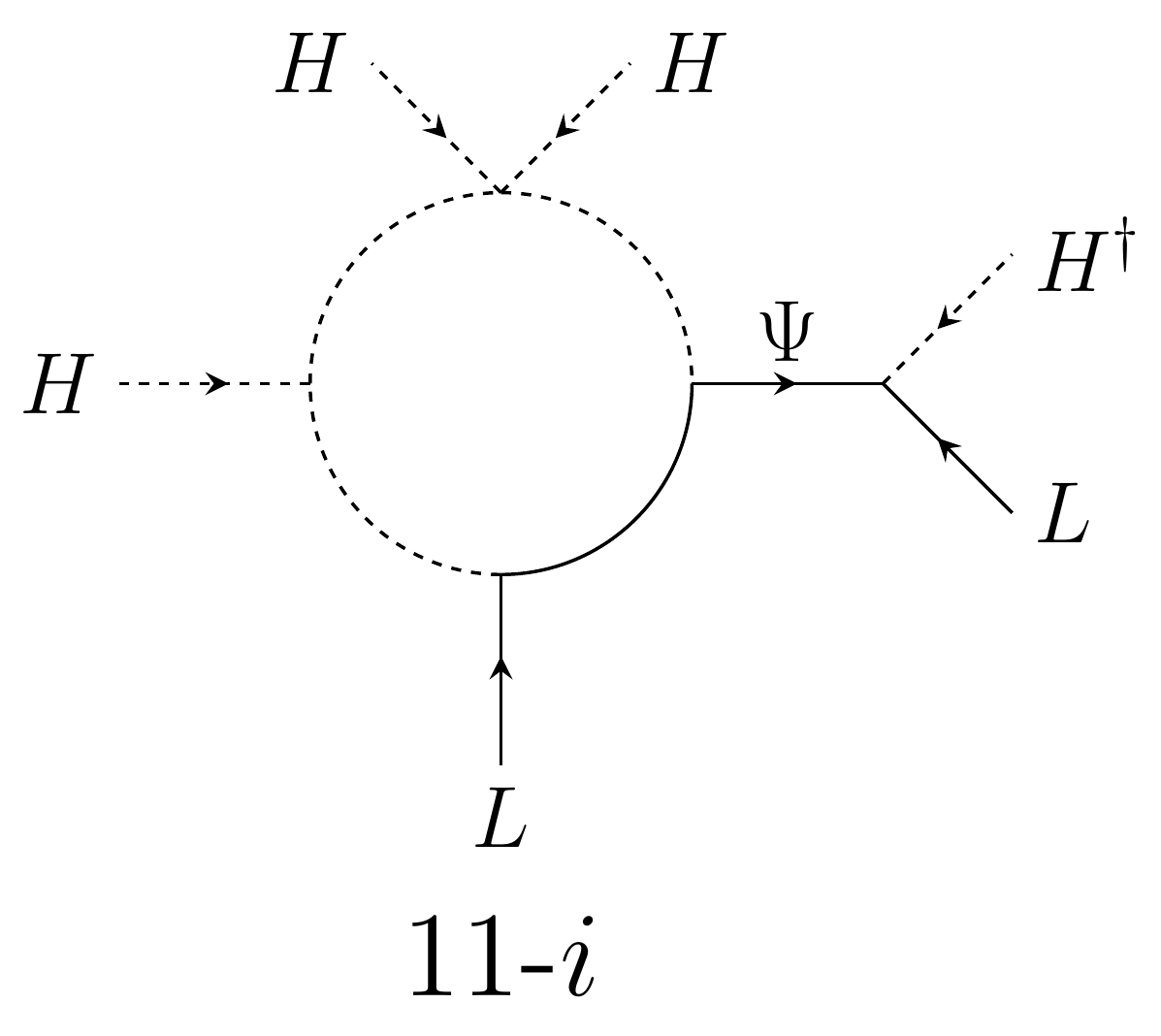}
    \caption{Topology $T_{11}$ to the left: The only topology for which a genuine model with no representations larger than triplets exist. The only genuine diagram for this topology is shown on the right.}
    \label{fig:dim7loop:Triplet}
\end{figure}

After giving all the diagrams that can be constructed with quadruplets as the highest representation (\figs{fig:app:topos:Diags4pletsIn}{fig:app:topos:Diags4pletsInOut}), the last case depicted in \fig{fig:dim7loop:Triplet} shows the only genuine diagram that can be constructed with no representation bigger than triplet. Despite its similarity to the structure given in \fig{fig:dim7loop:piece_4plet} (left), the 4-leg vertex prevents lower order neutrino masses already with triplets. The corresponding diagram \fig{fig:dim7loop:Triplet} (right) with a 4-legs vertex followed by a triplet $\Psi$ cannot be bridged to construct a one-loop $d=5$ contribution given the relation between the hypercharges of the fields running inside the loop.

To summarise, from the 8 genuine topologies, one can generate 23 diagrams. Among them, only one can be realised with no representation larger than triplets as the maximum $SU(2)_L$ representation, \fig{fig:dim7loop:Triplet}. The other 22 diagrams generated from the 7 topologies given in \fig{fig:app:topos:TopoGenuine_d7} can generate models with representations up to quadruplets. This whole set can be divided depending if the diagrams require one quadruplet running in the loop (\fig{fig:app:topos:Diags4pletsIn}), outside the loop (\fig{fig:app:topos:Diags4pletsS}) or two quadruplets both inside and outside the loop (\fig{fig:app:topos:Diags4pletsInOut}). Of course, models with larger representations can be constructed, and we will give one example in the next section.

%%%%%%%%%%%%%%%%%%%%%%%%%%%%%%%%%%%%%%%%%%%%%%%%%%%%%%%%%%%%%%%
%%%%%%%%%%%%%%%%%%%%%%%%%%%%%%%%%%%%%%%%%%%%%%%%%%%%%%%%%%%%%%%
\section{Example models}
\label{sec:dim7loop:models}

The complete list of diagrams, from which genuine $d=7$ one-loop models can be built is given in \app{app:topos}. Here, we will briefly discuss three example model-diagrams, which are among the simplest one can built from these diagrams. These models are: (i) The simplest $d=7$ model, which requires no representation larger than a triplet; (ii) one example model with an external quadruplet $S$; and (iii) an example model with an ${\rm SU(2)_L}$ quintuplet. The latter serves to show how models with larger representations can easily be constructed from our list of diagrams. The phenomenology of the first two example models will be studied in \ch{ch:Dim7_pheno}.

%%%%%%%%%%%%%%%%%%%%%%%%%%%%%%%%%%%%%%%%%%%%%%%%%%%%%%%%%%%%%%%
\subsection*{Triplet model}

\begin{figure}
    \centering
    \includegraphics[width=0.5\textwidth]{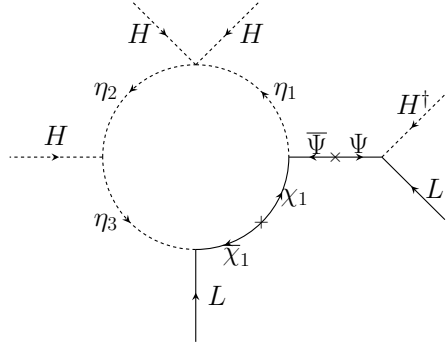}
    \caption{The most minimal model that one can construct at one-loop $d=7$ order with no ${\rm SU(2)_L}$ representations larger than triplet. The model is generated from the diagram $D_{11}^{(i)}$ in \fig{fig:dim7loop:Triplet}.}
    \label{fig:dim7loop:3plet_model}
\end{figure}

As discussed above, there is only one possible diagram that has a triplet as the largest ${\rm SU(2)_L}$ representation, see \fig{fig:dim7loop:Triplet}. The model requires the fermionic triplet $\Psi=\textbf{3}^F_1$, that also appears in the BNT model \cite{Babu:2009aq}. A priori, for the particles inside the loop hypercharge is not fixed. However, not all choices of hypercharge will lead to genuine models, since lower order contributions might appear. If we use only doublets and triplets inside the loop, the smallest hypercharge assignments that lead to a genuine model are,
\begin{center}
\begin{tabular}{ c c c }

$\Psi=\left(
\begin{matrix}
\Psi^{++} \\ 
\Psi^{+} \\
\Psi^{0}
\end{matrix} 
\right) \sim \textbf{3}_1^F$

&\qquad

$\eta_1=\left(
\begin{matrix}
\eta_1^{++} \\ 
\eta_1^{+}
\end{matrix} 
\right) \sim \textbf{2}_{3/2}^S$

&\qquad

$\eta_2=\left(
\begin{matrix}
\eta_2^{+++} \\ 
\eta_2^{++}
\end{matrix} 
\right) \sim \textbf{2}_{5/2}^S$

\end{tabular}

\vspace*{0.5cm}

\begin{tabular}{ c c }

$\eta_3=\left(
\begin{matrix}
\eta_3^{++++} \\ 
\eta_3^{+++} \\
\eta_3^{++}
\end{matrix} 
\right) \sim \textbf{3}_3^S$

&\qquad

$\chi_1=\left(
\begin{matrix}
\chi_1^{+++} \\ 
\chi_1^{++}
\end{matrix} 
\right) \sim \textbf{2}_{5/2}^F$.

\end{tabular}
\end{center}
The model generates neutrino masses via the diagram depicted in \fig{fig:dim7loop:3plet_model}. $\eta_1$ has the smallest hypercharge of the particles in the loop. For colourless particles it is not possible to find a smaller hypercharge assignment that leads to a genuine model. For example, choosing $\eta_1 = {\bf 2}_{1/2}^S$ instead would result in a model for which diagram T-3 at $d=5$ level is the dominant contribution to neutrino masses.

One interesting aspect of this model is that the scalar triplet inside the loop has a component which carries 4 units of electric charge. This would lead to a very clean signal in colliders that will be analysed in \ch{ch:Dim7_pheno}. Note that any other particle content that leads to a genuine model would have, at least, one field with 4 or more units of electric charge.

%%%%%%%%%%%%%%%%%%%%%%%%%%%%%%%%%%%%%%%%%%%%%%%%%%%%%%%%%%%%%%%
\subsection*{Quadruplet model}

While for triplets as the maximal representation there is only one diagram, for quadruplets three distinct groups of model exist, as discussed above. We choose an example based on an external quadruplet $S={\bf 4}^S_{3/2}$. The example model we choose is based on diagram $D_{16}^{(ii)}$.

As in the triplet case, hypercharge and ${\rm SU(2)_L}$ representation are not uniquely fixed. The minimal model, again in the sense of using the smallest possible hypercharge assignment for colourless internal fields, has the following particle content,
\begin{center}
\begin{tabular}{ c c c }

$S=\left(
\begin{matrix}
S^{+++} \\
S^{++} \\ 
S^{+} \\
S^{0}
\end{matrix} 
\right) \sim \textbf{4}_{3/2}^S$

\quad & \quad

$\chi_1=\left(
\begin{matrix}
\chi_1^{++} \\ 
\chi_1^{+}
\end{matrix} 
\right) \sim \textbf{2}_{3/2}^F$

\quad & \quad

$\chi_2=\left(
\begin{matrix}
\chi_2^{++++} \\ 
\chi_2^{+++} \\ 
\chi_2^{++}
\end{matrix} 
\right) \sim \textbf{3}_3^F$

\end{tabular}
\begin{tabular}{ c c }

$\eta_1=\eta_1^{++} \sim \textbf{1}_2^S$.

\qquad & \qquad

$\eta_2=\left(
\begin{matrix}
\eta_2^{+++} \\ 
\eta_2^{++}
\end{matrix} 
\right) \sim \textbf{2}_{5/2}^S$.

\end{tabular}
\end{center}

\begin{figure}
\centering
\includegraphics[width=0.5\textwidth]{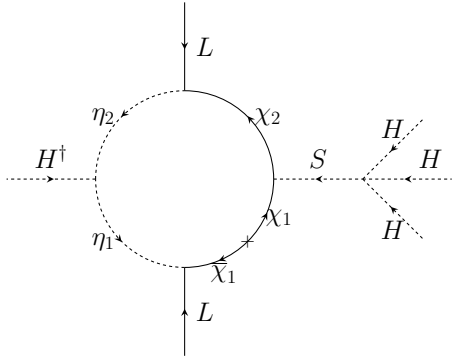}
\caption{Example of a $d=7$ one-loop model with an external quadruplet $S={\bf 4}^S_{3/2}$, generated from the diagram $D_{16}^{(ii)}$ in \fig{fig:dim7loop:Diags4plets}. This model contains only doublet and triplet representations inside the loop, see text.}
\label{fig:dim7loop:4plet_model}
\end{figure}

This model generates neutrino masses via the diagram of \fig{fig:dim7loop:4plet_model}. In this example, lower order contributions can be avoided due to the hypercharge of $S$. A model with smaller hypercharges, for example $\eta_1$ chosen to be $\eta_1={\bf 1}_1^S$, would again not be genuine, since it would necessarily have a $d=5$ one-loop contribution, i.e. the well-known Zee model \cite{Zee:1980ai}. Note that, while the triplet model contains a scalar with 4 units of electric charge, in the quadruplet model it is an internal fermion that has such a large electric charge.

%%%%%%%%%%%%%%%%%%%%%%%%%%%%%%%%%%%%%%%%%%%%%%%%%%%%%%%%%%%%%%%
\subsection*{Quintuplet model}

\begin{figure}
    \centering
	\includegraphics[width=0.5\textwidth]{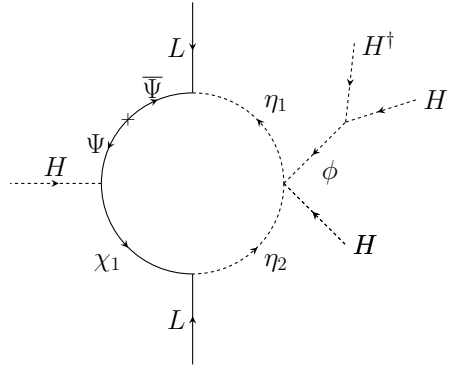}
    \caption{Example of one of the most minimal models that one can construct at $d=7$ one-loop order with ${\rm SU(2)_L}$ representations up to 5-plets generated from the diagram $D_{13}^{(i)}$ in \fig{fig:dim7loop:Diags4plets}.}
\label{fig:dim7loop:5plet_model}
\end{figure}

Finally, our last example is a model based on diagram $D_{13}^{(i)}$ in \fig{fig:dim7loop:Diags4plets}. It contains the field $\phi={\bf 3}_0^S$ and a quintuplet in the loop. The diagram for the generation of the neutrino masses is shown in \fig{fig:dim7loop:5plet_model}. The minimal particle content having a 5-plet is given by,

\begin{center}
\begin{tabular}{ c c c }

$\phi=\left(
\begin{matrix}
\phi^{+} 	\\
\phi^{0}	\\
\phi^{-}
\end{matrix} 
\right) \sim \textbf{3}_0^F$

&\qquad

$\Psi=\left(
\begin{matrix}
\Psi^{++} \\ 
\Psi^{+} \\
\Psi^{0}
\end{matrix} 
\right) \sim \textbf{3}_1^F$

&\qquad

$\chi_1=\left(
\begin{matrix}
\chi_1^{+++} \\ 
\chi_1^{++} \\
\chi_1^{+}  \\
\chi_1^{0}
\end{matrix} 
\right) \sim \textbf{4}_{3/2}^F$
\end{tabular}

\vspace*{0.5cm}

\begin{tabular}{ c c }

$\eta_1=\left(
\begin{matrix}
\eta_1^{++} \\ 
\eta_1^{+} 
\end{matrix} 
\right) \sim \textbf{2}_{3/2}^S$

&\qquad

$\eta_2=\left(
\begin{matrix}
\eta_2^{+++} \\ 
\eta_2^{++} \\
\eta_2^{+}  \\
\eta_2^{0}  \\
\eta_2^{-}  
\end{matrix} 
\right) \sim \textbf{5}_1^S$.

\end{tabular}
\end{center}

The maximum representation in this model is a quintuplet, $\eta_2$. Since it couples to the Higgs, $\phi$ and $\eta_1$, $\eta_1$ could be either a ${\bf 2}$, ${\bf 4}$, ${\bf 6}$ or a ${\bf 8}$. However, only for the case of $\eta_1$ being a doublet, a genuine model results. This is because, once the coupling $\eta_1 H \eta_2^\dagger$ is allowed, one can construct again a $d=5$ one-loop diagram with the particle content of the model.

It is worth noting that from three example models we have discussed, the quintuplet model is the only one, in which the representations in the loop contain a neutral component. For this model, one can thus follow the idea of the scotogenic model \cite{Ma:2006km}, i.e. add a discrete $Z_2$ symmetry to the model, under which the internal particles are odd, and the lightest neutral particle can be a cold dark matter candidate.

%%%%%%%%%%%%%%%%%%%%%%%%%%%%%%%%%%%%%%%%%%%%%%%%%%%%%%%%%%%%%%%
%%%%%%%%%%%%%%%%%%%%%%%%%%%%%%%%%%%%%%%%%%%%%%%%%%%%%%%%%%%%%%%
\section{Summary}
\label{sec:dim7loop:sum}

We have discussed neutrino masses at one-loop $d=7$ order. We have identified all possible topologies that can lead to genuine models, i.e. models that are not accompanied by either a $d=5$ or $d=7$ tree-level mass term nor by a $d=5$ one-loop neutrino mass. We have found that only 8 out of a total of 48 topologies can lead to genuine models.

We then ordered the remaining, possibly genuine, diagrams into different groups, depending on the minimal field content necessary to construct a model. There is only one possible diagram for which the largest necessary representation is a triplet. The remaining 7 topologies yield 22 diagrams, with the largest representation being at least a quadruplet. We then briefly discussed three example models, starting from the triplet model, with one additional example for a quadruplet and one for a quintuplet each. 

To avoid lower order neutrino masses, the ``genuine'' models we discussed always have to introduce five new multiplets, usually with quite a large hypercharge for at least one of them. Thus, these $d=7$ models are necessarily more complicated constructions than the classical seesaw. From a theoretical point of view this might make these models less attractive. However, in particular due to the large electrical charges in these models, one can expect interesting signatures for them at colliders. We reiterate that the $d=7$ one-loop contribution can only be dominant, if at least some of the new particles have masses below roughly 2 TeV. This implies an upper bound that can be easily tested in colliders given the low background expected for these kinds of signals.

\pagebreak
\fancyhf{}

%% file: Chapters/Dim5_3loop/Chapter_3loop.tex
\fancyhf{}
\fancyhead[LE,RO]{\thepage}
\fancyhead[RE]{\slshape{Chapter \thechapter. Classification of three-loop realisations}}
\fancyhead[LO]{\slshape\nouppercase{\rightmark}}

\chapter{Systematic classification of three-loop realisations of the Weinberg operator}
\label{ch:3loop}
\graphicspath{ {Chapters/Dim5_3loop/} }

In this chapter, based on \cite{Cepedello:2018rfh}, we study systematically the decomposition of the Weinberg operator at three-loop order. There are more than four thousand connected topologies. However, the vast majority of these are infinite corrections to lower order neutrino mass diagrams and only a very small percentage yields models for which the three-loop diagrams are the leading order contribution to the neutrino mass matrix, i.e. genuine, with scalars and fermions. Further subclasses can be found among the latter, as certain topologies can lead to genuine diagrams only for very specific choices of fields. This special genuine diagrams appears at the two-loop level and were not considered in previous classifications \cite{Cepedello:2019zqf}. 
\\

The chapter is organised as follows. In the first section, we explain our classification scheme and how our results are obtained. All topologies which we classify as \textit{genuine} have finite three-loop integrals and thus do not need lower order counter terms for renormalisation. We discuss that there is a further class of genuine topologies with finite three-loop integrals, which correspond to the loop generation of some 3- or 4-point vertices. We call these the \textit{special genuine} topologies. We then show in \sect{sec:3loop:diag} that the 73 genuine topologies are associated to 374 diagrams in the weak basis, which get reduced to only 30 diagrams in the mass basis. We will discuss the dichotomy between normal and special topologies/diagrams in detail. We would like to point out that we are mostly interested in diagrams with new scalars (and/or new fermions). However, there could also be diagrams with vector particles, either the Standard Model W-boson or some exotic vector. While many of our results apply also to diagrams with vectors, we stress that due to a particular loophole in our procedure for finding genuine diagrams our list of genuine diagrams is incomplete for vectors. This is discussed in \sect{sec:3loop:topos} in more detail. \Sect{sec:3loop:gen_loops} contains a discussion of the relation between incompressible loops and genuineness of a diagram. It is clear that genuine diagrams have incompressible loops, but, even if expected to be true, it is not straightforward to claim that incompressibility implies genuineness. In \sect{sec:3loop:examples} we show some example models, discussing them briefly. For two of them we perform also numerical calculations of the expected neutrino mass scale; they can easily reproduce the observed neutrino masses. 

We shall only give some topologies and diagrams in this chapter. The complete list of genuine topologies is given in \app{app:topos}, while the full list of topologies and diagrams, due to its extension, can be found in \cite{extraData}. For details about the reduction and calculation of the loop integrals, we refer to \app{app:loops}.

%%%%%%%%%%%%%%%%%%%%%%%%%%%%%%%%%%%%%%%%%%%%%%%%%%%%%%%%%%%%%%%
%%%%%%%%%%%%%%%%%%%%%%%%%%%%%%%%%%%%%%%%%%%%%%%%%%%%%%%%%%%%%%%
\section{Generating and classifying topologies}
\label{sec:3loop:topos}

We found all topologies and diagrams after a series of steps. First, using known algorithms from graph theory (see for example \cite{Read:1981}) we generate a list of all three-loop connected topologies with 3- and 4-point vertices, and four external lines. This list contains a total of 4367 topologies. Only 3269 of them can accommodate 2 external fermion lines plus 2 external scalars using renormalisable interactions only.

Still, at the level of topologies, we can already exclude a large number of these by applying the following straightforward criteria. We eliminate all cases with tadpoles (i.e., self-connecting vertices) and self-energies (i.e., 2-point subdiagrams with one or more loops), since these have always infinite parts in their loop integrals. This cut leaves us with 370 topologies.

\begin{figure}
    \centering
    \includegraphics[scale=0.90]{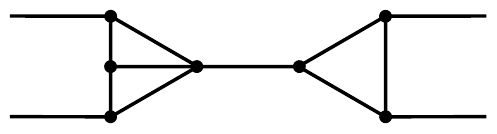}
    \caption{A 1-particle reducible three-loop topology. The line in the middle must correspond to one of the standard tree-level seesaw mediators, hence the topology is not genuine. The middle line is a fermion with the quantum numbers of $\nu_R\equiv(\boldsymbol{1},\boldsymbol{1},0)$ or $\Sigma\equiv(\boldsymbol{1},\boldsymbol{3},0)$ if it splits the external fields as $LH|LH$, or it is the scalar $\Delta=(\boldsymbol{1},\boldsymbol{3},-1)$ if the splitting is $LL|HH$.}
    \label{fig:3loop:discon}
\end{figure}

We then eliminate the 1-particle reducible topologies, that is those topologies which become disconnected by cutting one of its lines. These cases can be discarded because the line which would disconnect the topology must have the quantum numbers to mediate type-I, type-II or type-III seesaws (see \fig{fig:3loop:discon}).

This leaves us with 160 potentially genuine topologies, which can be further divided in three classes: \textit{normal genuine} topologies, \textit{special genuine} topologies and \textit{non-genuine} topologies.

%%%%%%%%%%%%%%%%%%%%%%%%%%%%%%%%%%%%%%%%%%%%%%%%%%%%%%%%%%%%%%%
\subsection{Normal genuine topologies}
\label{sec:3loop:normgen}

\begin{figure}
    \centering
    \includegraphics[scale=0.90]{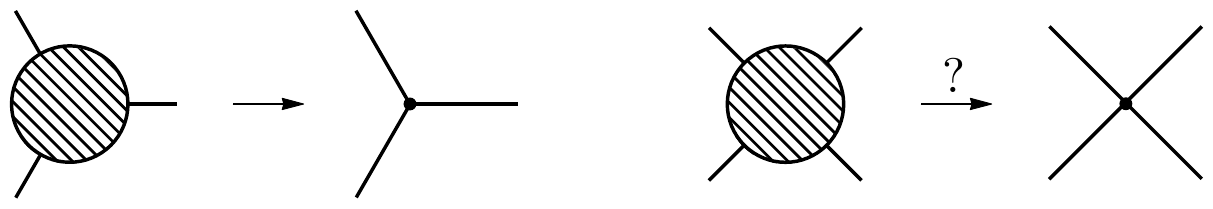}
    \caption{Subparts of diagrams with loops and three external lines can be compressed into a 3-line vertex, reducing the number of loops of the diagram. For subparts with four external lines, this will only generate a renormalisable interaction if all external lines are scalars.}
    \label{fig:3loop:loopcontraction}
\end{figure}

Consider first 3-point vertices. No matter what is the particle content of a model, if a loop with 3 external legs is allowed by symmetry, so is the trilinear vertex without the loop (see \fig{fig:3loop:loopcontraction}). Since this reasoning applies equally to fermion-fermion-scalar and to scalar-scalar-scalar vertices, this criterion can be defined at the level of topologies. For loops with four external legs, on the other hand, it is only possible to compress it to a renormalisable vertex if all external lines are scalars. Thus, this criterion needs to be used on diagrams, not topologies. The important point is that if a diagram has compressible subdiagrams (with 3 or 4 external legs), it cannot be genuine. We note that there is also the expectation of the converse: diagrams with incompressible loops are genuine, as there will be a choice of quantum numbers for the internal scalars and fermions such that there will be no other neutrino mass contribution with less loops.\footnote{In \sect{sec:3loop:gen_loops} we present an argument why we believe this is always true.}

We identify those topologies for which an internal loop (or loops) can be compressed to a 3-point vertex (of the type fermion-fermion-scalar or scalar-scalar-scalar). For the remaining topologies, we find all valid diagrams, labelling internal lines as scalars or fermions in all possible ways, keeping externally exactly two scalars and two fermions, and forbidding non-renormalisable vertices. In this list we identify all diagrams with internal loops which can be compressed to a 4-scalar interaction. All diagrams without 3-point nor scalar 4-point loop subdiagrams fall into one of 44 topologies. These we consider \textit{normal genuine} diagrams and topologies. Their complete list of topologies is given in \app{app:topos}, while the lists of diagrams are shown in \cite{extraData} due to their extension. A summary of the counting and the cuts applied can be found on \tab{tab:3loop:normalGenuineCounting}.

\begin{table}
    \centering
    \begin{tabular}{cc}
        \toprule 
        All topologies (connected, with 3 loops and 4 legs) & 4367\tabularnewline
        Allow two external fermion lines & 3269\tabularnewline
        No tadpoles & 1056\tabularnewline
        No self-energies & 370\tabularnewline
        1-particle irreducible & 160\tabularnewline
        No 3-point loop subgraphs & 70\tabularnewline
        No unavoidable 4-point scalar loop subgraphs & 44\tabularnewline
        \bottomrule
    \end{tabular}
    \caption{Number of topologies, after the cumulative application of a series of cuts described in the text. Out of the initial 4367, only 44 have all the properties listed above: these are the \textit{normal genuine topologies}. We would like to point out that dropping the requirement that a topology does no become disconnected by the cut of a single internal line (1-particle irreducibility), the final number of topologies would still be 44.}
    \label{tab:3loop:normalGenuineCounting}
\end{table}

%%%%%%%%%%%%%%%%%%%%%%%%%%%%%%%%%%%%%%%%%%%%%%%%%%%%%%%%%%%%%%%
\subsection{Special genuine topologies}
\label{sec:3loop:specgen}

\begin{figure}
    \centering
    \includegraphics[scale=0.90]{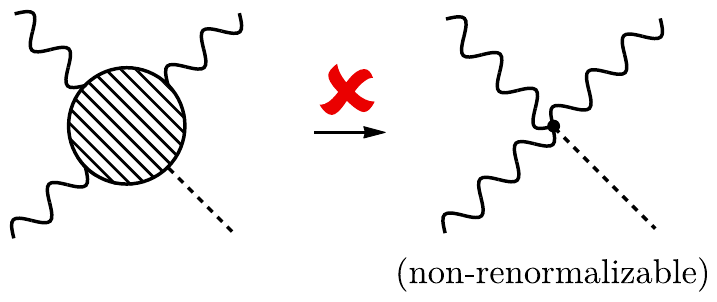}
    \caption{Our procedure to obtain genuine three-loop neutrino mass diagrams and topologies may not be valid in the presence of vector fields. In this example, a one-loop fragment of a larger diagram cannot be contracted into a renormalisable point interaction, see text.}
    \label{fig:3loop:exception1}
\end{figure}

The usage of the word ``normal'' in this context is explained by the existence of exceptions to the above arguments. First of all, strictly speaking, the cut on 4-point vertices (see the last row in \tab{tab:3loop:normalGenuineCounting}) is only valid for diagrams without vector fields. Consider the diagram shown in \fig{fig:3loop:exception1}: It has three external vector fields ($V$) and one scalar ($S$). Yet a term $VVVS$ is not Lorentz invariant, hence such a loop cannot be compressed into a renormalisable interaction (the effective interaction is $\partial VVVS$, dimension 5). As a consequence of this, some otherwise non-genuine topologies might be classified as genuine if vector fields are used. We are mentioning this exception only for completeness, since we are interested in diagrams with fermions and scalars only.

However, there are two subtle exceptions to the procedure used to obtain the previously mentioned 44 genuine diagrams with fermions and scalars only. The strategy discussed earlier admits a loophole which was overlooked in previous classifications, enlarging the set of \textit{genuine} topologies. We found that some topologies can generate neutrino mass diagrams which under normal circumstances can be redrawn with less loops, unless some particular quantum numbers are assigned to some of the particles in the internal lines. More specifically, these special diagrams contain fermion-fermion-scalar, $\left(\textrm{scalar}\right)^{3}$ and/or $\left(\textrm{scalar}\right)^{4}$ effective interactions generated through loops which cannot be compressed to a point, as is ordinarily the case. That is because these effective couplings involve derivatives of the fields, making them non-renormalisable, so there exist exceptional cases in which there is no corresponding tree-level realisation of the effective vertex. We call this class of diagrams \textit{special genuine} as they required very specific choices of fields in order to be genuine, as it will be discussed in the following sections. Out of the 160 topologies mentioned above, 44 are normal genuine ones, and of the remaining 116 there are 55 which fall into this class. The complete list is shown in \app{app:topos}, where we also classify them according to which particular particle combination is needed to make the corresponding model genuine.

%%%%%%%%%%%%%%%
\subsubsection*{${\rm SU(2)_L}$ antisymmetric contractions}

\begin{figure}
    \centering
    \includegraphics[width=0.9\textwidth]{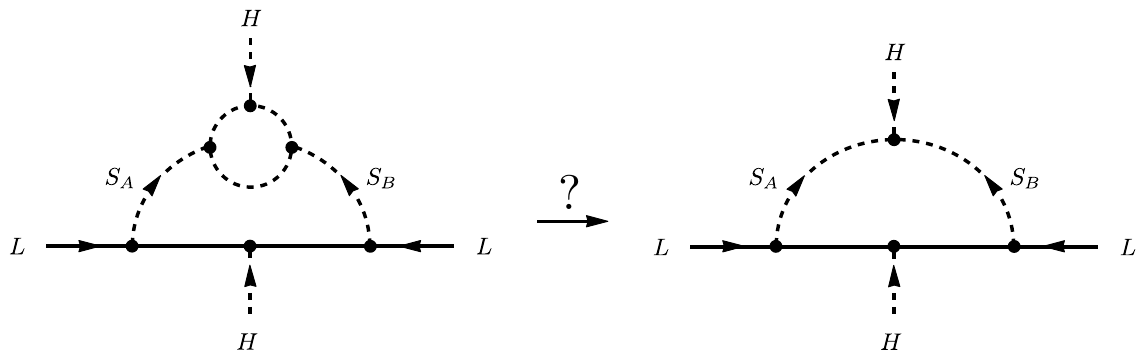}
    \caption{This two-loop realisation of the dimension five Weinberg operator illustrates a loophole in our automatised algorithm for finding genuine $n$-loops neutrino mass diagrams and topologies (we therefore track these special cases manually). In particular, if the scalar $S_A$ (or $S_B$) is the Higgs fields $H$, and $S_B$ (or $S_A$) is an ${\rm SU(2)}$ singlet with the correct hypercharge, then there is no point interaction $H S_A S_B$. Hence, the existence of the left diagram does not imply that one can build the diagram on the right, with one loop less.}
    \label{fig:3loop:exception2}
\end{figure}

We shall start with the first loophole. To understand it, consider the two-loop diagram in \fig{fig:3loop:exception2}. The diagram appears to be non-genuine because it requires fields with quantum numbers such that they would have a renormalisable interaction $H S_A S_B$, which could be used to remove one-loop from the diagram. In a sense, this is indeed always true: such trilinear combination of fields must be gauge invariant. Yet, $H S_A S_B$ might be identically zero for specific choices of $S_A$ and $S_B$. Take the case where $S_A$ is the Higgs field $H$, and $S_B$ is a scalar singlet with hypercharge -1. Then, $H H S_B\equiv 0$ because the ${\rm SU(2)}$ singlet contraction of two doublets is antisymmetric. For particular choices of the quantum numbers of the remaining fields, one can in fact check that no $d=5$ one-loop model is generated, hence the two-loop diagram/topology in \fig{fig:3loop:exception2} is genuine. This construction involving the use of repeated fields to forbid point-interactions (which otherwise would be allowed by their
quantum numbers) can obviously be extended to three-loop diagrams. These genuine three-loop diagrams (topologies) which lead to non-genuine model-diagrams, unless very special choices of quantum numbers are made, are inside the category of \textit{special genuine} diagrams (topologies). Out of the 160 topologies, 29 topologies shown in \app{app:topos} are genuine due to the arguments discussed above.

Note that, if we break down the fields into their components, the neutrino mass obtained from these special topologies arises from a difference of diagram amplitudes, with the negative sign(s) coming from the antisymmetry of ${\rm SU(2)_L}$ (and/or colour) contractions. This is very clear, for example, in the one-loop subdiagram in \fig{fig:3loop:exception2} (on the left), which must correspond to an $HHS_B$ interaction, as mentioned earlier. In the limit where the momenta flowing into these critical subdiagrams is small, the difference of amplitudes will approach zero. However, the momenta flowing into these subdiagrams is a loop momenta, hence the overall neutrino mass obtained from special genuine diagrams does not need to be small when compared to the mass obtained from normal genuine diagrams.

%%%%%%%%%%%%%%%
\subsubsection*{Massless fermion fields}

\begin{figure}
    \centering
    \includegraphics[width=1\textwidth]{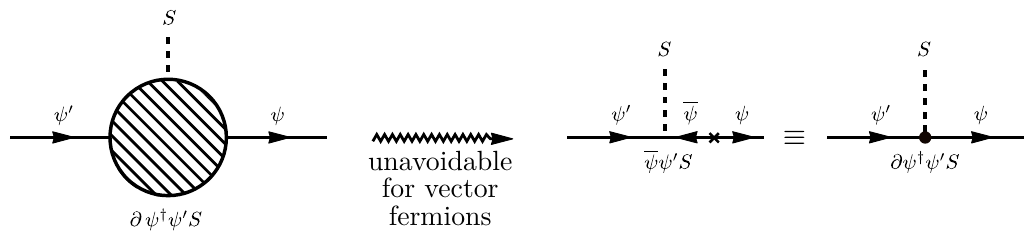}
    \caption{Consider a loop induced coupling of the left-handed fermions $\psi$ and $\psi'$ to a scalar $S$ as indicated on the left. Given the chirality of the fermions, the effective coupling is $\partial \psi^\dagger \psi' S$. If $\psi$ has a vector-like partner $\overline{\psi}$ (which we may consider to be left-handed as well), then the tree-level coupling $\overline{\psi} \psi' S$ exists, and together with a mass insertion $\overline{\psi} \psi$ it can be used to generate the effective interaction $\partial \psi^\dagger \psi' S$ without loops (this is the leading order interaction; extra pairs of derivatives appear at higher order). This argument fails if both $\psi$ and $\psi'$ are Standard Model fermions.}
    \label{fig:3loop:explanation}
\end{figure}

For the \textit{special genuine} topologies discussed before, the existence of derivatives can be traced to the antisymmetric nature of some ${\rm SU(2)_L}$ contractions, which makes some loop interactions non-compressible for appropriate choices of the quantum numbers of the diagram lines. However, there is a second possible reason why such derivative terms might be unavoidable: fermion-fermion-scalar couplings may contain a derivative due to the chirality of the Standard Model fermions. In particular, two left-handed Weyl fermions $\psi$ and $\psi'$ may interact with a scalar $S$ through a $\partial\psi^\dagger \psi' S$ effective coupling (the number of derivatives can be higher, as long as it is an odd number).\footnote{From a symmetry argument, one can see that the number of derivatives is odd, and therefore there is at least one of them. It goes as follows: the complexified algebra of the Lorentz group is the same as the complexified algebra of ${\rm SU(2)} \times {\rm SU(2)}$, so it's representations can be labelled by a pair of spins $(j_L,j_R)$. Left-handed fermions $\psi$ and their conjugates $\psi^\dagger$ transform as $(1/2,0)$ and $(0,1/2)$ (conjugation flips $j_L$ with $j_R$) respectively, while vector-like fields, such as derivatives, are bi-doublets $(1/2,1/2)$. Therefore, the bilinear $\psi \psi^\dagger$ transforms as a vector, and Lorentz invariance can only be obtained by adding to this fermion combination an odd number of derivatives.} However, if $\psi$ is a vector-like field, its left-handed partner $\overline{\psi}$ has the same gauge quantum numbers as $\psi^\dagger$ (the conjugate of $\psi$) and opposite chirality, hence there is no symmetry forbidding the renormalisable interaction $\overline{\psi} \psi' S$, with no derivatives.\footnote{The same argument follows interchanging $\psi$ with $\psi'$.} Using this coupling and a mass insertion $\overline{\psi}\psi$ one can then always generate $\partial \psi^\dagger \psi' S$ without loops, as depicted in \fig{fig:3loop:explanation}. However, if neither $\psi$ nor $\psi'$ has a vector-like partner, this argument fails and the vertex $\partial \psi^\dagger \psi' S$ may not be realisable at tree-level. Thus, if $\psi$ and $\psi'$ are fixed to be Standard Model fermions, the loop might not be compressible and our general arguments fail. Let us discuss this with one particular example. 

\begin{figure}
    \centering
    \includegraphics[width=1\textwidth]{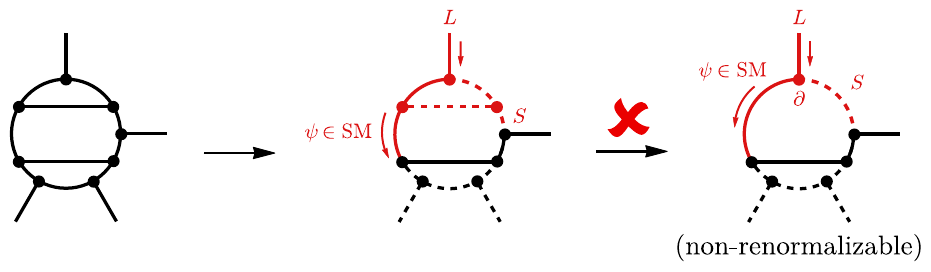}
    \caption{Example of a diagram with topology 89, containing a one-loop fermion-fermion-scalar effective interaction (in red). This loop is removable unless $\psi$ is a Standard Model fermion.}
    \label{fig:3loop:example}
\end{figure}

Take for instance topology 89 and one of its diagrams as an example, shown in \fig{fig:3loop:example}. This diagram contains a one-loop realisation of the vertex $\partial L \psi^\dagger S$, with $L$ the Standard Model lepton doublet and $\psi$ and $S$ an arbitrary left-handed fermion and scalar, respectively. If $\psi$ is not a Standard Model fermion, it is necessary to add its vector-like partner $\overline{\psi}$ to the model, in order to generate a bare mass term $M \overline{\psi}\psi$. From the argument in \fig{fig:3loop:explanation}, it is then possible to rewrite the diagram with one less loop. On the other hand, it might not be possible to do so if $\psi$ is a Standard Model fermion, in which case the diagram is genuine. 26 topologies shown in \app{app:topos} are classified as \textit{special genuine} due to this loophole.

Note that, since this loophole to our general argument exists only for Standard Model fermions, the list of all possible genuine models generated from the diagram in \fig{fig:3loop:example} will be quite constrained, due to the limited number of choices of $\psi \in \{L,e^c,Q,u^c,d^c\}$.

%%%%%%%%%%%%%%%%%%%%%%%%%%%%%%%%%%%%%%%%%%%%%%%%%%%%%%%%%%%%%%%
\subsection{Non-genuine topologies}
\label{sec:3loop:nongen}

\begin{figure}
    \centering
    \includegraphics[scale=1]{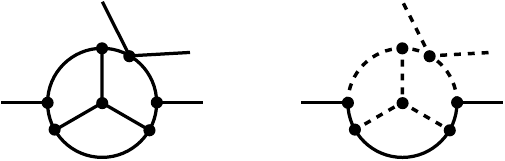}
    \caption{One example of a non-genuine but finite topology (to the left), generating the diagram on the right. See text.}
    \label{fig:3loop:ngfin}
\end{figure}

The other 61 remaining topologies ($160=44+29+26+61$) generate non-genuine diagrams. Even so, it is important to note that some of these topologies (9 of them) may lead to \textit{non-genuine finite diagrams}. These are diagrams for which an additional (broken) symmetry is always needed to forbid the otherwise allowed $\ell$-loop diagrams ($\ell<3$) that result from compressing one or more loops to a renormalisable vertex. We show an example of such a topology and corresponding non-genuine finite diagram in \fig{fig:3loop:ngfin}. In this diagram, the inner loop on the fermion line is an example of a compressible 3-point vertex. However, if this fermion is of Majorana type, one can add to the corresponding model an extra symmetry, for example a global ${\rm U(1)}$ as in \cite{Hatanaka:2014tba}. A Standard Model singlet scalar can then be coupled to the Majorana fermion and assigned a charge under the ${\rm U(1)}$. The tree-level coupling of the compressible inner loop could be forbidden in this way. Spontaneous breaking of this ${\rm U(1)}\to Z_2$ by the vacuum expectation value of the singlet generates a Majorana mass term for the fermion and allows then this three-loop diagram to exist.

We stress again that we do not consider this class to be genuine, as these models require extra symmetries (which need to be broken). Note that the symmetries can not be exact, otherwise the compressible loop is also forbidden by the symmetry. For the remaining 52 topologies in this non-genuine class all diagrams have infinite loop integrals. Due to the large number of topologies in this sub-class, we do not show them here; they can be can found in \cite{extraData}.

%%%%%%%%%%%%%%%%%%%%%%%%%%%%%%%%%%%%%%%%%%%%%%%%%%%%%%%%%%%%%%%
%%%%%%%%%%%%%%%%%%%%%%%%%%%%%%%%%%%%%%%%%%%%%%%%%%%%%%%%%%%%%%%
\section{Constructing diagrams}
\label{sec:3loop:diag}

\begin{figure}
    \centering
    \includegraphics[width=\textwidth]{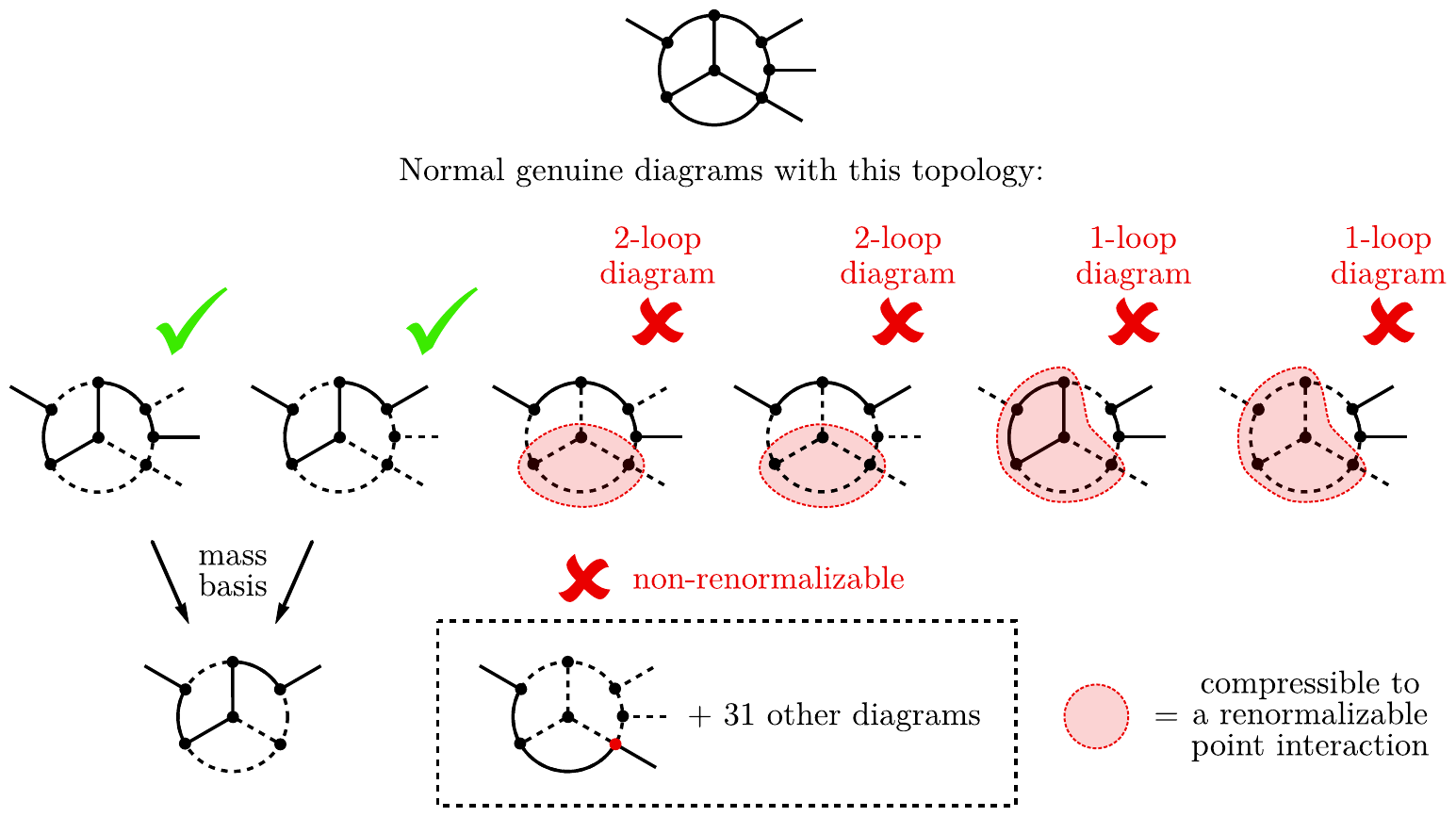}
    \caption{There are two normal genuine diagrams associated to the topology shown in the top. There is also a total of 32 diagrams which can be drawn with a non-renormalisable fermion-fermion-scalar-scalar interaction. Finally, there are 4 diagrams which are also not normal genuine ones because it is possible to shrink a subpart of them into a renormalisable point interaction. (Note however that under some very specific circumstances, the third, the fifth and sixth diagrams in the top row can be genuine, hence they are considered \textit{special} genuine diagrams.)}
    \label{fig:3loop:example-topology-to-diagrams}
\end{figure}

We now return to the construction of the genuine diagrams. From the 44 normal genuine topologies, a total of 228 genuine diagrams can be built \cite{extraData}. In \fig{fig:3loop:example-topology-to-diagrams} we show for one particular topology the possible diagrams, explaining graphically why several of them are not genuine. Diagrams with compressible loops are discarded in this set, as well as diagrams with non-renormalisable interactions. In this particular example, the topology has only two normal genuine diagrams. The remaining genuine diagrams, of the special kind, must be found carefully in a non-automated way. We show the procedure adopted to find the genuine diagrams generated from the special genuine topologies with some examples. We consider the case where the antisymmetry of ${\rm SU(2)_L}$ contractions forbids the lower order diagrams, as there are several possibilities that are worth to be studied.

\begin{figure}
    \centering
    \includegraphics[width=1\textwidth]{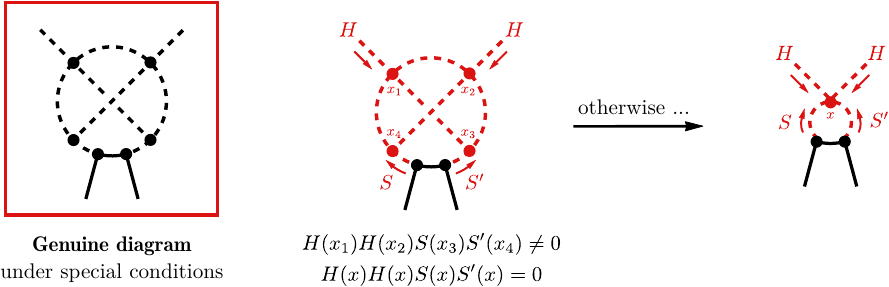}
    \caption{Topology 54 has only one diagram associated to it (shown here inside the box on the left). This diagram is only genuine under special conditions, in particular the two scalars interacting with the fermion line ($S$ and $S'$) must be a Higgs $H$ and a scalar $\phi_D\equiv \left(\boldsymbol{1},\boldsymbol{2},-3/2\right)$.}
    \label{fig:3loop:topologies_special_example1}
\end{figure}

Consider topology 54 (see \fig{fig:app:topos:topologies_special}), there is only one way of making a fermion chain connecting the two external $L$'s hence there is a single diagram to be considered, shown in \fig{fig:3loop:topologies_special_example1}. One can identify in it a two-loop subdiagram with 4 external scalar lines, shown in red in the middle of \fig{fig:3loop:topologies_special_example1}. Two of the external scalars are the Higgs fields of the Weinberg operator, while the others ($S$ and $S^\prime$) are unknown a priori, hence the subdiagram is associated to the operator $HHSS'$. This means that, for most assignments of quantum numbers to the internal fields, one can write down such an interaction directly in a renormalisable Lagrangian, in which case neutrino masses can be generated via the one-loop diagram shown in \fig{fig:3loop:topologies_special_example1} on the right.

However, strictly speaking the two-loop subdiagram generates the non-local operator $H(x_1)H(x_2)S(x_3)S'(x_4)$ which we may rewrite as,
\begin{eqnarray} %\label{eq:}
     H\left(x_{1}\right)H\left(x_{2}\right)S\left(x_{3}\right)S'\left(x_{4}\right) &=& H\left(x\right)H\left(x\right)S\left(x\right)S'\left(x\right)
     \\ \nn
     &+&\sum_{n}c_{n}\frac{\partial^{2n}}{\Lambda^{2n}}H\left(x\right)H\left(x\right)S\left(x\right)S'\left(x\right)\,,
\end{eqnarray}
where $x$ is some space-time point close to the $x_i$, $\Lambda$ is some mass scale and the $c_n$ are dimensionaless parameters. If $H(x)H(x)S(x)S'(x)$ is nullified, the corresponding Weinberg one-loop neutrino mass will not exist. Keeping in mind that in the Weinberg operator the two Higgs doublets contract as an ${\rm SU(2)}$ triplet, there is only one possibility of avoiding a renormalisable $HHSS'$ interaction: setting $S=H$ and $S'=\phi_D\equiv \left(\boldsymbol{1},\boldsymbol{2},-3/2\right)$, or vice-versa.\footnote{If $S^\prime$ was a ${\rm SU(2)}$ quadruplet, $S'= \left(\boldsymbol{1},\boldsymbol{4},-3/2\right)$, the local operator $H(x)H(x)H(x)S'(x)$ would not vanish. Another idea to avoid the $HHSS'$ point interaction is to have $S=S'=\left(\boldsymbol{1},\boldsymbol{R},-1/2\right)$ for some ${\rm SU(2)}$ representation $\boldsymbol{R}$, such that the two $S$'s contract antisymmetrically. This happens for odd-dimensional ${\rm SU(2)}$ representations (other than the trivial one): $\boldsymbol{R}=\boldsymbol{3},\boldsymbol{5}, \boldsymbol{7}, \cdots$.
The problem is that, for hypercharge $-1/2$, such $\boldsymbol{R}$'s lead to fractionally charged particles, the lightest of which would be stable and therefore pose a cosmological problem \cite{Langacker:2011db} (adding a non-trivial colour quantum number would not change this). Therefore we discarded such scenarios altogether.} With this very special setup, the three-loop diagram in \fig{fig:3loop:topologies_special_example1} is genuine, and that is why the corresponding topology is included in \fig{fig:app:topos:topologies_special}.

\begin{figure}
    \centering
    \includegraphics[width=1\textwidth]{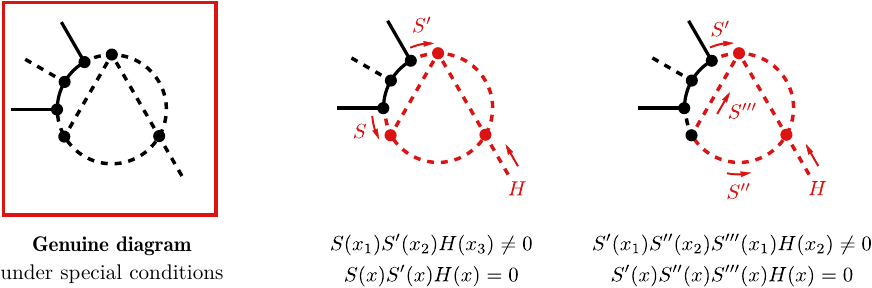}
    \caption{Topology 71 has one genuine diagram, shown here. The quantum numbers of the scalars must be special, otherwise one could build the Weinberg operator from a one- or two-loop diagram. In particular, this scenario is avoided only if $S=\phi_S\equiv \left(\boldsymbol{1},\boldsymbol{1},-1\right)$, $S^\prime=H$, $S^{\prime\prime}=\phi_{y}\equiv\left(\boldsymbol{1},\boldsymbol{1},y\right)$, and $S^{\prime\prime\prime}=\widetilde{\phi}_{y}\equiv\left(\boldsymbol{1},\boldsymbol{1},-1-y\right)$ for some hypercharge $y\neq 0, \pm 1, \pm 2$.}
    \label{fig:3loop:topologies_special_example2}
\end{figure}

As a more involved example, we will now discuss topology 71, for which there is a single genuine diagram, shown in \fig{fig:3loop:topologies_special_example2}. In this same figure, we indicate in red two subdiagrams (with 3 and 4 external lines) which should not be shrinkable to point interactions, otherwise the diagram becomes non-genuine. In particular, the internal scalars must be such that the local operators $HSS^\prime$ and $HS^\prime S^{\prime\prime} S^{\prime\prime\prime}$ are zero, while still allowing their non-local counterparts.

To nullify the first interaction, $HSS^\prime$, we must have either $S^\prime=S$, $S=H$, or $S^\prime=H$. But the first possibility ($HSS$) is no good, as there is no ${\rm SU(2)}$ representation $\boldsymbol{R}$ such that $\boldsymbol{R}\times \boldsymbol{R} \times \boldsymbol{2}$ is gauge invariant. The second and third possibilities ($HHS$ and $HHS'$), on the other hand, imply that either $S'=\phi_S\equiv \left(\boldsymbol{1},\boldsymbol{1},-1\right)$ or $S=\phi_S$, respectively.

We consider now the other interaction, $HS^\prime S^{\prime\prime} S^{\prime\prime\prime}$, which also needs to be zero in its point-like realisation. Given the two possible quantum number assignments for $S'$, we might have $H\phi_S S'' S'''$ or $HH S'' S'''$. However, it is not complicated to check that $H\phi_S S'' S'''$ would require either $S''$ or $S'''$ to be a gauge singlet $\left(\boldsymbol{1},\boldsymbol{1},0\right)$, so one could make a two-loop realisation of the Weinberg operator by removing this scalar line from the three-loop diagram.

We then proceed with the only viable hypothesis, i.e. $HS^\prime S^{\prime\prime} S^{\prime\prime\prime}=HH S'' S'''$. Again, we are faced with two scenarios: (a) one of the undetermined scalars ($S^{\prime\prime}$ and $S^{\prime\prime\prime}$) is equal to $H$, or (b) both $S^{\prime\prime}$ and $S^{\prime\prime\prime}$ are different from $H$. Scenario (a) implies that $\left(S^{\prime\prime},S^{\prime\prime\prime}\right)=\left(H,\phi_D\right)$, while scenario (b) leads to $S^{\prime\prime}=\phi_{y}\equiv\left(\boldsymbol{1},\boldsymbol{1},y\right)$, and $S^{\prime\prime\prime}=\widetilde{\phi}_{y}\equiv\left(\boldsymbol{1},\boldsymbol{1},-1-y\right)$ for some $y \in \mathbb{Z}$. In the last case, both scalars must be ${\rm SU(2)}$ singlets in order to ensure that the field product $S^{\prime\prime}S^{\prime\prime\prime}$ does not have a triplet component which would be responsible for coupling the two $H$'s symmetrically.

\begin{figure}
    \centering
    \includegraphics[scale=1.5]{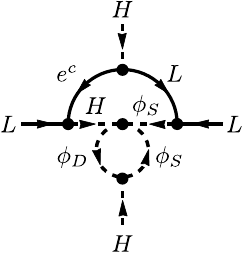}
    \caption{Two-loop diagram which can be built in a model with the scalar fields $\phi_S\equiv \left(\boldsymbol{1},\boldsymbol{1},-1\right)$ and $\phi_D\equiv \left(\boldsymbol{1},\boldsymbol{2},-3/2\right)$. Note that because $H(x)H(x)\phi_S(x)=0$, it is not possible to remove the bottom loop in the diagram.}
    \label{fig:3loop:phiD-phiS-diagram}
\end{figure}

Taking into consideration everything said so far, it might then seem that there are two possibilities for topology 71 with the labelling as indicated on \fig{fig:3loop:topologies_special_example2}: $\left(S,S',S'',S'''\right)=\left(\phi_{S},H,H,\phi_{D}\right)$ and $\left(\phi_{S},H,\phi_{y},\widetilde{\phi}_{y}\right)$ (possibly switching the quantum number of $S''$ and $S'''$). However, a model with both the scalar $\phi_S$ and the scalar $\phi_D$ will inevitable generate the two-loop diagram shown in \fig{fig:3loop:phiD-phiS-diagram}, so the diagram in \fig{fig:3loop:topologies_special_example2} is genuine only if $\left(S,S',S'',S'''\right)=\left(\phi_{S},H,\phi_{y},\widetilde{\phi}_{y}\right)$. It is worth mentioning that although the scalar loop with $\phi_{S}$ and $\phi_{D}$ in \fig{fig:3loop:phiD-phiS-diagram} seems to diverge, the loop is finite. This is because such special diagrams involve differences of two diagrams due to the ${\rm SU(2)_L}$ contractions, removing the divergences. This is the same contractions that makes precisely $H(x)H(x)\phi_S(x)=0$.

For all the topologies in \fig{fig:app:topos:topologies_special}, we performed a similar analysis as in the previous examples, identifying the loop or loops at the diagram level that can exploit the loophole and, therefore, the specific field content needed for the diagram to be genuine. With this analysis, we, not only, build the diagrams, but further classify the topologies in three groups such that all the diagrams generated by a certain topology require the same fields to be genuine. In \fig{fig:app:topos:topologies_special}, the first two rows of topologies (from topology 45 to 60) generate diagrams which contain one or two 4-point loop scalar vertices with at least one external Higgs. The models descending from these topologies necessarily have the fields $\phi_y$, $\widetilde{\phi}_y$ and/or $\phi_D$ in order to be genuine. Topologies 61 to 70 generate diagrams with one 3-point internal loop, i.e. with no leg being a external leptons or Higgs. This can be either a 3-point scalar or fermion-fermion-scalar vertex. Note that in both cases the tree-level should be zero, so one cannot have more than one copy of these fermions or scalars. Finally, the diagrams coming from the three topologies 71, 72 and 73, contain at least one reducible loop with two scalars and an external Higgs. In principle, one can avoid the corresponding two-loop diagram with the recipe described in \fig{fig:3loop:topologies_special_example2}, thus making the topology genuine. Nevertheless, we mention that this diagram alone generates models which are not able fit neutrino data as they contain a structure identical to the simplest realisation of the Zee model \cite{Zee:1980ai}.\footnote{Note that unlike the Zee model where another copy of the Higgs can be added to fit neutrino data, in the case under discussion this is not a viable solution, because the resulting model will be a correction to a dominant two-loop model generated by reducing the 3-point scalar loop with the copy of the Higgs.}

We explained how to generate genuine diagrams from the set of special topologies in \fig{fig:app:topos:topologies_special} exploiting the antisymmetry of ${\rm SU(2)_L}$ contractions. A similar analysis can be done for the special topologies in \fig{fig:app:topos:derivative}, which required at least a massless fermion running in the loop to be genuine (see \sect{sec:3loop:specgen}). The full set of genuine diagrams is much easier to build, as the possible massless fermions are constrained to the Standard Model fermions $\{L,e^c,Q,u^c,d^c\}$. For this reason, we do not enter into detail.

\begin{figure}
    \centering
    \includegraphics[width=\textwidth]{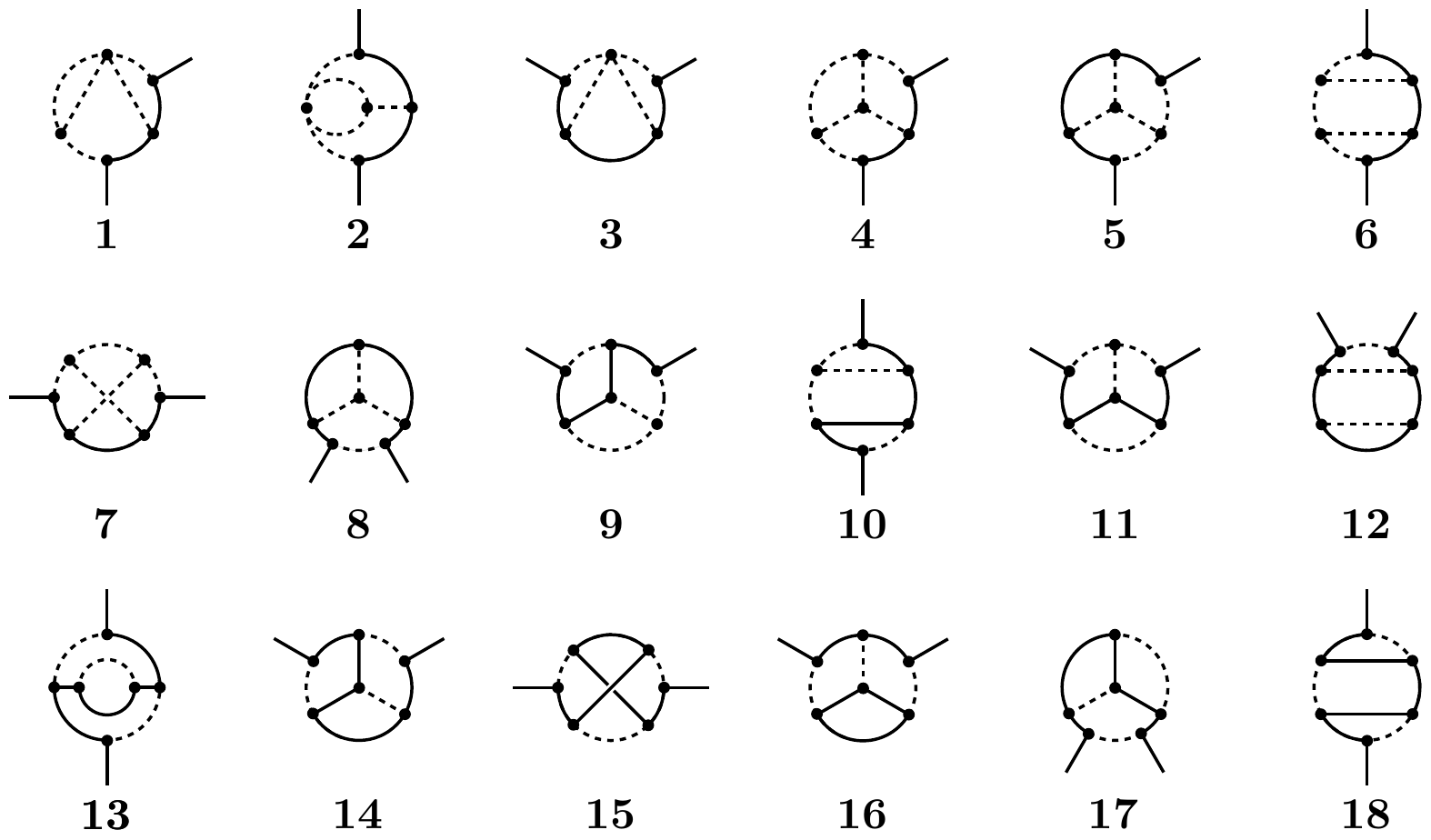}
    \caption{List of \textit{normal} genuine diagrams in the mass basis. Note that the two external Higgs lines were removed: in general there is a many-to-one relation between the original diagrams and the amputated ones shown here. Diagrams in this list are referred to in the text as $D^M_i$, where $i$ is the number of the diagram shown here.}
    \label{fig:3loop:normalgenuinediagrams}
\end{figure}

\begin{figure}[h]
    \centering
    \includegraphics[width=\textwidth]{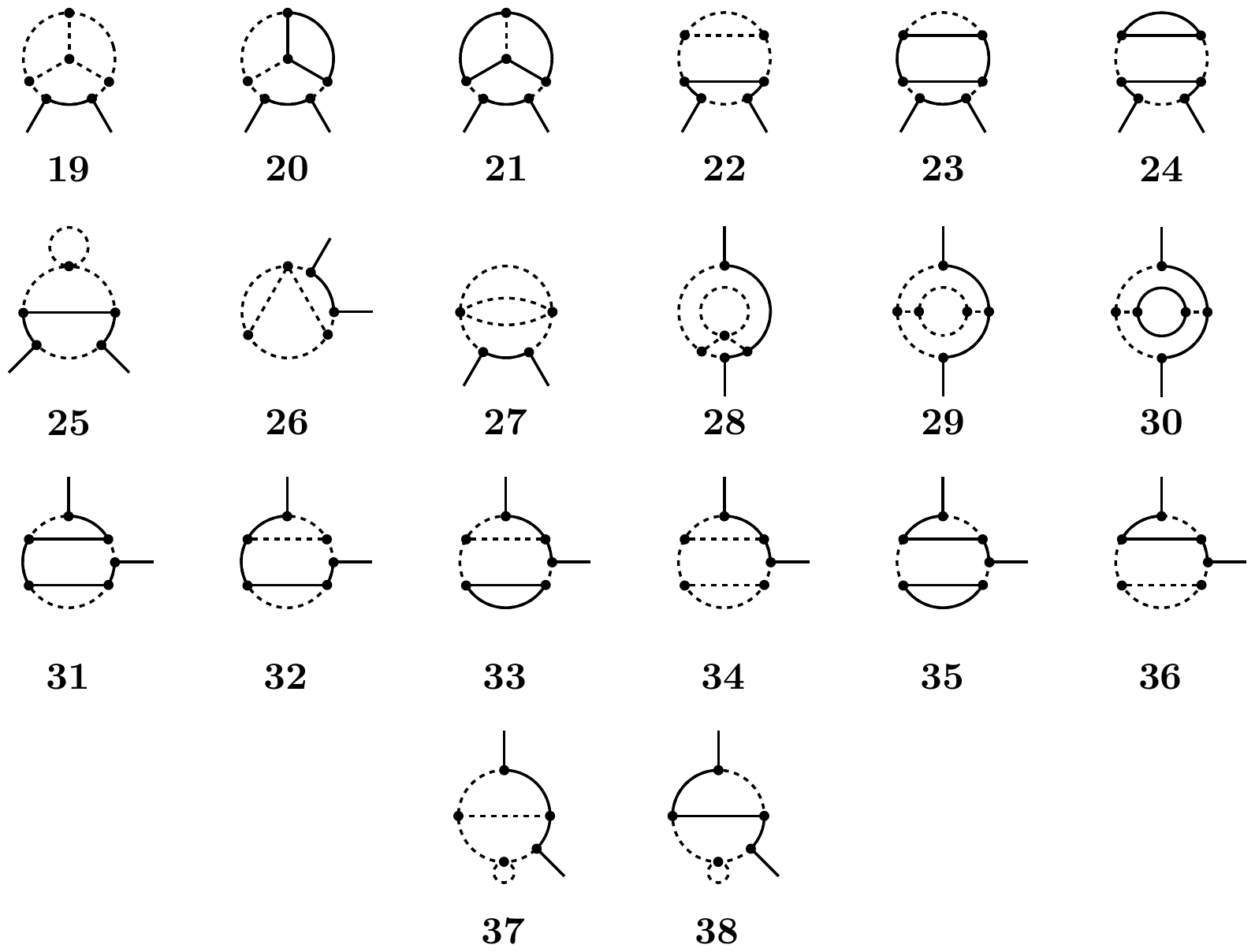}
    \caption{List of \textit{special} genuine diagrams, with the external Higgs lines removed. Diagrams in this list are referred to in the text as $D^M_i$, where $i$ is the number of the diagram shown here.}
    \label{fig:3loop:specialgenuinediagrams}
\end{figure}

\begin{figure}
    \centering
    \includegraphics[width=1\textwidth]{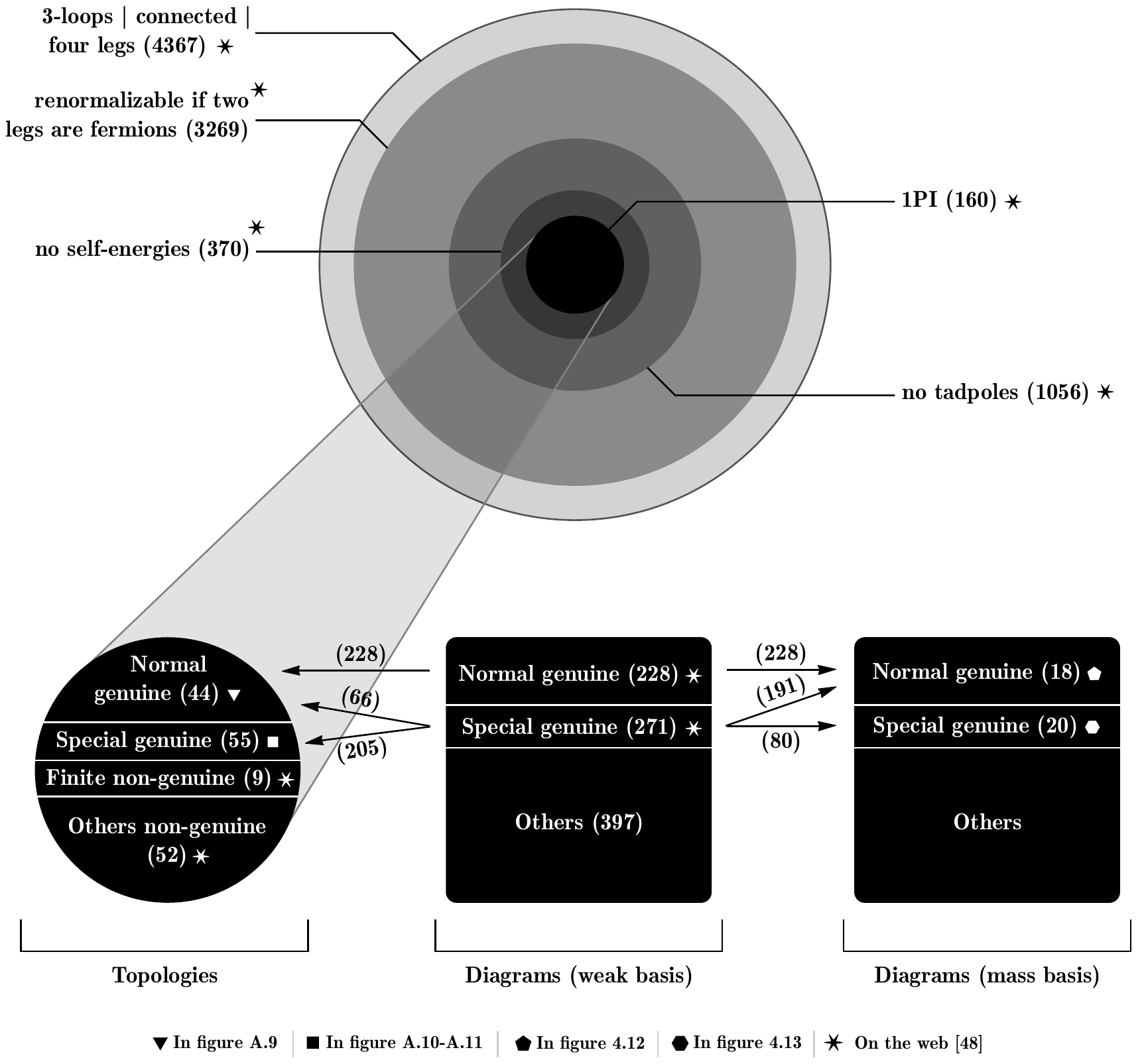}
    \caption{Summary of the different types of diagrams and topologies. Out of thousands of topologies, only 160 are potentially interesting. They correspond to a total of 896 diagrams: 228 can provide dominant neutrino mass contributions without special considerations (\textit{normal genuine diagrams}), and a further 271 can do so only with very special setups (\textit{special genuine diagrams}). We call \textit{normal genuine topologies} to those associated to at least one normal genuine diagram (there are 44); the \textit{special genuine topologies} are the remaining cases which are associated to at least one special genuine diagram (there are 55). The remaining topologies are \textit{non-genuine} but some of them (9) have at least one finite diagram. Once the external Higgs fields are removed, the 228 \textit{normal genuine diagrams} become 18 amputated diagrams, while the remaining genuine diagrams in the weak basis yield 20 more amputated diagrams.}
    \label{fig:3loop:summary}
\end{figure}

As a final step, the two external scalars standing for Higgs VEV insertions are removed, and a list of 18 genuine (amputated) diagrams is obtained. These are shown in \fig{fig:3loop:normalgenuinediagrams}. In other words, the 228 diagrams in the electroweak basis can be reduced to 18 diagrams in the mass eigenstate basis. To these one has to add the 20 mass diagrams in \fig{fig:3loop:specialgenuinediagrams} which are obtained from special genuine diagrams. A visual summary of the steps described so far, as well as a counting of the genuine diagrams and topologies, can be found in \fig{fig:3loop:summary}. We state again, that while most of our results apply also to diagrams with vector bosons, our lists are not complete for vectors, due to the loophole discussed above in \fig{fig:3loop:exception1}.

We close this section by noting that the amplitudes of the 18+20 diagrams from \fig{fig:3loop:normalgenuinediagrams} and \fig{fig:3loop:specialgenuinediagrams} can be decomposed as linear combination of five master integrals \cite{Martin:2016bgz}. Some of these master integrals admit an analytical solution, while others can only be solved numerically. A more detailed discussion is given in \app{app:loops}.

%%%%%%%%%%%%%%%%%%%%%%%%%%%%%%%%%%%%%%%%%%%%%%%%%%%%%%%%%%%%%%%
%%%%%%%%%%%%%%%%%%%%%%%%%%%%%%%%%%%%%%%%%%%%%%%%%%%%%%%%%%%%%%%
\section{Incompressible loops and genuineness of a diagram}
\label{sec:3loop:gen_loops}

We have mentioned in \sect{sec:3loop:normgen} that those diagrams for which it is possible to compress one or more loops into a renormalisable vertex $v$ are not genuine\footnote{We have also discussed in detail an exception to this rule, due to the potential presence of repeated fields. Hence, we introduced the concept of special genuine diagrams and topologies, which are genuine even though they have compressible loops.}, as one can then use the interaction $v$ to construct a similar diagram with less loops. In other words,
\begin{equation}
    \textrm{loop compressibility } \Rightarrow \textrm{non-genuineness}\,.
\end{equation}
Obviously, this is equivalent to the statement that genuine diagrams have incompressible loops ($\textrm{genuineness}\Rightarrow\textrm{loop incompressibility}$). However, this is not the same as,
\begin{equation} \label{eq:3loop:implication} 
    \textrm{loop incompressibility }\Rightarrow\textrm{genuineness} \, ,
\end{equation}
and yet it was stated before that we expect this to be true. Indeed, our analysis relies on this important assumption, so in this section we discuss why we believe it to be true. We think that the argument presented here is compelling, but we stop short of calling it a proof.

First, consider the following intuitive/informal explanation for the implication \eq{eq:3loop:implication}. For particular assignments of quantum numbers to the internal lines of a diagram, there might be extra interactions between some of the diagram's fields which are completely unrelated to the interactions used in the diagram. If that is the case, it might be possible to construct the Weinberg operator $LLHH$ with less loops by using the additional interactions. However, there will be a choice of quantum numbers of the internal lines such that this does not happen: no extra ``non-trivial interactions'' (see below) between the fields is possible, hence the operator $LLHH$ cannot be realised by a simpler diagram, with less loops.

\begin{figure}
    \centering
    \includegraphics[scale=0.7]{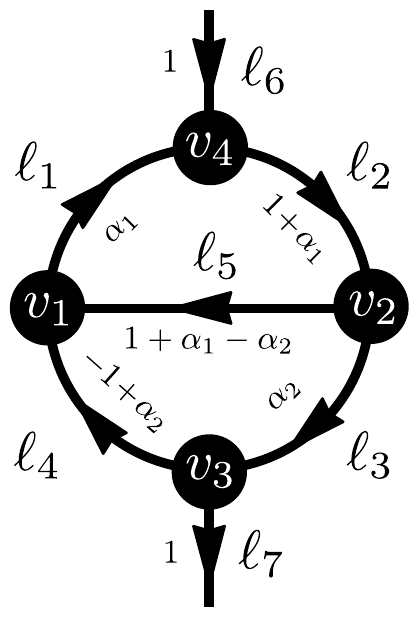}
    \caption{Two-loop diagram with oriented lines $\ell_{i}$ and vertices $v_{j}$. The hypercharge of each line is indicated as a function of two free numbers: $\alpha_{1}$ and $\alpha_{2}$. The hypercharge of the external lines was fixed to 1 in this example. }
    \label{fig:3loop:Two-loop-diagram}
\end{figure}

In order to formalise this idea, consider only the abelian ${\rm U(1)_{Y}}$ symmetry. In a $n$-loop diagram where the hypercharge of the external particles is fixed, the hypercharges $y_{i}$ of the internal lines depend on $n$ free numbers $\alpha_{j}$. More specifically, the $y_{i}$ are linear functions of these $n$ parameters,
\begin{equation} \label{eq:3loop:yi}
    y_{i}=c_{0}^{i}+\sum_{j=1}^{n}c_{j}^{i}\alpha_{j}\,,
\end{equation}
where the $c_{0}^{i}$ and $c_{j}^{i}$ are numbers which depend on the hypercharge of the external particles and on the topology.\footnote{Given that the $\alpha_{j}$ are free numbers, there is some arbitrariness in the choice of $c_{j}^{i}$: for any invertible matrix $X$ one can replace $c_{j}^{i}$ by $\sum_{j'}X_{jj'}c_{j'}^{i}$.} \fig{fig:3loop:Two-loop-diagram} shows an example where $y_{6}=y_{7}=1$ by choice, and $y_{1}=\alpha_{1}$, $y_{2}=1+\alpha_{1}$, $y_{3}=\alpha_{2}$, $y_{4}=-1+\alpha_{2}$, $y_{5}=1+\alpha_{1}-\alpha_{2}$.

A crucial question is then the following: what is the full list of interactions between the fields used in the diagram? From the point of view of the ${\rm U(1)_{Y}}$ symmetry, any hypothetical interaction beyond those used in the diagram will either be (a) forbidden, (b) allowed for particular values of the $\alpha_{j}$ or (c) allowed for all values of the $\alpha_{j}$. Referring to \fig{fig:3loop:Two-loop-diagram}, $\ell_{3}\ell_{4}^{*}$, $\ell_{1}\ell_{2}\ell_{3}$ and $\ell_{1}^{*}\ell_{2}\ell_{7}^{*}$ (respectively) are examples of each of these interactions. We will only be interested in those interaction of type (c) because we can choose the $\alpha_{j}$ in order to build a model where all interactions
of type (a) and (b) are forbidden.

For the rest of this discussion, it is important to keep in mind that the ${\rm U(1)_{Y}}$ symmetry is blind to combinations of a field and its conjugate, $\ell_{i}\ell_{i}^{*}$, hence, one can add/remove them at will from any allowed vertex. Now, note that the hypercharges $y_{i}$ in \eq{eq:3loop:yi} are the most general solutions to a linear system of equations,
\begin{equation} \label{eq:3loop:matrixC}
    \sum_{j}C_{ij}y_{j}=0\,,
\end{equation}
where the rows of the matrix $C$ represent each vertex, and its columns stand for each line in the diagram: if line number $j$ enters(leaves) vertex $i$, then $C_{ij}=1$($-1$), otherwise this entry is null. For the example in \fig{fig:3loop:Two-loop-diagram} we would have the following matrix,
\begin{equation}
    C = \left(
        \begin{array}{ccccccc}
            -1 &  0 &  0 &  1 &  1 & 0 &  0 \\
             0 &  1 & -1 &  0 & -1 & 0 &  0 \\
             0 &  0 &  1 & -1 &  0 & 0 & -1 \\
             1 & -1 &  0 &  0 &  0 & 1 &  0
        \end{array}
    \right) \, .
\end{equation}
For each external line in the diagram, since its hypercharge is fixed, one must add that constraint as well.

The important point is that any new vertex would correspond to adding a new row to matrix $C$. This operation will not change the solution space if and only if the new row is a linear combination of the existing rows (the solutions of $C\cdot y=0$ and $C'\cdot y=0$ are the same only if the lines of $C'$ are linear combinations of those of $C$). Additions and subtractions of rows of $C$ translates into making new vertices $v$ which are the product of existing ones (addition) or its conjugates (subtraction): $v=v_{i}^{(*)}v_{j}^{(*)}v_{k}^{(*)}\cdots$.\footnote{The fact that the hypercharge of external lines is fixed introduces a complication: the previous statement is true, but one can also add zero hypercharge combinations of the external fields. In the case of external Higgses $H$ and $L$'s, that would correspond to the combination $HL$, but one can invoke additionally Lorentz invariance to allow only the addition of pairs of this combination, i.e. $HHLL$. However, the sum of all vertices $v_{1}v_{2}v_{3}\cdots$ in our diagrams, by construction, yield the Weinberg operator (times irrelevant combinations of the internal lines of the form $I_{i}I_{i}^{*}$), so the original statement stands: the only extra vertices which do not spoil the solution space in \eq{eq:3loop:yi} are the trivial ones formed from the product of the vertices in the diagram and their conjugates.} The vertices obtained in this way account for all \textit{unavoidable interactions} (considering the ${\rm U(1)_Y}$ group only).

For example, there are vertices $v_{1}=\ell_{1}^{*}\ell_{4}\ell_{5}$ and $v_{2}=\ell_{2}\ell_{3}^{*}\ell_{5}^{*}$ and in \fig{fig:3loop:Two-loop-diagram}, so the interactions $v_{2}v_{2}=\ell_{2}\ell_{2}\ell_{3}^{*}\ell_{3}^{*}\ell_{5}^{*}\ell_{5}^{*}$ or $v_{1}^{*}v_{2}=\ell_{1}\ell_{2}\ell_{3}^{*}\ell_{4}^{*}\ell_{5}^{*}\ell_{5}^{*}$ cannot be forbidden by any choice of $\alpha_{1},\alpha_{2}$. Nevertheless, these are non-renormalisable interactions: to reduce the number of fields in these new interactions, one can only use the fact that combinations of the form $\ell_{i}\ell_{i}^{*}$ are irrelevant for the ${\rm U(1)_{Y}}$ symmetry, hence for example $v_{1}v_{2}=\ell_{1}^{*}\ell_{2}\ell_{3}^{*}\ell_{4}$ (modulo $\ell_{5}\ell_{5}^{*}$). Graphically, it is very easy to follow what is happening: we remove the line $\ell_{5}$ connecting the vertices $v_{1}$ and $v_{2}$, condensing them into a quartic interaction. Applying this process repeatedly for the internal lines $\ell_{3}$, $\ell_{4}$ and $\ell_{5}$, one generates the new interaction $\ell_{1}^{*}\ell_{2}\ell_{7}^{*}$ ($=v_{1}v_{2}v_{3}$ modulo $\ell_{i}\ell_{i}^{*}$'s) which graphically is obtained by collapsing the lower loop of the diagram into a point. 

Another interesting example are those cases where the same field appears in more than one line in the diagram. For the present discussion what is important are those situations where this is \textit{unavoidable}, rather than just possible. According to the previous discussion, two lines $\ell$ and $\ell^{\prime}$ must have the same hypercharge (and therefore potentially represent the same field) if and only if the bilinear interaction $\ell^{*}\ell^{\prime}$ is \textit{unavoidable}, i.e.  one must be able to merge various of the diagram vertices into such interaction. Lines $\ell_{6}$ and $\ell_{7}$ in \fig{fig:3loop:Two-loop-diagram} constitute an example of such a scenario (they are external lines, hence their hypercharge was fixed to 1 in our example, but even if they were internal lines of a bigger diagram, ${\rm U(1)_{Y}}$ invariance could not forbid the coupling $\ell_{6}^{*}\ell_{7}$).\\

In summary,
\begin{enumerate}
    \item For a $n$-loop diagram, the hypercharge of the internal lines depends on $n$ free numbers $\alpha_{i}$.
    
    \item The only interactions between the various lines which cannot be forbidden for any choice of $\alpha_{i}$ (the \textit{$U(1)_{Y}$ unavoidable interactions}) are those for which the sum of hypercharges is identically 0, i.e. $0+0\alpha_{1}+0\alpha_{2}+\cdots0\alpha_{n}$. (Only a subset of these interactions are truly unavoidable as one should take into account the full Standard Model symmetry, as well as Lorentz invariance.)
    
    \item The full list of \textit{${\rm U(1)_{Y}}$ unavoidable interactions} can be obtained by merging together the diagram vertices (and/or their conjugates). In this merging process, $\left(\textrm{line X}\right)\left(\textrm{line X}\right)^{*}$ combinations can be added or removed.
    
    \item Most of these \textit{unavoidable interactions} are non-renormalisable. Renormalisable new vertices can be formed only if there are those removable $\left(\textrm{line X}\right)\left(\textrm{line X}\right)^{*}$ combinations mentioned earlier, otherwise by merging vertices the number of lines constantly increasing. Graphically, this corresponds to coalescing adjacent vertices in the diagram, and removing the line(s) connecting them.\footnote{Even though it is not important for the present discussion, we mention here that the deleted lines can be the external $L$'s and/or $H$'s: graphically one would connect the diagram with a conjugated copy of itself (that is, a copy of the diagram with the orientation of all lines flipped) via the external $L$ and/or $H$ lines, which would become internal lines. For example, consider a diagram with Higgs interactions $v_{1}=H\ell_{1}\ell_{2}$ and $v_{2}=H\ell_{3}\ell_{4}$. Then $v_{1}v_{2}^{*}=\ell_{1}\ell_{2}\ell_{3}^{*}\ell_{4}^{*}$ is a new, unavoidable interaction which can be obtained graphically in the way just described.}
\end{enumerate}
There are 4 external lines ($LLHH$) indirectly connected to each other through a web of vertices and internal lines. The \textit{unavoidable} alternative ways of connecting these 4 lines must be through an alternative web of \textit{unavoidable} vertices. Graphically, this new web of lines and vertices must be obtained from the original one by the vertex-merging process described above. So, if it is not possible to remove one or more loops from a diagram by collapsing them into a point, the diagram is genuine.

%%%%%%%%%%%%%%%%%%%%%%%%%%%%%%%%%%%%%%%%%%%%%%%%%%%%%%%%%%%%%%%
%%%%%%%%%%%%%%%%%%%%%%%%%%%%%%%%%%%%%%%%%%%%%%%%%%%%%%%%%%%%%%%
\section{Some examples of models}
\label{sec:3loop:examples}

From the complete set of 228+271 genuine diagrams one can generate models by assigning quantum numbers to the internal fields following some basic rules. However, not all will lead to genuine three-loop neutrino mass models. For that, one should guarantee the absence of fields that generate lower order contributions. For example $\nu_R$, $\Delta$ and $\Sigma$ of the basic three tree-level seesaws, or the scalar $S\equiv(\mathbf{1},\mathbf{4})_{3/2}$ together with the fermion $\Psi\equiv(\mathbf{1},\mathbf{3})_{1}$ from the $d=7$
tree-level BNT model \cite{Babu:2009aq}. Here, we introduced the notation $(\mathbf{\rm SU(3)_C},\mathbf{\rm SU(2)_L})_{Y}$ for the quantum numbers of internal particles. We will use this notation mostly in the figures. Note that we shortened this to $\mathbf{SU(2)_L}_{Y}$ when all particles in the model are colourless. Thus, for example, a fermionic $\mathbf{1}_0$ corresponds to a right-handed neutrino $\nu_R$. As discussed above, it is however possible in many cases to construct models that avoid lower order diagrams, despite the use of particles such as $\nu_R$, by adding additional symmetries by hand to the model. We will show one example of such a model below. In that case, we add a superscript $\omega$ to the particle quantum numbers to indicate which particles are charged under the new symmetry. The simplest possibility is usually a $Z_2$.

\begin{figure}
    \centering
    \includegraphics[width=\textwidth]{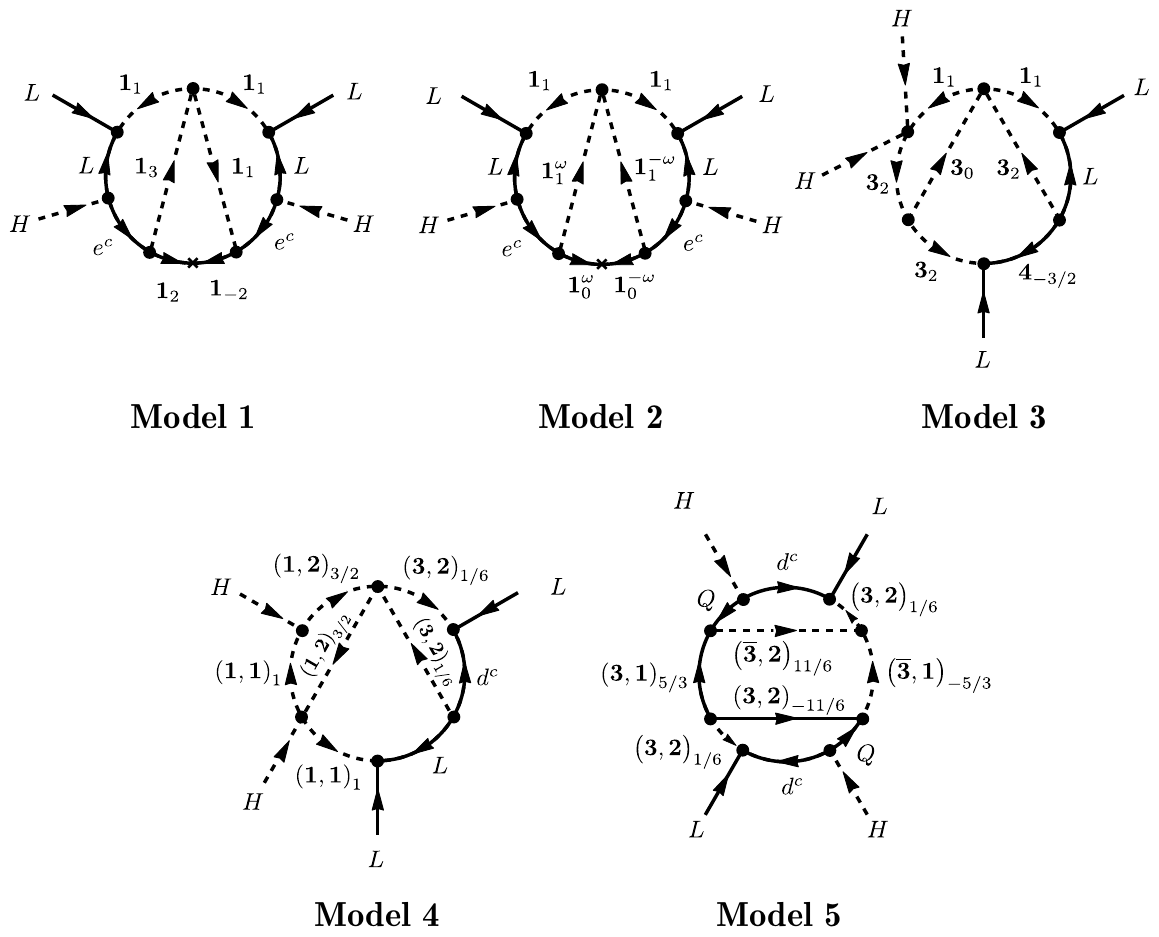}
    \caption{Five examples of three-loop $d=5$ genuine neutrino mass models. See the text for comments on the notation.}
    \label{fig:3loop:model_examples}
\end{figure}

Since there are endless possibilities for the quantum numbers of the internal fermions and scalars, the number of genuine models is infinite. Here, we will just show a few basic examples: Five comparatively simple models are shown in \fig{fig:3loop:model_examples}. Let us discuss them briefly.

\textbf{Model 1}, based on the same diagram as the KNT model \cite{Krauss:2002px}, can be considered as one of the simplest genuine three-loop models possible. The diagram needs only three singlets (two different scalars and one vector-like fermion) and no additional symmetry to produce a non-zero neutrino mass. All other models that we found need either (i) larger ${\rm SU(2)_L}$ representations and/or (ii) a larger number of beyond SM fields and/or (iii) an additional symmetry to avoid lower order diagrams.

\textbf{Model 2} is the simplest realisation of the KNT model. It also contains only three different singlets, as is the case for model 1. However, the KNT model needs an extra symmetry to avoid a tree-level seesaw contribution from the fermionic singlet (recall that $\mathbf{1}_0\equiv \nu_R$). We indicate the particles transforming non-trivially under the new symmetry, by writing their charge $\omega$ as a superscript. Note that for the simplest case of a $Z_2$ this simply reduces to the particles in the innermost loop to being odd, while the rest of the diagram contains only particles transforming even. There is a number of variations of this diagram in the literature containing larger ${\rm SU(2)_L}$ representations in the loops and coloured particles as well.

We have chosen \textbf{model 3} to show how larger ${\rm SU(2)_L}$ representations can also play a natural role in three-loop neutrino mass models. This model is associated to topology $T_{3}$, being the first three-loop model to do so in the literature, as far as we know. There are three new scalars, ${\bf 3}_0$, ${\bf 3}_2$, ${\bf 1}_1$ and one new fermion ${\bf 4}_{-3/2}$ (plus its vector partner). The model contains a triply-charged ``leptonic'' fermion as well as a triply charged scalar, and thus it should lead to a very rich accelerator phenomenology.

The second row in \fig{fig:3loop:model_examples} shows two models with coloured fields. Here, we have chosen the simplest possibilities for colour, i.e. we use only triplets. Variants with larger colour representations could be created in a straightforward manner. In the diagram of \textbf{model 4}, colour runs only in one of the loops. This model has again only three new fields. However, in contrast to models 1, 2 and 3, here all new fields are scalars. The scalar $(\mathbf{3,2})_{1/6}$ is a leptoquark, thus standard LHC searches for these particles should put bounds on this model. Note that model 4 descends from our topology $T_1$ (we have not found any model with this topology in the literature). \textbf{Model 5} is a second example with coloured particles: it needs 5 exotic fields, but no additional symmetry. Note again that the exotic fermions in this model, $({\bf 3},{\bf 2})_{-11/6}$ and $({\bf 3},{\bf 1})_{5/3}$, both must be vector-like.

In the following we will discuss models 1 and 5 in more detail, including a numerical calculation of the relevant three-loop integrals. We will only consider the unrealistic case where one neutrino is massive, adding just one generation of every new field for simplicity. Therefore, our results should be understood as estimates of the typical scale of the neutrino masses and not as a prediction for their exact values. Note, however, that it is possible to fit all neutrino oscillation data, including the mixing angles, in radiative models. Usually, adding more copies of the exotic fermions is enough (we discuss this briefly at the end of this section).

Also, unless we say otherwise, in the following all dimensionless couplings are set to one and, in this simplified setup, we will not put a hierarchy nor flavour structure in the indices of the Yukawa couplings. (This is done for simplicity; it is not a requirement/constraint on the models.) When there are no analytical solutions, the calculations for the three-loop integrals have been done numerically with the code \texttt{pySecDec} \cite{Borowka:2017idc}. For detailed definitions of the loop integrals see the \app{app:loops}.

%%%%%%%%%%%%%%%%%%%%%%%%%%%%%%%%%%%%%%%%%%%%%%%%%%%%%%%%%%%%%%%
\subsection{Model 1}\label{subsec:3loop:model1}

\begin{table}
    \centering
    \begin{tabular}{cccc}
        \hline \hline
        Fields & ${\rm SU(3)_C}$ & ${\rm SU(2)_L}$ & ${\rm U(1)_Y}$  \\
        \hline
        $S_1$ & 1 & 1 & 1
        \\
        $S_2$ & 1 & 1 & 3
        \\
        $F$ & 1 & 1 & 2
        \\
        \hline \hline
    \end{tabular}
    \caption{Quantum number assignments for the beyond-the-SM fields of model 1 (compare to \fig{fig:3loop:model_examples}).}
    \label{table:3loop:model1}
\end{table}

Model 1 contains the Standard Model fields plus the ones given in \tab{table:3loop:model1}. The fermion $F$ has a vector partner $\overline{F}$, which is not explicitly shown in the table. The neutrino mass in model 1 is generated from the following terms in the Lagrangian,
\begin{align} \label{eq:3loop:lagrangian_model1}
  \mathscr{L} & = \mathscr{L}_{SM} + Y_1 \overline{L^c} L S_1
        +  Y_2 \overline{F} e^c S_1
        +  Y_3 F e^c S_2^\dagger  +  \lambda_S S_2 (S_1^\dagger)^3  +  \hc
	\nonumber
	\\
	&+  m_1^2 S_1^\dagger S_1   +  m_2^2 S_2^\dagger S_2
        +  M \overline{F} F + \cdots.
\end{align}
Other quartic terms in the scalar potential (such as $H^\dagger H S^\dagger S$) are not explicitly given here, as they will only result in uninteresting corrections to the scalar masses. It is worth mentioning that $Y_1$ in \eq{eq:3loop:lagrangian_model1}, in principle, is a $3 \times 3$ antisymmetric matrix. This fact is important if one wants to fit the complete neutrino oscillation data (see the discussion at the end of this section).

The mass diagram of model 1 in \fig{fig:3loop:model_examples} shows that the neutrino mass is proportional to the product of two masses of Standard Model charged leptons. Considering the dominant contribution with two $\tau$ leptons running in the loop, the neutrino mass matrix is then calculated straightforwardly as,
\begin{equation} \label{eq:3loop:NuMassModel1}
    \small
    (M_\nu)_{\alpha \beta} = -\frac{3!}{(16\pi^2)^3} \lambda_S \frac{ m_\tau^2 }{M} 
    \left[ (Y_1)_{\alpha \tau} (Y_2)_\tau (Y_3)_\tau (Y_1)_{\tau \beta} \; +  \; (\alpha \leftrightarrow  \beta) \right]
    \; F_{loop} \! \left( x_1, x_2 \right) \, .
\end{equation}

After EWSB, in the mass eigenbasis, model 1 generates the diagram $D_3^M$ in \fig{fig:3loop:normalgenuinediagrams} with a mass insertion in each of the three internal fermions. From the diagram, and assigning momenta to the internal fields, we get the following expression for the loop function $F_{loop}$, which is given by a dimensionless integral,
\begin{equation} \label{eq:3loop:Floopmodel1}
\medmath{
    F_{loop} \! \left( x_1, x_2 \right) = \!\!\!\!\!\! \iiint\limits_{(k_1,k_2,k_3)} \!\!\! \frac{ 1 }{ [k_1^2] [k_1^2-x_1] [k_2^2] [k_2^2-x_1] [k_3^2-1] [(k_2-k_3)^2-x_1] [(k_3-k_1)^2-x_2] } \,}.
\end{equation}
Here $m_\tau$ was neglected, while the other masses were normalised to the vector-like mass $M$ of the field $F$,
\begin{equation}
    x_1 = \frac{m_1^2}{M^2},\; x_2 = \frac{m_2^2}{M^2} \, .
\end{equation}
We also used the short-hand notation,
\begin{equation} \label{eq:3loop:intnotation}
    \int\limits_k \equiv (16 \pi^2) \int \frac{ d^4 k }{(2\pi)^4} \, .
\end{equation}
For the full decomposition of $F_{loop} \! \left( x_1, x_2 \right)$, in terms of master integrals, suitable for numerical evaluation, see \app{app:loops}.

\begin{figure}[t]
    \centering
    \includegraphics[width=0.7\textwidth]{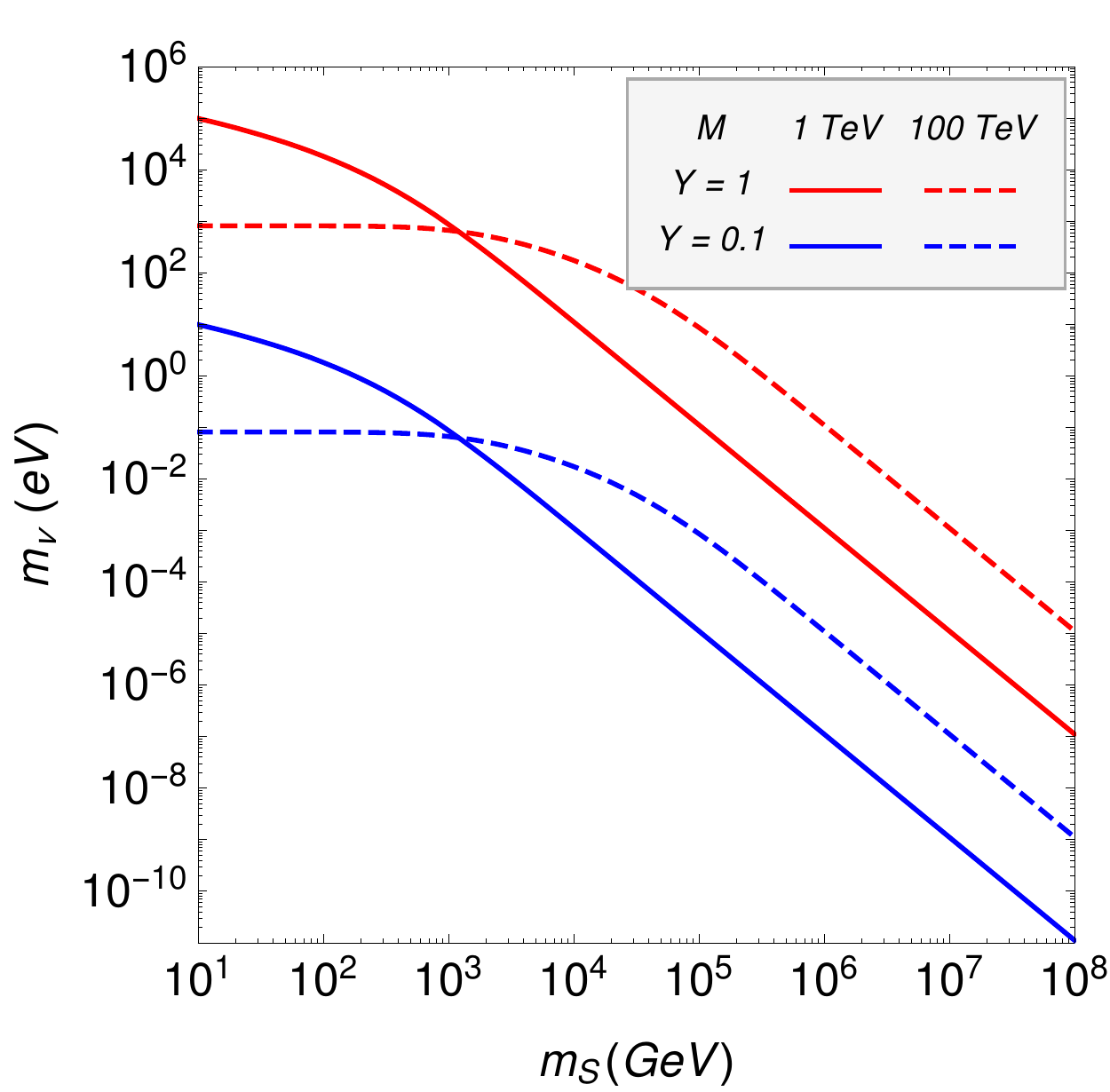}
    \caption{The neutrino mass scale in model 1 for a few sample choices of parameters, see text for details.}
    \label{fig:3loop:NumassModel1}
\end{figure}

In \fig{fig:3loop:NumassModel1} we show the neutrino mass scale for different choices of parameters. For the calculation we have taken all masses of the new scalar singlets equal, i.e. $m_1 = m_2 = m_S$. As can be seen from \eq{eq:3loop:NuMassModel1}, $M_\nu$ is proportional to the fourth power of the Yukawas. For masses of order $\mathcal{O}(1)$ TeV one can reproduce the neutrino atmospheric scale ($\sim 0.05$ eV) with Yukawas $\mathcal{O}(10^{-2}-10^{-1})$.

The dependence of the neutrino mass on the masses of the fields in the loop is also understood straightforwardly. From the diagram of model 1 in the mass eigenbasis, it is straightforward to see that the neutrino mass should scale as,
\begin{equation}
    M_\nu \sim m_{\tau}^2 \frac{M}{\Lambda^2}, 
\end{equation}
where $\Lambda$ is some characteristic energy scale. As the loop function contains only two mass scales, i.e. $m_S$ and $M$ (neglecting $m_\tau$), for $m_S \gg M$ the neutrino mass decreases with $1/m_S^2$, while for small scalar masses one obtains a constant value (for a fixed $M$).\footnote{This is true only if $M \gg m_{\tau}$. It can be easily checked that for $M \rightarrow 0$ the integral vanishes and neutrinos remain massless.}

In summary, as \fig{fig:3loop:NumassModel1} shows, the correct neutrino mass scale is obtained in this model for a wide range of masses. In one extreme case, the new physics scale can be as high as $10^3$ TeV if all Yukawas are order one. On the other hand, even for masses $M$ and $m_S$ of the order of $1$ TeV, Yukawa couplings can be as large as ${\cal O}(0.1)$.

%%%%%%%%%%%%%%%%%%%%%%%%%%%%%%%%%%%%%%%%%%%%%%%%%%%%%%%%%%%%%%%
\subsection{Model 5}\label{subsec:model5}

We have performed an analogous study for model 5 in \fig{fig:3loop:model_examples}. In the mass eigenbasis, the neutrino diagram corresponds to diagram 10 in \fig{fig:3loop:normalgenuinediagrams} with a mass insertion on both d-quark internal lines.

\begin{table}
    \centering
    \begin{tabular}{ccccc}
        \hline \hline
        Fields & ${\rm SU(3)_C}$ & ${\rm SU(2)_L}$ & ${\rm U(1)_Y}$  \\
        \hline
        $S_Q$ & 3 & 2 & 1/6
        \\
        $S_1$ & 3 & 1 & 5/3
        \\
        $S_2$ & 3 & 2 & -11/6
        \\
        $F_1$ & 3 & 1 & 5/3
        \\
        $F_2$ & 3 & 2 & -11/6
        \\
        \hline \hline
    \end{tabular}
    \caption{Quantum numbers of the new fields given in model 5 (see \fig{fig:3loop:model_examples}).}
    \label{table:3loop:model5}
\end{table}

The new fields present in the model are listed in \tab{table:3loop:model5}. Among others, the Lagrangian contains the following interactions,
\begin{align} \label{eq:3loop:lagrangian_model5}
    \mathscr{L} = \mathscr{L}_{SM} &+ Y_1 L d^c S_Q + Y_2 Q F_1 S_2 + Y_3 Q F_2 S_1 + \hc
    \\ \nn
     & + Y^L_4 \overline{F_1}\, \overline{F_2} S_Q^\dagger + Y^R_4 F_1 F_2 S_Q + \mu_S S_Q^\dagger S_1^\dagger S_2^\dagger +   \hc
    \\ \nn
    &+  M_{F_1} \overline{F_1} F_1 + M_{F_2} \overline{F_2} F_2 + m_{S_Q}^2 S_Q^\dagger S_Q + m_{S_1}^2 S_1^\dagger S_1 + m_{S_2}^2 S_2^\dagger S_2  + \cdots.
\end{align}
Additional quartic terms in the scalar potential coupling the new scalars and the higgs field are not written down explicitly.

Similarly to model 1, the neutrino mass in model 5 is proportional to the product of two d-quark masses. Thus, one expects the dominant contribution will be proportional to the mass of the bottom quark squared,
\begin{align} \label{eq:3loop:NumassModel5}
    (M_\nu)_{\alpha \beta} &= - \frac{12 \mu_S}{(16\pi^2)^3} \frac{m_b^2}{ m_{S_Q}^2 }
    \left[ (Y_1)_{\alpha b} (Y_2)_b (Y_3)_b (Y_1)_{b \beta} \; +  \; (\alpha \leftrightarrow  \beta) \right]  
    \\ \nn
    & \times \left[ Y^L_4 F_L( x_1,x_2,x_3,x_4 ) + Y^{R*}_4 F_R( x_1,x_2,x_3,x_4 ) \right] \, ,
\end{align}
where
\begin{equation} \label{eq:3loop:Floop_model5}
    F_L( x_1,x_2,x_3,x_4 ) = \iiint\limits_{(k_1,k_2,k_3)} \!\!\! \frac{ \sqrt{x_1 x_3} }{\mathcal{D}} \, , \quad 
    F_R( x_1,x_2,x_3,x_4 ) = \iiint\limits_{(k_1,k_2,k_3)} \!\!\! \frac{ \slashed{k_3}(\slashed{k_2}-\slashed{k_3} }{\mathcal{D}} \, ,
\end{equation}
are two dimensionless loop integrals normalised to the mass of the new scalar $S_Q$,
\begin{equation}
    x_1 = \frac{M_{F_1}^2}{m_{S_Q}^2},\; x_2 = \frac{m_{S_1}^2}{m_{S_Q}^2},\; x_3 = \frac{M_{F_2}^2}{m_{S_Q}^2},\; x_4 = \frac{m_{S_2}^2}{m_{S_Q}^2} \, ,
\end{equation}
with the common denominator,
\begin{equation} %\label{eq:}
    \mathcal{D} = [k_1^2] [k_1^2-1] [k_2^2] [k_2^2-1] [k_3^2-x_1] [k_3^2-x_2] [(k_2-k_3)^2-x_3] [(k_3-k_1)^2-x_4] \, .
\end{equation}
For the decomposition of both integrals in terms of master integrals we refer again to \app{app:loops}.

\begin{figure}[t]
    \centering
    \includegraphics[width=0.7\textwidth]{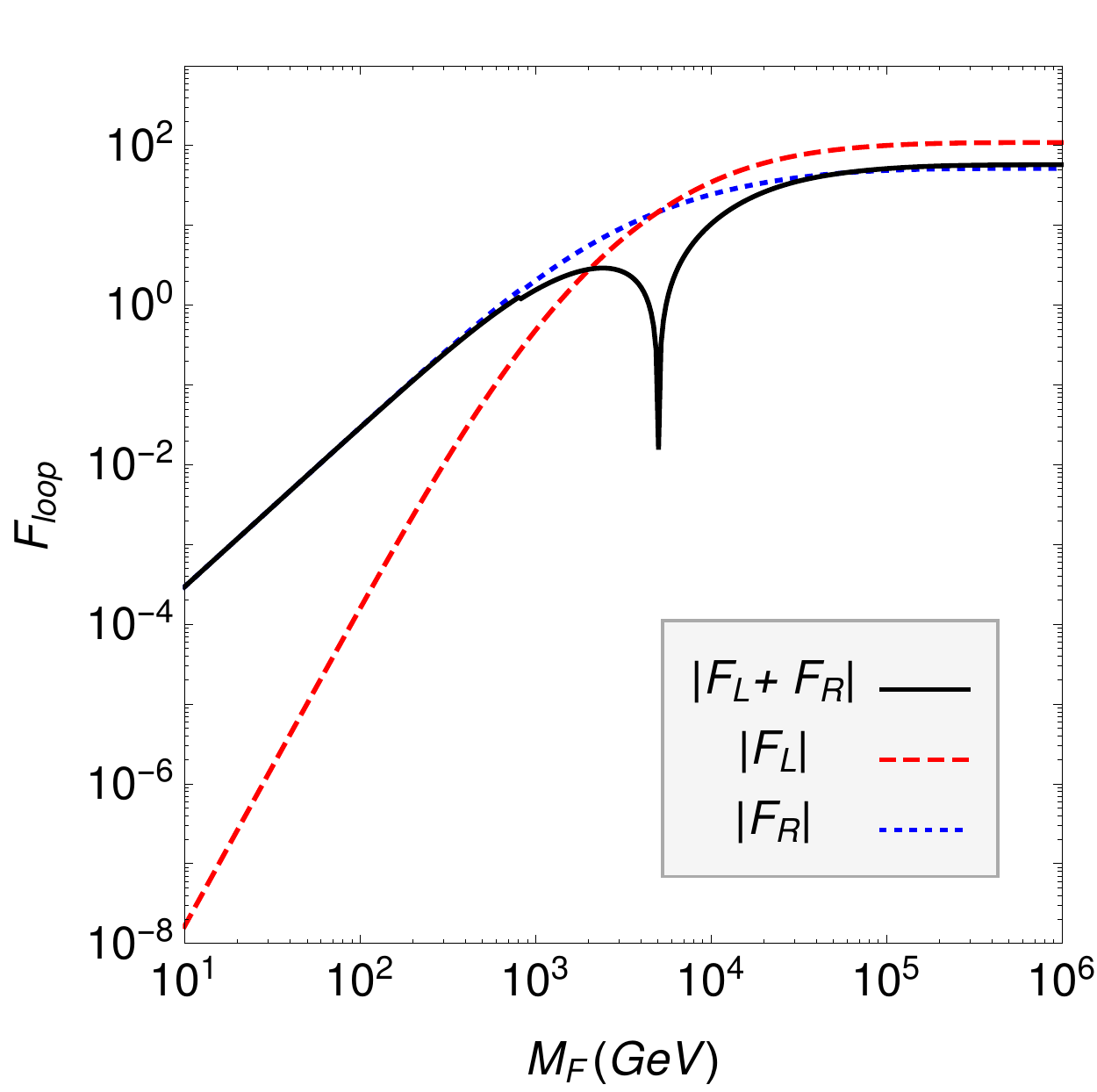}
    \caption{Loop functions, see \eq{eq:3loop:Floop_model5}, that enter the neutrino mass \eq{eq:3loop:NumassModel5} generated by model 5, see \fig{fig:3loop:model_examples}. For this plot, the masses of $S_1$ and $S_2$ are taken to be $1$ TeV, while both fermion masses $M_{F_1} =  M_{F_2} = M_F$.}
    \label{fig:3loop:Loops model5}
\end{figure}

There are two different integrals contributing to the neutrino mass, as can be seen in \eq{eq:3loop:NumassModel5}, due to the fact that one may flip the chirality of the internal fermions with mass insertions. $F_1$ and $F_2$ must have vector-like masses\footnote{New fermions beyond the Standard Model fields must have vector-like mass terms for phenomenological reasons. For instance, a fourth chiral family is excluded by the Higgs production measurements at the LHC.}, thus there are two possible choices for the chiral structure of the vertex, i.e. either one uses $Y_4^L \overline{F_1}\, \overline{F_2}$ or $Y_4^{R*} (F_1)^*(F_2)^*$. This yields one loop integral with $M_{F_1} M_{F_2}$ in the numerator and another one with the loop momenta of both fermions instead of their masses. This fact is important because, unlike in model 1, for model 5 a cancellation can occur between both contributions, as shown in \fig{fig:3loop:Loops model5}. This cancellation occurs when all the masses in the diagram are of the same order. For example, $\mathcal{O}(1)$ TeV in the case shown in \fig{fig:3loop:Loops model5}.

\begin{figure}[t]
    \centering
    \includegraphics[width=0.7\textwidth]{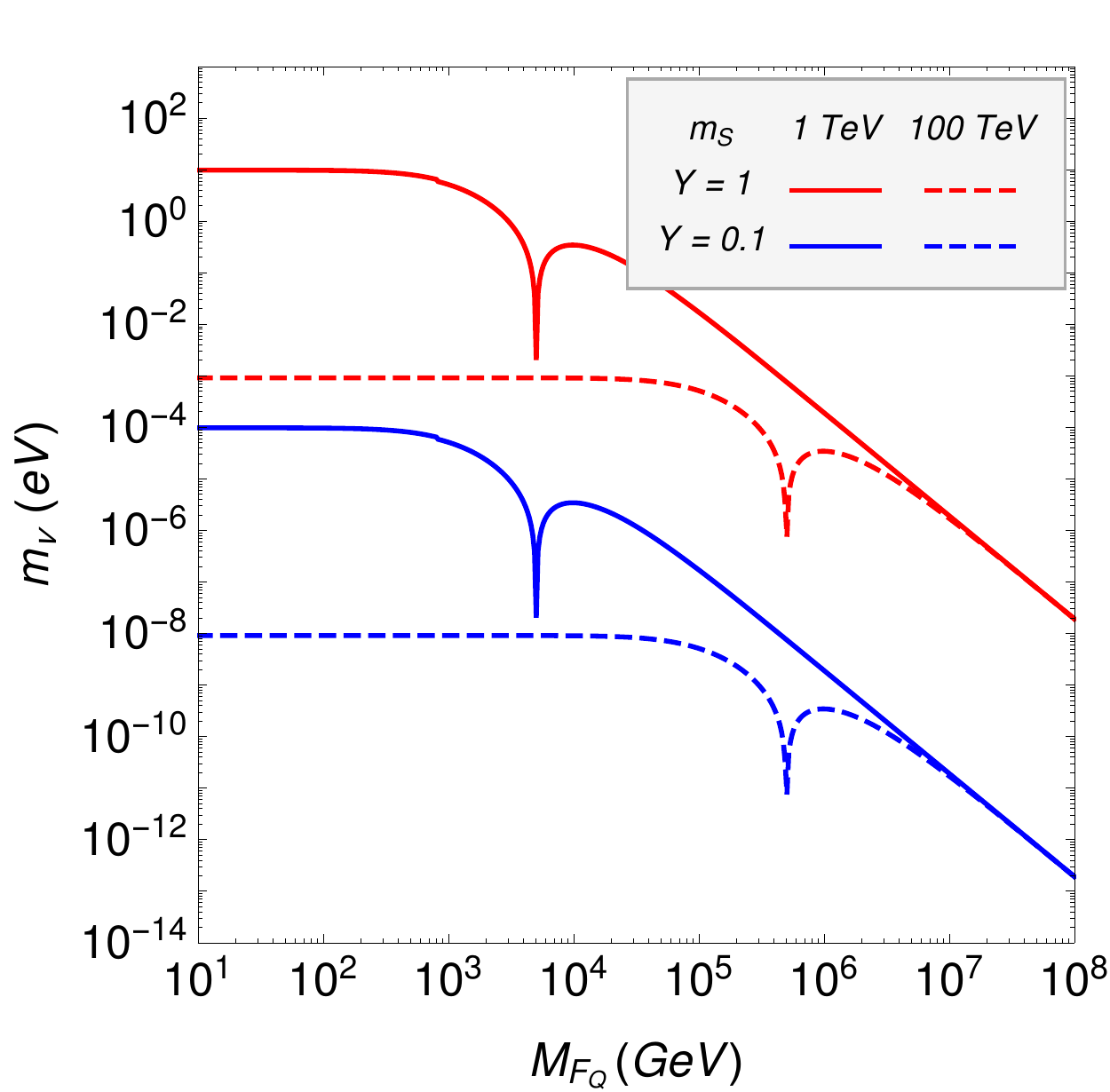}
    \caption{The neutrino mass scale for model 5 in \fig{fig:3loop:model_examples}, for some example choices of parameters. See text for details.}
    \label{fig:3loop:NumassModel5}
\end{figure}

In \fig{fig:3loop:NumassModel5} we show some examples for the neutrino mass scale for specific but arbitrary choices of parameters. Taking all the masses of the new scalars equal for simplicity, i.e. $m_{S_1} = m_{S_2} = m_S$, for masses of $\mathcal{O}(1)$ TeV Yukawas around ${\cal O}(0.5)$ are needed to generate the atmospheric scale. The difference, compared to the previous case of model 1, arises from the fact that the loop integral in model 5 contains one extra propagator compared to model 1, as well as one extra Yukawa coupling. Thus, the neutrino mass scales differently in model 5. The dependence on the masses is again easily understood, considering that for model 5 one has,
\begin{equation} \label{eq:3loop:Mnu_approx_model5}
    M_\nu \sim \mu_S m_b^2 \left( \frac{M_{F}^2}{\Lambda^4} + \frac{1}{\Lambda^2} \right) \, .
\end{equation}
For $M_{F} \ll m_S$ one has a plateau whose height scales as $1/\Lambda^2$, instead of $1/\Lambda$ as in model 1, see \eq{eq:3loop:Mnu_approx_model5}, while for large fermion masses both models have the same behaviour.

It is worth mentioning that model 5 contains an extra mass scale, i.e. the coupling $\mu_S$ in \eq{eq:3loop:lagrangian_model5}. In \fig{fig:3loop:NumassModel5} we have chosen $\mu_S = m_S$. Increasing its value will smoothen the differences between both models, making it possible to reach the measured neutrino mass scale with smaller Yukawas couplings. On the other hand, the need of at least one neutrino mass of the order of $0.05$ eV can be interpreted as a lower limit on this parameter.

We have chosen to discuss models 1 and 5 in more detail because they span the typical range of three-loop neutrino mass models. By direct inspection of the genuine diagrams, listed in \fig{fig:3loop:normalgenuinediagrams}, it can be seen that every integral contains 7 or 8 propagators, leading to the same behaviour as in either model 1 or model 5, respectively, in the limit of large masses. For small scalar and fermion masses, the scale of $M_\nu$ depends on the numerator of the integral, i.e. the number of fermions inside the loop along with the chiral structure of each vertex, as well as the presence of Standard Model mass insertions.

Finally, we should point out that obviously any realistic neutrino mass model should be able to reproduce all neutrino oscillation data, i.e. the two neutrino mass squared differences along with three neutrino mixing angles and phases. The aim of our simplified examples was to show the typical neutrino mass scales in three-loop models; it was not to make a detailed neutrino flavour fit. However, going beyond the simplified scenario where there is just one non-zero charged lepton (or down-quark) mass the neutrino mass matrices given in \eq{eq:3loop:NuMassModel1} and \eq{eq:3loop:NumassModel5} have rank-2. This makes it possible to fit normal or inverted hierarchical neutrino spectra, including a correct fit for angles and phases, in both model 1 and model 5. In order to fit a degenerate neutrino spectrum, a rank-3 neutrino mass matrix is needed. This can be achieved easily in model 5 just by adding extra copies of the new fields, for instance having two copies of $F_1$ and $F_2$. However, fitting a degenerate spectrum is not possible for the case of model 1, disregarding the number of copies of the fields. This is due to the antisymmetry of the Yukawa $Y_1$.\footnote{This can be understood recalling that the rank of any $n \times n$ antisymmetric matrix is at most $n-1$ for odd $n$'s, together with the identity $\text{rank}(AB) \leq \text{min}( \text{rank}(A), \text{rank}(B) )$, for two arbitrary matrices $A$ and $B$.} Again, as with the overall mass scale, our two example models represent the two typical kind of models, that can be found at three-loop order.

%%%%%%%%%%%%%%%%%%%%%%%%%%%%%%%%%%%%%%%%%%%%%%%%%%%%%%%%%%%%%%%
%%%%%%%%%%%%%%%%%%%%%%%%%%%%%%%%%%%%%%%%%%%%%%%%%%%%%%%%%%%%%%%
\section{Summary}

In this chapter we have discussed the complete decomposition of the Weinberg operator at three-loop order. Our analysis concentrates on finding those topologies and diagrams
that can give the dominant contribution to the neutrino mass matrix, without the use of additional symmetries beyond those of the Standard Model. We call such topologies/diagrams genuine. We considered models with scalars and fermions only.

The requirement of ``genuineness'' eliminates the large majority of possible topologies: From more than four thousands, there are 99 topologies which satisfy this criteria. We have discussed how to identify these cases, and we listed them in \app{app:topos}. Those genuine topologies were sub-divided into two classes: Normal ones (44 topologies) and special ones (55 topologies). While the former can be found systematically by our selection criteria, the latter topologies form an exception to our general rules, as explained in detail in \sect{sec:3loop:specgen}. This exception is related to the fact that usually, if any three fields (or four scalars) can interact through a loop, then they can also do so through a renormalisable local interaction. However, for special combinations of fields this is is not true: for example, the Higgs-Higgs-singlet local interaction is null, but a  loop with these 3 external scalars does not need to have a zero amplitude.

The 44 topologies we have found are associated to a total of 228 diagrams in the electroweak basis, from which one can get three-loop leading order neutrino masses contributions. Going to the mass eigenstate basis, this list is reduced to only 18 diagrams (they are shown in \fig{fig:3loop:normalgenuinediagrams}). To these normal genuine diagrams one has to add 271 special ones, in the electroweak basis, which give another 20 mass eigenstate diagrams (see \fig{fig:3loop:specialgenuinediagrams}). All those diagrams can be calculated with only five master integrals which where analysed in the literature previously \cite{Martin:2016bgz}. We give them in \app{app:loops}, where we also show how the loop integrals for specific neutrino mass models can be constructed with two examples.

We have then also shown in \sect{sec:3loop:examples}, how our general results can be easily used to build genuine three-loop neutrino mass models. A few examples are briefly mentioned, and for two of them we have calculated the neutrino mass scale in more detail. This allows us to estimate the typical parameter range (couplings and masses), for which these three-loop models can explain the measured neutrino oscillation data. We find that dimension 5 three-loop models will give a good fit to data if the new particles have masses roughly in the range $1-10^3$ TeV. Such a low scale is partially testable at current and future colliders, as well as in experiments searching for lepton flavour violation. Thus, three-loop models are interesting constructions, since they are experimentally testable. We hope that model builders will find our results useful.

\pagebreak
\fancyhf{}

%% file: Chapters/Dirac_2loop/Chapter_Dirac2loop.tex
\fancyhf{}
\fancyhead[LE,RO]{\thepage}
\fancyhead[RE]{\slshape{Chapter \thechapter. Classification of two-loop $d = 4$ Dirac mass models}}
\fancyhead[LO]{\slshape\nouppercase{\rightmark}}

\chapter{Systematic classification of two-loop $d = 4$ Dirac neutrino mass models}
\label{ch:dirac2l}
\graphicspath{ {Chapters/Dirac_2loop/} }

In this chapter, based on \cite{CentellesChulia:2019xky}, we will build on previous works on classifications of Dirac neutrino models \cite{Ma:2016mwh, Yao:2017vtm, CentellesChulia:2018gwr, Yao:2018ekp} (see \sect{subsec:numass:dirac_class}), and give a systematic classification of dimension four Dirac neutrino mass models at the two-loop level.
%For the two-loop models to provide the leading order contribution to neutrino masses, one needs to ensure that the dimension 4 tree-level and one-loop contributions are absent. This can happen in a variety of scenarios involving flavour symmetries \cite{Chulia:2016ngi, Chulia:2016giq, CentellesChulia:2017koy} or right-handed neutrinos with chiral lepton number charges \cite{Ma:2014qra, Ma:2015mjd, Bonilla:2018ynb}. Furthermore, one has to ensure that Majorana mass terms are absent at all orders. This can be easily accomplished by, for example, requiring that lepton number (or $B-L$) is conserved exactly \cite{Farzan:2012sa}, or at least an appropriate subgroup of it \cite{Ma:2015raa, Hirsch:2017col, Fonseca:2018ehk, Calle:2018ovc, Bonilla:2019hfb, Dasgupta:2019rmf}, or also by invoking the presence of additional new symmetries \cite{Ma:2019yfo}.

We consider that neutrino masses are generated from the dimension four operator $\bar{L} H^c \nu_R$. Starting from this operator, in general Dirac neutrino mass $m_\nu$ can roughly be written as,
\begin{equation} \label{eq:mnu estimate}
    m_\nu \sim C \left( \frac{1}{16 \pi^2} \right)^n \vev{H^0} \, ,    
\end{equation}
where $n$ is the number of loops needed to generate the Dirac neutrino masses, $\vev{H^0} \sim v$ is the Standard Model Higgs vacuum expectation value and $C$ is a dimensionless constant containing all the information of the couplings involved in the neutrino mass. The aim of going to radiative models is that one can explain the smallness of neutrino masses naturally without requiring extremely small couplings. For instance, focusing on the $n = 2$ case and in accordance with the cosmological constraints \cite{Aghanim:2018eyx}, we can take the neutrino masses to be of the order of the atmospheric mass scale $\mathcal{O}(0.05)$ eV \cite{deSalas:2017kay}, which by means of \eq{eq:mnu estimate} implies couplings of order $0.1$-$0.01$.
\\

The chapter is organised as follows: the main body is presented in the first section. We start by classifying all the possible topologies realising the dimension 4 operator $\bar{L} H^c \nu_R$ that can give a dominant Dirac neutrino mass at two-loop level. We then generate all possible diagrams and arrange them in three differentiated classes according to their symmetry and field restrictions. In \sect{sec:dirac2l:genmodels} we explain how to generate models within each class of diagrams, and we then illustrate how it can be implemented, studying in detail two examples in \sect{sec:dirac2l:models}.
%Finally, in \sect{sec:dirac2l:dm} we shortly discuss the connection between the symmetry that ensures the Dirac nature of neutrinos and the stability of dark matter, topic that will be later analysed in detail in \ch{ch:dm_DiracMajo}.

%%%%%%%%%%%%%%%%%%%%%%%%%%%%%%%%%%%%%%%%%%%%%%%%%%%%%%%%%%%%%%%
%%%%%%%%%%%%%%%%%%%%%%%%%%%%%%%%%%%%%%%%%%%%%%%%%%%%%%%%%%%%%%%
\section{Classification: from topologies to models} \label{sec:dirac2l:class}

We start our discussion by first introducing certain key concepts and setting up the notation in generic terms. Our aim is to consider all possible decompositions of the dimension four operator $\bar{L} H^c \nu_R$ at two-loop level for Dirac neutrino masses. One of the first key requirements is that the Dirac nature of neutrinos should be protected by some symmetry. This symmetry should remain exact and be such that it forbids the Majorana masses at all loop orders. Such feature can be easily achieved by the global lepton number ${\rm U(1)_L}$ symmetry already present in the Standard Model or one of its appropriate unbroken subgroups. However, on general grounds this symmetry protection can in principle have different origins and need not be related with lepton number.

Given an appropriate symmetry protecting the Dirac nature of neutrinos, the next issue is regarding the leading contribution to neutrino masses. Since we are interested in two-loop UV completions of the operator $\bar{L} H^c \nu_R$, the tree-level and one-loop contributions should be absent. This can also happen naturally in many models involving an additional $Z_2$ symmetry \cite{Chulia:2016ngi}, a flavour symmetry \cite{Chulia:2016giq, CentellesChulia:2017koy} or chiral ${\rm U(1)_L}$ charges for $\nu_R$ \cite{Ma:2014qra, Ma:2015mjd, Ma:2015raa, Bonilla:2018ynb, Bonilla:2019hfb}. In fact, it has been recently shown that an appropriate residual subgroup of the lepton number (or equivalently $B-L$ symmetry) alone is enough to guarantee the Dirac nature of neutrinos to all loops and ensure that the leading contribution to neutrino mass only arises at higher loops \cite{Bonilla:2018ynb}. Since all the requirements to have Dirac neutrino masses with a leading contribution at two-loops can always be met, henceforth we will take a different approach and not bother about the details of the symmetries required for an specific model. Instead, we will rather focus on the classification of all such possible models in general.

In this section we begin by looking at how one can systematically organise and analyse all the two-loop realisations of $\bar{L} H^c \nu_R$. We make use of the usual definitions for topology, diagram and model-diagram, already given in \sect{sec:numass:class}. An important concept in all the classifications is the \textit{genuineness} of a topology or diagram. We identify as \textit{genuine} those model-diagrams (and consequently the topologies or diagrams which generate them) for which the main contribution to the neutrino masses arises at two-loops. Contrary to the Majorana case, for Dirac neutrinos one always needs a symmetry argument to forbid the Yukawa coupling $\bar{L} H^c \nu_R$ at tree-level, as discussed before. For this reason, every finite 1 Particle Irreducible (1PI) topology\footnote{Note that, in general, loop integrals have finite and divergent parts. In any consistent renormalisation scheme infinite integrals always require a lower order counter term to absorb the infinities, so they are never genuine.} is genuine in our sense enforcing the correct symmetry and transformations of the fields in order to avoid lower order contributions. 

Although in principle the choice of the symmetries used to forbid tree-level masses and ensuring the Dirac nature of neutrinos is model-dependent, some general conclusions can be given in order to establish a useful classification for model builders.

It is important to clarify that if one imposes a symmetry that forbids the tree-level Dirac mass term, all the realisations of the operator $\bar{L} H^c \nu_R (H^\dagger H)^n$ with $n\in\mathbb{N}$ will automatically vanish. For this reason, one must break such a symmetry, either softly or spontaneously, in order to allow radiative or higher-dimensional Dirac neutrino mass models such that the tree-level term is still absent \cite{Ma:2014qra, Chulia:2016ngi, Bonilla:2018ynb}. Keep in mind that any model with a softly-broken symmetry can be replaced by one with spontaneous breaking by just adding a VEV-carrying scalar with the adequate charge in all the soft-breaking terms. Since this would increase the dimensionality of the UV complete Dirac mass operator, we will not study it further.

%%%%%%%%%%%%%%%%%%%%%%%%%%%%%%%%%%%%%%%%%%%%%%%%%%%%%%%%%%%%%%%
\subsection{Topologies} \label{subsec:dirac2l:topos}

\begin{figure}
\centering
    \includegraphics[width=0.3\textwidth]{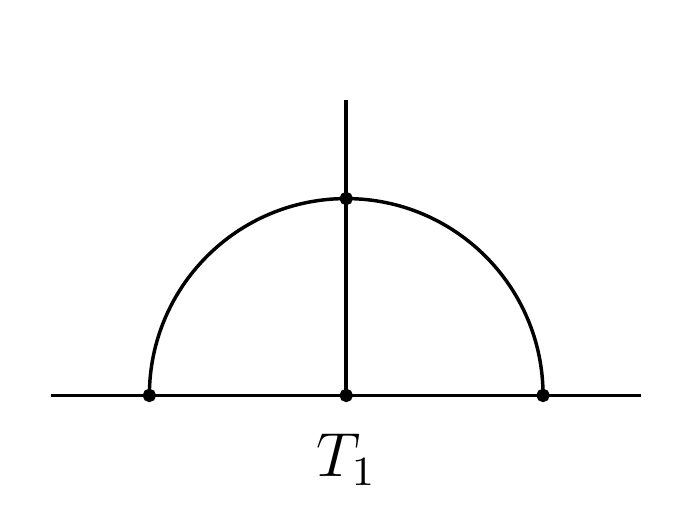}
    \includegraphics[width=0.3\textwidth]{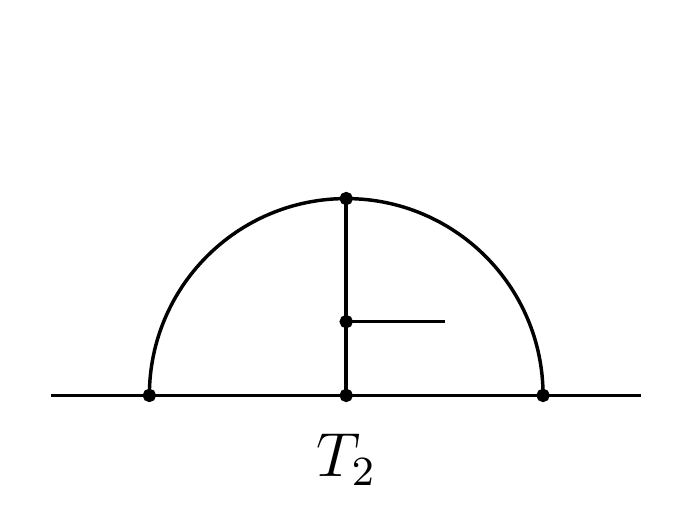}
    \\
    \includegraphics[width=0.3\textwidth]{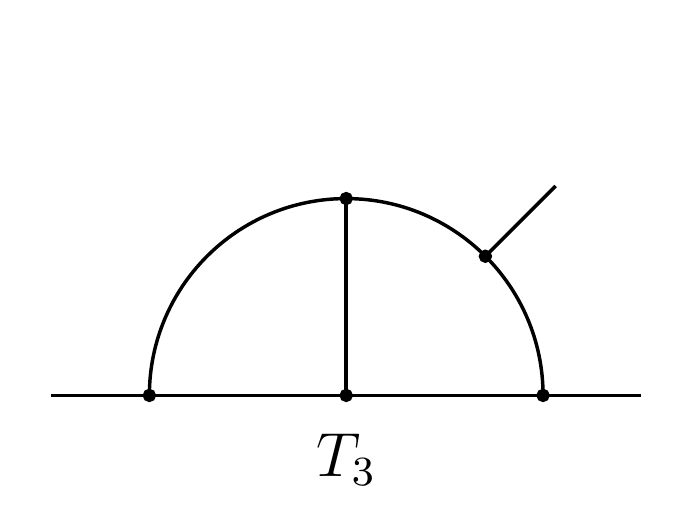}
    \includegraphics[width=0.3\textwidth]{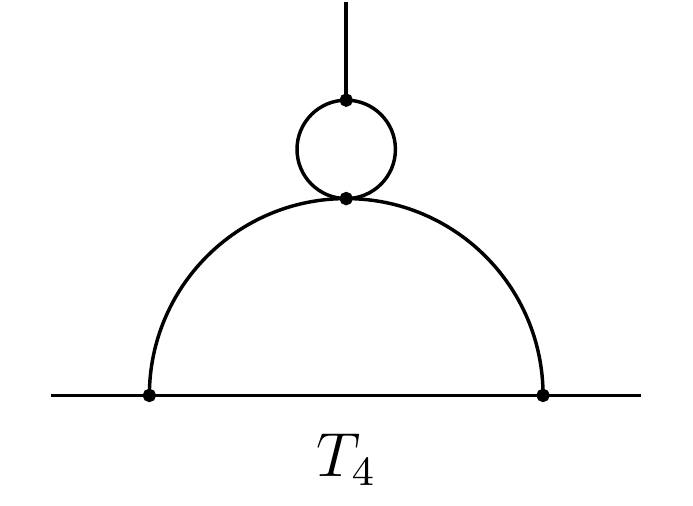}
    \includegraphics[width=0.3\textwidth]{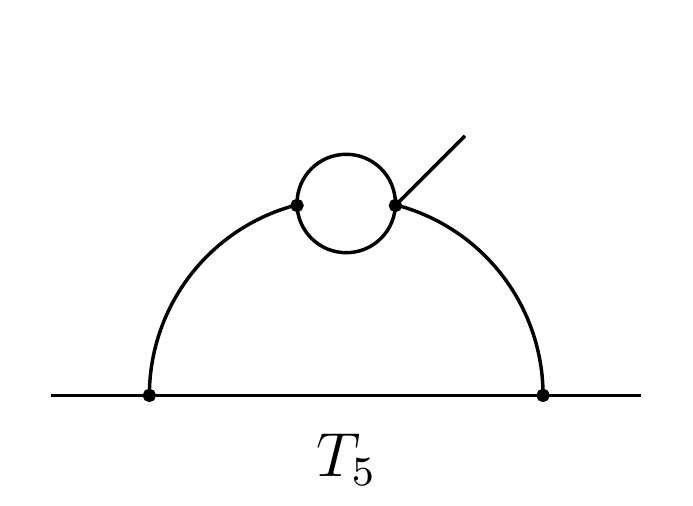}
    
    \caption{Finite two-loops 1PI topologies with 3- and 4-point vertices and three external lines. $T_1$ and $T_2$ generate in general genuine models. Topologies $T_3$, $T_4$ and $T_5$ are a special kind of genuine topologies, diagrams generated from them are finite but contain a 3-point or 4-point renormalisable vertex which can be in principle reducible, generating a lower order contribution. See the text for details.}
    \label{fig:dirac2l:topologies} 
\end{figure}

We generate all possible connected topologies with two-loops, 3- and 4-point vertices and three external lines. This gives a total number of 70 topologies. From these 70 topologies we remove all the topologies corresponding to tadpoles and self-energy diagrams as they always imply infinite parts in the loop integral. Furthermore, since at the topology level we have not specified the Lorentz nature of the lines, there are also non-renormalisable diagrams, for instance, 3-point vertices with only fermions or 4-point vertices with a fermion insertion. After removing all these topologies a small set of 5 1PI topologies remains, shown in \fig{fig:dirac2l:topologies}.

These 5 1PI topologies can be divided into two differentiated sets of topologies. Topologies $T_3$, $T_4$ and $T_5$ contain an internal loop that can be compressed to a 3-point vertex, whereas the two remaining topologies $T_1$ and $T_2$ do not. One can argue that the latter are genuine, while the former are corrections to their corresponding one-loop topologies, as they contain a loop realisation of a renormalisable vertex. Nevertheless, there are various ways to address this reducibility, leading to differentiated classes at the diagram level.

\begin{figure}
\centering
    \includegraphics[width=\textwidth]{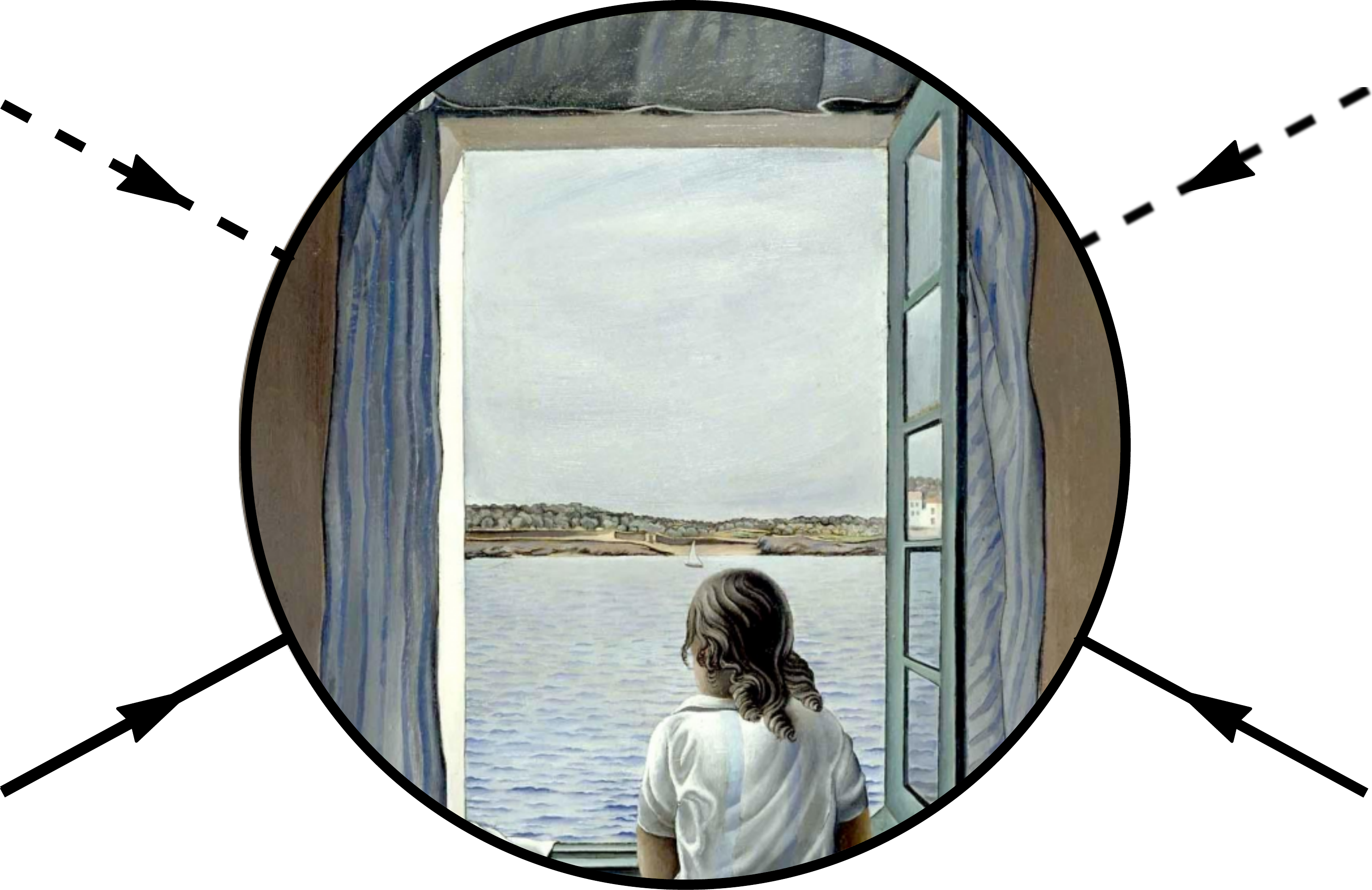}
    \caption{Set of renormalisable diagrams generated from topology $T_3$. We found that the topology is not genuine at the diagram level as all the diagrams contain a reducible 3-point loop vertex, coloured in red for each case. The same happens with $T_4$ and $T_5$. Among the diagrams two differentiated sets are given regarding the type of reducible loop vertex. This will be important at the model-diagram level, in order to promote this a priori non-genuine diagrams to genuine in \sect{subsec:dirac2l:class2} and \sect{subsec:dirac2l:class3}.}
    \label{fig:dirac2l:diagramsT3}
\end{figure}

As an example, one can construct all possible diagrams of topology $T_3$ and see whether there is a reducible (renormalisable) loop interaction. In \fig{fig:dirac2l:diagramsT3}, we can see the procedure to follow. Given a topology one should introduce scalar and fermion lines such that there are two external fermions and one external scalar, as required for the diagram to generate effectively $\bar{L} H^c \nu_R$. Then, one should check if any diagram contains a loop realisation of a renormalisable vertex (marked in red in \fig{fig:dirac2l:diagramsT3}). If so, a priori, the diagram is not genuine. If all of the diagrams generated from a certain topology are not genuine, then that topology too is considered not genuine. This is the standard way to address genuineness. In general, one can classify topologies as genuine checking reducibility on any loop at the diagram level. Nevertheless, loopholes to this procedure can be found, and we will exploit them in the subsequent sections.

We now move to the construction and classification of genuine diagrams. From the set of 5 topologies, a total of 18 diagrams can be built with two external fermion lines, one external scalar and containing only renormalisable vertices. One immediately finds three classes among the 18 diagrams by looking at the compressibility of one-loop vertices in the diagram, as well as the Lorentz nature of these vertices.

Note that all these conclusions so far are derived taking into account only fermions and scalars, but they can be directly generalised to include vectors. No new topology or diagram appear if vectors are considered. To extend our classification to vectors one just has to replace one or more scalars with vectors, provided that the resulting diagram is still renormalisable\footnote{Note that some vertices with vectors cannot be built such as vector-vector-vector-scalar.}.

%%%%%%%%%%%%%%%%%%%%%%%%%%%%%%%%%%%%%%%%%%%%%%%%%%%%%%%%%%%%%%%
\subsection{Completely genuine diagrams} \label{subsec:dirac2l:class1}

\begin{figure}
\centering
    \includegraphics[width=0.3\textwidth]{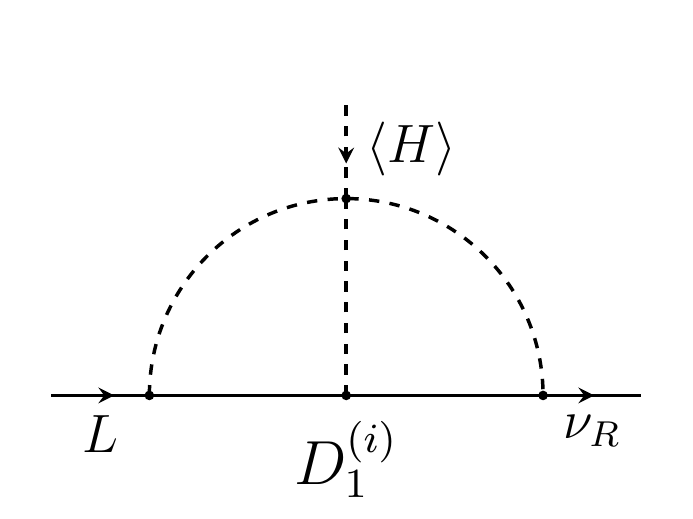}
    \includegraphics[width=0.3\textwidth]{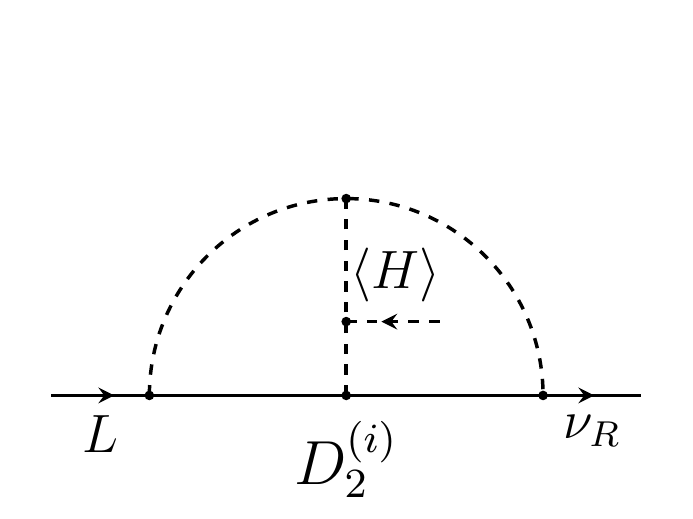}
    \includegraphics[width=0.3\textwidth]{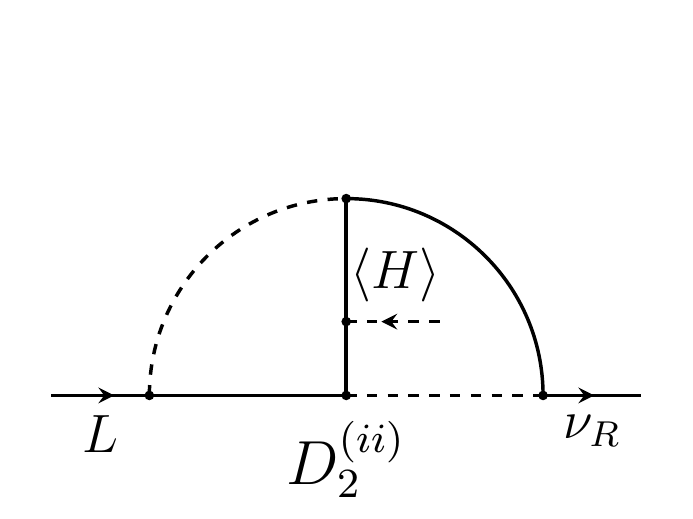}
    \caption{Set of diagrams which, in general, do not generate a lower order contribution, i.e. they contain no reducible 3- or 4-point renormalisable vertex. However, the tree-level Dirac mass term should be forbidden by some symmetry.}
    \label{fig:dirac2l:diagrams1} 
\end{figure}

The topologies $T_1$ and $T_2$ do not contain compressible renormalisable sub-parts. All the diagrams generated from these topologies will be genuine in our sense. The complete list of diagrams for these two topologies is given in \fig{fig:dirac2l:diagrams1}.

Note that the concept of genuineness stated above implies that these diagrams will generate at least one model at leading two-loop order. Of course, if the symmetries and particle content are not well-chosen, one can find specific sets of fields for these diagrams that generate a lower order contribution.

As said before, diagrams generated from the rest of the topologies ($T_3$, $T_4$, $T_5$), will not be straightforwardly genuine, due to the presence of a compressible 3-point loop vertex. Nevertheless, genuineness can be addressed in various ways and general conclusions can be drawn, as we will discuss in the next sections.

%%%%%%%%%%%%%%%%%%%%%%%%%%%%%%%%%%%%%%%%%%%%%%%%%%%%%%%%%%%%%%%
\subsection{Diagrams with a fermion-fermion-scalar loop vertex} \label{subsec:dirac2l:class2}

Naively, we could be tempted to think that, in general, the diagrams of this class are a set of corrections to the corresponding one-loop diagrams obtained by shrinking the fermion-fermion-scalar loop vertex. It is trivial to see that if a renormalisable loop vertex is allowed by the symmetries, so should be the vertex without the loop. This means that, a priori, this diagrams will not be genuine in general, as they always generate a one-loop diagram. However, there are ways to avoid this issue. One can try to forbid the tree-level vertex which generates the lower order one-loop diagram by adding an extra symmetry to the Standard Model. Then this symmetry can be softly broken to allow the loop vertex leading to a two-loop genuine diagram.

\begin{figure}[h!]
\centering
    \includegraphics[width=0.3\textwidth]{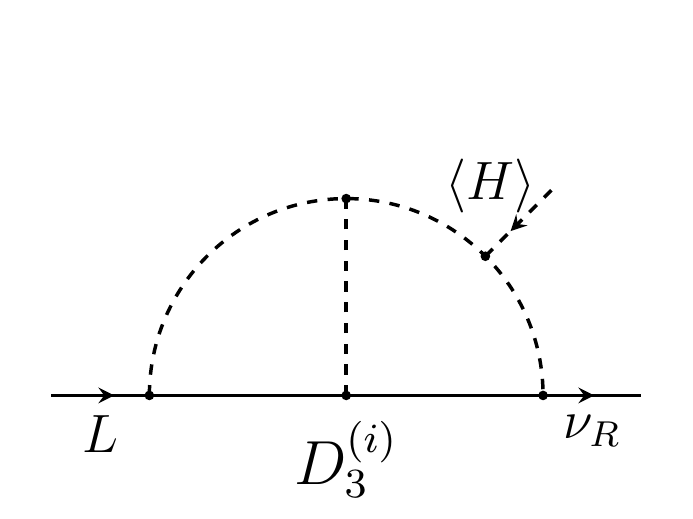}
    \includegraphics[width=0.3\textwidth]{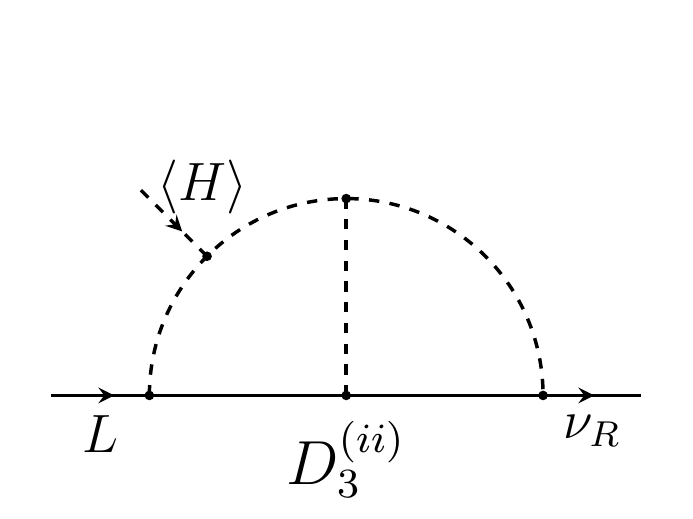}
    \includegraphics[width=0.3\textwidth]{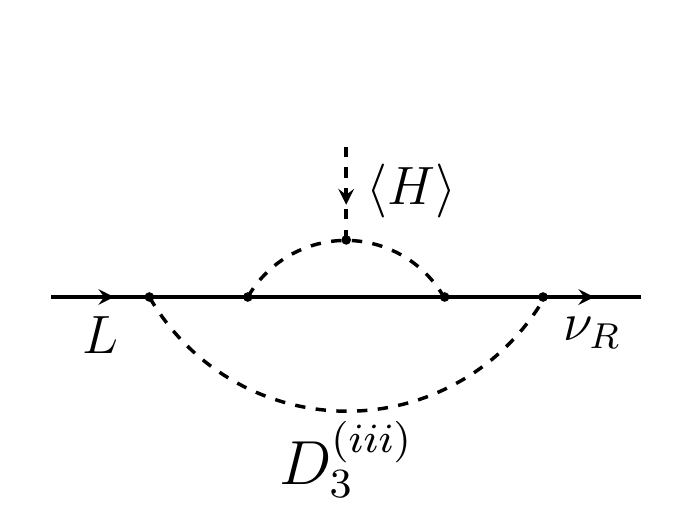}
    \\
    \includegraphics[width=0.3\textwidth]{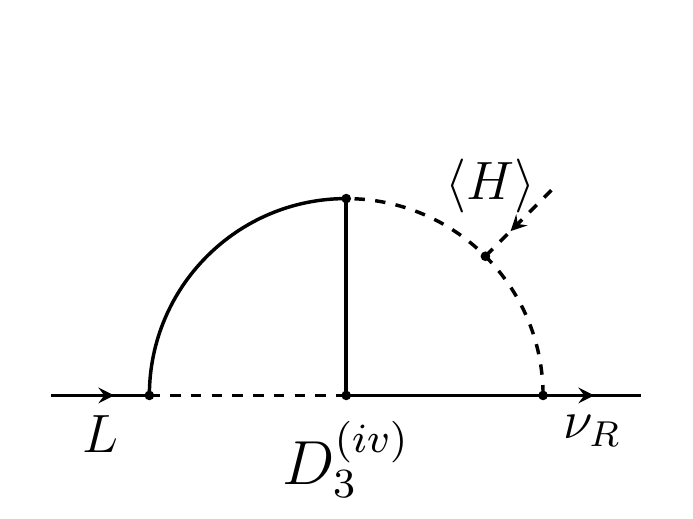}
    \includegraphics[width=0.3\textwidth]{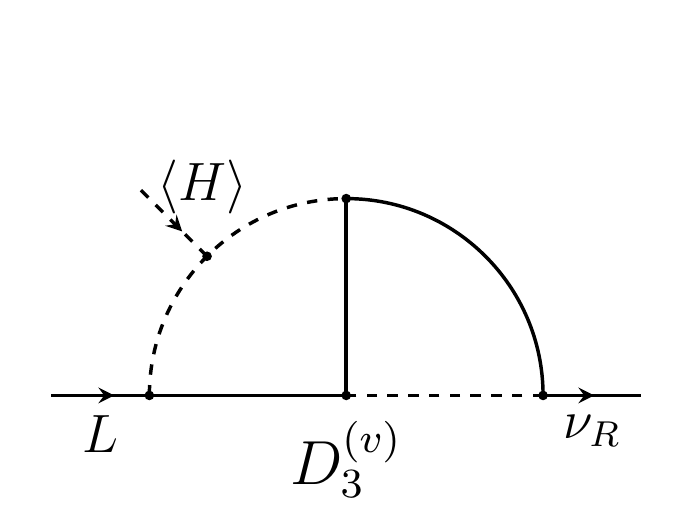}
    \includegraphics[width=0.3\textwidth]{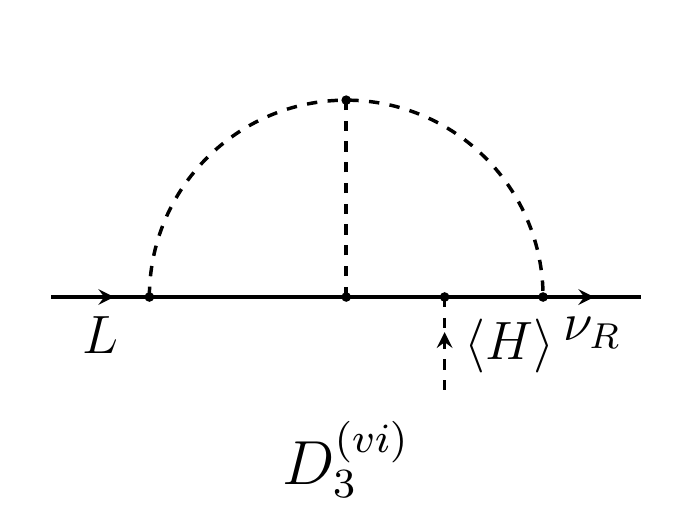}
    \\
    \includegraphics[width=0.3\textwidth]{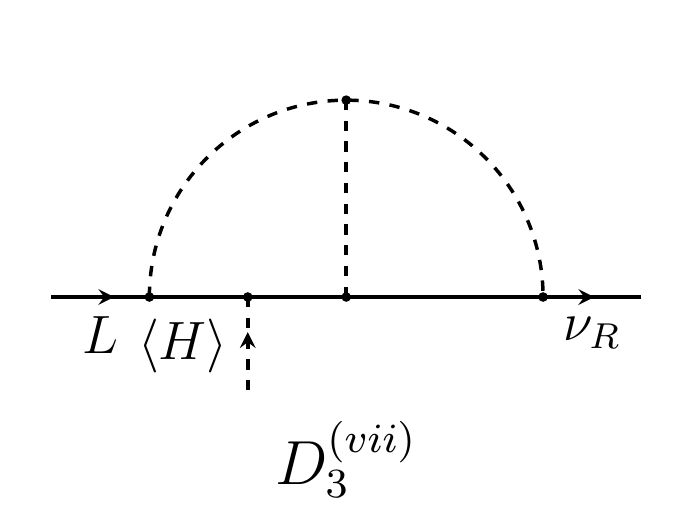}
    \includegraphics[width=0.3\textwidth]{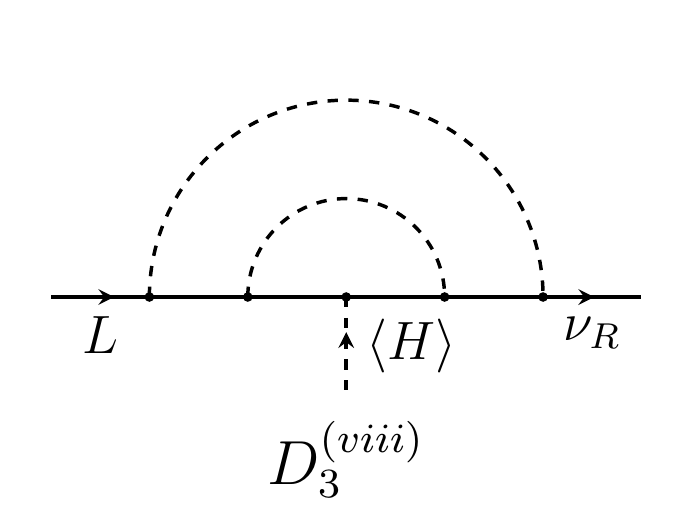}
    \\
    \includegraphics[width=0.3\textwidth]{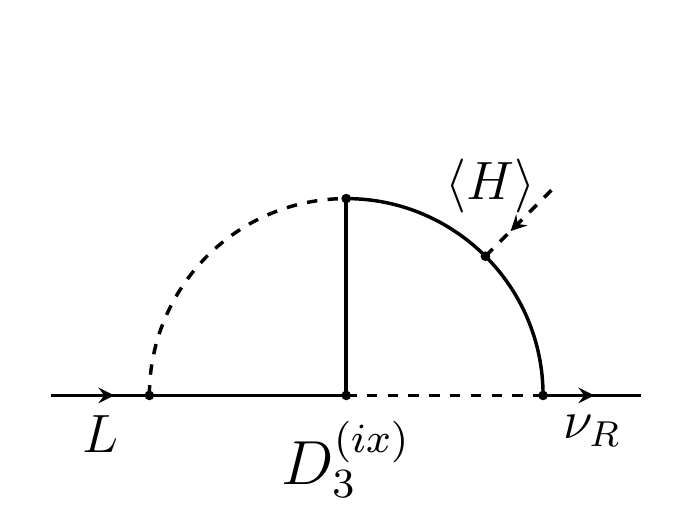}
    \includegraphics[width=0.3\textwidth]{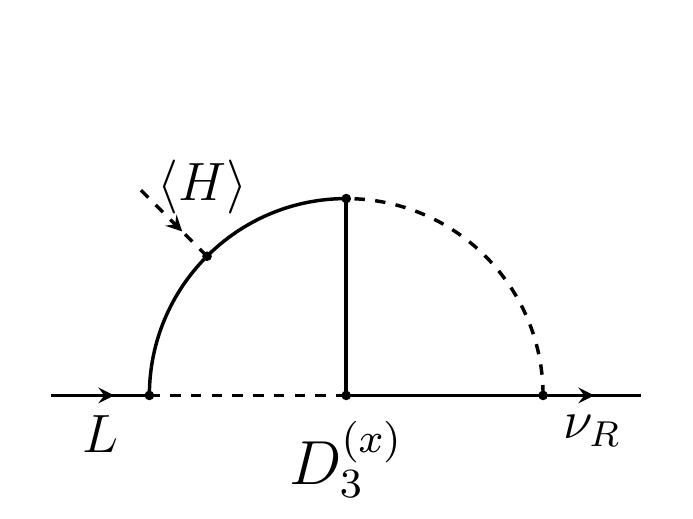}
    \caption{Set of finite diagrams with a compressible fermion-fermion-scalar vertex. These two-loop diagrams can be the dominant contribution to neutrino masses, provided the lower order contributions are forbidden by a softly broken symmetry. Note that each diagram contains a \textit{soft vertex} which breaks the symmetry forbidding the lower order diagrams.}
    \label{fig:dirac2l:diagrams2}
\end{figure}

Given the correct extra symmetry and transformation of the fields, one can use the same symmetry which forbids the Dirac tree-level mass term to forbid the tree-level vertex and then break it softly to allow the loop vertex. Since in this case the forbidden fermion-fermion-scalar coupling is a hard vertex, the tree-level realisation is still absent after the soft breaking, while the model remains completely renormalisable. In this sense, the soft breaking term should be contained in a soft coupling or mass participating in the fermion-fermion-scalar loop vertex. This procedure thus forbids the lower order loop diagram but allows the two-loop diagram making use of the same symmetry that protects Diracness. The diagrams in this class are given in the \fig{fig:dirac2l:diagrams2}.

It is worth mentioning that there is another way to avoid the compressibility of a fermion-fermion-scalar loop vertices. Instead of using a softly broken symmetry, one can force the fermion-fermion-scalar loop vertex to contain a derivative choosing the correct chirality of the fermions. This would make the effective tree-level coupling non-renormalisable (dimension 5 or beyond) \cite{Cepedello:2018rfh,Cepedello:2019zqf}. This way to address genuineness was already explained in \sect{sec:3loop:specgen}.

%%%%%%%%%%%%%%%%%%%%%%%%%%%%%%%%%%%%%%%%%%%%%%%%%%%%%%%%%%%%%%%
\subsection{Diagrams with a scalar-scalar-scalar loop vertex} \label{subsec:dirac2l:class3}

Analogous to the class of diagrams discussed in \sect{subsec:dirac2l:class2}, the diagrams with compressible three scalar vertex are also, in general, corrections to a one-loop neutrino mass diagrams. Nevertheless, in this case the procedure with softly broken symmetries (or derivatives) does not work. The problem arises due to the fact that a three scalar vertex is a soft term, so the tree-level vertex needs to be included in order to have a consistent renormalisable model. This makes the procedure useless as the one-loop diagram cannot be absent.

The solution, first pointed out in \cite{Cepedello:2018rfh} and explained in \sect{sec:3loop:specgen}, is to introduce a scalar $S$ transforming as $(\mathbf{1},\mathbf{1},-1)$ under the Standard Model gauge group. The idea is that the antisymmetric ${\rm SU(2)_L}$ contractions makes the vertex $H H S$ exactly zero at tree-level, while the one-loop (non-local) realisation of the same operator is in general non zero, see \fig{fig:3loop:exception2}. This can only be applied to scalar-scalar-scalar vertices with just one copy of the Higgs and the new singlet $S$. 

%\begin{figure}
%\centering
%    \includegraphics[scale=0.8]{Figures/example_HHS}
%    \caption{For the SM Higgs $H$ and the scalar singlet $S$ transforming as $(\mathbf{1},\mathbf{1},-1)$ under the Standard Model gauge group, the local operator $H(x)H(x)S(x)$ (right) is automatically zero, as the contraction under ${\rm SU(2)_L}$ of two doublets to a singlet is antisymmetric. On the contrary, the non-local operator $H(x_1)H(x_2)S(x_3)$ (left) does not vanish in general, but implies the difference of two diagrams \cite{Cepedello:2018rfh}.}
%    \label{fig:dirac2l:exampleHHS}
%\end{figure}

\begin{figure}
\centering
    \includegraphics[width=0.3\textwidth]{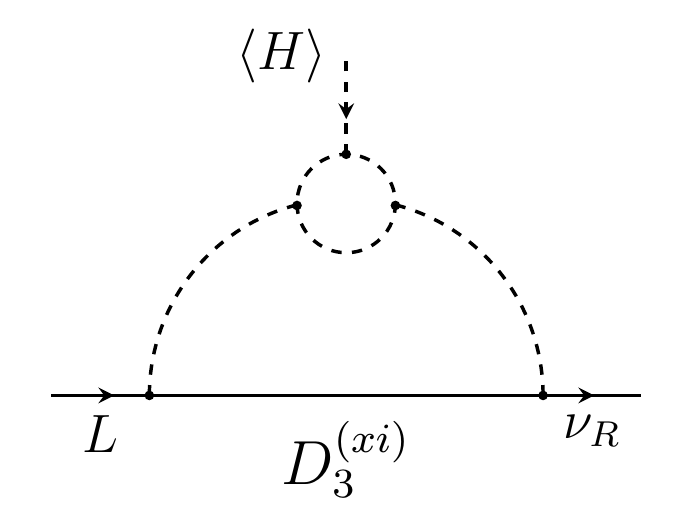}
    \includegraphics[width=0.3\textwidth]{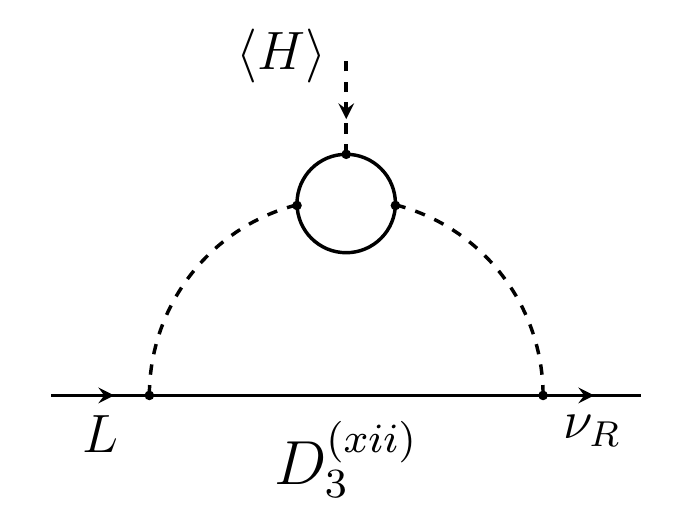}
    \includegraphics[width=0.3\textwidth]{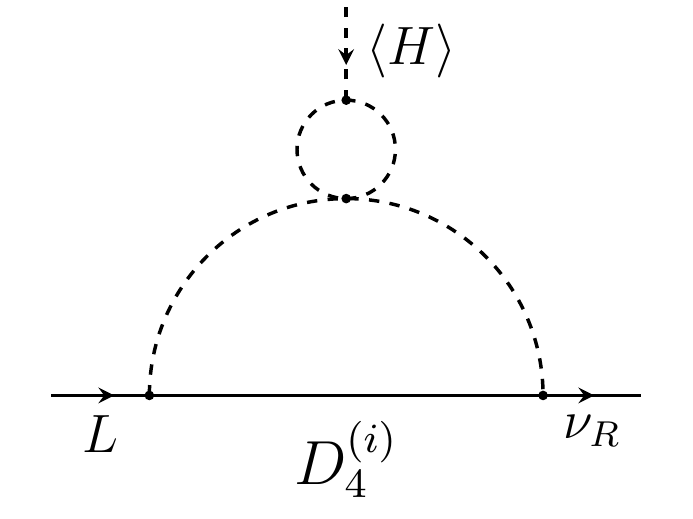}
    \\
    \includegraphics[width=0.3\textwidth]{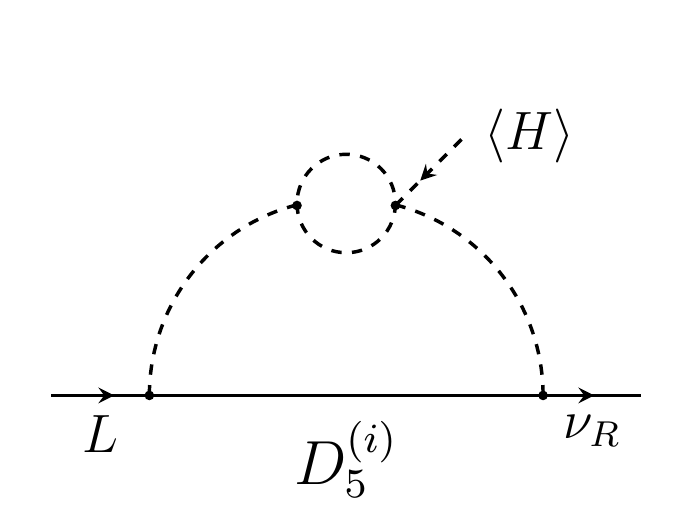}
    \includegraphics[width=0.3\textwidth]{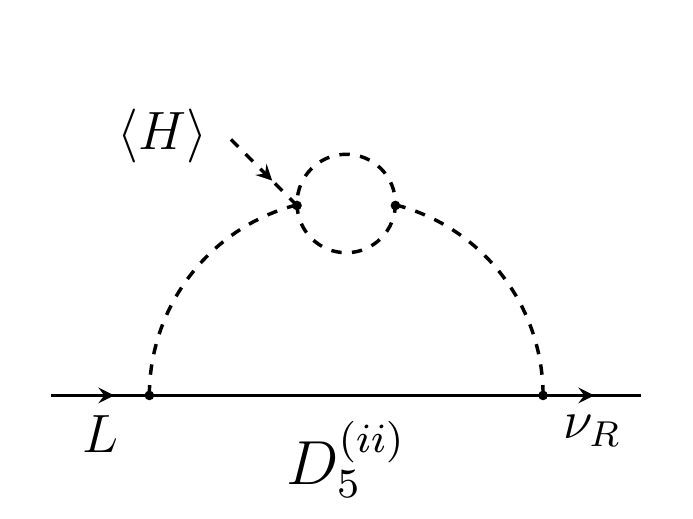}
    
    \caption{Diagrams with a compressible scalar-scalar-scalar vertex. All these diagrams contain a one-loop (i.e. non-local) realisation of a 3-point scalar vertex. In each case, the tree-level vertex $H(x)H(x)S(x)$ is exactly zero, thanks to the fact that the antisymmetric nature of the ${\rm SU(2)_L}$ contraction of the two doublets to a singlet. See \fig{fig:3loop:exception2} for details.}
    \label{fig:dirac2l:diagrams3}
\end{figure}

In \fig{fig:dirac2l:diagrams3} we give all the diagrams which fall into this class. All the genuine models generated from these diagrams should contain just one Higgs and one scalar $S\equiv(\mathbf{1},\mathbf{1},-1)$. In order to fit the neutrino mass spectra, all such diagrams should also contain at least two copies of a new vector-like pair of fermions with exactly the same Standard Model charges as those of either $e_R$ or $L$.

In general, the field content of the models generated from this class of diagrams is extremely constrained. Contrary to other two-loop diagrams, here there is only one free choice for the colour, ${\rm SU(2)_L}$ representation or hypercharge of the particles running in the loops.

%%%%%%%%%%%%%%%%%%%%%%%%%%%%%%%%%%%%%%%%%%%%%%%%%%%%%%%%%%%%%%%
\subsection{Diagrams in the mass basis} \label{subsec:dirac2l:massdiagrams}

\begin{figure}
\centering
    \includegraphics[width=0.3\textwidth]{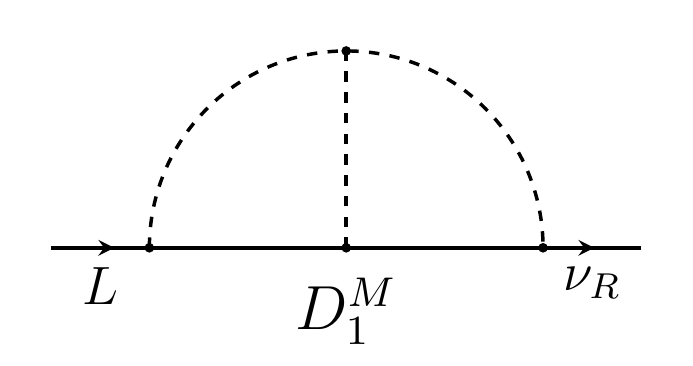}
    \includegraphics[width=0.3\textwidth]{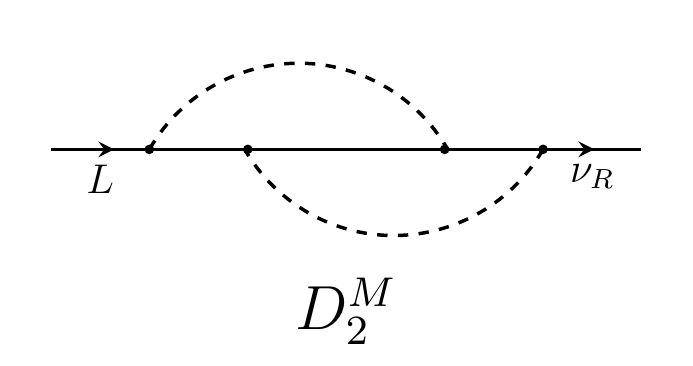}
    \includegraphics[width=0.3\textwidth]{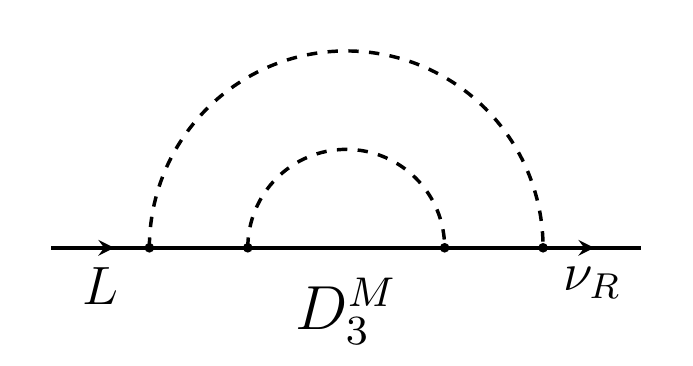}
    \\
    \includegraphics[width=0.3\textwidth]{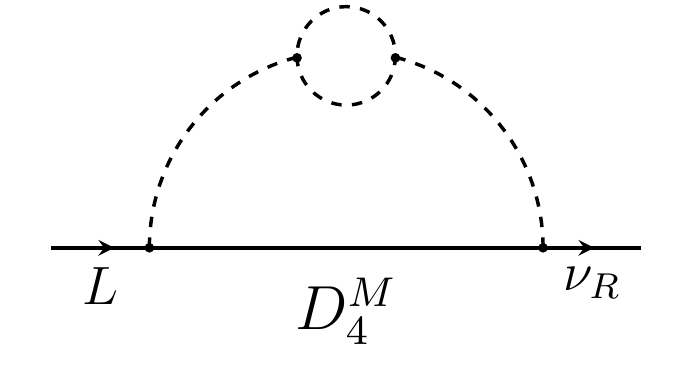}
    \includegraphics[width=0.3\textwidth]{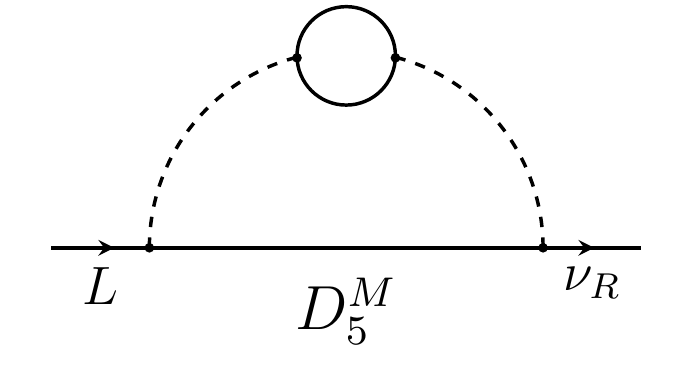}
    \includegraphics[width=0.3\textwidth]{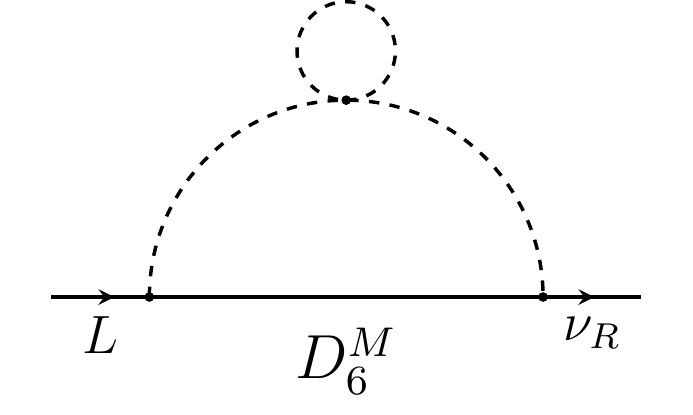}
    \caption{List of diagrams in the mass basis. Note that after removing the external Higgs line, there is no one-to-one correspondence between the diagrams in the gauge basis (\figs{fig:dirac2l:diagrams1}{fig:dirac2l:diagrams3}) and the mass diagrams given here.}
    \label{fig:dirac2l:massdiagrams}
\end{figure}

Finally, after spontaneous symmetry breaking the Higgs gets a VEV generating the neutrino masses. Thus, the external scalar denoting the Higgs insertion is removed from the diagrams in the mass basis. The initial set of 18 genuine diagrams obtained in the electroweak basis is reduced to 6 diagrams. In \fig{fig:dirac2l:massdiagrams} we show the list of the genuine mass diagrams.

All the mass diagrams can be computed analytically. Following the results of \cite{vanderBij:1983bw}, one can easily decompose any two-loop integral in \fig{fig:dirac2l:massdiagrams} in terms of just two master integrals. A more detailed discussion is given in \app{app:loops}.

%%%%%%%%%%%%%%%%%%%%%%%%%%%%%%%%%%%%%%%%%%%%%%%%%%%%%%%%%%%%%%%
%%%%%%%%%%%%%%%%%%%%%%%%%%%%%%%%%%%%%%%%%%%%%%%%%%%%%%%%%%%%%%%
\section{Generating models} \label{sec:dirac2l:genmodels}

In this section we will discuss how to assign quantum numbers to the internal fields of the loops to obtain the \textit{model-diagrams}. It should be noted that on top of the gauge group of the Standard Model, an extra symmetry is needed to forbid the tree-level Dirac mass term $\bar{L} H^c \nu_R$, as well as to protect the Dirac nature of neutrinos. This can always be achieved by just one symmetry, which can be a residual subgroup of the global $B-L$ symmetry of the Standard Model \cite{Bonilla:2018ynb}. For now, we will only consider the Standard Model quantum numbers. The issue of the extra symmetry and its charge assignment will be discussed in the next section.

\begin{figure}
\centering
    \includegraphics[width=0.3\textwidth]{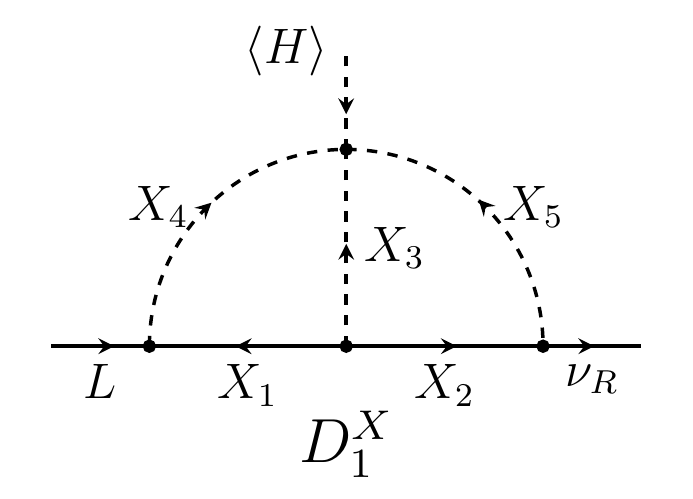}
    \includegraphics[width=0.3\textwidth]{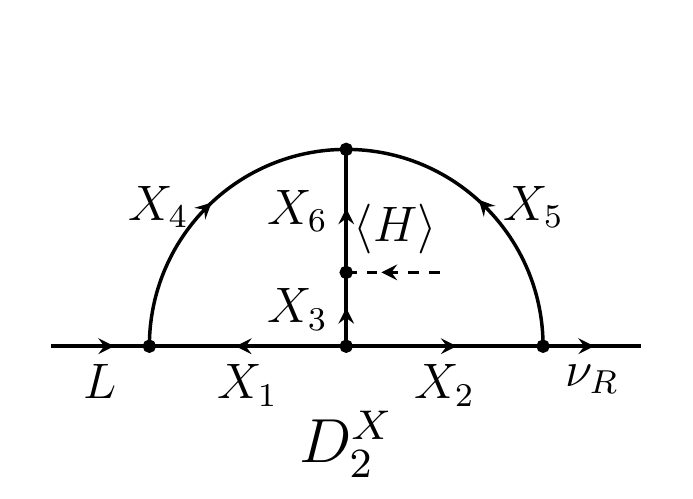}
    \includegraphics[width=0.3\textwidth]{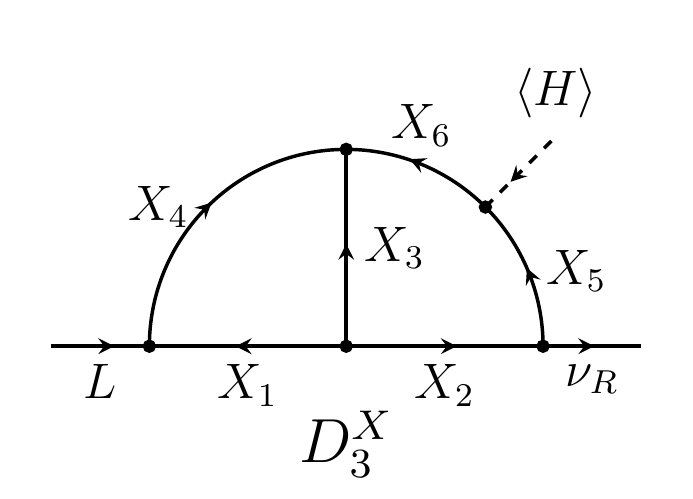}
    \vspace{-0.5cm}
    \\
    \includegraphics[width=0.3\textwidth]{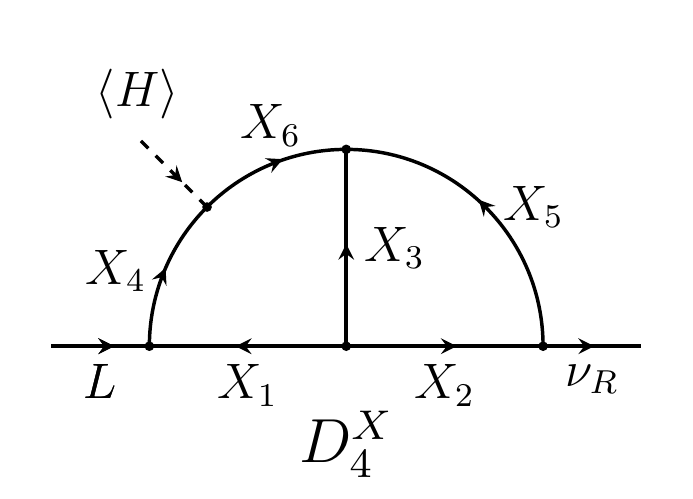}
    \includegraphics[width=0.3\textwidth]{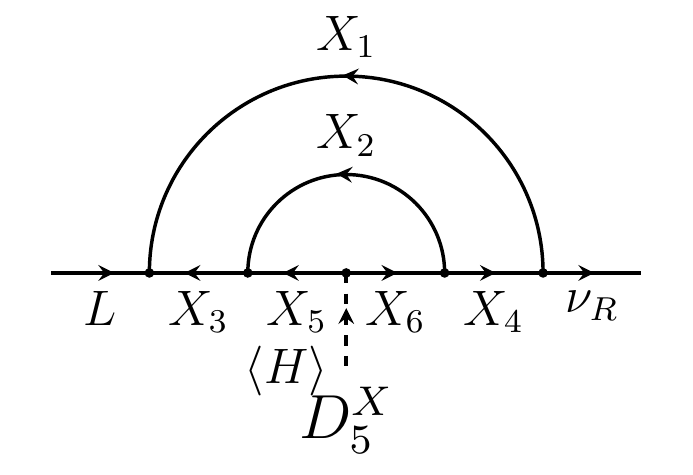}
    \caption{Auxiliary diagrams with a symbolic assignment of the internal fields at topology level. The external lines are assigned to $L$, $H$ or $\nu_R$, while the internal fields $X_i$ if represented by solid lines, can be either scalar or fermions. The arrows indicate the flowing of the quantum numbers.}
    \label{fig:dirac2l:diagrams-models-naming}
\end{figure}

\begin{figure}[t!]
\centering
    \includegraphics[width=1\textwidth]{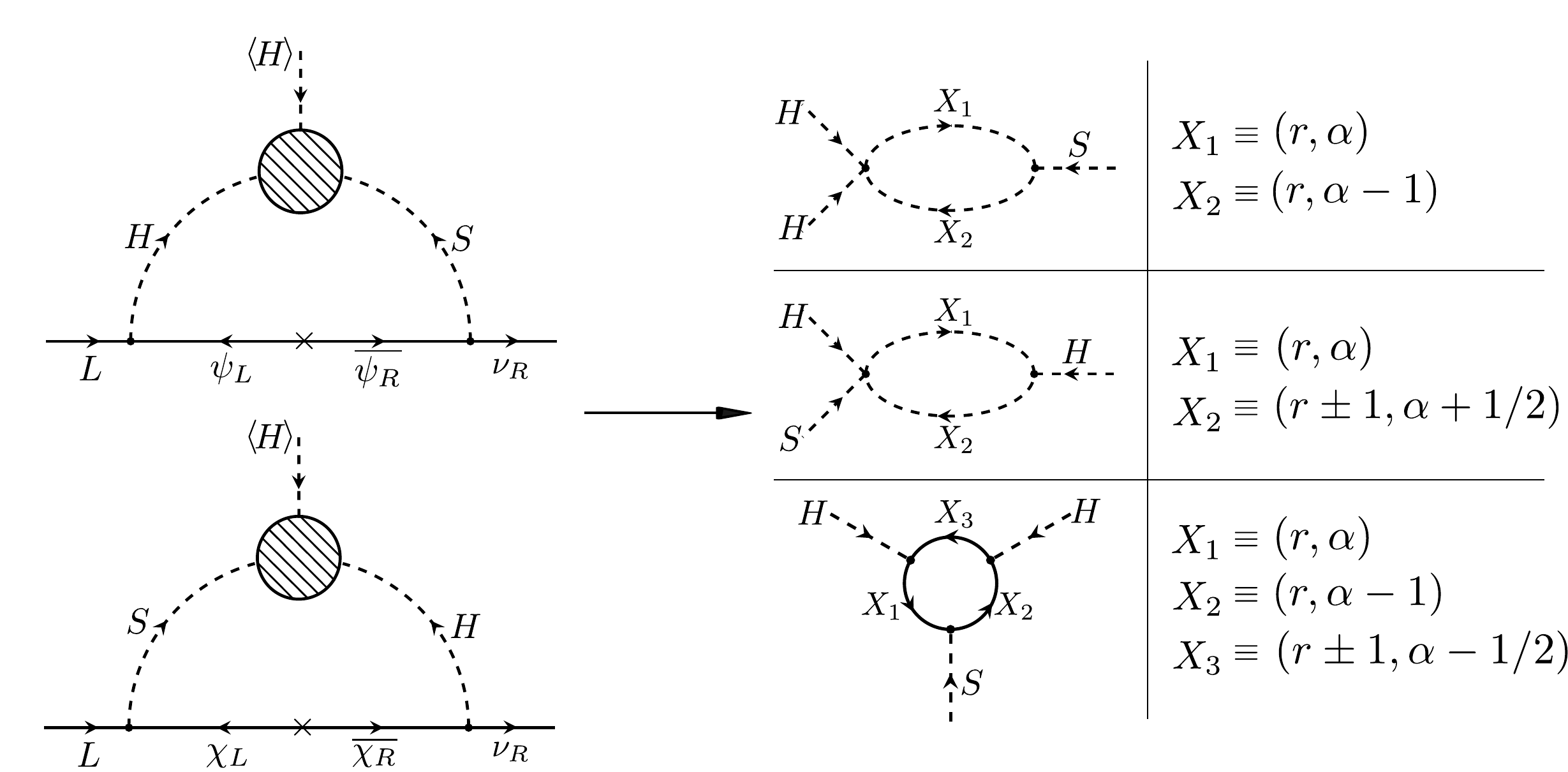}
    \caption{Auxiliary diagrams corresponding to those of \fig{fig:dirac2l:diagrams3}. The particle content depicted is the SM Higgs $H \equiv (\mathbf{2},1/2)$, $S\equiv(\mathbf{1},-1)$, $\psi_{L/R} \equiv (\mathbf{1},1)$, $\chi_{L/R} \equiv (\mathbf{2},-1/2)$ with charges under ${\rm SU(2)_L} \times {\rm U(1)_Y}$, while the unknown fields $X$ if solid lines, can be either scalars or fermions. We only consider colour singlets for simplicity. All the quantum numbers of the fields are determined once an input field $X_1 \equiv (r,\alpha)$ is given with $r>1$ (for $r=1$ only $r+1$ holds). See text for details.}
    \label{fig:dirac2l:models-loophole}
\end{figure}

Due to the large number of diagrams, it is more convenient to assign the quantum numbers at the topology level, while fixing the external fields, i.e. $L$, $H$ and $\nu_R$. This leaves us with the seven diagrams given in \fig{fig:dirac2l:diagrams-models-naming} and \fig{fig:dirac2l:models-loophole}. The separation into two distinct sets is done because the latter always requires the field $S\equiv(\mathbf{1},\mathbf{1},-1)$ in order to be genuine, which considerably constraints the possible fields running in the loop. See \sect{subsec:dirac2l:class3} for details.

In \fig{fig:dirac2l:diagrams-models-naming}, the internal fields $X_i$ depicted as solid lines, can be either scalars or fermions. The only exception is in the $D_1^X$ diagram where, due to the quartic coupling, $X_3$, $X_4$ and $X_5$ can only be scalars and, in consequence, they are drawn with dashed lines. This diagram corresponds directly to diagram $D_1^{(i)}$. In the rest of the cases the correspondence with the diagrams in \figs{fig:dirac2l:diagrams1}{fig:dirac2l:diagrams2} depends on whether the fields $X_i$ are scalars or fermions. For example, $D^X_2$ corresponds to diagrams $D_2^{(i)}$ or $D_2^{(ii)}$ depending on if $X_2$, $X_3$, $X_5$ and $X_6$ are scalars or fermions. $D_3^X$ corresponds to diagram $D_3^{(i)}$, $D_3^{(iii)}$, $D_3^{(vi)}$ and $D_3^{(ix)}$; $D_4^X$ to $D_3^{(ii)}$, $D_3^{(iv)}$, $D_3^{(vii)}$ and $D_3^{(x)}$; while diagrams $D_3^{(v)}$ and $D_3^{(viii)}$ are generated from $D_5^X$.

Since in all diagrams there are two-loops, there are two independent sets of quantum numbers (colour, ${\rm SU(2)_L}$ representation and hypercharge), that need to be chosen in order to determine the rest of the fields. To fix the particle content, we start by assigning quantum numbers to $X_1$ and $X_2$. Once the gauge charges of these two fields are chosen, all the hypercharges of all other fields are automatically fixed. For the ${\rm SU(2)_L}$ and ${\rm SU(3)_C}$ representations, though, no general straightforward relation can be found since a product of two fields contains several irreducible representations. In spite of this, once the representations of $X_1$ and $X_2$ are chosen, the freedom of the other fields get severely restricted. For simplicity, we will work with colour singlets, because colour assignments can be trivially added taking into account that external fields are colour-blind. We will explicitly omit this quantum number for the internal fields $X_i$. As a side remark, note that every new fermion should have its corresponding vector-like partner to provide mass to them, as a fourth chiral family is excluded by Higgs production measurements and direct searches.

In \tab{tab:dirac2l:QN1}, we give all possible fields assignments for a general hypercharge and up to ${\rm SU(2)_L}$ triplets for the diagram $D_1^X$ of \fig{fig:dirac2l:diagrams-models-naming}.

\begin{table}[t!]
\centering
    \begin{tabular}{|*{5}{c|}}
        \hline\hline
        \multicolumn{5}{c}{\textbf{Hypercharge for $D_1^X$}} \\
        \hline
        \makebox[3em]{$X_1$} & \makebox[3em]{$X_2$} & $X_3$ & \makebox[3em]{$X_4$} & \makebox[3em]{$X_5$} \\
        \hline
        $\alpha_1$ & $\alpha_2$ & $-\alpha_1-\alpha_2$ & $\alpha_1-1/2$ & $\alpha_2$ \\
        \hline
    \end{tabular}
 \\[3ex]
    \begin{tabular}{|c||*{3}{c|}|*{3}{c|}|*{3}{c|}}
        \hline\hline
        \multicolumn{10}{c}{\textbf{${\rm SU(2)_L}$ representations for $D_1^X$}} \\
        \hline
        \backslashbox[3em]{$X_2$}{$X_1$} & \multicolumn{3}{c||}{1} & \multicolumn{3}{c||}{2} & \multicolumn{3}{c|}{3} \\
        \hline\hline
        & \makebox[2em]{$X_3$} & \makebox[2em]{$X_4$} & \makebox[2em]{$X_5$} & \makebox[2em]{$X_3$} & \makebox[2em]{$X_4$} & \makebox[2em]{$X_5$} & \makebox[2em]{$X_3$} & \makebox[2em]{$X_4$} & \makebox[2em]{$X_5$} \\
        \hline\hline
        1 & 1 & 2 & 1 & 2 & \slashbox[2.4em]{1}{3} & 1 & 3 & 2 & 1 \\
        \hline
        2 & 2 & 2 & 2 & \slashbox[2.4em]{1}{3} & \slashbox[2.4em]{1}{3} & 2 & 2 & 2 & 2 \\
        \hline
        3 & 3 & 2 & 3 & 2 & \slashbox[2.4em]{1}{3} & 3 & \slashbox[2.4em]{1}{3} & 2 & 3 \\
        \hline
    \end{tabular}
\caption{The ${\rm SU(2)_L}$ and ${\rm U(1)_Y}$ quantum numbers for the diagram $D_1^X$ (T1-i) of \fig{fig:dirac2l:diagrams-models-naming}. Fixing the charges of the two fields $X_1$ and $X_2$, fixes the possible charges of all the other $X_{3-5}$ fields.  The possible ${\rm SU(2)_L}$ representations (up to triplets) of the fields $X_1$ and $X_2$ are given in the first row and column of the second table. Their hypercharges are denoted by $\alpha_1$ and $\alpha_2$, respectively, in the first table. For the rest of the fields $X_{3-5}$ we give all the possible hypercharges and ${\rm SU(2)_L}$ representations (up to triplets). For simplicity, all the fields are colour singlets.}
\label{tab:dirac2l:QN1}
\end{table}

\Tab{tab:dirac2l:QN1} is divided in two panels: hypercharge and ${\rm SU(2)_L}$ representation. Hypercharges can be given in general by solving the system of equations for each vertex in terms of two input values $\alpha_1$ and $\alpha_2$, which are the hypercharges of $X_1$ and $X_2$ respectively, as shown in the upper panel of \tab{tab:dirac2l:QN1}. In the lower panel, we show all the possible ${\rm SU(2)_L}$ representations for the internal fields of $D_1^X$ up to triplets for different values of the quantum numbers of $X_1$ and $X_2$ in the first row and column, respectively. In the cases when several representations are possible (for example, $2 \otimes 2 = 1 \oplus 3$), the cell is subdivided to indicate that any of the two representations can be chosen.

Note that certain particular choices of the fields can generate lower order masses, i.e. tree or one-loop neutrino masses. In contrast to the Majorana case, here an additional model dependent symmetry is needed such that it forbids the lower order contributions. A judicious choice of the transformation of the fields under this symmetry and its appropriate breaking pattern is sufficient to ensure the genuineness of any two-loop model generated from \fig{fig:dirac2l:diagrams-models-naming} (see \sect{sec:dirac2l:class} for details).

For the diagrams $D_{2-4}^X$, for simplicity we do not give one set of tables for every diagram, but we unified them with \tab{tab:dirac2l:QN1} for $D_1^X$. From \fig{fig:dirac2l:diagrams-models-naming} it can be seen that the diagram $D_1^X$ is obtained by shrinking the field $X_6$. This means that for the diagrams $D_2^X$, $D_3^X$ and $D_4^X$ the fields are identical to those of $D_1^X$, except for $X_6$. For each assignment of ${\rm SU(2)_L}$ representation and hypercharge of the fields $X_1$ and $X_2$, the quantum numbers of $X_6$ for each diagram in $D_{2-4}^X$ are depicted \tab{tab:dirac2l:QN2-4}, completing the charge assignment for fields $X_{1-5}$ in \tab{tab:dirac2l:QN1}, identical for all the diagrams.

\begin{table}
\centering
    \begin{tabular}{|*{4}{c|}}
        \hline\hline
        \multicolumn{4}{c}{\textbf{Hypercharge of $X_6$ for $D_{2-4}^X$}} \\
        \hline
         & \makebox[7em]{$D_2^X$} & \makebox[5em]{$D_3^X$} & \makebox[3em]{$D_4^X$} \\
        \hline
       \makebox[3em]{$X_6$} & $-\alpha_1-\alpha_2-1/2$ & $\alpha_2+1/2$ & $\alpha_1$ \\
        \hline
    \end{tabular}
    \\[3ex]
    \begin{tabular}{|c||*{3}{c|}|*{3}{c|}|*{3}{c|}}
        \hline\hline
        \multicolumn{10}{c}{\textbf{${\rm SU(2)_L}$ representations of $X_6$ for $D_{2-4}^X$}} \\
        \hline
        \backslashbox[3em]{$X_2$}{$X_1$} & \multicolumn{3}{c||}{1} & \multicolumn{3}{c||}{2} & \multicolumn{3}{c|}{3} \\
        \hline\hline
        & \multicolumn{3}{c||}{$X_6$} & \multicolumn{3}{c||}{$X_6$} & \multicolumn{3}{c|}{$X_6$} \\
        \hline
        & $D_2^X$ & $D_3^X$ & $D_4^X$ & $D_2^X$ & $D_3^X$ & $D_4^X$ & $D_2^X$ & $D_3^X$ & $D_4^X$ \\
        \hline\hline
        1 & 2 & 2 & \slashbox[2.4em]{1}{3} & \slashbox[2.4em]{1}{3} & 2 & 2 & 2 & 2 & \slashbox[2.4em]{1}{3} \\
        \hline
        2 & \slashbox[2.4em]{1}{3} & \slashbox[2.4em]{1}{3} & \slashbox[2.4em]{1}{3} & 2 & \slashbox[2.4em]{1}{3} & 2 & \slashbox[2.4em]{1}{3} & \slashbox[2.4em]{1}{3} & \slashbox[2.4em]{1}{3} \\
        \hline
        3 & 2 & 2 & \slashbox[2.4em]{1}{3} & \slashbox[2.4em]{1}{3} & 2 & 2 & 2 & 2 & \slashbox[2.4em]{1}{3} \\
        \hline
    \end{tabular}
\caption{Standard Model quantum numbers for the field $X_6$ of the diagrams $D_2^X$, $D_3^X$ and $D_4^X$ in \fig{fig:dirac2l:diagrams-models-naming}. Two input fields are needed $X_1$ and $X_2$ with hypercharges $\alpha_1$ and $\alpha_2$, respectively, and ${\rm SU(2)_L}$ representations explicitly given in the first row and column of the right table. These tables should be completed with \tab{tab:dirac2l:QN1} which contains the quantum numbers for the fields $X_3$, $X_4$ and $X_5$, common for all the diagrams. For simplicity, all the fields are colour singlets.}
\label{tab:dirac2l:QN2-4}
\end{table}

The only diagram that do not shrink to $D_1^X$ is $D_5^X$, for which a specific set of tables is needed. For $D_5^X$ the quantum numbers are given in \tab{tab:dirac2l:QN5}, in the same fashion as the example already discussed: (1) two tables are given for ${\rm SU(2)_L}$ representations and hypercharge, (2) we consider representations up to triplets, and (3) two input fields are needed $X_1$ and $X_2$ with hypercharges $\alpha_1$ and $\alpha_2$, respectively, while the ${\rm SU(2)_L}$ representations are explicitly given in the first row and column of each table.
\\

\begin{table}
\centering
    \begin{tabular}{|*{6}{c|}}
        \hline\hline
        \multicolumn{6}{c}{\textbf{Hypercharge for $D_5^X$}} \\
        \hline
        \makebox[3em]{$X_1$} & \makebox[3em]{$X_2$} & $X_3$ & \makebox[3em]{$X_4$} & \makebox[3em]{$X_5$} & \makebox[3em]{$X_6$} \\
        \hline
        $\alpha_1$ & $\alpha_2$ & $-\alpha_1+1/2$ & $\alpha_1$ & $-\alpha_1-\alpha_2+1/2$ & $\alpha_1+\alpha_2$ \\
        \hline
    \end{tabular}
    \\[3ex]
    \begin{tabular}{|c||*{4}{c|}|*{4}{c|}|*{4}{c|}}
        \hline\hline
        \multicolumn{13}{c}{\textbf{${\rm SU(2)_L}$ representations for $D_5^X$}} \\
        \hline
        \backslashbox[3em]{$X_2$}{$X_1$} & \multicolumn{4}{c||}{1} & \multicolumn{4}{c||}{2} & \multicolumn{4}{c|}{3} \\
        \hline\hline
        & \makebox[1.3em]{$X_3$} & \makebox[1.3em]{$X_4$} & \makebox[1.3em]{$X_5$} & \makebox[1.3em]{$X_6$} & \makebox[1.3em]{$X_3$} & \makebox[1.3em]{$X_4$} & \makebox[1.3em]{$X_5$} & \makebox[1.3em]{$X_6$} & \makebox[1.3em]{$X_3$} & \makebox[1.3em]{$X_4$} & \makebox[1.3em]{$X_5$} & \makebox[1.3em]{$X_6$} \\
        \hline\hline
        1 & 2 & 1 & 2 & 1 & \slashbox[2.4em]{1}{3} & 2 & \slashbox[2.4em]{1}{3} & 2 & 2 & 3 & 2 & 3 \\
        \hline
        2 & 2 & 1 & \slashbox[2.4em]{1}{3} & 2 & \slashbox[2.4em]{1}{3} & 2 & 2 & \slashbox[2.4em]{1}{3} & 2 & 3 & \slashbox[2.4em]{1}{3} & 2 \\
        \hline
        3 & 2 & 1 & 2 & 3 & \slashbox[2.4em]{1}{3} & 2 & \slashbox[2.4em]{1}{3} & 2 & 2 & 3 & 2 & \slashbox[2.4em]{1}{3} \\
        \hline
    \end{tabular}
\caption{Standard Model quantum numbers for the diagram $D_5^X$ ($D_5^{(vii)}$ and $D_5^{(x)}$) in \fig{fig:dirac2l:diagrams-models-naming}. All the fields are colour singlets.}
\label{tab:dirac2l:QN5}
\end{table}

The diagrams that require the scalar field $S\equiv(\mathbf{1},\mathbf{1},-1)$ in order to be genuine, are shown in \fig{fig:dirac2l:models-loophole}. We show all possible fields, along with their ${\rm SU(2)_L} \times {\rm U(1)_Y}$ charges, that close the diagrams in \sect{subsec:dirac2l:class3}. Like before, we take all the fields to be colour singlets. As already explained, the main difference from the diagrams of the previous class (\fig{fig:dirac2l:diagrams-models-naming}), is the necessity of a Higgs and the scalar $S$ running in the loop. In these diagrams, there is only one free set of quantum numbers, i.e. the quantum numbers of one of the fields running in the loop that generates the effective vertex $HH S$. As the external fields are fixed to be one ${\rm SU(2)_L}$ singlet and two doublets, the quantum numbers of all the fields in the loop can be determined in general, once we pick the quantum numbers of any one of the remaining fields. For example, choosing the ${\rm SU(2)_L} \times {\rm U(1)_Y}$ charges of $X_1$ field in \fig{fig:dirac2l:models-loophole} as $X_1 \equiv (r,\alpha)$, $r$ denoting the ${\rm SU(2)_L}$ representation and $\alpha$ the hypercharge, automatically fixes the possible charges of the remaining fields. Note that unlike the previous class of models, here the coloured particles can only run in the small loop, see \fig{fig:dirac2l:models-loophole} (right). All the internal fields in this loop need to have the same ${\rm SU(3)_C}$ representation since all the external fields are colour singlets.

As stated earlier, the set of models following from the topologies of \fig{fig:dirac2l:models-loophole} are phenomenologically constrained. In all of them, one of the vector-like internal fermions must always have the Standard Model quantum numbers similar to either the quantum numbers of $L$ or $e^c$, i.e. $\chi \equiv (\mathbf{1},\mathbf{2},-1/2)$ and $\psi \equiv (\mathbf{1},\mathbf{1},1)$, respectively. Consequently, limits on their masses can be derived from collider searches \cite{Thomas:1998wy, Kumar:2015tna} and lepton flavour violating processes \cite{Falkowski:2013jya}, forcing their mass to be at least a  few TeV. Nevertheless, these constraints do not run in conflict with neutrino masses, since even for $\mathcal{O}(1)$ values of the couplings and the internal fermion masses in TeV range, the neutrino masses can easily be  $\mathcal{O}(0.1)$ eV scale \cite{Sierra:2014rxa}, as required by the current data.

%%%%%%%%%%%%%%%%%%%%%%%%%%%%%%%%%%%%%%%%%%%%%%%%%%%%%%%%%%%%%%%
%%%%%%%%%%%%%%%%%%%%%%%%%%%%%%%%%%%%%%%%%%%%%%%%%%%%%%%%%%%%%%%
\section{Example models} \label{sec:dirac2l:models}

We construct two example models to show in action the ideas discussed before. We have already presented the basic features and gauge charge requirements for the internal particles in the two-loop models. However, so far we have not explicitly discussed the role and nature of the symmetry or symmetries forbidding the tree-level coupling and/or protecting the Dirac nature of neutrinos. Since, as mentioned before, there are various options for such symmetries, a completely model independent approach is not possible. Let us now finally address the role of these symmetries by means of some example models.
  
There are many ways to arrange the additional symmetries of the model in such a way that all the necessary features are satisfied, namely neutrinos are Dirac particles and the leading contribution to its mass comes from the two-loop level. Another interesting feature that has been noticed before \cite{Chulia:2016ngi, Chulia:2016giq, Bonilla:2018ynb, Bonilla:2019hfb} is the connection between the Dirac nature of neutrinos and dark matter stability. If chosen correctly, the symmetry protecting the Diracness of neutrinos can also forbid the decay of the dark matter, ensuring its stability. Thus, the additional symmetry can play multiple roles. Furthermore, as has been discussed in \cite{Bonilla:2018ynb}, this symmetry can also forbid the lower order mass terms. Additionally, it need not be a new symmetry and can just be a residual subgroup of the global ${\rm U(1)_{B-L}}$ symmetry already present in the Standard Model. We will discuss the Diracness-dark matter stability connection in more details in \ch{ch:dm_DiracMajo}.

The two  examples we show in this section employ the discrete abelian cyclic $Z_4$ group as the symmetry protecting the Dirac nature of neutrinos. Although not necessary, this  symmetry can be a residual subgroup of the ${\rm U(1)_{B-L}}$ symmetry of the Standard Model \cite{Chulia:2016ngi, Chulia:2016giq, CentellesChulia:2017koy, Dasgupta:2019rmf} or of some other ${\rm U(1)_X}$ symmetry \cite{Ma:2019yfo}. The choice of the $Z_4$ symmetry is done keeping in mind the Diracness connection to dark matter stability to be discussed in the next chapter. It is worth to notice that if this symmetry is taken as $Z_2$ then neutrinos will be Majorana fields \cite{Hirsch:2017col}. Taking  it to be $Z_3$ will necessarily lead either to decaying dark matter or to the existence of an accidental symmetry that stabilises dark matter \cite{Bonilla:2018ynb,Bonilla:2019hfb}. Therefore $Z_4$ is the smallest group that achieves simultaneously the stability of the dark matter while protecting the Dirac nature of neutrinos. 

For both models the lepton doublets $L_i $ and right-handed neutrinos $\nu_{R,i}$ transform as ``$Z_4$ odd'' particles i.e. $z^1 = e^{i \pi/2}=i$ under the $Z_4$ symmetry with $z^4=1$.\footnote{We call $Z_4$ odd the fields that transform as odd powers, i.e. the fields transforming as $z^1 \equiv i$ or as $z^3 \equiv -i$ under the $Z_4$ symmetry. Similarly, $Z_4$ even are the fields transforming as even powers i.e. $z^0 \equiv 1$ or as $z^2 \equiv -1$ under the $Z_4$ symmetry.} This automatically forbids all Majorana mass terms for the neutrinos at all dimensions and loop orders, and ensures they are Dirac particles. We also add a $Z_2$ symmetry whose role is to forbid the tree-level mass term for neutrinos. This symmetry will be softly broken to allow for the two-loop realisation of the operator $\overline{L} H^c \nu_R$ \cite{Chulia:2016ngi}. We would like to remark that this additional $Z_2$ symmetry is not always necessary to forbid the tree-level mass term~\cite{Ma:2014qra, Ma:2015mjd,Bonilla:2018ynb,Bonilla:2019hfb}, however we have added it in order to keep the discussion simple. Further note that, as shown in \cite{Bonilla:2018ynb}, all these features can be obtained using only the $B-L$ symmetry without the need of extra symmetries. Although this construction is appealing because of its economic symmetry inventory, it is conceptually a bit more involved than the simple one we choose here as an example.

%%%%%%%%%%%%%%%%%%%%%%%%%%%%%%%%%%%%%%%%%%%%%%%%%%%%%%%%%%%%%%%
\subsection{A genuine two-loop Dirac neutrino mass model} \label{subsec:dirac2l:genmodel}

From the diagrams given in \sect{subsec:dirac2l:class1}, we choose $D_1^{(i)}$ in \fig{fig:dirac2l:diagrams1} to illustrate how a simple genuine model can be built. As described before, in contrast to the diagrams of \sect{subsec:dirac2l:class2} and \sect{subsec:dirac2l:class3}, the main characteristic of the models generated from these diagrams is that, a priori, there is no restriction on the possible internal fields or the position of the soft symmetry breaking terms. One should only be careful about choosing the charges of the internal fermions in such a way that the leading contribution comes at the two-loop order.\footnote{For instance, avoid new fermions $F$ with quantum numbers that allow the vertex $\overline{L}H^c F$.}

\begin{figure}
\centering
    \includegraphics[width=0.5\textwidth]{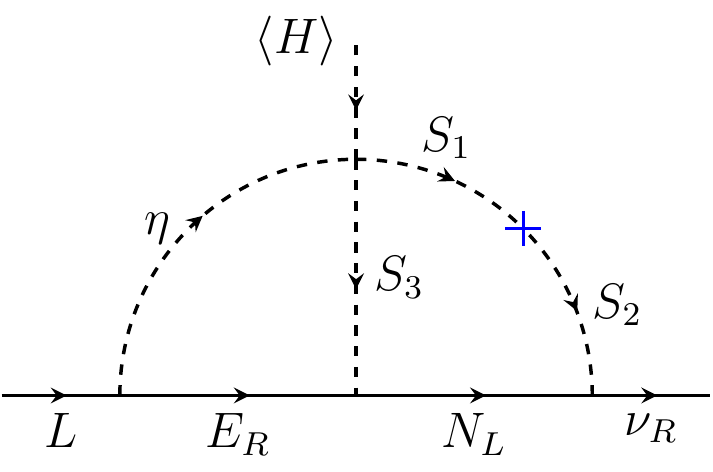}
    \caption{Completely genuine two-loop diagram that gives mass to neutrinos. The blue cross marks the soft breaking term of the $Z_2$ symmetry that allows the loop realisation of the operator $\overline{L} H^c \nu_R$ forbidding the tree-level.}
    \label{fig:dirac2l:softly} 
\end{figure}

Following \tab{tab:dirac2l:QN1}, for the simplest case when  $X_1$ and $X_2$ are ${\rm SU(2)_L}$ singlets, we construct the model of \fig{fig:dirac2l:softly}, whose particle content and relevant quantum numbers are given in \tab{tab:dirac2l:modelgenuine}. Two extra symmetries, apart from those of the Standard Model gauge group are added, a $Z_4$ and a $Z_2$. The former ensures the Dirac nature of neutrinos, at the same time it also provides the stability of dark matter, while the latter is related to the smallness of neutrino masses, forbidding the tree-level mass operator $\overline{L} H^c \nu_R$. The $Z_2$ symmetry is softly broken in order to allow the loop realisation of \fig{fig:dirac2l:softly}. Including all the soft-breaking terms to the Lagrangian means that we have to add the mass term $S_2^\dagger S_1 + \hc$, depicted as a blue cross on the diagram.\footnote{The soft term $H\eta^\dagger S_2$ should be added too for consistency, although it plays no role in the neutrino mass generation or the dark matter stability.}

\begin{table}
\begin{center}
\begin{tabular}{| c || c | c | c | c |}
  \hline
&   \hspace{0.1cm}  Fields     \hspace{0.1cm}       &  \hspace{0.4cm}  ${\rm SU(2)_L} \times {\rm U(1)_Y}$     \hspace{0.4cm}    & \hspace{0.4cm}   $Z_4$            \hspace{0.4cm}   &  \hspace{0.4cm}  $Z_2$            \hspace{0.4cm}                            \\
\hline \hline
\multirow{6}{*}{ \begin{turn}{90} Fermions \end{turn} } 
&   $L$        	     &   ($\mathbf{2}, {-1/2}$)      &   $i$ & $+$\\
&   $\nu_{R}$        &   ($\mathbf{1}, {0}$)         &   $i$ & $-$ \\
&   $e_{R}$          &   ($\mathbf{1}, {-1}$)        &   $i$ & $+$  \\
&   $E_{R}$         &   ($\mathbf{1}, {-1}$)        &   $1$ & $+$   \\
&   $E_{L}$    	 &   ($\mathbf{1}, {-1}$)        &   $1$ & $+$    \\
&   $N_{L}$      	 &   ($\mathbf{1}, {0}$)         &   $1$ & $+$     \\
\hline \hline                                                                               
\multirow{5}{*}{ \begin{turn}{90} Scalars \end{turn} } 
& $H$  	       	 &  ($\mathbf{2}, {1/2}$)        & $1$   & $+$        \\
& $S_1$          	 &  ($\mathbf{1}, {0}$)          & $i$   & $+$         \\
& $S_2$              &  ($\mathbf{1}, {0}$)          & $i$   & $-$          \\
& $S_3$              &  ($\mathbf{1}, {1}$)          & $1$   & $+$           \\
& $\eta$             &  ($\mathbf{2}, {1/2}$)        & $i$   & $+$            \\
    \hline
  \end{tabular}
\end{center}
\caption{Particle content of the completely genuine example model. The gauge charges along with the $Z_4 \times Z_2$ charges are also shown. All the fields listed in the table are ${\rm SU(3)_C}$ singlets. The role of $Z_4$ is to protect Diracness and to stabilise the dark matter candidate. The lightest particle among $S_1$, $S_2$, the neutral component of $\eta$ and the Majorana fermion $N_{L}$. The $Z_2$ symmetry forbids the neutrino tree-level mass $\overline{L} H^c \nu_R$ and it is  softly broken to allow the two-loop realisation of such operator.}
 \label{tab:dirac2l:modelgenuine}
\end{table}

The $Z_4$ charges are chosen to forbid the Majorana mass term for $\nu_R$. It also prevents the mixing of the internal fermions with the Standard Model fermions. This avoids undesirable  terms that may mix new fermions with the charged leptons or the neutrinos. Moreover, $Z_4$ ensures the stability of the dark matter candidate, in our case the lightest of the ``$Z_4$ odd'' scalars and ``$Z_4$ even'' fermions, i.e. the lightest among ($S_1$, $S_2$, $\eta^0$, $N_{L}$).
%, as we will show in \sect{sec:dirac2l:dm}.

The new fermions in the loop are of two types. $E_{L,R}$ is a massive, ${\rm SU(2)_L}$ singlet vector-like fermion carrying hypercharge. Although its quantum numbers are the same as those of right-handed charged leptons, the $Z_4$ symmetry forbids their mixing. Since it is electrically charged, it cannot be the dark matter candidate. Therefore, its mass has to be taken sufficiently high. The fermion $N_{L}$ is also a ${\rm SU(2)_L}$ singlet fermion but carries no hypercharge. Owing to its quantum charges, one can write down a Majorana mass term for it and hence it's right-handed partner is not needed to give it mass. Being a neutral $Z_4$ even fermion, it can be a good dark matter candidate.

The scalars running in the loop, $\eta$ and $S_i$, must have exactly zero VEV in order to avoid breaking the $Z_4$ symmetry. Moreover, given their charges under $Z_4$, the lightest can be a good dark matter candidate, except $S_3$ which decays to the Standard Model.\footnote{Specifically, $S_3$ decays via the operator $H^\dagger H^\dagger S_3$, which can be generated at the one-loop level with the particle content of the model.}

The neutrino mass of the diagram in \fig{fig:dirac2l:softly} is generated from the following terms of the Lagrangian,
\begin{eqnarray} \label{eq:dirac2l:lagrangian1}
\mathcal{L}_\nu & = & (Y_1)_{\alpha i} \, \overline{L}_\alpha E_{R_i} \eta 
\, + \, (Y_2)_{ i \alpha}\, \overline{N}_{L_i} \nu_{R_\alpha}  S_2^\dagger 
\, +\, (Y_{12})_{ij}\, \overline{N}_{L_i} E_{R_j} S_3 
\,+ \, \hc
    \nonumber
    \\
&+& (\mathcal{M}_E)_{ij} \, \overline{E}_{L_i} E_{R_j} 
\,+\, (\mathcal{M}_N)_{ij} \, \overline{N}^c_{L_i} N_{L_j} \,+\, \hc
    \\
    \nonumber
    &+& \left[ \lambda \, \eta H S_1^\dagger S_3^\dagger \,+\, \mu_{12}^2 \, S_2^\dagger S_1  \,+\, \hc \right] 
    \,+\, m_\eta^2 \, \eta^\dagger \eta \,+\, \sum\limits_{k=1}^3 m_{S_k}^2 S_k^\dagger S_k
    \,+\, ...\;,
\end{eqnarray}
with $\alpha=1,2,3$ and where the term $\mu_{12}^2$ breaks softly the $Z_2$ symmetry. Other terms of the Lagrangian are not explicitly given, as they are not relevant for the neutrino mass generation. At this point, there is no need to fix the number of internal fermion copies. Nevertheless, given the fact that at least two neutrinos should have mass, the minimal choice in order to fit neutrino data would be $i,j=1,2$. Consequently, the effective Yukawa is given by
\begin{equation} \label{eq:dirac2l:yukawanu1}
    \left( Y_\nu \right)_{\alpha\beta} \approx \frac{1}{(16\pi^2)^2}\, \lambda\, \frac{\mu_{12}^2}{m_{S_3}^2} \left[ \frac{M_{Ei} M_{Nj}}{m_{S_3}^2} F^{(1)}_{ij} + F^{(2)}_{ij} \right]\, (Y_1)_{\alpha i} (Y_{12})_{ij} (Y_2)_{\beta j} ,
\end{equation}
with $M_{Ei}$ and $M_{Ni}$ the mass eigenstates of the $i$-copy of the fermions $E$ and $N_L$, respectively. The dimensionless loop integrals $F_{ij}$ are obtained directly in the mass insertion approximation assigning momenta to the internal fields as,
\begin{subequations} \label{eq:dirac2l:Fintegrals1}
    \begin{equation} %\label{eq:}
    \medmath{
        F^{(1)}_{ij} = m_{S_3}^4 \iint\limits_{(k,q)} \frac{ 1 }{ (k^2-m_\eta^2) (k^2-M_{Ei}^2) (q^2-M_{Nj}^2) (q^2-m_{S_1}^2) (q^2-m_{S_2}^2) ((k+q)^2-m_{S_3}^2) }
        },
    \end{equation}
\vspace*{-0.5cm}
    \begin{equation} %\label{eq:}
    \medmath{
        F^{(2)}_{ij} = m_{S_3}^2 \iint\limits_{(k,q)} \frac{ k \cdot q }{ (k^2-m_\eta^2) (k^2-M_{Ei}^2) (q^2-M_{Nj}^2) (q^2-m_{S_1}^2) (q^2-m_{S_2}^2) ((k+q)^2-m_{S_3}^2) }
        },
    \end{equation}
\end{subequations}
with the shorthand $\int\limits_k \equiv (16\pi^2)\int d^4\!k / (2\pi)^4$. Both integrals can be written in terms of simple one-loop and two-loop integrals, for which analytical expressions can be found. The decomposition of the integrals $F^{(a)}$ is done as an example of how to compute two-loop radiative masses in \app{app:loops}.

In order to fit the neutrino oscillation data \cite{deSalas:2017kay}, we need at least two copies of $E$ and $N_L$, so that $Y_\nu$ is a rank-2 matrix, giving masses to two neutrinos. We have then three rank-2 matrices ($Y_1$, $Y_2$, $Y_{12}$) with enough freedom to fit the two neutrino mass squared differences, along with the three mixing angles and phases. However, here we will only consider the case with one massive neutrino, by assuming no hierarchy or flavour structure in the Yukawas and just one copy of the new fermions, i.e. $Y_1 = Y_2 = Y_{12} = Y$. This is done in order to simplify our analysis and show the characteristic neutrino mass scale $m_\nu$ in this type of models. 

\begin{figure}[t!]
\centering
    \includegraphics[width=0.75\textwidth]{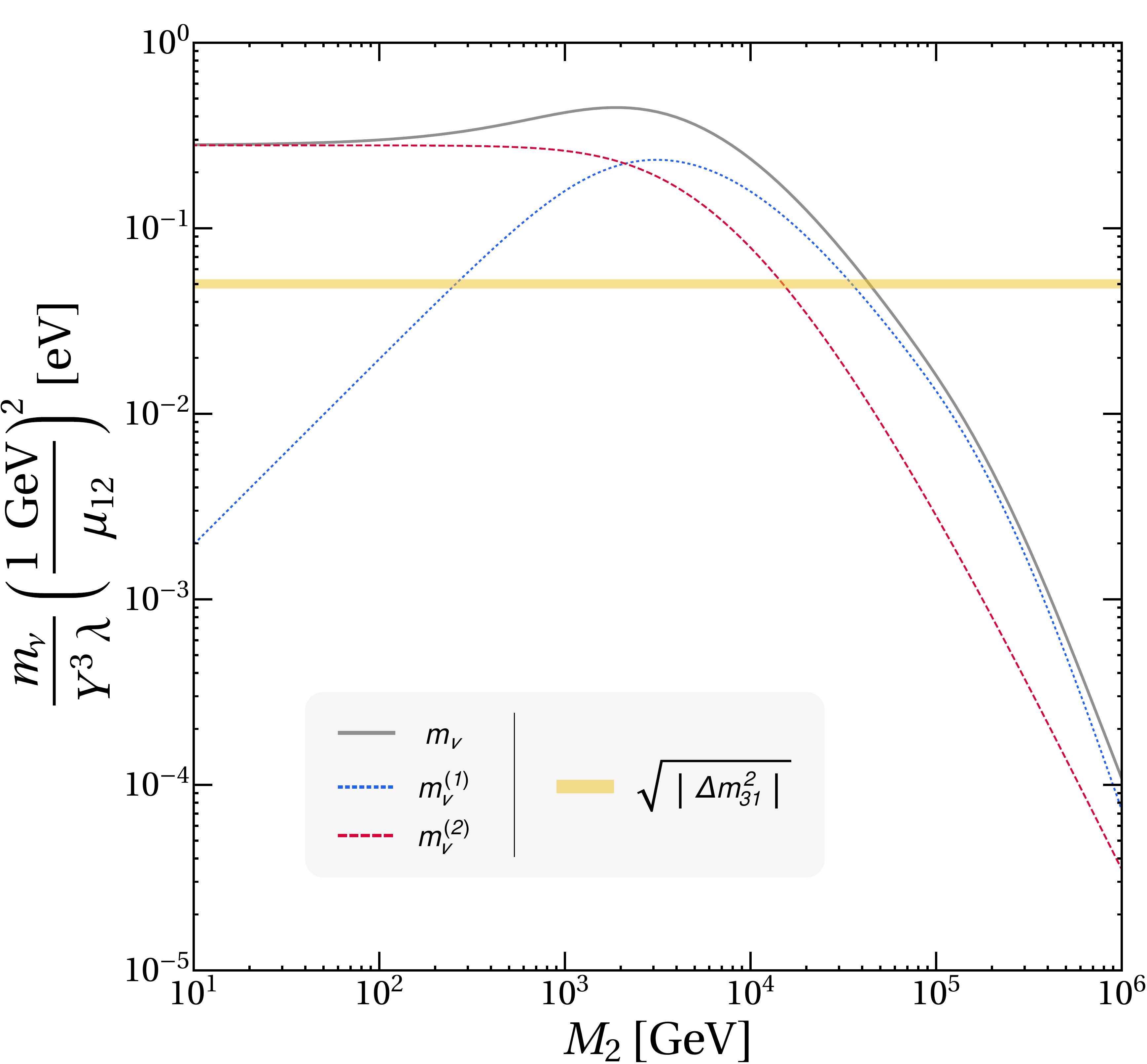}
    \caption{Neutrino mass scale $m_\nu$ (solid line) for the diagram of \fig{fig:dirac2l:softly} with respect to the mass of $N_L$. The rest of the masses are $\mathcal{O}(1)$ TeV. The contributions coming from both $F^{(a)}$, see \eq{eq:dirac2l:yukawanu1}, has been separated (dashed lines). The atmospheric mass scale (yellow line) is plotted for comparison. See text for details.}
    \label{fig:dirac2l:NuMassplot1} 
\end{figure}

The behaviour of the neutrino mass is given in \fig{fig:dirac2l:NuMassplot1} in terms of the couplings of the model and the mass of $N_{L}$ field. Here, we consider that the masses of the rest of the internal fields are of order $1$ TeV. The atmospheric scale $\sqrt{|\Delta m_{13}^2 |} \approx 0.05$ eV is plotted for comparison. The dashed lines $m_\nu^{(a)}$ represent the neutrino mass scale when only the loop integral $F^{(a)}$ is considered, while the solid line is for the complete mass equation \eq{eq:dirac2l:yukawanu1}.

In \fig{fig:dirac2l:NuMassplot1} we see the distinct behaviour of $F^{(1)}$ and $F^{(2)}$ due to the different numerators. Also, notice how the function decreases when $M_N$ becomes larger than the rest of the masses of $\mathcal{O}(1)$ TeV and its propagator starts dominating the integral. If all couplings are taken to be  $\mathcal{O}(1)$, the mass scale should be about $100$ TeV. Nevertheless, the cubic dependence of the neutrino mass with the Yukawas can lower this scale considerably, allowing masses of order $1$ TeV or below accessible at colliders.

%%%%%%%%%%%%%%%%%%%%%%%%%%%%%%%%%%%%%%%%%%%%%%%%%%%%%%%%%%%%%%%
\subsection{Model exploiting the non-local realisation \\of $HH S$} \label{subsec:dirac2l:nonlocal}

Now we move to a different class of diagrams, those depicted in \fig{fig:dirac2l:diagrams2}. Models generated from these diagrams need certain fields in order to be genuine. As explained in \sect{subsec:dirac2l:class3}, they contain a one-loop three scalar vertex with one external Higgs. Such loops are reducible unless the other scalars are another Higgs and the charged singlet $S\equiv(\mathbf{1},\mathbf{1},-1)$, realising the loop effective coupling $HH S$ (see \fig{fig:3loop:exception2}).

As a simple example of how these models work, we will take the diagram $D_3^{(xi)}$ in \fig{fig:dirac2l:diagrams3} and add to the Standard Model the particle content given in \tab{tab:dirac2l:modelHHS}. The role of the cyclic $Z_4$ and $Z_2$ symmetries is analogous to the previous model.

\begin{table}[t!]
\begin{center}
\begin{tabular}{| c || c | c | c | c |}
  \hline 
&   \hspace{0.1cm}  Fields     \hspace{0.1cm}       &  \hspace{0.4cm}  ${\rm SU(2)_L} \times {\rm U(1)_Y}$     \hspace{0.4cm}    & \hspace{0.4cm}   $Z_4$            \hspace{0.4cm}   &  \hspace{0.4cm}  $Z_2$            \hspace{0.4cm}                            \\
\hline \hline
\multirow{5}{*}{ \begin{turn}{90} Fermions \end{turn} }
&   $L$              &   ($\mathbf{2}, {-1/2}$)      &   $i$ & $+$ \\	
&   $\nu_{R}$        &   ($\mathbf{1}, {0}$)         &   $i$ & $-$  \\
&   $e_{R}$          &   ($\mathbf{1}, {-1}$)        &   $i$ & $+$   \\
&   $E_{R}$          &   ($\mathbf{1}, {-1}$)        &   $i$ & $+$    \\
&   $E_{L}$    	     &   ($\mathbf{1}, {-1}$)        &   $i$ & $+$     \\
\hline \hline                                                                              
\multirow{5}{*}{ \begin{turn}{90} Scalars \end{turn} } 
& $H$  	       	 &  ($\mathbf{2}, {1/2}$)        & $1$   & $+$        \\		
& $S_1$          	 &  ($\mathbf{1}, {1}$)          & $1$   & $-$         \\
& $S_0$              &  ($\mathbf{1}, {0}$)          & $i$   & $+$          \\	
& $S_1'$             &  ($\mathbf{1}, {1}$)          & $i$   & $+$           \\	
& $\eta$             &  ($\mathbf{2}, {1/2}$)        & $i$   & $+$            \\
    \hline
  \end{tabular}
\end{center}
\caption{Particle content of the example $HH S$ model. The gauge charges along with the $Z_4 \times Z_2$ charges are also shown. All the fields listed in the table are ${\rm SU(3)_C}$ singlets. Again, the role of $Z_4$ is to protect Diracness and to stabilise the dark matter candidate: the lightest out of $S_0$ and the neutral component of $\eta$. The $Z_2$ symmetry forbids the neutrino tree-level mass $\overline{L} H ^c \nu_R$ and it is softly broken to allow the two-loop realisation of such operator.}
\label{tab:dirac2l:modelHHS}
\end{table}

Note that the new fermions have the same gauge charges as the right-handed charged leptons and, therefore, they mix. This mixing has to be controlled in order for the model to be phenomenologically viable. Looking at the relevant Lagrangian terms,
\begin{eqnarray}
 \mathcal{L}_{\nu N} &=& (Y_e)_{\alpha\beta} \, \overline{L}_\alpha H \,  e_{R\beta} \,+\, (Y_1)_{\alpha i}  \, \overline{L}_\alpha H E_{Ri} \,+\, \hc 
 \\ \nn &+& (\mathcal{M}_E)_{ij} \, \overline{E}_{Li} E_{Rj} \,+\, \mathcal{X}_{\alpha i} \, \overline{E}_{Li} \, e_{R\alpha} \,+\, \hc \, ,
\end{eqnarray}
which can be written in matrix form as,
\begin{equation}
    \left (\begin{matrix} \overline{L} & \overline{E}_L \end{matrix} \right)
                                    \left (\begin{matrix}
                                    Y_e v & Y_1 v \\
                                    \mathcal{X} & \mathcal{M}_E
                                   \end{matrix} \right)
                                   \left (\begin{matrix}
                                   e_R \\
                                   E_R
                                 \end{matrix} \right).
\end{equation}
Here, $\alpha, \beta = 1,2,3$ and $(i,j)$ the number of copies of $E$. Taking the Yukawa matrices of order $1$, it is easy to see that if the elements of matrices $\mathcal{X}$ and $\mathcal{M}_E$ are bigger than the Standard Model vacuum expectation value $v$, then the mixing in the left-handed sector will be sufficiently small to avoid collider constraints. The phenomenology of vector-like singlet leptons has been extensively studied in the literature, setting limits on their masses around $\sim 100$ GeV \cite{Kumar:2015tna, Falkowski:2013jya}. Therefore it is safe to consider the mass of the charged fermions to be around or above the TeV scale. Lepton flavour violating processes can set also stringent limits on the mass depending on the value of $Y_1$. Nevertheless, given our model, lepton flavour violation can be hidden, as there are enough free parameters between $Y_1$ and $Y_2$ in order to fit neutrino data while suppressing significantly any flavour violating signal.

\begin{figure}
\centering
    \includegraphics[width=0.53\textwidth]{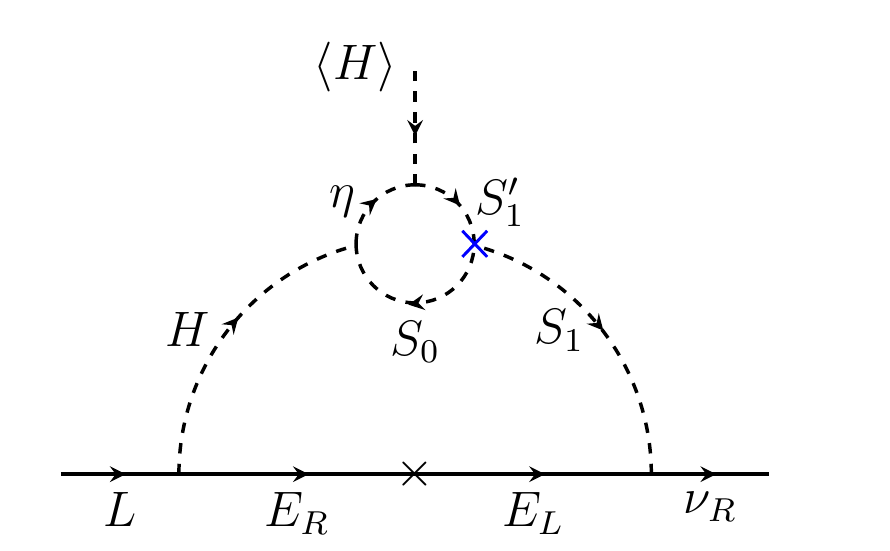}
    \hspace*{-1.3cm}
    \includegraphics[width=0.53\textwidth]{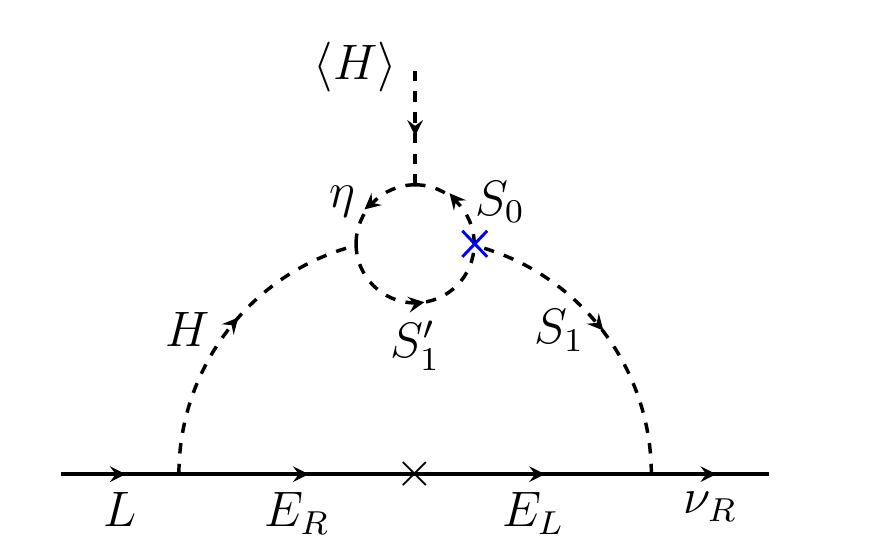}
    \caption{Leading contributions to neutrino masses. The blue cross marks the soft breaking term of the $Z_2$ symmetry that allows the two-loop realisation of the operator $\overline{L} H^c \nu_R$. Note that the small scalar loop cannot be reduced into a tree-level vertex.}
    \label{fig:dirac2l:loopHHS}
\end{figure}

In the neutrino sector, as explained before, the tree-level mass term $\overline{L} H ^c \nu_R$ is forbidden by the $Z_2$ symmetry. Indeed, this symmetry will forbid all the loop realisations of such operators unless it is softly broken. Once we allow the soft breaking of $Z_2$, we have to add only one extra term to the Lagrangian, $S_1^\dagger S_1' S_0^\dagger$. This term is essential and leads to the neutrino mass diagrams depicted in \fig{fig:dirac2l:loopHHS}.

One could be tempted to add also the tree-level coupling $H H S_1$, which is allowed by the gauge symmetry and $Z_4$ and breaks $Z_2$ only softly. However, this term vanishes because the contraction $HH$ to a singlet is completely antisymmetric. Therefore the leading contribution to neutrino masses will be the two-loop diagrams shown in \fig{fig:dirac2l:loopHHS} (see \sect{subsec:dirac2l:class3} for details).

The lightest among the neutral component of $\eta$, $S_1'$ and $S_0$ will be stable and thus a good dark matter candidate. In this model all the dark matter candidates are scalars.

\begin{figure}
\centering
    \includegraphics[width=0.75\textwidth]{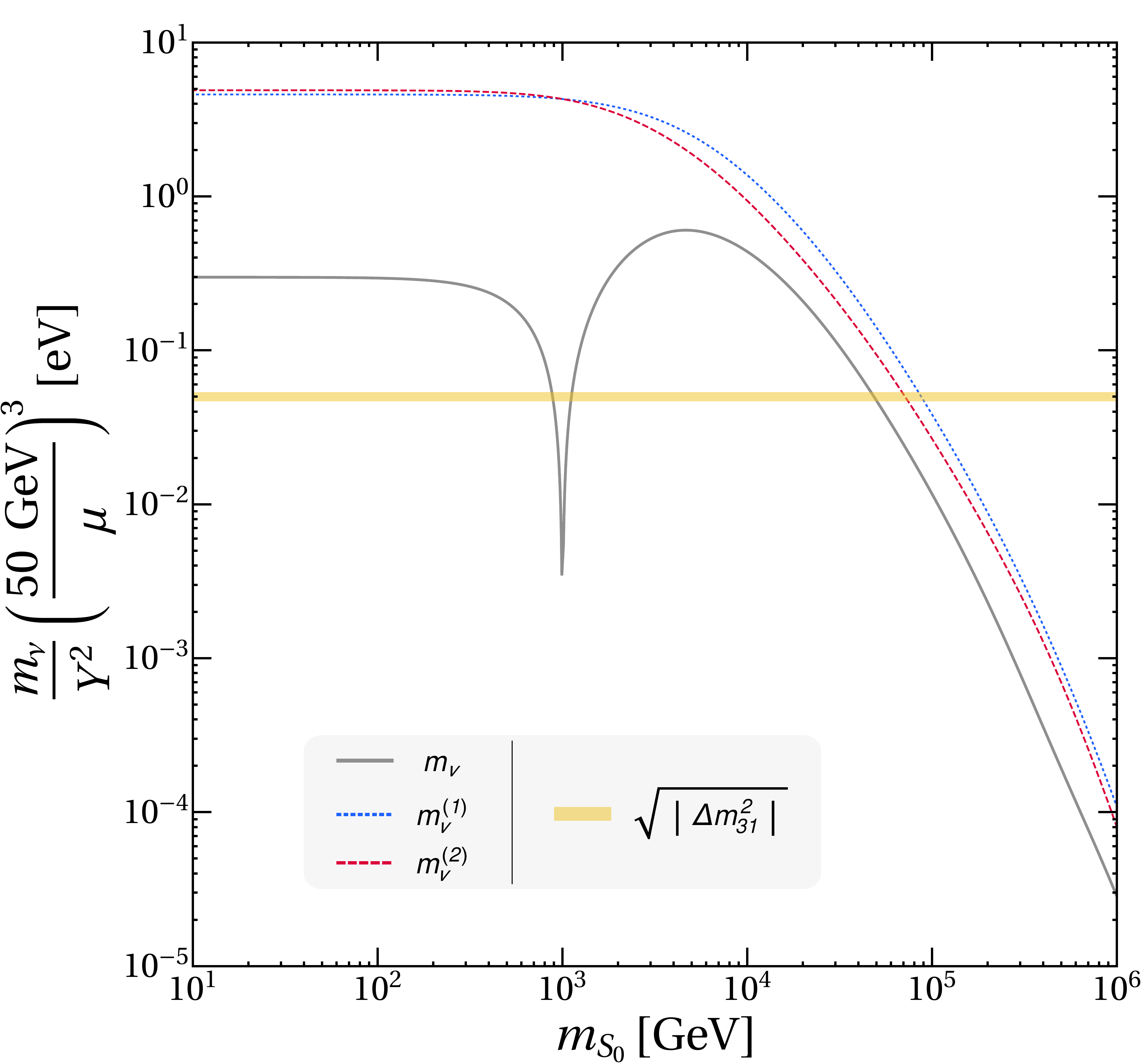}
    \caption{Neutrino mass scale $m_\nu$ (solid line) for the diagrams of \fig{fig:dirac2l:loopHHS} with respect to the mass of the neutral scalar $S_0$. The rest of the masses are $\mathcal{O}(1)$ TeV. Both contributions are separated and represented as dashed lines. The atmospheric mass scale (yellow line) is plotted for comparison. See text for details.}
    \label{fig:dirac2l:NuMassplot2} 
\end{figure}

The main feature of this class of models is that given the loop vertex with two identical Higgs, there are always two contributions simply interchanging both Higgs. Moreover, both contributions have a relative minus sign due to the antisymmetric nature of ${\rm SU(2)_L}$ contractions. Precisely, the minus sign comes from the coupling $H \eta S_1^{'\dagger} \equiv \epsilon^{ij} H_i \eta_j S_1^{'\dagger} = (H^+ \eta^0 - H^0 \eta^+) S_1^{'\dagger}$. This can produce a cancellation between both diagrams, leading to a suppression of the neutrino mass scale as can be seen in \fig{fig:dirac2l:NuMassplot2}.

The corresponding terms of the Lagrangian that appear in the diagrams of \fig{fig:dirac2l:loopHHS} are,
\begin{eqnarray} \label{eq:dirac2l:lagrangian2}
    \mathcal{L} &=& (Y_1)_{\alpha i}\, L_\alpha \overline{E}_{Ri} H^\dagger \,+\, (Y_2)_{\alpha i}\, \overline{\nu_R}_\alpha E_{Li} S_1 \,+\, (\mathcal{M}_E)_{ij} \, \overline{E}_{Ri} E_{Lj} \,+\, \hc
    \nonumber
    \\
    &+& \mu_S S_1' S_1^\dagger S_0^\dagger \,+\, \mu_1 \eta H  S_1^{'\dagger} \,+\, \mu_2 \eta^\dagger H S_0 \,+\, \hc
    \\
    \nonumber
    &+& m_\eta^2 \eta^\dagger \eta \,+\, m_{S_0}^2 S_0^\dagger S_0 \,+\, m_{S_1}^2 S_1^\dagger S_1 \,+\, m_{S_1'}^2  S_1^{'\dagger} S_1' \,+\, ...\;,
\end{eqnarray}
where the term $\mu_S$ breaks softly the $Z_2$ symmetry. The rest of the scalar potential is omitted for simplicity.

The effective Yukawa associated to the neutrino masses is given by,
\begin{equation} \label{eq:dirac2l:yukawanu2}
    \left( Y_\nu \right)_{\alpha\beta} \approx \frac{1}{(16\pi^2)^2} \frac{\mu_S \mu_1 \mu_2}{M_i^3} \left[ \left( \Delta m_0^2 c^2_0 + \Delta m_+^2 s^2_+ \right) F^{(1)}_i + F^{(2)}_i \right]\, (Y_1)_{\alpha i} (Y_2)_{\beta i}.
\end{equation}
Here, $M_i$ are the mass eigenvalues of the vector-like fermions $E$, $\Delta m_0^2$ is the mass difference between the neutral eigenstates coming from the mixing of ($\eta^0$,$S_0$) with mixing angle $\cos \theta_0 \equiv c_0$ and $\Delta m_+^2$ the same for the charged eigenstates of ($\eta^+$,$S_1'$) with mixing angle $\sin \theta_+ \equiv s_+$. 

$F_{i}$ are the dimensionless loop integrals of the form,

\begin{subequations} \label{eq:dirac2l:Fintegrals2}
    \begin{equation} %\label{eq:}
        F^{(1)}_{i} = M_i^6 \iint\limits_{(k,q)} \frac{ 1 }{ \mathcal{D}_0 } \times \frac{1}{ (q^2-m_{\eta^+}^2) ((k+q)^2-m_{\eta^0}^2) } \, 
    \end{equation}
    \begin{equation}
        F^{(2)}_{i} = M_i^4 \iint\limits_{(k,q)} \frac{ 1 }{ \mathcal{D}_0 } \times \left[ \frac{1}{(q^2-m_{\eta^+}^2)} - \frac{1}{((k+q)^2-m_{\eta^0}^2)} \right] \,
    \end{equation}
\end{subequations}
with $\mathcal{D}_0^{-1} = (k^2-M_i^2) (k^2-M_W^2) (k^2-m_{S_1}^2) (q^2-m_{S_1}^2) ((k+q)^2-m_{S_0}^2)$. Both integrals can be written in terms of simple one-loop and two-loop integrals, for which analytical expressions can be found, see \app{app:loops}.

In the same fashion as before, the neutrino mass scale is given in \fig{fig:dirac2l:NuMassplot2} in terms of the couplings of the model and the mass of $S_0$, keeping other the masses of order $1$ TeV. Here, we consider no hierarchy in the Yukawas, obtaining only a characteristic mass scale for neutrinos. The contributions for both integrals \eq{eq:dirac2l:Fintegrals2} are considered separately, plotted as dashed lines with labels $m_\nu^{(a)}$. The combination \eq{eq:dirac2l:yukawanu2} is depicted as a solid line.

The overall behaviour is similar to \fig{fig:dirac2l:NuMassplot1}, but with a cancellation among diagrams. For small masses there is a visible suppression of the neutrino mass scale that even vanishes when $m_{S_0}^2 \approx m_{S_1'}^2$, leading to a lower neutrino mass compared to the previous example.

%%%%%%%%%%%%%%%%%%%%%%%%%%%%%%%%%%%%%%%%%%%%%%%%%%%%%%%%%%%%%%%
%%%%%%%%%%%%%%%%%%%%%%%%%%%%%%%%%%%%%%%%%%%%%%%%%%%%%%%%%%%%%%%
\section{Summary} \label{sec:dirac2l:summary}

We have discussed the complete decomposition of the Dirac neutrino mass operator $\bar{L} H^c \nu_R$ at two-loop order. We have identified all the 1PI topologies and diagrams with 3 external legs, two-loops and 3, 4-point vertices which give the dominant contribution to the neutrino mass. We call such diagrams \textit{genuine}. From an initial set of 70 topologies, only 5 satisfy the genuineness criteria (\fig{fig:dirac2l:topologies}), obtained after removing tadpoles, self-energy diagrams and non-renormalisable contributions.

A set of 18 renormalisable diagrams are generated straightforward from the 5 genuine topologies. We classify them in three different classes depending on the requirements imposed on their possible particle content, in order to generate a genuine two-loop model. The three diagrams generated from topologies $T_1$ and $T_2$ given in \fig{fig:dirac2l:diagrams1} are genuine in general. This means that there is no particular field or symmetry breaking requirement for these diagrams to be the dominant contribution to neutrino masses. Meanwhile, the other 15 diagrams contain a one-loop realisation of either a fermion-fermion-scalar vertex (\fig{fig:dirac2l:diagrams2}) or a three scalar vertex (\fig{fig:dirac2l:diagrams3}). The former is genuine if one provides a symmetry transformation that forbids not only the tree-level but also the one-loop diagram and, then, breaks it softly allowing the two-loop mass diagram. The latter always requires that the three scalars of the loop vertex are $H$, $H$ and $S\equiv(\mathbf{1},\mathbf{1},-1)$. The antisymmetric nature of ${\rm SU(2)_L}$ contractions makes the local tree-level operator $H(x) H(x) S(x)$ zero but not its loop realisations, consequently forbidding the reduction of this class of two-loop diagrams into their corresponding one-loop diagrams, see \fig{fig:3loop:exception2} and \sect{sec:3loop:specgen} for details. Finally, we have found that every neutrino mass generated from the operator $\bar{L} H^c \nu_R$ at two-loop order can be written in terms of 6 mass diagrams or integrals (\fig{fig:dirac2l:massdiagrams}). These integrals can be decomposed in terms of two master integrals for which analytical expressions already exist \cite{Sierra:2014rxa, Martin:2016bgz}.

We have shown how one can generate models from our classification, listing all possible Standard Model quantum numbers with ${\rm SU(2)_L}$ representations up to triplets. Although, for simplicity, we only discussed the cases with colour singlet fields, nevertheless as explained before, introducing non-trivial representations of ${\rm SU(3)_C}$ is straightforward as the external fields are colour-blind. To illustrate how our classification can be used to generate genuine models, we have constructed and discussed in detail two different Dirac neutrino mass models. Each of the models is built from two characteristic sets of diagrams explained in the previous section. One of the example models uses a completely genuine topology, so that the two-loop contribution is guaranteed to be the leading order contribution to neutrino masses. The second example model makes use of the non-locality of the operator $H H S$ in order to be non-reducible. We have shown that these types of models are able to fit neutrino oscillation data for reasonable values of the masses and parameters. Such models may be testable, while a part of the parameter space is already excluded by collider searches. In this direction, a more involved analysis of the phenomenology of these models would be needed for detailed quantitative results.

Finally, we have not discussed in detail the connection between the symmetry that protects the Dirac nature of neutrinos and the stability of dark matter. We have given two example models where a $Z_4$ symmetry forbids a Majorana mass term for neutrinos at all orders, and, at the same time, this same symmetry stabilises a dark matter candidate. For both cases, the exact $Z_4$ symmetry ensures the stability of the lightest particle belonging to the \textit{dark sector}, form of $Z_4$-odd scalars and $Z_4$-even fermions. This is due to the interplay between the $Z_2$ subgroup of $Z_4$ and the Lorentz symmetry, although other possibilities exist, as it will be discussed in detail in the following chapter. The relationship between the Diracness of neutrinos and dark matter stability is an attractive possibility intimately connecting the neutrino and dark matter physics through the same symmetry.

\pagebreak
\fancyhf{}

%% file: Chapters/Dirac_Majorana_DM/Chapter_DM.tex
\fancyhf{}
\fancyhead[LE,RO]{\thepage}
\fancyhead[RE]{\slshape{Chapter \thechapter. Dark matter stability and the nature of neutrinos}}
\fancyhead[LO]{\slshape\nouppercase{\rightmark}}

\chapter{Dark matter stability and the nature of neutrino masses}
\label{ch:dm_DiracMajo}
\graphicspath{ {Chapters/Dirac_Majorana_DM/} }

There are particularly attractive scenarios that connect dark matter to neutrino physics in an intimate manner. The scotogenic model is one such model where the ``dark sector'' participates in the loop responsible for neutrino mass generation \cite{Ma:2006km}. It is also possible to find scenarios where the dark matter stability is related to the Dirac or Majorana nature of neutrinos \cite{Chulia:2016ngi, CentellesChulia:2018gwr}. The main idea of these works is to use the Standard Model lepton number ${\rm U(1)_L}$ symmetry (equivalently, the anomaly free ${\rm U(1)_{B-L}}$ symmetry), or its appropriate $Z_n$ subgroup, to enforce either Dirac or Majorana neutrino masses, as well as to stabilise dark matter. In this approach, the nature of neutrinos and the stability of dark matter are intimately connected, having their origins in the same lepton number symmetry.
\\

In this chapter, based on \cite{Bonilla:2018ynb, CentellesChulia:2019gic}, we study in general how the breaking pattern of lepton number determines whether neutrinos are Dirac or Majorana, depending on the residual $Z_n$ symmetry. We aim to relate this same symmetry to the stability of a dark matter candidate that participates in the radiative generation of the neutrino masses. Then in \sects{sec:dm:dirac}{sec:dm:majo}, we describe in detail, as general as possible, the scenarios where neutrinos are Dirac or Majorana, showing some examples for both cases.

%%%%%%%%%%%%%%%%%%%%%%%%%%%%%%%%%%%%%%%%%%%%%%%%%%%%%%%%%%%%%%%
%%%%%%%%%%%%%%%%%%%%%%%%%%%%%%%%%%%%%%%%%%%%%%%%%%%%%%%%%%%%%%%
\section{Lepton number and the nature of neutrinos} \label{sec:dm:lepton}

From a theoretical point of view, the issue of the Dirac/Majorana nature of neutrinos is intimately connected to lepton number or, equivalently, ${\rm U(1)_{B-L}}$ symmetry of the Standard Model and its possible breaking pattern \cite{Hirsch:2017col}. If the ${\rm U(1)_{B-L}}$ symmetry is conserved in nature, then the neutrinos will be Dirac fermions. However, if it is broken to a residual $Z_m$ subgroup with $m\in \mathbb{Z}^+$ and $m \geq 2$,\footnote{ $\mathbb{Z}^+$ being the set of all positive integers.} then the Dirac/Majorana nature will depend on the residual $Z_m$ symmetry provided that the Standard Model lepton doublets $L_i = (\nu_{L_i}, l_{L_i})^T$  do not transform trivially under it. Thus, we have
\begin{eqnarray} \label{eq:dm:oddzn}
    {\rm U(1)_{B-L}}   & \, \to  \, &   Z_m \equiv Z_{2n+1} \, \text{with} \,  n \in \mathbb{Z}^+ 
    \nn \\
    && \, \Longrightarrow \, \text{neutrinos are Dirac particles} \, ,
    \nn \\
    && 
    \nn \\
    {\rm U(1)_{B-L}}  & \, \to  \,  & Z_m \equiv Z_{2n} \, \text{with} \,  n \in \mathbb{Z}^+  
    \\ \nn
    && \, \Longrightarrow \, \text{neutrinos can be Dirac or Majorana } \, .
\end{eqnarray}
If the ${\rm U(1)_{B-L}}$ is broken to a $Z_{2n}$ subgroup, then one can make a further classification depending on how the $L_i$ transform,
\begin{eqnarray} \label{eq:dm:evenzndir}
    L_i \; \left\{ 
        \begin{array}{llc}
            \nsim  \, \omega^{n} \ \ \text{under $Z_{2n}$} &  \Longrightarrow & \text{Dirac neutrinos}\\
            \sim  \, \omega^{n} \ \ \text{under $Z_{2n}$} &  \Longrightarrow  & \text{Majorana neutrinos}
        \end{array} \right. \, ,
\end{eqnarray}
where $\omega^{2n}=1$. Note that $\{ 1, \, \omega^n \}$ form a $Z_2$ subgroup of $Z_{2n}$. Indeed, we can conclude that if the Standard Model lepton doublet transforms either trivially under $Z_m$, or belongs to a $Z_2$ subgroup of the residual $Z_m$ symmetry, neutrinos can be Majorana particles. Else, neutrinos will be Dirac particles.

%%%%%%%%%%%%%%%%%%%%%%%%%%%%%%%%%%%%%%%%%%%%%%%%%%%%%%%%%%%%%%%
\section{Dark matter stability} \label{sec:dm:dm}

As pointed out before, if we start with a ${\rm U(1)_{B-L}}$ symmetry, by controlling its breaking and the charges of the leptons under it, we can have either Majorana or Dirac neutrinos protected by a remnant $Z_m$ symmetry. We can take advantage of this symmetry to stabilise a possible dark matter candidate, without adding a new \textit{ad hoc} symmetry.

Here, we are interested in the case where the residual $Z_m$ symmetry, and consequently ${\rm U(1)_{B-L}}$, is responsible for the stability of dark matter. By this we mean that we want to avoid the appearance of accidental symmetries that may be ultimately responsible for the stability of dark matter. A well-known case of the latter is the \textit{Minimal dark matter} \cite{Cirelli:2005uq}, where the dark matter candidate is stable because no Standard Model particles have the quantum numbers to couple to it. The accidental nature of the symmetry is reflected in the fact that, although every operator is invariant at dimension 4, at higher dimensions operators appear which violate the symmetry. This is for instance the case of lepton number in the Standard Model, though a symmetry of the dimension 4 theory, it is broken by the dimension five Weinberg operator.

\begin{figure}
    \centering
    \includegraphics[width=0.5\textwidth]{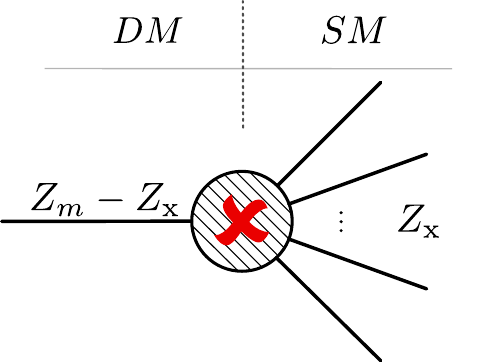}
    \caption{The decay of the dark matter to the Standard Model is forbidden by the residual symmetry $Z_m$ and protected by the subgroup $Z_{\rm x} \subset Z_{m}$. In this scenario, the Standard Model is charged under the subgroup $Z_{\rm x}$, while the dark matter candidate will be the lightest among the particles transforming as an element of $Z_{m}-Z_{\rm x}$. Here, the solid lines denote either scalars or fermions. For details, see point (I) in \sect{sec:dm:dm}.}
    \label{fig:dm:dm1}
\end{figure}

\begin{figure}
    \centering
    \begin{subfigure}[t]{0.47\textwidth}
        \centering
        \includegraphics[width=1\textwidth]{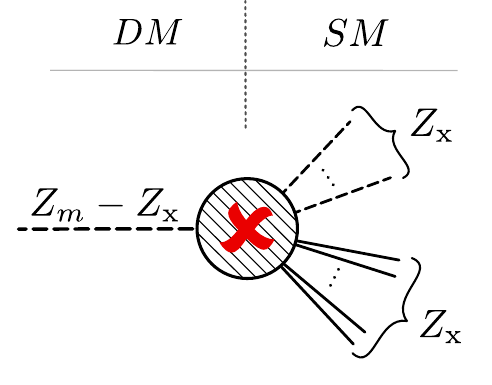}
        \caption{Stable scalar dark matter candidate.}
    \end{subfigure}%
    \hfill
    \begin{subfigure}[t]{0.47\textwidth}
        \centering
        \includegraphics[width=1\textwidth]{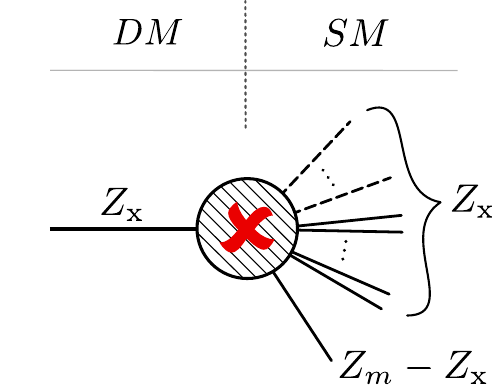}
        \caption{Stable fermion dark matter candidate.}
    \end{subfigure}
    \caption{Dark matter stability according to point (II) in \sect{sec:dm:dm}. Here, the Standard Model fermions (scalars) transform as an element of $Z_{m}-Z_{\rm x}$ ($Z_{\rm x}$). Providing that every Lorentz invariant combination of SM fields (denoted for fermions as pairs of lines in the figures) transforms under $Z_{\rm x}$; the lightest among fermions and scalars charged under $Z_{\rm x}$ and $Z_m - Z_{\rm x}$, respectively, will be stable.}
    \label{fig:dm:dm2}
\end{figure}

Without invoking new exact or accidental symmetries, the idea behind dark matter stability by means of a $Z_m$ symmetry is very simple: $Z_m$ should have at least a subgroup $Z_{\rm x}$, this implies that $m$ cannot be a prime number. We shall denote as $Z_m - Z_{\rm x}$ the elements in $Z_m$ that are not in $Z_{\rm x}$.\footnote{Note that this is related to the quotient group $Z_m / Z_{\rm x}$.} We have then two possible scenarios for the stability of dark matter from ${\rm U(1)_{B-L}} \rightarrow Z_m \supset Z_{\rm x}$:
\begin{enumerate}[(I).]
    \item All the Standard Model fields are charged under $Z_{\rm x}$:
    
        As $Z_{\rm x}$ is a subgroup and, consequently, it is closed, every operator build from Standard Model fields will be charged under $Z_{\rm x}$. So any particle transforming as an element of $Z_m - Z_{\rm x}$ will not decay only to Standard Model particles at any order. i.e. will be stable. This is depicted schematically in \fig{fig:dm:dm1}.
    
    \item All the Standard Model fermions are charged under $Z_m - Z_{\rm x}$:
    
        In this case the stability of dark matter is not straightforward as before. Lorentz's symmetry provides that fermions should appear in pairs in the operators. With the Standard Model Higgs charged under $Z_{\rm x}$, any scalar transforming as an element of $Z_m - Z_{\rm x}$ will be stable. Similarly, a fermion charged under the subgroup $Z_{\rm x}$ will not decay solely to the Standard Model. Note that this is true only if the product of any two elements in $Z_m - Z_{\rm x}$ always belong to $Z_{\rm x}$. Moreover, this relation holds only for even cyclic groups, i.e. $Z_m \equiv Z_{2n}$, and considering the subgroup $Z_{\rm x} \equiv Z_n$. See \fig{fig:dm:dm2}.
\end{enumerate}

In this section, we have focused on the possible model-independent scenarios where we can have a completely stable dark matter candidate considering only the residual $Z_m$ symmetry. It is also possible that in some models, the stability of dark matters comes from the interplay between the Standard Model symmetries and the discrete symmetry $Z_m$ \cite{Ma:2019yfo}. The other possibility is that the effective decay operators cannot be UV completed in a model due to its limited particle content. If this happens, then a new accidental symmetry will appear in such a model, protecting the dark matter against decaying. We have not considered such cases as they depend on the details of the model. Furthermore, we have always been looking for a completely stable dark matter candidate, it is possible that the dark matter candidate is not absolutely stable, but it decays slowly enough so that its half-life is much larger than age of the Universe, being then a phenomenologically viable dark matter candidate.

%%%%%%%%%%%%%%%%%%%%%%%%%%%%%%%%%%%%%%%%%%%%%%%%%%%%%%%%%%%%%%%
%%%%%%%%%%%%%%%%%%%%%%%%%%%%%%%%%%%%%%%%%%%%%%%%%%%%%%%%%%%%%%%
\section{Dirac neutrinos} \label{sec:dm:dirac}

As we already discussed, to explain dark matter, the particle content of the Standard Model needs to be extended. Furthermore, to account for dark matter stability new explicit \cite{Silveira:1985rk, Ma:2006km} or accidental symmetries \cite{Cirelli:2005uq} beyond those of the Standard Model are also invoked. On the other hand, the understanding of the tiny, yet non-zero, masses of neutrinos also requires extending the Standard Model in one way or another \cite{Ma:1998dn, CentellesChulia:2018gwr}.

In this section, we aim to develop a general formalism where the following conditions are satisfied:
\begin{enumerate}[(I).]
    \item Neutrinos are Dirac in nature.
    \item Naturally small neutrino masses are generated through finite loops, forbidding the tree-level neutrino Yukawa couplings.
    \item The dark sector participates in the loop. The lightest particle being stable is a good dark matter candidate.
\end{enumerate}
Usually one needs at least three different symmetries besides those of the Standard Model to achieve this \cite{Bonilla:2016diq}. However, we show that all of these requirements can be satisfied with just lepton number, without adding any extra explicit or accidental symmetries. In our formalism, we employ a chiral realisation of ${\rm U(1)_{B-L}}$, spontaneously broken to a residual $Z_n$ symmetry. This ${\rm U(1)_{B-L}}$ can be anomaly free.

Before going into the details of the formalism, let us briefly discuss the possibility of chiral solutions to ${\rm U(1)_{B-L}}$ anomaly cancellation conditions. It is well-known that the accidental ${\rm U(1)_B}$ and ${\rm U(1)_L}$ symmetries of the Standard Model are anomalous, but the ${\rm U(1)_{B-L}}$ combination can be made anomaly free by adding three right-handed neutrinos $\nu_{R_i}$ with $(-1,-1,-1)$ vector charges under ${\rm U(1)_{B-L}}$. However, chiral solutions to ${\rm U(1)_{B-L}}$ anomaly cancellation conditions are also possible. The particularly attractive feature of chiral solutions is that by using them one can automatically satisfy conditions (I) and (II), as shown in \cite{Ma:2014qra, Ma:2015mjd}, using for instance the chiral solution $\nu_{R_i} \sim (-4,-4,5)$ under ${\rm U(1)_{B-L}}$ symmetry.

\begin{figure}
    \centering
    \begin{subfigure}[b]{0.43\textwidth}
        \includegraphics[width=1.\textwidth]{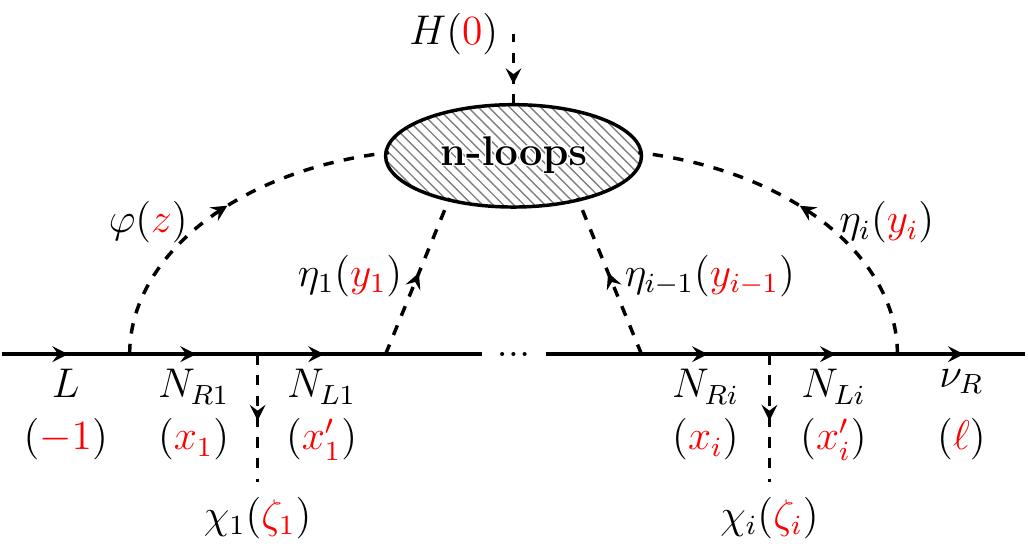}
        \caption{General ${\rm U(1)_{B-L}}$ charge assignment.}
        \label{fig:dm:genU1}
    \end{subfigure}
    \begin{subfigure}[t]{0.1\textwidth}
        \vspace*{-4.15cm}
        \hspace*{-0.45cm}
        \includegraphics[width=1.5\textwidth]{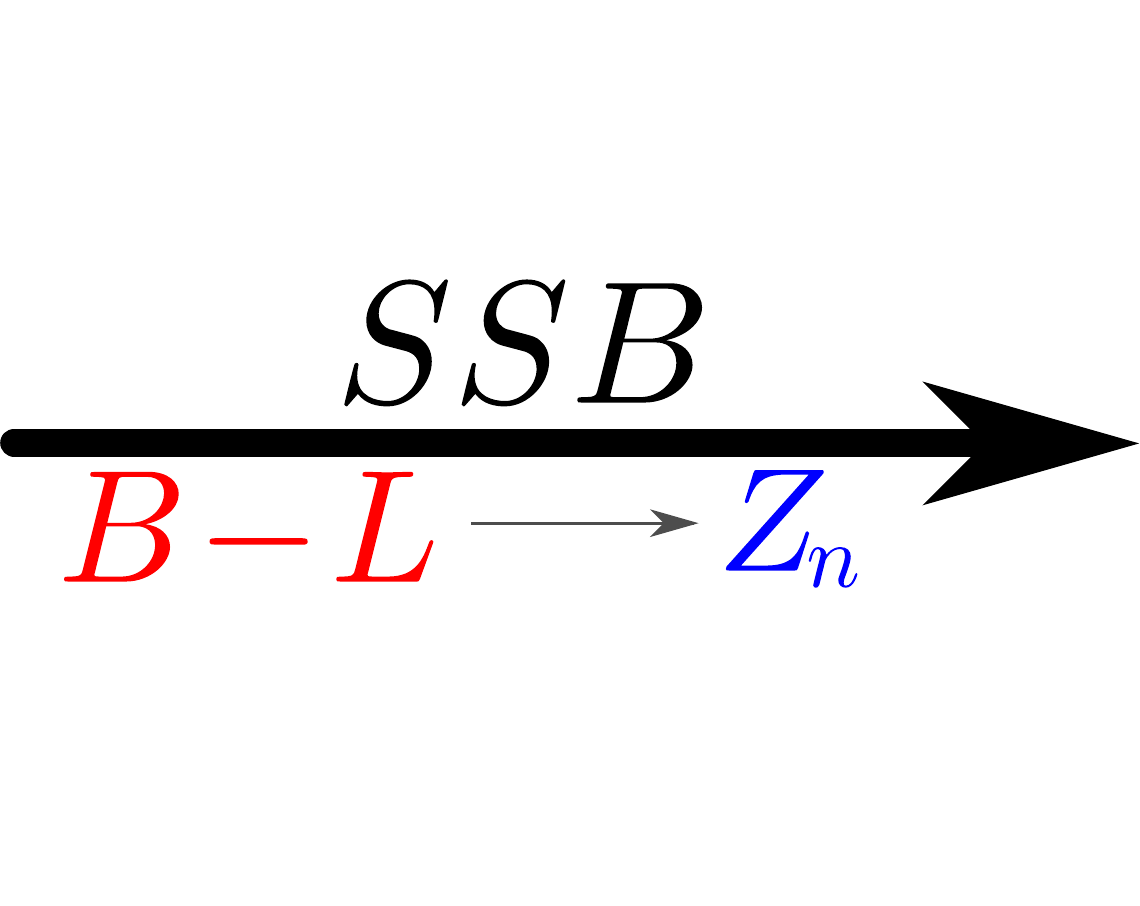}
    \end{subfigure}
    \begin{subfigure}[b]{0.43\textwidth}
        \includegraphics[width=1.\textwidth]{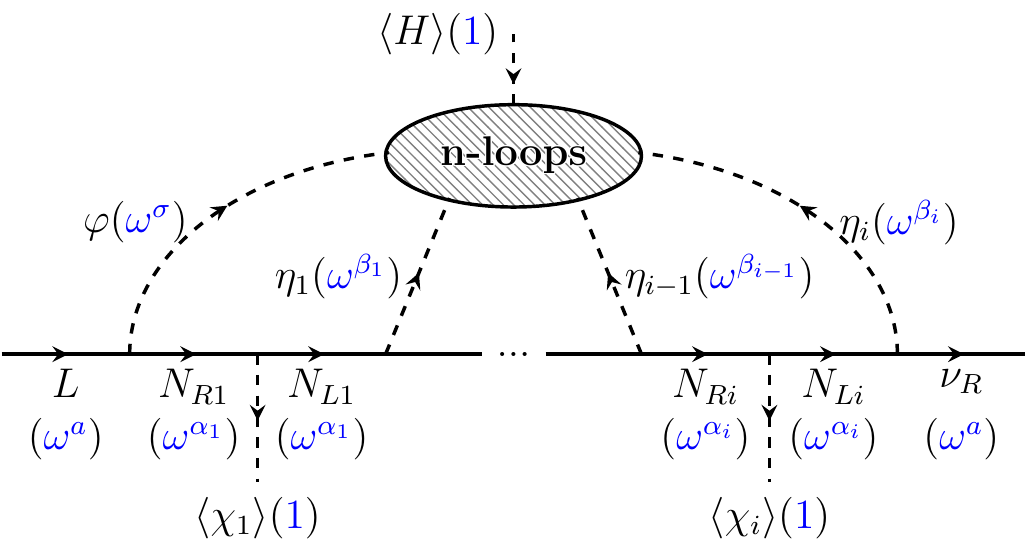}
        \caption{General residual $Z_n$ charge assignment.}
        \label{fig:dm:genZn}
    \end{subfigure}
    \caption{General $n$-loop Dirac neutrino mass diagram, along with the general charge assignment and its spontaneous
symmetry breaking pattern. For details, see text.}
    \label{fig:dm:gencase}
\end{figure}

Our general strategy is to use chiral anomaly free solutions of the ${\rm U(1)_{B-L}}$ symmetry to generate loop masses for Dirac neutrinos and also have a stable dark matter particle mediating the aforementioned loop. Then, after symmetry breaking, once all the scalars get a VEV, the ${\rm U(1)_{B-L}}$ symmetry will be broken down to one of its $Z_n$ subgroups, such that the dark matter stability and Dirac nature of neutrinos remain protected. This scheme is shown diagrammatically in \fig{fig:dm:gencase}. 

In \fig{fig:dm:gencase} the Standard Model singlet fermions $N_{Li}, N_{Ri}$, as well as the right-handed neutrinos $\nu_{R}$, have non-trivial chiral charges under ${\rm U(1)_{B-L}}$ symmetry.\footnote{It is not necessary that all fermions $N_{Li}, N_{Ri}$ be chiral under ${\rm U(1)_{B-L}}$ symmetry.} In order to generate the masses of these chiral fermions we have also added Standard Model singlet scalars $\chi_i$ which carry ${\rm U(1)_{B-L}}$ charges. To complete the radiative neutrino mass generation, additional scalars $\varphi, \eta_i$ are required. After the spontaneous symmetry breaking of the ${\rm U(1)_{B-L}}$ symmetry, all the scalars $\chi_i$ will acquire VEVs breaking ${\rm U(1)_{B-L}} \to Z_n$ residual symmetry. The fermions $N_{Li}, N_{Ri}$ get masses through the VEVs of the scalars $\chi_i$, while the neutrinos acquire a naturally small $n$-loop mass as shown in \fig{fig:dm:gencase}.

In order to satisfy all the requirements listed above, several conditions must be applied. First of all, the model should be anomaly-free:
\begin{itemize}
    \item The chiral charges of the fermions must be taken in such a way that the anomalies are cancelled.
\end{itemize}
In order to obtain non-zero but naturally small Dirac neutrino masses we impose the following conditions:
\begin{itemize}
    \item The tree-level Yukawa coupling $\bar{L} \tilde{H} \nu_R$ should be forbidden. This implies that apart from the Standard Model lepton doublets $L_i$ no other fermion can have ${\rm U(1)_{B-L}}$ charge of $\pm 1$. Furthermore, to ensure that the desired loop diagram gives the dominant contribution to the neutrino masses, all lower loop diagrams should also be forbidden by an appropriate choice of the charges of the fields. The latter can be done easily with the help of systematic classifications of radiative neutrino masses.
    \item The neutrino mass operator, i.e. $\bar{L} H^c \chi_1 \dots \chi_i \nu_{R}$, has to be invariant under the Standard Model gauge symmetries as well as under ${\rm U(1)_{B-L}}$. Following the charge convention of \fig{fig:dm:gencase}, the charges $\zeta_i$ of the VEV carrying scalars $\chi_i$ should be such that $\sum_i \zeta_i = -1 - \ell$, with $\ell$ the charge of $\nu_R$ (see \fig{fig:dm:genU1}).
    \item All the new fermions and scalars participating in the neutrino mass loop must be massive. Since the fermions will be in general chiral, this mass can only be generated via the coupling with a VEV carrying scalar. For example, in the diagram in \fig{fig:dm:gencase} we should have $-x_i + x'_i + \zeta_i = 0$.
    \item To protect the Diracness of neutrinos, all the Majorana mass terms for the neutrino fields at all loops must be forbidden in accordance with \eq{eq:dm:evenzndir}.
\end{itemize}

Additionally, for dark matter stability, we impose the following conditions:
\begin{itemize}
    \item After spontaneous symmetry breaking, the ${\rm U(1)_{B-L}}$ symmetry is broken down to a $Z_n$ subgroup. We shall consider only $Z_n$ subgroups, with $n$ a non-prime integer, to protect dark matter stability.\footnote{For a $Z_n$ subgroups with $n$ a prime number, there will always be an effective dark matter decay operator allowed by the residual $Z_n$ symmetry. Even then it is possible that such an operator cannot be closed within a particular model, thus pinpointing the existence of an accidental symmetry that stabilises dark matter. Another possibility is that the dark matter candidate decays at a sufficiently slow rate.} The symmetry breaking pattern can be extracted as follows. First all the ${\rm U(1)}$ charges must be re-scaled in such a way that all the charges are integers and the least common multiple (LCM) of all the re-scaled charges is $1$. Defining $n$ as the least common multiple of the charges of the scalars $\chi_i$, it is easy to see that the ${\rm U(1)}$ will break to a residual $Z_n$. This $n$ is taken to be even as explained before, i.e. $n \equiv {\rm LCM} (\zeta_i) \in 2\mathbb{Z}$. 
    \item Dark sector particles should neither mix with nor decay to Standard Model particles or to VEV carrying scalars. This can be accomplished by means of one of the two viable dark matter scenarios explained in \sect{sec:dm:dm}:
\end{itemize}

Given the long list of requirements, most of the possible solutions that lead to anomaly cancellation fail to satisfy some or most of them. Still, we have found some simple one-loop and several two-loop solutions that can satisfy all the conditions.

We demonstrate the idea for a simple example in which the ${\rm U(1)_{B-L}}$ symmetry is broken down to a residual $Z_6$ symmetry. However, in general, many other examples with different residual even $Z_n$ symmetries can be found by applying the given framework.

%%%%%%%%%%%%%%%%%%%%%%%%%%%%%%%%%%%%%%%%%%%%%%%%%%%%%%%%%%%%%%%
\subsection{Realistic example} \label{subsec:dm:diracexamp}

\begin{figure}
    \centering
    \begin{subfigure}[b]{0.4\textwidth}
        \includegraphics[width=\textwidth]{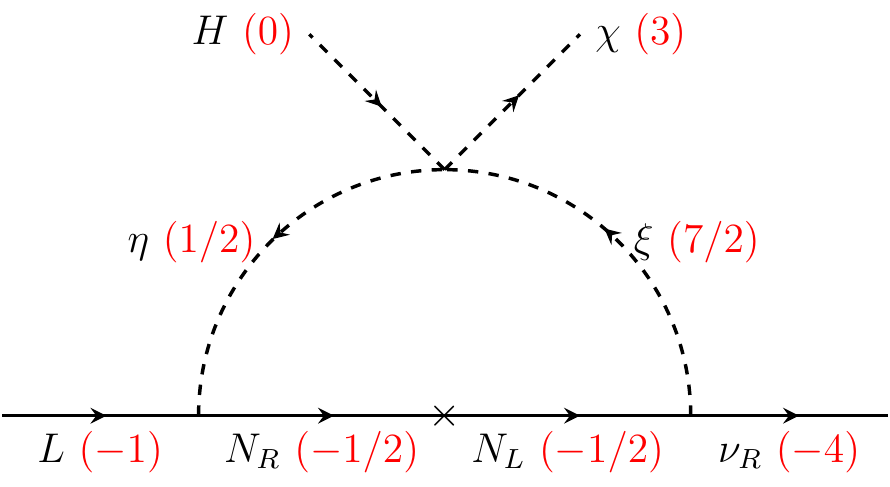}
        \caption{Charge assignment under ${\rm U(1)_{B-L}}$.}
        \label{fig:dm:gull}
    \end{subfigure}
    ~
    \begin{subfigure}[t]{0.1\textwidth}
    \vspace*{-3.5cm}
    \hspace*{-0.4cm}
    \includegraphics[width=1.5\textwidth]{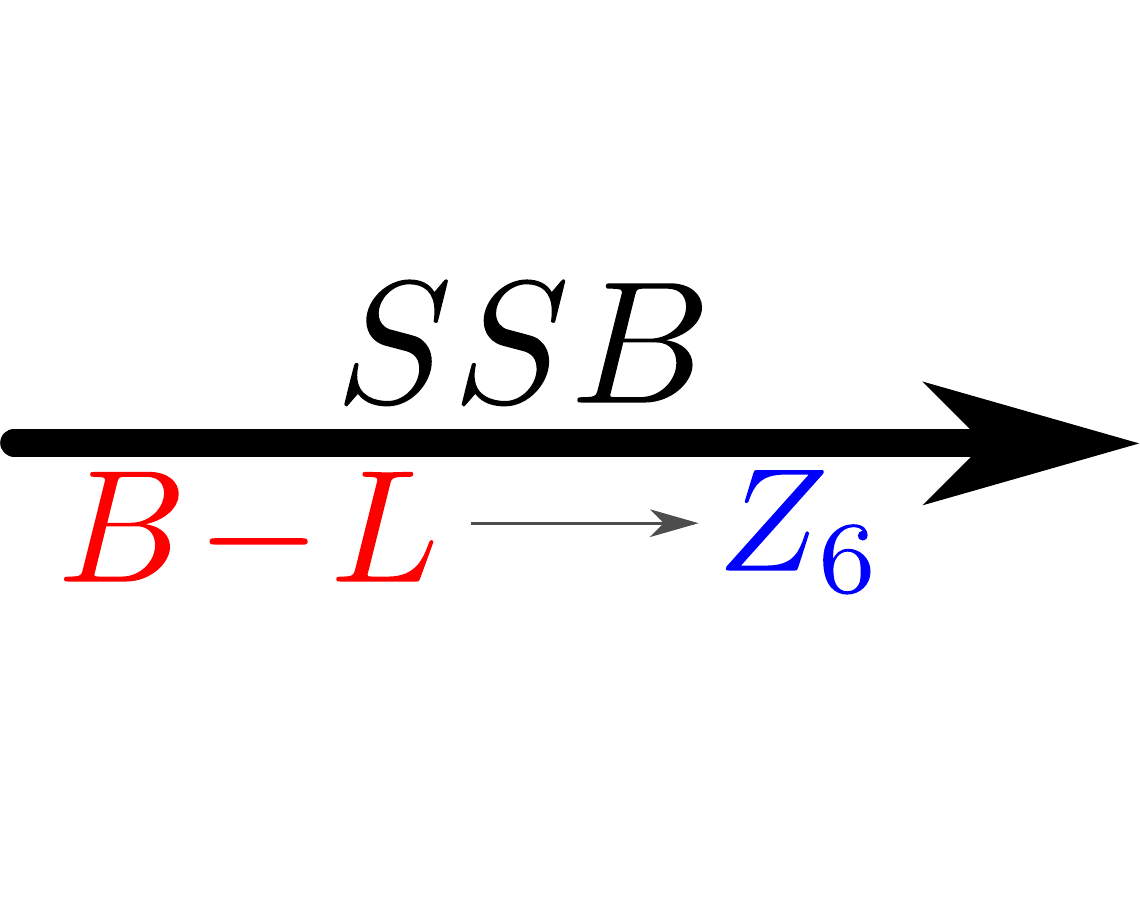}      \end{subfigure}
    ~
    \begin{subfigure}[b]{0.4\textwidth}
        \includegraphics[width=\textwidth]{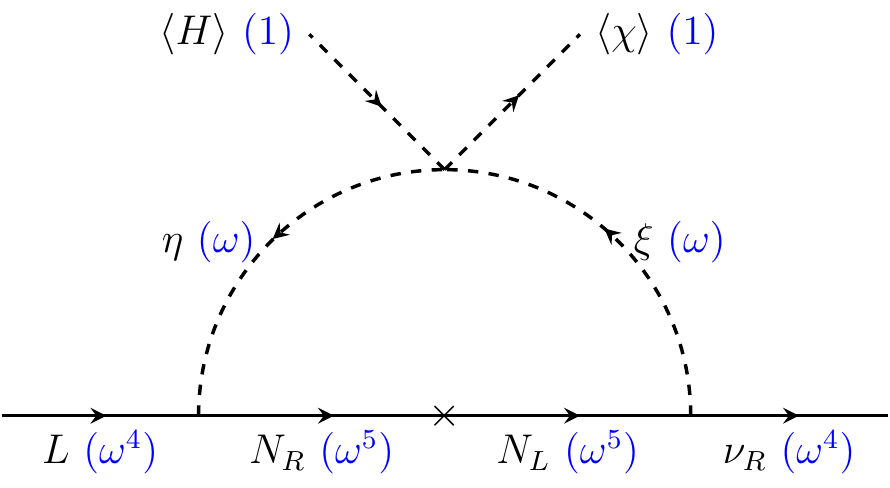}
        \caption{Residual $Z_6$ charge assignment.}
        \label{fig:dm:mouse}
    \end{subfigure} \\
    \caption{Diagram generating neutrino masses at one-loop order for the example model.}
    \label{fig:dm:z6}
\end{figure}

Let us consider an extension of the Standard Model by adding an extra Higgs singlet $\chi$ with a ${\rm U(1)_{B-L}}$ charge of $3$, along with a scalar doublet $\eta$, a singlet $\xi$ and two vector-like fermions $N_{L_l}$ and $N_{R_l}$, with $l=1,2$, all carrying non-trivial ${\rm U(1)_{B-L}}$ charges as shown in \tab{tab:dm:modelZ6} and depicted in \fig{fig:dm:gull}.

\begin{table}[t!]
    \centering
    \begin{tabular}{| c || c | c | c || c |}
  \hline 
&   Fields            &    ${\rm SU(2)_L} \times {\rm U(1)_Y}$            &     ${\rm U(1)_{B-L}}$                       & 
  $Z_{6}$                              \\
\hline \hline
\multirow{4}{*}{ \begin{turn}{90} Fermions \end{turn} } &
 $L_i$        	  &    ($\mathbf{2}, {-1/2}$)       &   {\color{red}${-1}$}    	  &	 {\color{blue}$\omega^4$}                     \\	
&   $\nu_{R_i}$       &   ($\mathbf{1}, {0}$)      & {\color{red} $({-4},{-4},\,{5})$ }   &  	 {\color{blue}($\omega^4, \omega^4, \omega^4)$}\\
&   $N_{L_l}$    	  &   ($\mathbf{1}, {0}$)      & {\color{red}${-1/2}$ }   &    {\color{blue} $\omega^5$}     \\
&  $N_{R_l}$     	  &  ($\mathbf{1}, {0}$) 	     & {\color{red} ${-1/2}$ } &  {\color{blue}$\omega^5$}     \\
\hline \hline
\multirow{4}{*}{ \begin{turn}{90} Scalars \end{turn} } &
 $H$  		 &  ($\mathbf{2}, {1/2}$)      &  {\color{red}${0}$ }    & {\color{blue} $1$}    \\
& $\chi$          	 &  ($\mathbf{1}, {0}$)        &  {\color{red}${3}$ }  &  {\color{blue} $1$}     \\		
& $\eta$          	 &  ($\mathbf{2}, {1/2}$)      &  {\color{red}${1/2}$}    &  {\color{blue}$\omega$}       \\
& $\xi$             &  ($\mathbf{1}, {0}$)        &  {\color{red}${7/2}$}      &	{\color{blue}$\omega$} \\	
    \hline
  \end{tabular}
    \caption{Charge assignment for all the fields. $Z_6$ is the residual symmetry in this example, with $\omega^6=1$. All the fields are colour singlets. All the generations have the same quantum number, except the right-handed neutrinos, for which ${\rm U(1)_{B-L}}$ and $Z_{6}$ charges are given for each copy.}
    \label{tab:dm:modelZ6} 
\end{table}

The neutrino interactions are described by the following Lagrangian,
\begin{equation}\label{eq:dm:YukInt}
    {\cal L}_\nu = y_{il} \, \bar{L}_i\tilde{\eta}N_{R_l} + y'_{li} \, \bar{N}_{L_l}\nu_{R_i}\xi + M_{lm} \, \bar{N}_{R_l}N_{L_m} + \hc \, ,
\end{equation}
where $\tilde{\eta} = i \tau_2 \eta^*$, with the indices $i = 1,2,3$ and $l,m = 1,2$. The relevant part of the scalar potential for generating the Dirac neutrino mass is given by,
\begin{equation} \label{eq:dm:Veps}
    {\cal V}\supset m_{\eta}^2 \, \eta^\dagger \eta + m_\xi^2 \, \xi^\dagger \xi + (\lambda_D \, H^\dagger \eta \chi \xi^* + \hc) \, ,
\end{equation}
where $\lambda_D$ is a dimensionless quartic coupling. 
 
After the spontaneous symmetry breaking of ${\rm U(1)_{B-L}}$, the scalar $\chi$ gets a VEV $\vev{\chi}=u$, giving mass to two neutrinos through the loop depicted in \fig{fig:dm:z6}. Note that only $\nu_{R_1}$ and $\nu_{R_2}$ can participate in this mass generation due to the chiral charges $(-4, -4, \, 5)$, i.e. $y'_{l3}=0$ in \eq{eq:dm:YukInt}. The third right-handed neutrino $\nu_{R_3}$ remains massless and decouples from the rest of the model, although it is trivial to extend this simple model to generate its mass. 

The neutral component of the gauge doublet $\eta$ and the singlet $\xi$ are rotated into the mass eigenbasis with eigenvalues $m^2_i$ in the basis of ($\xi$,\,$\eta^0$). The neutrino mass matrix is then given in the mass insertion approximation as,
 \begin{equation} \label{eq:dm:numass}
    (M_\nu)_{\alpha\beta} \approx \frac{1}{16\pi^2} \frac{\lambda_D \, v  \, u}{m^2_\xi-m^2_\eta} \, y_{\alpha k} \, y'_{k\beta} \, M_k \, \sum\limits_{i=1}^2 (-1)^i B_0(0,m_i^2,M_k^2) \, ,
\end{equation}
where $M_k$ ($k=1,2$) are the masses of the Dirac fermions $N_k$ and $\vev{H}=v$ the Standard Model VEV. $B_0$ is one of the Passarino-Veltman functions \cite{Passarino:1978jh}.

As a benchmark point, we can take the internal fermion to be heavier than the scalars running in the loop, one of which will be the dark matter candidate. Then, the neutrino mass scale can be roughly approximated by,
\begin{equation}
    m_\nu \sim \frac{1}{16 \pi^2} \, \frac{v \, u}{M} \, y \, y' \, \lambda_D \, ,
\end{equation}
where generation indices have been omitted for simplicity.

For comparison, we can take the Yukawa couplings to be of order $10^{-2}$ and the quartic coupling $\lambda_D \sim 10^{-4}$, like in the original scotogenic model \cite{Ma:2006km}. We can also take neutrino masses to be of order $0.1$ eV and $u \sim v$. With these choices, we can find the mass scale of the neutral fermions,
\begin{equation}
    M \sim \frac{1}{16 \pi^2} \, \frac{v \, u}{m_\nu} \, y \, y' \, \lambda_D \sim 10^{4} \, \text{GeV} \, .
\end{equation}

Compared with the type-I seesaw scale $M \approx y^2 \frac{v^2}{m_\nu} \sim 10^{10}$ GeV we can see a five order of magnitude suppression coming from the loop and the possibility of a broader parameter space.

It is worth mentioning that since the ${\rm U(1)_{B-L}}$ is anomaly free, it can be gauged. Then the physical Nambu-Goldstone boson associated to the dynamical generation of the Dirac neutrino mass \cite{Bonilla:2016zef} is absent.

Regarding dark matter stability in this particular model, we can see that the lightest particle inside the loop is stable. This is true for both the fermionic and scalar dark matter candidates. As can be seen in \fig{fig:dm:mouse}, all the internal loop particles are odd under the remnant $Z_6$, while all the Standard Model particles are even. Translating to the language used in \sect{sec:dm:dm}, the Standard Model belongs to the $Z_3 \equiv \{ 1, \omega^2, \omega^4 \}$ subgroup of $Z_6$, while the internal particle in the loop transform as elements of $Z_6 - Z_3 \equiv \{ \omega, \omega^3, \omega^5 \}$, being the lightest among them our dark matter candidate. Therefore any combination of Standard Model fields will always transform under $Z_3$, forbidding all effective operators leading to dark matter decay as shown graphically in \fig{fig:dm:dm1}.

%%%%%%%%%%%%%%%%%%%%%%%%%%%%%%%%%%%%%%%%%%%%%%%%%%%%%%%%%%%%%%%
%%%%%%%%%%%%%%%%%%%%%%%%%%%%%%%%%%%%%%%%%%%%%%%%%%%%%%%%%%%%%%%
\section{Majorana neutrinos} \label{sec:dm:majo}

As pointed out before in \eq{eq:dm:oddzn} and \eq{eq:dm:evenzndir}, in order to have Majorana neutrinos one has to break ${\rm U(1)_{B-L}}$ symmetry into an even subgroup $Z_{2n}$. In addition, the lepton doublets $L_i$ should also belong to the subgroup $Z_2 \subset Z_{2n}$, i.e. $L_i$ either transform trivially or as $\omega^n$ with $\omega^{2n}=1$. A connection between these symmetries and the stability of dark matter can be found, as first stated in \cite{Bonilla:2018ynb} for Dirac neutrinos. In this section, we follow an analogous approach linking the generation of naturally small Majorana neutrino masses with the stability of dark matter providing the appropriate symmetry breaking pattern ${\rm U(1)_{B-L}}\rightarrow Z_{2n}$. This further implies that neutrino masses arise at loop level, as the tree-level Majorana and Dirac masses are forbidden by the symmetry.

In order to do this, new fields with exotic $B-L$ charges are required. Since in the Standard Model lepton doublets $L_i$ have $B-L$ charge $-1$, in order to avoid all possible tree-level Dirac mass terms, no new fermion can carry $\pm 1$ charges under ${\rm U(1)_{B-L}}$ symmetry. Furthermore, the lowest order Majorana mass term, i.e. the Weinberg operator $\bar{L}^cLHH$, is not invariant under ${\rm U(1)_{B-L}}$, so it is automatically absent. To generate neutrino masses we should go to higher dimensional operators,
\begin{equation} \label{eq:dm:operator}
    \bar{L}^cLHH \chi_1 ...\chi_k,
\end{equation}
where the $\chi_i$ ($i = 1, ... k$) are scalar fields transforming non-trivially under ${\rm U(1)_{B-L}}$. The operator in \eq{eq:dm:operator} should be invariant under the Standard Model symmetries including ${\rm U(1)_{B-L}}$. This means that the $B-L$ charges of the fields $\chi_i$ must sum up to $2$. Although in principle some of them can also have non-trivial transformations under ${\rm SU(2)_L} \times {\rm U(1)_Y}$, for sake of simplicity we will take all $\chi_i$ to be Standard Model gauge singlets. Since the $\chi_i$ are charged under that ${\rm U(1)_{B-L}}$, once they acquire a VEV, the ${\rm U(1)_{B-L}}$ symmetry will break down to a residual $Z_{2n}$ subgroup, with $n$ depending on the charges of the particles in the model.

%%%%%%%%%%%%%%%%%%%%%%%%%%%%%%%%%%%%%%%%%%%%%%%%%%%%%%%%%%%%%%%
\subsection{One-loop realisations of the operator $\bar{L}^c LHH \chi$} \label{subsec:dm:1loop}

\begin{figure}[t!]
    \centering
    \includegraphics[width=0.89\textwidth]{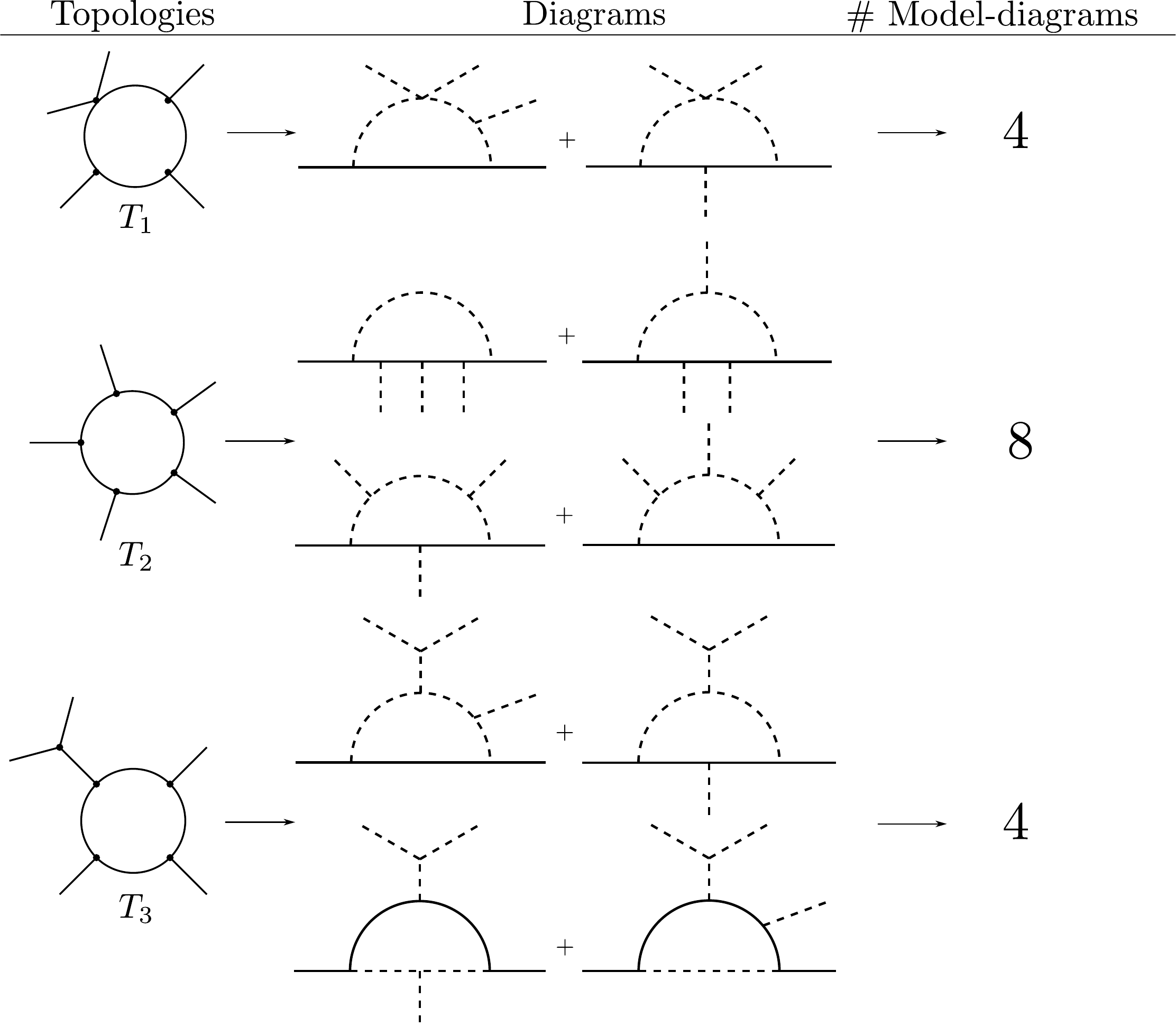}
    \caption{Renormalisable genuine topologies that generate the operator $\bar{L}^c LHH \chi$. For each topology, all the diagrams are given along with the number of model-diagrams. Each model-diagram can be generated by arranging in all possible ways the $\chi$ and the two $H$ in the external scalar legs.}
    \label{fig:dm:topos}
\end{figure}

Following this framework, in the simplest scenario, one can realise the operator \eq{eq:dm:operator} at the one-loop level with only one field $\chi$ with $B-L$ charge $2$, i.e. the dimension 6 operator $\bar{L}^c LHH \chi$. The possible one-loop realisations of the operator can be classified, following the philosophy of previous classifications, into three renormalisable genuine topologies which lead to 10 different diagrams as can be seen in \fig{fig:dm:topos}. The concept of genuineness is then attributed to those models for which the main contribution to neutrino masses comes from the one-loop level realisation of the operator $\bar{L}^c LHH \chi$. We call topologies or diagrams that generate at least one of these models genuine by inference. For example, diagrams which unavoidably contain the vertices ($L H$ $+$ fermion) or ($\bar{L}^c L$ $+$ scalar) are not genuine as they would generate a dominant type-I/III or type-II seesaw contribution, respectively.
 
The ten different diagrams depicted in \fig{fig:dm:topos} generate 16 model-diagrams. Each model-diagram is generated from a given diagram by the different arrangements of the two Higgs doublets and the Higgs singlet $\chi$ in the external scalar lines. For instance, take topology $T_1$, each of its diagrams generate two model-diagrams inserting $\chi$: (1) in the quartic scalar coupling or (2) in the trilinear coupling with scalars or fermions. In the case of $T_3$ the arrangement of $\chi$ and both Higgses is unique, as a trilinear vertex with two $H$ is not allowed because it would generate a dominant type-II seesaw contribution. Note that for each model-diagram there is an infinite series of possible models as there is always a free set of charges running in the loop.

We shall stop here the discussion about the possible one-loop realisations of $\bar{L}^c LHH \chi$. It is not our intention to make an exhaustive classification, but to show in a systematic way the wide range of possibilities, yet unexplored. We will now choose one of the simplest diagrams to build a particular, consistent and complete model as an example of how this general method works.

%%%%%%%%%%%%%%%%%%%%%%%%%%%%%%%%%%%%%%%%%%%%%%%%%%%%%%%%%%%%%%%
\subsection{A simple explicit model} \label{subsec:dm:majoexa}

In this section, we construct an explicit UV complete model realisation of the dimension 6 operator $\bar{L}^c LHH \chi$, in order to further describe the application of the formalism developed in this chapter to Majorana neutrinos. We add a new vector-like fermion pair $F_L$ and $F_R$ with charge $1/2$ under ${\rm U(1)_{B-L}}$, but singlet under the Standard Model gauge symmetry group. Since the field breaking the ${\rm U(1)_{B-L}} $ symmetry, $\chi$, transforms as $2$, the fractional charges of the new fields will imply that the breaking pattern is ${\rm U(1)_{B-L}} \rightarrow Z_4$. Note that, given the fractional charges of $F_L$ and $F_R$, there will be no tree-level Dirac mass term for neutrinos.

Additional scalars $\eta_i$ ($i =1,2,3$) are also needed to generate a one-loop contribution to neutrino masses. The relevant matter fields and their transformation under ${\rm SU(2)_L} \times {\rm U(1)_Y} \times {\rm U(1)_{B-L}}$ are given in \tab{tab:dm:tab1}, as well as the charges under the residual $Z_4$ subgroup that survives after spontaneous symmetry breaking.

\begin{table}
\begin{center}
\begin{tabular}{| c || c | c | c | c |}
  \hline 
&   \hspace{0.1cm}  Fields     \hspace{0.1cm}       &    ${\rm SU(2)_L} \times {\rm U(1)_Y}$            &    \hspace{0.2cm}   ${\rm U(1)_{B-L}}$        \hspace{0.2cm}               & 
   \hspace{0.4cm} $Z_{4}$ \hspace{0.4cm}                             \\
\hline \hline
\multirow{4}{*}{ \begin{turn}{90} Fermions \end{turn} } &
 $L_i$        	  &    ($\mathbf{2}, {-1/2}$)       &   {\color{red}$-1$}    	  &	 {\color{blue}$\omega^2$}                     \\	
&   $e_{R_i}$       &   ($\mathbf{1}, {-1}$)      & {\color{red} $-1$}   &  	 {\color{blue}$\omega^2$}\\
&   $F_{R}$       &   ($\mathbf{1}, {0}$)      & {\color{red} $1/2$}   &  	 {\color{blue}$\omega$}\\
&   $F_{L}$    	  &   ($\mathbf{1}, {0}$)      & {\color{red}$1/2$}   &    {\color{blue}$\omega$}     \\
\hline \hline
\multirow{5}{*}{ \begin{turn}{90} Scalars \end{turn} } &
 $H$  		 &  ($\mathbf{2}, {1/2}$)      &  {\color{red}$0$}    & {\color{blue}$1$}    \\
& $\chi$          	 &  ($\mathbf{1}, {0}$)        &  {\color{red}$2$ }  &  {\color{blue} $1$}     \\		
& $\eta_1$          	 &  ($\mathbf{2}, {-1/2}$)      &  {\color{red}${-3/2}$}    &  {\color{blue}$\omega$}       \\
& $\eta_2$             &  ($\mathbf{2}, {-1/2}$)        &  {\color{red}${-1/2}$}      &	{\color{blue}$\omega^3$} \\
& $\eta_3$             &  ($\mathbf{2}, {-1/2}$)        &  {\color{red}${3/2}$}      &	{\color{blue}$\omega^3$} \\	
    \hline
  \end{tabular}
\end{center}
\caption{Particle content of the model with $i\in\{e, \mu, \tau\}$. All the fields listed in the table are colour singlets. The field $\chi$ acquires a VEV, breaking the ${\rm U(1)_{B-L}}$ symmetry into its $Z_4$ subgroup given the half-integer charges running in the loop (see text for details).}
 \label{tab:dm:tab1}
\end{table}%

It is clear that the ${\rm U(1)_{B-L}}$ symmetry given in \tab{tab:dm:tab1} is anomalous. The canonical solution to make ${\rm U(1)_{B-L}}$ anomaly free is to add three right-handed fermions $N_R$ with $(-1,-1,-1)$ charges under $B-L$ symmetry. However, as noted before, these charges are not allowed as they lead to tree-level Dirac coupling between the Standard Model lepton doublets $L_i$. Instead, to cancel the anomalies, one can simply add three new neutral right-handed fermions $\nu_R$ with charges $(-4,-4,5)$ under ${\rm U(1)_{B-L}}$. This charge assignment also leads to anomaly free ${\rm U(1)_{B-L}}$ symmetry \cite{Montero:2007cd,Ma:2014qra,Ma:2015mjd,Ma:2015raa}. Other anomaly free solutions with several additional chiral fermions carrying exotic $B-L$ charges, can also be found as discussed in \cite{Ho:2016aye,Ma:2016nnn,Patra:2016ofq, Wang:2016lve, Wang:2017mcy, Nanda:2017bmi, Han:2018zcn, Kang:2018lyy}. However, the $(-4,-4,5)$ solution seems to be minimal.\footnote{Some of the subsequent works on $(-4,-4,5)$ can be found in \cite{Modak:2016ung, Singirala:2017see, DeRomeri:2017oxa, Nomura:2017jxb, Das:2017deo, Okada:2018tgy, Calle:2018ovc}.} These right-handed neutrinos can be given Majorana masses through VEV of singlet Higgses, $\chi_8$ and $\chi_{10}$, with charges $8$ and $10$ under $B-L$.\footnote{Note that VEV to these Higgses is also consistent with the ${\rm U(1)_{B-L}} \to Z_4$ breaking.} The $\nu_R$ will not play a role in the light neutrino mass generation, but they could be relevant in colliders, particularly if one gauges the ${\rm U(1)_{B-L}}$ symmetry. The dark matter phenomenology will also be influenced by the addition of these neutral fields. Note that the residual charges of these fields are $(1, 1, \omega^2)$, i.e. they are even fermions and therefore there are effective decay operators allowed by the symmetry. However, since they would be disconnected from the rest of the model they would be accidentally stable. This implies that the dark matter would be multi-component. Another option is to extend the model in such a way that these neutral fermions decay. The minimal content we found is the addition of a doublet $H_6$ with charge $6$ under $B-L$ and with a small induced VEV via $H_6 H^\dagger \chi \chi_8^\dagger$, along with a singlet scalar with charge $1$.

\begin{figure}[h!t]
    \centering
    \begin{subfigure}[b]{0.4\textwidth}
        \includegraphics[width=\textwidth]{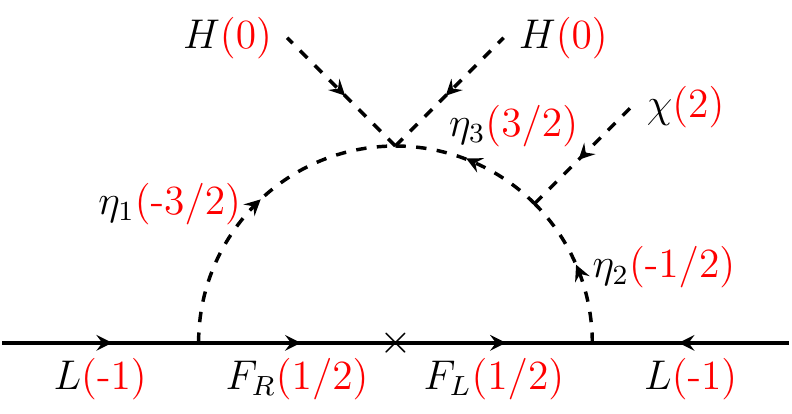}
    \end{subfigure}
    \begin{subfigure}[t]{0.15\textwidth}
        \vspace*{-2.2cm}
        \hspace*{-0.1cm}
        \includegraphics[width=\textwidth]{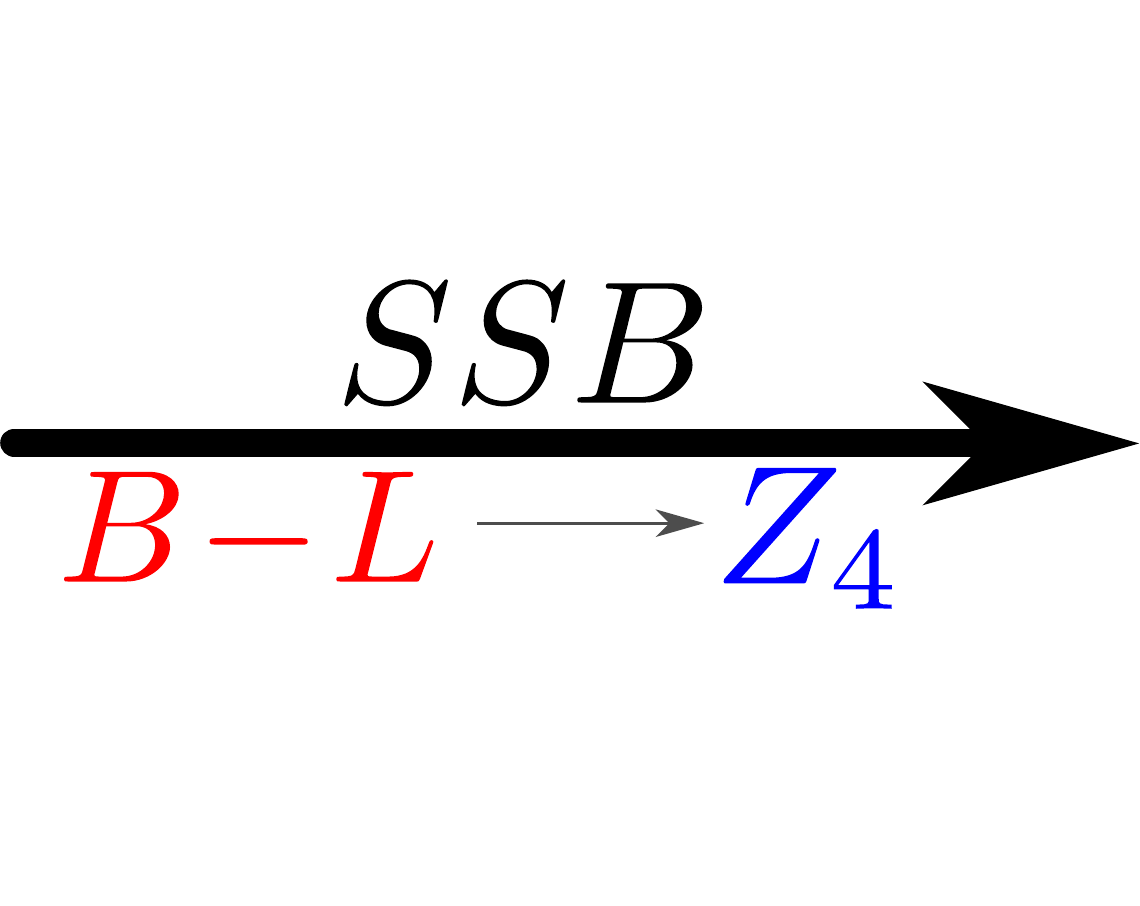}
    \end{subfigure}
    \begin{subfigure}[b]{0.4\textwidth}
        \includegraphics[width=\textwidth]{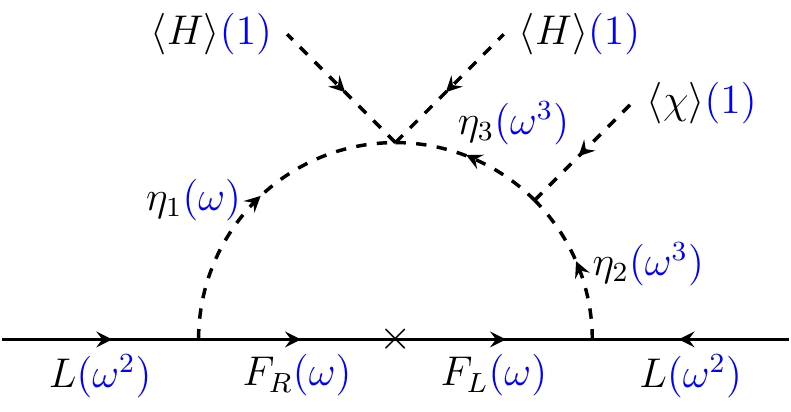}
    \end{subfigure}
    \caption{Leading order neutrino mass diagram with $B-L$ charges (left). After spontaneous symmetry breaking, as $\chi$ has charge $2$ under $B-L$ and there are half-integer charged fields, ${\rm U(1)_{B-L}}$ is broken to its subgroup $Z_4$ (right).}
    \label{fig:dm:diagloop}
\end{figure}

With this setup, the anomaly free $B-L$ will forbid the tree-level mass term for the neutrinos, but the new field content can accommodate the one-loop neutrino mass diagram of \fig{fig:dm:diagloop} in the \textit{scotogenic spirit}, thus explaining the smallness of neutrino masses and dark matter stability in a natural way. We will now write down the complete Lagrangian in several pieces for a better understanding. The Lagrangian of the model consist of the following parts:

\begin{itemize} 
    \item Yukawa terms that participate in the one-loop neutrino mass, along with the vector-like mass of $F_{L,R}$:
    \begin{eqnarray} \label{eq:dm:lag5}
        \mathcal{L} \, \in \, Y_1 \, \bar{L} F_{R} \eta_1 \, + \, Y_2 \bar{L}^c F_{L} \eta_2^\dagger \, + \, M \, \bar{F}_{L} F_{R} \, +  \, \hc \, ,
    \end{eqnarray}
    where we have omitted generation indices. In the following expression, we will use $m_F$ for the eigenvalues of $M$.

    \item  The scalar terms relevant for neutrino masses are given by:
    \begin{eqnarray} \label{eq:dm:scal}
        \mathcal{L}_{Scalar} \, \in \, \kappa \, \eta_2^\dagger \eta_3 \chi \, + \, \lambda \, \eta_3 \eta_1 H H \, + \hc \, .
    \end{eqnarray}
\end{itemize}

Apart from the standard kinetic and gauge terms, the scalar potential consists of 37 extra different terms which we do not write for simplicity. ${\rm SU(2)}$ contractions have been suppressed for brevity.

Regarding neutrino masses, as we pointed out before there is no tree-level mass term for the neutrinos, since the exotic charges of the new fermions forbid the Standard Model-like coupling with the Higgs. Moreover, note that the Weinberg operator $\bar{L}^c L H H$ is also forbidden by the same ${\rm U(1)_{B-L}}$ charges. The leading contribution to neutrino masses will arise at the radiative level coming from the allowed operator $\bar{L}^c LHH \chi$ as shown in \fig{fig:dm:diagloop}. 

A rough estimation for neutrino masses coming from the diagram in \fig{fig:dm:diagloop} is given by
\begin{equation} %\label{eq:}
    m_\nu \sim \frac{1}{16\pi^2} \, Y^2 \, \lambda \, \kappa \, u \, v ^2 \, \frac{1}{\Lambda^3} \, ,
\end{equation}
where $u$ is the VEV of $\chi$ and $\Lambda$ is the characteristic scale of the loop. The mass of the dark matter candidate will necessarily be lower than this scale. Note that in order to have two massive neutrinos only one generation of $F$ is needed, while two generations of $F$ can generate three non-zero neutrino masses. This is due to, as usual, the sum of two contributions: one coming from the diagram depicted in \fig{fig:dm:diagloop} and another coming from its transpose.

An estimate of the neutrino mass scale can be obtained if one considers that $u \sim \mathcal{O}(10)$ GeV. With $\kappa$ order $1$ GeV, $Y \sim \mathcal{O}(0.1)$ and $\lambda \sim \mathcal{O}(1)$, one can fit the atmospheric scale of $0.05$ eV with masses of order $10$ TeV.

Moreover, as can be seen from \fig{fig:dm:diagloop} all the particles running in the neutrino mass loop are \textit{odd} under the residual $Z_4$ symmetry. Thus, they all belong to the dark sector with the lightest among them, i.e. the lightest out of $\eta_i$ and $F_{L,R}$, being a good candidate for stable dark matter. As mentioned before the stability of the dark matter is owed to the fact that all the dark sector particles have charges that are odd under the residual $Z_4$ symmetry, while all the Standard Model particles are even under $Z_4$. Hence, for the lightest dark sector particle there is no possible effective decay operator at any order allowed by the remnant $Z_4$, see \fig{fig:dm:dm1}.

%%%%%%%%%%%%%%%%%%%%%%%%%%%%%%%%%%%%%%%%%%%%%%%%%%%%%%%%%%%%%%%
%%%%%%%%%%%%%%%%%%%%%%%%%%%%%%%%%%%%%%%%%%%%%%%%%%%%%%%%%%%%%%%
\section{Summary} \label{sec:dm:summary}

To summarise, neutrino masses and dark matter remain two of the most important shortcomings of the Standard Model. \textit{Scotogenic-like} models, where the dark sector particles run in the neutrino mass loop, provide a particularly attractive scenario to address both these shortcomings in a Standard Model extension. In this chapter, we have shown that the symmetry responsible for the dark matter stability can be obtained as a residual $Z_{n}$ subgroup of the ${\rm U(1)_{B-L}}$ symmetry of the Standard Model.

We have listed the general conditions required to have either Dirac or Majorana neutrinos depending on the residual $Z_{n}$ subgroup, as well as the different scenarios where the dark matter stability is ensured by this same symmetry. We showed that our framework can be applied broadly to many different cases, yet unexplored. 

For Dirac neutrinos, we have described a general framework in which Diracness and dark matter stability are realised by exploiting the anomaly free chiral solutions of a global ${\rm U(1)_{B-L}}$. This framework can be utilised in a wide variety of scenarios. We have presented a particular simple realisation of this idea where neutrino masses are generated at the one-loop level and the ${\rm U(1)_{B-L}}$ symmetry is broken spontaneously to a residual $Z_6$ symmetry. The framework can also be used in models with higher-order loops, as well as in cases where ${\rm U(1)_{B-L}}$ symmetry is broken to other even $Z_n$ subgroups. Since the ${\rm U(1)_{B-L}}$ is anomaly free, it can be gauged in a straightforward way, giving a richer phenomenology from the dark matter and collider point of view.

For Majorana neutrinos, we have shown that there are still many unexplored possibilities. We have particularised to a simple case with just one extra scalar Higgs singlet and discussed all the possible realisations at one-loop level. At the end, one simple realistic example with a remnant $Z_{4}$ symmetry is explained in more detail to illustrate how the spontaneous symmetry breaking of ${\rm U(1)_{B-L}}$ to an even $Z_{2n}$ can be easily accommodated, granting the stability of dark matter.

Before ending, we would like to remark that, although in this chapter we focus on one-loop models, this formalism can be implemented at higher loop orders and for any $Z_{n}$ symmetry, with $n$ a non-prime number.

\pagebreak
\fancyhf{}

%% file: Chapters/Loop_seesaw/Chapter_loopseesaw.tex
\fancyhf{}
\fancyhead[LE,RO]{\thepage}
\fancyhead[RE]{\slshape\nouppercase{\leftmark}}
\fancyhead[LO]{\slshape\nouppercase{\rightmark}}

\chapter{Radiative type-I seesaw}
\label{ch:loop_seesaw}
\graphicspath{ {Chapters/Loop_seesaw/} }

The simplest possibility to generate the Weinberg operator \eq{eq:nuphys:weinberg} is certainly the type-I seesaw mechanism \cite{Minkowski:1977sc, Yanagida:1979as, Mohapatra:1979ia}. In the classical type-I seesaw the Yukawa vertices are point-like $Y_{\nu} \bar{L} H \nu_{R}$ and the smallness of the neutrino masses is controlled by the large Majorana mass, $\Lambda\sim M_R$, of the right-handed neutrinos $\nu_{R}$.

After electroweak symmetry breaking with the Higgs VEV, $v\equiv\vev{H^0}$, the Weinberg operator leads to the light active neutrino Majorana mass terms. In one generation notation, the active neutrino mass is then given by the well-known relation,
\begin{equation} \label{eq:loopss:sstI}
    m_\nu \approx m_D^2/M_R \,
\end{equation}
with $m_{D} = Y_{\nu} \vev{H^{0}}$, see \sect{subsec:numass:maj_tree}.

Assuming that the Yukawas entering $m_D$ take values of order ${\cal O}(1)$ current neutrino oscillation data \cite{deSalas:2017kay} would then point to $M_R \sim 10^{(14-15)}$ GeV. This setup, apart from being able to explain neutrino oscillation data, leads to only one experimentally ``testable'' prediction: Neutrinoless double-$\beta$ decay should be
observed at some level.

Here, we instead discuss a simple idea, based on \cite{Arbelaez:2019wyz}, that allows for a much lower scale $M_R$, even for all involved Yukawa couplings of order ${\cal O}(1)$, by effectively generating the Dirac mass term corresponding to the Yukawa vertices.
\\

We start the chapter by introducing the main idea that we will develop, the effective type-I seesaw. This realisation of the type-I seesaw, relies on the radiative generation of neutrino Dirac couplings, discuss in \sect{sec:loopss:radiative} from a model-independent point of view. This allows us to estimate the typical scales for the Majorana mass of neutrinos as a function of the loop level, at which the Dirac couplings are generated. Afterwards, we study in detail two concrete example models at one- and two-loop level. We
estimate the neutrino masses, discuss possible constraints from lepton flavour violation and, finally, turn briefly to dark matter.

%%%%%%%%%%%%%%%%%%%%%%%%%%%%%%%%%%%%%%%%%%%%%%%%%%%%%%%%%%%%%%%
%%%%%%%%%%%%%%%%%%%%%%%%%%%%%%%%%%%%%%%%%%%%%%%%%%%%%%%%%%%%%%%
\section{Effective type-I seesaw} \label{sec:loopss:seesaw}

\begin{figure}
    \centering
    \includegraphics[width=0.6\textwidth]{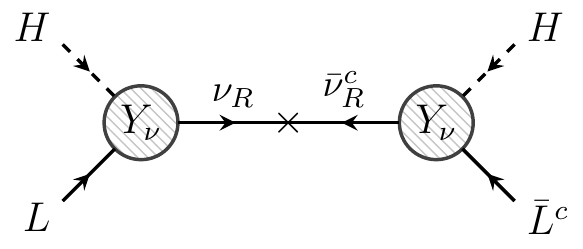}
    \caption{Effective type-I seesaw. The neutrino mass is suppressed by the Majorana mass of $\nu_R$ and by the square of the Dirac Yukawa term $Y_\nu$ which is generated effectively, see \eq{eq:loopss:Yuk}.}
    \label{fig:loopss:SeesawEff}
\end{figure}

As described before, we are interested on lowering the scale $M_R$ by generating the Dirac Yukawa term $m_D$, or correspondingly $Y_\nu$, effectively. This corresponds to generate the diagram depicted in \fig{fig:loopss:SeesawEff}. To this end one can claim that the elementary Yukawa coupling is forbidden by some symmetry, which being softly broken allows one to generate these vertices at a certain loop level directly or via higher dimension effective operators of the form,
\begin{equation}\label{eq:loopss:Eff-Operator-1}
    \frac{\kappa}{M^{2n}} \bar{L} H \nu_{R} \left( H^{\dagger} H \right)^{n} \, ,
\end{equation}
where $M$ is the scale of new physics underlying these operators, supposedly somewhere above the electroweak scale, and $\kappa$ is a loop suppression factor. The Dirac mass term is generated by the operator \eq{eq:loopss:Eff-Operator-1} after electroweak symmetry breaking. We assume that only the Standard Model Higgs acquires a VEV, though it is straightforward to generalise this to non-SM Higgses with VEVs as well. Then, the resulting effective Yukawa couplings would be suppressed as,
\begin{equation} \label{eq:loopss:Yuk}
    Y_\nu \sim \left( \frac{1}{16\pi^2} \right)^\ell  \left( \frac{v^{2}}{M^{2}} \right)^{n} \, ,
\end{equation}
where $\ell$ is the number of loops in the diagram generating the operator \eqref{eq:loopss:Eff-Operator-1}. As $Y_\nu$ is generated effectively, it can be naturally small, while all couplings arising in the UV complete theory can take values of order ${\cal O}(1)$.

As shown in the next section, right-handed neutrino masses of order the electroweak scale are easily possible in this setup. Such moderately heavy right-handed neutrinos could be searched for in accelerator based experiments via displaced vertices. The topic of ``long-lived light particles'' (LLLPs) has attracted much attention in the recent literature \cite{Alimena:2019zri}. A number of experimental proposals \cite{Curtin:2018mvb, Gligorov:2017nwh, Feng:2017uoz, Gligorov:2018vkc} could search for this signal. Sensitivity estimates for right-handed neutrinos for these experiments can be found in \cite{Helo:2018qej, Dercks:2018wum}, and those for the main LHC experiments can be found in \cite{Cottin:2018kmq, Liu:2019ayx, Drewes:2019fou}.

As we already mentioned, in order to forbid tree-level Dirac Yukawa couplings, it is necessary to postulate some additional symmetry beyond the Standard Model gauge group. This symmetry could be either gauged or discrete. For simplicity, in our model constructions we will concentrate on discrete symmetries, starting with a $Z_4$ symmetry, that gets softly broken to an exact remnant $Z_2$ symmetry. Thus, the same symmetry responsible for explaining the smallness of the neutrino mass is able to stabilise a dark matter candidate too.

In our setup neutrinos are Majorana particles. However, our constructions have some overlap with papers on Dirac neutrinos. The possibility that Dirac neutrino masses are small, because they are radiatively generated has been considered already in the pioneering works of \cite{Cheng:1977ir, Chang:1986bp, Mohapatra:1987nx, Branco:1987yg, Gu:2007ug}. Some general considerations on how to obtain small Dirac neutrino masses have been discussed in \cite{Ma:2016mwh}. Systematic studies of one-loop (and two-loop) Dirac neutrino masses were given in \cite{Yao:2017vtm}, (\cite{CentellesChulia:2019xky} and \ch{ch:dirac2l}).

Closer to this setup are the neutrino mass models presented in \cite{Nasri:2001ax, Kanemura:2011jj}. In these models neutrinos are Majorana particles with radiatively generated Dirac mass terms. However, both of these papers presented just one particular one-loop example model, while here we formulate general conditions for the implementation of the radiative seesaw type-I mechanism at any loop order. We also describe in more detail two specific models at one- and two-loop level. Moreover, different from the example model in \cite{Nasri:2001ax, Kanemura:2011jj}, our models also have a candidate for cold dark matter.

%%%%%%%%%%%%%%%%%%%%%%%%%%%%%%%%%%%%%%%%%%%%%%%%%%%%%%%%%%%%%%%
%%%%%%%%%%%%%%%%%%%%%%%%%%%%%%%%%%%%%%%%%%%%%%%%%%%%%%%%%%%%%%%
\section{Radiative Dirac Yukawa couplings} \label{sec:loopss:radiative}

In this brief section we discuss the radiative generation of neutrino Dirac Yukawa couplings from a model-independent point of view. Here, we consider only the $d=4$ Dirac mass operator $L H \bar \nu_R$ generated via loops. The mass of the light active neutrinos arises from the diagram depicted in \fig{fig:loopss:SeesawEff} and is given by \eq{eq:loopss:sstI}.

For simplicity, we restrict the discussion to the phenomenologically unrealistic case of one massive neutrino with no hierarchy or flavour structure for the Yukawas. This is sufficient for understanding the parameter dependence, and extending to three generations of active neutrinos is straightforward. The Dirac Yukawa, $Y_\nu$, can be parametrised in general in terms of five exponents, $(\ell,\alpha,\beta,\gamma,\delta) \in \mathbb{N}^+$, whose values will depend on the specific UV complete realisation of the operator $Y_\nu \, L H \bar \nu_R$, as
\begin{equation} \label{eq:paramYuk}
    Y_\nu \sim \left( \frac{1}{16 \pi^2} \right)^\ell \left( \frac{m_\tau}{v} \right)^\alpha \left( \frac{M_F}{\Lambda} \right)^\beta \left( \frac{\mu}{\Lambda} \right)^\gamma \epsilon^\delta \, .
\end{equation}
This corresponds to effectively generating the Yukawa via a diagram with the following features:
\begin{itemize}
    \item $\ell$ loops.
    \item $\alpha$ insertions of SM Yukawas. Unless the UV model allows for a top-quark in the loop, this corresponds to a suppression of typically $\sim 10^{-2}$, from $Y_{\tau}^{SM}$ (or $Y_{b}^{SM}$).
    \item $\beta$ mass insertions of new (vector-like) fermions that are not part of the SM, all set to $M_F$ for simplicity.
    \item $\gamma$ dimensionful couplings in the scalar sector, i.e. trilinear scalar couplings.
    \item $\delta$ dimensionless couplings, such as Yukawas or four-point scalar couplings.
\end{itemize}
For a UV complete model which is genuine, i.e., gives the dominant contribution to the neutrino mass, the possible sets of values of these exponents are limited. For example, for the simplest case of a one-loop Dirac mass term, there are only two genuine diagrams \cite{Yao:2017vtm} (see \sect{subsec:numass:dir_loop}) with one or two mass insertions and, at least, three couplings. So, for $\ell=1$ it is not possible to generate a genuine diagram with, for instance, $\alpha,\beta > 2$. The possible combinations of $(\alpha,\beta,\gamma,\delta)$ can be deduced from the systematic studies of radiative Dirac models given in \cite{Yao:2017vtm, CentellesChulia:2019xky}, already discussed in \sect{subsec:numass:dirac_class} and \ch{ch:dirac2l}.

For our numerical estimates, we assume that all couplings are in the perturbative regime, i.e. $\epsilon \lesssim 1$.\footnote{It is often argued that perturbativity only requires Yukawa couplings to be $Y \lesssim \sqrt{4\pi}$. However, saturating this limit would imply that higher order contributions are (at least) equally important than the leading order (that we consider), thus rendering estimates effectively inconsistent.} If $\mu$ is a trilinear coupling between some BSM scalar and the Higgs, it enters the calculation of the stability of the Higgs potential, i.e. it will induce a modification of the quartic Higgs coupling at the one-loop level. Thus, we also assume that $\mu \lesssim m_S \equiv \epsilon \, m_S$, in order not to run into problems with the Standard Model Higgs sector. With these considerations the light neutrino mass can be written in terms of the same five exponents, using the seesaw relation \eqref{eq:loopss:sstI},
\begin{equation} \label{eq:loopss:param}
    m_\nu \sim \left( \frac{1}{16 \pi^2} \right)^{2\ell} \frac{v^2}{M_R} \left( \frac{m_\tau}{v} \right)^{2\alpha} \left( \frac{M_F}{\Lambda} \right)^{2\beta} \left( \frac{m_S}{\Lambda} \right)^{2\gamma} \epsilon^{2(\gamma+\delta)} \, .
\end{equation}
As this equation shows, neutrino masses generated from this class of models will be very suppressed. If, for instance, the Dirac neutrino mass arises at two-loop order, then $m_\nu$ will effectively come from a four-loop diagram with an extra suppression due to the Majorana scale $M_R$. Thus, for relatively low masses of the order of TeV and couplings order one, a reasonable neutrino mass can be easily obtained.

A rough, but conservative limit on the Majorana mass scale can be obtained setting all masses in the loop to the same scale $\Lambda = M_F = m_S$. Conservatively taking $\epsilon = 1$, we find
\begin{equation} \label{eq:loopss:ParMnu2}
    m_\nu \sim \left( \frac{1}{16 \pi^2} \right)^{2\ell} \frac{v^2}{M_R} \left( \frac{m_\tau}{v} \right)^{2\alpha} \, .
\end{equation}
Note, that the scale $\Lambda$ does not appear in this simple case in the expression for $m_{\nu}$. This is to be expected, given the dimension of the neutrino Dirac coupling. Taking as a reference scale the atmospheric neutrino mass $\sqrt{|\Delta m_{31}^2|} \approx 0.05$ eV, we can set upper limits on $M_R$ as a function of the exponents $\ell$ and $\alpha$. The limits are given in \tab{tab:loopss:GenLim} up to three-loops and two Standard Model Yukawa insertions. The numbers given correspond to couplings of order one.

\begin{table}
    \begin{center}
        \begin{tabular}{ c|*{3}{|c} }
            \xrowht[()]{10pt}
            $M_R$ & $\alpha=0$ & $\alpha=1$ & $\alpha=2$ \\
            \hline\hline
            \xrowht[()]{14pt}
            $\ell=1$ & $2\times 10^{10}$ GeV & $2\times 10^{6}$ GeV & $2\times 10^2$ GeV \\
            \hline
            \xrowht[()]{14pt}
            $\ell=2$ & $10^{6}$ GeV & $10^2$ GeV & $9\times 10^{-3}$ GeV \\
            \hline
            \xrowht[()]{14pt}
            $\ell=3$ & $4\times 10^1$ GeV & $4\times 10^{-3}$ GeV & $4\times 10^{-7}$ GeV
        \end{tabular}
    \end{center}
    \caption{Estimated values for $M_R$ needed to fit a neutrino mass of $0.05$ eV with couplings of order one for different realisations of the Dirac mass operator $L H \bar \nu_R$, considering $\ell$ loops and $\alpha$ Standard Model Yukawa insertions. These mass scales constitute a rough, but conservative upper limit on $M_R$ for each class of models parametrised by the exponents $\ell$ and $\alpha$ in \eqref{eq:loopss:ParMnu2}.}
    \label{tab:loopss:GenLim}
\end{table}

Obviously, $M_R$ decreases very fast as $\alpha$ or $\ell$ increases. This is due to the fact that for Majorana neutrinos $m_\nu$ depends quadratically on $Y_\nu$, rather than linearly. For $\alpha=1$ and $l=2$ one finds a scale of $M_R \sim 10^2$ GeV, and similar values for $\alpha=2$ and $l=1$ or $\alpha=0$ and $l=3$. These are phenomenologically the most interesting cases.

Apart from the upper limit on the Majorana mass coming from the neutrino mass scale, lower limits on $M_R$ can be set from big bang nucleosynthesis (BBN) \cite{Deppisch:2015qwa} and the effective number of neutrinos in the early universe $\Delta N_{eff}$ \cite{Gariazzo:2019gyi}. These limits depend on the mixing angle between the right-handed and active neutrinos (as a function of the mass $M_R$). For our class of models, as for the ordinary type-I seesaw, one expects the bound $M_R \gtrsim (0.1-1)$ GeV from these considerations \cite{Deppisch:2015qwa,Gariazzo:2019gyi}. This significantly constrains the space of possible models to only those with three loops or less, and at most two Standard Model mass insertions (for the one-loop case). Therefore, in the next section, we will discuss two model examples in more detail: a one-loop and a two-loop model.

%%%%%%%%%%%%%%%%%%%%%%%%%%%%%%%%%%%%%%%%%%%%%%%%%%%%%%%%%%%%%%%
%%%%%%%%%%%%%%%%%%%%%%%%%%%%%%%%%%%%%%%%%%%%%%%%%%%%%%%%%%%%%%%
\section{Examples of models} \label{sec:loopss:examples}

We show two simple models where the Dirac mass is generated at one- and two-loops, both containing a stable dark matter candidate that participates in the loop. We give an estimate of the neutrino mass scale involved for a simplified benchmark, as well as insight into the phenomenological constraints coming from charged lepton flavour violating processes.

In both models, we assume a $Z_4$ symmetry, which is softly broken down to the preserved $Z_2$ symmetry, in order to guarantee that the Dirac neutrino mass matrix is generated
at the corresponding loop level. At the same time, the residual $Z_2$ symmetry stabilises the lightest of the $Z_2$-odd particles.

%%%%%%%%%%%%%%%%%%%%%%%%%%%%%%%%%%%%%%%%%%%%%%%%%%%%%%%%%%%%%%%
\subsection{One-loop Dirac mass} \label{subsec:loopss:oneloop}

The particle spectrum of the model and their assignments under the Standard Model gauge and $Z_4$ discrete symmetry are shown in \tab{tab:loopss:fields}. The scalar sector of the model is composed of the SM Higgs doublet $H$, the inert ${\rm SU(2)_L}$ scalar doublet $\eta$ and the electrically charged gauge singlet scalar $S^-$. In addition, the Standard Model fermion sector is extended by the inclusion of a right-handed Majorana neutrino $\nu_{R}$ \footnote{We repeat that here we are only interested in a rough estimate for the neutrino mass scale. For phenomenological reasons, one would need at least two right-handed neutrinos that generate the solar and atmospheric neutrino mass. Since fits of the type-I seesaw to neutrino data are straightforward and have been done many times in the literature, we do not repeat these details here.} and the vector-like charged leptons $\chi_L$ and $\chi_R$. The relevant terms for the neutrino mass are,
\begin{eqnarray} \label{eq:loopss:Lag-Y}
    - \mathcal{L}_{Y} &=& Y_e \, L H^{\dagger} e^c + Y_{L} \, L \eta^{\dagger} \chi_{L}
    + Y_{R} \, \overline{\chi_{R}} S^{+} \overline{\nu_{R}} + \hc \, ,
    \\
    \label{eq:loopss:Lag-M}
    \mathcal{L}_M &=& M_{R} \, \overline{\nu _{R}^{c}} \nu_{R}
    + M_{\chi} \, \overline{\chi_{R}} \chi _{L} + \hc \, ,
\end{eqnarray}
where flavour indices and ${\rm SU(2)}$ contractions have been suppressed for brevity.

\begin{figure}
    \centering
    \includegraphics[width=0.5\textwidth]{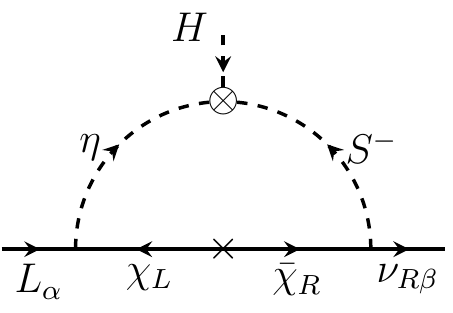}
    \caption{One-loop Dirac neutrino mass. The diagram is realised when the $Z_4$ is softly broken (denoted by the symbol $\otimes$). As the symmetry is broken in two units, the diagram is still invariant under a remnant $Z_2$ of $Z_4$.}
    \label{fig:loopss:model1}
\end{figure}

The terms above generate the Dirac neutrino mass matrix at the one-loop level through the diagram shown in \fig{fig:loopss:model1} provided the following $Z_4$ trilinear soft breaking term is added to the scalar potential,
\begin{equation} \label{eq:loopss:soft}
    \mathcal{V} \supset \mu_S \, H \eta S^{-} + \hc \, .
\end{equation}
The softly broken $Z_{4}$ guarantees that the Dirac mass term is forbidden at tree-level but generated by loops, i.e. that the diagram is genuine (non-reducible) \cite{Yao:2017vtm}.

\begin{table}[t!]
    \begin{center}
        \begin{tabular}{| c || c | c || c |}
            \hline
            \hspace{0.1cm} Fields \hspace{0.1cm}  &  ${\rm SU(3)_C} \times {\rm SU(2)_L} \times {\rm U(1)_Y}$  &  \hspace{0.2cm} $Z_4$ \hspace{0.2cm}  &  Residual $Z_2$  \\
            \hline \hline
                   $L$ & ($\mathbf{1}$, $\mathbf{2}$, -1/2) &  $1$ &  $1$ \\
                 $e^c$ & ($\mathbf{1}$, $\mathbf{1}$,    1) &  $1$ &  $1$ \\
               $\nu_R$ & ($\mathbf{1}$, $\mathbf{1}$,    0) & $-1$ &  $1$ \\
                   $H$ & ($\mathbf{1}$, $\mathbf{2}$,  1/2) &  $1$ &  $1$ \\
            \hline
  ($\chi_{L}$, $\chi_{R}$) & ($\mathbf{1}$, $\mathbf{1}$,    1) &  ($i$, $i$) &  ($-1$, $-1$) \\
                    $\eta$ & ($\mathbf{1}$, $\mathbf{2}$,  1/2) &         $i$ &          $-1$ \\
                     $S^-$ & ($\mathbf{1}$, $\mathbf{1}$,   -1) &         $i$ &          $-1$ \\
            \hline
        \end{tabular}
    \end{center}
    \caption{Particle content of the example model that generates the one-loop diagram of \fig{fig:loopss:model1} once the $Z_4$ is softly broken by the trilinear term $H \eta S^-$. After the breaking of $Z_4$ a remnant $Z_2$ is exactly conserved.}
    \label{tab:loopss:fields}
\end{table}

The Dirac mass term can be computed directly from the diagram in \fig{fig:loopss:model1} given the Lagrangians \eq{eq:loopss:Lag-Y} and \eq{eq:loopss:Lag-M}, and the soft-breaking term \eq{eq:loopss:soft}. In the mass insertion approximation and, for simplicity, setting all the masses of the internal scalars, as well as the soft-breaking parameter $\mu_S$, to $m_S$, one finds,
\begin{equation} %\label{eq:}
    m_D \approx \frac{1}{16\pi^2} \frac{v m_S}{M_\chi} Y_L Y_R \; \mathcal{I}_1(m_S^2/M_\chi^2) \, .
\end{equation}
The loop integral $\mathcal{I}_1(x)$ can be written in terms of the Passarino-Veltman $B_0$ function \cite{Passarino:1978jh} as,
\begin{equation} %\label{eq:}
    \mathcal{I}_1(x) = \frac{1}{1-x} \left[ B_0(0,1,x) - B_0(0,x,x) \right] \, .
\end{equation}

The mass scale of the lightest active neutrino can be directly estimated through the seesaw approximation as,
\begin{equation} \label{eq:loopss:mnu1}
    m_{\nu} \sim \left(\frac{1}{16\pi^2} \right)^2 Y_{L}^2Y_{R}^2 \, \frac{v^2 m^2_{S}}{M^2_{\chi}M_{R}} \, [\mathcal{I}_1(m_{S}^2/M_{\chi}^2)]^2 \, .
\end{equation}

\begin{figure}[t!]
    \centering
    \includegraphics[width=0.7\textwidth]{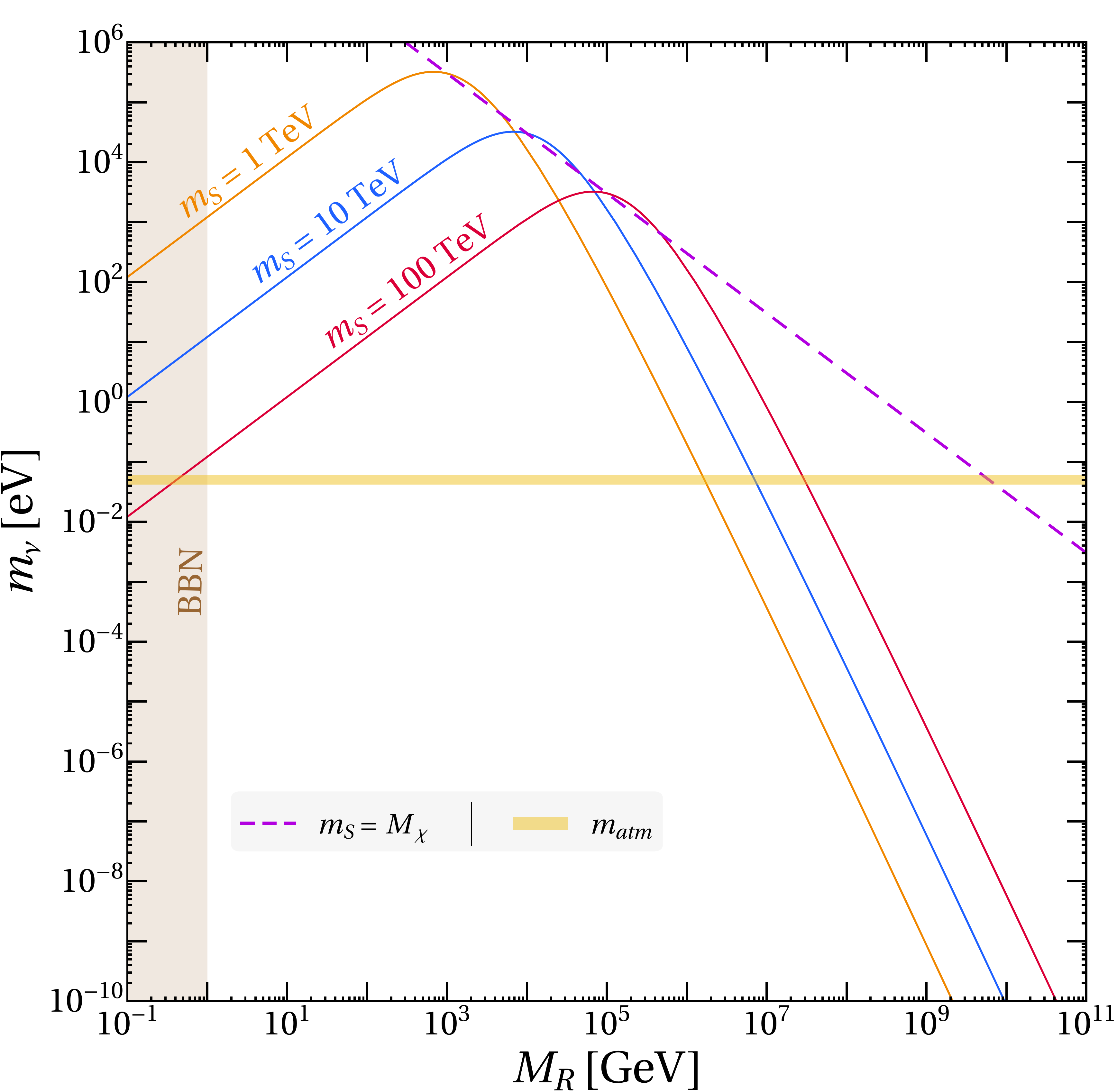} 
    \caption{One-loop neutrino mass scale. The dashed line corresponds to the case where $m_{S}=M_{\chi}$, while the solid lines depict the case where $M_{\chi}=M_{R}$ for different scalar masses. The Yukawas $Y_{L}$ and $Y_{R}$ are set to 1. BBN excludes $M_R >(0.1-1)$ GeV, depending on mixing, for this class of models \cite{Deppisch:2015qwa}.}
    \label{fig:loopss:oneloopmnu}
\end{figure}

This mass scale as a function of $M_{R}$ is plotted in \fig{fig:loopss:oneloopmnu}. Two different benchmarks with $M_{\chi}=M_{R}$ and $m_{S} = M_{\chi}$ are represented by the solid and dashed lines respectively. For both cases, we can observe that the neutrino mass is strongly suppressed even for small values of $M_{R}$. In the $m_{S} = M_{\chi}$ scenario, the neutrino mass falls as $\sim 1/M_{R}$ independently of the one-loop internal scalar masses. Moreover, in the $M_{\chi}=M_{R}$ scenario, the neutrino mass is a function of both mass scales $m_{S}$ and $M_{R}$. It behaves as $M_R$ or $1/M_R^3$ depending on which of these two scales dominates the loop.

The window of allowed $M_{R}$ values that could fit the neutrino oscillation scale $m_{\nu} \sim 0.05$ eV becomes narrower for larger masses $m_{S}$. Note that in \fig{fig:loopss:oneloopmnu} the neutrino mass is plotted for order-one couplings. Consequently, the points with a neutrino mass lying roughly below the atmospheric scale are phenomenologically non-viable, as they would require couplings larger than one (non-perturbative) to give a reasonable mass scale.

Current upper limits on lepton flavour violating (LFV) decays such as $\mu \rightarrow e\gamma$ can provide constraints on the parameters of our model. These depend on specific choices for the Yukawas $Y_L$ and $Y_R$. As \eq{eq:loopss:mnu1} shows, $m_{\nu}$ depends on the product of these couplings, while LFV decays are mostly sensitive to only $Y_L$. There are then two extreme cases: (i) choose $Y_L \simeq 1$ and fit $Y_R$ to $m_{\nu}$ as a function of the other model parameters, or (ii) choose $Y_R \simeq 1$  and fit $Y_L$. Case (i) is very similar to the situation in our two-loop model (see \sect{subsec:loopss:twoloop}), and thus we will discuss the details in the next section. For case (ii), on the other hand, we found that LFV limits do not impose interesting limits on our one-loop model.

The residual $Z_2$ symmetry ensures that the lightest of the fields running inside the loop will be stable. In order to not run into conflict with cosmology and to provide a good dark matter candidate, one should force the neutral component of the doublet $\eta$ to be the lightest of the loop particles. Similar dark matter candidates have been studied in the literature.\footnote{See for instance the well-known \textit{Inert Doublet Model} \cite{Deshpande:1977rw} or the \textit{Scotogenic model} \cite{Ma:1998dn}.} Considering $\eta$ as the only source of dark matter, the observed relic density, together with direct detection limits and the constraints on the invisible width of the Higgs boson severely limit its mass to lie either around $m_{h}/2 \simeq 62.5$ GeV, in a small region around $m_\eta\simeq 72$ GeV or above $m_\eta \gtrsim 500$ GeV \cite{Eiteneuer:2017hoh}.

%%%%%%%%%%%%%%%%%%%%%%%%%%%%%%%%%%%%%%%%%%%%%%%%%%%%%%%%%%%%%%%
\subsection{Two-loop Dirac mass} \label{subsec:loopss:twoloop}

Analogously to the first example, we build a two-loop radiative seesaw model that softly breaks a $Z_4$ discrete group to an exact $Z_2$ symmetry. The particle content and their transformation properties under the SM gauge and the $Z_4$ discrete symmetry are shown in \tab{tab:loopss:fields2loops}. We again include a right-handed Majorana neutrino $\nu_{R}$.

The relevant terms of the Lagrangian and the scalar sector invariant under $Z_4$ are,
\begin{eqnarray}
    -\mathcal{L}_{Y} &=& Y_e \, L H^{\dagger} e^{c} + Y_L \, \overline{F_L} \eta_{2} \overline{e^c} + Y_R \, \overline{\nu_{R}} \eta_{2} F_{L} + \hc \, ,
    \\
    \mathcal{L}_{M} &=& M_{R} \, \overline{\nu_{R}^{c}} \nu_{R} + M_{F} \, \overline{F_R} F_{L} + \hc \, ,
    \\
    \mathcal{V} &\supset& \lambda \, \eta_{1}^{\dagger}H \eta_{2}^{\dagger} H + \hc \, ,
\end{eqnarray}

\begin{table}
    \begin{center}
        \begin{tabular}{| c || c | c || c |}
            \hline
            \hspace{0.1cm} Fields \hspace{0.1cm}  &  ${\rm SU(3)_C} \times {\rm SU(2)_L} \times {\rm U(1)_Y}$  &  \hspace{0.2cm} $Z_4$ \hspace{0.2cm}  &  Residual $Z_2$  \\
            \hline \hline
                   $L$ &  ($\mathbf{1}$, $\mathbf{2}$, -1/2) &   $1$ &  $1$ \\
                 $e^c$ &  ($\mathbf{1}$, $\mathbf{1}$,    1) &   $1$ &  $1$ \\
               $\nu_R$ &  ($\mathbf{1}$, $\mathbf{1}$,    0) &  $-1$ &  $1$ \\
                   $H$ &  ($\mathbf{1}$, $\mathbf{2}$,  1/2) &   $1$ &  $1$ \\
            \hline
    ($F_{L}$, $F_{R}$) &  ($\mathbf{1}$, $\mathbf{2}$, -1/2) &  ($i$, $i$) &  ($-1$, $-1$) \\
              $\eta_1$ &  ($\mathbf{1}$, $\mathbf{2}$,  1/2) &        $-i$ &          $-1$ \\
              $\eta_2$ &  ($\mathbf{1}$, $\mathbf{2}$,  1/2) &         $i$ &          $-1$ \\
            \hline
        \end{tabular}
    \end{center}
    \caption{Particle content of the example model that generates the two-loop diagram of \fig{fig:loopss:model2} once the $Z_4$ is softly broken by the term $\eta_2^\dagger \eta_1$. After the breaking of $Z_4$ a remnant $Z_2$ is conserved.}
    \label{tab:loopss:fields2loops}
\end{table}

\begin{figure}[t!]
    \centering
    \includegraphics[width=0.5\textwidth]{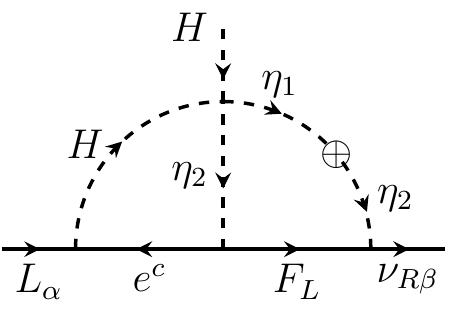}
    \caption{Two-loop Dirac neutrino mass. The diagram is realised when the $Z_4$ is softly broken (denoted by the symbol $\otimes$). As the symmetry is broken in two units, the diagram is still invariant under a remnant $Z_2$.}
    \label{fig:loopss:model2}
\end{figure}

An effective Dirac term is generated once the $Z_4$ symmetry is softly broken in the scalar sector by the term,
\begin{equation}
    -\mathcal{L}_{\text{soft}} = \mu_{12}^{2} \, \eta_{2}^{\dagger} \eta_{1} + \hc \, .
\end{equation}
A Dirac mass appears at the two-loop level, as depicted in \fig{fig:loopss:model2}, which can be expressed in the mass insertion approximation, assuming no flavour structure in the Yukawa couplings, as
\begin{equation}
    m_{D} \approx \left( \frac{1}{16 \pi^2} \right)^2 \lambda \, Y_{e} Y_{L} Y_{R} \frac{v \, \mu_{12}^2 }{ M^2_F } \, \mathcal{I}_2 ( m_{S}^{2} / M_{F}^{2} ) \, ,
\end{equation}
where $\mathcal{I}_2(x)$ is a dimensionless two-loop function. $\mu_{12}$ is the soft breaking mass term depicted by $\otimes$ in \fig{fig:loopss:model2}. For simplicity, we set all the masses of the new internal scalars to $m_{S}$. Taking into account that the main contribution of the Standard Model Yukawa $Y_{e}$ would be $m_{\tau}/v$, the mass scale of the lightest active neutrino is directly estimated through the seesaw approximation as,
\begin{equation} \label{eq:loopss:mnu2}
    m_{\nu} \sim \left( \frac{1}{16 \pi^2} \right)^4 \lambda^2 Y_L^2 Y_R^2 \, \frac{ m_{\tau}^2 m_{S}^4 }{ M_F^4 M_{R} } \, \left[ \mathcal{I}_2(m_{S}^2/M_{F}^2) \right]^2 \, ,
\end{equation}
where, as before, we have set $\mu_{12} = m_S$. $\mathcal{I}_2$ can be written in terms of simple two-loop integrals for which analytical solutions are known \cite{Martin:2016bgz}. We do not give its decomposition here for brevity, though it can be found in the literature \cite{Sierra:2014rxa}.

\begin{figure}[t!]
    \begin{center}
        \begin{tabular}{cc}
            \hspace{-1cm}
            \includegraphics[width=0.7\textwidth]{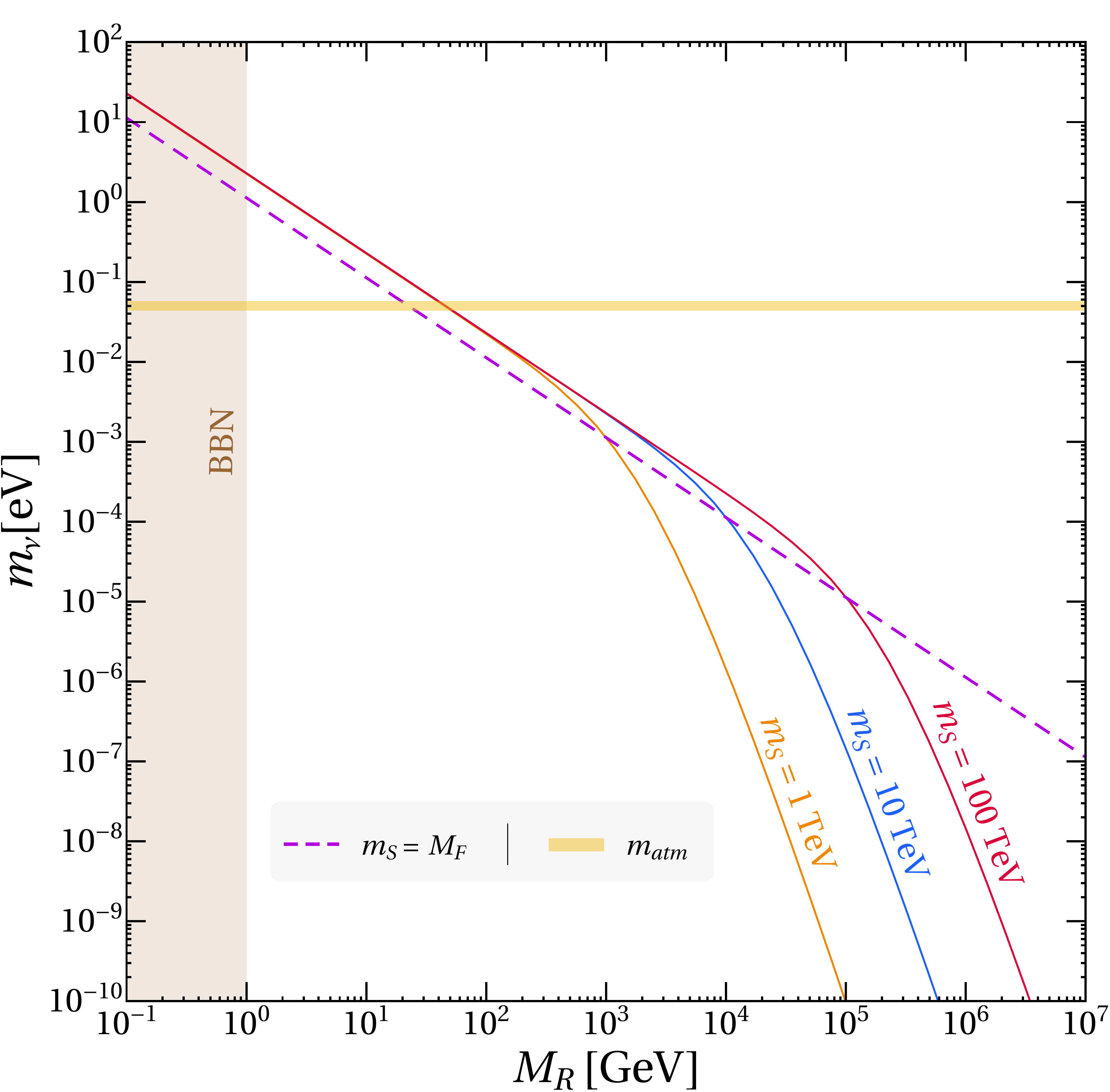}
        \end{tabular}
    \caption{Two-loop neutrino mass scale assuming that $m_{S}=M_{F}$ and $M_{F}=M_{R}$, depicted as dashed and solid lines respectively. All dimensionless couplings are set to 1 and the BBN exclusion region is indicated on the left.}
    \label{fig:loopss:massplot2}
    \end{center}
\end{figure}

The neutrino mass scale, \eq{eq:loopss:mnu2}, as a function of $M_{R}$ is plotted in \fig{fig:loopss:massplot2}. We consider two different approximations: $M_{F}=m_{S}$ and $M_{F}=M_{R}$, represented by the dashed and solid lines respectively. As expected from \tab{tab:loopss:GenLim}, the neutrino mass is more strongly suppressed compared to the one-loop model described previously. For the case $m_{S}=M_{F}$ the Dirac Yukawa is independent of the scale, and consequently the neutrino mass falls simply as $\sim 1/M_{R}$. On the other hand, in the scenario where $M_{F}=M_{R}$, this same behaviour is reproduced when $m_{S}$ dominates, while for values of $M_{R} > m_S$, the neutrino mass follows the curve $1/M_R^5$.

Given the suppression factor $(m_\tau/v)^2 \sim 10^{-4}$, and if we take into account the limit coming from cosmology (BBN), the range of allowed values of $M_{R}$ which can fit the neutrino oscillation scale $m_{atm} \sim 0.05$ eV is considerably limited. For $m_S>10^2$ GeV, $M_{R}$ has to be $M_{R} \lesssim 10^{2}$ GeV. This makes the model testable in future heavy neutral lepton searches.

We mention again that the remnant $Z_2$ symmetry stabilises the lightest of the fields that are odd under this symmetry. Fermionic dark matter coming from a doublet is ruled out by direct detection experiments \cite{Cirelli:2005uq}, while for the scalar inert doublet the same limits described in the previous section apply.

\begin{figure}[t!]
    \centering
    \includegraphics[width=0.7\textwidth]{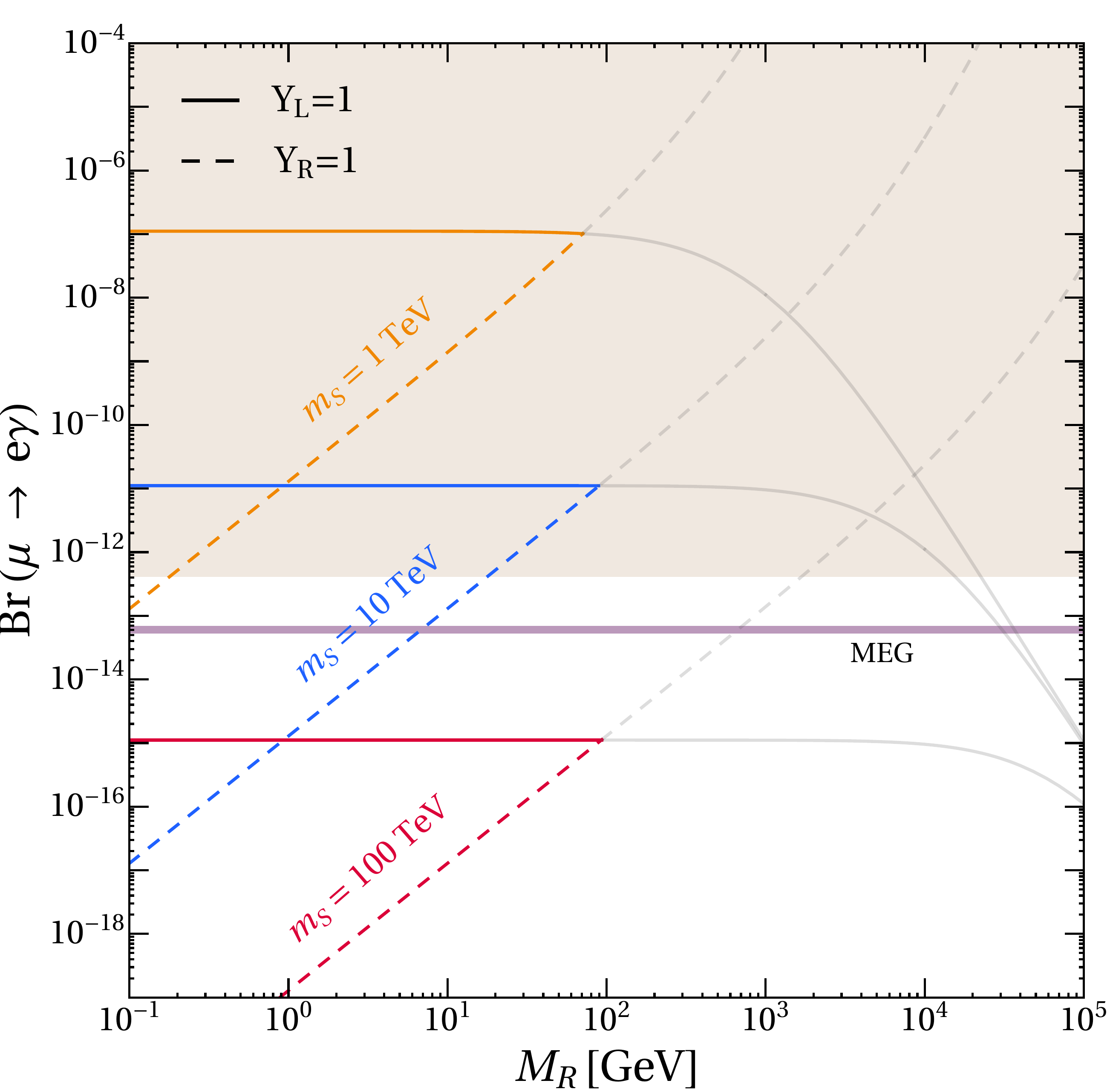}
    \caption{Estimate of the branching ratio of $\mu \rightarrow e \gamma$ as a function of $M_{R}$ for different values of $m_{S}$ fitting the neutrino mass to $m_{atm}$. The areas between the coloured lines are allowed in this model, see text. The grey lines represent the values of $M_{R}$ where one of the Yukawa couplings becomes non-perturbative in order to fit neutrino oscillation data. The shaded region represents the experimentally excluded area for $Br(\mu \rightarrow e \gamma) > 4.2 \times 10^{-13}$ \cite{TheMEG:2016wtm}, while the purple line corresponds to the future prospect limit from MEG Collaboration \cite{Baldini:2013ke}.}
    \label{fig:loopss:dwSST3}
\end{figure}

Turning to LFV processes, \fig{fig:loopss:dwSST3} shows $Br(\mu \rightarrow e \gamma)$ as a function of $M_{R}$ for two different scenarios, already mentioned in section \ref{subsec:loopss:oneloop}: (i) choose $Y_L \simeq 1$ and fit $Y_R$ to $m_{\nu}$, or (ii) choose $Y_R \simeq 1$ and fit $Y_L$. All other possible Yukawa choices lie between these extremes. The dominant (one-loop) contribution to $Br(\mu \rightarrow e \gamma)$ comes always from $Y_L$, which directly connects the new particles with the Standard Model leptons. For $M_F=M_R$ and $Y_R=1$ the branching is dominated by the fit of the neutrino mass, \eq{eq:loopss:mnu2}. The branching increases as a function of $M_R$ as $Y_L$ gets larger, counteracting the suppression of $1/M_R^5$ in the neutrino mass. We stop the calculation when $Y_L$ grows larger than 1. In contrast, for $Y_L=1$ there is no dependence from the neutrino mass fit, but rather a suppression of $1/M_R^4$ when this mass scale dominates over $m_S$ in the $\mu \rightarrow e \gamma$ loop function \cite{Lavoura:2003xp}. The regions in between these extremes are the allowed regions for this neutrino mass model.

%%%%%%%%%%%%%%%%%%%%%%%%%%%%%%%%%%%%%%%%%%%%%%%%%%%%%%%%%%%%%%%
%%%%%%%%%%%%%%%%%%%%%%%%%%%%%%%%%%%%%%%%%%%%%%%%%%%%%%%%%%%%%%%
\section{Summary} \label{sec:loopss:summary}

We have constructed a new realisation of the type-I seesaw mechanism based on radiatively generated Dirac neutrino masses. We showed that this class of models can naturally generate a small neutrino mass for order-one couplings and relatively low mass scales. Compared to the standard type-I seesaw mechanism, for which the Majorana mass scale should be of the order of the GUT scale, we found viable models even for $M_R$ below 100 GeV. Parametrising the neutrino mass in terms of five integers, we derived a conservative limit on $M_R$ for each set of models, requiring only that they should fit the atmospheric neutrino mass scale. The strong suppression of the light neutrino mass with the number of loops, i.e. $(1/16\pi^2)^{2\ell}$, along with the seesaw Majorana mass suppression allows remarkably low $M_{R}$ values. This fact makes models with a large number of loops (or Standard Model mass insertions) run into conflict with big bang nucleosynthesis and $\Delta N_{eff}$, which therefore significantly constrains the space of possible models.

To illustrate this idea in further detail, we presented two example models where the Dirac neutrino mass matrix is generated at the one- and two-loop level. The latter lies at the edge of the excluded models. An extra $Z_4$ symmetry is incorporated to forbid a tree-level Dirac mass, but it is broken softly in order to radiatively generate the Dirac Yukawa. A remnant exact $Z_2$ symmetry is kept stabilising the lightest of the $Z_2$ charged fields and providing a good dark matter candidate.

\pagebreak
\fancyhf{}

%% file: Chapters/Dim7_pheno/Chapter_dim7pheno.tex
\fancyhf{}
\fancyhead[LE,RO]{\thepage}
\fancyhead[RE]{\slshape{Chapter \thechapter. LNV phenomenology of $d = 7$ neutrino mass models}}
\fancyhead[LO]{\slshape\nouppercase{\rightmark}}

\chapter{Lepton number violating phenomenology of $d = 7$ neutrino mass models}
\label{ch:Dim7_pheno}
\graphicspath{ {Chapters/Dim7_pheno/} }

A Majorana mass term for neutrinos always implies also the existence of lepton number violating (LNV) processes. The best-known example is neutrinoless double-$\beta$ decay ($0\nu\beta\beta$), for reviews see \cite{Deppisch:2012nb, Avignone:2007fu}. A high-scale mechanism, such as the classical seesaw type-I \cite{Minkowski:1977sc, Yanagida:1979as, Mohapatra:1979ia}, however, will leave no other LNV signal than $0\nu\beta\beta$ decay. From this point of view, models in which the scale of LNV is around the electroweak scale are phenomenologically much more interesting.

Low-scale Majorana neutrino mass models need some suppression mechanism to explain the observed smallness of neutrino masses \cite{Cai:2017jrq}. This suppression could be due to loop factors \cite{Bonnet:2012kz, Sierra:2014rxa}, or neutrino masses could be generated by higher order operators \cite{Bonnet:2009ej, Babu:2009aq}, or both. In this chapter based on \cite{Cepedello:2017lyo}, we will study the phenomenology of a particular class of models, namely $d=7$ one-loop models \cite{Cepedello:2017eqf} already discussed in \ch{ch:Dim7_1loop}. Our main motivation is that $d=7$ one-loop contributions to neutrino masses can be dominant only, if new particles below approximately $2$ TeV exist. This mass range can be covered by the LHC experiments in the near future, if some dedicated search for the LNV signals we discuss in this chapter is carried out.

Apart from LNV signals, the parameter space of $d=7$ neutrino mass models can be constrained by a variety of searches. First, neutrino masses and angles should be correctly fitted. Since we now know that all three active neutrino mixing angles are non-zero, this fit leads to certain predictions for lepton flavour violating decays. We therefore discuss also current and future constraints coming from $\mu\to e \gamma$, $\mu \to 3 e$ and $(\mu-e)$ conversion in nuclei.

Constraints on our models come also from lepton number conserving LHC searches. The same-sign dilepton searches \cite{ATLAS:2014kca,ATLAS:2016pbt,ATLAS:2017iqw,CMS:2016cpz} can be recast into lower mass limits valid for our models. In addition, also multi-lepton searches \cite{Sirunyan:2017qkz}, motivated by the seesaw type-III, can be used to obtain interesting limits.
\\

The chapter is therefore organised as follows. In the next section, we discuss the basic setup of $d=7$ models and then present the Lagrangians and neutrino mass diagrams of our two example models. These models are taken from the classification done in \ch{ch:Dim7_1loop}. In \sect{sec:pheno:lowenergy} we calculate neutrino masses and constraints from low energy probes. We then move to the discussion of LHC phenomenology in \sect{sec:pheno:lhc}. We first derive constraints from existing searches, before discussing possible searches for LNV final state.

%%%%%%%%%%%%%%%%%%%%%%%%%%%%%%%%%%%%%%%%%%%%%%%%%%%%%%%%%%%%%%%
%%%%%%%%%%%%%%%%%%%%%%%%%%%%%%%%%%%%%%%%%%%%%%%%%%%%%%%%%%%%%%%
\section{Theoretical setup: $d=7$ models} \label{sec:pheno:dim7}

Before we discuss our example models, it may be useful to recapitulate some basics about Majorana neutrino masses in general and $d=7$ models in particular. Majorana neutrino masses can be generated from $d=5+2n$ operators,
\begin{equation} \label{eq:pheno:d2n}
  {\cal O}^{d=5+2n}= LLHH \times ({HH^{\dagger}})^n \, ,
\end{equation}
where the lowest order, $d=5$, is the well-known Weinberg operator ${\cal O}^W$
\cite{Weinberg:1979sa}.

Higher order contributions to neutrino masses are expected to be subdominant, unless the underlying model does not generate ${\cal O}^W$. This can be achieved essentially in two ways: Either via introducing a discrete symmetry \cite{Bonnet:2009ej} or simply because the particle content of the model does not allow to complete the lowest order operator \cite{Babu:2009aq, Cepedello:2017eqf}. We will not be interested in models with additional discrete symmetries here, since such models, although interesting theoretically, usually are based on additional Standard Model singlet states, which leave very little LHC phenomenology to explore.\footnote{``Sterile'' neutrino searches at the LHC provide of course constraints on these models.} Consider, instead, the BNT model \cite{Babu:2009aq} (\sect{subsec:numass:maj_tree}). This $d=7$ tree-level model introduces a vector-like fermion pair, $\Psi$ and ${\bar\Psi}$ with quantum numbers ${\bf 3}^F_1$ and a scalar quadruplet $S \equiv {\bf 4}^S_{3/2}$. (Here and elsewhere we will use a notation which gives the ${\rm SU(2)_L}$ representation and hypercharge in the form ${\bf R}_Y$ with a superscript $S$ or $F$, where necessary.) By construction, at tree-level the lowest order contribution to the neutrino masses is $d=7$. Being higher order, already at tree-level, two new particles are needed in order to generate a neutrino mass. This model has a rich LHC phenomenology \cite{Babu:2009aq, Ghosh:2017jbw} and, in particular, generates the LNV final state $W^{\pm}W^{\pm}W^{\pm}+W^{\mp}l^{\mp}l^{\mp}$.

The BNT model is unique at tree-level in the sense that no additional symmetries are required to make it the leading contribution to neutrino masses (we call such models ``genuine''). In \ch{ch:Dim7_1loop}, we have analysed systematically $d=7$ one-loop models. While there exists a large number of topologies, only a few of them can lead to genuine $d=7$ models. These topologies can still generate 23 different diagrams, but all models underlying these diagrams share the following common features: (i) five new multiplets must be added to the Standard Model particle content; and (ii) all models contain highly charged particles. In all cases, there is at least one triply charged state. Thus, one expects that all $d=7$ one-loop models have rather similar accelerator phenomenology. For this reason, we concentrate on only two of the simplest example models.\footnote{Strictly speaking this is true only for variants of the $d=7$ one-loop models for which the particles appearing in the loop are colour singlets. For a brief discussion for the case of coloured particles see \sect{sec:pheno:summary}.}

According to the classification shown in \ch{ch:Dim7_1loop}, one can classify the $d=7$ models w.r.t. increasing size of the largest ${\rm SU(2)_L}$ multiplet. There is one model, in which no representation larger than triplets is needed. All other models require at least one quadruplet. Our two example models, introduced below, are therefore just the simplest realisations of ${\cal O}^{d=7}$ at one-loop, but are expected to cover most of the interesting phenomenology.

Finally, let us recall an important fact about the higher order operator \eq{eq:pheno:d2n}. As discussed in \sect{sec:dim7loop:intro}, the operator ${\cal O}^{d=7}$ generates automatically also a $d=5$ neutrino mass contribution,
\begin{equation}\label{eq:pheno:nlp1}
    \frac{1}{\Lambda^3}LLHHHH^{\dagger} \rightarrow \frac{1}{16 \pi^2} \frac{1}{\Lambda}LLHH \, .
\end{equation}
It is easy to estimate that this loop contribution will become more important than the tree-level if $\Lambda \gtrsim 2$ TeV. Our main motivation for the present study is that the LHC can explore large parts of this parameter space.

%%%%%%%%%%%%%%%%%%%%%%%%%%%%%%%%%%%%%%%%%%%%%%%%%%%%%%%%%%%%%%%
\subsection{Triplet model} \label{subsec:pheno:3plet}

\begin{figure}
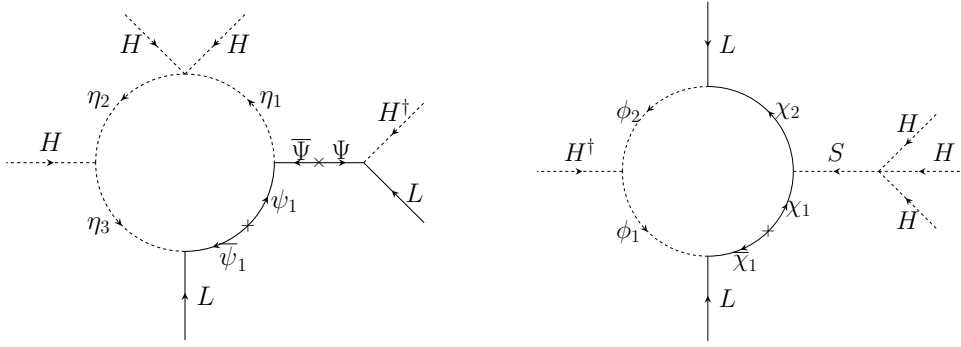

    \centering
    \includegraphics[width=0.48\linewidth]{./figures/model_T11-i.pdf}
    \hfill
    \includegraphics[width=0.48\linewidth]{./figures/model_T16-ii.pdf}
    \caption{One-loop dimension 7 neutrino mass diagrams for the triplet model (left) and for the quadruplet model (right). Diagrams are given in the gauge basis.}
    \label{fig:pheno:diag}
\end{figure}

Our first example model is the ``minimal'' one-loop $d=7$ model. This model, depicted in \fig{fig:pheno:diag} to the left, is minimal in the sense that it uses no multiplet larger than triplets. The model adds two new (vector-like) fermions and three scalars to the Standard Model particle content.
\begin{center}
    \begin{tabular}{ c c c }

        $\Psi=\left(
        \begin{matrix}
        \Psi^{++} \\ 
        \Psi^{+}  \\
        \Psi^{0}
        \end{matrix} 
        \right) \sim \textbf{3}_1^F$ ,

      &\hfill

        $\eta_1=\left(
        \begin{matrix}
        \eta_1^{++} \\ 
        \eta_1^{+}
        \end{matrix} 
        \right) \sim \textbf{2}_{3/2}^S$ ,

      &\hfill

        $\eta_2=\left(
        \begin{matrix}
        \eta_2^{+++} \\ 
        \eta_2^{++}
        \end{matrix} 
        \right) \sim \textbf{2}_{5/2}^S$ ,

    \end{tabular}

\vspace*{0.5cm}

    \begin{tabular}{ c c }

        $\eta_3=\left(
        \begin{matrix}
        \eta_3^{++++} \\ 
        \eta_3^{+++}  \\
        \eta_3^{++}
        \end{matrix} 
        \right) \sim \textbf{3}_3^S$ ,

      &\quad

        $\psi_1=\left(
        \begin{matrix}
        \psi_1^{+++} \\ 
        \psi_1^{++}
        \end{matrix} 
        \right) \sim \textbf{2}_{5/2}^F$ .

    \end{tabular}
\end{center}
Note that both, $\Psi$ and ${\bar\Psi}$ are needed. The Lagrangian of the model contains the following terms,
\begin{eqnarray} \label{eq:pheno:lagr_3plet}
    \mathscr{L} &\supset& \left[ Y_1 H^\dagger \Psi P_L L \,+\, Y_2 \overline{\psi}_1 P_L L \eta_3 \,+\, Y_3 \eta_1^\dagger \overline{\Psi} \psi_1 \right.
     \\ \nn
            && \,+\, \left. Y_4 \eta_1 \overline{\Psi} P_L L \,+\, Y_5\, e_R \eta_1^\dagger \psi_1 \,+\, \hc \right] 
     \\ \nn 
            && - M_\Psi \overline{\Psi}\Psi - M_{\psi_1} \overline{\psi}_1 \psi_1 - V_{scalar} \, ,
\end{eqnarray}
with the scalar potential given by,
\begin{eqnarray} \label{eq:pheno:pot_3plet}
    V_{scalar} &=& m_H^2 H^\dagger H \,+\, \frac 12 \lambda_1 (H^\dagger H)^2 \,+\, m_{\eta_1}^2 \eta_1^\dagger \eta_1 \,+\, m_{\eta_2}^2 \eta_2^\dagger \eta_2 \,+\, m_{\eta_3}^2 \eta_3^{\dagger} \eta_3
    \nn \\ \nn 
               &&+ \left[ \mu_1 \, H \eta_2 \eta_3^{\dagger} \,+\, \mu_2 \eta_1 \eta_1 \eta_3^{\dagger} \,+\, \lambda_2 \eta_2^\dagger H \eta_1 H \,+\, \lambda_3 \, \eta_1^\dagger \eta_2 \eta_1^\dagger H \,+\, \hc \right]
    \\ \nn
               &&+ \frac 12 \lambda_4 (\eta_1^\dagger \eta_1)^2 \,+\, \frac 12 \lambda_5 (\eta_2^\dagger \eta_2)^2 \,+\, \frac 12 \lambda_6 (\eta_3^{\dagger} \eta_3)^2 \,+\, \frac 12 \lambda_7 (\eta_3^{\dagger} \eta_3^\dagger)(\eta_3 \eta_3)
    \\
               &&+ \lambda_8 (H^\dagger H)(\eta_1^\dagger \eta_1) \,+\, \lambda_9 (H^\dagger H)(\eta_2^\dagger \eta_2) \,+\, \lambda_{10} (H^\dagger H)(\eta_3^{\dagger} \eta_3)
    \\ \nn
               &&+ \lambda_{11} (\eta_1^\dagger \eta_1)(\eta_2^\dagger \eta_2) \,+\, \lambda_{12} (\eta_1^\dagger \eta_1)(\eta_3^{\dagger} \eta_3) \,+\, \lambda_{13} (\eta_2^\dagger \eta_2)(\eta_3^{\dagger} \eta_3)
    \\ \nn 
               &&+ \lambda_{14} (H^\dagger \eta_1)(\eta_1^\dagger H) \,+\, \lambda_{15} (H^\dagger \eta_2)(\eta_2^\dagger H) \,+\, \lambda_{16} ( H^\dagger \eta_3 )( \eta_3^{\dagger} H ) 
    \\ \nn
               &&+ \lambda_{17} ( \eta_1^{\dagger} \eta_2 )( \eta_2^\dagger \eta_1 ) \,+\, \lambda_{18} ( \eta_1^\dagger \eta_3 )( \eta_3^{\dagger} \eta_1 ) \,+\, \lambda_{19} ( \eta_2^\dagger \eta_3 )(\eta_3^{\dagger}\eta_2 ) \, .
\end{eqnarray}
The model contains many charged scalars, but the only neutral scalar is in the Standard Model Higgs doublet.

From the Yukawa couplings only $Y_1$, $Y_2$, $Y_3$ enter the neutrino mass calculation directly, see next section. Similarly, from the scalar terms only the coupling $\lambda_2$ and mass term $\mu_1$ and the mass matrix of the doubly charged scalars play an important role. We therefore give here only the mass matrix for the $S_i^{++}$ states. In the basis ($\eta_1,\eta_2,\eta_3$) it is given as,
\begin{equation} \label{eq:pheno:etpp}
    {\cal M}_{\eta^{++}}^2 = 
    \begin{pmatrix}
                       m_{S_1}^2 &  -\frac{\lambda_2 v^2}{2} &                         0
        \\
        -\frac{\lambda_2 v^2}{2} &                 m_{S_2}^2 & -\frac{\mu_1 v}{\sqrt{2}}
        \\
                               0 & -\frac{\mu_1 v}{\sqrt{2}} &                 m_{S_3}^2
    \end{pmatrix} \, .
\end{equation}
Here, $v$ is the Higgs vacuum expectation value, and
\begin{eqnarray} \label{eq:pheno:defmsq}
    m_{S_1}^2 = m_{\eta_1}^2 \,+\, \frac{\lambda_7}{2}v^2 \, ,
    \\ \nn
    m_{S_2}^2 = m_{\eta_2}^2 \,+\, \frac{\lambda_8+\lambda_{14}}{2}v^2 \, , 
    \\ \nn
    m_{S_3}^2 = m_{\eta_3}^2 \,+\, \frac{\lambda_9+\lambda_{15}}{2}v^2 \, . 
    \\ \nn
\end{eqnarray}
The mass matrix of the doubly charged scalars \eq{eq:pheno:etpp} can be diagonalised by,
\begin{equation} \label{eq:pheno:diagetpp}
    {\hat{\cal M}}_{\eta^{++}}^2 = R_{\eta^{++}}^T {\cal M}_{\eta^{++}}^2 R_{\eta^{++}} \, .
\end{equation}
All other mass matrices of the model can be easily derived, and we do not give them here for brevity.

%%%%%%%%%%%%%%%%%%%%%%%%%%%%%%%%%%%%%%%%%%%%%%%%%%%%%%%%%%%%%%%
\subsection{Quadruplet model} \label{subsec:pheno:4plet}

Our second example model makes use of the quadruplet $S$. The corresponding neutrino mass diagram is given in \fig{fig:pheno:diag} to the right. The full new particle content of the model is,

\begin{center}
    \begin{tabular}{ c c c }

        $S=\left(
        \begin{matrix}
        S^{+++} \\
        S^{++} \\ 
        S^{+}  \\
        S^{0}
        \end{matrix} 
        \right) \sim \textbf{4}_{3/2}^S$ ,

      & \hfill

        $\chi_1=\left(
        \begin{matrix}
        \chi_1^{++} \\ 
        \chi_1^{+}
        \end{matrix} 
        \right) \sim \textbf{2}_{3/2}^F$ ,

      & \hfill

        $\chi_2=\left(
        \begin{matrix}
        \chi_2^{++++} \\ 
        \chi_2^{+++} \\ 
        \chi_2^{++}
        \end{matrix} 
        \right) \sim \textbf{3}_3^F$ ,

    \end{tabular}

    \begin{tabular}{ c c }

        $\phi_1=\phi_1^{++} \sim \textbf{1}_2^S$ ,

      & \quad

        $\phi_2=\left(
        \begin{matrix}
        \phi_2^{+++} \\ 
        \phi_2^{++}
        \end{matrix} 
        \right) \sim \textbf{2}_{5/2}^S$ .

    \end{tabular}
\end{center}
Again, fermions need to be vector-like. The Lagrangian of the model is given by,
\begin{eqnarray} \label{eq:pheno:lagr_4plet}
    \mathscr{L} &\supset& \left[ Y_1 \overline{\chi}_1 P_L L \phi_1 \,+\, Y_2 \phi^\dagger_2 P_L L \chi_2 \,+\, Y_3 \chi_1 S\, \overline{\chi}_2 \,+\, Y_4 e_R \overline{\chi}_1 \phi_2 \right.
    \\ \nn 
                &&+ \left. Y_5 e_R H^\dagger \chi_1 \,+\, Y_6 e_R e_R \phi_1 \,+\, \hc \right] 
     \\ \nn
                &&- M_{\chi_1} \overline{\chi}_1 \chi_1 - M_{\chi_2} \overline{\chi}_2 \chi_2 - V_{scalar} \, ,
\end{eqnarray}
where the scalar potential is defined as,
\begin{eqnarray} \label{eq:pheno:pot_4plet}
    V_{scalar} &=& m_H^2 H^\dagger H \,+\, \frac 1 2 \lambda_1 ( H^\dagger H )^2 \,+\, m_S^2 S^\dagger S \,+\, m_{\phi_1}^2 \phi_1^\dagger \phi_1 \,+\, m_{\phi_2}^2 \phi_2^\dagger \phi_2
   \nn  \\ \nn 
               &&+ \left[ \mu_1 \phi_1^\dagger H^\dagger \phi_2 \,+\, \lambda_2 S^\dagger H H H \,+\, \lambda_3 \phi_2^\dagger S H H \,+\, \lambda_4 \phi_2^\dagger S H^\dagger S \,+\, \hc \right]
    \\ \nn 
               &&+ \frac 1 2 \lambda_5 ( \phi_1^\dagger \phi_1 )^2 \,+\, \frac 1 2 \lambda_6 ( \phi_2^{\dagger} \phi_2 )^2 \,+\, \frac 1 2 \lambda_7 ( S^\dagger S )^2 \,+\, \frac 1 2 \lambda_8 ( S^\dagger S^\dagger )(S S)
    \\
               &&+  \lambda_9 ( H^\dagger H )( \phi_1^\dagger \phi_1 )  \,+\, \lambda_{10} ( H^\dagger H ) ( \phi_2^\dagger \phi_2 ) \,+\, \lambda_{11} ( H^\dagger H ) ( S^\dagger S )
    \\ \nn
               &&+ \lambda_{12} ( \phi_1^\dagger \phi_1 ) ( \phi_2^\dagger \phi_2 ) \,+\, \lambda_{13} ( \phi_1^\dagger \phi_1 ) ( S^\dagger S ) \,+\, \lambda_{14} ( \phi_2^\dagger \phi_2 ) ( S^\dagger S ) 
    \\ \nn
               &&+ \lambda_{15} ( H^\dagger \phi_2 ) ( \phi_2^\dagger H ) \,+\, \lambda_{16} ( H^\dagger S )( S^\dagger H ) \,+\, \lambda_{17} ( S^\dagger \phi_2 )( \phi_2^\dagger S ) \, .
\end{eqnarray}
Note that the term proportional to $\lambda_2$ will induce a non-zero value for the VEV of the neutral scalar $S$, even if $m_S^2$ is positive. One can thus take either $\lambda_2$ or $v_S$ as a free parameter and solve the tadpole equations for the other. In our numerical calculation we choose $v_S$ as input.

%%%%%%%%%%%%%%%%%%%%%%%%%%%%%%%%%%%%%%%%%%%%%%%%%%%%%%%%%%%%%%%
%%%%%%%%%%%%%%%%%%%%%%%%%%%%%%%%%%%%%%%%%%%%%%%%%%%%%%%%%%%%%%%
\section{Low energy constraints} \label{sec:pheno:lowenergy}

In this section we will discuss non-accelerator constraints on the parameters of our two example models. We consider first neutrino masses and angles and then turn to lepton flavour violating (LFV) decays. The LHC phenomenology is discussed in \sect{sec:pheno:lhc}.

We have implemented both of our example models in SARAH \cite{Staub:2012pb, Staub:2013tta}. Using Toolbox \cite{Staub:2011dp}, the implementation can be used to generate \texttt{SPheno} code \cite{Porod:2003um, Porod:2011nf}, for the numerical evaluation of mass spectra and observables, such as LFV decays ($\mu \to e \gamma$, $\mu \to 3 e$ etc) calculated using Flavour Kit \cite{Porod:2014xia}. The Toolbox subpackage SSP has then be used for our numerical scans.

%%%%%%%%%%%%%%%%%%%%%%%%%%%%%%%%%%%%%%%%%%%%%%%%%%%%%%%%%%%%%%%
\subsection{Neutrino masses} \label{subsec:pheno:numass}

Here we discuss the calculation of neutrino masses in our two example models. We first consider the triplet model, then only briefly summarise the calculation of the quadruplet model, since it is very similar in both cases. Note that \texttt{SPheno} calculates one-loop corrected masses numerically. We have checked that the description given below agrees very well with the numerical results from \texttt{SPheno}.

The triplet model is described by the Lagrangian given in \eq{eq:pheno:lagr_3plet} and generates $d=7$ one-loop neutrino masses via the diagram shown in \fig{fig:pheno:diag} to the left. Rotating the doubly charged scalars to the mass eigenstate basis, the diagram in \fig{fig:pheno:diag} results in a neutrino mass matrix given by,
\begin{eqnarray} \label{eq:pheno:mnu1}
    (m_{\nu})_{\alpha\beta} &=& \frac{1}{16 \pi^2} \frac{Y_3 v}{m_\Psi}m_{\psi_1} \times
    \\ \nn
    && \sum_i (R_{\eta^{++}})_{1i} (R_{\eta^{++}})_{3i}  B_0(0, m^2_{\psi_1}, m^2_{S_i}) 
    \left[ (Y_1)_{\alpha} (Y_2)_{\beta} \,+\, (Y_1)_{\beta} (Y_2)_{\alpha} \right] \, .
\end{eqnarray}
Here $(R_{\eta^{++}})$ is the rotation matrix defined in \eq{eq:pheno:diagetpp} and $m_{S_i}$ are the eigenvalues of \eq{eq:pheno:etpp}. $B_0(0, m^2_{\psi_1}, m^2_{S_i})$ is a Passarino-Veltman function \cite{Passarino:1978jh}. Note that \eq{eq:pheno:mnu1} is already an approximation: $\Psi_0$ mixes with the light active neutrinos. So, the total neutral fermion mass matrix is $4 \times 4$. However, this mixing should be small and is estimated here simply by the factor $\frac{Y_3 v}{m_\Psi}$.

In the numerical calculation we have used \eq{eq:pheno:mnu1} to fit the neutrino masses of the model to neutrino oscillation data. However, in order to have a better understanding of the dependence of \eq{eq:pheno:mnu1} on the different parameters of the Lagrangian, we also give the expression of the neutrino mass matrix in the mass insertion approximation. We replace the full diagonalisation matrices and eigenvalues of the doubly charged scalar mass matrix by their leading order ones. The resulting equation can be written simply as,
\begin{equation} \label{eq:pheno:mnu2}
    (m_{\nu})_{\alpha\beta} = {\cal F} \times  \left[ (Y_1)_{\alpha} (Y_2)_{\beta} \,+\, (Y_1)_{\beta} (Y_2)_{\alpha} \right] \, ,
\end{equation}
where
\begin{eqnarray} \label{eq:pheno:prefac}
    {\cal F} &=& \frac{1}{16 \pi^2} \frac{Y_3 v}{m_\Psi} \frac{v^2 \lambda_2}{m^2_{S_2}-m^2_{S_1}} \frac{v \mu_1}{m^2_{S_3}-m^2_{S_2}} \times 
    \\ \nn
    && m_{\psi_1} \left[\frac{m^2_{S_1}}{m^2_{\psi_1}-m^2_{S_1}} \ln\left(\frac{m^2_{S_1}}{m^2_{\psi_1}}\right) - \frac{m^2_{S_2}}{m^2_{\psi_1}-m^2_{S_2}} \ln\left(\frac{m^2_{S_2}}{m^2_{\psi_1}}\right) \right] \, .
\end{eqnarray}
\eq{eq:pheno:mnu2} shows that neutrino angles predicted by the model depend on ratios of Yukawa couplings, while the overall mass scale is determined by the prefactor ${\cal F}$. The model has the interesting feature that $\det(m_{\nu})=0$. Therefore it can fit only hierarchical neutrino mass spectra (normal or inverse), but not a degenerate spectrum.\footnote{In order to fit also a quasi-degenerate spectrum we would need to include more than one copy of $\Psi$ or/and $\psi_1$.} The eigenvalues of \eq{eq:pheno:mnu2} are,
\begin{equation} \label{eq:pheno:mnu-eigen}
    m_{\nu_{1(3)}} = 0,
    \quad 
    m_{\nu_{2,3(1,2)}} = 
    \left[
    \sum_{\alpha} (Y_1)_{\alpha} (Y_2)_{\alpha} 
    \mp 
    \sqrt{
    \sum_{\alpha}\left|(Y_1)_{\alpha}\right|^{2} 
    \sum_{\alpha}\left|(Y_2)_{\alpha} \right|^{2}
    }
    \right]
    {\cal F} \, ,
\end{equation}
for normal (inverted) hierarchy. From \eq{eq:pheno:mnu-eigen}, one can estimate the constraints from neutrino masses on the size of the Yukawa couplings. In order to reproduce the neutrino mass suggested by atmospheric neutrino oscillations ($m_{\nu_{3}} \sim 0.05$ eV), keeping the mass scale of the new particles $M \sim 1$ TeV, the scalar coupling $\lambda_2 \sim 1$ and mass term $\mu \sim 1$ TeV, the Yukawa couplings $Y_{1}$, $Y_{2}$, $Y_{3}$ must be set typically to $\mathcal{O}(10^{-2})$. Note, however, that this is only a rough estimate and in our numerical calculations we scan over the free parameters of the model. As discussed in the next subsection, LFV produces upper limits on these Yukawa couplings very roughly of this order.

In our numerical fits to neutrino data, we do not only fit to solar and atmospheric neutrino mass differences, but also to the observed neutrino angles \cite{Forero:2014bxa}. This is done in the following way. First, we choose all free parameters appearing in the prefactor ${\cal F}$. These leaves us with the six free parameters in the two vectors $Y_1$ and $Y_2$. Two neutrino masses and three neutrino angles give us five constraints. We arbitrarily choose $(Y_1)_e$ as a free parameter, the remaining five entries are then fixed. Since $\det(m_{\nu})=0$, finding the solutions for those five parameters implies solving coupled quadratic equations, which can be done numerically.
\\

For the quadruplet model we show the neutrino mass diagram in \fig{fig:pheno:diag} to the right. The Lagrangian of this model is given in \eq{eq:pheno:lagr_4plet}. The calculation of the neutrino mass matrix for this model gives,
\begin{eqnarray}\label{eq:pheno:mnu3}
    (m_{\nu})_{\alpha\beta} &=& \frac{1}{16 \pi^2} \sum_j\sum_im_{\chi^{++}_j} (R_{S^{++}})_{1i} (R_{S^{++}})_{3i} (R_{\chi^{++}})_{1j} (R_{\chi^{++}})_{j2}^*
    \nn \\
                       &\times&  B_0(0, m^2_{\chi_j^{++}}, m^2_{S^{++}_i}) \left[ (Y_1)_{\alpha} (Y_2)_{\beta} \,+\, (Y_1)_{\beta} (Y_2)_{\alpha} \right] \, .
\end{eqnarray}
Here $R_{S^{++}}$ and $R_{\chi^{++}}$ are the matrices which diagonalise the doubly charged scalar and fermion mass matrices in the quadruplet model. As in the triplet model, $\det(m_{\nu})=0$. Thus, the fit of neutrino data is analogous to the one described above for the triplet model. Recall, however, that in the numerical calculation we use the induced VEV $v_s$ as a free parameter.

%%%%%%%%%%%%%%%%%%%%%%%%%%%%%%%%%%%%%%%%%%%%%%%%%%%%%%%%%%%%%%%
\subsection{Lepton flavour violating decays} \label{subsec:pheno:lfv}

As is well-known, experimental upper limits on lepton flavour violating decays provide important constraints on TeV-scale extensions of the Standard Model, see for example \cite{Vicente:2015cka, Cai:2017jrq} and references therein. Flavour Kit \cite{Porod:2014xia} implements a large number of observables into \texttt{SPheno} \cite{Porod:2011nf}. In the following we will concentrate on \mueg, \mueee and $(\mu-e)$ conversion in Ti.

Currently, \mueg \cite{TheMEG:2016wtm} and $\mu\to 3 e$ \cite{Bellgardt:1987du} provide the most stringent constraints. There is also a limit on $(\mu-e)$ conversion in Ti \cite{Dohmen:1993mp}. However, while there will be only some improvement in the sensitivity in \mueg \cite{Baldini:2013ke}, proposals to improve \mueee \cite{Blondel:2013ia} and $(\mu-e)$ conversion on both Ti \cite{prime2003} and Al \cite{Pezzullo:2017iqq} exist, which claim current bounds can be improved by 4-6 orders of magnitude. Constraints involving $\tau$'s also exist, but are much weaker. Thus, while we routinely calculate constraints also for the $\tau$ sector, we will not discuss the results in detail.

Again, let us first discuss the triplet model. The Lagrangian \eq{eq:pheno:lagr_3plet} of the model contains five different Yukawa couplings. We can divide them into two groups: $Y_1$, $Y_2$ and $Y_3$ enter the neutrino mass calculation, while $Y_4$ and $Y_5$ are parameters with no relation to $m_\nu$. This implies that for the former, neutrino physics imposes a lower bound on certain products of these Yukawas (as a function of the other parameters), while the latter could, in principle, be arbitrarily small.

\begin{figure}[t!]
    \centering
    \includegraphics[width=0.48\linewidth]{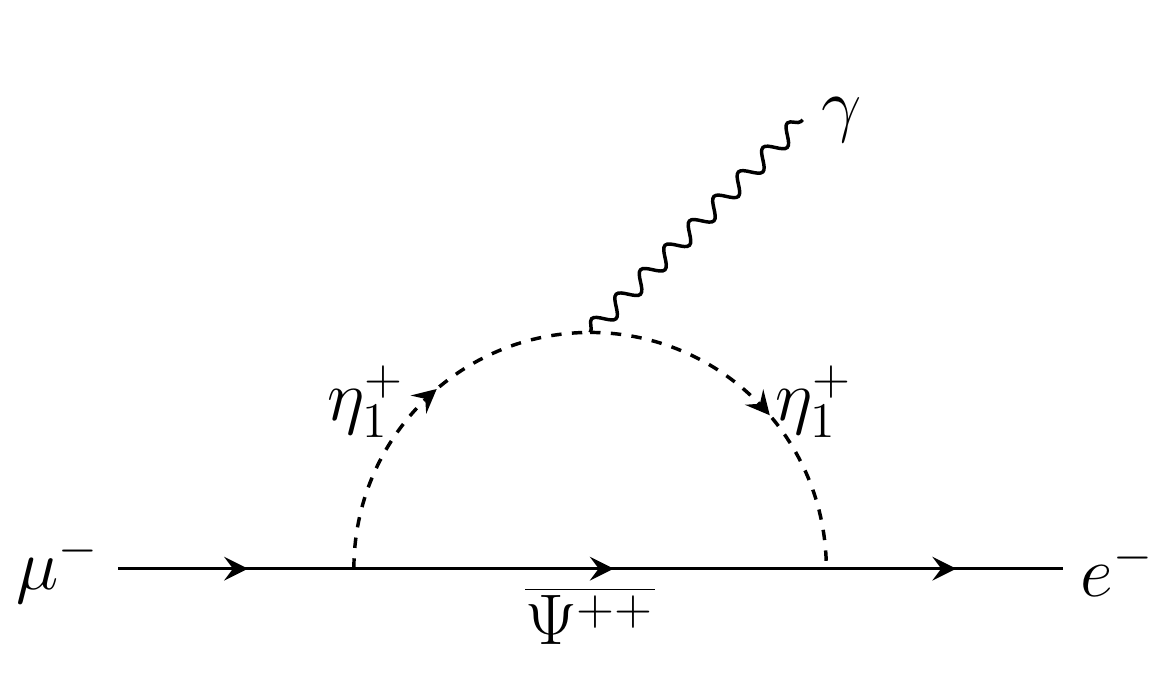}
    \hfill
    \includegraphics[width=0.48\linewidth]{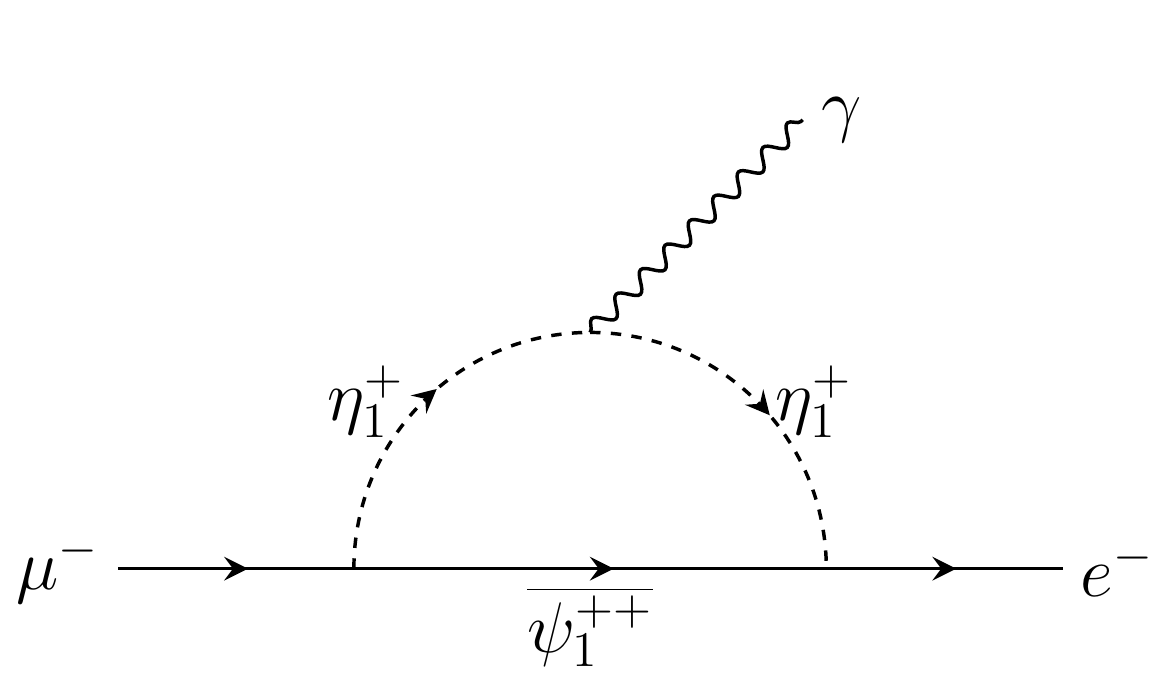}
    \caption{Example diagrams for $\mu\to e \gamma$ in the triplet model, proportional to $(Y_{4})_e(Y_{4})_{\mu}$ (left) and $(Y_{5})_e(Y_{5})_{\mu}$ (right).}
    \label{fig:pheno:diagmueg}
\end{figure}

Consider first the simpler case of $Y_4$ and $Y_5$. The diagrams in \fig{fig:pheno:diagmueg} show contributions to \mueg due to these couplings. The current upper limit on Br(\mueg) then puts a bound on both, $Y_4$ and $Y_5$, of roughly $(Y_{4/5})_e (Y_{4/5})_{\mu} \lesssim 10^{-4}$ for masses of $\eta_1$ and $\Psi$ or $\chi$ of the order ${\cal O}(1)$ TeV.

\begin{figure}[t!]
    \centering
    \includegraphics[width=0.48\linewidth]{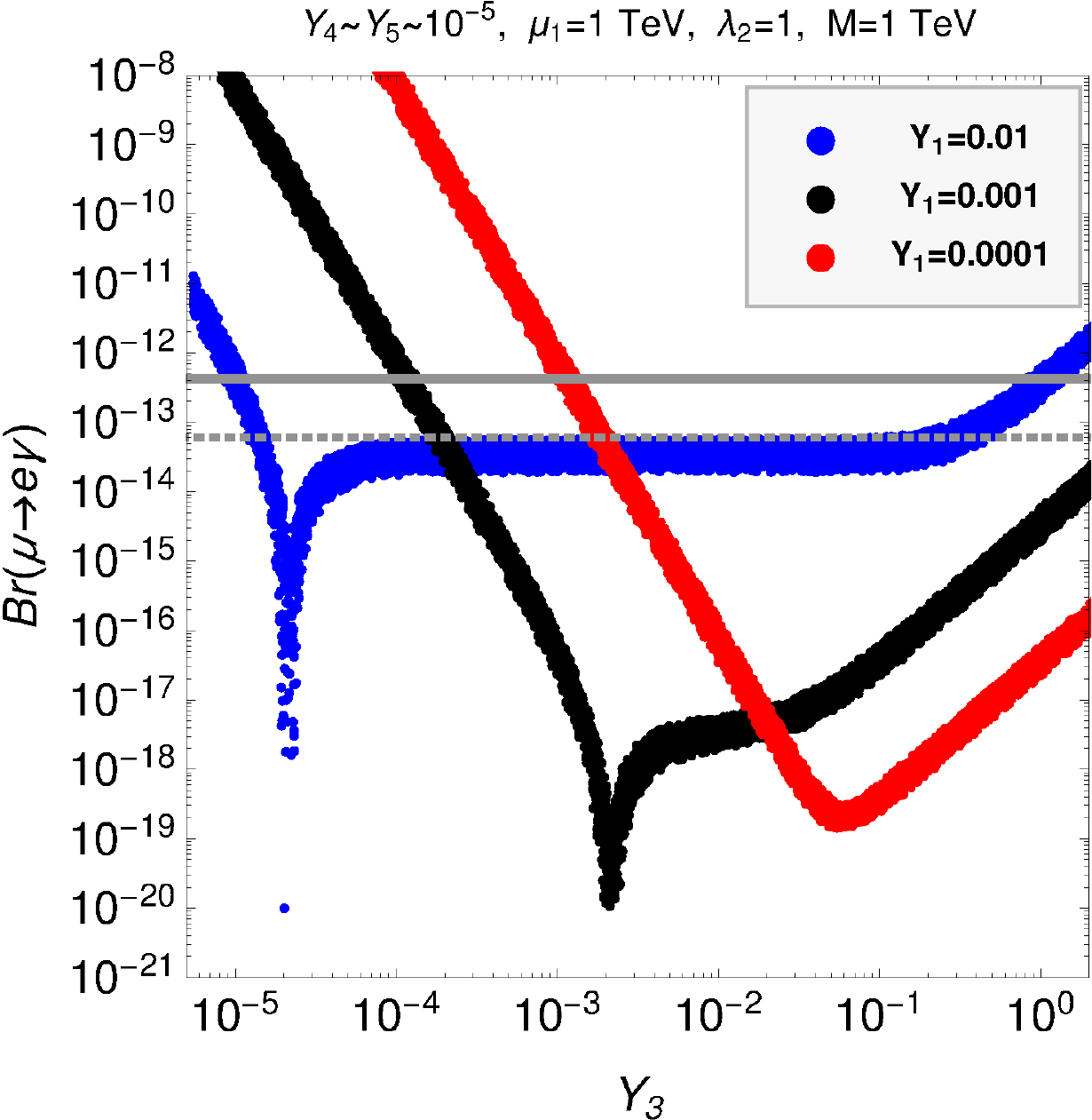}
    \hfill
    \includegraphics[width=0.48\linewidth]{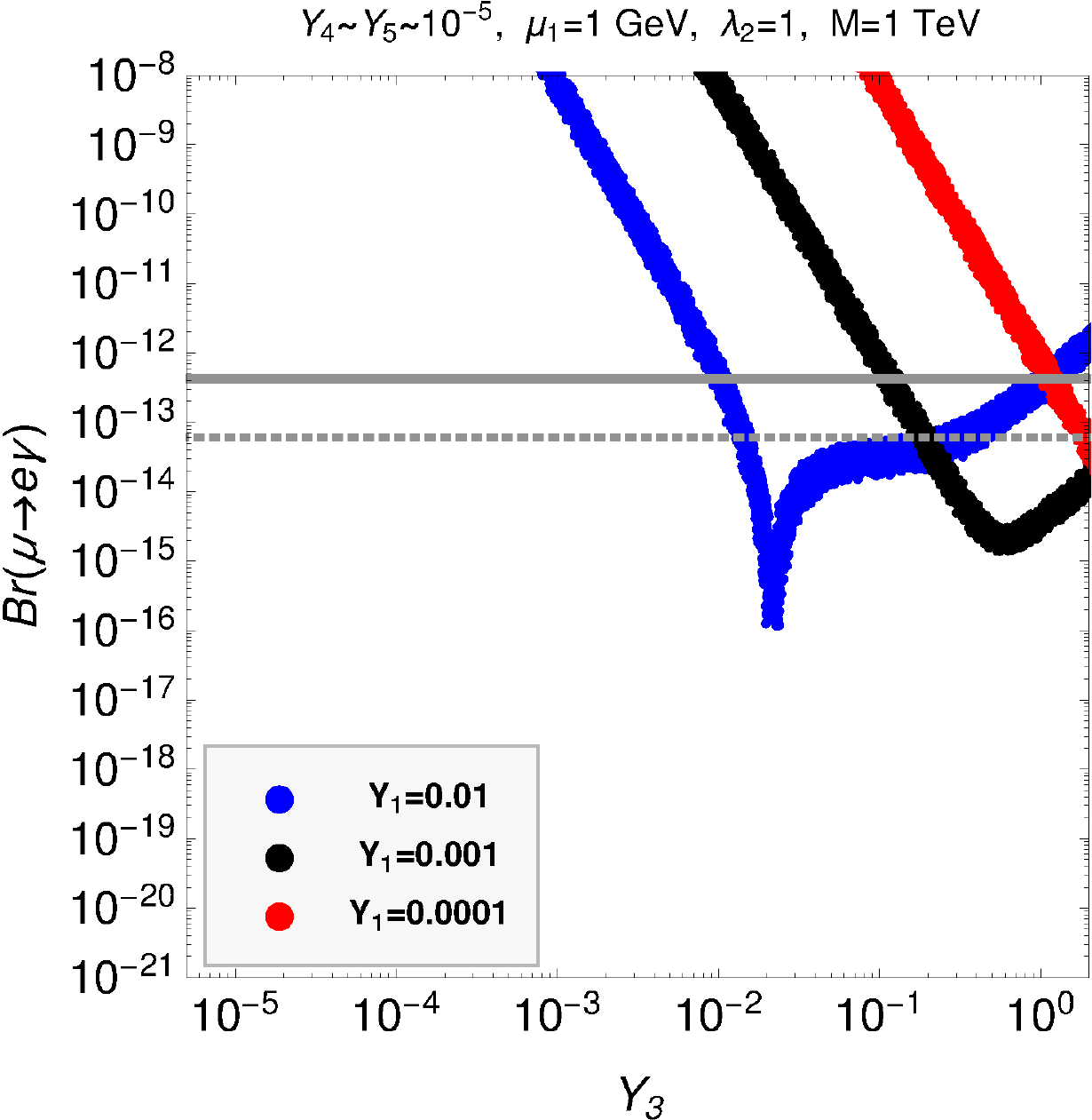}
    \\
    \includegraphics[width=0.48\linewidth]{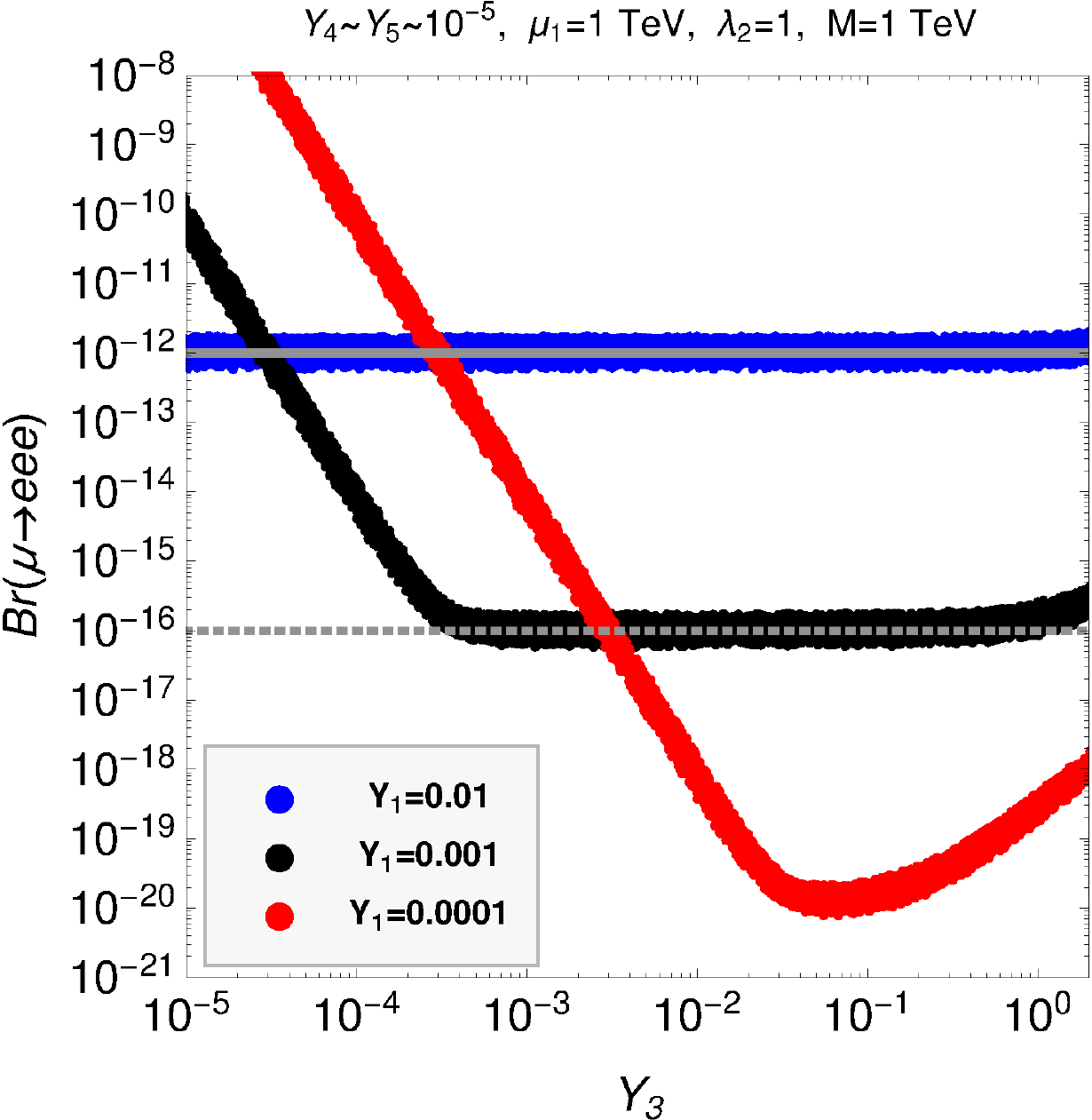}
    \hfill
    \includegraphics[width=0.48\linewidth]{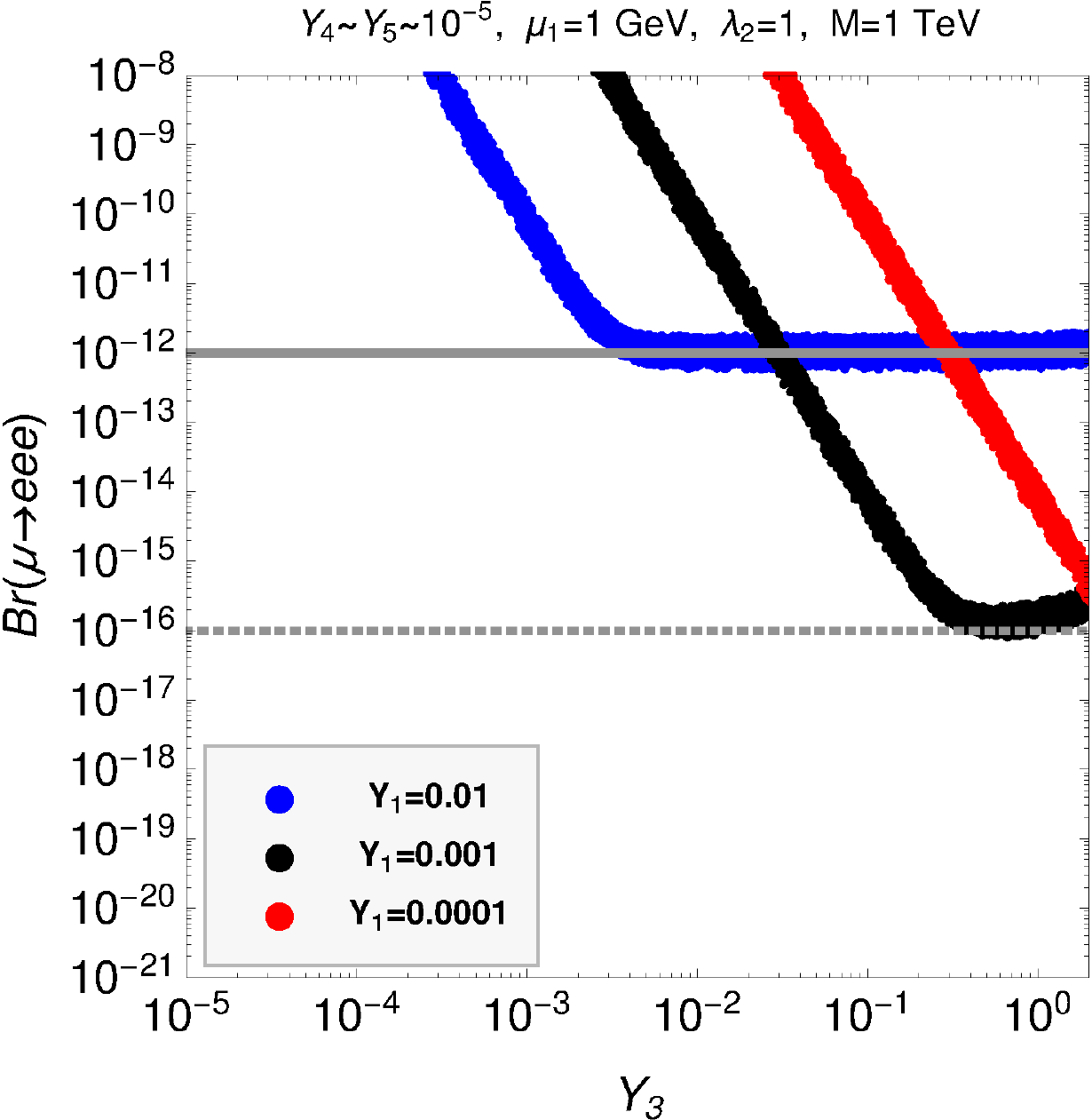}
\caption{(Continues in \fig{fig:pheno:LFVT}).}
    \label{fig:pheno:LFVT1}
\end{figure}

\begin{figure}[t!]
    \centering
    \includegraphics[width=0.48\linewidth]{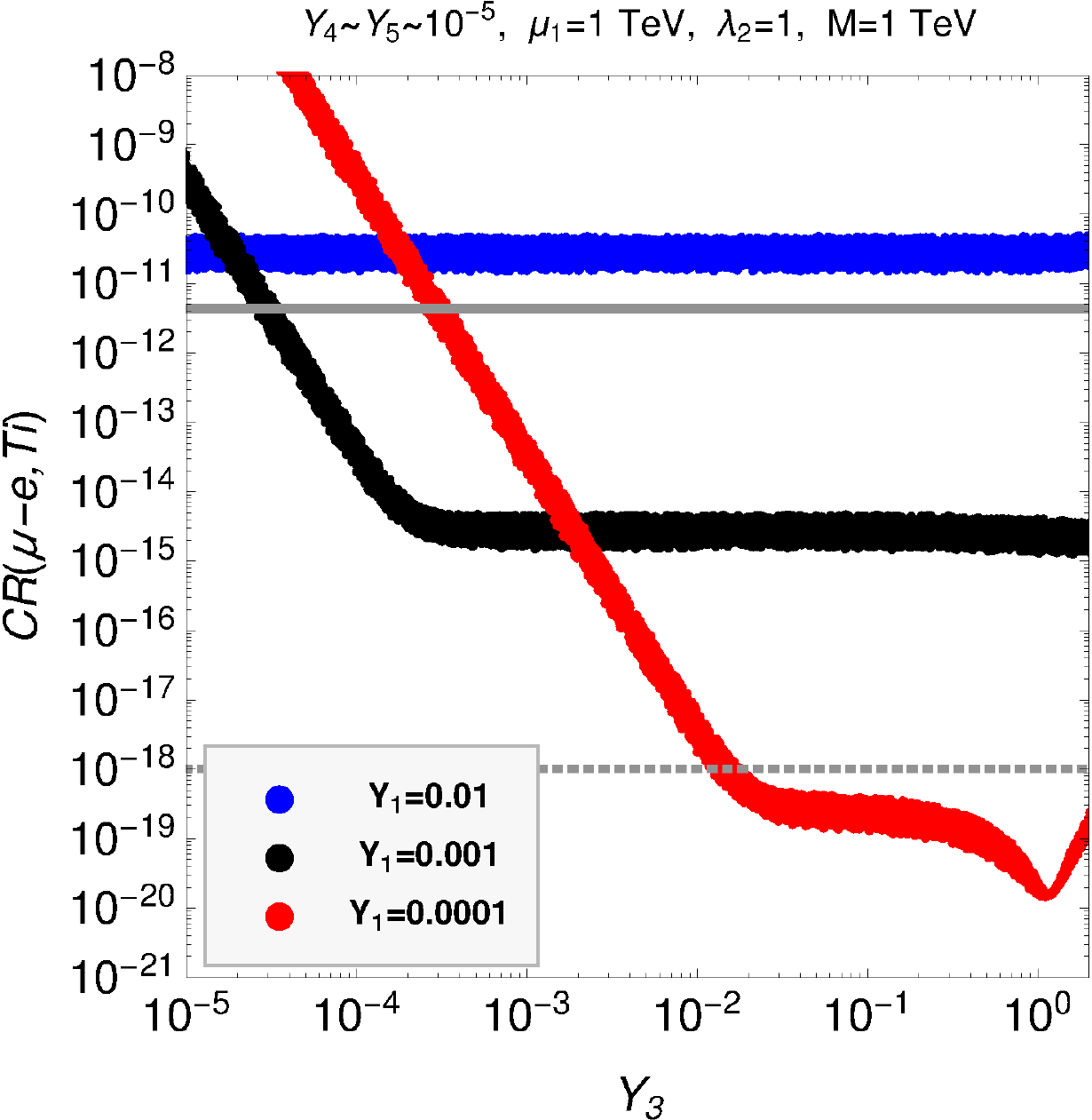}
    \hfill
    \includegraphics[width=0.48\linewidth]{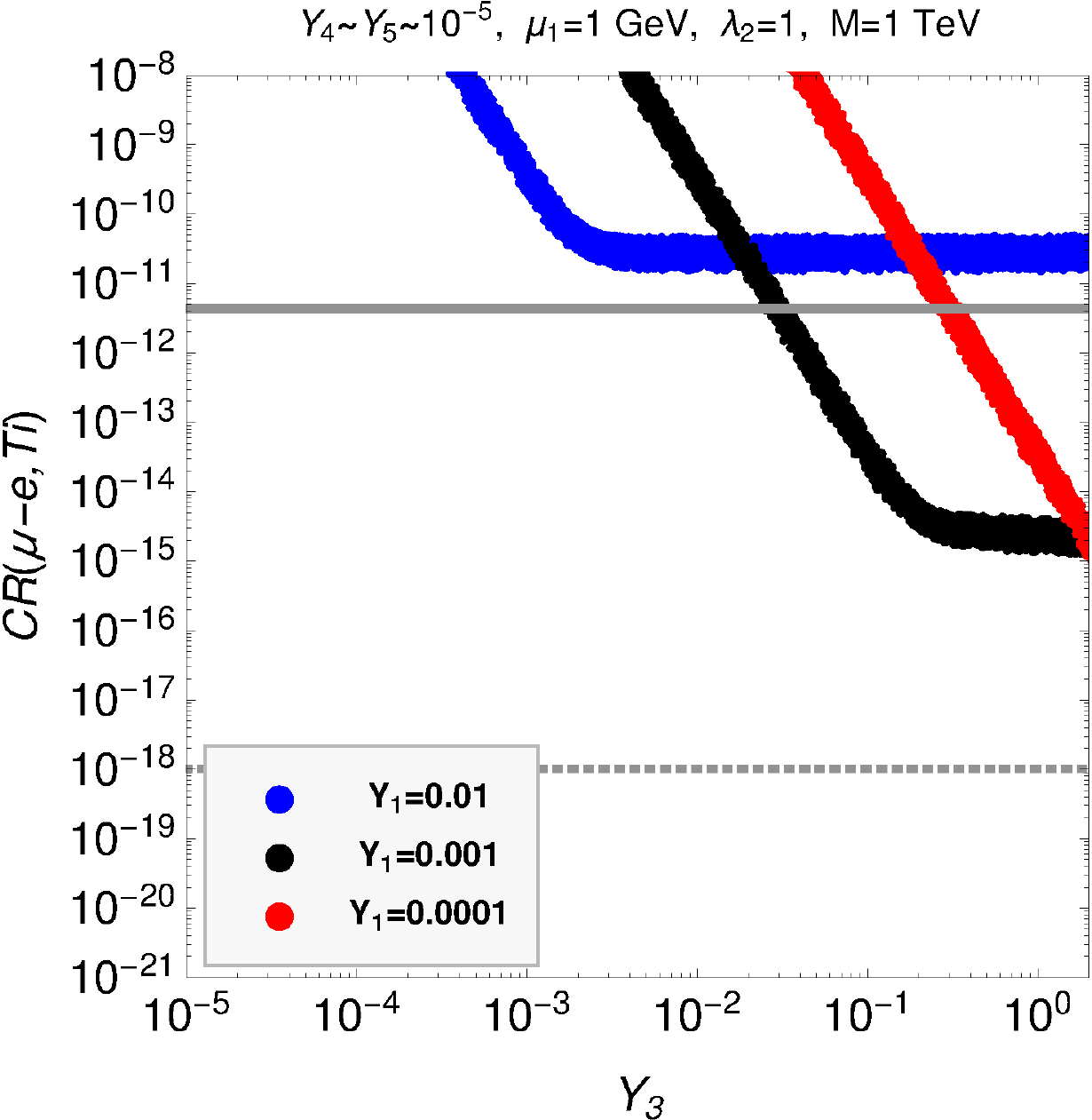}
    \caption{Lepton flavour violating decays calculated in the triplet model. Top panel: Br(\mueg); middle panel: Br(\mueee); bottom: $(\mu-e)$ conversion in Ti. Rates are plotted versus the coupling $Y_3$, for discussion see text. Left row: $\mu_1=1$ TeV, right row $\mu_1=1$ GeV. The full (dashed) horizontal lines are the current limits (and future expected sensitivities). This figure is a continuation of \fig{fig:pheno:LFVT1}.}
    \label{fig:pheno:LFVT}
\end{figure}

The fit to neutrino data imposes relations among the parameters $Y_1$, $Y_2$ and $Y_3$, see the discussion in the previous section. Thus, the dependence of LFV decays on these parameters is slightly more subtle. \Fig{fig:pheno:LFVT} shows results for calculated branching ratios of \mueg, \mueee and $(\mu-e)$ conversion in Ti, for several different choices of parameters, as function of $Y_3$. The horizontal lines show current experimental limits (full lines) and future expected sensitivities (dashed lines). Note that $Y_3$ has no lepton flavour indices and, thus, by itself can not generate a LFV diagram. Instead, for fixed values of masses and the parameters $\lambda_2$ and $\mu_1$, the prefactor ${\cal F}$ determining the size of the calculated neutrino masses \eq{eq:pheno:prefac} depends linearly on $Y_3$. Keeping neutrino masses constant while varying $Y_3$, thus leads to a corresponding change in (the inverse of) $Y_1 \times Y_2$. For this reason, for small values of $Y_3$ the branching ratios in \fig{fig:pheno:LFVT} decrease with increasing $Y_3$. For the largest values of $Y_3$, diagrams with additional $Y_3 v/m_{\Psi}$ insertions can become important and branching ratios start to rise again as a function of $Y_3$. Note that in all calculations in \fig{fig:pheno:LFVT}, we have chosen $Y_4$ and $Y_5$ small enough, such that their contribution to the LFV decays is negligible.

Both, $Y_1$ and $Y_2$, generate LFV decays. Whether diagrams proportional to $(Y_1)_e (Y_1)_{\mu}$ or to $(Y_2)_e (Y_2)_{\mu}$ give the more important contribution to \mueg depends on the (mostly) arbitrary choice of $(Y_1)_e$. In \fig{fig:pheno:LFVT} we plot results for three different choices of $(Y_1)_e$. For $(Y_1)_e = 10^{-2}$ there is a large range of $Y_3$, for which \mueg and \mueee remain constant. In this case, diagrams proportional to $(Y_1)_e (Y_1)_{\mu}$ dominate the partial width.

We also show in \fig{fig:pheno:LFVT} two different choices of the parameter $\mu_1$. To the left: $\mu_1 = 1$ TeV, to the right $\mu_1=1$ GeV. Smaller values of $\mu_1$ require again larger values of the Yukawa coupling $Y_2$, and thus lead to larger LFV decays. While for $\mu_1 = 1$ TeV nearly all points in the parameter space are allowed with current constraints, once $(Y_1)_e$ is smaller than roughly (few) $10^{-3}$, for $\mu_1 = 1$ GeV large parts of the parameter space are already ruled out. For $\mu_1 \simeq 10^{-2}$ GeV and masses below 2 TeV there remain already now no valid points in the parameter space which, at the same time, can obey upper limits from \mueg and explain neutrino masses, except in the small regions where different diagrams cancel each other exactly accidentally.

It is worth to mention that for the triplet model the branching ratio of \mueee is higher than the corresponding of \mueg. Naively one would expect the former to be two orders (an order of $\alpha_{EM}$) lower than the latter. However, \mueg occurs at one-loop level, while in this model there exists a tree-level diagram for \mueee mediated by a $Z^0$, due to the mixing between leptons and $\overline{\Psi^+}$, proportional to $(Y_1)_e (Y_1)_{\mu}$. Other tree-level contributions mediated by doubly charged scalars are also possible due to this mixing. These are proportional to $(Y_4)_{\mu} (Y_4)_e (Y_1)_e (Y_1)_e$, so the upper limit given by \mueg is still dominant.

The plots in \fig{fig:pheno:LFVT} also show the discovery potential of future \mueee and $(\mu-e)$ conversion experiments. In particular, an upper bound on $(\mu-e)$ conversion of the order $10^{-18}$ would require both, very small Yukawas (for example: $(Y_1)_e \lesssim 10^{-5}$) and a large value of $\mu_1 \gtrsim 1$ TeV at the same time. All other points in the parameter space of the triplet model (assuming they explain neutrino data) with masses below $2$ TeV, should lead to the discovery of $(\mu-e)$ conversion. This is an interesting constraint, since such small values of the Yukawa couplings would imply very long-lived particles at the LHC. We will come back to this discussion in the next section.

\begin{figure}[t!]
    \centering
    \includegraphics[width=0.48\linewidth]{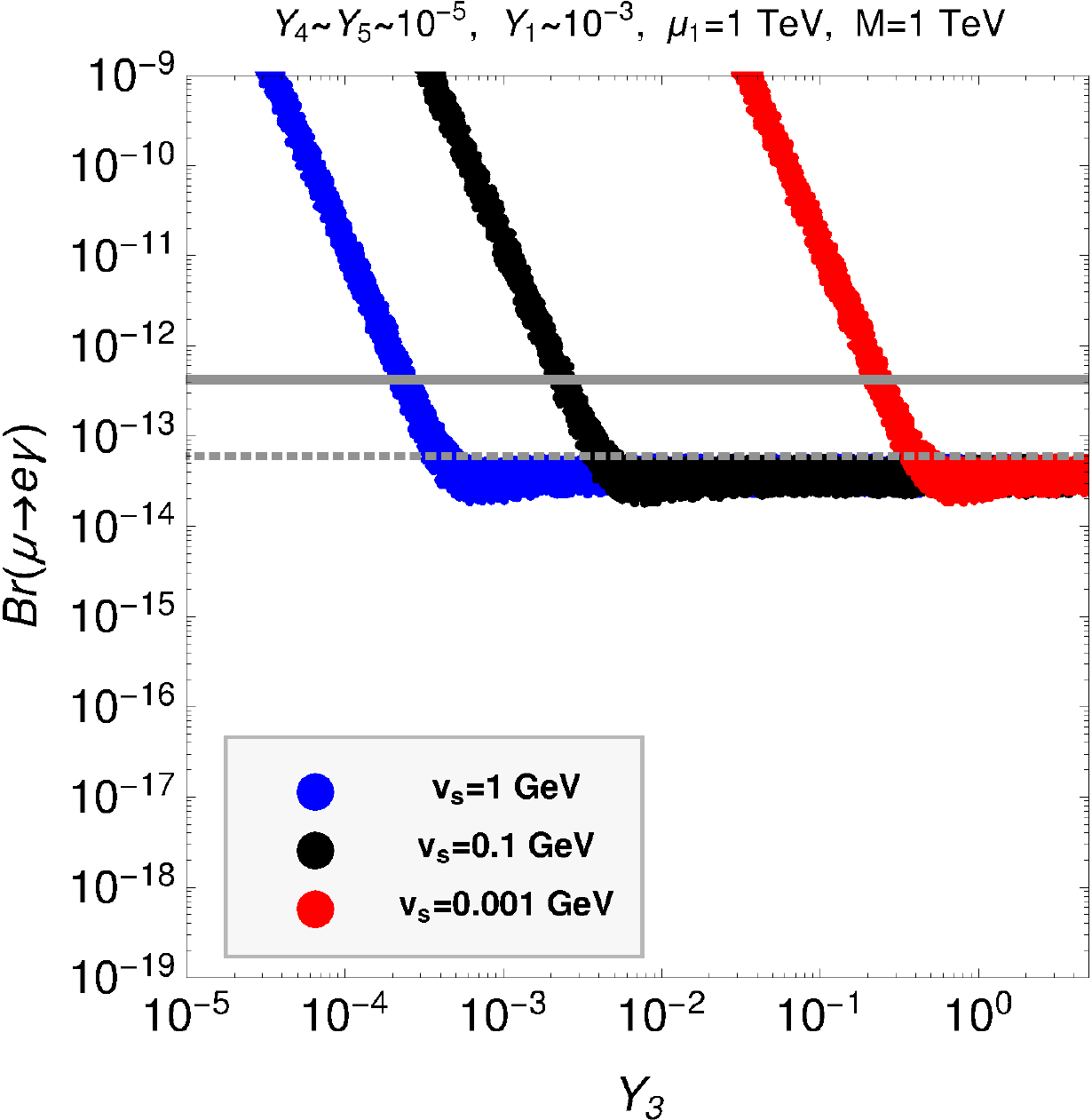}
    \hfill
    \includegraphics[width=0.48\linewidth]{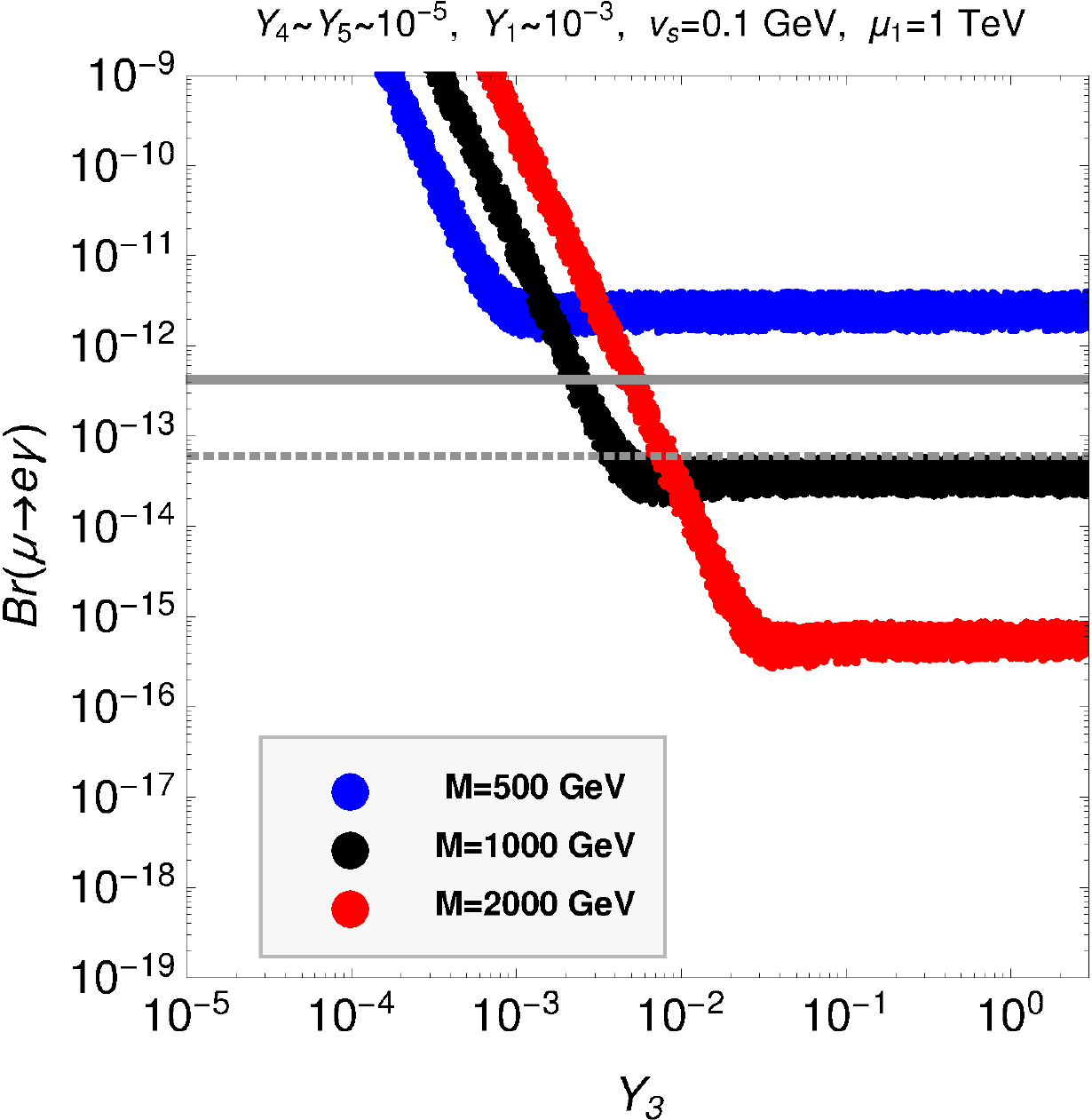}
    \\
    \includegraphics[width=0.48\linewidth]{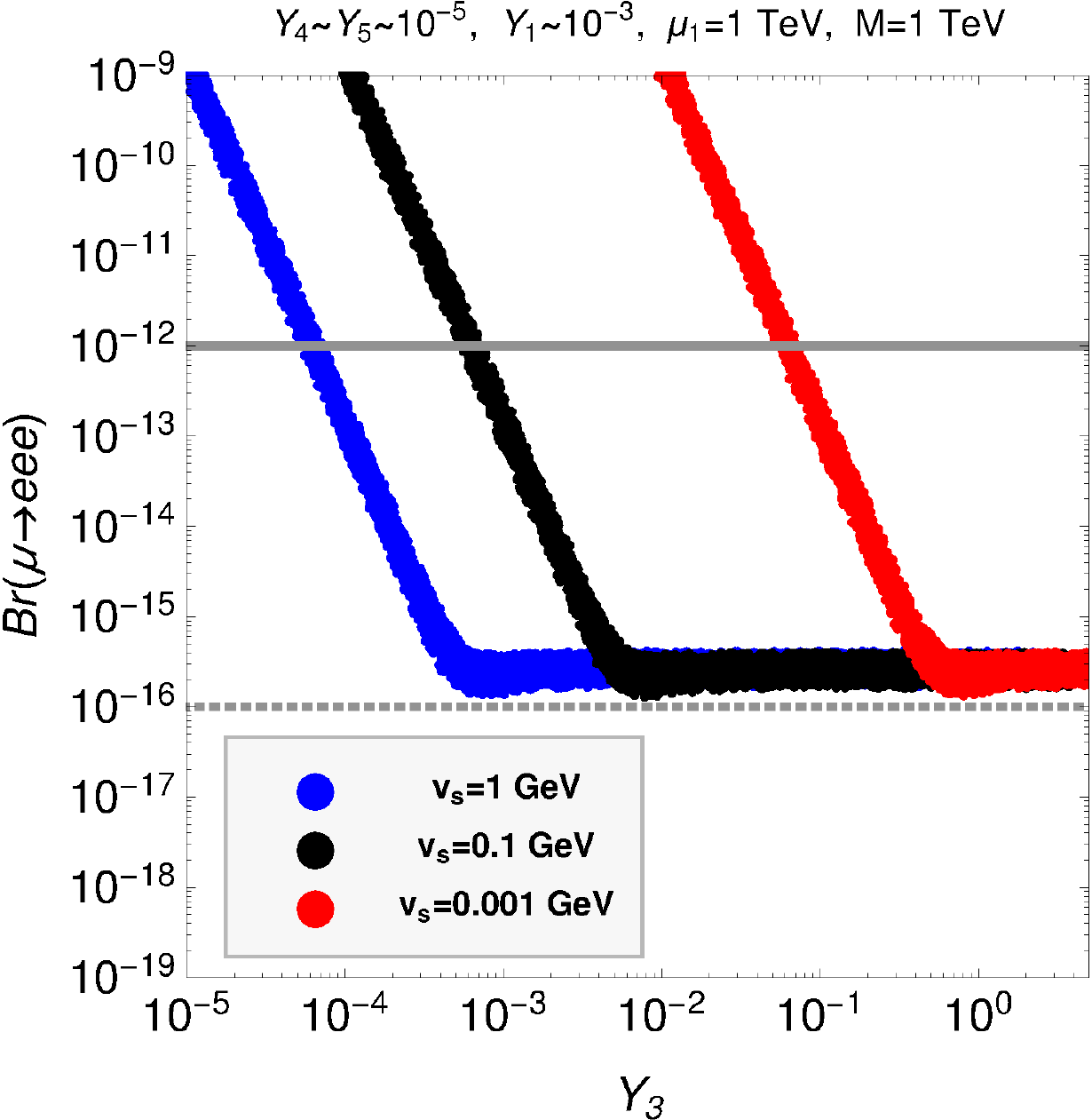}
    \hfill
    \includegraphics[width=0.48\linewidth]{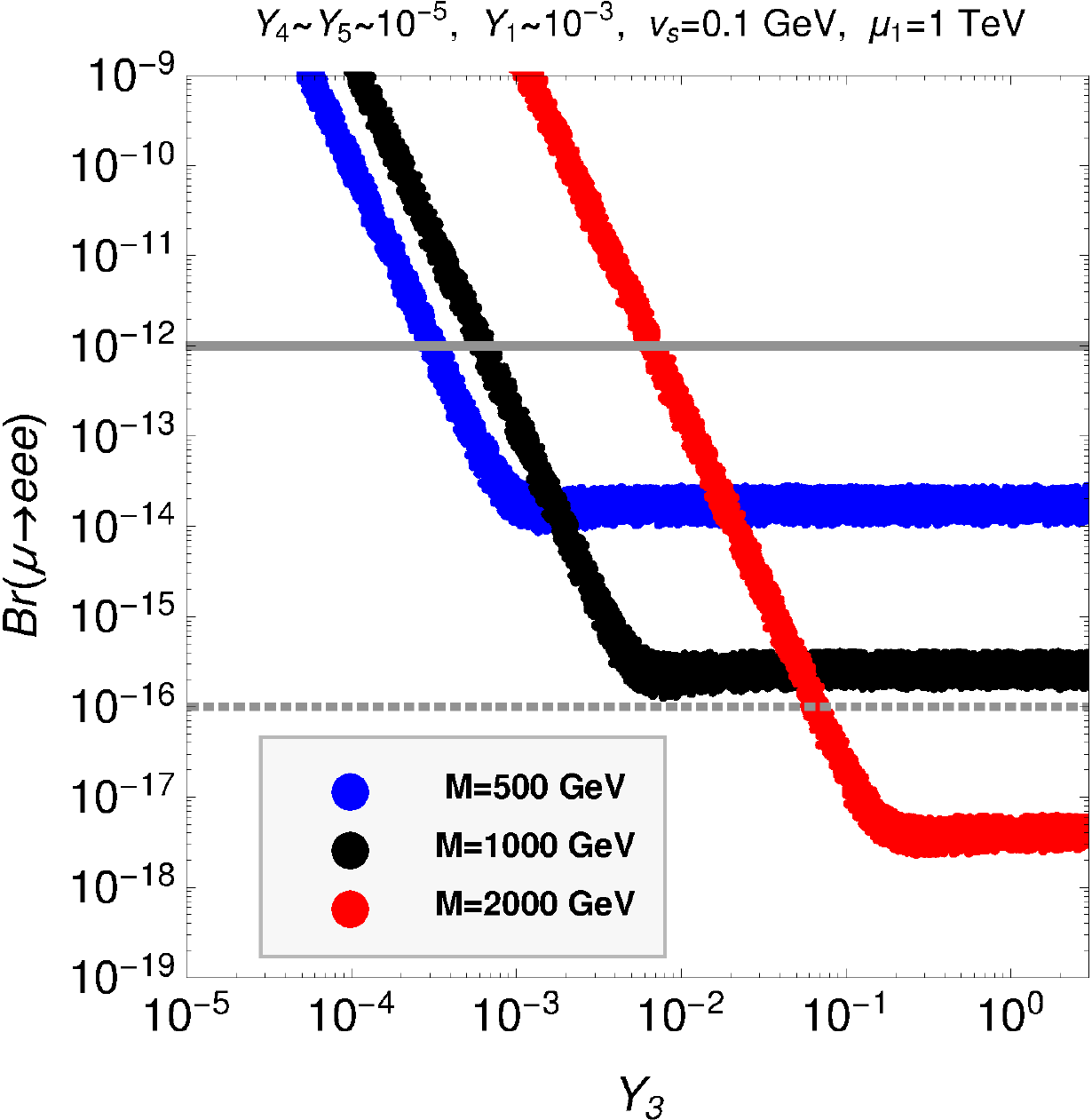}
\caption{(Continues in \fig{fig:pheno:LFVQ}).}
    \label{fig:pheno:LFVQ1}
\end{figure}

\begin{figure}[t!]
    \includegraphics[width=0.48\linewidth]{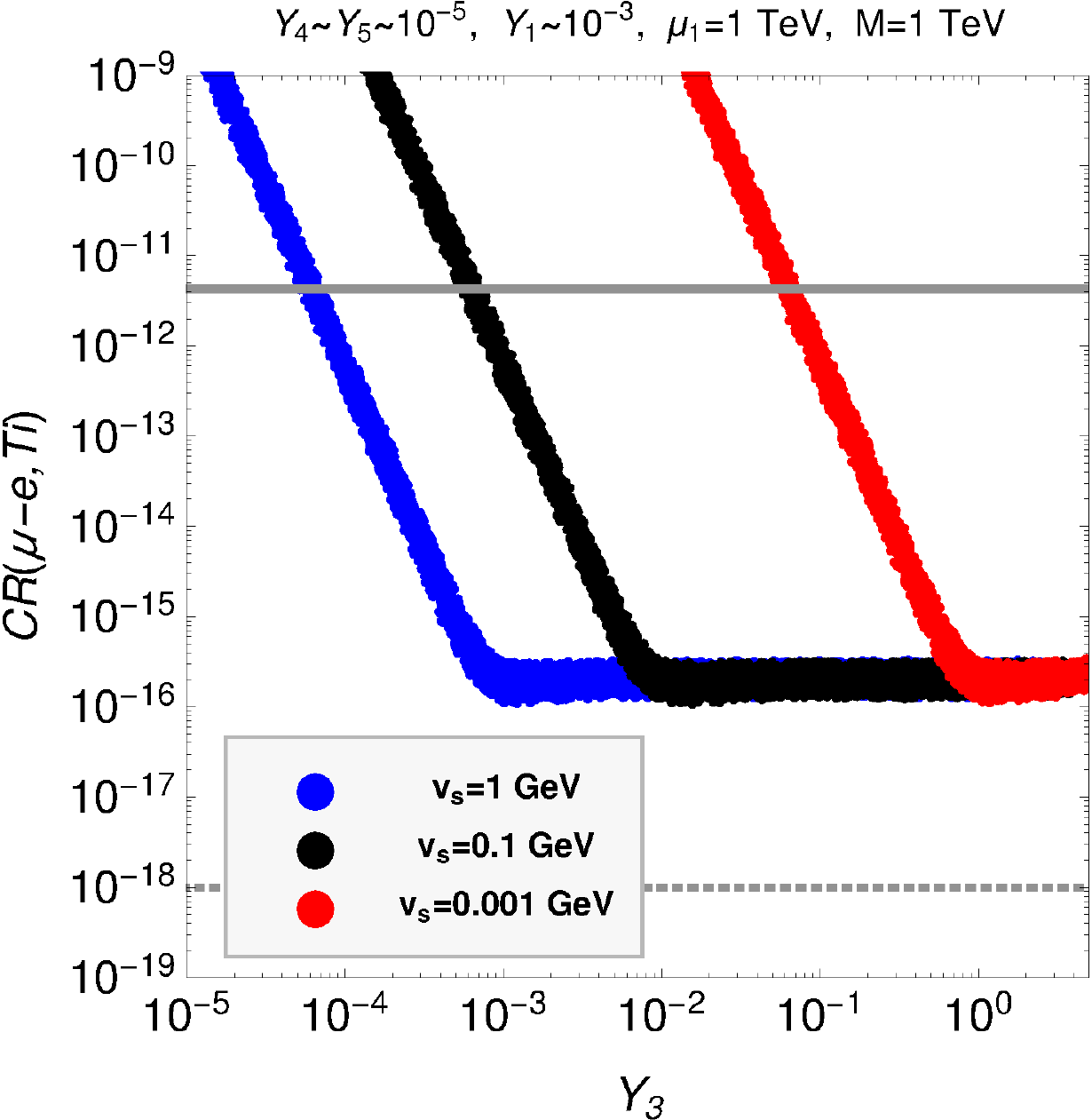}
    \hfill
    \includegraphics[width=0.48\linewidth]{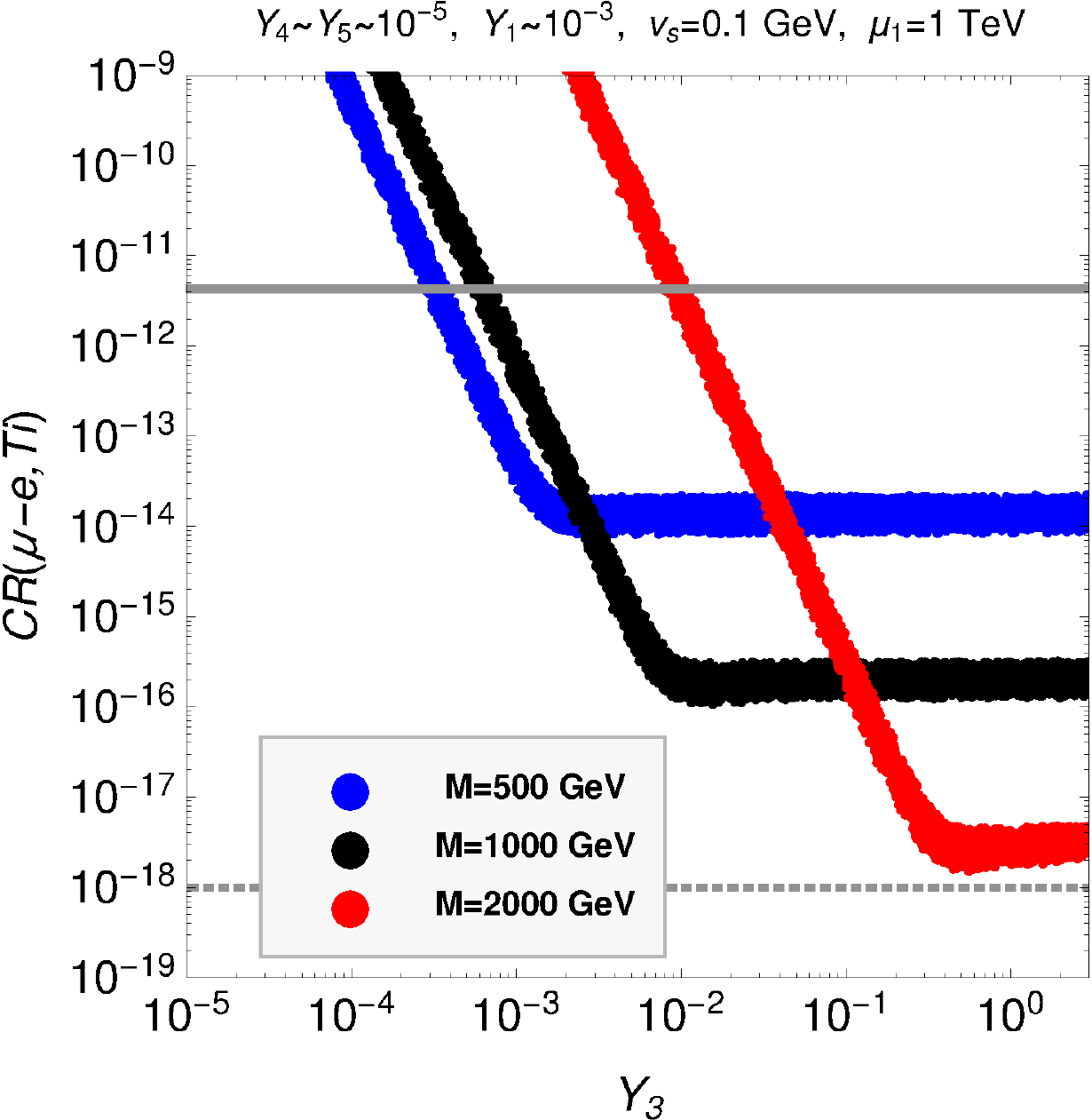}
    \caption{Lepton flavour violating decays calculated in the quadruplet model. Plots to the left show results for different choices of the quadruplet VEV $v_S$, while the ones to the right use different masses for the new scalars and fermions. Here, we assume that all new particles have similar masses of the order indicated in the figure panels. This figure is a continuation of \fig{fig:pheno:LFVQ1}}
    \label{fig:pheno:LFVQ}
\end{figure}

We now turn to a discussion of LFV in the quadruplet model. Similarly to the triplet model, we can divide parameters into two groups: $Y_1$-$Y_3$ depend on the neutrino mass fit, while $Y_4$-$Y_6$ are unconstrained parameters. Constraints on $Y_4$ and $Y_5$ from LFV are very similar to those found in the triplet model. The constraints on $Y_6$ are somewhat more stringent, $(Y_6)_{\mu e}(Y_6)_{ee} \lesssim 10^{-5}$, since there exists a tree-level diagram via doubly charged scalar exchange contributing to the decay \mueee.

Turning to $Y_1$-$Y_3$, \fig{fig:pheno:LFVQ} shows some sample calculations of LFV decays as function of $Y_3$ in the quadruplet model. The plots to the left show \mueg, \mueee and $(\mu-e)$ conversion in Ti, for several different choices of the quadruplet VEV $v_S$. Smaller values of $v_S$ need larger values of the Yukawa couplings $Y_2$ for constant neutrino masses. Thus, LFV decays are larger at the same values of $Y_3$ for smaller values of $v_S$.

The plots on the right of \fig{fig:pheno:LFVQ} show the same LFV decays, for a fixed value of $v_S=0.1$ GeV, but different values of the new scalar and fermion masses. As simplification in this plot we assume that all new scalars and fermions have roughly the same mass, $M$, as indicated in the plot panels. Larger values of masses lead to smaller LFV decay widths, as expected. As also is the case for the triplet model, future bounds from \mueee and $(\mu-e)$ conversion will test most of the relevant parameter space of the quadruplet model up to masses of order $2$ TeV.

In fact, even for masses as large as $2$ TeV, non-observation of $(\mu-e)$ conversion would put an interesting lower limit on the value of $v_S$, which we roughly estimate to be around $v_S=0.1$ GeV. Note that there is an upper limit on $v_S$ from the Standard Model $\rho$ parameter of the order of $v_S \lesssim 2.5$ GeV \cite{Babu:2009aq}.

In summary, the non-observation of LFV decays can be used to put upper bounds on the Yukawa couplings of our models. At the same time the observed neutrino masses require lower bounds on these Yukawa couplings and the combination of both constraints result in a very restricted range of allowed parameters. We have shown this explicitly only for our two example models, but the same should be true for any of the possible (genuine) $d=7$ one-loop models.

%%%%%%%%%%%%%%%%%%%%%%%%%%%%%%%%%%%%%%%%%%%%%%%%%%%%%%%%%%%%%%%
%%%%%%%%%%%%%%%%%%%%%%%%%%%%%%%%%%%%%%%%%%%%%%%%%%%%%%%%%%%%%%%
\section{Phenomenology at the LHC} \label{sec:pheno:lhc}

One of the main motivations to study neutrino mass models at dimension 7 and one-loop is that they can be dominant only if new particles below approximately $2$ TeV exist. This mass range can be covered by the LHC experiments in the near future, if some dedicated search for the LNV signals we discuss is carried out.

Lepton number violation has been searched for at the LHC so far using the final state of same-sign dileptons plus jets, $l^{\pm}l^{\pm}jj$. Many different LNV extensions of the Standard Model can lead to this signal \cite{Helo:2013dla, Helo:2013ika}. However, ATLAS and CMS searches usually concentrate on only two theoretical scenarios, left-right symmetry \cite{Keung:1983uu} and the Standard Model extended with ``sterile neutrinos''. Note that these two models lead to the same final state signal, but rather different kinematical regions are explored in the corresponding experimental searches. CMS has published first results from searches at run-II \cite{CMS:2017uoz} and run-I \cite{Khachatryan:2014dka}, both for $eejj$ and $\mu\mu jj$ final states, concentrating on the left-right symmetric model.\footnote{CMS has searched also for $\tau\tau jj$ \cite{Sirunyan:2017yrk}.  However, that search is not a test for LNV, since one $\tau$ is assumed to decay hadronically.} There is also a CMS search for sterile Majorana neutrinos, based on ${\cal L}=19.7/fb$ at $\sqrt{s}=8$ TeV \cite{Khachatryan:2016olu}. ATLAS published a search for $lljj$ based on $8$ TeV data, for both Standard Model with steriles and for the Left-Right model \cite{Aad:2015xaa}. However, only like-sign lepton data was analysed in \cite{Aad:2015xaa} and no update for $\sqrt{s}=13$ TeV has been published so far from ATLAS. No signal has been seen in any of these searches so far and thus lower (upper) limits on masses (mixing angles) have been derived.

Other final states that can test LNV have been discussed in the literature. For example, in the seesaw type-II \cite{Schechter:1980gr} the doubly charged component of the scalar triplet $\Delta$ can decay to either $\Delta^{++}\to l^+l^+$ or $\Delta^{++}\to W^+W^+$ final states. If the branching ratios to both of these final states are of similar order, LNV can be established experimentally \cite{Azuelos:2004mwa, Perez:2008ha, Melfo:2011nx, Babu:2016rcr}. No such search has been carried out by the  LHC experiments so far. Instead, ATLAS \cite{ATLAS:2014kca, ATLAS:2016pbt, ATLAS:2017iqw} and CMS \cite{CMS:2016cpz} have searched for invariant mass peaks in the same-sign dilepton distributions. Assuming that the branching ratios for $ee$ and/or $\mu\mu$ are large, i.e. ${\cal O}(1)$, lower limits on the mass of the $\Delta^{\pm\pm}$ up to $850$ GeV \cite{ATLAS:2017iqw}, depending on the flavour, have been derived. Note that, if only one of the two channels are observed, LNV can not be established at the LHC but the type of scalar multiplet could be still determined \cite{delAguila:2013yaa}.

Dimension 7 neutrino mass models can lead to new LNV final states at the LHC. The prototype tree-level model of this kind is the BNT model \cite{Babu:2009aq}. As pointed out in \cite{Babu:2009aq} the model predicts the final state $W^{\pm}W^{\pm}W^{\pm}+W^{\mp}l^{\mp}l^{\mp}$. The LHC phenomenology of the BNT model has been studied recently in detail in \cite{Ghosh:2017jbw}. As in the case of $W^{\pm}W^{\pm}+l^{\mp}l^{\mp}$ predicted by the seesaw type-II, no experimental search for this particular LNV final state has been published so far. We note in passing that we have also checked that the LNV searches for $lljj$ \cite{Aad:2015xaa,CMS:2017uoz} are currently not competitive for the models we consider in this chapter.

As already explained in \ch{ch:Dim7_1loop}, at tree-level the BNT model is unique, i.e. the only $d=7$ genuine model \cite{Bonnet:2009ej, Cepedello:2017eqf}, while the possibilities explode when considering one-loop realisations. These $d=7$ one-loop models, while necessarily richer in their particle content than simple $d=5$ (or $d=7$) tree-level neutrino mass models, offer a variety of interesting LNV signals at the LHC, so far not discussed in the literature. As we show below, depending on the unknown mass spectrum, several different multi-lepton final states with gauge bosons up to $W^{\pm}W^{\pm}l^{\mp}l^{\mp}+l^{\pm}l^{\pm}l^{\mp}l^{\mp}$ can occur. Note that for such high multiplicity final states one can expect very low Standard Model backgrounds.

%%%%%%%%%%%%%%%%%%%%%%%%%%%%%%%%%%%%%%%%%%%%%%%%%%%%%%%%%%%%%%%
\subsection{Constraints from LHC searches} \label{subsec:pheno:lhc}

We have calculated the production cross-sections for the different scalars and fermions of our example models using MadGraph \cite{Alwall:2014hca}. Pair production is usually calculated via s-channel photon and $Z^0$ exchange, while associated production, such as $\eta^{--}\eta^{+++}$, proceeds via $W^+$ diagrams. However, as pointed out in \cite{Ghosh:2017jbw}, for large masses the pair production cross-section of charged particles via photon-photon fusion can give the dominant contribution to the cross-section, despite the small photon density in the proton. In our calculation we use the NNPDF23$\_$nlo$\_$as$\_$0119 parton distribution function, which contains NLO corrections, necessary for inclusion of the photon-photon fusion contributions. We have checked numerically and find that at the largest masses cross-sections can be enhanced up to one order of magnitude for multiply charged particles. For this reason we concentrate on pair production of particles in the following. Note, however, that for lower masses (up to roughly $1$ TeV), associated production is large enough to produce additional signals, not discussed here.

\begin{figure}
    \centering
    \includegraphics[width=0.48\textwidth]{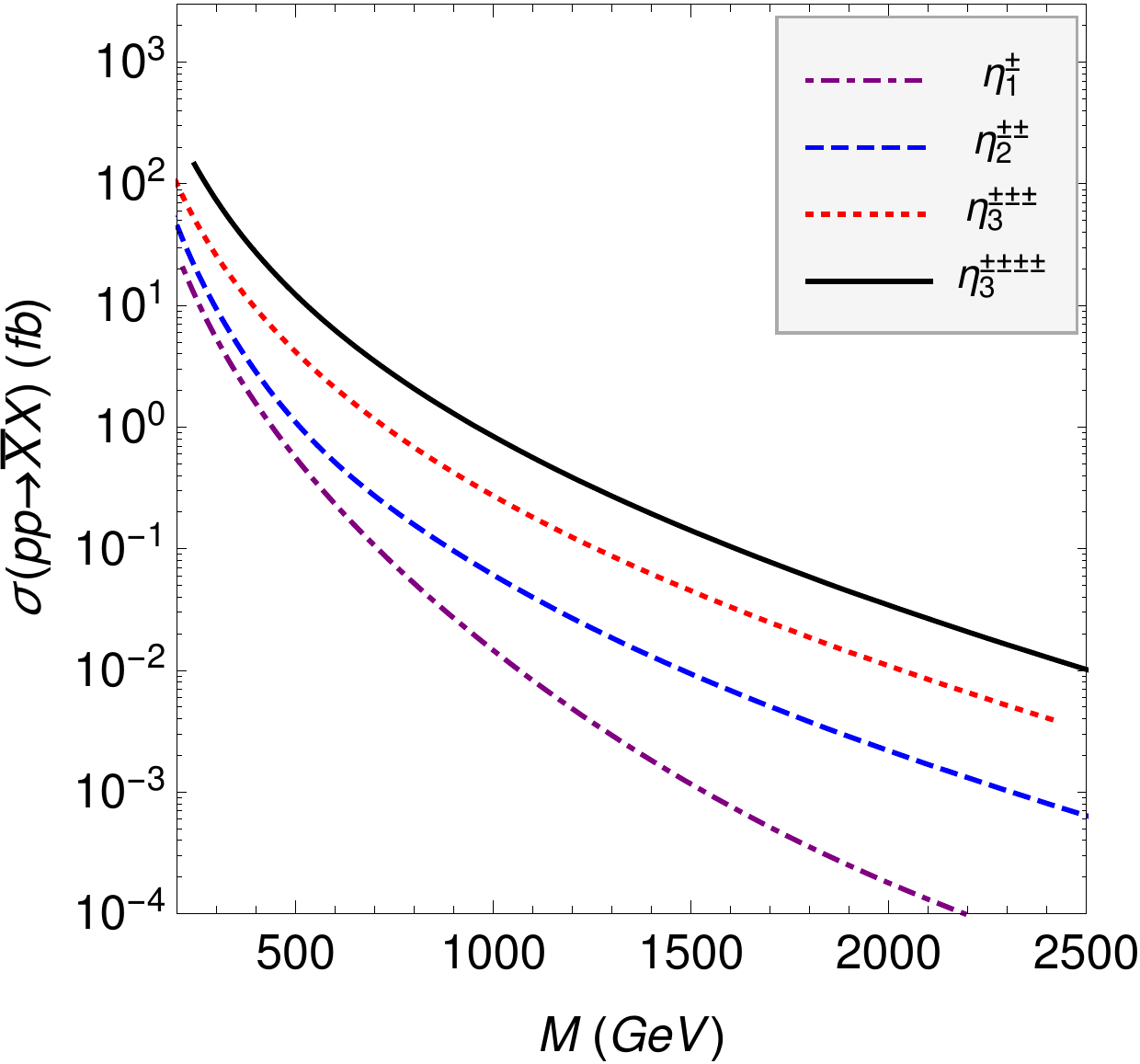}
    \hfill
    \includegraphics[width=0.48\textwidth]{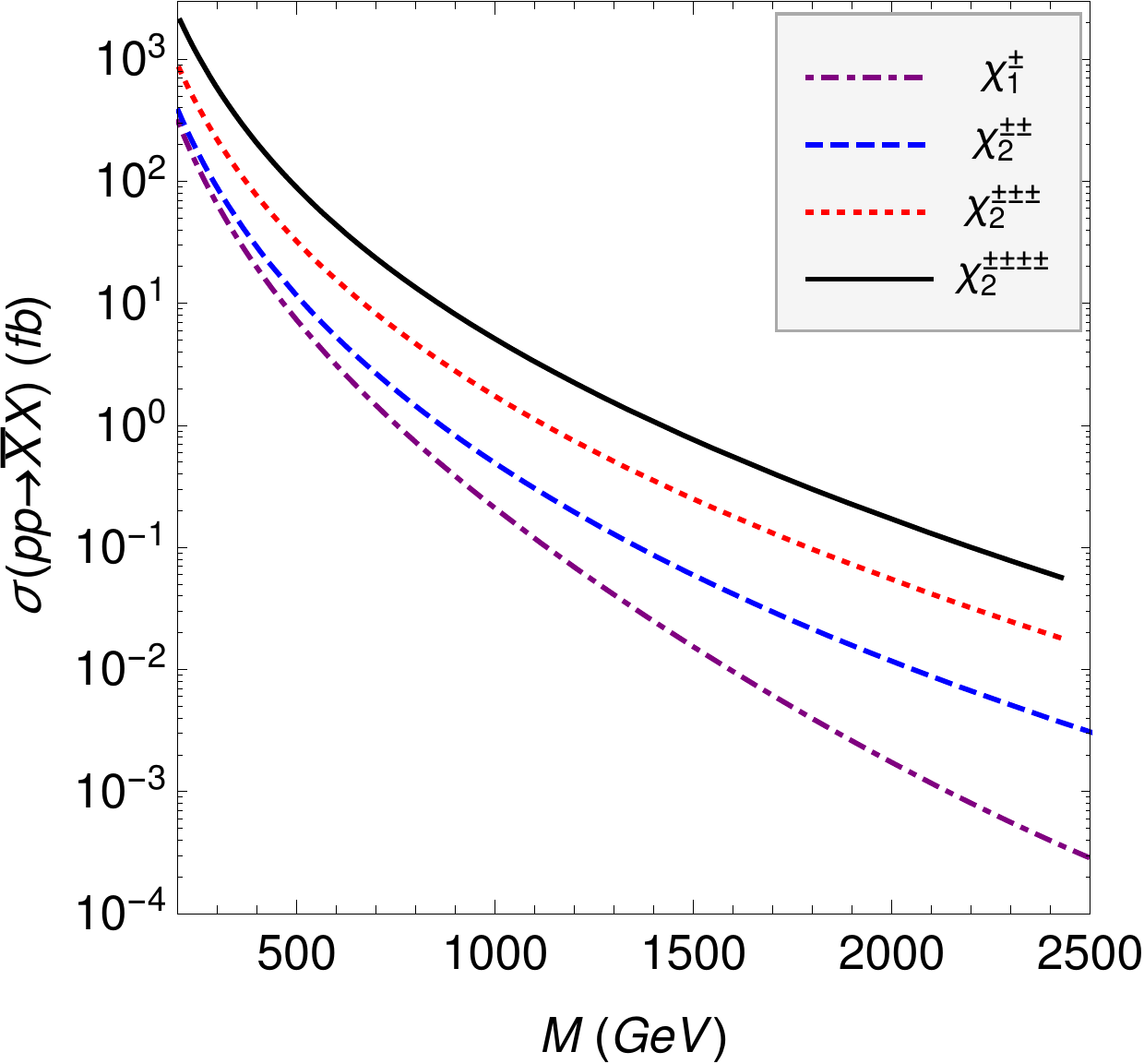}
    \caption{Pair production cross-sections for the different scalars (left) and fermions (right) of the two example models. For discussion see text.}
    \label{fig:pheno:Prod}
\end{figure}

Results for the cross-sections are shown in \fig{fig:pheno:Prod} for $\sqrt{s}=13$ TeV. To the left we show results for scalars, to the right the cross-sections for fermions. The scalar cross-sections (to the left) were calculated for the scalars of the triplet model. The fermion cross-section (to the right) corresponds to the fermions of the quadruplet model. The underlying Lagrangian parameters were chosen such that the corresponding gauge states (index shown in the figure) are the lightest mass eigenstate of the corresponding charge. Cross-sections do also depend, to some extent, on the hypercharge of the particle. However, since photon-fusion dominates the cross-section at large values of the masses, all mass eigenstates with the same electric charge have similar cross-sections. We therefore do not repeat those plots for all the particles in our models.

For the quadruply charged particles of the models cross-sections larger than $10^{-2}$ fb are obtained, even for masses up to $2.5$ TeV. Note that at the largest value of masses pair production cross-section ratios for different charged particles simply scale as the ratio of the charges to the 4th power. We will come back to this in the discussion of the LNV signals in the next subsection.

\begin{figure}[t!]
    \centering
    \includegraphics[width=0.65\textwidth]{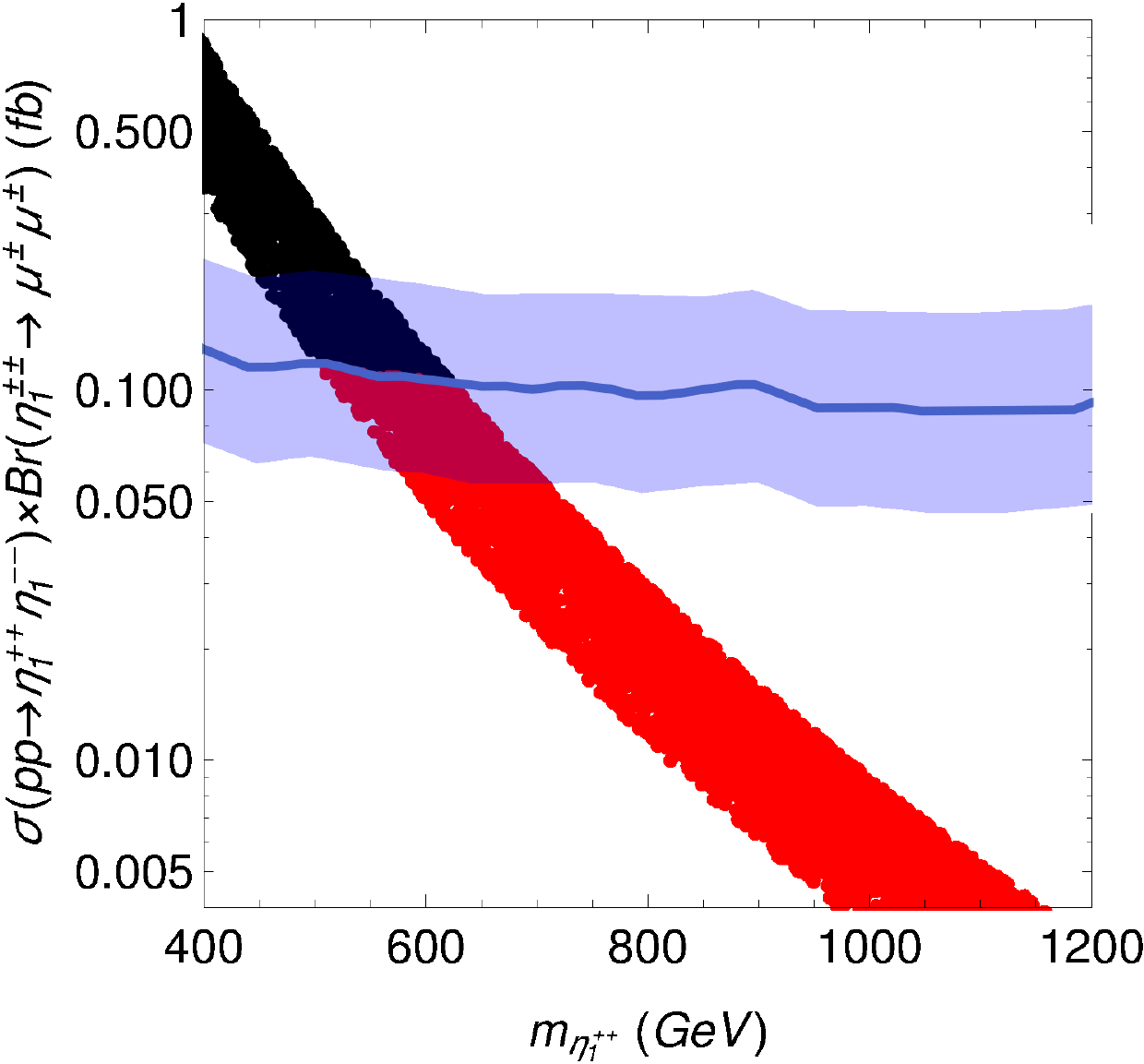}
    \caption{Constraints on doubly charged scalars, using the recent search by ATLAS \cite{ATLAS:2017iqw}. The blue line is the limit quoted in \cite{ATLAS:2017iqw}, the light blue region the 95\% c.l. region. Points are our calculation, scanning over the allowed ranges of neutrino angles. Red points are allowed by this search.}
    \label{fig:pheno:Limpp}
\end{figure}

A number of different LHC searches can be used to set limits on the various particles of our example models. The simplest search, and currently the most stringent LHC limit for our models, comes from a recent ATLAS search for doubly charged particles decaying to either $e^{\pm}e^{\pm}$, $e^{\pm}\mu^{\pm}$ or $\mu^{\pm}\mu^{\pm}$ final states \cite{ATLAS:2017iqw}. Results of our calculation, compared to the experimental limit are shown in \fig{fig:pheno:Limpp} for the $\mu^{\pm}\mu^{\pm}$ final state.

The two-body decay with of the doubly charged scalar $\eta_1^{++}$ is approximately given by,
\begin{equation}\label{eq:pheno:GamPP}
    \Gamma(\eta_1^{++}\to l^{+}_{\alpha}l^{+}_{\beta}) \simeq
    \frac{1}{8\pi} \left( \frac{v}{m_{\Psi}} \right)^2
    \left[ (Y_4)_{\alpha}(Y_1)_{\beta}+(Y_4)_{\beta}(Y_1)_{\alpha} \right]^2 m_{\eta^{++}_1} \, .
\end{equation}
Since the Yukawa coupling $Y_4$ does not enter the neutrino mass calculation, the exact value and flavour composition of this decay can not be predicted. However, $Y_1$ enters our neutrino mass fit. The observed large neutrino angles require that all entries in the vector $Y_1$ are different from zero and of similar order. Typically, from the fit we find numerically ratios in the range $(Y_1)_e:(Y_1)_{\mu}:(Y_1)_{\tau} \sim ([1/4,1/2]:[1,3]:1)$, but the exact ratios depend on the allowed range of neutrino angles. Scanning over the allowed neutrino parameters then leads to a variation of the branching ratios of the $\eta_1^{++}$ into the different lepton generations. This explains the spread of the numerically calculated points in \fig{fig:pheno:Limpp}. Combined with the experimental limit from ATLAS, lower mass limits in the range of ($500-650$) GeV result. Note that in this plot, we allow all three neutrino angles to float within the 3 $\sigma$ regions of the global fit \cite{Forero:2014bxa}.

\begin{figure}[t!]
    \centering
    \includegraphics[width=0.48\textwidth]{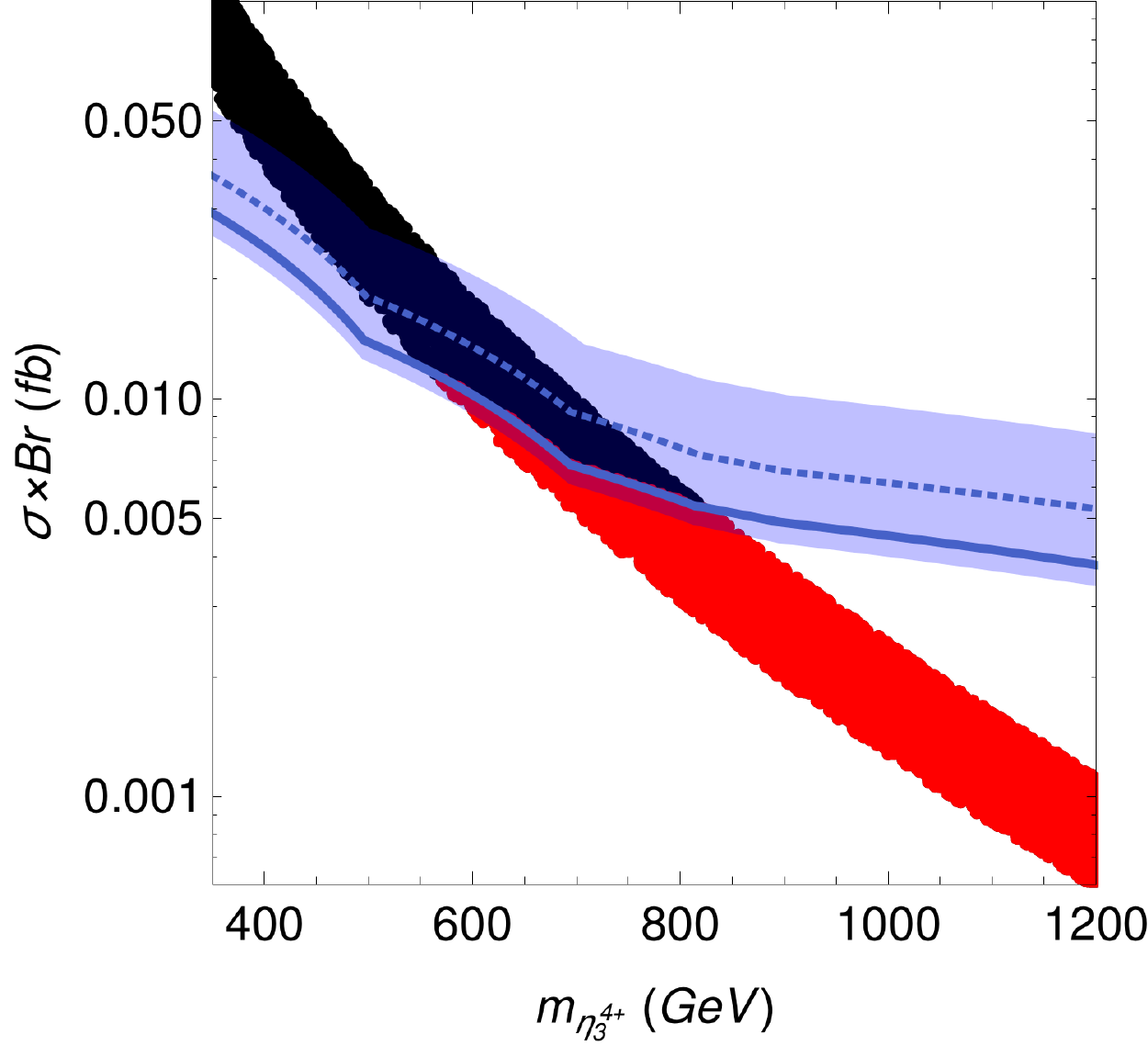}
    \hfill
    \includegraphics[width=0.48\textwidth]{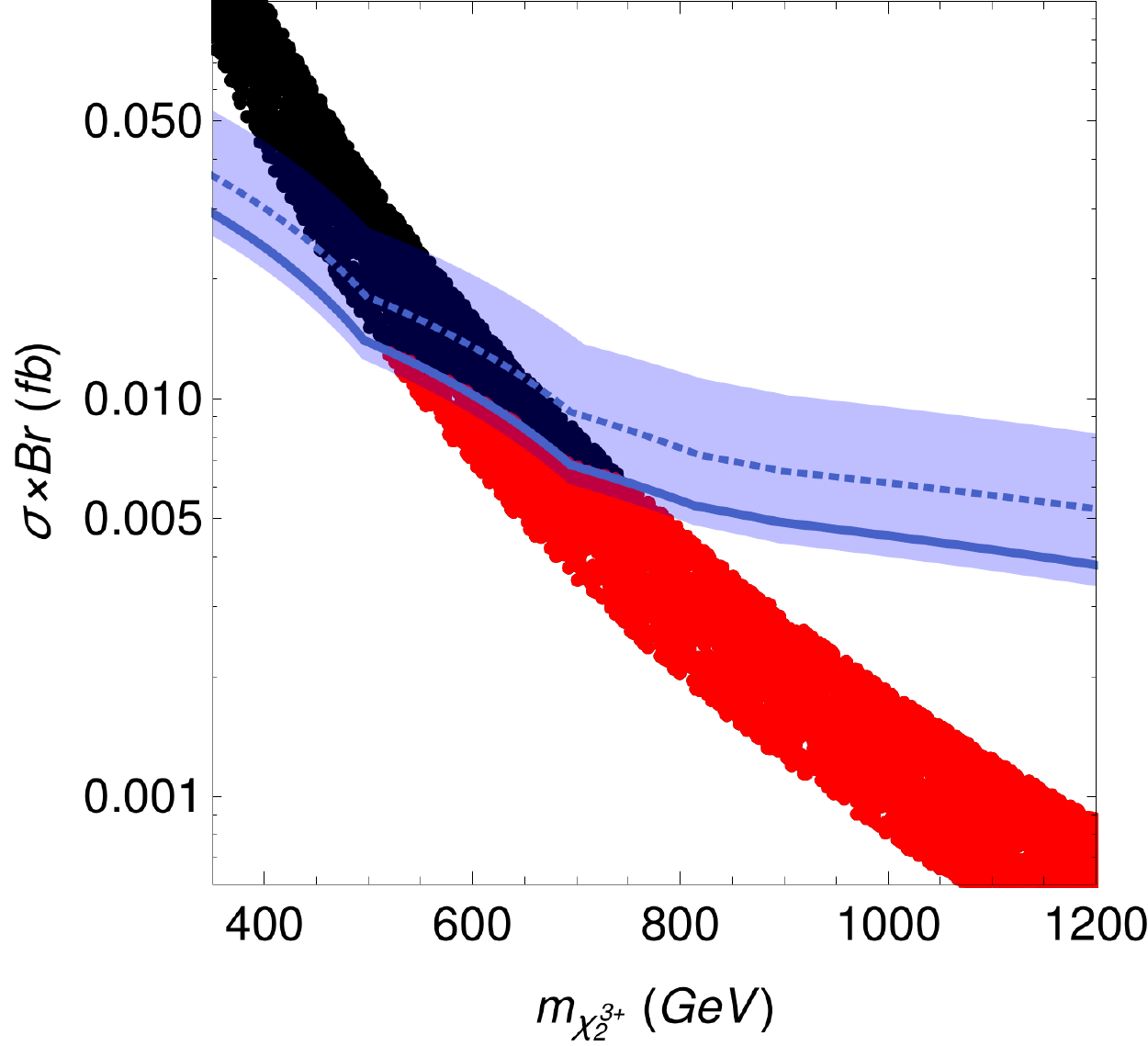}
  \\
    \includegraphics[width=0.48\textwidth]{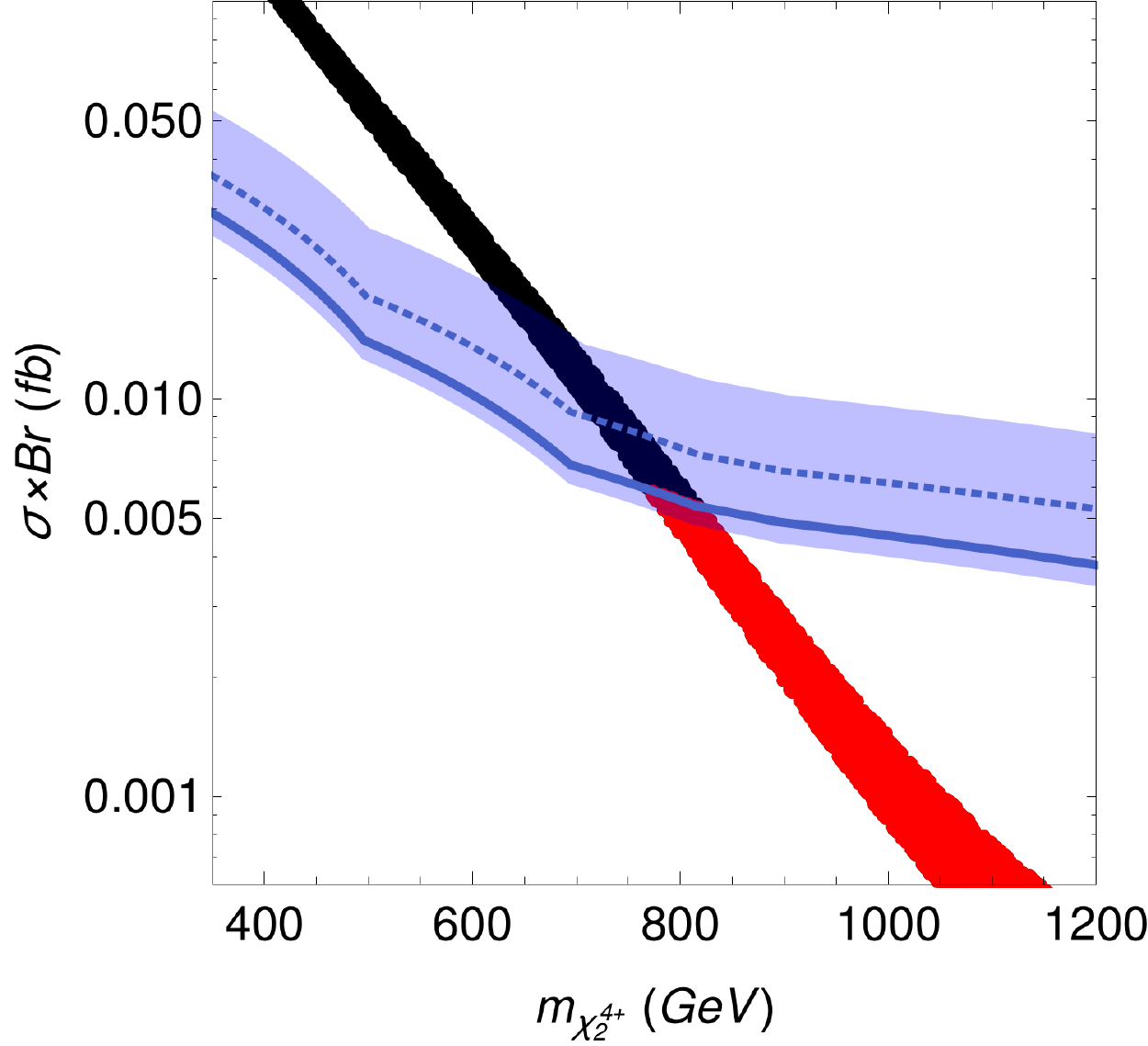}
    \caption{Constraints on charged scalars and fermions using the multi-lepton search \cite{Sirunyan:2017qkz}. Points are our numerical results, the bands are  experimental limits, see also \fig{fig:pheno:Limpp}.}
    \label{fig:pheno:MultiLep}
\end{figure}

The CMS collaboration has recently published a search based on multi-lepton final states \cite{Sirunyan:2017qkz}. The original motivation for this search is the expectation that the fermions of the seesaw type-III lead to final states containing multiple charged leptons and missing momentum. For example, $\Sigma^{\pm}\Sigma^0 \to W^{\pm}\nu W^{\pm}l^{\mp}$ from the associated production of the fermionic triplet $\Sigma=(\Sigma^{+},\Sigma^0,\Sigma^{-})$. The analysis \cite{Sirunyan:2017qkz} requires than at least three charged leptons plus missing energy and takes into account both, electrons and muons.

In our models, these final states can occur in various decay chains. Consider for example $\chi_2^{4+}$. Once produced, it can decay into a $\chi_1^{3+}+W^+$, which further decays to a doubly charged scalar and $l^+$. The doubly charged scalar decays to either leptons or $W$'s. The missing energy is then produced in the leptonic decays of the $W$'s. Here, all intermediate particles can be either on-shell or off-shell, depending on the unknown mass hierarchies. Constraints can then be derived from the results of \cite{Sirunyan:2017qkz}, scanning over the allowed ranges of the branching ratios, which lead to a least three charged leptons plus at least one $W$ in the final state. 

In \fig{fig:pheno:MultiLep} we show results of this procedure for the examples of $\eta_3^{4+}$, $\chi_2^{3+}$ and $\chi_2^{4+}$. The lower limits, derived from this exercise, have a rather large uncertainty, due to the unknown branching ratios. For example, the lower mass limit for $\eta_3^{4+}$ is in the range of ($550-850$) GeV. Note that $\eta_3^{4+}$ could decay, in principle to four charged leptons with a branching ratio close to 100\%. The final state from pair production of $\eta_3^{4+}$ would then contain eight charged leptons and missing momentum would appear only from the decays of the $\tau$'s. In this case, our simple-minded recasting of the multi-lepton search \cite{Sirunyan:2017qkz} ceases to be valid and the lower limit on the mass of $\eta_3^{4+}$, mentioned above does not apply. As \fig{fig:pheno:MultiLep} shows, the lower limit on the mass of $\chi_2^{4+}$ is more stringent than the one for $\eta_3^{4+}$. This simply reflects the larger production cross-sections for fermions, compare to \fig{fig:pheno:Prod}.

%%%%%%%%%%%%%%%%%%%%%%%%%%%%%%%%%%%%%%%%%%%%%%%%%%%%%%%%%%%%%%%
\subsection{New LNV searches} \label{subsec:pheno:lnv}

We now turn to a discussion of possible LNV signals at the LHC. \Tab{tab:pheno:Tablelnv} shows examples of different LNV final states from pair production of scalars or fermions in the two models under consideration. This list is not complete since (i) associated production of particles is not considered; (ii) the table gives only ``symmetric'' LNV states, see below, and (iii) we do not give LNV final states with neutrinos, since such states do not allow to establish LNV experimentally. 

\begin{table}[t!]
    \begin{center}
        \small\addtolength{\tabcolsep}{-5pt}
        \begin{tabular}{|c|c|c|c|c|c||c|c|c|c|}
            \hline
            Multiplicity  &LNV Signal & Particles& Model & Mass range  \\
            \hline
            4 (6) & $l^{\pm} l^{\pm}  +  W^{\mp} W^{\mp}$ & $S^{{\pm} {\pm}}$, $\phi_{1}^{{\pm}{\pm}}$, $\phi_{2}^{\pm\pm}$& Q & $m < 1.4$ TeV  \\
            \hline
            6 (8) & $l^{\pm} l^{\pm} l^{\pm}  +   W^{\mp} W^{\mp} l^{\mp}$ & $\chi_{2}^{3+}$& Q & $m < 2.6$  TeV \\
            \hline
             6 (10) & $l^{\pm} l^{\pm} W^{\pm} +  W^{\mp} W^{\mp} W^{\mp}$ & $S^{3+}$, $\phi_{2}^{3+}$& Q &   $m < 2.0$  TeV\\
            \hline
             8 (10) & $l^{\pm} l^{\pm} l^{\pm} l^{\pm}+  l^{\mp} l^{\mp} W^{\mp} W^{\mp}$ & $\eta_{3}^{4+}$& T & $m < 2.5$ TeV \\
            \hline
            8 (12) & $l^{\pm} W^{\pm} W^{\pm} W^{\pm}+   l^{\mp} l^{\mp} l^{\mp} W^{\mp}$ & $\chi_2^{4+}$ & Q & $m < 3.2$ TeV\\
            \hline
            8 (14) & $l^{\pm} l^{\pm} W^{\pm} W^{\pm}+   W^{\mp} W^{\mp} W^{\mp} W^{\mp}$ & -- & -- & -- \\
            \hline
            \hline
        \end{tabular}
    \end{center}
    \caption{List of \textit{symmetric} LNV final states in $d=7$ models. The first column counts the number of final state particles, the second column gives the LNV signal. The multiplicity is given twice, the value without the bracket gives the number counting $W$'s, while the number in brackets counts each $W$ as two jets. This is done, since only the hadronic decays of the $W$ can be used for establishing LNV, see text. Here, we have separated the total final state into the two sets of particles, coming from the pair produced states listed in column 3. The invariant masses of the quoted subsystems should peak at the mass of the particle quoted in column 3. Column four gives the model in which this signal could be found. The last column gives our simple estimate for the mass range, which can be probed at the LHC with ${\cal L}\simeq 300$/fb. For a discussion see text.}
    \label{tab:pheno:Tablelnv}
\end{table}

The table gives in column 1 the multiplicity of the final state and in column 2 the LNV signal. The multiplicity in column 1 is given twice, once counting directly the number of leptons and $W$'s (value without bracket) and second, counting each $W$ as two jets in the final state (number in bracket). We stress again that the leptonic decays of the $W$ can be used in searches to derive lower mass limits on exotic particles, but can not be used to establish LNV experimentally. This is because the neutrinos from the leptonic $W$ decays show up only as missing energy, i.e. their lepton number can never be tagged. In the following, we will discuss final states as leptons plus $W$, but one should always bear in mind that we assume the $W$ to decay hadronically. 

In the 2nd column, the two possible final states from the decay of the particle given in column 3 are given separately. The invariant masses of both separate subsystems in column 2, should therefore give peaks in the mass of the particle in column 3.

\begin{figure}
    \centering
    \includegraphics[width=0.48\textwidth]{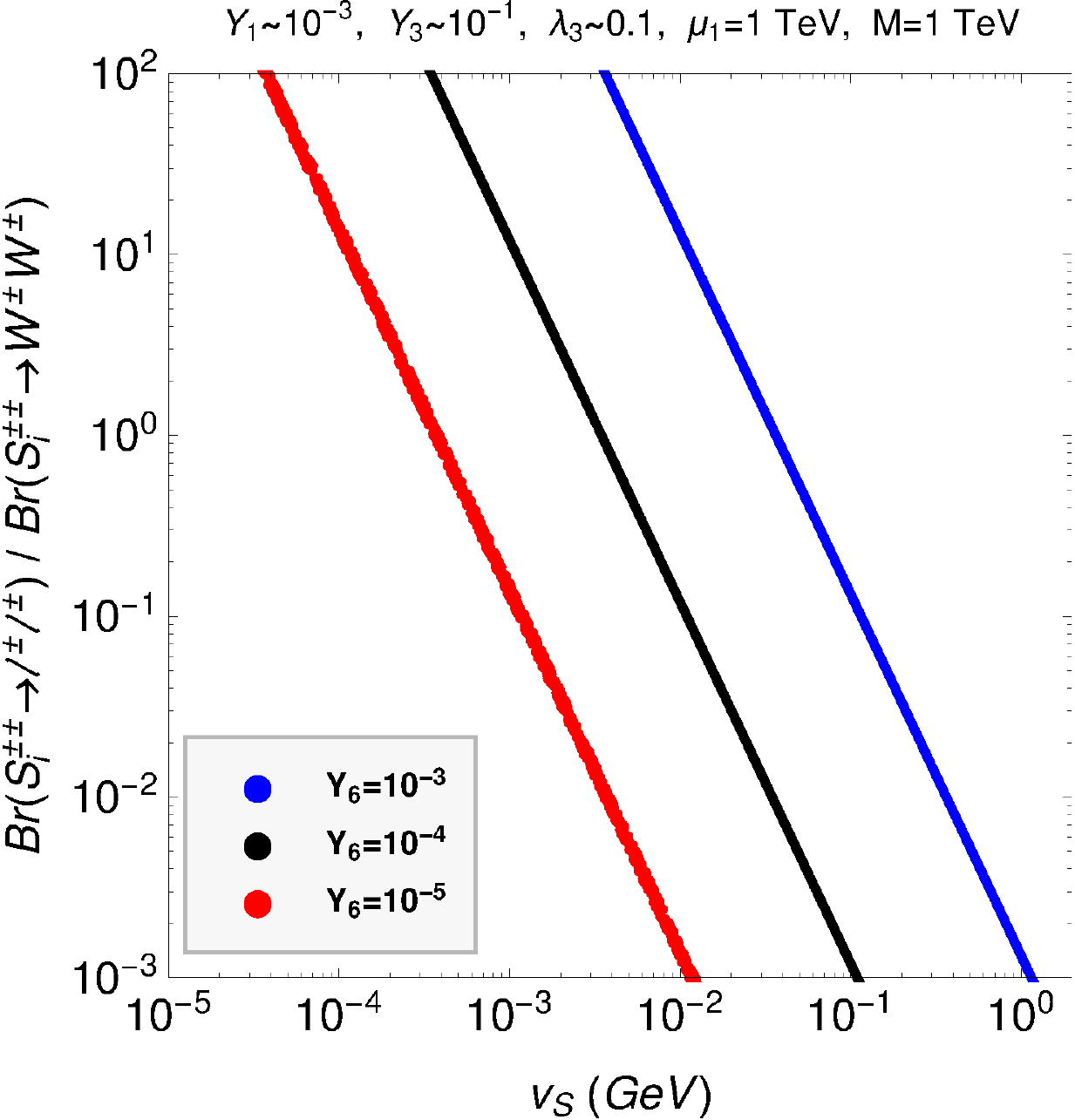}
    \hfill
    \includegraphics[width=0.48\textwidth]{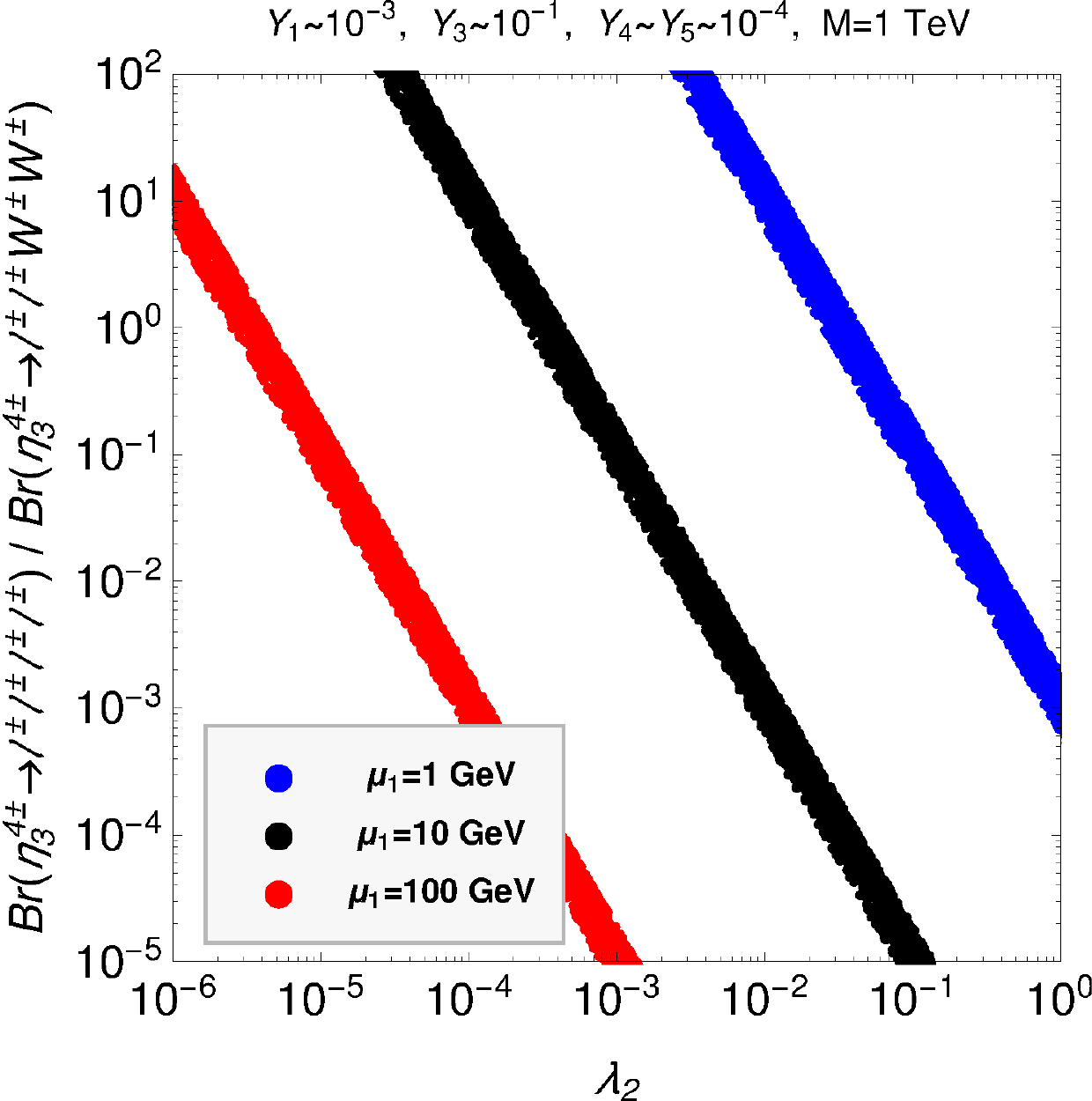}
    \caption{To the left: Ratio of branching ratios of the doubly charged scalar, $S_1^{++}$ decaying to $l^{\pm}l^{\pm}$ divided by $W^{\pm}W^{\pm}$ as a function of $v_S$ for some fixed choice of the other model parameters and three different values of $Y_6$. This plot assumes that the lightest doubly charged scalar $S_1^{++}$ is mostly the gauge state $S^{++}$. Results for the other cases are qualitatively very similar and thus not repeated. To the right: Ratio of branching ratios for $\eta_3^{4+}$ decays. As in the case of $S_1^{++}$, LNV will be observable only if this ratio is of order ${\cal O}(1)$.}
    \label{fig:pheno:LNV}
\end{figure}

Particles in column 3 are quoted as gauge eigenstates. However, scalars in our models are, in general, admixtures of different gauge eigenstates. Consider, for example, the simplest final state $l^{\pm}l^{\pm} +W^{\mp}W^{\mp}$. $\phi_1^{\pm\pm}$ can decay to $l^{\pm}l^{\pm}$, via the coupling $Y_6$, while $S^{\pm\pm}$ can decay to $W^{\pm}W^{\pm}$ via the induced VEV $v_S$ (or, equivalently proportional to $\lambda_2$). The doubly charged scalars mix via the entries in the mass matrices proportional to $\mu_1$, $\lambda_3$ (and $\lambda_4$), see \eq{eq:pheno:pot_4plet}. Whether the lightest doubly charged mass eigenstate is mostly $\phi_1$, $\phi_2$ or $S$ depends on the choice of parameters, but the results are qualitatively very similar in all cases. We therefore show in \fig{fig:pheno:LNV} only the results for the case where $S_1^{++}$ is mostly $S$. 

\Fig{fig:pheno:LNV} (left) shows the ratio of branching ratios of the doubly charged scalar, $S_1^{++}$ decaying to $l^{\pm}l^{\pm}$ divided by the decay to $W^{\pm}W^{\pm}$ as a function of $v_S$ for some fixed choice of the other model parameters and three different values of $Y_6$. Observation of LNV is only possible, if $\Gamma(S_i^{\pm\pm} \to l^{\pm}l^{\pm})$ is of similar order than $\Gamma(S_i^{\pm\pm} \to W^{\pm}W^{\pm})$, since both final states are needed to establish that LNV is indeed taking place. One can see from the figure that equality of partial widths is possible for different choices of parameters. However, since the decay to two charged leptons is proportional to (the square of) a Yukawa coupling that is not fixed by our neutrino mass fit, the relative ratio of branching ratios can not be predicted from current data.

Similarly, also for all other decays to LNV final states, the two competing final states have to have similar branching ratios. \Fig{fig:pheno:LNV} to the right show results for the decay of $\eta_3^{4+}$ of the triplet model. Depending on the parameter $\mu_1$ equality of branching ratio can occur in a large range of values of the parameter $\lambda_2$. Note that the rate of LNV final states is not suppressed by the smallness of neutrino masses. Neutrino masses require the product of ${\cal F}\times Y_1Y_2$ to be small, see \eq{eq:pheno:mnu2}. For a fixed neutrino mass, smaller values of $\mu_1\lambda_2$ require larger Yukawa couplings $Y_1Y_2Y_3$. Depending on the ratio between $\mu_1\lambda_2$ and $Y_1Y_2Y_3$, either the final state $4l$ or the final state $2l+2W$ can dominate. Whether LNV rates are observable, therefore, does not depend so much on absolute values of some (supposedly small) parameters, but on certain ratios of these parameters.

\Tab{tab:pheno:Tablelnv} is ordered with respect to increasing multiplicity of the final state. Note that, as discussed in the last subsection, cross-sections at the LHC increase with electric charge and decrease (strongly) with increasing mass. Which of the possible signals has the largest rate, can not be predicted because of the unknown mass spectrum. However, if the different members of the scalar (or fermion) multiplets have similar masses, final states with larger multiplicities have actually larger rates at the LHC. Since large multiplicity final states also have lower backgrounds. Searches for such states should give stronger bounds.

The last column in \tab{tab:pheno:Tablelnv} gives our estimate for the reach of the LHC. The numbers for the mass reach quoted in that column are simply based on the cross-section calculation, discussed in the last subsection. We assume here that in particular for the high multiplicity final states Standard Model backgrounds are very low (order of one event or less). Then we simply take the cross-section for which 3 events for a luminosity of 300 $fb^{-1}$ are produced as the approximate limit, that maybe achieved in a dedicated search. In fact, with supposedly no backgrounds even slightly lower masses than those quoted in the table would lead to 5 or more events, which maybe sufficient for a discovery. 

However, we need to mention that our calculation, using MadGraph, calculates the cross-sections at leading order only. Also our calculation does not include any cuts and thus, should be taken only as a rough estimate. Thus, the numbers in the table should probably come with an uncertainty of the order of ($100-150$) GeV or so for the larger multiplicity states. On the other hand, for the simpler signal $ pp \rightarrow l^{+} l^{+} W^- W^- $, the number given in the table should be taken with a grain of salt. Currently, for dilepton searches with luminosity of 36 $fb^{-1}$, there are no background events in the bins above $1$ TeV in the invariant mass distribution $m (ll)$, see \cite{ATLAS:2017iqw}.  This in turns implies for a luminosity of 300 $fb^{-1}$ in the most pessimistic case an upper limit of roughly $8$ background events for the signal $pp \rightarrow l^{+} l^{+} W^-
W^-$. Our estimate of 3 signal events would then correspond only a 1 $\sigma$ c.l. limit.

We mention that the final state with 2 $l$ and $6$ $W$'s and LNV signals with 10 or more particles are also possible in $d=7$ one-loop models, but do not occur within our two example models. This is simply due to the fact that scalars or fermions with 5 units of charge are needed for such states. Thus, such signals can appear in versions of the $d=7$ one-loop type models, that include larger ${\rm SU(2)_L}$ representations, such as quintuplets, or with particles with a larger hypercharge.

Finally, the table considers only ``symmetric'' LNV final states. Here, by symmetric we define that both branches of the decay contain the same number of final states particles. For example, for the quadruplet model, we have included the LNV signal with symmetric final states $ p p \rightarrow \chi_2^{3+} \chi_2^{3-}, \chi_2^{3+} \rightarrow l^{+} l^{+} l^{+} , \chi_2^{3-} \rightarrow W^- W^- l^{-} $, but we have not considered the possible LNV signal with asymmetric final states $ p p \rightarrow \chi_2^{++} \chi_2^{- - }, \chi_2^{++} \rightarrow l^{+} W^+ , \chi_2^{- - } \rightarrow W^- W^- W^- l^{+} $. The reason for this choice is simply that we consider ``asymmetric'' LNV signals, although in principle possible, are less likely to occur. This can be understood from phase space considerations: A two-body final state has a prefactor of $\frac{1}{8 \pi}$ in the partial width, while a four-body phase space is smaller by a factor $3072 \pi^4$. Naturally one then expects that the ratio of branching ratios for these asymmetric cases is never close to one, unless there is a corresponding hierarchy in the couplings involved.

Decay widths for the lightest particle in our models are often very small numerically. This opens up the possibility that some particle decays might occur with a charged track. Charged tracks are more likely to occur in the triplet model, so we concentrate in our discussion on this case. The two-body decay width of $\eta_1^{++}$ is estimated in \eq{eq:pheno:GamPP}. For the decay of $\eta_3^{3+}$, assuming $\eta_3^{3+}$ is the lightest particle, one can estimate,
\begin{equation}\label{eq:pheno:Gam3P}
    \Gamma(\eta_3^{+++} \to W^+l^+l^+) \sim \frac{1}{32\pi^2}
    \Big(\frac{\mu_1}{m_{\eta_2^{++}}^2}\Big)^2
    \frac{m_{\eta_3^{++}}^3}{m_{\eta_1^{++}}}\theta_{\eta_1\eta_2}^2
    \Gamma(\eta_1^{++}\to l^+l^+) \, .
\end{equation}
Here, $\theta_{\eta_1\eta_2}$ is the mixing angle between the states $\eta_1$ and $\eta_2$. The decay width \eq{eq:pheno:Gam3P} contains three parameters related to the smallness of the observed neutrino masses: $\mu_1$, $\theta_{\eta_1\eta_2}$ and $Y_1$. Assuming all mass parameters roughly equal $\mu_1\simeq m_{\eta_3^{++}} \simeq m_{\eta_2^{++}}\simeq m_{\eta_1^{++}}=M$ this leads to the estimate,
\begin{equation}\label{eq:pheno:Len3P}
    L_0(\eta_3^{3+} \to W^+l^+l^+) \sim 0.3 
    \Big(\frac{10^{-1}}{\theta_{\eta_1\eta_2}}\Big)^2
    \Big(\frac{10^{-2}}{|Y_1|}\Big)^2
    \Big(\frac{10^{-2}}{|Y_4|}\Big)^2
    \Big(\frac{m_{\psi}}{\rm TeV}\Big)^2
    \Big(\frac{\rm TeV}{M}\Big) {\rm mm} \, .
\end{equation}
Here, the choice for the Yukawa couplings being order $10^{-2}$ is motivated by the upper limits on the CLFV branching ratios, discussed in the last section. The decay length \eq{eq:pheno:Len3P} represents only a very rough estimate, but it is worth pointing out that more stringent upper limits from charged LFV would result in smaller values for the Yukawa couplings, leading to correspondingly large decay lengths. Note also that smaller values of $\mu_1$ would lead to quadratically large lengths.

Similarly, one can estimate roughly the order of magnitude of the decay length for $\eta_3^{4+}$. The result is
\begin{equation}\label{eq:pheno:Len4P}
    L_0(\eta_3^{4+} \to W^+W^+l^+l^+) \sim 4
    \Big(\frac{1}{\lambda_2}\Big)^2
    \Big(\frac{10^{-2}}{|Y_1|}\Big)^2
    \Big(\frac{10^{-2}}{|Y_4|}\Big)^2
    \Big(\frac{m_{\psi}}{\rm TeV}\Big)^2
    \Big(\frac{\rm TeV}{M}\Big) {\rm cm} \, .
\end{equation}
The width of $\eta_3^{4+}$ is smaller than the corresponding one for $\eta_3^{3+}$ due to the phase space suppression for a 4-body final state. \eq{eq:pheno:Len4P} shows that within the triplet model a charged track for the decay of $\eta_3^{4+}$ is actually expected.

%%%%%%%%%%%%%%%%%%%%%%%%%%%%%%%%%%%%%%%%%%%%%%%%%%%%%%%%%%%%%%%
%%%%%%%%%%%%%%%%%%%%%%%%%%%%%%%%%%%%%%%%%%%%%%%%%%%%%%%%%%%%%%%
\section{Summary} \label{sec:pheno:summary}

In this chapter we have discussed the phenomenology of $d=7$ one-loop neutrino mass models. Models in this class are far from the simplest variants of BSM models that can fit existing neutrino data, but are interesting on their own right, since they predict that new physics must exist below roughly 2 TeV. If neutrino masses were indeed generated by one of the models in this class, one can thus expect that the LHC will find signatures of new resonances. Searches for doubly charged scalars and multi-lepton final states already put some bounds on these models. However, for the most interesting aspect of $d=7$ one-loop models, namely lepton number violating final states, no LHC search exists so far. In particular, final states with large multiplicities are predicted to occur (multiple $W$ and multiple leptons) for which we expect Standard Model backgrounds to be negligible.

In our discussion, we have limited ourselves to just two simple example models. Our motivation to do so is that all $d=7$ one-loop neutrino mass models, which are genuine in the sense that they give the leading contribution to neutrino mass without invoking new symmetries, predict similar LHC signals. The two models which we considered have either an ${\rm SU(2)_L}$ triplet or a quadruplet as the largest representations. Other $d=7$ models will contain even larger ${\rm SU(2)_L}$ multiplets and thus also particles with multiple electric charges, to which very similar constraints than those analysed here will apply.

Finally, we mention that there exist variants of $d=7$ one-loop models, in which the internal scalars and fermions carry non-trivial colour charges. These variants are not fully covered by our analysis. While the neutrino mass fit and the constraints from LFV searches will be qualitatively very similar to what we have discussed here, additional colour factors in the calculations will lead to some quantitative changes. The resulting bounds will, in general be somewhat more stringent than the numbers we give in this thesis. More important, however, are the changes in the LHC phenomenology. For example, in the colour-singlet models, which we analysed in this chapter, the lightest doubly charged scalar will decay to two charged leptons. In the coloured variants of the model, the corresponding lightest scalar will behave like a leptoquark, decaying to $l+j$, instead. Thus, different LHC searches will apply to the coloured $d=7$ models. More interesting, however, is that for coloured models also the LNV final states, which we discussed, will change, since at the end of the decay chain instead of two charged lepton, one lepton plus jet will appear. Although this variety of signals will be interesting on their own rights, we have concentrated here on the colour singlet variants of the model, because dileptons are cleaner (and thus easier to probe) in the challenging experimental environment that is the LHC.

\pagebreak
\fancyhf{}

%% file: Chapters/cLFV_3loop/Chapter_clfv.tex
\fancyhf{}
\fancyhead[LE,RO]{\thepage}
\fancyhead[RE]{\slshape{Chapter \thechapter. Minimal three-loop neutrino mass models and CLFV}}
\fancyhead[LO]{\slshape\nouppercase{\rightmark}}

\chapter{Minimal three-loop neutrino mass models and charged lepton flavour violation}
\label{ch:clfv}
\graphicspath{ {Chapters/cLFV_3loop/} }

In this chapter, based on \cite{Cepedello:2020lul}, we will study how upper limits on charged lepton flavour violating (CLFV) observables constrain three-loop neutrino mass models. We will focus on some particular, well-known models, which we consider ``minimal'' models. The term ``minimal'' here refers to the fact that for models at three-loop level at least three different {\em types} of particles beyond the Standard Model particle content are needed, in order to avoid lower order diagrams.\footnote{{\em Types} of particles refers to the fact, that in case one of the new particles is a fermion, usually at least two copies (``families'') of fermions are needed for a realistic neutrino mass matrix.} The three models that we will study in this chapter are the so-called cocktail \cite{Gustafsson:2012vj}, Krauss-Nasri-Trodden (KNT) \cite{Krauss:2002px} and Aoki-Kanemura-Seto (AKS) \cite{Aoki:2008av} models.

These three models are probably the best-known three-loop models in the literature, and a number of other papers have studied them (or some variations thereof). The cocktail model, for example, has been studied also in \cite{Geng:2014gua}. There are also versions of the cocktail model in which the $W$ bosons are replaced by scalars \cite{Kajiyama:2013lja, Hatanaka:2014tba, Alcaide:2017xoe}. For the AKS model, one can find some discussion on phenomenology and vacuum stability constraints in \cite{Aoki:2009vf, Aoki:2010aq, Aoki:2011zg}, while a variant of the AKS model with doubly charged vector-like fermions and a scalar doublet with hypercharge $Y=3/2$ (plus the singlets of the AKS model) can be found in \cite{Okada:2015hia}. Other variants of the AKS model in which the exotic particles are all electroweak singlets can be found in \cite{Gu:2016xno, Ho:2016aye}. Finally, for the KNT model, different phenomenological and theoretical aspects were studied in \cite{Cheung:2004xm, Ahriche:2014cda, Ahriche:2014oda, Ahriche:2015loa, Ahriche:2015taa, Ahriche:2015lqa, Ahriche:2014xra, Ahriche:2015wha}. There are also variations of the KNT model, like the coloured KNT \cite{Gu:2012tn, Nomura:2016ezz, Cheung:2016frv, Hati:2018fzc}, or a model with vector-like fermions added to the KNT model \cite{Okada:2016rav}. Other variants can be found in \cite{Ng:2013xja, Chen:2014ska}.

Common to all the three minimal models is that their neutrino mass diagrams are proportional to two powers of Standard Model lepton masses. Together with the three-loop suppression of $1/(16 \pi^2)^3$, this results in the prediction of rather small neutrino mass eigenvalues, unless the new Yukawa couplings of the models take very large values.  However, in all models off-diagonal entries for these new Yukawa couplings are required, since neutrino oscillation experiments have measured large neutrino angles \cite{deSalas:2017kay}. Therefore, one expects that CLFV limits will put severe constraints on these minimal models. This simple observation forms the motivation for the current chapter.
\\

The chapter is organised as follows. In \sect{sec:clfv:notation} we will set up the notation and briefly discuss two scalar extensions of the Standard Model. In \sect{sec:clfv:cocktail} we will discuss the cocktail model. We will first introduce the model and the neutrino mass generation mechanism, and then we will present our numerical results for this model. We start with the cocktail model, since the flavour structure of the neutrino mass matrix, in this case, is the simplest of the three models. We then discuss in a similar way the KNT model in \sect{sec:clfv:KNT} and the AKS model in \sect{sec:clfv:AKS}. A number of technical aspects on the calculation of the loop integrals are relegated to \app{app:loops}.

%%%%%%%%%%%%%%%%%%%%%%%%%%%%%%%%%%%%%%%%%%%%%%%%%%%%%%%%%%%%%%%
%%%%%%%%%%%%%%%%%%%%%%%%%%%%%%%%%%%%%%%%%%%%%%%%%%%%%%%%%%%%%%%
\section{Notation and conventions} \label{sec:clfv:notation}

In order to make the discussion more transparent for the reader, it is convenient to adopt a common notation and use the same conventions for the three models considered here. This is the aim of this section.

The three minimal three-loop neutrino mass models studied in this chapter are based on the Standard Model gauge group, ${\rm SU(3)_C} \times {\rm SU(2)_L} \times {\rm U(1)_Y}$. This local symmetry is supplemented by a global $Z_2$ parity, which is introduced to forbid the tree-, one- and two-loop contributions to the neutrino mass matrix, as explained below. All the Standard Model fields are assumed to be even under the global $Z_2$ symmetry. The particle spectrum of the three-loop models explored here may contain new fermions, and these will be fully specified for each model in the next sections. In what concerns their scalar sectors, they can be regarded as extensions of three well-known scenarios: the Standard Model scalar sector, the Two Higgs Doublet Model (2HDM) scalar sector and the Inert Doublet Model (IDM) scalar sector. We now describe the scalar fields in the 2HDM and IDM scenarios, the way the electroweak symmetry gets broken in each case and the Yukawa interactions with the Standard Model fermions.

%\begin{itemize}
%    \item {\bf Standard Model scalar sector}
%\end{itemize}

%\begin{table}
%    \centering
%    \begin{tabular}{| c c c c c |}
%        \hline  
%         & generations & $\mathrm{SU(3)}_c$ & $\mathrm{SU(2)}_L$ & $\mathrm{U(1)}_Y$ \\
%        \hline
%        \hline
%        $H$ & 1 & ${\bf 1}$ & ${\bf 2}$ & $1/2$ \\
%        \hline
%        \hline
%    \end{tabular}
%    \caption{Standard Model scalar sector, containing only the usual Higgs doublet $H$.}
%    \label{tab:clfv:SMscalar}
%\end{table}

%The Standard Model scalar sector contains only the usual Higgs doublet, $H$, as shown in \tab{tab:clfv:SMscalar}. This doublet can be decomposed in terms of its $\rm SU(2)_L$ components as,
%%
%\begin{equation}
%    H = \left( \begin{array}{c}
%    H^+ \\
%    H^0 \end{array} \right) \, .
%\end{equation}
%%
%The Yukawa couplings of the Standard Model are
%%
%\begin{equation}  \label{eq:clfv:SMYuk}
%    - \mathcal L_Y^{\rm SM} = y_e \, \overline{L} \, H \, e_R + y_u \, \overline Q \, \widetilde H \, u_R + y_d \, \overline Q \, H \, d_R + \hc \, ,
%\end{equation}
%%
%where we have defined $\widetilde H = i \tau_2 H^\ast$, with $\tau_2$ the second Pauli matrix. We have omitted flavour and $\mathrm{SU(2)}_L$ indices in the previous expression to simplify the notation. The electroweak symmetry gets spontaneously broken by the Higgs VEV,
%%
%\begin{equation}
%    \langle H \rangle = \frac{1}{\sqrt{2}} \left( \begin{array}{c}
%    0 \\
%    v \end{array} \right) \, ,
%\end{equation}
%%
%with $v \simeq 246$ GeV. In the three models discussed below, $H$ will be even under the $Z_2$ parity.

\begin{itemize}
    \item {\bf 2HDM scalar sector}
\end{itemize}

\begin{table}
    \centering
    \begin{tabular}{| c c c c c |}
        \hline  
         & generations & $\mathrm{SU(3)_C}$ & $\mathrm{SU(2)_L}$ & $\mathrm{U(1)_Y}$ \\
        \hline
        \hline
        $\Phi_1$ & 1 & ${\bf 1}$ & ${\bf 2}$ & $1/2$ \\
        $\Phi_2$ & 1 & ${\bf 1}$ & ${\bf 2}$ & $1/2$ \\
        \hline
        \hline
    \end{tabular}
    \caption{2HDM scalar sector, composed of the two $\mathrm{SU(2)_L}$ scalar doublets $\Phi_1$ and $\Phi_2$.}
    \label{tab:clfv:2HDMscalar}
\end{table}

The 2HDM scalar sector is composed of two scalar doublets, $\Phi_1$ and $\Phi_2$, with identical quantum numbers under the Standard Model gauge symmetry, as shown in \tab{tab:clfv:2HDMscalar}. They can be decomposed in terms of their $\rm SU(2)_L$ components as,
\begin{equation}
    \Phi_1 = \left( \begin{array}{c}
    \Phi_1^+ \\
    \Phi_1^0 \end{array} \right) \quad , \quad \Phi_2 = \left( \begin{array}{c}
    \Phi_2^+ \\
    \Phi_2^0 \end{array} \right) \, .
\end{equation}
Since both scalar doublets have exactly the same quantum numbers, and in particular since they will both be assumed to be even under the $Z_2$ symmetry, flavour changing neutral current interactions are in principle present. This dangerous feature can be fixed by introducing a second (softly broken) $Z_2$ symmetry, under which one of the two doublets and some of the Standard Model fermions are charged. There are several possibilities, and here we will just assume that this symmetry makes $\Phi_1$ leptophilic, and $\Phi_2$ leptophobic.\footnote{This is the choice of the authors of the AKS model \cite{Aoki:2008av}, and we will stick to it although the more common possibilities of a type-I or type-II 2HDM are equally valid.} Under this assumption, the 2HDM Yukawa interactions are given by,
\begin{equation}  \label{eq:clfv:2HDMYuk}
    - \mathcal L_Y^{\rm 2HDM} = y_e \, \overline L \, \Phi_1 \, e_R + y_u \, \overline Q \, \widetilde \Phi_2 \, u_R + y_d \, \overline Q \, \Phi_2 \, d_R + \hc \, .
\end{equation}
Again, flavour and $\mathrm{SU(2)_L}$ indices have been omitted for the sake of clarity. We see that, as explained above, $\Phi_1$ only couples to leptons, while $\Phi_2$ only couples to quarks. In the 2HDM, both scalar doublets are assumed to take VEVs,
\begin{equation}
    \langle \Phi_1 \rangle = \frac{1}{\sqrt{2}} \left( \begin{array}{c}
    0 \\
    v_1 \end{array} \right) \quad , \quad \langle \Phi_2 \rangle = \frac{1}{\sqrt{2}} \left( \begin{array}{c}
    0 \\
    v_2 \end{array} \right) \, ,
\end{equation}
such that the usual electroweak VEV $v$ is given by,
\begin{equation}
    v^2 = v_1^2 + v_2^2 \, .
\end{equation}
We also define the ratio $\tan \beta = v_2 / v_1$.

\begin{itemize}
    \item {\bf IDM scalar sector}
\end{itemize}

\begin{table}
    \centering
    \begin{tabular}{| c c c c c c |}
        \hline  
         & generations & $\mathrm{SU(3)_C}$ & $\mathrm{SU(2)_L}$ & $\mathrm{U(1)_Y}$ & $Z_2$ \\
        \hline
        \hline
        $H$ & 1 & ${\bf 1}$ & ${\bf 2}$ & $1/2$ & $+$ \\
        $\eta$ & 1 & ${\bf 1}$ & ${\bf 2}$ & $1/2$ & $-$ \\
        \hline
        \hline
    \end{tabular}
    \caption{IDM scalar sector, containing the standard Higgs doublet $H$ as well as a second inert doublet $\eta$ charged under the $Z_2$ parity.}
    \label{tab:clfv:IDMscalar}
\end{table}

In the IDM, a second scalar doublet denoted as $\eta$ is introduced. In contrast to the 2HDM, this doublet is odd under the $Z_2$ parity, as shown in \tab{tab:clfv:IDMscalar}. The inert doublet $\eta$ can be decomposed in terms of its $\mathrm{SU(2)_L}$ components as,
\begin{equation}
    \eta = \left( \begin{array}{c}
    \eta^+ \\
    \eta^0 \end{array} \right) \, .
\end{equation}
Since the Standard Model fermions are even under $Z_2$, $\eta$ does not couple to them and the IDM Yukawa interactions are exactly the same as those in the Standard Model. The scalar potential of the IDM is assumed to be such that only the Standard Model Higgs doublet takes a VEV,
\begin{equation}
    \langle H \rangle = \frac{1}{\sqrt{2}} \left( \begin{array}{c}
    0 \\
    v \end{array} \right) \quad , \quad \langle \eta \rangle = 0 \, .
\end{equation}
Therefore, electroweak symmetry breaking takes place in the standard way and the $Z_2$ parity remains exactly conserved. Finally, one can split the neutral component of the $\eta$ doublet as,
\begin{equation}
    \eta^0 = \frac{1}{\sqrt{2}} (\eta_R + i \, \eta_I) \, ,
\end{equation}
so that $\eta_R$ and $\eta_I$ are, respectively, the real and imaginary parts of $\eta^0$. Under the assumption of CP conservation in the scalar sector, these two states are mass eigenstates, since the $Z_2$ symmetry forbids their mixing with the Standard Model neutral scalar.

%%%%%%%%%%%%%%%%%%%%%%%%%%%%%%%%%%%%%%%%%%%%%%%%%%%%%%%%%%%%%%%
%%%%%%%%%%%%%%%%%%%%%%%%%%%%%%%%%%%%%%%%%%%%%%%%%%%%%%%%%%%%%%%
\section{Cocktail model} \label{sec:clfv:cocktail}

We begin with the so-called cocktail model, introduced in \cite{Gustafsson:2012vj}, since the neutrino mass matrix in this model has the simplest flavour structure.
\\

The cocktail model can be regarded as an extension of the IDM. In addition to the IDM fields, the particle content of the cocktail model includes the two $\mathrm{SU(2)_L}$ singlet scalars $S$ and $\rho$, singly and doubly charged. Interestingly, the model does not have any new fermion, just scalars. The $\eta$ and $S$ scalar fields are taken to be odd under the $Z_2$ parity, while the rest of the fields in the model are even.\footnote{$S$ and $\eta$ need to be odd under the $Z_2$ symmetry, in order to forbid Yukawa couplings with the Standard Model leptons. These couplings would otherwise generate a one-loop neutrino mass diagram, as in the Zee model \cite{Zee:1980ai}.} The quantum numbers $S$ and $\rho$ are given in \tab{tab:clfv:cocktail}. Counting also $\eta$, there are then three new multiplets in the cocktail model, with respect to the Standard Model.

\begin{table}
    \centering
    \begin{tabular}{| c c c c c c |}
        \hline  
         & generations & $\mathrm{SU(3)_C}$ & $\mathrm{SU(2)_L}$ & $\mathrm{U(1)_Y}$ & $Z_2$ \\
        \hline
        \hline    
        $S$ & 1 & ${\bf 1}$ & ${\bf 1}$ & $1$ & $-$ \\
        $\rho$ & 1 & ${\bf 1}$ & ${\bf 1}$ & $2$ & $+$ \\
        \hline
        \hline
    \end{tabular}
    \caption{New particles in the cocktail model with respect to the IDM.}
    \label{tab:clfv:cocktail}
\end{table}

The Lagrangian of the cocktail model contains only one additional Yukawa term with respect to the Standard Model,
\begin{equation} \label{eq:clfv:YukCocktail}
    - \mathcal L \supset h \, \overline{e_R^c} \, e_R \, \rho + \hc \, ,
\end{equation}
where $h$ is a symmetric $3 \times 3$ matrix. Flavour indices have been omitted in this expression for the sake of clarity. In addition, the new scalar potential couplings are given by,
\begin{align} \label{eq:clfv:PotCocktail}
    \mathcal V &\supset M_{S}^2 |S|^2 + M_{\rho}^2 |\rho|^2 + M_{\eta}^2
    |\eta|^2 + \frac{1}{2} \lambda_S \, |S|^4 + \frac{1}{2} \lambda_\rho
    \, |\rho|^4 + \frac{1}{2} \lambda_\eta \, |\eta|^4 \nn \\ &+
    \lambda_{S\rho} \, |S|^2 |\rho|^2 + \lambda_{S\eta} \, |S|^2 |\eta|^2
    + \lambda_{\rho\eta} \, |\rho|^2 |\eta|^2 \nn \\
    &+ \lambda_{\rho H} |\rho|^2 |H|^2 + \lambda_{SH} |S|^2 |H|^2
    + \lambda_{\eta H}^{(1)} |\eta|^2 |H|^2
    + \lambda_{\eta H}^{(3)} H^{\dagger}\eta^{\dagger}H\eta
    \nn \\
    &+ \Big[ \mu_1 \,  H \eta S^\ast  + \frac 12 \mu_2  \rho \, S^\ast S^\ast  
       + \kappa  H \eta S \rho^\ast 
    + \frac{1}{2} \, \lambda_5  (H \eta^\ast)^2 + \hc \Big] \, .
\end{align}
We have omitted $\mathrm{SU(2)_L}$ indices to simplify the notation. The parameters $\mu_1$ and $\mu_2$ are trilinear couplings with dimensions of mass, $\kappa$ and all $\lambda$'s are dimensionless. Most important is the term proportional to $\lambda_5$, see discussion below.

The singly charged scalars $S^+$ and $\eta^+$ mix after electroweak symmetry breaking, due to the term proportional to $\mu_1$. This leads to two $\mathcal{H}^+_i$ mass eigenstates, with mixing angle $\beta$, see \app{app:loops}. The model also includes the doubly charged scalar $\rho^{++}$, with mass
\begin{equation}
    m_{\rho^{++}}^2 = M_{\rho}^2 + \frac{1}{2} \, \lambda_{\rho H} \, v^2 \, .
\end{equation}

The cocktail model has other interesting features that will not be discussed in any detail here. For instance, the $Z_2$ parity of the model is conserved after electroweak symmetry breaking, so that the lightest $Z_2$-odd state is stable and can in principle constitute a good dark matter candidate.

%%%%%%%%%%%%%%%%%%%%%%%%%%%%%%
\subsubsection*{Neutrino masses}

\begin{figure}
    \centering
    \includegraphics[width=0.65\textwidth]{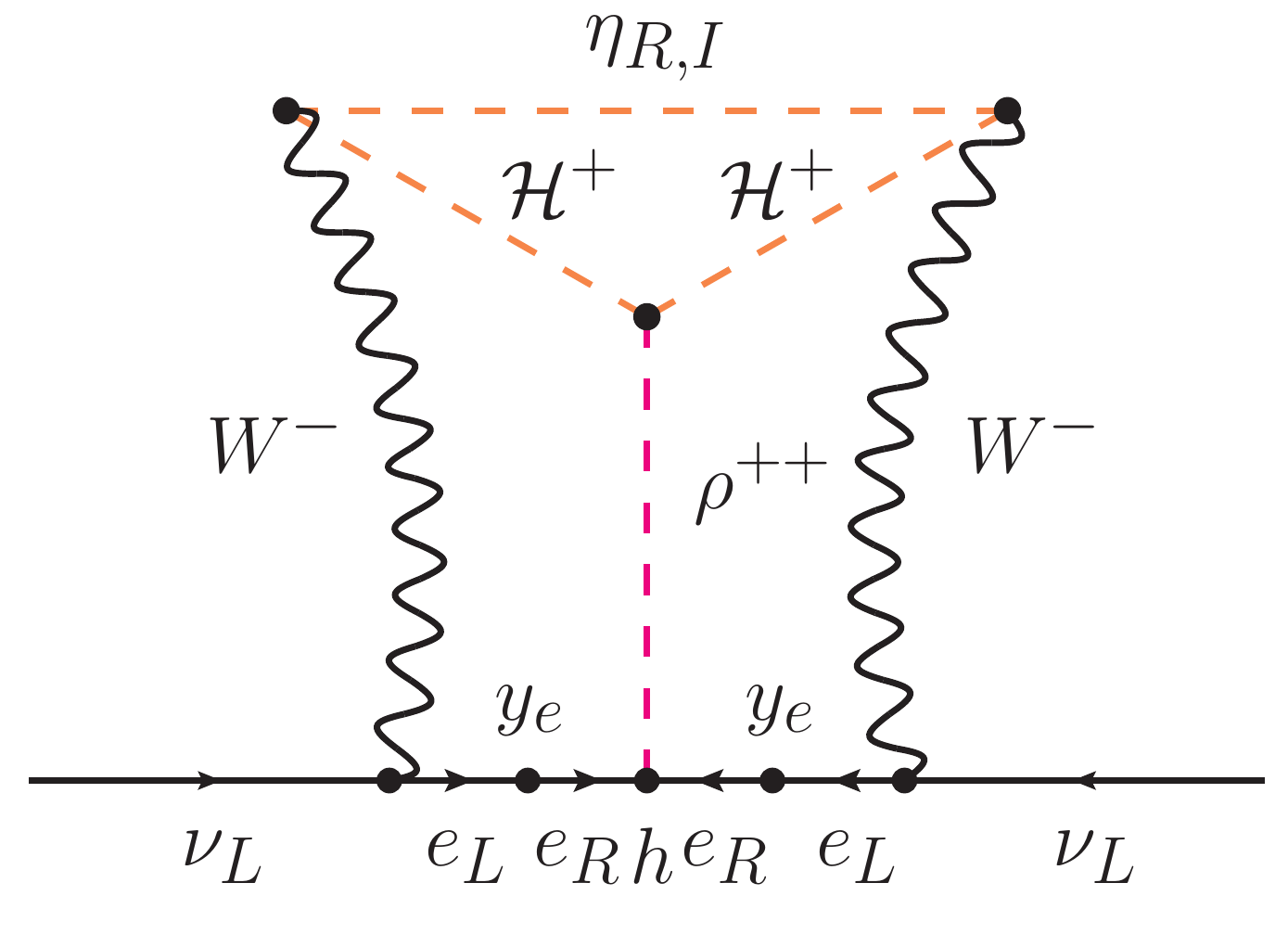}
    \caption{Three-loop neutrino masses in the cocktail model. The inert doublet $\eta$ is split into its real and imaginary parts, $\eta^0 = \frac{1}{\sqrt{2}} (\eta_R + i \, \eta_I)$, due to the scalar potential terms proportional to $\lambda_5$. $\mathcal{H}^+ \equiv \mathcal{H}^+_{1,2}$ represent the singly charged scalars in the model, obtained after diagonalising the mass matrix of the $\left\{S^+ , \eta^+ \right\}$ states.}
    \label{fig:clfv:cocktail}
\end{figure}

The three-loop diagram leading to neutrino masses in the cocktail model is shown in \fig{fig:clfv:cocktail}. In the unitary gauge this diagram is the only diagram contributing to the neutrino mass matrix. However, in order to understand how to maximise the contribution of this diagram to the neutrino mass matrix, it is more useful to calculate all diagrams in Feynman-'t Hooft gauge. This is discussed in detail in \app{app:loops}.

In an analogous way to the well-known scotogenic model \cite{Ma:2006km}, the diagram shown in \fig{fig:clfv:cocktail} vanishes in the limit $m_{\eta_R}^2 - m_{\eta_I}^2 \propto \lambda_5 \to 0$, since in this limit the model conserves lepton number. We can then write the neutrino mass matrix in the cocktail model as,
\begin{align} \label{eq:clfv:mnucocktail}
    \left(\mathcal{M}_\nu\right)_{ij} = \frac{\lambda_5}{(16\pi^2)^3}
    \frac{m_i \, h_{ij} \, m_j}{m_{\rho^{++}}} \, F_{\rm Cocktail} \, ,
\end{align}
where $m_i$ and $m_j$ are charged lepton masses. Here we have hidden all the complexities of the calculation in the dimensionless factor $F_{\rm Cocktail}$. This factor contains the loop integrals, depending on the masses of the scalars, and prefactors containing coupling constants, etc, see \app{app:loops}.

%%%%%%%%%%%%%%%%%%%%%%%%%%%%%%%%%%%%%%%%%%%%%%%%%%%%%%%%%%%%%%%
\subsection{Results} \label{subsec:clfv:results-cocktail}

The cocktail model is an example of a \textit{type-II-seesaw-like} model. In this class of models, the neutrino mass matrix is proportional to a symmetric Yukawa matrix,
\begin{equation} \label{eq:clfv:typeII}
    \mathcal{M}_\nu \sim Y \, \frac{v^2}{\Lambda} \, ,
\end{equation}
with $Y_{ij} = Y_{ji}$ and $\Lambda$ some generic mass scale. This allows one to fit the observed neutrino masses and mixing angles in a trivial way. Furthermore, the tight relation given in \eq{eq:clfv:typeII} implies very specific predictions for ratios of CLFV observables and strongly reduces the number of free parameters in the model. As we will discuss now, this has very important consequences.

From the experimental data we can reconstruct the neutrino mass matrix in the flavour basis as,
\begin{equation} \label{eq:clfv:Mnu}
    {\cal M}_{\nu} = U^\ast \, \widehat{\cal M}_\nu \, U^\dagger \, .
\end{equation}
This allows us to calculate the Yukawa $h$ necessary to fit the experimental data using the expression in \eq{eq:clfv:mnucocktail}. We find,
\begin{equation} \label{eq:clfv:hfit}
    h = (16\pi^2)^3 \, \frac{m_{\rho^{++}}}{\lambda_5 \, F_{\rm Cocktail}} \,
          {\widehat{\cal M}_{e}}^{-1} \, {\cal M}_\nu \, {\widehat{\cal M}_{e}}^{-1} \, .
\end{equation}
where $\widehat{\cal M}_{e}$ is the diagonal matrix with the measured charged lepton masses.

Let us first make a rough numerical estimate. Choosing normal hierarchy ($m_{\nu_1}\to 0$) and $\delta=0$ for simplicity and inserting $m_{\rho^{++}} = 800$ GeV, which is roughly the current experimental bound from LHC data \cite{Aaboud:2017qph, CMS:2017pet, Aaboud:2018qcu}, we obtain,
\begin{equation} \label{eq:clfv:hexa}
      h \simeq
     \begin{pmatrix}
     46000 & 450 & 5.7 \\
     450   & 8.1 & 0.36 \\
     5.7   & 0.36 & 0.026
     \end{pmatrix}
       \Big(\frac{m_{\rho^{++}}}{\rm 800\hskip1mm GeV}\Big)
       \Big(\frac{1}{\lambda_5}\Big)
       \Big(\frac{1}{F_{\rm Cocktail}}\Big) \, .
\end{equation}
These values are obviously much too large to be realistic. We therefore searched the parameter space, intending to identify regions, in which $h$ can fulfill the bounds from perturbativity and lepton flavour violation searches. This search was done in two steps.

First, we maximise $F_{\rm Cocktail}$ and $\lambda_5$. For $\lambda_5$ we use $\lambda_5=4\pi$, the largest value allowed by perturbativity. We then scanned all free mass parameters entering in $F_{\rm Cocktail}$, for details see \app{app:loops}. Generally speaking, $F_{\rm Cocktail}$ is maximised when $\mu_1$, $\mu_2$ and $\kappa$ take the largest values allowed, while the remaining free mass eigenvalues of the model take the lowest possible values allowed by experimental searches. The maximal value of $F_{\rm Cocktail}$ found in this numerical scan is $F_{\rm Cocktail}^{\rm max} \simeq 192$. We will use this number in all plots below. This choice is conservative in the sense that the Yukawa couplings $h_{ij}$ will be larger for all other choices, thus constraints from charged lepton flavour violation searches will only be more stringent in other parts of the parameter space.

\begin{figure}[t!]
    \centering
    \includegraphics[width=0.48\textwidth]{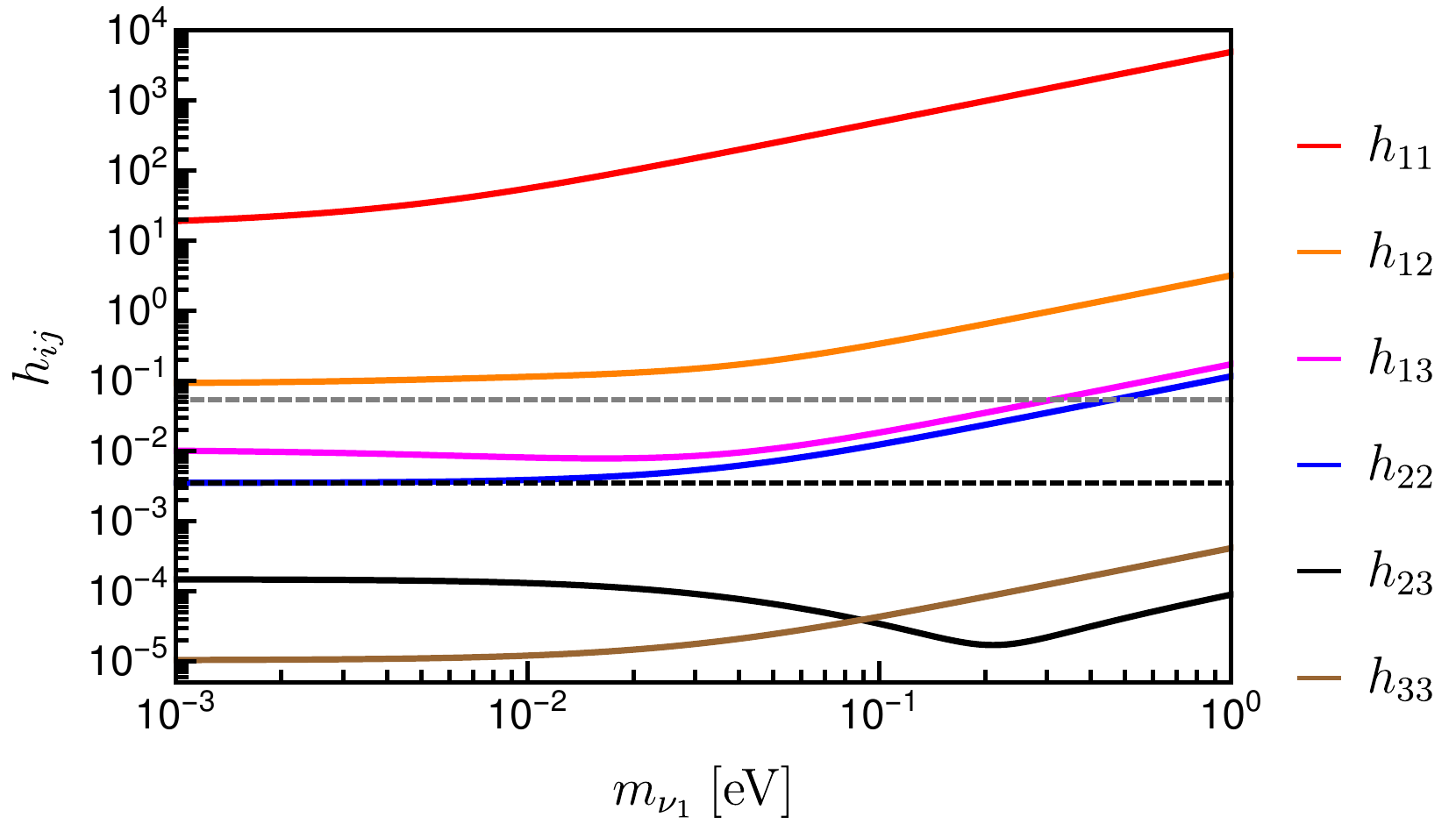}
    \hfill
    \includegraphics[width=0.48\textwidth]{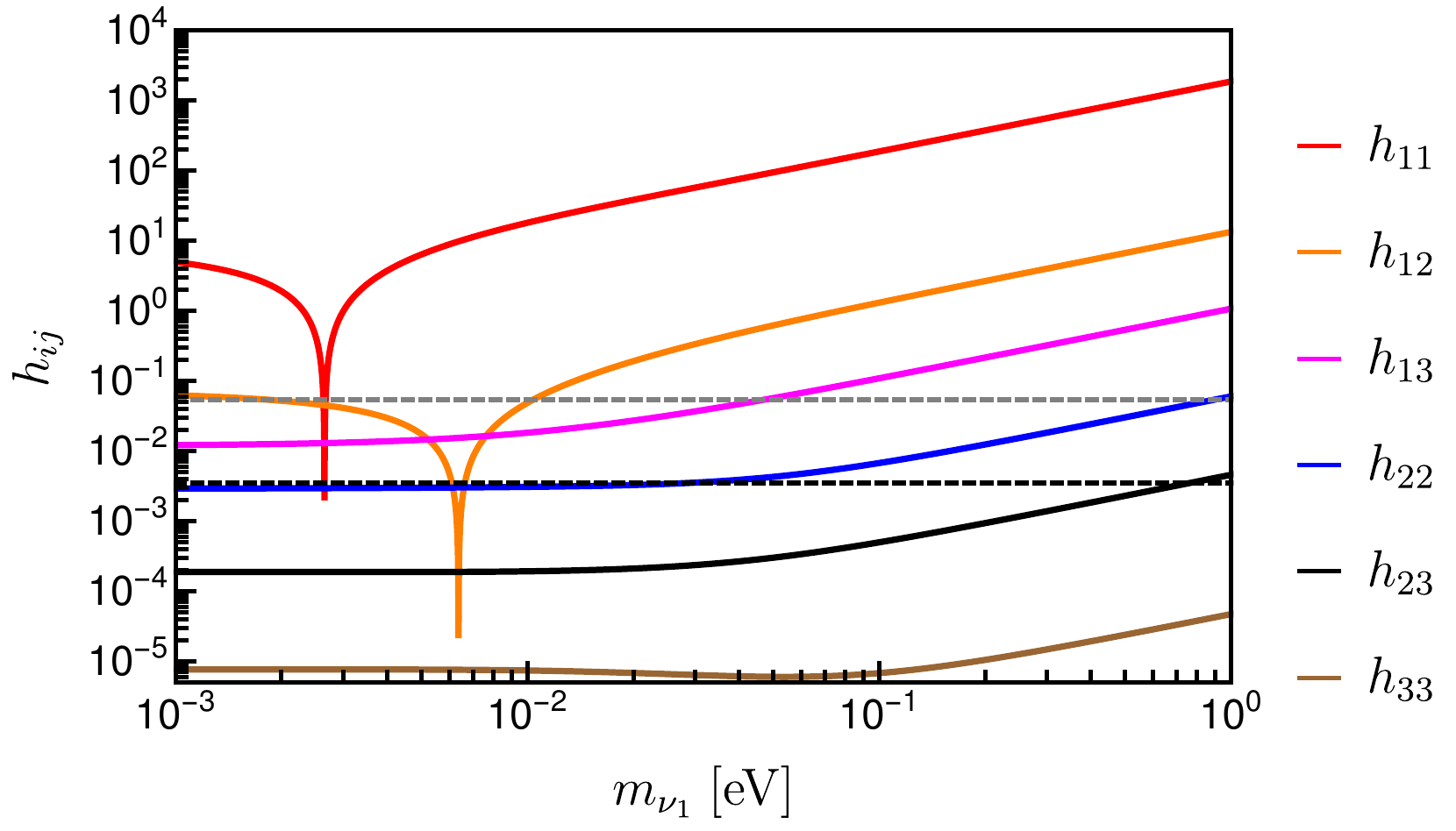}
    \caption{Yukawa couplings $h_{ij}$ as function of the lightest neutrino mass, calculated with $F_{\rm Cocktail}^{\rm max}$. These Yukawas should therefore be understood as {\em lower limits}. In both plots we have used the best fit point data from the global oscillation fit \cite{deSalas:2017kay}, except $\delta=0$. The plot to the left shows the case $(\alpha_{12}, \alpha_{13})=(0,0)$, the plot on the right $(\alpha_{12}, \alpha_{13})=(\pi,0)$. The dashed grey (black) lines in the background are rough estimates for the typical size that the Yukawa couplings should have, in order to satisfy limits from muon (tau) CLFV decays. These lines are only for orientation.}
    \label{fig:clfv:Yuks}
\end{figure}

\begin{figure}[t!]
    \centering
    \includegraphics[width=0.55\textwidth]{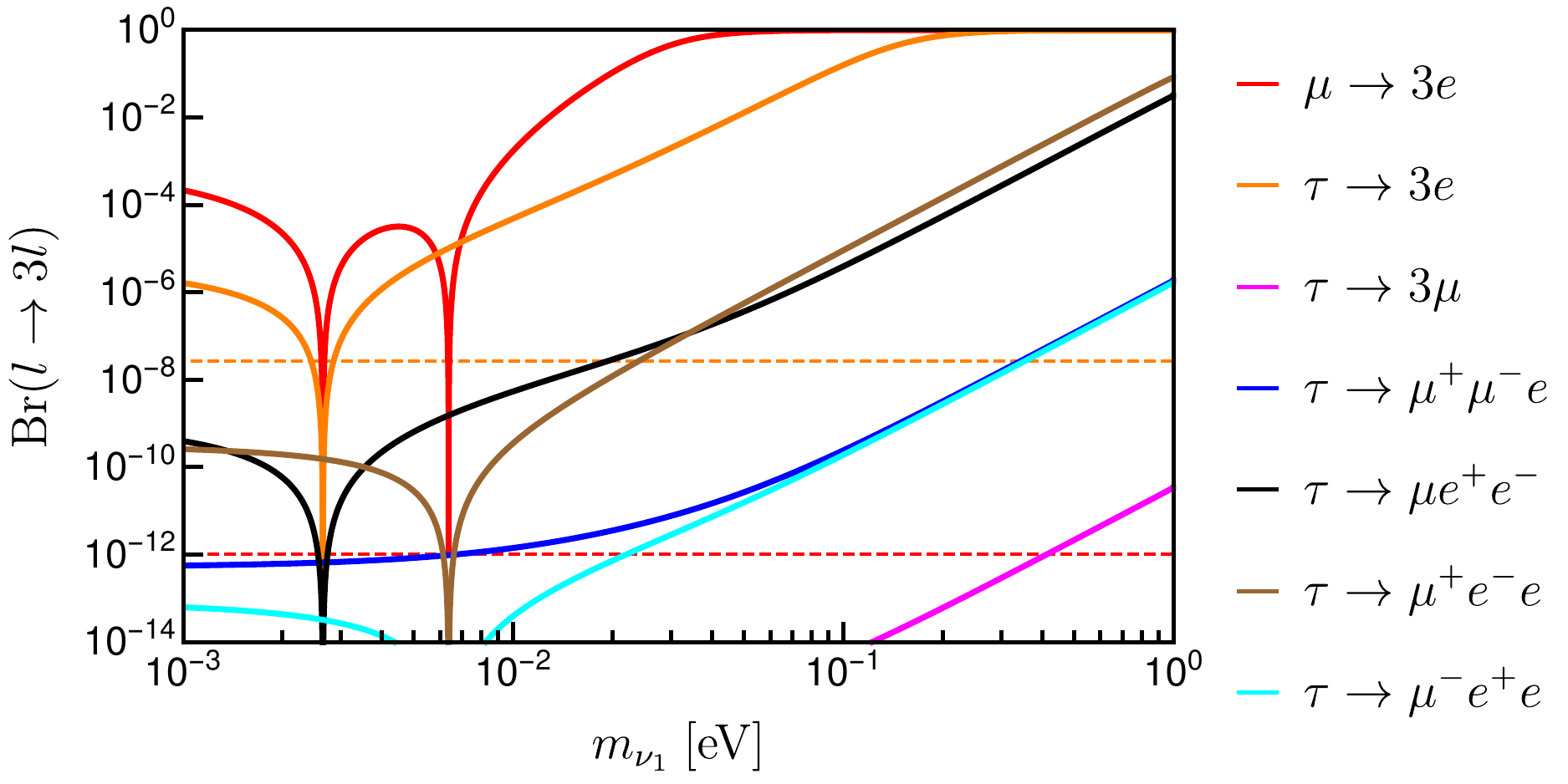}
    \hfill
    \includegraphics[width=0.43\textwidth]{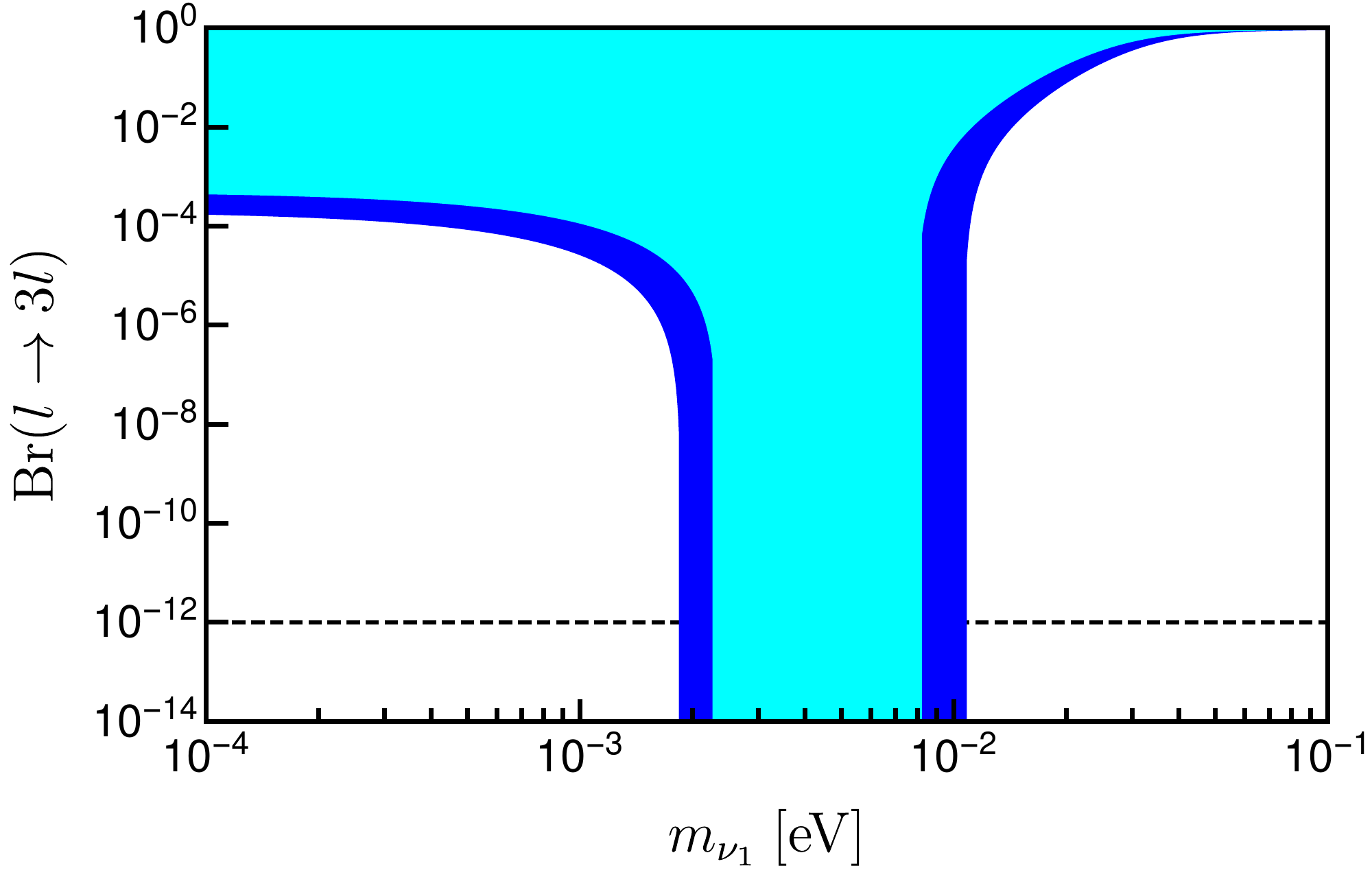}
    \caption{Br($l_i \to l_j l_k l_m$) as function of the lightest neutrino mass. To the left, all different combinations of lepton generations are considered using the b.f.p. of oscillation data and $\alpha_{12}=\pi$ and $2\delta-\alpha_{13}=0$. To the right, Br(\mueee) scanned over the uncertainty in neutrino oscillation data is shown. The light and dark blue areas correspond to the $1 \, \sigma$ and $3 \, \sigma$ uncertainties, respectively. This plot scans over the Majorana phases.}
    \label{fig:clfv:Br}
\end{figure}

Once $F_{\rm Cocktail}^{\rm max}$ is fixed, one can scan over the free parameters in the neutrino sector. Oscillation data \cite{deSalas:2017kay} fixes rather well $\Delta m^2_{\rm atm}$, $\Delta m^2_{\odot}$ and all three mixing angles; there is also an indication for a non-zero value of $\delta$. This leaves us with three essentially free parameters, the two Majorana phases and $m_{\nu_1}$, equivalent to the overall neutrino mass scale, for which there are only upper limits from neutrinoless double-$\beta$ ($0\nu\beta\beta$) decay \cite{KamLAND-Zen:2016pfg,Agostini:2018tnm} and cosmology \cite{Aghanim:2018eyx}. We note that there is a slight preference in the data for normal hierarchy (NH, also called normal ordering) over inverted hierarchy (IH).

In \fig{fig:clfv:Yuks} we plot the absolute values of the 6 independent entries in the Yukawa matrix $h$ as a function of $m_{\nu_1}$. The oscillation data have been fixed at their best fit point (b.f.p.) values, except $\delta=0$ for simplicity. The plot on the left was obtained with vanishing Majorana phases, whereas the one on the right takes $(\alpha_{12}, \alpha_{13})=(\pi,0)$. Given that we used $F_{\rm Cocktail}^{\rm max}$ in this plot, the numerical values of $h_{ij}$ are much smaller than in \eq{eq:clfv:hexa}, but $h_{11}$ is still in the non-perturbative region everywhere in the left plot. In the right plot, however, there are two special points, where cancellations among different contributions of the neutrino mass eigenstates lead to a vanishing value for either $h_{11}$ or $h_{12}$. Such cancellations are well-known in studies of $0\nu\beta\beta$ decay. The effective Majorana mass, $m_{ee}$,\footnote{$m_{ee}$, also sometimes called $\langle m_{\nu}\rangle$, is defined as $m_{ee}=\sum_j U_{ej}^2m_j$.} depends on the Majorana phases in the same way as $h_{11}$. As in $m_{ee}$, one can therefore not obtain a cancellation for the cases (i) NH without Majorana phases, and (ii) IH for any choice of parameters. The cocktail model can therefore explain neutrino data only for normal hierarchy and some particular combination of Majorana phases, as we are going to discuss now in some more detail.

As \fig{fig:clfv:Yuks} demonstrates, only in some exceptional points can $h_{11}$ be small enough to enter the perturbative region. We therefore scanned over $(\alpha_{12}, \alpha_{13})$ and $m_{\nu_1}$, in the full $3 \, \sigma$ range of oscillation data. In this scan, we calculate the CLFV observable Br($l_i \to l_j l_k l_m$), with different combinations of lepton flavours, for the minimal value of $m_{\rho^{++}}$ allowed by LHC data. \Fig{fig:clfv:Br} to the left shows Br($l_i \to l_j l_k l_m$) for all different combinations of $i,j,k,m$ using the b.f.p. of neutrino oscillation data. The most stringent constraint on the model comes from the experimental upper limit on Br(\mueee)$\le 10^{-12}$ \cite{Bellgardt:1987du}. The plot to the right then shows the allowed regions in parameter space, scanning over the complete range of oscillation parameters and phases. All acceptable points lie in the range $m_{\nu_1} = (2-10)$ meV.

\begin{figure}[t!]
    \centering
    \includegraphics[width=0.495\textwidth]{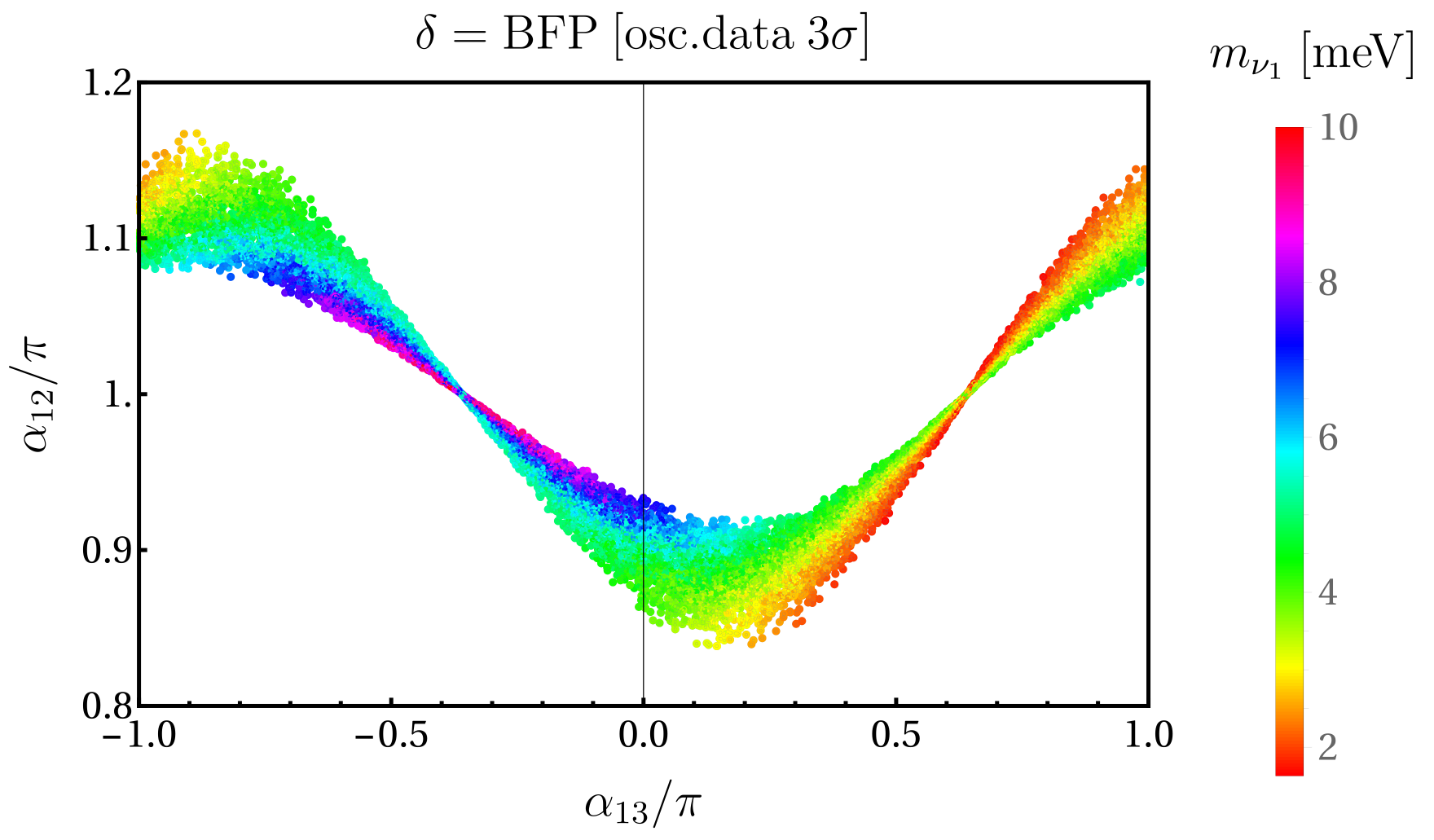}
    \hfill
    \includegraphics[width=0.47\textwidth]{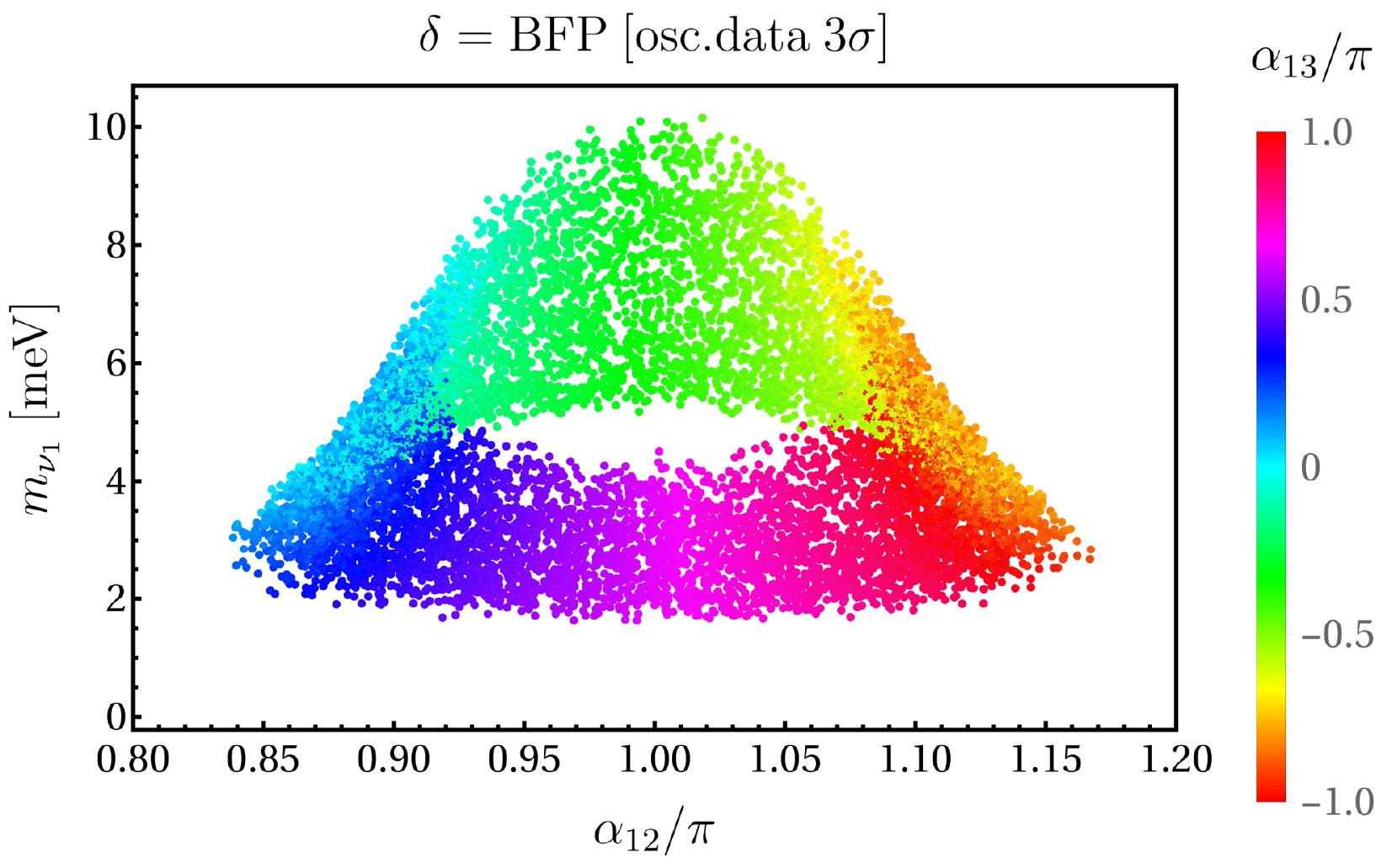}
    \caption{Allowed parameter space for $\alpha_{12}$, $\alpha_{13}$ and $m_{\nu_1}$. Note that $m_{\nu_1}$ is shown in units of meV. Neutrino oscillation data was scanned over the $3 \, \sigma$ uncertainties, except $\delta$ which is taken at its best fit value for simplicity.}
    \label{fig:clfv:ph}
\end{figure}

\Fig{fig:clfv:ph} shows a scan over the allowed range of Majorana phases and the lightest neutrino mass. The plot to the left shows the plane ($\alpha_{13},\alpha_{12}$), the one to the right ($\alpha_{12},m_{\nu_1}$). The model can fulfill the constraint from Br(\mueee) only in a very narrow range of phases. In particular, $\alpha_{12}$ has to be close to $\pi$ in all points, while also $m_{\nu_1}$ is fixed in a rather narrow interval.

\begin{figure}
    \centering
    \includegraphics[width=0.65\textwidth]{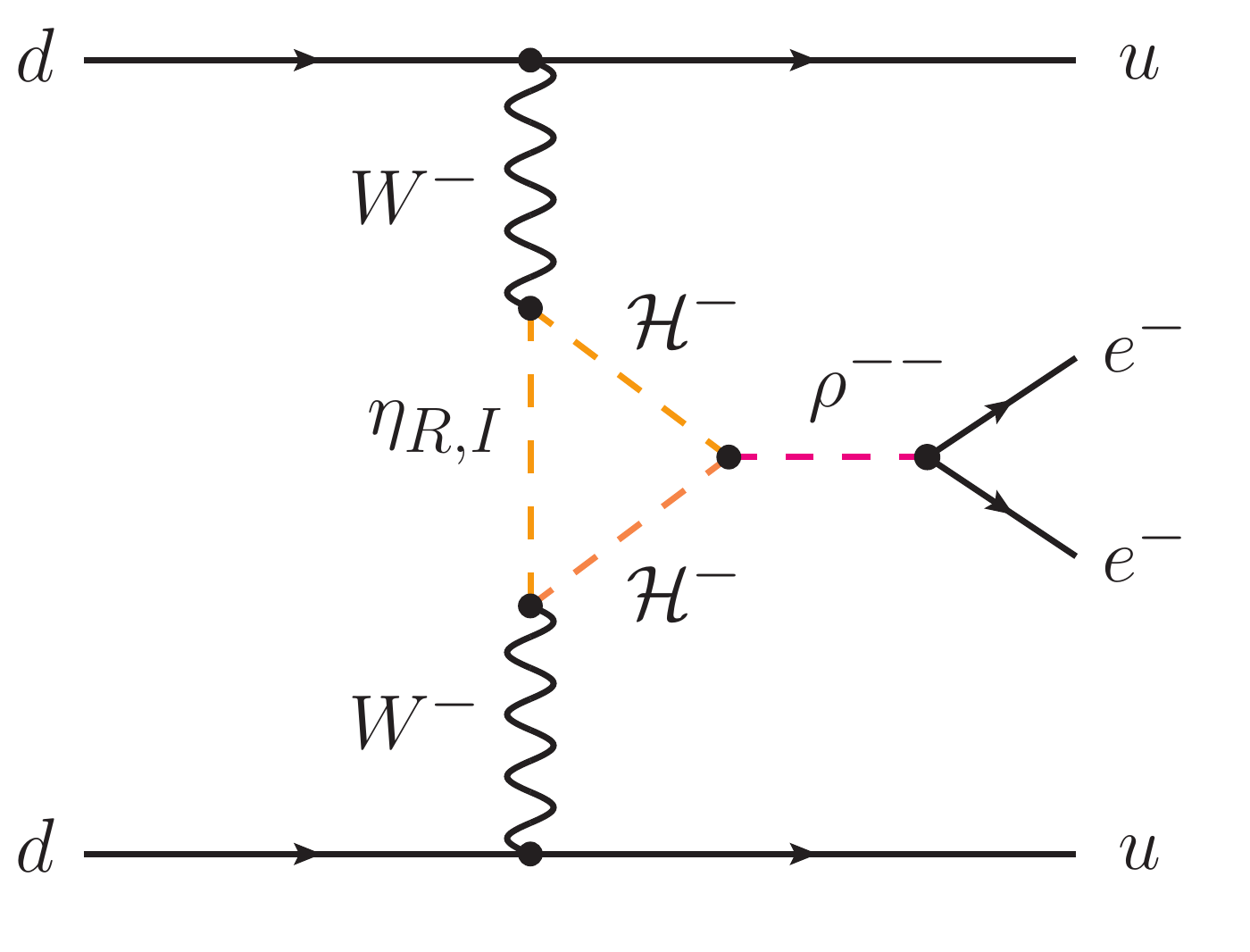}
    \caption{One-loop neutrinoless double-$\beta$ decay diagram in the cocktail model.}
    \label{fig:clfv:cocktail_0nbb}
\end{figure}

Finally we note that the acceptable points of the model lie in regions of parameter space where the $0\nu\beta\beta$ decay observable $m_{ee}$ is unmeasurably small. There is, however, a one-loop short-range diagram contributing to $0\nu\beta\beta$ decay in the cocktail model \cite{Gustafsson:2014vpa}, see \fig{fig:clfv:cocktail_0nbb}. This diagram depends on the same parameters as the three-loop neutrino mass diagram in \fig{fig:clfv:cocktail}. In particular, note that the sum over $\eta_{R,I}$ generates the same dependence on $\lambda_5$ as for the neutrino mass.

We have calculated this diagram and estimated its contribution to the $0\nu\beta\beta$ decay half-life, including the QCD running of the short-range operator \cite{Gonzalez:2015ady, Arbelaez:2016uto}. Using the same mass parameters that maximise the three-loop diagram, in particular $m_{\rho^{++}}=800$ GeV, the current limit on the half-life of $^{136}$Xe \cite{KamLAND-Zen:2016pfg} imposes a limit on $h_{11}$ of roughly $|h_{11}|\lesssim 5 \times 10^{-4}$. This limit is around a factor $\sim 7$ more stringent than the one obtained from the upper limit on Br(\mueee).\footnote{See \cite{Alcaide:2017xoe} for a variant of the cocktail model inducing a different one-loop short-range $0\nu\beta\beta$ decay diagram.}

We can conclude that the cocktail model is severely constrained from perturbativity arguments and from searches for CLVF. The model has acceptable points only within a narrow window of $m_{\nu_1}$ and for particular combinations of the Majorana phases.

%%%%%%%%%%%%%%%%%%%%%%%%%%%%%%%%%%%%%%%%%%%%%%%%%%%%%%%%%%%%%%%
%%%%%%%%%%%%%%%%%%%%%%%%%%%%%%%%%%%%%%%%%%%%%%%%%%%%%%%%%%%%%%%
\section{KNT model} \label{sec:clfv:KNT}

We continue with the KNT model \cite{Krauss:2002px}. This was the first radiative neutrino mass model at three-loop order proposed.
\\

In addition to the Standard Model particles, the KNT model contains three copies of the fermionic singlet $N$ and two singly charged singlet scalars $X$ and $S$. A discrete $Z_2$ symmetry is imposed, under which $S$ and $N$ are odd and the rest of particles in the model are even. The quantum numbers of the new particles in the KNT model are given in \tab{tab:clfv:KNT}.

\begin{table}[t!]
    \centering
    \begin{tabular}{| c c c c c c |}
        \hline  
         & generations & $\mathrm{SU(3)_C}$ & $\mathrm{SU(2)_L}$ & $\mathrm{U(1)_Y}$ & $Z_2$ \\
        \hline
        \hline    
        $X$ & 1 & ${\bf 1}$ & ${\bf 1}$ & $1$ & $+$ \\
        $S$ & 1 & ${\bf 1}$ & ${\bf 1}$ & $1$ & $-$ \\
        \hline
        \hline    
        $N$ & 3 & ${\bf 1}$ & ${\bf 1}$ & $0$ & $-$ \\  
        \hline
        \hline
    \end{tabular}
    \caption{New particles in the KNT model with respect to the Standard Model.}
    \label{tab:clfv:KNT}
\end{table}

The Lagrangian of the model contains the following pieces,
\begin{align} \label{eq:clfv:YukKNT}
    -\mathcal L &\supset f \, \overline{L^c} \, L \, X
    + g^\ast \, \overline{N^c} \, e_R \, S
    + \frac{1}{2} M_N \overline{N^c} N + \hc \, ,
\end{align}
where we have omitted $\mathrm{SU(2)_L}$ and flavour indices to simplify the notation. We note that $f$ is an antisymmetric $3 \times 3$ Yukawa matrix, while $M_N$ is a symmetric $3 \times 3$ Majorana mass matrix, which we take to be diagonal without loss of generality. The scalar potential of the model also contains additional terms besides those in the Standard Model. These are given by,
\begin{align} \label{eq:clfv:PotKNT}
    \mathcal V &\supset M_{X}^2 |X|^2 + M_{S}^2 |S|^2 + \frac{1}{2} \lambda_1 \, |X|^4 + \frac{1}{2} \lambda_2 \, |S|^4 + \lambda_{12} \, |X|^2 |S|^2 
    \nn \\
               &+ \lambda_H^{(1)} \, |H|^2 |X|^2 + \lambda_H^{(2)} \, |H|^2 |S|^2 + \frac{1}{4} \, \left[ \lambda_S \, (X S^\ast)^2 + \hc \right] \, .
\end{align}
The presence of the $\lambda_S$ quartic coupling precludes the definition of a conserved lepton number. Indeed, one can easily see that the simultaneous presence of the Lagrangian terms in \eq{eq:clfv:YukKNT} and \eq{eq:clfv:PotKNT} breaks lepton number in two units. The masses of the physical scalar states in the KNT model are given by,
\begin{eqnarray}
    m_H^2&=&\lambda \, v^2 \, ,
    \\
    m_{s_1}^2&=& M_{X}^2 + \frac{1}{2} \, \lambda_H^{(1)} \, v^2 \, ,
    \\
    m_{s_2}^2&=& M_{S}^2 + \frac{1}{2} \, \lambda_H^{(2)} \, v^2 \, .
\end{eqnarray}

We also note that the lightest $Z_2$-odd state in the KNT model is completely stable. Assuming the hierarchy $M_{N_1} < m_{s_1} < m_{s_2}$, this state is the lightest fermion singlet, which then constitutes a good dark matter candidate. In fact, the KNT model is historically the first radiative neutrino mass theory with a stable dark matter candidate running in the loop.

%%%%%%%%%%%%%%%
\subsubsection*{Neutrino masses}

\begin{figure}[t!]
    \centering
    \includegraphics[width=0.78\textwidth]{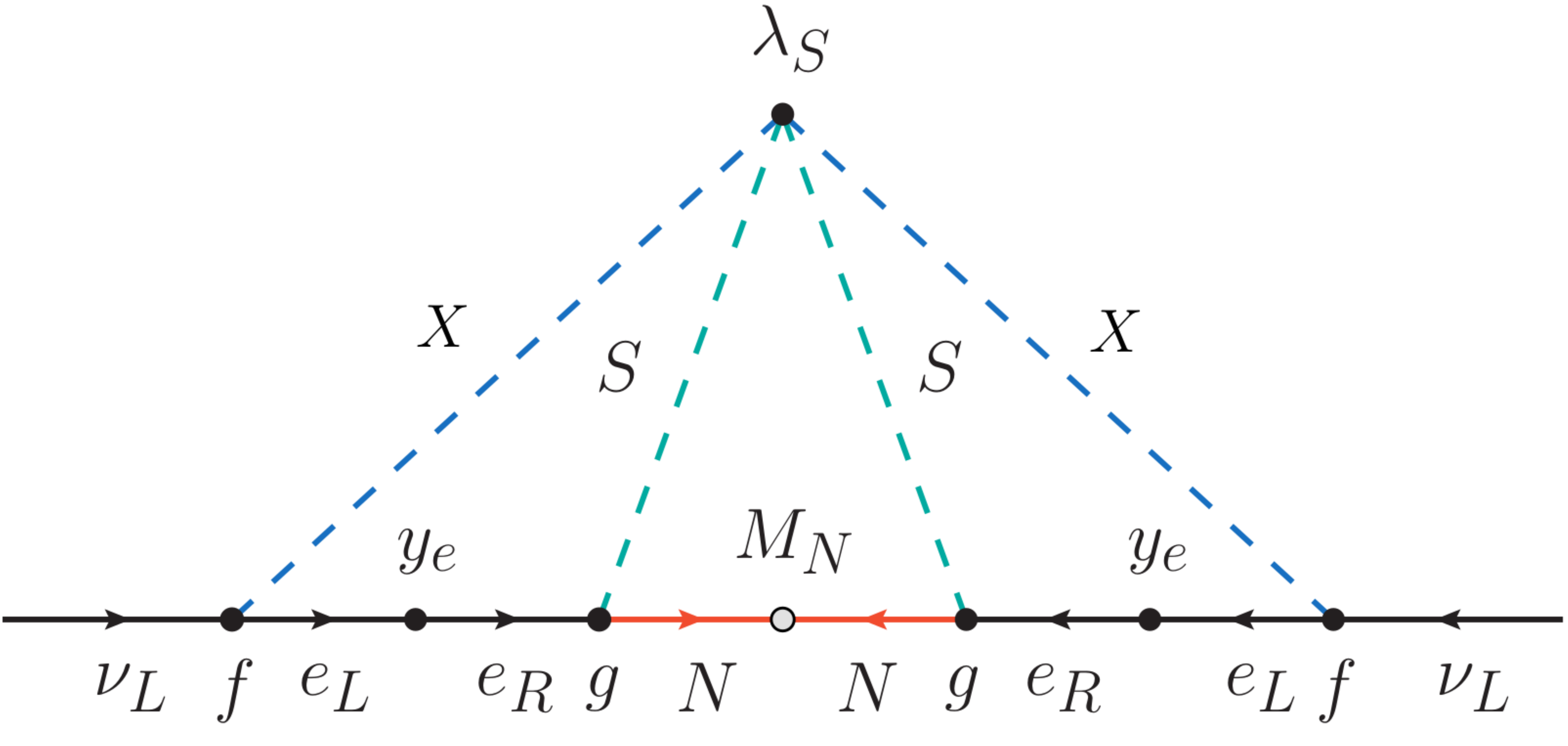}
    \caption{Three-loop neutrino masses in the KNT model.}
    \label{fig:clfv:KNT}
\end{figure}

The $Z_2$ symmetry forbids the standard Higgs Yukawa coupling with the lepton doublet $L$ and the $N$ singlets. Therefore, the usual type-I seesaw contribution at tree-level is absent. Instead, neutrino masses are generated at three-loop order as shown in \fig{fig:clfv:KNT}. The neutrino mass matrix is given by,
\begin{equation} \label{eq:clfv:MnuKNT}
    \left(\mathcal{M}_\nu\right)_{ij} = \frac{2 \lambda_S}{(16 \pi^2)^3} \sum_{\alpha \beta a}
    \frac{m_\alpha m_\beta}{M_{N_a}} f_{i\alpha} f_{j\beta}
    g_{\alpha a} g_{\beta a} \, F_{\rm KNT} \, ,
\end{equation}
where $m_\alpha$ is the mass of the $\ell_\alpha$ charged lepton and $F_{\rm KNT}$ is a loop function that depends on the masses of the scalars and fermions running in the loops. More information about this function can be found in \app{app:loops}.

It is important to stress that in the KNT model, each entry in $(\mathcal{M}_\nu)_{ij}$ contains the sum over the Standard Model charged lepton masses. Therefore, different from the other models discussed in this chapter, the suppression of the entries in $\mathcal{M}_\nu$ is at most $m_\mu^2$. The neutrino fit can then reproduce experimental data with Yukawas which are considerably smaller than in the cocktail or AKS models.

%%%%%%%%%%%%%%%%%%%%%%%%%%%%%%%%%%%%%%%%%%%%%%%%%%%%%%%%%%%%%%%
\subsection{Results} \label{subsec:clfv:results-KNT}

We start this section again with a discussion of the neutrino mass fit. The coupling $f$ in \eq{eq:clfv:YukKNT} is antisymmetric, thus the determinant of the neutrino mass matrix in \eq{eq:clfv:MnuKNT} is zero, implying that one neutrino is massless. This is reminiscent of the two-loop Babu-Zee model of neutrino mass \cite{Zee:1985id,Babu:1988ki}, where the same singly charged scalar is used. In our fitting procedure we use therefore an adapted version of the solution found in \cite{Babu:2002uu, Herrero-Garcia:2014hfa} for the Babu-Zee model.

The procedure consists of two steps. First, because $\det(f)= 0$, the matrix has one eigenvector ${\bf a}= (f_{23},-f_{13},f_{12})$, which is also an eigenvector of $\mathcal{M}_\nu$,
\begin{equation} \label{eq:clfv:defa}
    \widehat{\cal M}_\nu \, U^T{\bf a}=0 \, .
\end{equation}
This implies three equations, one of which is trivial, while the other two allow to express the ratios $(f_{13}/f_{12}, f_{23}/f_{12})$ as functions of the neutrino angles and phases only. These solutions depend on the neutrino mass hierarchy.

Next, we can write the neutrino mass matrix as,
\begin{equation} \label{eq:clfv:Maux}
    \mathcal{M}_\nu = -c \, f \, M_{\rm aux} \, f \, .
\end{equation}
$c$ contains all global constants, we have used $f^T=-f$ and $M_{\rm aux}$ is an auxiliary matrix, which is complex symmetric. This defines a set of 6 complex equations relating the entries in $M_{\rm aux}$ to neutrino data. With three independent entries $f_{ij}$, we can use three of the six equations to express three entries in $M_{\rm aux}$ as a function of the remaining ones, neutrino data and $f_{ij}$. The resulting equations are very lengthy and not at all illuminating, so we do not present them here.

The definition of $M_{\rm aux}$ in \eq{eq:clfv:Maux} shows that,
\begin{equation} \label{eq:clfv:Maux2}
    M_{\rm aux} = {\widehat{\cal M}_{e}} \, g \left({\widehat M^{\rm eff}}\right)^{-1}
    g^T {\widehat{\cal M}_{e}} \, ,
\end{equation}
where
\begin{equation} \label{eq:clfv:MNeff}
    \left({\widehat M^{\rm eff}}\right)^{-1} =
    \begin{pmatrix}
        \frac{F_{KNT}(r^X_1,r^S_1)}{M_{N_1}} & 0 & 0
        \\
        0 & \frac{F_{KNT}(r^X_2,r^S_2)}{M_{N_2}} & 0
        \\
        0 & 0 & \frac{F_{KNT}(r^X_3,r^S_3)}{M_{N_3}} 
    \end{pmatrix}
    \, ,
\end{equation}
and $r^X_i=(m_{s_1}/M_{N_i})^2$, $r^S_i=(m_{s_2}/M_{N_i})^2$. With $M_{\rm aux}$ being complex symmetric, we can use a suitably modified \cite{Cordero-Carrion:2018xre, Cordero-Carrion:2019qtu} Casas-Ibarra parametrisation \cite{Casas:2001sr} to express the matrix $g$ as,
\begin{equation}
    g = \sqrt{{\widehat M^{\rm eff}}}{\cal R} \, \sqrt{\hat M_{\rm aux}} \, U_{\rm aux}^T
    \left({\widehat{\cal M}_{e}}\right)^{-1} \, .
\end{equation}
${\hat M_{\rm aux}}$ and $U_{\rm aux}$ are the eigenvalues and eigenvectors of the auxiliary matrix $M_{\rm aux}$. Although in principle it would be possible to determine ${\hat M_{\rm aux}}$ and $U_{\rm aux}$ in terms of the input neutrino data analytically, in practice we find these two matrices numerically for any input point of experimental data and choice of free parameters. Finally, ${\cal R}$ is a $3 \times 3$ orthogonal matrix.

In summary, neutrino oscillation data provides 6 constraints: $\Delta m_{\rm atm}^2$, $\Delta m_{\odot}^2$, three angles and the CP-phase $\delta$. A number of free parameters can then be scanned over, using the above procedure. In the neutrino sector we still have $\alpha_{12}$.\footnote{Since one neutrino is massless, only one of the two Majorana phases, i.e. $\alpha_{12}$, is physical.} The matrix $f$ is fixed from experimental data, up to the overall scale of the matrix. We choose $f_{12}$ as the free parameter. The matrix ${\cal R}$, in the most general case, contains 3 complex angles. There are three right-handed neutrino masses, $M_{N_i}$, and two scalar masses, $m_{s_{1,2}}$. And, finally, we can use 3 of the 6 equations for $M_{\rm aux}$ to eliminate some particularly chosen $(M_{\rm aux})_{ij}$. This leaves as free inputs the remaining 3 entries in $M_{\rm aux}$.

Up to now, we have been completely general in our discussion. However, there is still certain freedom as to which 3 entries in $(M_{\rm aux})_{ij}$ we fix via 3 of the equations defined by \eq{eq:clfv:Maux}. In practice, we choose to solve for $(M_{\rm aux})_{22}$, $(M_{\rm aux})_{23}$ and $(M_{\rm aux})_{33}$ and assume $(M_{\rm aux})_{1k} = 0$. This particular choice is motivated by the observation that in this limit all terms in $g \propto 1/m_e$ disappear. In other words, this solution guarantees that the contribution to \mueg and \taueg from loops involving $s_2$ and $N_i$ are automatically absent in our scans, due to $g_{1k}=0$ $\forall k$. Our ansatz is therefore the optimal choice for minimising fine-tuning on the other parameters in $g$.\footnote{We have explored other ans\"atze but concluded that this is indeed the optimal one. For instance, we can generate textures with either the 2nd or 3rd column of $g$ vanishing. These will make the \taueg or \taumug branching ratio vanish, but at the cost of a large \mueg branching ratio. We have also considered a solution with any or all of $(M_{\rm aux})_{1k}\ne 0$. However, in this case we did not find any configuration for the remaining free parameters that induces a cancellation that suppresses the \mueg branching ratio.}

Before we explore the remaining parameter space of the model, we must consider lower limits on the masses of the charged scalars from accelerator searches. LEP provides a lower limit on charged particles decaying to leptons plus missing momentum, which will essentially rule out all values of $m_{s_1}$ below $100$ GeV \cite{Tanabashi:2018oca} and similarly for $m_{s_2}$, unless $M_{N_1}$ is close to $m_{s_2}$.\footnote{There are no accelerator limits on $N$, since the $Z_2$ symmetry prohibits their mixing with the active neutrinos.} At the LHC there is currently no specific search for particles with the quantum numbers of $S$ and $X$. However, slepton pair production with their subsequent decays to a lepton plus a neutralino provide the same signal and thus, we can make a reinterpretation of the corresponding searches at CMS \cite{Sirunyan:2018nwe} and ATLAS \cite{Aad:2019vnb}. The CMS slepton search \cite{Sirunyan:2018nwe} is based on $35.9/$fb, while ATLAS' chargino and slepton search \cite{Aad:2019vnb} uses $139/$fb. The ATLAS limits are correspondingly more stringent, and we will therefore discuss these. We have implemented the KNT model in \texttt{SARAH} \cite{Staub:2012pb,Staub:2013tta} and generated \texttt{SPheno} routines \cite{Porod:2003um,Porod:2011nf} and model files for \texttt{MadGraph} \cite{Alwall:2007st,Alwall:2011uj,Alwall:2014hca}. We have then calculated cross sections with \texttt{MadGraph} to recast the results of \cite{Aad:2019vnb}. For $s_1$ the mass range between (very roughly) $m_{s_1}=(250-400)$ GeV is excluded by this search. The range $m_{s_1}=(100-250)$ GeV is currently unconstrained, due to large backgrounds in \cite{Aad:2019vnb}. For $s_2$ the limits are even weaker, unless $m_{s_2}-M_{N_1}$ is larger than ($50-70$) GeV, depending on $m_{s_2}$.
\\

Let us now turn to the discussion of CLFV. Consider first the antisymmetric Yukawa coupling $f$. Neutrino data requires all three elements of $f$ to be non-zero, thus there will always be a non-zero value for the three possible decays \mueg, \taueg and \taumug. The constraint from \mueg is the most stringent one. However, since we still have the overall scale of $f$ as a free parameter, in our choice the value of $f_{12}$, we can use it to fix Br(\mueg) to the upper limit (present or future) for any point in the parameter space. Since the neutrino mass matrix is proportional to the square of the matrices $f$ and $g$ however, once this choice is made there is no longer any overall scaling freedom in the coupling $g$. Putting the calculated Br(\mueg) to equal the experimental bound will generate the smallest values for the entries of $g$ allowed in the model parameter space. A smaller upper limit on Br(\mueg) will lead to larger $g$ and thus more stringent constraints from \taumug.

\begin{figure}[t!]
    \centering
    \includegraphics[width=0.49\textwidth]{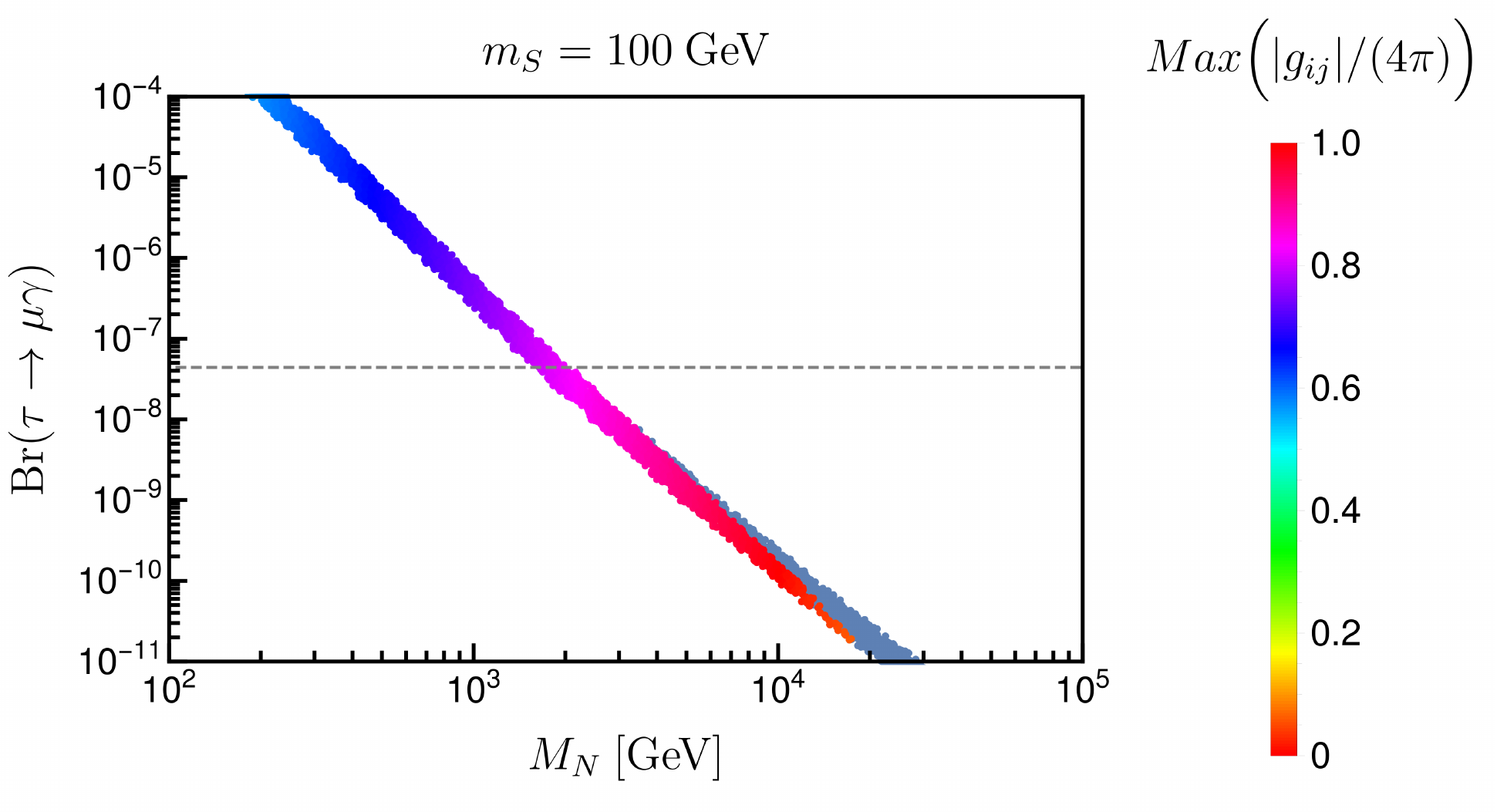}
    \hfill
    \includegraphics[width=0.49\textwidth]{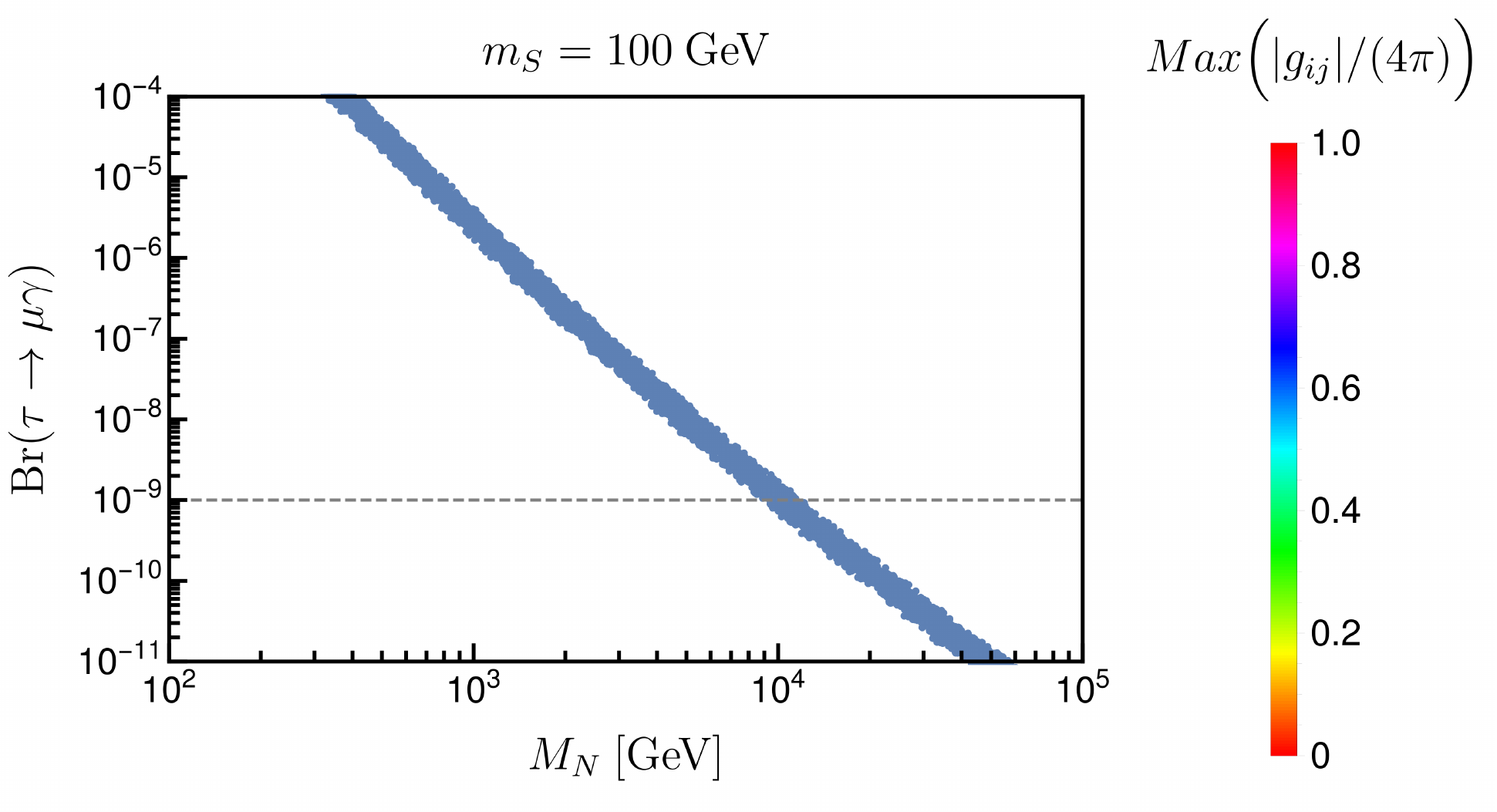}
    \caption{Calculated values for Br(\taumug) as function of the common fermion mass $M_N$, for the current lower limit on $m_{s_1}=m_{s_2}=100$ GeV. To the left, with current experimental limits; to the right for the expected future experimental limits. The required size of the Yukawas is colour-coded. Bluish-grey points mean that at least one entry in $g$ is larger than $4\pi$.}
    \label{fig:clfv:KNTTauMuG}
\end{figure}

We then scanned over the remaining parameters of the model numerically. Consider first the case of NH. Some examples for Br(\taumug) are shown in \fig{fig:clfv:KNTTauMuG}. In this plot we have chosen the fixed value $m_{s_{1,2}}=100$ GeV, corresponding to the experimental lower limit, and the three right-handed neutrino masses all equal to a common $M_N$.\footnote{For degenerate $M_{N_i}$ ${\cal R}$ becomes unphysical and drops out of the calculation.} The points are scanned over the allowed $3 \, \sigma$ ranges for the neutrino data for NH. The size of the largest entry in $g$ is colour-coded in the points. The plot to the left has been calculated for the current experimental limit on Br(\mueg)$< 4.2 \times 10^{-13}$ \cite{TheMEG:2016wtm}, the plot to the right is for the expected future limit Br(\mueg)$< 6 \times 10^{-14}$ \cite{Baldini:2013ke}. For the choice of $m_{s_{1,2}}=100$ GeV no valid point with $g_{ij}\le 4\pi$~$\forall ij$ remains in the parameter space. Constraints are more stringent for IH, and therefore the same conclusion is reached.

\begin{figure}[t!]
    \centering
    \includegraphics[width=0.49\textwidth]{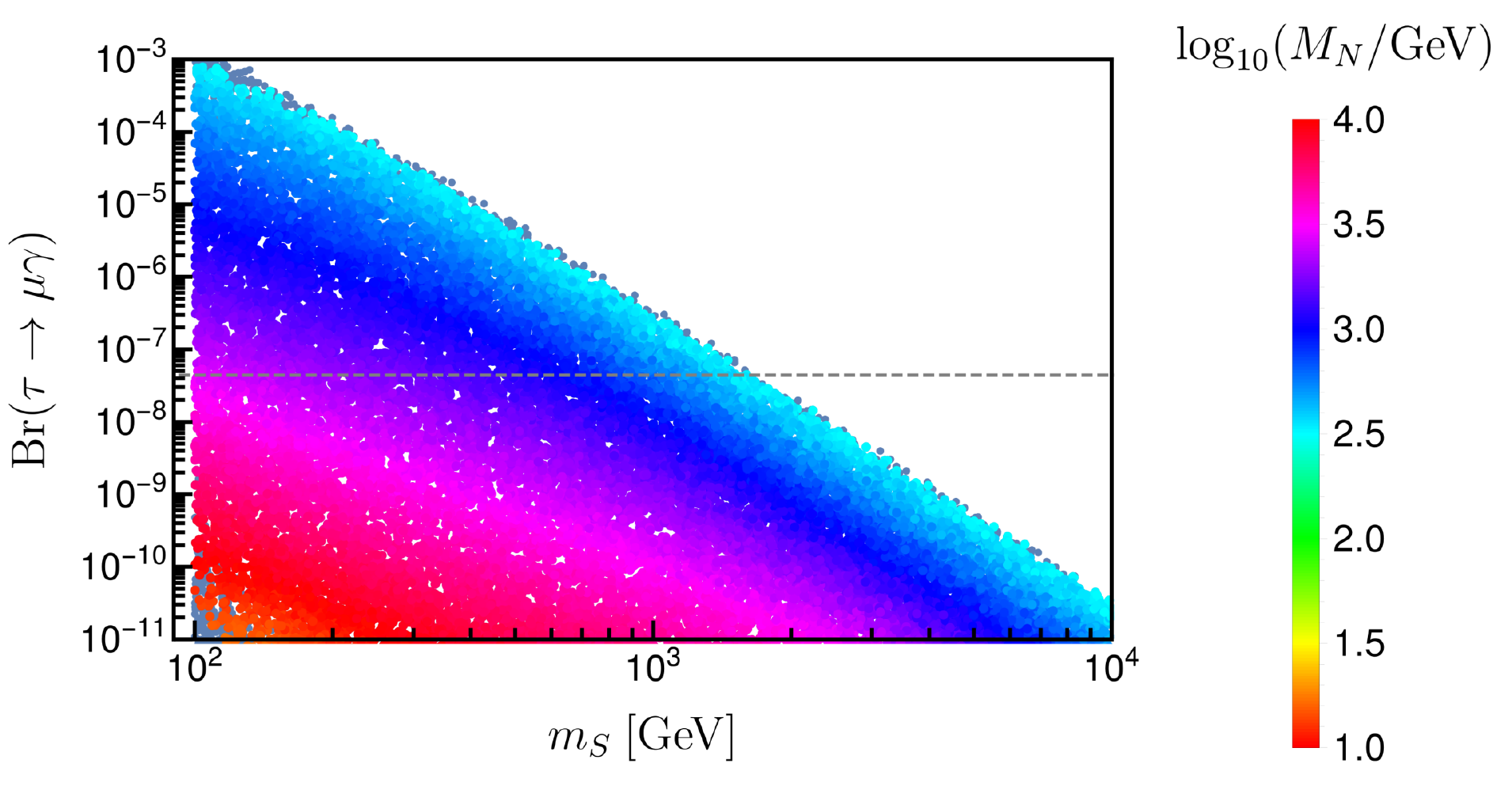}
    \hfill
    \includegraphics[width=0.49\textwidth]{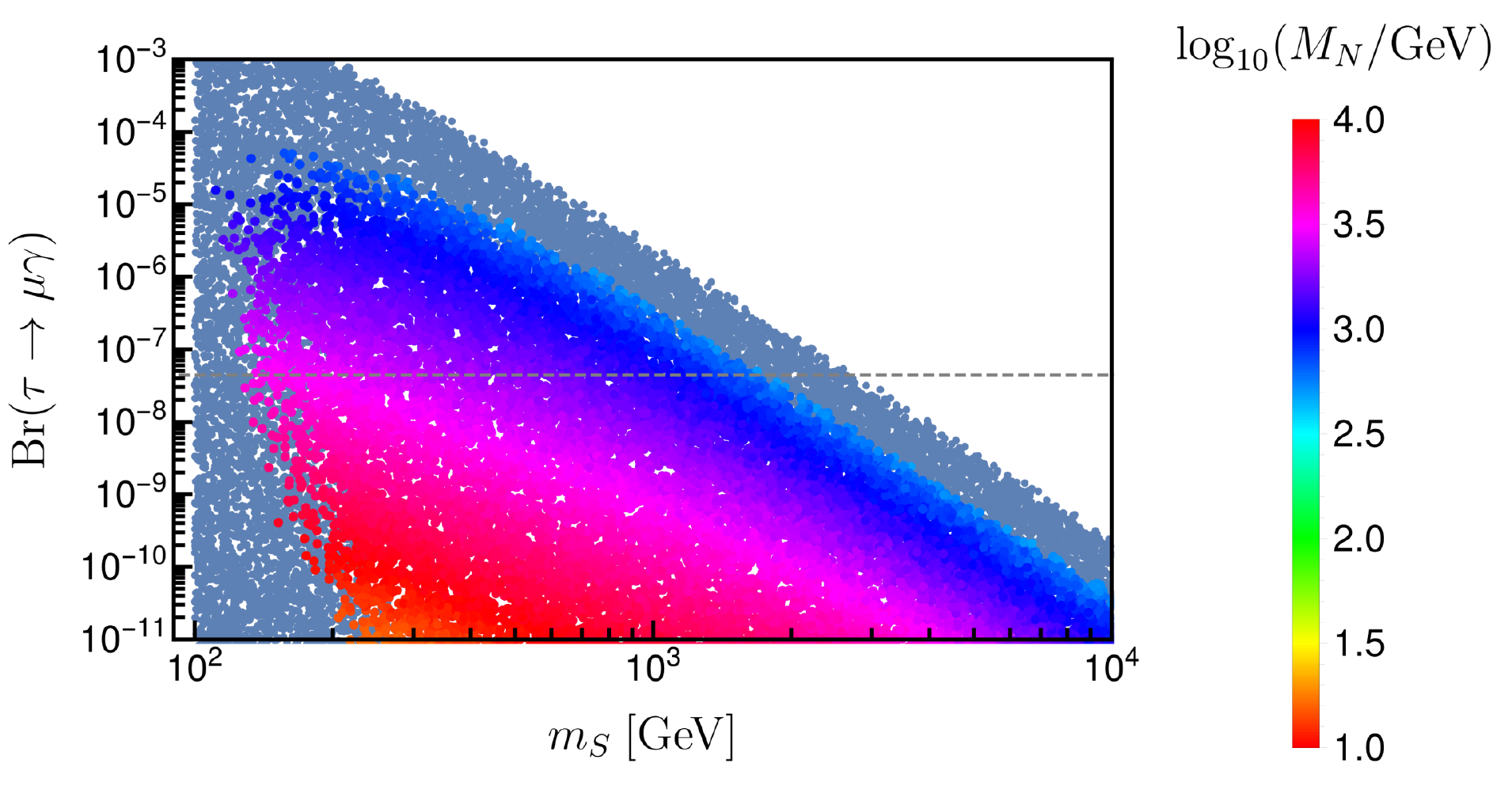}
    \caption{Calculated values for Br(\taumug) as function of the scalar mass $m_s$ for different values of $M_N$ (colour-coded). Bluish points are ruled out by non-perturbative couplings. To the left, with current experimental limits; to the right for the expected future experimental limit on Br(\mueg).}
    \label{fig:clfv:KNTTauMuG2}
\end{figure}

We therefore scanned over $m_{s_{1,2}} \equiv m_S$ and $M_{N_i}$ simultaneously. The results are shown in \fig{fig:clfv:KNTTauMuG2}. Here, $M_{N_i}$ are varied within $20\%$ of a common $M_N$. The range of $M_N$ is colour-coded in the points. Again, the plot to the left is for the current bound on Br(\mueg), while the plot to the right is for the future bound. In these plots, points with non-perturbative couplings are shown in bluish colour. This bound eliminates all points below roughly $M_N = {\cal O}(100)$ GeV already with the current experimental bound on Br(\mueg), see however the discussion below. We show only the cases with a trivial ${\cal R}$ matrix. For non-zero angles in ${\cal R}$ the results look similar, although fewer points lie in the perturbative regime.

The combined constraints of perturbativity and future limits from CLFV searches would put a lower bound on $m_S$ roughly of order ($180-200$) GeV. This limit becomes stronger for lower values of $M_N$, as the plots shows.

\begin{figure}[t!]
    \centering
    \includegraphics[width=0.49\textwidth]{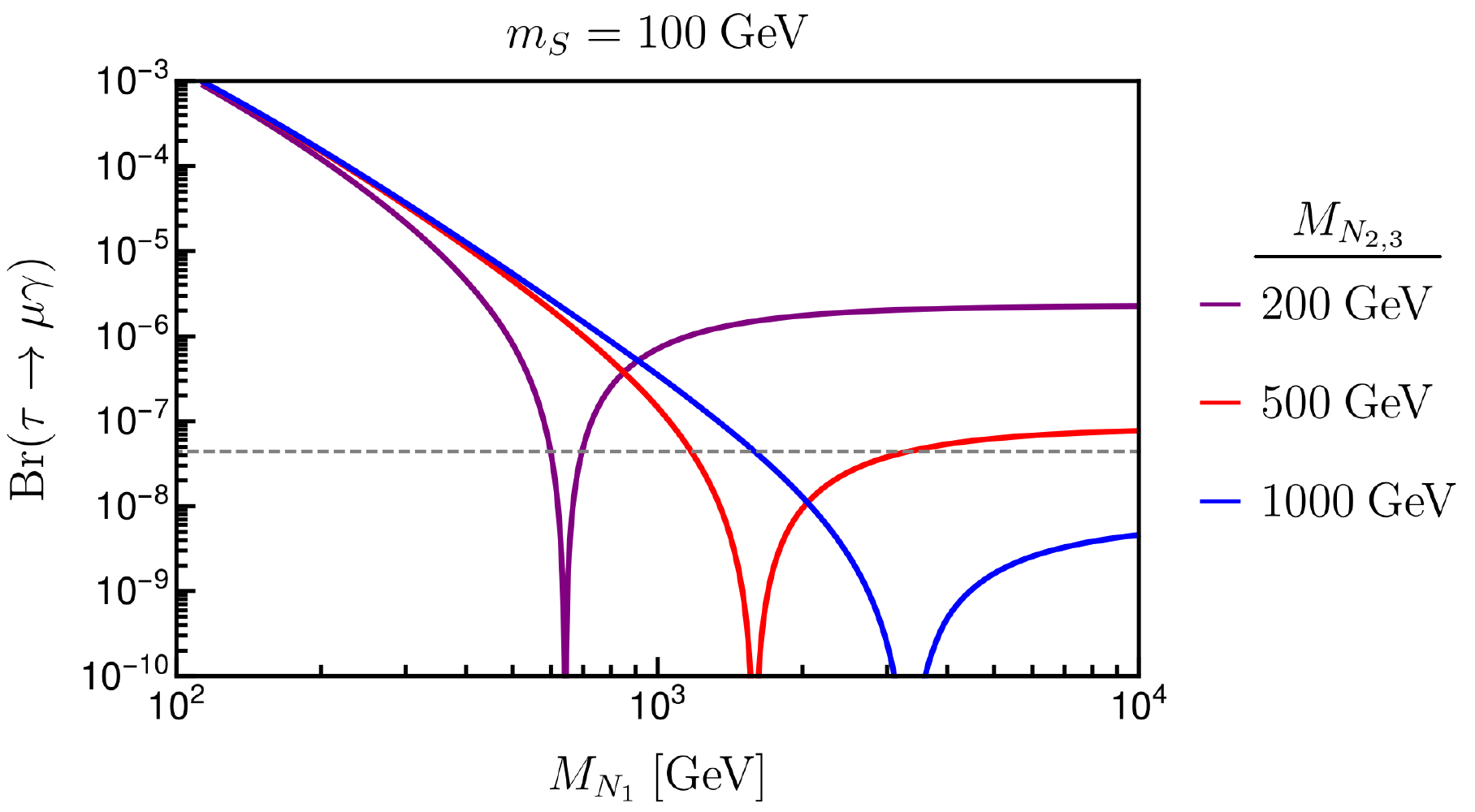}
    \hfill
    \includegraphics[width=0.49\textwidth]{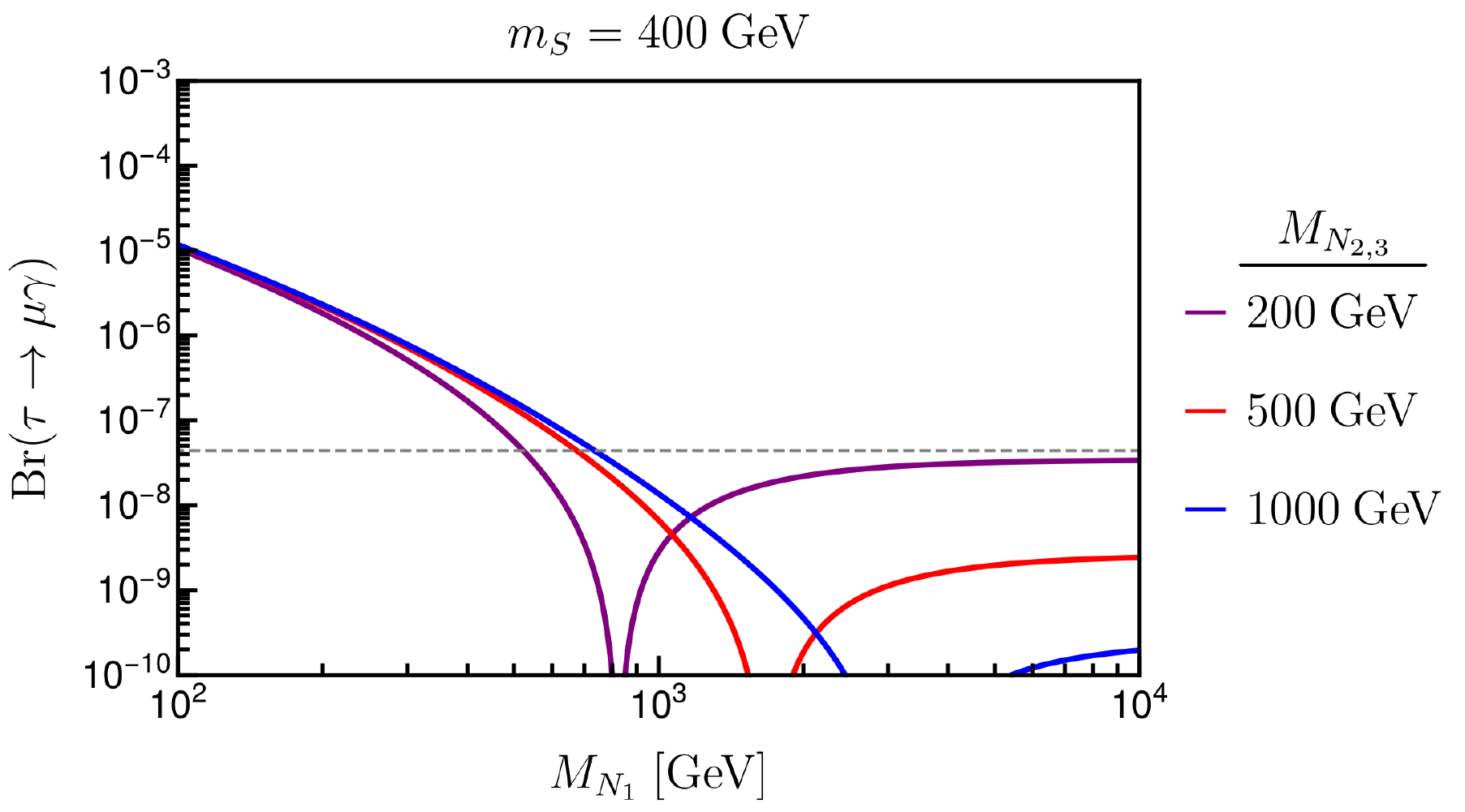}
    \caption{Calculated values for Br(\taumug) as function of the lightest right-handed neutrino mass $M_{N_1}$ for different choices of $M_{N_{2,3}}$ and two different scalar masses $m_S$ (left and right). The calculation uses for simplicity the b.f.p. for neutrino data.}
    \label{fig:clfv:KNTTauMuG3}
\end{figure}

The above discussion is strictly valid only for the case where the three right-handed neutrinos have similar masses. For hierarchical right-handed neutrinos the constraints are usually dominated by the lightest of these. There exist, however, exceptional points in the parameter space, where the contributions to Br(\mueg) from the three different neutrinos conspire to (nearly) cancel each other. This is shown in \fig{fig:clfv:KNTTauMuG3}. The figure shows Br(\taumug) as a function of the ``lightest'' right-handed neutrino mass, for different choices of $M_{N_{2,3}}$. Br(\taumug) is dominated by the lightest mass eigenstate, except in some particular points, where cancellations occur. \Fig{fig:clfv:KNTTauMuG} and \fig{fig:clfv:KNTTauMuG2} do not cover these exceptional combinations of parameters.

\begin{figure}[t!]
    \centering
    \includegraphics[width=0.49\textwidth]{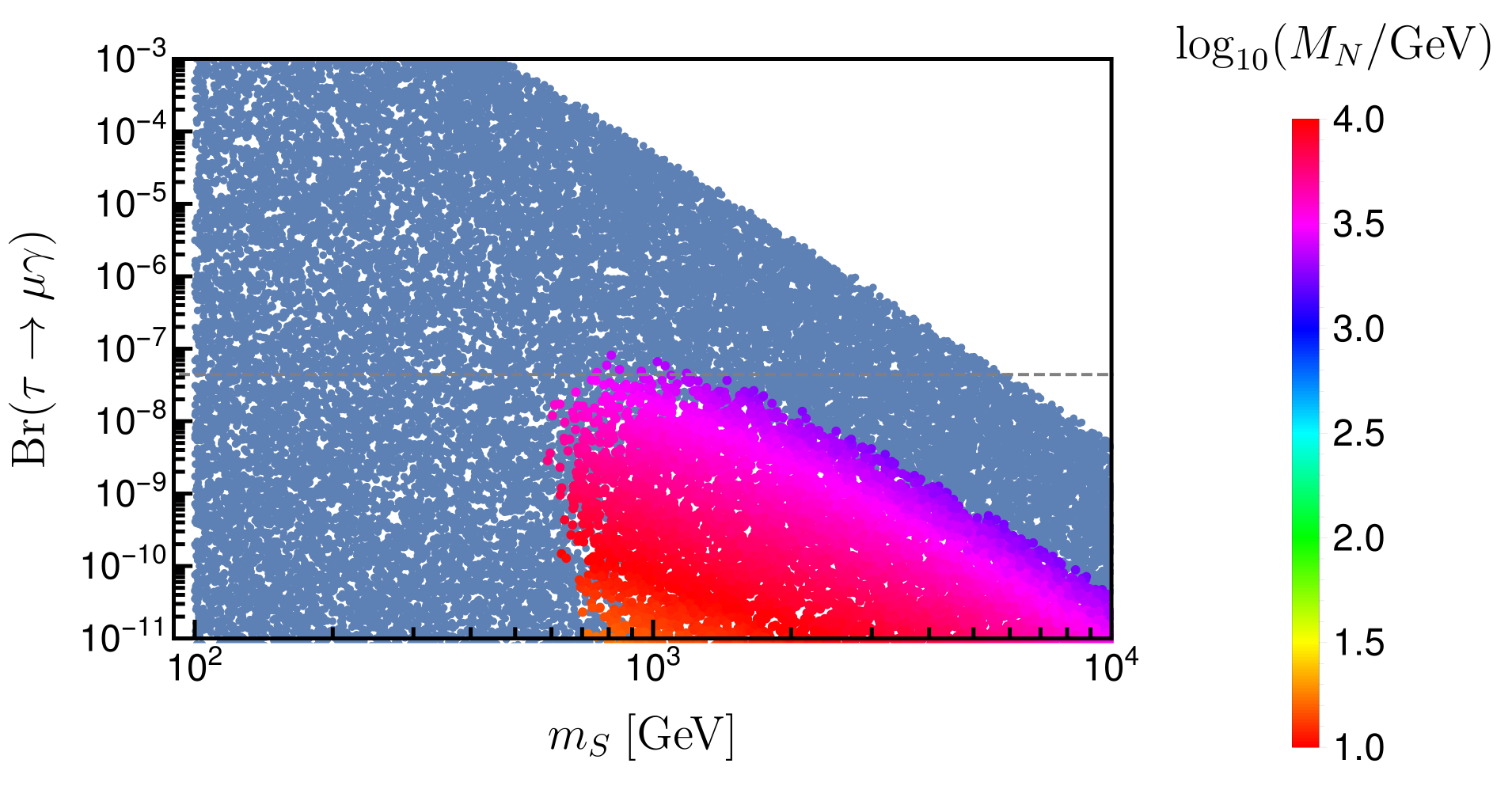}
    \hfill
    \includegraphics[width=0.49\textwidth]{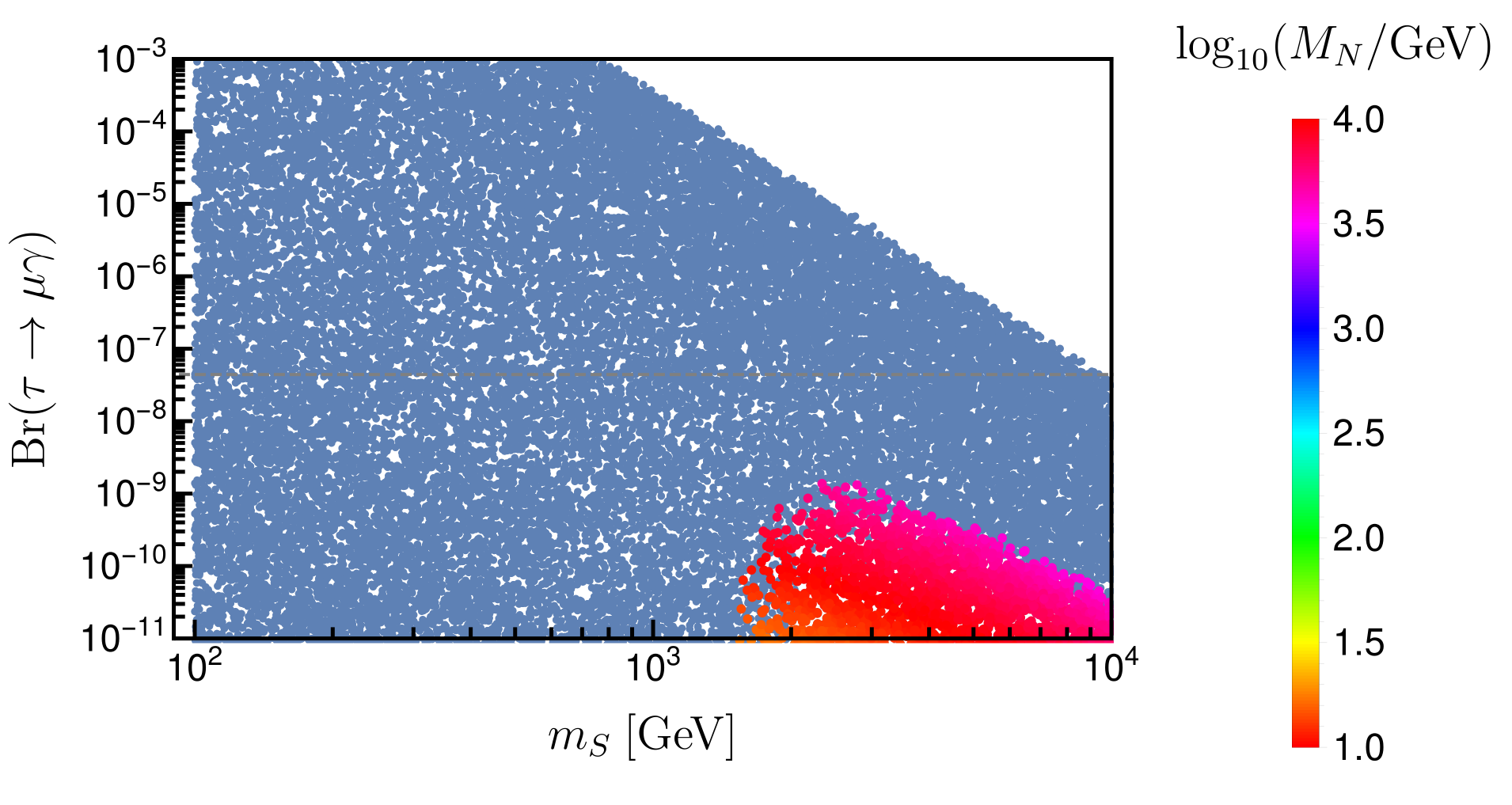}
    \caption{Same as \fig{fig:clfv:KNTTauMuG2} but for IH for the neutrino masses. Bluish points are excluded due to perturbativity arguments.}
    \label{fig:clfv:KNTTauMuG2IH}
\end{figure}

We have repeated the scans discussed above also for the case of IH. An example is shown in \fig{fig:clfv:KNTTauMuG2IH}. IH requires larger Yukawa couplings, since now two neutrino have masses of order $\sqrt{\Delta m^2_{\rm atm}}$. Thus, many more points in the parameter space are ruled out due to the perturbativity constraint. This pushes both fermion and scalar masses to larger values. Indeed, already with current constraints there are no points with $m_S$ below roughly $600$ GeV.

Finally, let us mention that in the KNT model there is no short-range diagram contributing to $0\nu\beta\beta$ decay. Given that the KNT model predicts one (nearly)\footnote{A tiny lightest neutrino mass will be generated at higher loop order.} massless neutrino, it predicts both, an upper and a lower limit for $0\nu\beta\beta$ decay. For normal [inverted] hierarchy the allowed range is roughly $m_{ee} \sim (1-5)$ meV [$(20-50)$ meV]. Observing $0\nu\beta\beta$ decay outside this range would rule out the KNT model as an explanation for the experimental neutrino oscillation data.

%%%%%%%%%%%%%%%%%%%%%%%%%%%%%%%%%%%%%%%%%%%%%%%%%%%%%%%%%%%%%%%
%%%%%%%%%%%%%%%%%%%%%%%%%%%%%%%%%%%%%%%%%%%%%%%%%%%%%%%%%%%%%%%
\section{AKS model} \label{sec:clfv:AKS}

A general class of models is represented by the AKS model \cite{Aoki:2008av}. In this case the particle content is extended to include new scalars and fermions.
\\

The AKS model extends the usual 2HDM with the real scalar singlet $\varphi$, the singly charged scalar $S$ and three generations of singlet fermions $N$. Even though a more minimal version with only two generations of $N$ is possible, we will consider three in the following. The fields $S$, $\varphi$ and $N$ are assumed to be odd under the $Z_2$ parity, while the rest of the particles are even. The quantum numbers of the new particles in the AKS model are given in \tab{tab:clfv:AKS}.

\begin{table}[t!]
    \centering
    \begin{tabular}{| c c c c c c |}
        \hline  
         & generations & $\mathrm{SU(3)_C}$ & $\mathrm{SU(2)_L}$ & $\mathrm{U(1)_Y}$ & $Z_2$ \\
        \hline
        \hline 
        $\varphi$ & 1 & ${\bf 1}$ & ${\bf 1}$ & $0$ & $-$ \\
        $S$ & 1 & ${\bf 1}$ & ${\bf 1}$ & $1$ & $-$ \\
        \hline
        \hline    
        $N$ & 3 & ${\bf 1}$ & ${\bf 1}$ & $0$ & $-$ \\  
        \hline
        \hline
    \end{tabular}
    \caption{New particles in the AKS model with respect to the 2HDM.}
    \label{tab:clfv:AKS}
\end{table}

As explained in \sect{sec:clfv:notation}, an additional softly-broken $Z_2$ symmetry is introduced to avoid dangerous flavour changing neutral currents. We choose to follow \cite{Aoki:2008av} and use this symmetry to couple one of the scalar doublets ($\Phi_1$) only to leptons, and the other ($\Phi_2$) only to quarks. Due to this choice, the Yukawa couplings of the model are given in \eq{eq:clfv:2HDMYuk}, along with the Yukawa,
\begin{equation} %\label{eq:}
    -\mathcal L \supset Y^\ast \, \overline{N^c} \, e_R \, S + \hc \, .
\end{equation}
One can also write Majorana masses for the $N$ singlets,
\begin{equation}
    - \mathcal L_N = \frac{1}{2} M_N \overline{N^c} N + \hc \, ,
\end{equation}
with $M_N$ a symmetric matrix. The scalar potential of the model is given by
\begin{align} \label{eq:clfv:PotAKS}
    \mathcal V &\supset m_{1}^2 |\Phi_1|^2 + m_{2}^2 |\Phi_2|^2 + \left( \mu_{12}^2 \, \Phi_1^\dagger \Phi_2 + \hc \right) + \frac{1}{2} \lambda_1 \, |\Phi_1|^4 + \frac{1}{2} \lambda_2 \, |\Phi_2|^4
    \nn \\
    &+ \lambda_{3} \, |\Phi_1|^2 |\Phi_2|^2 + \lambda_4 \, |\Phi_1^\dagger \Phi_2|^2  + \frac{1}{2} \left[ \lambda_5 \, (\Phi_1^\dagger \Phi_2)^2 + \hc \right]
    \\ \nn 
    &+ \lambda_{\Phi S}^{(1)} \, |\Phi_1|^2 |S|^2 + \lambda_{\Phi S}^{(2)} \, |\Phi_2|^2 |S|^2 + \frac{1}{2} \lambda_{\Phi \varphi}^{(1)} \, |\Phi_1|^2 \varphi^2 + \frac{1}{2} \lambda_{\Phi \varphi}^{(2)} \,
     |\Phi_2|^2 \varphi^2
     \\ \nn
    &+ \left[ \kappa \, \Phi_1 \, \Phi_2 \, S^\ast \, \varphi + \hc \right] + \frac{M_{\varphi}^2}{2} \varphi^2 + M_{S}^2 |S|^2 + \frac{1}{2} \lambda_S \, |S|^4 + \frac{1}{4!} \lambda_{\varphi} \varphi^4 + \frac{1}{2} \xi \, \varphi^2 |S|^2 \, .
\end{align}
As usual, we have omitted $\mathrm{SU(2)_L}$ indices in the previous expression. We point out that lepton number would be restored in the limit $\kappa \to 0$. The presence of this coupling breaks lepton number in one unit.

After electroweak symmetry breaking, the doublet scalars $\Phi_1$ and $\Phi_2$ get mixed. The mass eigenstates resulting from this mixing are the Standard Model Higgs, another Higgs, a new charged scalar, and a pseudoscalar. The charged and neutral Goldstone bosons are absorbed by the $Z$ and $W$ gauge bosons. The mass matrix for the CP-even neutral states in the basis $\mathcal{H}^0 = \text{Re} \,(\Phi_1^0, \Phi_2^0)^T$ is given by,
\begin{equation}
    \mathcal{M}^2_{\mathcal{H}^0} = \left( 
    \begin{array}{cc}
        \lambda_1 v_1^2 - \mu_{12}^2 \, \tan \beta  &   v_1 v_2 \left( \lambda_3 + \lambda_4 + \lambda_5\right) + \mu_{12}^2
        \\
        v_1 v_2 \left( \lambda_3 + \lambda_4 + \lambda_5\right) + \mu_{12}^2 & \lambda_2 v_2^2 - \mu_{12}^2 \, \cot\beta 
    \end{array}
    \right) \, .
\end{equation}
The CP-odd neutral scalar mass matrix in the basis $\mathcal{A}^0 = \text{Im} \,(\Phi_1^0, \Phi_2^0)^T$ is
\begin{equation}
    \mathcal{M}^2_{\mathcal{A}^0} = \left(
    \begin{array}{cc}
        - v_2^2 \lambda_5 - \mu_{12}^2 \, \tan \beta  &  v_1 v_2 \lambda_5 + \mu_{12}^2
        \\
        v_1 v_2 \lambda_5 + \mu_{12}^2 & - v_1^2 \lambda_5 - \mu_{12}^2 \, \cot\beta 
    \end{array}
    \right) \, .
\end{equation}
One finds a massless state, the Goldstone boson that becomes the longitudinal component of the $Z$ boson. The other state has a mass
\begin{equation}
    m^2_{\mathcal{A}^0}= - \left(v_1 v_2 \lambda_5 + \mu_{12}^2\right) \, \frac{v^2}{v_1 v_2} \, ,
\end{equation}
while the mass of the $Z$ boson is $m_Z^2=\frac{1}{4} v^2 (g_1^2 + g_2^2)$. The mass matrix for the charged states in the $\mathcal{H}^\pm = (\Phi_1^\pm, \Phi_2^\pm)^T$ basis is,
\begin{equation}
    \mathcal{M}^2_{\mathcal{H}^\pm} = \left(
    \begin{array}{cc}
        -\frac{1}{2} v_2^2 \, \left(\lambda_4 + \lambda_5 \right) - \mu_{12}^2 \, \tan \beta  &   \frac{1}{2} v_1 v_2 \, \left(\lambda_4 + \lambda_5 \right)+ \mu_{12}^2
        \\
        \frac{1}{2} v_1 v_2 \, \left(\lambda_4 + \lambda_5 \right)+ \mu_{12}^2 & -\frac{1}{2} v_1^2 \, \left(\lambda_4 + \lambda_5 \right) - \mu_{12}^2 \, \cot\beta 
    \end{array}
    \right) \, .
\end{equation}
Again, after diagonalisation one obtains a massless state, identified with the Goldstone boson that becomes the longitudinal part of the $W$ boson, and a massive physical charged scalar with mass
\begin{equation}
    m^2_{\mathcal{H}^\pm}=-\left(\frac{\mu_{12}^2}{v_1 v_2}+\frac{\lambda_4+\lambda_5}{2} \right) v^2 \, .
\end{equation}
The mass of the $W$ boson is given by the standard expression $m_{W^\pm}^2=\frac{1}{4} g_2^2 v^2$. Finally, the masses of the singlet scalars $\varphi$ and $S$ are
\begin{align}
    m_{\varphi}^2 &= M_{\varphi}^2 + \frac{1}{2}\left(\lambda_{\Phi \varphi}^{(1)} \, v_1^2+\lambda_{\Phi \varphi}^{(2)} \, v_2^2 \right) \, ,
    \\
    m_{S^+}^2 &= M_{S}^2 + \frac{1}{2}\left(\lambda_{\Phi S}^{(1)} \, v_1^2+\lambda_{\Phi S}^{(2)} \, v_2^2\right) \, .
\end{align}

As in the cocktail model, the lightest $Z_2$-odd state in the AKS model is stable and can constitute a dark matter candidate.

\subsubsection*{Neutrino masses}

\begin{figure}[t!]
    \centering
    \includegraphics[width=0.48\textwidth]{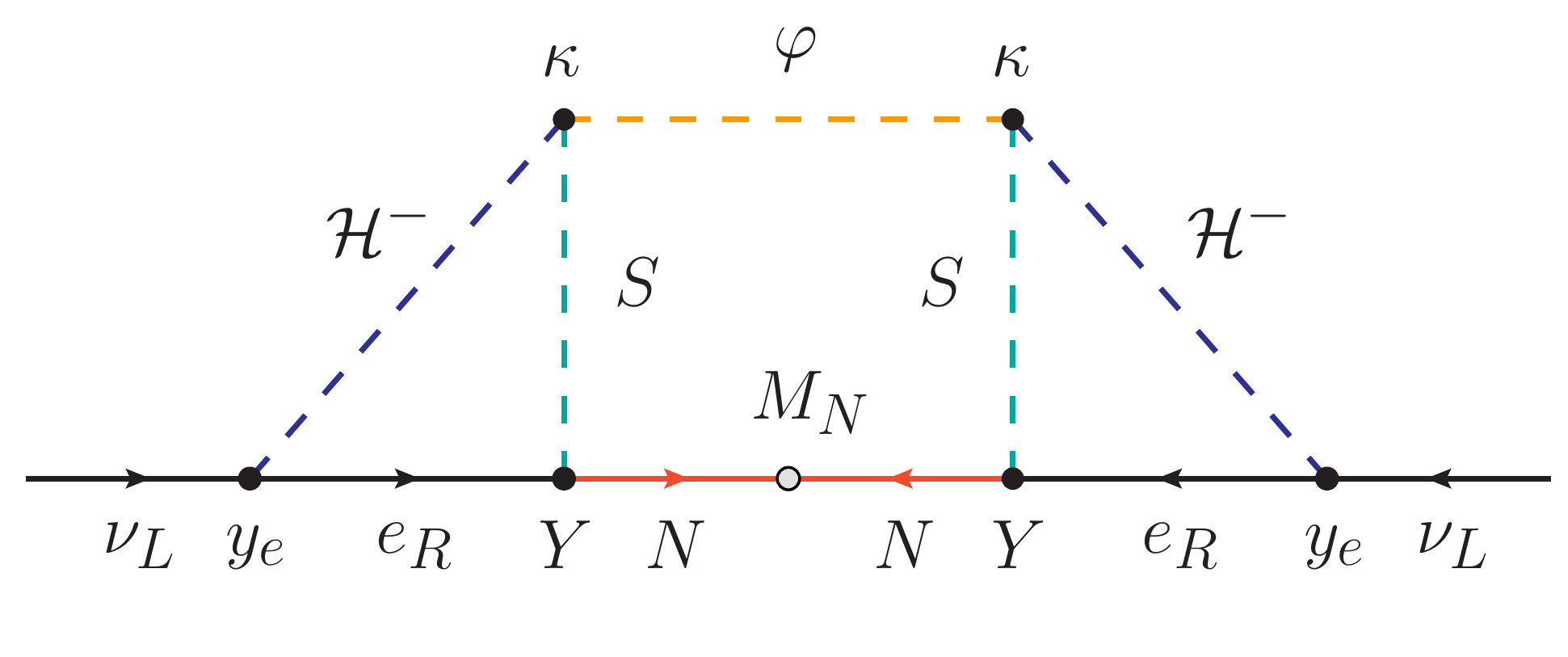}
    \hfill
    \includegraphics[width=0.48\textwidth]{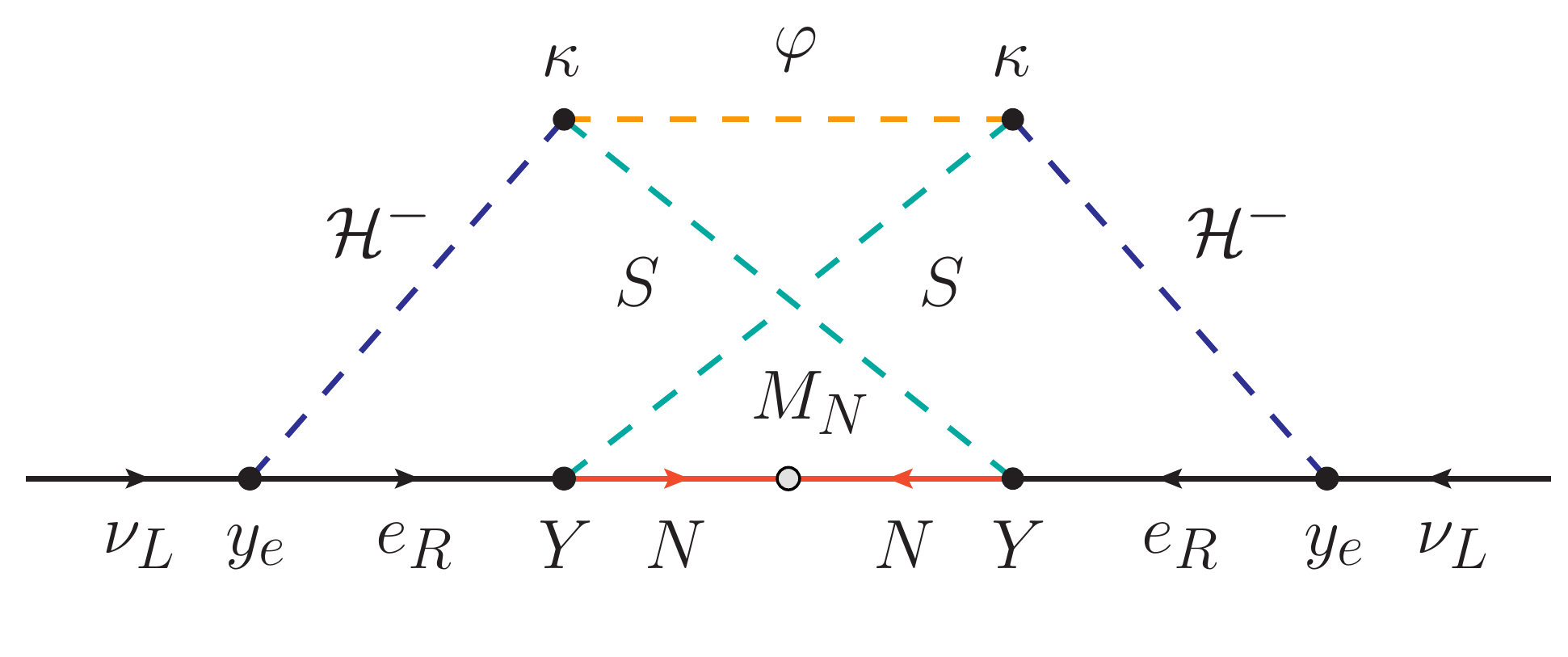}
    \caption{Three-loop neutrino masses in the AKS model. $\mathcal{H}^- \equiv \mathcal{H}^-_{1,2}$ represent the singly charged scalars in the model, obtained after diagonalising the mass matrix of the $\left\{ \Phi_1^- , \Phi_2^- \right\}$ states.}
  \label{fig:clfv:AKS}
\end{figure}

In the AKS model, neutrino masses are induced at three-loop order, as shown in the diagrams in \fig{fig:clfv:AKS}. The resulting neutrino mass matrix is given by,
\begin{align} \label{eq:clfv:MnuAKS}
    \left(\mathcal{M}_\nu\right)_{ij} = \frac{\kappa^2 \, \tan^2 \beta}{(16\pi^2)^3} \, \sum_{\alpha \beta} \frac{m_i \, Y_{i \alpha} Y_{j \beta} \, m_j}{(M_N)_{\alpha \beta}} \, F_{\rm AKS} \,,
\end{align}
where $m_i$ is the mass of the $i$-th charged lepton and $F_{\rm AKS}$ is a dimensionless loop function that depends on the masses of the scalars and fermions in the loop. More details about the calculation of this loop function can be found in \app{app:loops}.

The Yukawa matrix $Y$ in the AKS model does not have any specific symmetry. Therefore, this model represents the general class of models in which the Yukawa matrices can be described by using a generalisation of the Casas-Ibarra parametrisation \cite{Casas:2001sr} (see also \cite{Cordero-Carrion:2018xre, Cordero-Carrion:2019qtu}).

%%%%%%%%%%%%%%%%%%%%%%%%%%%%%%%%%%%%%%%%%%%%%%%%%%%%%%%%%%%%%%%
\subsection{Results} \label{subsec:clfv:results-AKS}

The Yukawa structure of the neutrino mass matrix shown in \eq{eq:clfv:MnuAKS} resembles that of the type-I seesaw. In order to fit the experimental oscillation data, we use the Casas-Ibarra parametrisation introducing the neutrino mass matrix in the flavour basis given in \eq{eq:clfv:Mnu}. We find that,
\begin{equation} \label{eq:clfv:YAKS_fit}
    Y = \frac{i\, (16 \pi^2)^{3/2}}{\kappa \, \tan \beta} \, {\cal R} \, \sqrt{M_N / F_{\rm AKS}} \, \sqrt{\widehat{\cal M}_\nu} \, U^\dagger \, \widehat{\cal M}^{-1}_e \, ,
\end{equation}
where $M_N$ has been taken to be diagonal and ${\cal R}$ is an arbitrary complex $3\times 3$ orthogonal matrix. We include $F_{\rm AKS}$ as it is, in general, a function of the eigenvalues of $M_N$, see \app{app:loops}. Similar to the KNT model, the presence of $\widehat{\cal M}^{-1}_e$ in the fit, implies the enhancement of each column of the Yukawa matrix in terms of the charged lepton masses, i.e. $Y_{\alpha i} \propto 1/m_i$. This leads to unacceptably large Yukawa entries in the first column. For instance, choosing NH with $m_{\nu_1}=0.1$ eV, setting all the phases to $0$ for simplicity, ${\cal R} = \mathbb{I}$ and $(M_N)_{ii} = m_{N}$, we find
\begin{equation} \label{eq:clfv:Y_example}
    Y \simeq
    \begin{pmatrix}
        320  &  -0.88  &   0.038 \\
        220  &   0.93  &  -0.074 \\
        160  &   1.45  &   0.079
    \end{pmatrix}
    \Big(\frac{1}{\kappa \, \tan \beta}\Big)
    \Big(\frac{m_{N}}{100\,\text{GeV}}\Big)^{1/2}
    \Big(\frac{1}{F_{\rm AKS}}\Big)^{1/2} \, ,
\end{equation}
clearly in the non-perturbative regime. Insisting on perturbative Yukawa couplings thus calls for cancellations, especially in the first column, proportional to $1/m_e$. Moreover, even if for a choice of parameters, the Yukawa lives at the edge of perturbativity, one should take care of the constraints coming from CLFV. Especially \mueg, given the hierarchy among the entries of the Yukawa matrix $Y$.

In order to avoid non-perturbativity and CLFV constraints, first we exploit the freedom in ${\cal R}$. We fix two of the complex angles to make two entries of the Yukawa matrix zero or close to zero. We choose $Y_{21}$ and $Y_{31}$. With this, we find that the third free angle of ${\cal R}$ is not enough to cancel another entry in the Yukawa matrix. Therefore, we can only fix the values of the phases and $m_{\nu_1}$ to minimise or cancel $Y_{11}$, similarly to the cocktail model, or $Y_{12}$, to live below the experimental limit on Br(\mueg), proportional to $|Y_{k1} Y_{k2}^\ast|^2$. From now on, we also consider $\kappa = 4\pi$, at the edge of perturbativity, and $\tan \beta = 1$.

\begin{figure}[t!]
    \centering
    \includegraphics[width=0.47\textwidth]{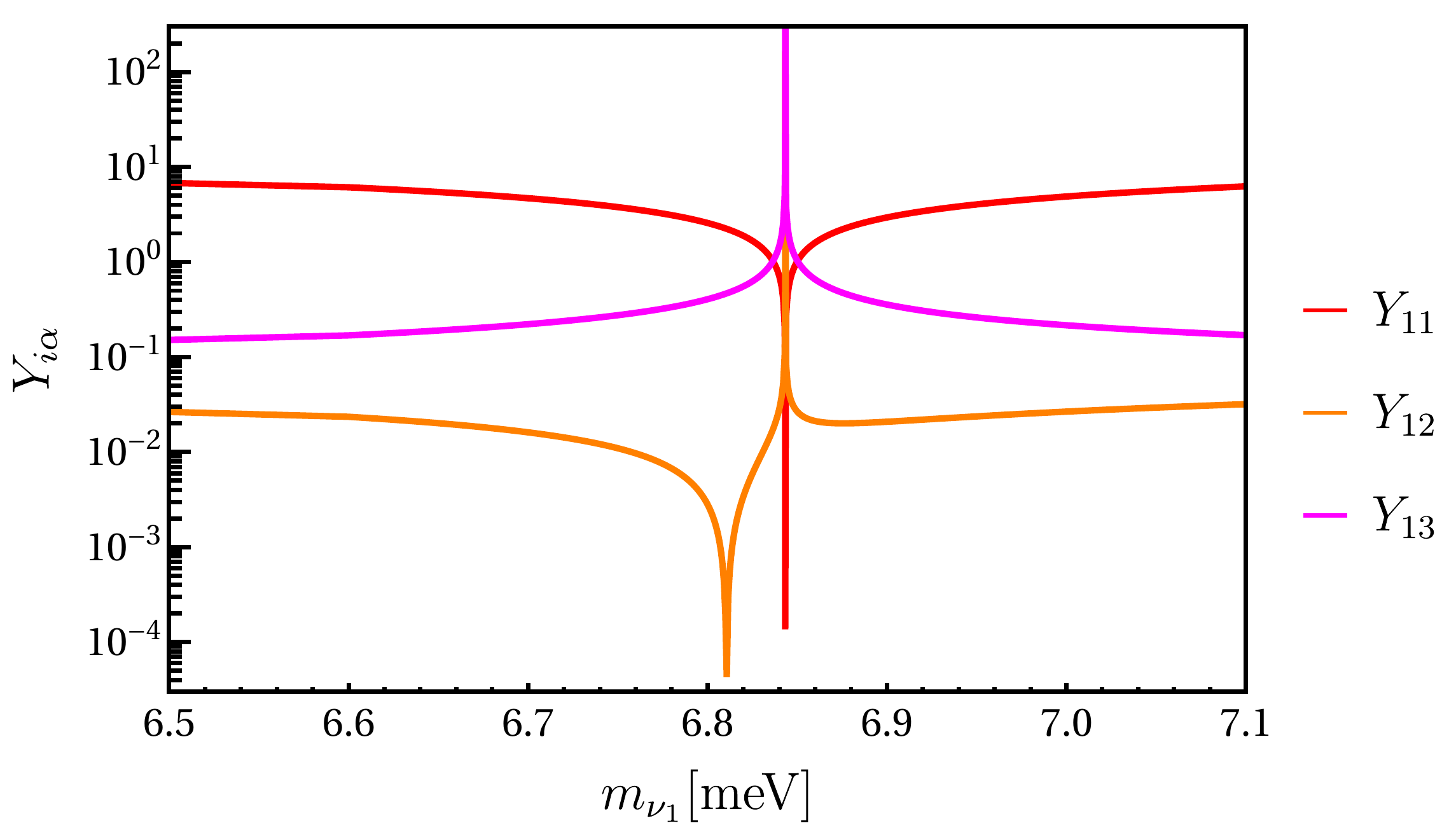}
    \hfill
    \includegraphics[width=0.51\textwidth]{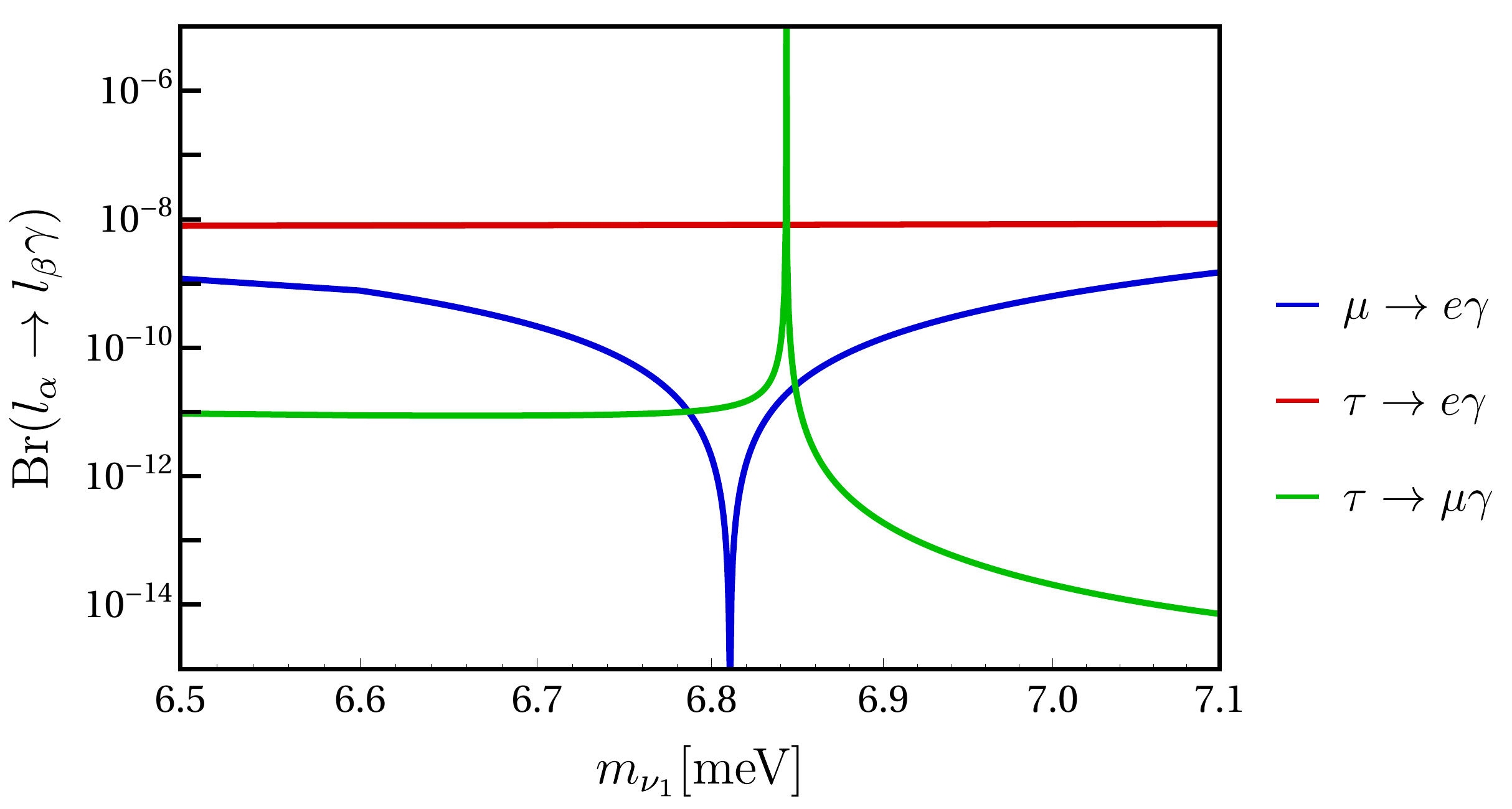}
    \caption{To the left, the entries of the first row of the Yukawa coupling matrix $Y$ \textit{zoomed} around the poles for $Y_{11}$ and $Y_{12}$ for $\alpha_{12}=\alpha_{13}=\delta=\pi$. To the right, the calculated Br($l_\alpha \to l_\beta \gamma$). Both computed fixing ${\cal R}$ for $Y_{21} = Y_{31} = 0$ and at the minimum allowed value of $m_N / F_{\rm AKS}$. While a pole for $Y_{11}$ exists, no pole for Br(\mueg) or Br(\taueg) is associated to it due to the divergence of $Y_{12}$ and $Y_{13}$ on the pole, i.e. the product $Y_{11} \times Y_{1j}$ with $j=2,3$ has no divergence. Note that only near the pole for $Y_{12}$ Br(\mueg) is below the experimental limit.}
    \label{fig:clfv:AKS_yuks}
\end{figure}

In \fig{fig:clfv:AKS_yuks} to the left, we show the behaviour of the first row of $Y$ for $\alpha_{12}=\alpha_{13}=\delta=\pi$. We considered for simplicity that all the scalar masses are equal to $m_{S^+} = m_\varphi = m_{\mathcal{H}^\pm} \equiv m_S$ and all the $N$ singlet fermion masses to be degenerate, $m_N$, and minimise $m_N / F_{\rm AKS}$ to find the lowest value of $Y$, see \eq{eq:clfv:YAKS_fit}. We found this minimum for $m_N = 272$ GeV and $m_{\mathcal{S}} = 100$ GeV, where $F_{\rm AKS} \approx 0.44$, compatible with the limit on scalar masses from LEP \cite{Tanabashi:2018oca}. Here we do not show the other four non-zero Yukawas for simplicity. They are nearly constant and of order $0.1$. Similar to the cocktail model (\fig{fig:clfv:Yuks}), poles exist in the different Yukawa entries for particular values of the phases and $m_{\nu_1}$. The main difference lies in the divergence that appears when $Y_{11}=0$. This is caused by our choice of ${\cal R}$ matrix, such that $Y_{21} = Y_{31} = 0$. In this case, the pole in $Y_{11}$ does not imply a pole in Br(\mueg) or Br(\taueg), as it can be seen in \fig{fig:clfv:AKS_yuks} to the right. In fact, the product of $|Y_{11} Y_{13}^*|$ remains constant over the pole and very close to the current experimental limit of $3.3 \times 10^{-8}$ \cite{Aubert:2009ag}. Only the region around the pole in $Y_{12}$ is allowed by the experimental limit Br(\mueg)$< 4.2 \times 10^{-13}$ \cite{TheMEG:2016wtm}.

To sum up, the parameter space of the AKS model is constrained mainly by perturbativity and Br(\mueg). The former can be addressed with the freedom in ${\cal R}$ to set $Y_{21}$ and $Y_{31}$ to zero. As well as by fixing the Majorana and Dirac phases, and the lightest neutrino mass, to be near the pole of $Y_{11}$, where its value is lower than $4\pi$. On the other hand, to be below the experimental limit on Br(\mueg), a similar fine-tuning of the phases and $m_{\nu_1}$ should be done to be around the \textit{narrow} pole of $Y_{12}$. The parameter space is then restricted to those values of the phases and $m_{\nu_1}$ where the poles of $Y_{11}$ and $Y_{12}$ exist, and they are close enough to each other to avoid the limit on Br(\mueg) while $Y_{11}$ is still perturbative.

\begin{figure}[t!]
    \centering
    \includegraphics[width=0.7\textwidth]{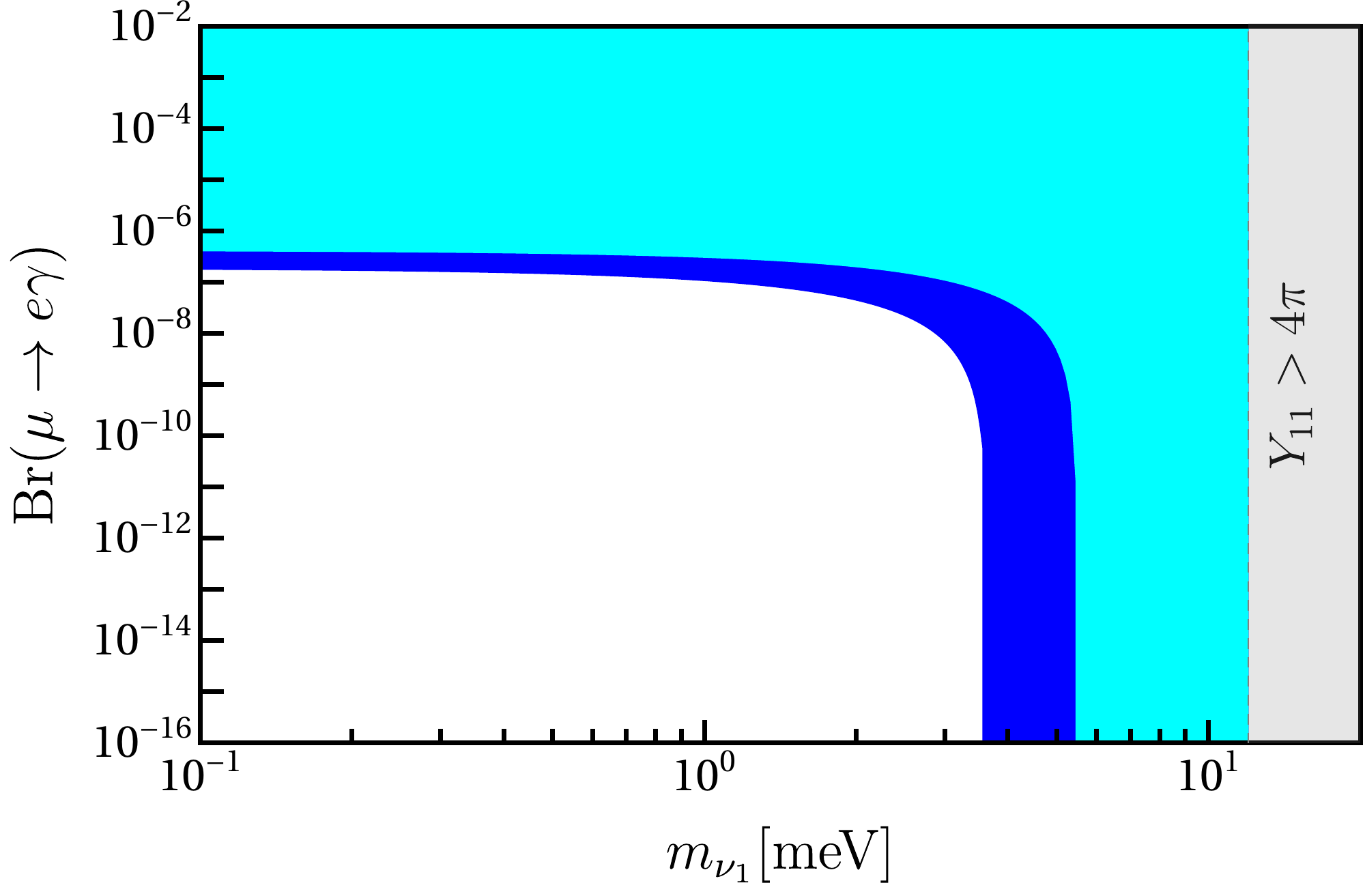}
    \caption{Br(\mueg) scanned over neutrino oscillation data in $1\sigma$ (light blue) and $3\sigma$ (dark blue) ranges. This plot scans over the Majorana phases. The shaded grey area corresponds to the most conservative limit to non-perturbative Yukawas.}
    \label{fig:clfv:AKS_meg}
\end{figure}

In \fig{fig:clfv:AKS_meg} we show the value of Br(\mueg) scanning over the complete range of oscillation parameters (NH) and phases. On the right, we give the limit due to perturbativity of $Y_{11}$, reducing the parameter space to a small window of $m_{\nu_1} = (4.5 - 20)$ meV. Note that like in the cocktail model, $Y_{11}$ behaves as $m_{ee}$, and for $m_{\nu_1} \gtrsim 10$ meV, $m_{ee}$ has no pole, so $Y_{11}$ is in the non-perturbative region. Moreover, the cancellation of $Y_{11}$ and $Y_{12}$ only occurs for NH, so the model can only explain neutrino data with this neutrino mass ordering. In the following, we shall consider only NH.

\begin{figure}[t!]
    \centering
    \includegraphics[width=0.7\textwidth]{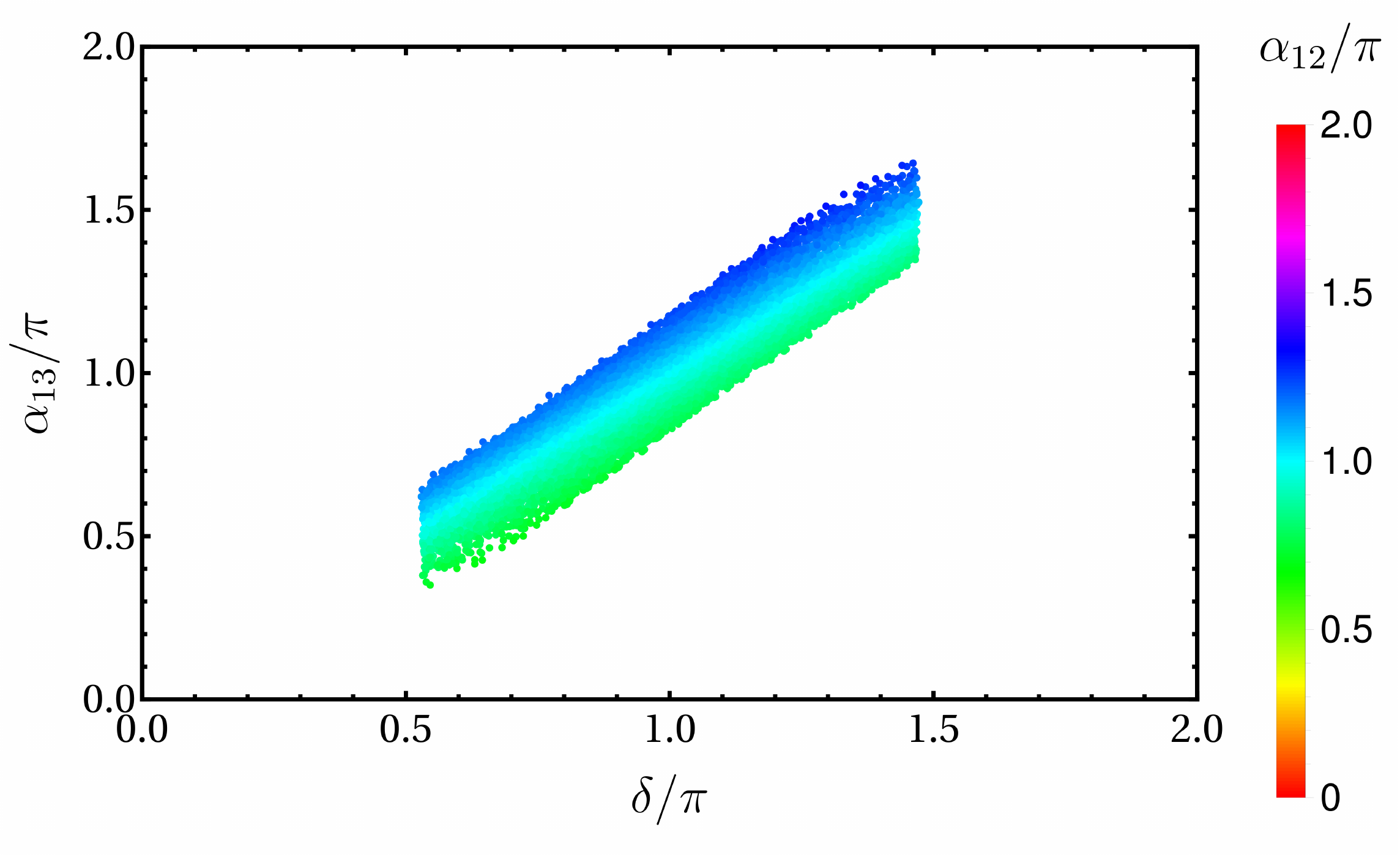}
    \caption{Allowed parameter space for $\alpha_{12}$, $\alpha_{13}$ and $\delta$. Neutrino oscillation data was scanned over the $3\sigma$ uncertainties, except $\delta$ which was left free. For points outside this region, either $Y_{11}$ is non-perturbative or Br(\mueg) is above the experimental limit.}
    \label{fig:clfv:AKS_phases}
\end{figure}

\Fig{fig:clfv:AKS_meg} not only implies a constraint on $m_{\nu_1}$, but also on the phases. In \fig{fig:clfv:AKS_phases} we show the points allowed by perturbativity and the experimental limits on Br(\mueg), Br(\taueg) and Br(\taumug), for the values of the three phases. We scanned over the phases and masses, with $m_S > 100$ GeV, allowing oscillation data to vary in $3\sigma$. As can be seen, $\alpha_{12}$ should lie around $\pi$, while $\delta$ is constrained to values between roughly $\pi/2$ and $3\pi/2$. For $\delta$ outside this window, there is no cancellation of $Y_{12}$. In the following, we restrict the results shown to the region where Br(\mueg)$< 4.2 \times 10^{-13}$.

\begin{figure}[t!]
    \centering
    \includegraphics[width=0.48\textwidth]{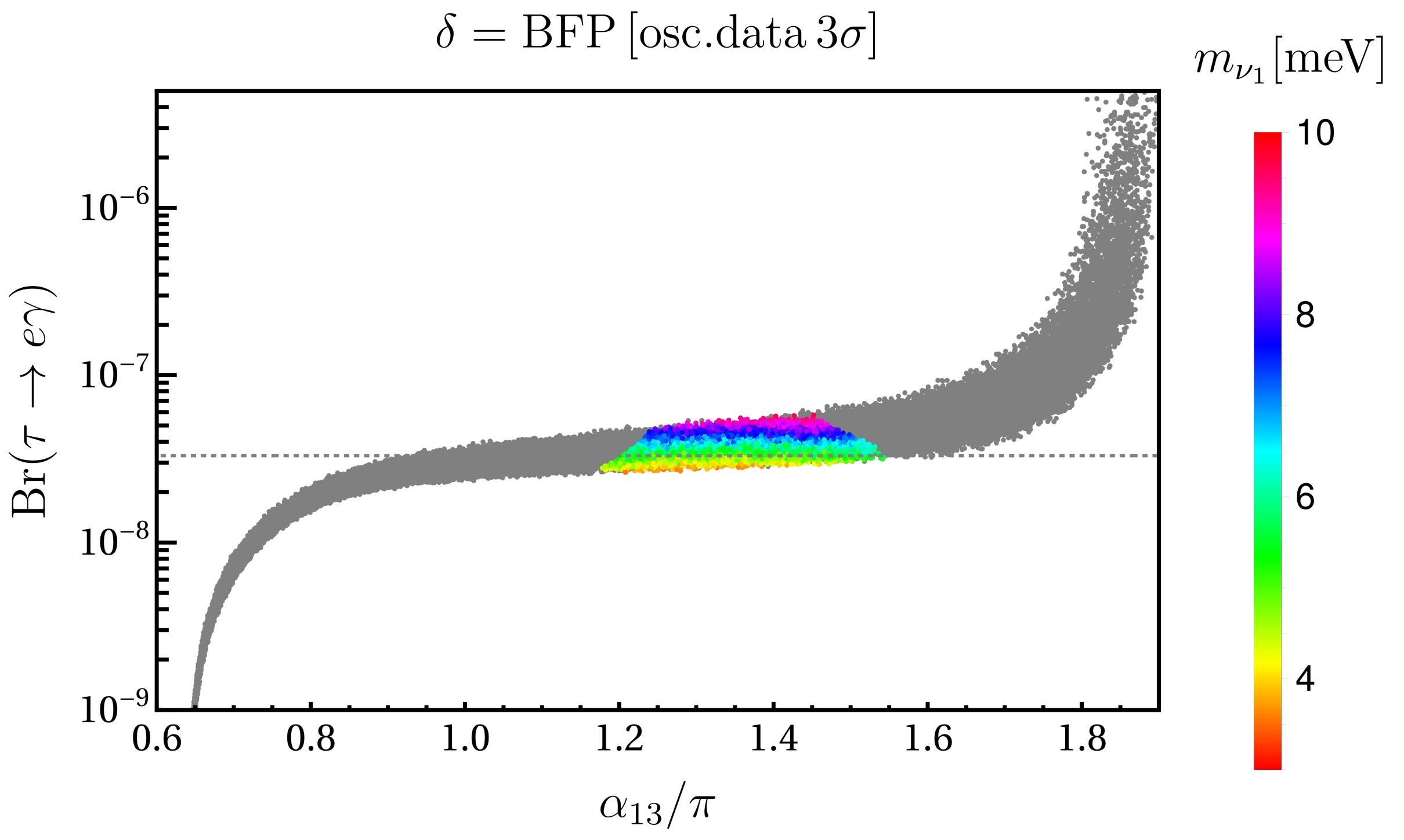}
    \hfill
    \includegraphics[width=0.48\textwidth]{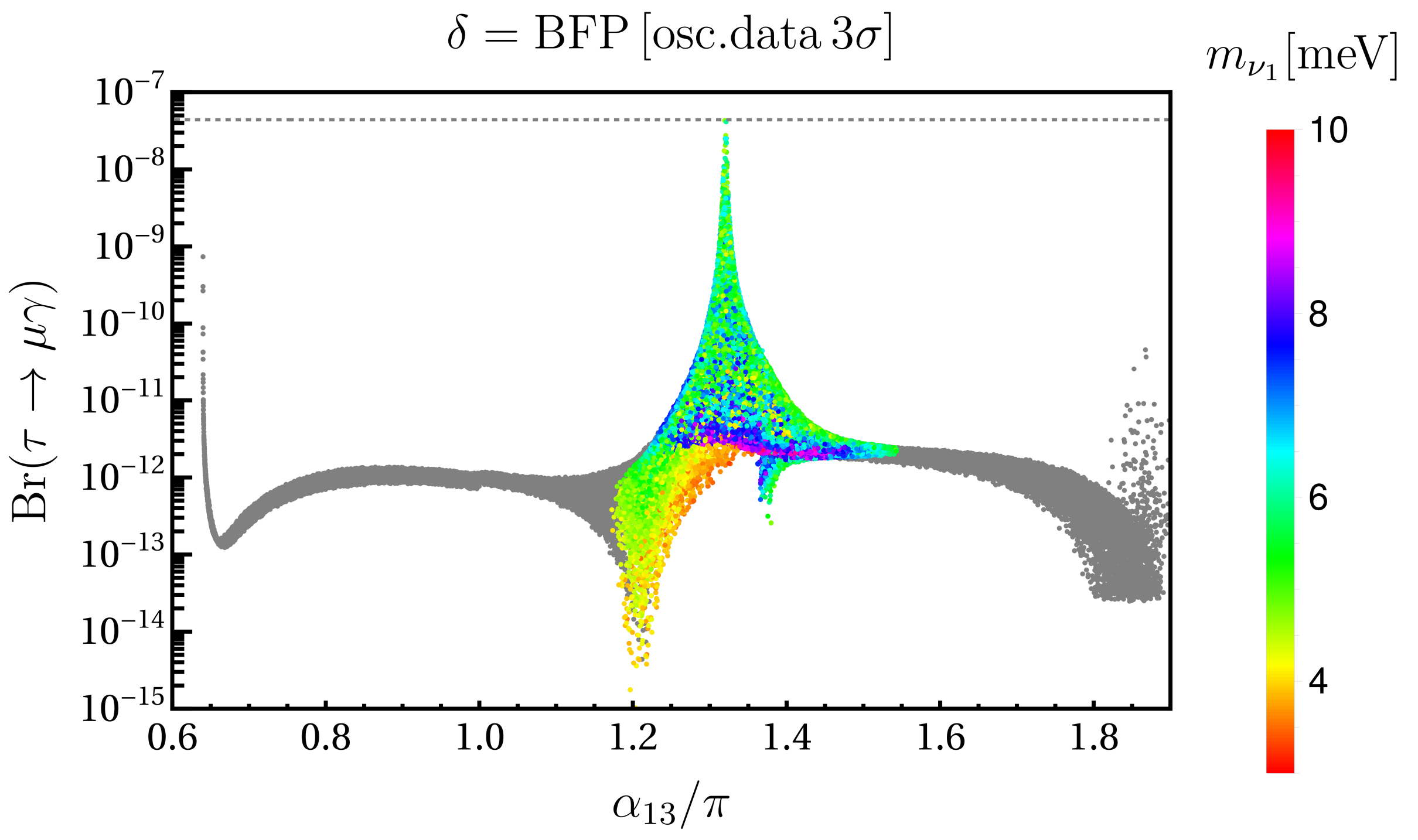}
    \caption{Br(\taueg) and Br(\taumug) as functions of $\alpha_{13}$ for different values of the lightest neutrino mass (colour-coded) along with the current experimental limits (dotted line). We scanned over $3\sigma$ uncertainties of the oscillation data, except for $\delta$ which was fixed to the b.f.p. Grey points are excluded due to perturbativity arguments.}
    \label{fig:clfv:AKS_tau}
\end{figure}

\begin{figure}[t!]
    \centering
    \includegraphics[width=0.48\textwidth]{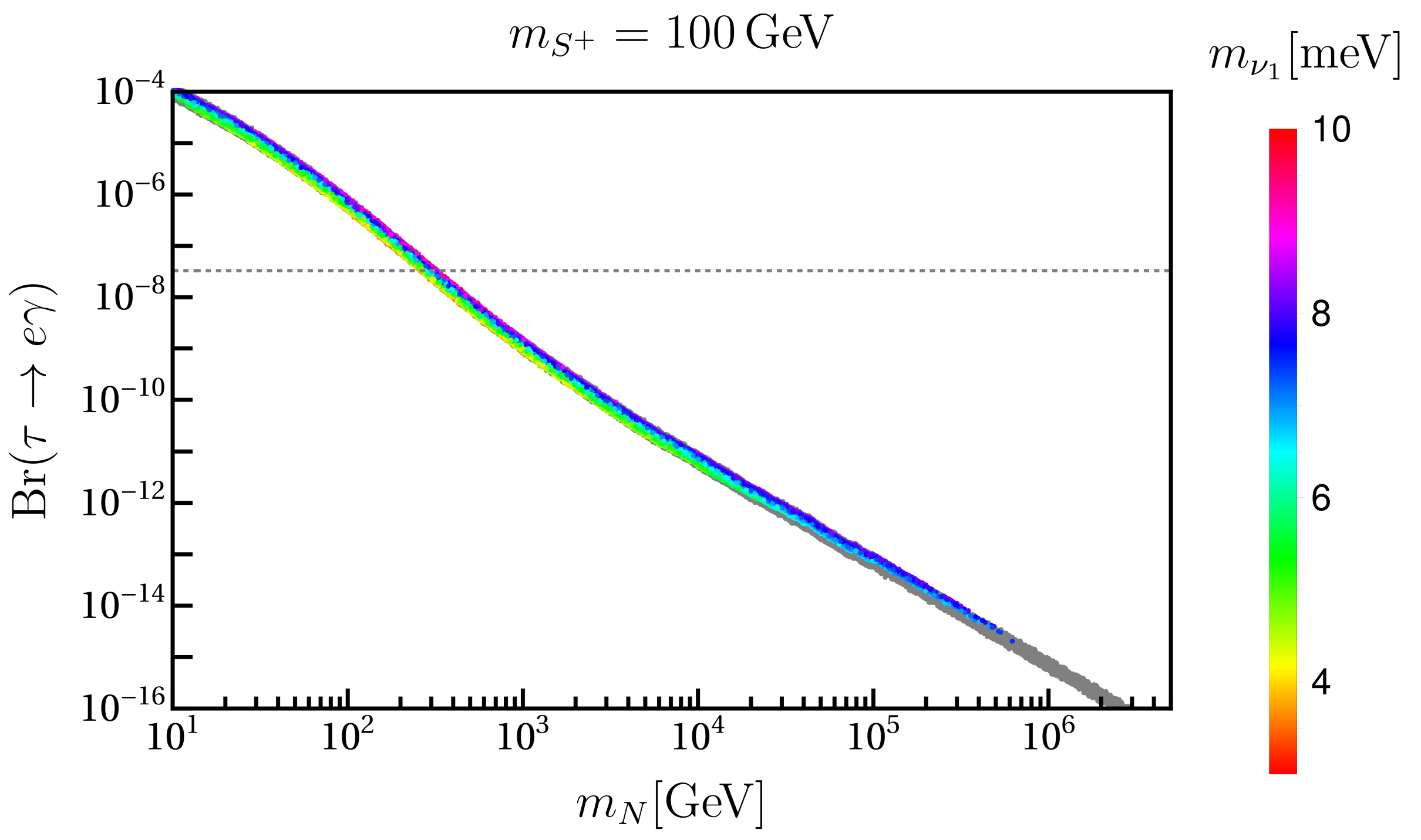}
    \hfill
    \includegraphics[width=0.48\textwidth]{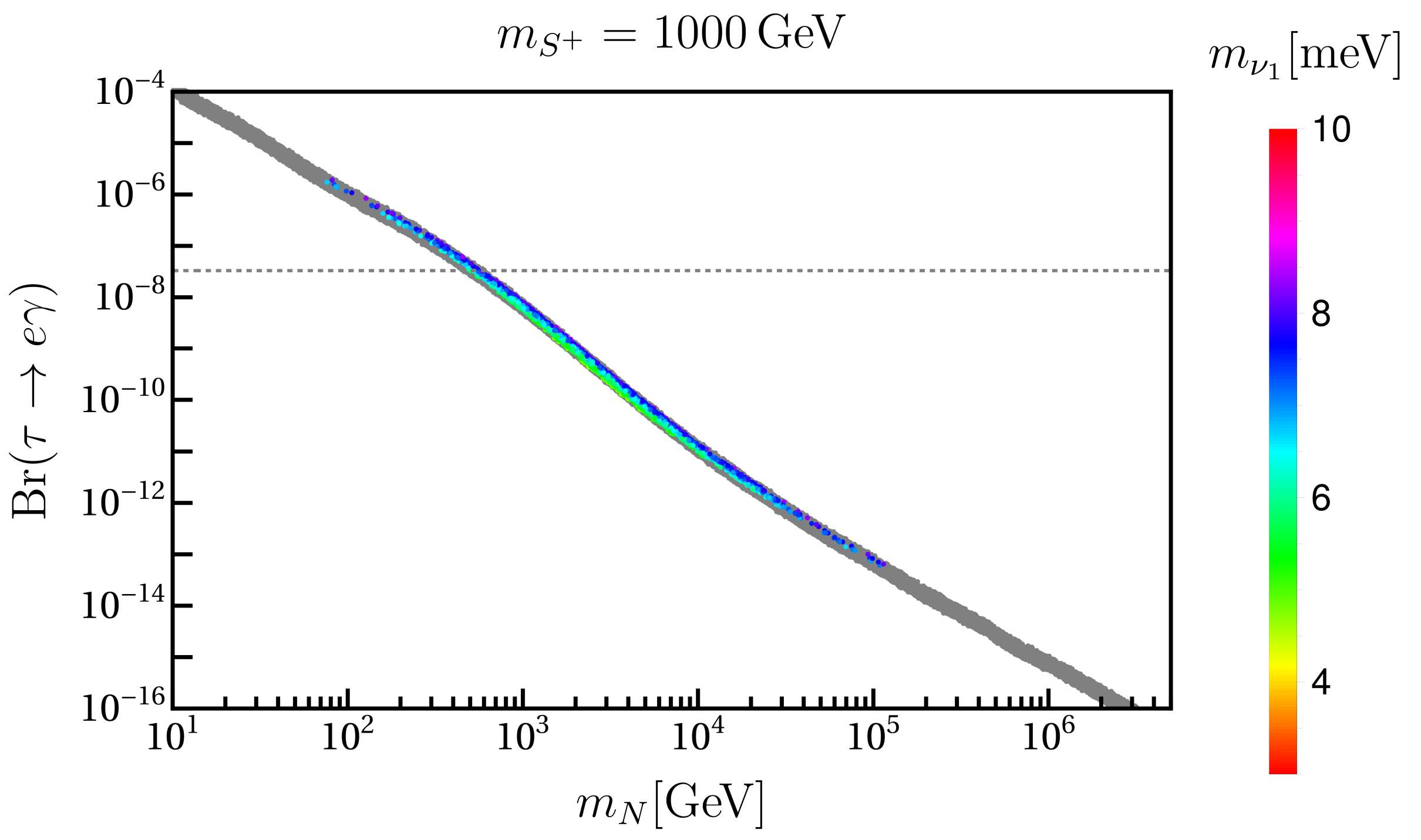}
    \caption{Calculated values for Br(\taueg) as function of $m_N$ for different values of $m_{S^+}$. Here, we are maximising the allowed parameter space in terms of $\alpha_{13}$. For the grey points, at least one entry in $Y$ is larger than $4\pi$.}
    \label{fig:clfv:AKS_mnms}
\end{figure}

Now we move to analyse \taueg and \taumug. As shown in \fig{fig:clfv:AKS_yuks} (right), while Br(\taumug) is below the experimental limit, except on the pole of $Y_{11}$, Br(\taueg) is mainly constant and close to the experimental limit. In \fig{fig:clfv:AKS_tau} we give both branching ratios fixing $\delta$ to the b.f.p. and scanning over the uncertainties in the rest of the oscillation parameters. We consider $(m_N / F_{\rm AKS})_{min}$ with $m_S = 100$ GeV and $m_N = 272$ GeV. Points coloured in grey correspond to non-perturbative Yukawas. We see that while Br(\taumug) is \textit{safe}, the allowed region on the left plot is severely constrained by the experimental limit Br(\taueg)$<3.3 \times 10^{-8}$. This tension can be mitigated by raising the masses, see \fig{fig:clfv:AKS_mnms}. For the AKS model, the dominant contribution to Br($l_\alpha \to l_\beta \gamma$) is approximately proportional to $1/M^4$, with $M$ the dominant scale \cite{Lavoura:2003xp}. On the other hand, $m_N / F_{\rm AKS}$ is minimal for masses around $m_S = 100$ GeV and $m_N = 272$ GeV. So for masses away from these values, $m_N / F_{\rm AKS}$ increases and, consequently, the absolute scale of the Yukawas increases as well (see \eq{eq:clfv:YAKS_fit}), hence narrowing the region where the Yukawas are perturbative. For $m_N (m_{S^+}) \sim 10^6$ GeV, we found no points allowed by perturbativity and the experimental limit on Br(\mueg). In \fig{fig:clfv:AKS_mnms}, in order to minimise the Yukawas, we fixed $m_\varphi = m_{\mathcal{H}^{\pm}} = 100$ GeV and change $m_{S^+}$ and
$m_N$, which enter the calculation of Br($l_\alpha \to l_\beta \gamma$).

\begin{figure}[t!]
    \centering
    \includegraphics[width=0.56\textwidth]{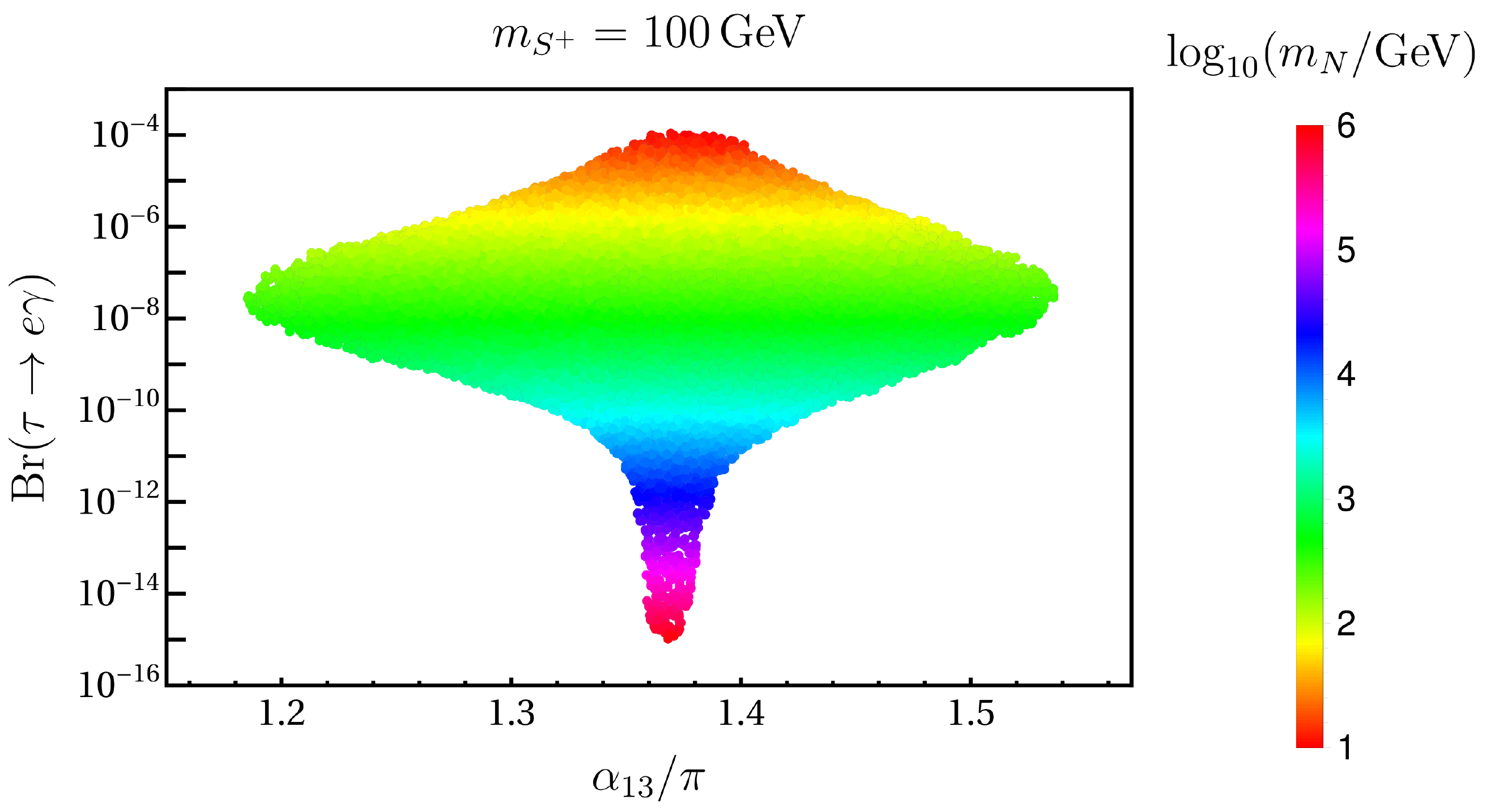}
    \hfill
    \includegraphics[width=0.423\textwidth]{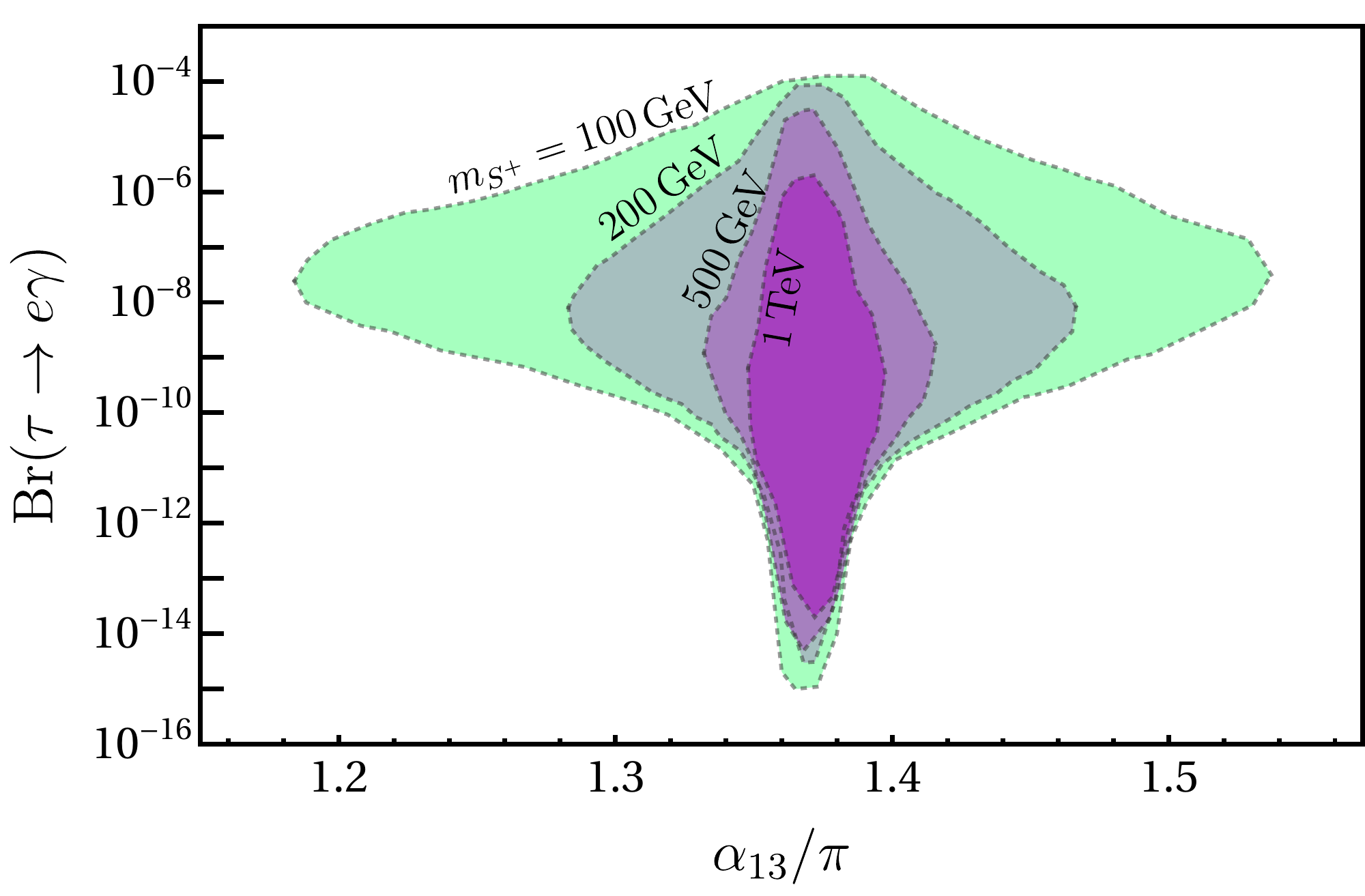}
    \caption{Br(\taueg) for different values of $m_N$ and $m_{S^+}$. To the left, we fixed $m_{S^+}$ and see how by modifying $m_N$ the parameter space widens or narrows due to perturbativity arguments. A similar behaviour can be observed for the plot on the right, where we show contour lines for different values of $m_{S^+}$ scanning over $m_N$.}
    \label{fig:clfv:AKS_alp13_m}
\end{figure}

A similar analysis can be done scanning over the Majorana phases too. \Fig{fig:clfv:AKS_alp13_m} shows Br(\taueg) as a function of $\alpha_{13}$ for different fermion and scalar masses. The allowed parameter space is \textit{bigger} for $m_N$ around $272$ GeV, where $m_N / F_{\rm AKS}$ is minimal. For different masses the parameter space \textit{narrows}, because $m_N / F_{\rm AKS}$ increases, as explained before. The upper limit is due to the phenomenological limit $m_{S^+} > 100$ GeV, as for $m_N \ll m_{S^+}$, Br(\taueg) is dominated by $m_{S^+}$. On the other side, while going to larger $m_N$ reduces considerably Br(\taueg), a lower limit always exists due to perturbativity.
\\

We close this discussion with a short comment on $0\nu\beta\beta$ decay. There is no short-range diagram for $0\nu\beta\beta$ decay in the AKS model. Since, as discussed above, the AKS model survives only for normal hierarchy and in the part of parameter space where $m_{ee}$ is largely cancelled, observation of  $0\nu\beta\beta$ decay in the next round of experiments would definitely rule out AKS as an explanation of neutrino masses.

%%%%%%%%%%%%%%%%%%%%%%%%%%%%%%%%%%%%%%%%%%%%%%%%%%%%%%%%%%%%%%%
%%%%%%%%%%%%%%%%%%%%%%%%%%%%%%%%%%%%%%%%%%%%%%%%%%%%%%%%%%%%%%%
\section{Summary and discussion} \label{sec:clfv:summary}

In this chapter we have considered the cocktail, KNT and AKS models and studied their CLFV phenomenology. In these models, Majorana neutrino masses are generated at the three-loop order, which naturally implies that large Yukawa couplings are required in order to reproduce the mass scales observed in neutrino oscillation experiments. As a result of this, perturbativity is typically lost. We have shown that one can decrease the Yukawa couplings by tuning some of the free parameters of these scenarios, such as the lightest neutrino mass $m_{\nu_1}$ or the Dirac and Majorana phases contained in the leptonic mixing matrix $U$. However, even after these parameters are tuned to recover perturbativity, the resulting CLFV branching ratios tend to largely exceed the existing bounds. In order to reduce the CLFV rates further tuning is needed. Our main conclusion is that the three models survive only in tiny correlated regions of their parameter spaces.

One should note that CLFV alone cannot exclude any of these models. The reason is that one can always reduce the CLFV rates as much as necessary by tuning the parameters of the model more finely. However, additional experimental handles exist. First, perturbativity imposes upper limits on the masses of some of the particles running in the loops. The reason is simple: larger mediator masses would imply a stronger suppression of the loop functions and then require larger Yukawa couplings.  Thus, also future searches at the LHC in the high-luminosity phase would further restrict the available parameter space. An important experimental handle on the models is $0\nu\beta\beta$ decay. Since $m_{\nu_1}$ and the Majorana phases must be tuned for the models to survive, the effective $0\nu\beta\beta$ neutrino mass $m_{ee}$ becomes strongly constrained and definite predictions for the $0\nu\beta\beta$ rates are obtained for the AKS and KNT models. Any observation of $0\nu\beta\beta$ decay with the next generation of experiments would definitely rule out the AKS model. For the KNT model, because $m_{\nu_1}\simeq 0$, $m_{ee}$ has to be either in the range $m_{ee} \simeq (2-6)$ meV or $(15-50)$ meV for normal hierarchy or inverted hierarchy. Only the cocktail model is more flexible in its predictions for $0\nu\beta\beta$ decay, due to additional contributions from a sizeable short-range diagram.

We mention also that only the KNT model can explain neutrino data for both hierarchies. Neither the cocktail nor the AKS model has any acceptable point in all of their parameter space in the case of inverse hierarchy.

A crucial ingredient in our analysis is the allowed size for the quartic scalar potential couplings that play a role in the neutrino mass generation mechanism, for example $\lambda_5$ in the cocktail model, $\lambda_S$ in the KNT model and $\kappa$ in the AKS model. Since neutrino masses are proportional to (some power of) these couplings, the larger they are, the smaller the Yukawa couplings can be. In our analysis, scalar couplings as large as $4 \pi$ have been allowed. A more restrictive choice, with couplings at most of $\mathcal{O}(1)$, would alter the conclusions dramatically. In fact, all three models would already be ruled out, if all their couplings are restricted to be not larger than $\mathcal{O}(1)$.

Finally, we emphasise again that our strong claims only apply to the three minimal models considered here. There are several ways to modify these models so that they can evade the perturbativity and flavour constraints. For instance, one can introduce new exotic states in order to get rid of the proportionality to the charged lepton masses, at the origin of the problems discussed in the chapter. Also, one may enhance the contributions to the neutrino mass matrix by using coloured states. Nevertheless, we also note that there may be many other three-loop (or four-loop) neutrino mass models with the same issues.

\pagebreak
\fancyhf{}

%% file: Chapters/0vBB_majoron/Chapter_majoron.tex
\fancyhf{}
\fancyhead[LE,RO]{\thepage}
\fancyhead[RE]{\slshape{Chapter \thechapter. $0\nu\beta\beta$ decay with non-standard Majoron emission}}
\fancyhead[LO]{\slshape\nouppercase{\rightmark}}

\chapter{Neutrinoless double-$\beta$ decay with non-standard Majoron emission}
\label{ch:0vbb}
\graphicspath{ {Chapters/0vBB_majoron/} }

Double-$\beta$ decay processes are sensitive probes of physics beyond the Standard Model. The Standard Model process of two-neutrino double-$\beta$ ($2\nu\beta\beta$) decay is the rarest process ever observed with half-lives of order $T_{1/2}^{2\nu\beta\beta} \sim 10^{21}~\text{y}$. Neutrinoless double-$\beta$ ($0\nu\beta\beta$) decay, with no observation of any missing energy, is clearly the most important mode beyond the Standard Model as it probes the Majorana nature and mass $m_\nu$ of light neutrinos, with current experimental sensitives of $T_{1/2}^{0\nu\beta\beta} \sim (0.1~\text{eV}/m_\nu)^2 \times 10^{26}~\text{y}$. In general, $0\nu\beta\beta$ is a crucial test for any new physics scenario that violates lepton number by two units.

On the other hand, one or more exotic neutral particles may also be emitted, with a signature of anomalous missing energy beyond that expected in $2\nu\beta\beta$ decay. A well studied set of theories involve the emission of a scalar particle, called \textit{Majoron} $J$. The first such proposed Majoron was a Goldstone boson associated with the spontaneous breaking of lepton number symmetry \cite{Chikashige:1980ui, Gelmini:1980re}, coupling to a neutrino $\nu$ as $g_J \, \nu\nu J$. Current searches have a sensitivity of the order $T_{1/2}^{0\nu\beta\beta J} \sim (10^{-5}/g_J)^2 \times 10^{24}~\text{y}$. The term Majoron has been used in a wider sense, implying just a charge-neutral scalar particle (Goldstone boson or not) or vector particle \cite{Carone:1993jv}. Originally considered to be massless, it may also be a light particle \cite{Bamert:1994hb, Hirsch:1995in, Blum:2018ljv} that can potentially be a dark matter candidate \cite{Berezinsky:1993fm, Garcia-Cely:2017oco, Brune:2018sab}. Searches for extra particles in double-$\beta$ decay are crucial in understanding neutrinos. Most importantly, violation of lepton number by two units and thus the Majorana nature of neutrinos can only be firmly established in the case of $0\nu\beta\beta$ decay.

Not all such emission modes have been discussed in the literature yet. Existing experimental searches so far focus on the emission of one or two Majorons originating from the intermediate neutrino exchanged in the process. The different Majoron scenarios have been classified into several categories, all of which assume Standard Model $(V-A)$ charged currents with the electrons and quarks. In this chapter based on \cite{Cepedello:2018zvr}, we instead consider $0\nu\beta\beta\phi$ decay with emission of a light neutral scalar $\phi$ from a single effective dimension-7 operator of the form $\Lambda_{\rm NP}^{-3}(\bar u\mathcal{O} d)(\bar e\mathcal{O}\nu)\phi$, with the fermion currents having a different chiral structure from that in the Standard Model. In the following, we will refer to the light scalar as ``Majoron'', independent of its origin. We determine the sensitivity to $\Lambda_{\rm NP}$ and analyse the effect on the energy and angular distributions in comparison with $2\nu\beta\beta$ decay.

%%%%%%%%%%%%%%%%%%%%%%%%%%%%%%%%%%%%%%%%%%%%%%%%%%%%%%%%%%%%%%%
%%%%%%%%%%%%%%%%%%%%%%%%%%%%%%%%%%%%%%%%%%%%%%%%%%%%%%%%%%%%%%%
\section{Effective Long-Range Interactions} \label{sec:0vbb:effect}

\begin{figure}
    \centering
    \includegraphics[width=0.29\textwidth]{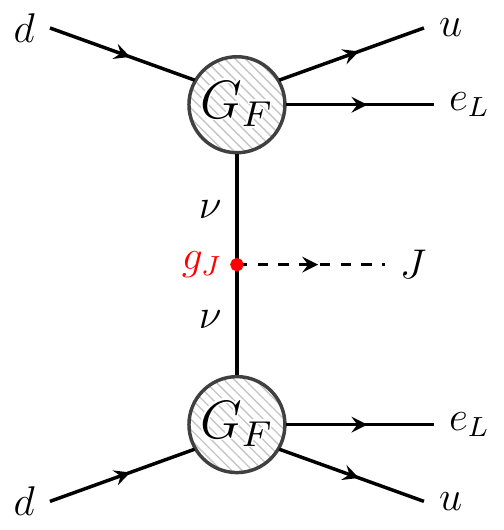}
    \hfill
    \includegraphics[width=0.29\textwidth]{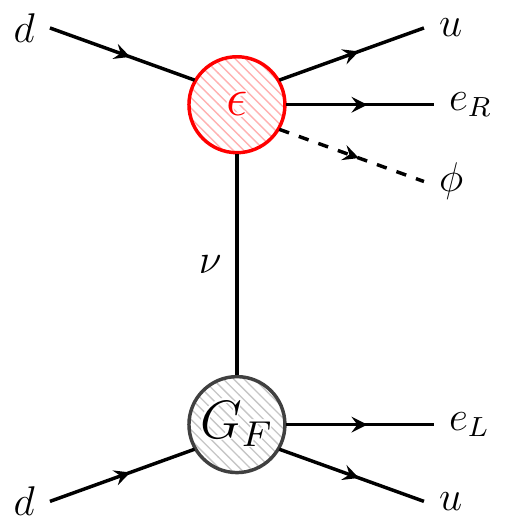}
    \hfill
    \includegraphics[width=0.35\textwidth]{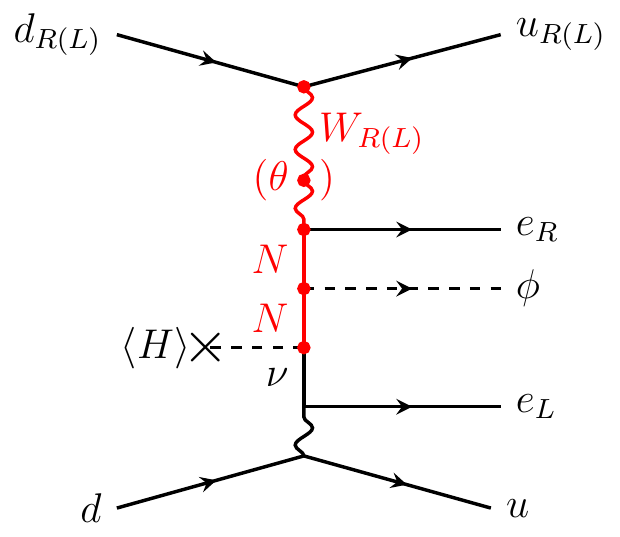}
    \caption{Feynman diagrams for ordinary $0\nu\beta\beta J$ Majoron decay (left), $0\nu\beta\beta\phi$ decay triggered by an effective operator of the form $\Lambda^{-3}_{\rm NP}(\bar u\mathcal{O} d)(\bar e\mathcal{O}\nu)\phi$ (centre) and possible ultraviolet completion of the latter in a Left-Right symmetric model (right).}
    \label{fig:0vbb:diagram} 
\end{figure}

We are interested in processes where right- and left-handed electrons are emitted along with a scalar $\phi$ considering as a first approach, only $(V+A)$ and $(V-A)$ currents. The effective Lagrangian can then be written as,
\begin{align}
\label{eq:0vbb:lagrangian}
    \mathcal{L}_{0\nu\beta\beta\phi} &=
    \frac{G_F \cos\theta_C}{\sqrt{2}}\left(
          j^\mu_L J^{\phantom{\mu}}_{L\mu} 
        + \frac{\epsilon^\phi_{RL}}{m_p} j_R^\mu J^{\phantom{\mu}}_{L\mu}\phi 
        + \frac{\epsilon^\phi_{RR}}{m_p} j_R^\mu J^{\phantom{\mu}}_{R\mu}\phi
    \right) + \hc \, ,
\end{align}
with the Fermi constant $G_F$, the Cabbibo angle $\theta_C$, and the leptonic and hadronic currents $j_{L,R}^\mu = \bar e\gamma^\mu(1\mp\gamma_5)\nu$ and $J^\mu_{L,R} = \bar u\gamma^\mu(1\mp\gamma_5)d$, respectively. Here, $\nu$ is a 4-spinor field of the light electron neutrino, either defined by $\nu = \nu_L + \nu_L^c$ (i.e. a Majorana spinor constructed from the Standard Model active left-handed neutrino $\nu_L$) or $\nu = \nu_L + \nu_R$ (a Dirac spinor constructed from the Standard Model $\nu_L$ and a new sterile right-handed neutrino $\nu_R$). Whether the light neutrinos are of Majorana or Dirac type and whether total lepton number is broken or conserved is of crucial importance for an underlying model, determined by the chosen lepton numbers for $\nu_R$ and $\phi$, but as far as the effective interactions in \eq{eq:0vbb:lagrangian} are concerned, this does not play a role in our calculations. The proton mass $m_p$ is introduced in the exotic interactions as normalisation to make the effective coupling constants $\epsilon^\phi_{RL}$ and $\epsilon^\phi_{RR}$ dimensionless, in analogy to the effective operator treatment of $0\nu\beta\beta$ decay \cite{Pas:1999fc,Deppisch:2012nb}. In \eq{eq:0vbb:lagrangian}, we omit exotic operators with left-handed lepton currents; as in the standard long-range case, such contributions will be additionally suppressed by the small neutrino masses \cite{Deppisch:2012nb}. We instead focus on the process depicted in \fig{fig:0vbb:diagram}~(centre), where the Standard Model $(V-A)$ Fermi interaction, the first term in \eq{eq:0vbb:lagrangian}, meets one of the exotic operators. In this case, the momentum part in the numerator of the neutrino propagator contributes, rather than the mass. In \eq{eq:0vbb:lagrangian} we consider the first generation electron and neutrino only. Generalising to three flavours amounts to promoting the $\epsilon^\phi_{RX}$ couplings to $3\times 3$ matrices in generation space, $(\epsilon^\phi_{RX})_{\alpha i}$ ($\alpha = e, \mu, \tau$, $i = \nu_1, \nu_2, \nu_3$). The final decay rate will then be proportional to $|\epsilon^\phi_{RX}|^2 \to |\sum_i (\epsilon^\phi_{RX})_{ei} U_{ei}|^2$, where $U$ is the Standard Model lepton mixing matrix.

%%%%%%%%%%%%%%%%%%%%%%%%%%%%%%%%%%%%%%%%%%%%%%%%%%%%%%%%%%%%%%%
%%%%%%%%%%%%%%%%%%%%%%%%%%%%%%%%%%%%%%%%%%%%%%%%%%%%%%%%%%%%%%%
\section{Calculation of the exotic Majoron process} \label{sec:0vbb:calculation}

Here, we detail the computation of the amplitude and differential decay rate of the $0\nu\beta\beta\phi$ process. We follow the calculation of the standard long-range contributions presented in \cite{Doi:1985dx} and start from the effective Lagrangian,
\begin{align} \label{eq:0vbb:lagrangianapp}
    \mathcal{L}_{0\nu\beta\beta\phi} = \mathcal{L}_{\rm SM} + \mathcal{L}_{R\phi} \, ,
\end{align}
with the Standard Model charged current,
\begin{align}
    \mathcal{L}_{\rm SM} = 
    \frac{G_F\cos\theta_C}{\sqrt{2}}
    j^\mu_L J^{\phantom{\mu}}_{L\mu} + \hc \, ,
\end{align}
and the exotic 7-dimensional operators incorporating right-handed lepton currents and the Majoron $\phi$,
\begin{align}
    \mathcal{L}_{R\phi} = \frac{G_F\cos\theta_C}{\sqrt{2} m_p}
    \left(
          \epsilon_{RL}^\phi j_R^\mu J^{\phantom{\mu}}_{L\mu}\phi 
        + \epsilon_{RR}^\phi j_R^\mu J^{\phantom{\mu}}_{R\mu}\phi  
    \right)
    + \hc \, .
\end{align}
Here, $G_F$ is the Fermi constant, $\theta_C$ is the Cabbibo angle and the leptonic and hadronic currents are defined as,
\begin{align}
    j_{L,R}^\mu = \bar e\gamma^\mu (1\mp\gamma_5) \nu \, ,
    \quad
    J_{L,R}^\mu = \bar u\gamma^\mu (1\mp\gamma_5) d \, ,
\end{align}
respectively.

To the lowest order in perturbation theory, the amplitude for the process of $0_I^+ \to 0_F^+$ $0\nu\beta\beta\phi$ decay depicted in \fig{fig:0vbb:diagram}~(centre) is,
\begin{align}
    \mathcal{M} = -\int d^4x d^4y \langle F|\mathcal{T} \left\lbrace \mathcal{L}_{\rm SM}(x)\mathcal{L}_{R\phi}(y) 
    \right\rbrace |I \rangle \, .
\end{align}
The time-ordered product can be expanded as,
\begin{align}
    \mathcal{T} \left\lbrace \mathcal{L}_{\rm SM}(x) \mathcal{L}_{R\phi}(y) \right\rbrace 
    &= 2 \, \epsilon_{RX} \frac{(G_F\cos\theta_C)^2}{m_p}
    \nn\\ 
    &\times\mathcal{T} \lbrace J_L^\mu(x) J_X^\nu(y) 
    \underbrace{\bar e(x)\gamma_\mu P_L\nu(x)\bar\nu(y)\gamma_\nu P_L e^c(y)}_{ \Xi^L_{\mu\nu}(x,y)} 
    \phi(y) \rbrace \, ,
\end{align}
with the chiral projectors defined as $P_{L,R} = \frac{1}{2}(1\mp \gamma_5)$. Using the neutrino propagator with momentum $q$ and mass $m_\nu$, the highlighted term $\Xi^L_{\mu\nu}(x,y)$ can be expressed as,
\begin{align}
    \Xi^L_{\mu\nu}(x,y) &= 
    \int\frac{d^4 q}{(2\pi)^4} \frac{e^{-iq(x-y)}}{q^2-m_\nu^2 + i\varepsilon} 
    \bar e(x) \gamma_\mu P_L(\slashed{q} + m_\nu)\gamma_\nu P_L e^c(y)
    \nn\\
    &= \int\frac{d^4 q}{(2\pi)^4} q^\alpha 
    \frac{e^{-iq(x-y)}}{q^2 - m_\nu^2 + i\varepsilon} 
    \bar e(x)\gamma_\mu \gamma_\alpha \gamma_\nu P_L e^c(y) \, .
\end{align}
The amplitude needs to be antisymmetric under the exchange of the electrons $e_1$ and $e_2$, and thus we generalise,
\begin{align}
    \Xi^{L/R}_{\mu\nu}(x,y) = 
    \frac{1}{\sqrt{2}}\int\frac{d^4 q}{(2\pi)^4} \frac{e^{-iq(x-y)}}{q^2-m_\nu^2+i\varepsilon}
    \left(u^{L/R}_{\mu\nu}(E_1x,E_2y) - u^{L/R}_{\mu\nu}(E_2x,E_1y)\right) \, ,
\end{align}
with $u^{L/R}_{\mu\nu}(E_1x, E_2y) = q^\alpha \bar e(E_1,x)\gamma_\mu \gamma_\alpha \gamma_\nu P_{L/R} \, e^c(E_2,y)$ and $E_i$ is the energy of each electron.

We now perform the integral over the temporal variables. The integration over $q_0$ is straightforward by means of the residue theorem,
\begin{align}
    \int\frac{dq_0}{2\pi} \frac{1}{q_0^2-\omega^2} f(q_0) 
    = \frac{i}{2 \omega} f(\omega) \, ,
\end{align}
with $\omega^2 = \mathbf{q}^2 + m_\nu^2$.

On the other hand, expanding the time-ordered product as,
\begin{eqnarray}
    \mathcal{T} \left\lbrace 
    \mathcal{L}_{\rm SM}(x)\mathcal{L}_{R\phi}(y) \right\rbrace 
    &=& \Theta(x^0-y^0) \mathcal{L}_{\rm SM}(x) \mathcal{L}_{R\phi}(y)
    \\ \nn
    &+& \Theta(y^0-x^0) \mathcal{L}_{R\phi}(y)\mathcal{L}_{\rm SM}(x) \, ,
\end{eqnarray}
and using the operator $e^{iHt}$ to extract the temporal dependence from the different wave functions, for example $\phi(y) = e^{iE_\phi y^0} \phi(\mathbf{y})$, one can directly integrate over $x^0$ and $y^0$ obtaining the analogous expression to equation~(C.2.19) in \cite{Doi:1985dx},
\begin{align} \label{eq:0vbb:M}
    \mathcal{M} = \epsilon_{RX}^{\phi}\frac{(G_F\cos\theta_C)^2}{\sqrt{2}m_p} &
    \sum_N \int d^3x d^3y \int\frac{d^3q}{2\pi^2\omega} J_{LX}^{\rho\sigma}(\mathbf{x},\mathbf{y})\phi(\mathbf{y})
    \\ \nn
     &\times\left\lbrace\quad\,  e^{i\mathbf{q}(\mathbf{x} - \mathbf{y})}
      \left[     
           \frac{u_{\rho\sigma}^L(E_1\mathbf{x},E_2\mathbf{y})}{\omega+A_2+\frac 12 E_\phi} 
         - \frac{u_{\sigma\rho}^R(E_1\mathbf{y},E_2\mathbf{x})}{\omega+A_1+\frac 12 E_\phi}
      \right]\right.
      \\ \nn
      &\quad\quad\! -e^{i\mathbf{q}(\mathbf{y}-\mathbf{x})} 
      \left.\left[
            \frac{u_{\rho\sigma}^L(E_1\mathbf{x},E_2\mathbf{y})}{\omega+A_1-\frac 12 E_\phi} 
          - \frac{u_{\sigma\rho}^R(E_1\mathbf{y},E_2\mathbf{x})}{\omega+A_2-\frac 12 E_\phi} 
      \right]\right\rbrace \, ,
\end{align}
where $A_{1/2} = E_N - E_I + \frac 12 Q_{\beta\beta} + m_e \pm \frac{1}{2}(E_1 - E_2)$. We anticipate the closure approximation and define the matrix element of the hadronic currents as,
\begin{align}
    J_{LX}^{\rho\sigma}(\mathbf{x},\mathbf{y}) 
    = \frac 12 \left[ \bra{F} J_L^\rho(\mathbf{x}) \ket{N}\bra{N} J_X^\sigma(\mathbf{y}) \ket{I} + \bra{F} J_X^\sigma(\mathbf{y}) \ket{N}\bra{N} J_L^\rho(\mathbf{x}) \ket{I} \right] \, .
\end{align}
In addition, the following properties under the exchange of position and electron energies were used in \eq{eq:0vbb:M},
\begin{equation}
    u^{L/R}_{\rho\sigma}(E_1\mathbf{x},E_2\mathbf{y}) = u^{R/L}_{\sigma\rho}(E_1\mathbf{x},E_2\mathbf{y}), \qquad J_{LX}^{\rho\sigma}(\mathbf{x},\mathbf{y}) = J_{XL}^{\sigma\rho}(\mathbf{y},\mathbf{x}) \, .
\end{equation}
The integration over $x^0$ and $y^0$ in \eq{eq:0vbb:M} also provides the overall energy conservation condition $\delta(Q_{\beta\beta} + 2m_e - E_1 - E_2 - E_\phi)$ with $Q_{\beta\beta} = E_I - E_F - 2m_e$. It is included in the phase space, \eq{eq:0vbb:decayrate} below, by requiring $E_\phi = Q_{\beta\beta} + 2m_e - E_1 - E_2$. We additionally assume that the Majoron $\phi$ is emitted predominantly in an $S$-wave configuration, $\phi(\mathbf{y}) \approx 1$.

Considering the term between braces in \eq{eq:0vbb:M}, one can write everything under the same exponential by interchanging $\mathbf{x}$ and $\mathbf{y}$,
\begin{align} \label{eq:0vbb:y}
    e^{i\mathbf{q}(\mathbf{x}-\mathbf{y})} 
    &\left\lbrace \left[
          \frac{J^{\rho\sigma}_{LX}(\mathbf{x},\mathbf{y})  
          u^L_{\rho\sigma}(E_1\mathbf{x},E_2\mathbf{y})}{\omega+A_2+\frac{1}{2}E_\phi} 
        + \frac{J^{\rho\sigma}_{XL}(\mathbf{x},\mathbf{y})  
            u^R_{\rho\sigma}(E_1\mathbf{x},E_2\mathbf{y})}{\omega+A_2-\frac{1}{2}E_\phi} 
      \right]\right.
    \\ \nn
    &\left.\quad\! -\left[ 
          \frac{J^{\rho\sigma}_{LX}(\mathbf{x},\mathbf{y})    
          u^L_{\rho\sigma}(E_2\mathbf{x},E_1\mathbf{y})}{\omega+A_1+\frac{1}{2}E_\phi} 
        + \frac{J^{\rho\sigma}_{XL}(\mathbf{x},\mathbf{y}) \; 
          u^R_{\rho\sigma}(E_2\mathbf{x},E_1\mathbf{y})}{\omega+A_1-\frac{1}{2}E_\phi}
    \right]\right\rbrace \, .
\end{align}
It is furthermore useful to split the leptonic $u^{L,R}_{\rho\sigma}$ functions by separating out the part containing $\gamma_5$ as $u^{L/R}_{\rho\sigma} = \frac{1}{2} \left[u_{\rho\sigma} \mp u^5_{\rho\sigma}\right]$. We then define,
\begin{align} \label{eq:0vbb:fmunu}
    F^\pm_{\rho\sigma} &= u_{\rho\sigma}(E_1\mathbf{x},E_2\mathbf{y}) 
        \pm u_{\sigma\rho}(E_1\mathbf{y},E_2\mathbf{x}),
    \\
    F^{5\pm}_{\rho\sigma} &= u^5_{\rho\sigma}(E_1\mathbf{x},E_2\mathbf{y}) 
        \pm u^5_{\sigma\rho}(E_1\mathbf{y},E_2\mathbf{x}),
    \\
    J^\pm_{\rho\sigma} &= J^{LX}_{\rho\sigma}(E_1\mathbf{y},E_2\mathbf{x}) 
        \pm J^{XL}_{\rho\sigma}(E_1\mathbf{y},E_2\mathbf{x}) \, .
    \label{eq:0vbb:fmunu3}
\end{align}
These definitions become useful if one recalls that in the non-relativistic impulse approximation, the $J^L$ part of $J^{LX}_{\rho\sigma}$ acts on the $n$-th nucleon whereas the $J^X$ part acts on the $m$-th when performing the sum over all neutrons in the initial nucleus. The superscript $\pm$ in $J^\pm_{\rho\sigma}$ thus indicates if the combination of currents is symmetric or antisymmetric under the interchange of $m\leftrightarrow n$. The same applies to $F^\pm_{\rho\sigma}$ and $F^{5\pm}_{\rho\sigma}$.

The closure approximation implies that the sum over all possible intermediate states is performed analytically using the completeness of all intermediate states and by replacing the intermediate state energies $E_N$ with a common average $\langle E_N \rangle$. This means that the antisymmetric combinations under the interchange of the nucleons $m$ and $n$ will vanish, as the sum is performed over all possible configurations. From \eq{eq:0vbb:M} and \eq{eq:0vbb:y}, the non-vanishing terms are,
\begin{eqnarray}
    \mathcal{M} &=& \epsilon_{RX} \frac{(G_F\cos\theta_C)^2}{2\sqrt{2}m_p} \sum_N (H_{\omega 2} - H_{\omega 1})
     \\\nn
    &&\times\left\lbrace 
          J_{\mu\nu}^+ F^{+,\mu\nu}   
        - J_{\mu\nu}^- F^{5-,\mu\nu}
        + \frac{E_\phi}{E_{12}}
          \left(J_{\mu\nu}^+ F^{5+,\mu\nu} - J_{\mu\nu}^-F^{-,\mu\nu}
          \right)\right\rbrace \, ,
\end{eqnarray}
where $H_{\omega i}$ are neutrino potentials defined as,
\begin{align} \label{eq:0vbb:Hw}
    H_{\omega i} = \int \frac{d^3q}{2\pi^2\omega} \frac{\omega}{\omega + A_i} e^{i\mathbf{q}(\mathbf{x}-\mathbf{y})} \, .
\end{align}
Now, the connection with the results of \cite{Doi:1985dx} can be done by contracting the leptonic and nuclear currents within the impulse approximation. The only change in our case is in the $\omega$ term,
\begin{align} 
\label{eq:0vbb:M2}
    \mathcal{M}_\omega &\propto (H_{\omega 2} - H_{\omega 1})
    \left\lbrace 
          (X_3 + X_{5 R})\left[F_+^0 + \frac{E_{\phi}}{E_{12}}F_{5 +}^0\right] 
        + Y_{3R}\left[F_{5 -}^0 + \frac{E_{\phi}}{E_{12}}F_{-}^0\right]\right. 
    \nn\\
     &+ \left.(X_{4 R}^l + X_{5}^l)\left[F_+^l +\frac{E_{\phi}}{E_{12}}F_{5 +}^l \right] 
      + (Y_4^l -Y_{5R}^l)\left[F_{5-}^l 
      + \frac{E_{\phi}}{E_{12}}F_{-}^l\right]\right\rbrace \, ,
\end{align} 
where the $X$ and $Y$ terms are functions of nuclear parameters and operators defined in the Appendix~C in \cite{Doi:1985dx}. The $F^{\alpha}_{(5)\pm}$-terms are generated by the contraction of the hadronic and leptonic parts in \eqs{eq:0vbb:fmunu}{eq:0vbb:fmunu3} factorising out the dependence with the momentum $q_{\alpha}$ from the leptonic part (see  equation~(C.2.25) in \cite{Doi:1985dx}). One trivially recovers the $\omega$ term in the expression~(C.2.23) in \cite{Doi:1985dx} for $E_\phi \to 0$.

Comparing \eq{eq:0vbb:M2} with the results from \cite{Doi:1985dx}, one can track the dependence with $E_\phi$ in the decay rate down to equation~(C.3.9) of \cite{Doi:1985dx}. The main change for $0^+_I\to 0^+_F$ transitions is in the terms $N_3$ and $N_4$ where a contribution proportional to $E_{\phi}$ appears explicitly, 
\begin{align} \label{eq:0vbb:N12}
    \begin{pmatrix}
        N_1 \\
        N_2
    \end{pmatrix}
    &=
    \begin{pmatrix}
        \alpha_{-1-1}^* \\
        \alpha_{11}^*
    \end{pmatrix}
    \left[ \frac{4}{3}Z_6
    \mp \frac{4}{m_e R}\left(Z_{4R}-\frac{1}{6} \zeta Z_6\right)\right] \, , \\
    \begin{pmatrix}
        N_3 \\
        N_4
    \end{pmatrix}
    &=
    \begin{pmatrix}
        \alpha_{1-1}^* \\
        \alpha_{-11}^*
    \end{pmatrix}
    \left[ -\frac{2}{3}Z_5 
    \mp \frac{E_{12}}{m_e}
    \left(Z_3 + \frac{1}{3}Z_5\right)+\frac{E_{\phi}}{m_e}Z_3 \right] \, .
    \label{eq:0vbb:N34}
\end{align}
Here, $\alpha_{jk} = \tilde{A}_j(E_1) \tilde{A}_k(E_2)$ describe the Coulomb-corrected relativistic electron wave functions and $\zeta = 3\alpha Z + (Q_{\beta\beta} + 2m_e)R$ the correction of the electron $P$ wave, with the fine structure constant $\alpha$ and the radius $R$ and charge $Z$ of the final state nucleus. The information about the electron wave functions is encoded in 
\begin{align}
    \tilde{A}_{\pm k}(E) = \sqrt{\frac{E \mp m_e}{2E} F_{k-1}(Z,E)} \, ,
\end{align}
with the Fermi factor,
\begin{align}
    F_{k-1}(Z,E) = \left[\frac{\Gamma(2k+1)}{\Gamma(k)\Gamma(2\gamma_k+1)}\right]^2(2pR)^{2(\gamma_k -k)}|\Gamma(\gamma_k+ iy)|^2e^{\pi y} \, ,
\end{align}
where $\gamma_k = \sqrt{k^2 + (\alpha Z)^2}$, $y = \alpha Z E/p$ and $p = \sqrt{E^2 - m_e^2}$.

In order to arrive at \eq{eq:0vbb:M2} one should neglect the higher order terms $E_{12}^2$, $E_{12}E_\phi$ and $E_\phi^2$ as they are suppressed with an extra denominator $(\omega+A_i)$ compared to \eq{eq:0vbb:Hw}.

\begin{table}
    \centering
    \small\setlength{\tabcolsep}{3.2pt}
    \begin{tabular}{ccccccccccc}
        \hline
        Isotope & $Q_{\beta\beta}$ & $M_{GT}$ & $\chi_F$ & $\chi_{GT\omega}$ & $\chi_{F\omega}$ & $\chi_{GT}'$ & $\chi_F'$ & $\chi_T$ & $\chi_R$ & $\chi_P$
        \\
        \hline
        $^{82}\text{Se}$ & 2.99 & 2.993 & $-0.134$ & 0.947 & $-0.131$ & 1.003 & $-0.103$ & $\phantom{-}0.004$ & 1.086 & 0.430
        \\
        $^{136}\text{Xe}$ & 2.46 & 1.770 & $-0.158$ & 0.908 & $-0.149$ & 1.092 & $-0.167$ & $-0.031$ & 0.955 & 0.256
        \\
        \hline
    \end{tabular}
    \caption{Energy release $Q_{\beta\beta}$ [MeV] and relevant nuclear matrix elements for $^{82}\text{Se}$ and $^{136}\text{Xe}$ used in the calculation of the $0\nu\beta\beta\phi$ decay rate and distributions. The nuclear matrix elements were taken from the shell model calculations \cite{Horoi:2015gdv} ($^{82}$Se) and \cite{Caurier:1996bu} ($^{136}$Xe), except for $M_{GT}$ in $^{136}\text{Xe}$ where we use an updated value from the same group \cite{Caurier:2007qn}.}
    \label{tab:0vbb:NMEs}
\end{table}

The $Z_i$ terms are given in \eqs{eq:0vbb:Zs}{eq:0vbb:Zs4} below, and they contain the nuclear matrix elements and effective particle physics couplings. The $Z_i$ terms are the same as in \cite{Doi:1985dx}, with the relevant couplings $\lambda \to \epsilon_{RR}^{\phi}$ and $\eta \to \epsilon_{RL}^{\phi}$ substituted. Note that the term with $Z_1$ in equation~(C.3.9) in \cite{Doi:1985dx} related to the standard $0\nu\beta\beta$ decay disappears from \eq{eq:0vbb:N12}, as we are not considering the interaction $\mathcal{L}_{SM}(x) \mathcal{L}_{SM}(y)$.
\begin{align} 
\label{eq:0vbb:Zs}
    Z_3 &= \left[-\epsilon_{RR}^{\phi}(\chi_{GT\omega}-\chi_{F\omega}) + \epsilon_{RL}^{\phi}(\chi_{GT\omega}+\chi_{F\omega}) \right] M_{GT} \, ,
    \\
    Z_{4R} &= \epsilon_{RL}^{\phi} \chi_R M_{GT} \, ,
    \\
    Z_5 &= \frac 13 \left[ \epsilon_{RR}^{\phi}(\chi'_{GT}-6\chi_T+3\chi'_F) - \epsilon_{RL}^{\phi}(\chi'_{GT}-6\chi_T-3\chi'_F) \right] M_{GT} \, ,
    \\
    Z_6 &= \epsilon_{RL}^{\phi} \chi_P M_{GT} \, .
   \label{eq:0vbb:Zs4}
\end{align}
These equations are valid when both $\epsilon^\phi_{RL}$ and $\epsilon^\phi_{RR}$ are present. For our numerical calculations, we use the $Q_{\beta\beta}$ values and nuclear matrix elements $M_{GT}$, $\chi_F$, etc. presented in \tab{tab:0vbb:NMEs} for $^{82}\text{Se}$ and $^{136}\text{Xe}$. We use the following values for the remaining parameters: $G_F = 1.2 \times 10^{-5}~\text{GeV}^{-2}$, $\alpha = 1/137$, $g_A = 1.27$, $R = 1.2 A^{1/3}$~fm with the mass number $A$ of the isotope in question. The factors $N_1$, $N_2$, $N_3$ and $N_4$ in \eq{eq:0vbb:N12} and \eq{eq:0vbb:N34} are then fully described and the energy-dependent coefficients are,
\begin{align}
    a(E_1, E_2, E_\phi) &= |N_1|^2 + |N_2|^2 + |N_3|^2 + |N_4|^2 \, ,
    \\
    b(E_1, E_2, E_\phi) &= -2\,\text{Re}\left(N_1^* N_2 + N_3^* N_4\right) \, .
\end{align}
The differential decay rate for the $0^+\to 0^+$ $0\nu\beta\beta\phi$ decay can then be written as \cite{Doi:1985dx},
\begin{align}
\label{eq:0vbb:decayrate}
    d\Gamma = C
    \left[a(E_1,E_2,E_\phi) + b(E_1,E_2,E_\phi) \cos\theta\right] 
    w(E_1, E_2, E_\phi) \, dE_1 \, dE_2 \, d\!\cos\theta  \, ,
\end{align}
with 
\begin{gather}
    C = \frac{(G_F\cos\theta_C g_A)^4 m_e^{9}}{256\pi^7 (m_p R)^2} \, ,
    \\
    w(E_1, E_2, E_\phi) = m_e^{-7} p_1 p_2 E_1 E_2 E_\phi \, .
\end{gather}
Here, $g_A$ is the axial coupling of the nucleon and $R$ is the radius of the nucleus. The magnitudes of the electron momenta are given by $p_i = \sqrt{E_i^2 - m_e^2}$ and $0 \leq \theta \leq \pi$ is the angle between the emitted electrons. Throughout the above expressions, the Majoron energy is implicitly fixed by the electron energies as $E_\phi = Q_{\beta\beta} + 2m_e - E_1 - E_2$ due to overall energy conservation.

From the fully differential decay rate \eq{eq:0vbb:decayrate}, we can extract all the necessary information for the different distributions.

%%%%%%%%%%%%%%%%%%%%%%%%%%%%%%%%%%%%%%%%%%%%%%%%%%%%%%%%%%%%%%%
%%%%%%%%%%%%%%%%%%%%%%%%%%%%%%%%%%%%%%%%%%%%%%%%%%%%%%%%%%%%%%%
\section{Decay rate and distributions} \label{sec:0vbb:rate}

From \eq{eq:0vbb:decayrate}, the total decay rate and thus the half-life is calculated by performing the integration of $a(E_1,E_2,E_\phi)$ over all energies within the allowed phase space limits $E_1$, $E_2 \geq 0$ and $E_1 + E_2 \leq Q_{\beta\beta} + 2m_e$. The total decay rate $\Gamma$ and the half-life $T_{1/2}$ are then calculated as,
\begin{align}
\label{eq:0vbb:TotalRate}
    \Gamma = \frac{\ln 2}{T_{1/2}} = 2 C
    \int_{m_e}^{Q_{\beta\beta} + m_e} dE_1 \int_{m_e}^{Q_{\beta\beta} + 2m_e - E_1} dE_2 \, 
    a(E_1, E_2, E_\phi) w(E_1, E_2, E_\phi) \, .
\end{align}

In addition, we determine and discuss several distributions below. We will show results for the $\epsilon^\phi_{RL}$ and $\epsilon^\phi_{RR}$ versions of the effective operators where we consider only one of these to be present at a time. We assume the exotic $\phi$ Majoron to be massless in our calculations and comment on massive $\phi$ in the discussion below. For our numerical evaluation we focus on two isotopes: (i) $^{136}\text{Xe}$, for which the KamLAND-Zen collaboration \cite{KamLAND-Zen:2016pfg} currently provides the most stringent constraints; (ii) $^{82}\text{Se}$ used by NEMO-3 and the upcoming SuperNEMO experiments \cite{Arnold:2018tmo} that can measure the detailed electron topology.
\\

\begin{figure}
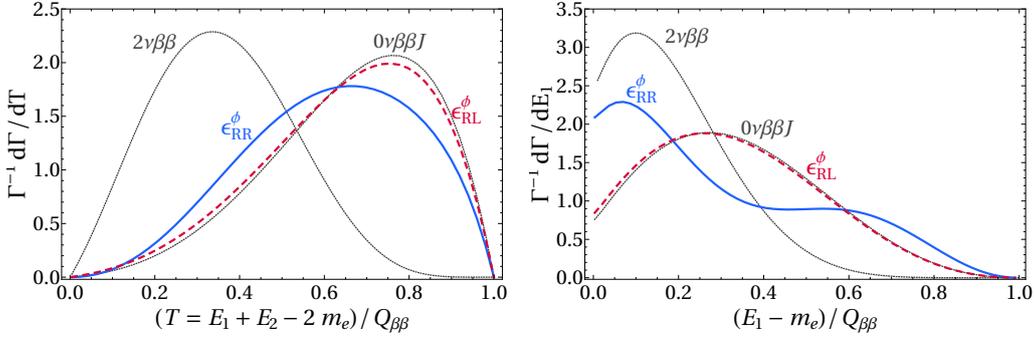

    \centering
    \includegraphics[clip,trim={10 10 0 0},width=0.49\textwidth]{figures/TotalDistributions_Xe}
    \hfill
    \includegraphics[clip,trim={10 10 0 0},width=0.49\textwidth]{figures/Single_Se_0vBBJ}
    \caption{Left: Normalised $0\nu\beta\beta\phi$ decay distributions in the total kinetic energy of the electrons for $^{136}\text{Xe}$. Right: Normalised $0\nu\beta\beta\phi$ decay distribution in the single electron kinetic energy distribution for $^{82}\text{Se}$. The blue solid and red dashed lines correspond to the $\epsilon^\phi_{RR}$ and $\epsilon^\phi_{RL}$ cases, respectively. The corresponding distributions for the Standard Model $2\nu\beta\beta$ decay and ordinary $0\nu\beta\beta J$ Majoron decay (spectral index $n = 1$) are given for comparison.}
    \label{fig:0vbb:energydist}
\end{figure}

For all experimental searches, the crucial distribution is with respect to the sum of the kinetic energies of the detected electrons. With the Standard Model $2\nu\beta\beta$ decay as irreducible background to any exotic signal, it is important to calculate it precisely. In \fig{fig:0vbb:energydist}~(left), we compare the normalised total electron kinetic energy distribution of $0\nu\beta\beta\phi$ decay with that of $2\nu\beta\beta$ decay and ordinary $0\nu\beta\beta J$ Majoron decay (with spectral index $n = 1$) for the isotope $^{136}\text{Xe}$. The distribution associated with $\epsilon^\phi_{RL}$ is very similar to ordinary $0\nu\beta\beta J$ decay, while the introduction of a hadronic right-handed current in the $\epsilon^\phi_{RR}$ term changes considerably the shape of the distribution. In both cases, the spectral index still corresponds to $n=1$ with the characteristic onset near the kinematic endpoint. We emphasise that because of the different shape, a dedicated signal over background analysis is required to determine the experimental sensitivity on the effective parameters $\epsilon^\phi_{RL}$ and $\epsilon^\phi_{RR}$ precisely.

NEMO-3 and SuperNEMO are able to measure the individual electron energies. In right-handed current scenarios without emission of a Majoron, the single energy distribution exhibits a distinctive valley-type shape. This occurs as the dominant term is proportional to $(E_1 - E_2)$ for the corresponding $\epsilon_{RR}$ term, as a result of the antisymmetry with respect to electron exchange.\footnote{For the $\epsilon_{RL}$ term with a left-handed hadronic current, $P$-wave and nuclear recoil contribute constructively, giving a dominant contribution proportional to $(E_1 + E_2)$ \cite{Doi:1985dx}.} In our case, depicted in \fig{fig:0vbb:energydist}~(right), part of the energy is being carried away by the Majoron, shifting the distribution towards lower electron energies and softening the characteristic valley-type distribution for $\epsilon^\phi_{RR}$. The distribution does not vanish for $E_1 - m_e = \frac{1}{2}Q_{\beta\beta}$ (as in the ordinary right-handed current case), but is still significantly different from that of ordinary Majoron emission.

\begin{figure}[t!]
    \centering
    \includegraphics[clip,trim={70 20 50 20},width=\textwidth]{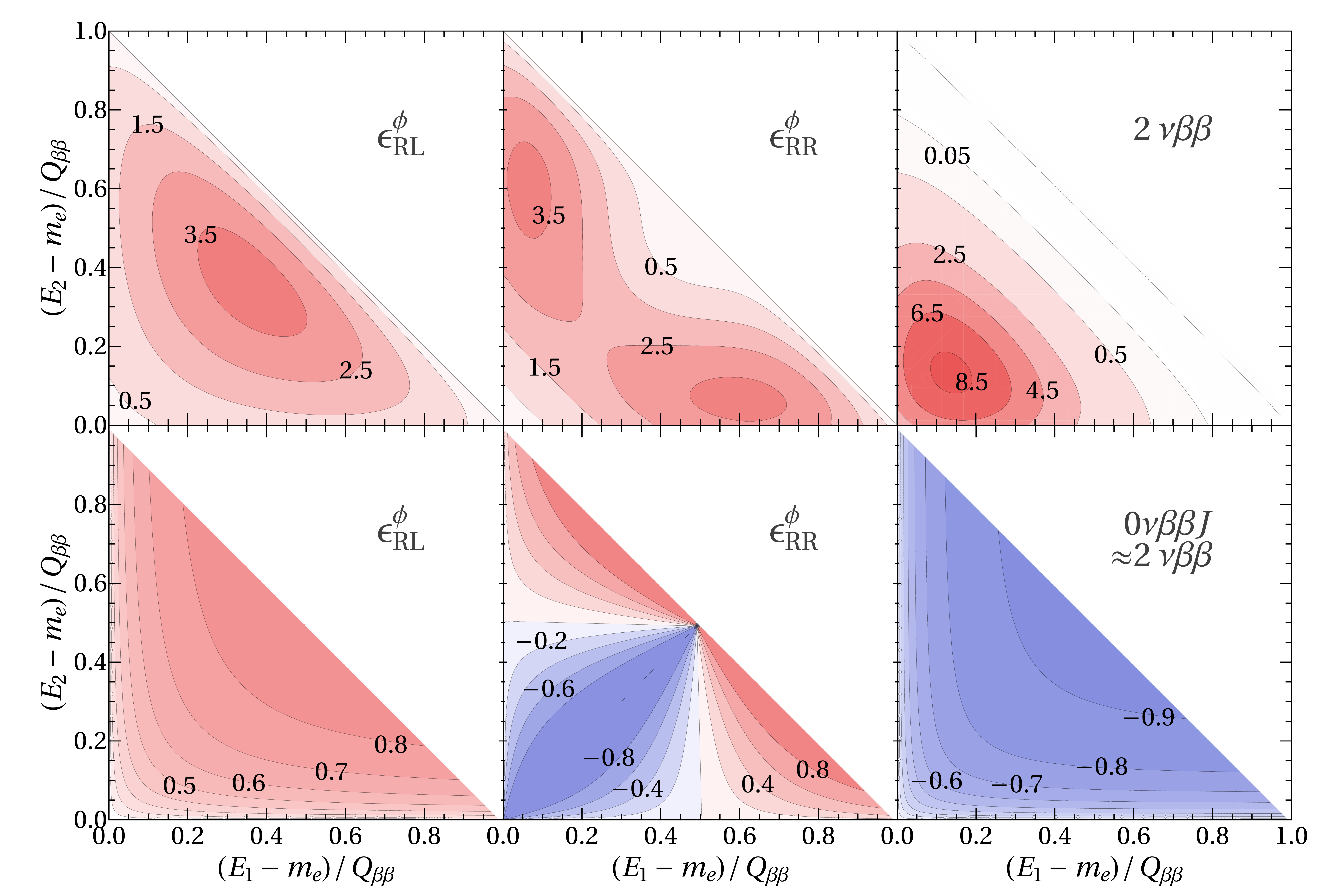}
    \caption{Double electron energy distribution $\frac{d\Gamma}{dE_1dE_2}$ (top row) and electron angular correlation $\alpha$ (bottom row) as function of the individual electron kinetic energies for $^{82}\text{Se}$. Each column is for a specific scenario: $0\nu\beta\beta\phi$ Majoron emission through $\epsilon^\phi_{RL}$ (left) and $\epsilon^\phi_{RR}$ (centre); Standard Model $2\nu\beta\beta$ decay (right). The angular correlation of the latter is approximately identical to ordinary Majoron emission $0\nu\beta\beta J$.}
    \label{fig:0vbb:angular2}
\end{figure}

We can analyse in detail the distributions for the individual electron energies integrating the angular dependence in \eq{eq:0vbb:decayrate}. The fully differential energy information is encoded in the normalised double energy distribution,
\begin{align}
    \Gamma^{-1}\frac{d\Gamma}{dE_1 dE_2} 
    = \frac{2 C}{\Gamma} a(E_1, E_2, E_\phi) w(E_1, E_2, E_\phi) \, .
\end{align}
This function, in terms of the kinetic energies normalised to the $Q$ value, $(E_i - m_e)/Q_{\beta\beta}$, is plotted in the top row of \fig{fig:0vbb:angular2} for the case of $0\nu\beta\beta\phi$ Majoron emission through $\epsilon^\phi_{RL}$ (left) and $\epsilon^\phi_{RR}$ (centre) as well as for the Standard Model $2\nu\beta\beta$ decay (right). The plots are for the isotope $^{82}\text{Se}$ but would be qualitatively similar for $^{136}\text{Xe}$. As can be seen, the shapes depicted as contours are different between all three modes. Especially the $\epsilon^\phi_{RR}$ exhibits an asymmetry in that one of the electrons takes the majority of the visible energy. If the individual electron energies can be measured, as e.g. in the NEMO-3 or SuperNEMO experiments, this can be exploited to enhance the signal over the $2\nu\beta\beta$ background. The distribution with respect to both electron energies depicted in \fig{fig:0vbb:angular2}~(top panel) exhibits an even more pronounced difference between the $\epsilon^\phi_{RR}$ mode and $2\nu\beta\beta$. This may be used experimentally to improve the sensitivity through kinematic selection criteria, counteracting the effect of the less peaked total energy distribution, cf. \fig{fig:0vbb:energydist}~(left). As an illustrating example, requiring that any one of the electrons in a signal event has a kinetic energy $E_i - m_e > Q_{\beta\beta} / 2$ would reduce the $0\nu\beta\beta\phi$-$\epsilon^\phi_{RR}$ rate only by a factor of 2 but would suppress the $2\nu\beta\beta$ rate by a factor of 20.

Note that \fig{fig:0vbb:angular2} provides the full kinematical information in each mode; all measurable quantities can be constructed from these distributions. For example, the distributions in \fig{fig:0vbb:energydist} can be easily determined by appropriately integrating the first row over $\frac{d\Gamma}{dE_1 dE_2}$.

In addition to the energies, the angle between the electron momenta also contains useful information. The so-called angular correlation defined by,
\begin{align}
    \alpha(E_1, E_2) 
    = \frac{b(E_1, E_2, E_\phi)}{a(E_1, E_2, E_\phi)} \, ,
\end{align}
is a function of the individual electron energies which can take values between $(-1)$ (the two electrons are dominantly emitted back-to-back) and $(+1)$ (the two electrons are dominantly emitted collinearly). For $^{82}\text{Se}$ it is plotted in the bottom row of \fig{fig:0vbb:angular2} in the three modes of interest. As expected from angular momentum considerations, the electrons are dominantly emitted back-to-back in the Standard Model $2\nu\beta\beta$ decay with $(V-A)$ lepton currents, $\alpha < 0$ for all energies. For $\epsilon^\phi_{RL}$, they are dominantly emitted collinearly, $\alpha > 0$ for all energies. In the case of $\epsilon^\phi_{RR}$, the behaviour is complex due to the asymmetry of the amplitude under the exchange of electrons and nuclear recoil effects. The correlation $\alpha$ is positive when one of the electrons has a kinetic energy $E_i - m_e > Q_{\beta\beta} / 2$, but changes sign ($\alpha <0$) when the kinetic energy of both electrons is below $Q_{\beta\beta} / 2$.

One can use the angular correlations to distinguish between left-handed and right-handed currents \cite{Doi:1983wv,Arnold:2010tu}, see \fig{fig:0vbb:angular2}~(bottom panel). Integrating over the electron energies one obtains the average angular distribution which takes the simple form,
\begin{equation} %\label{eq:}
    \frac{d\Gamma}{d\!\cos\theta} = \frac{\Gamma}{2}(1 +k\cos\theta) \, .
\end{equation}
The coefficient $k$ is $k^\phi_{RL} = + 0.70$ (electrons are dominantly emitted collinearly) and $k^\phi_{RR} = -0.05$ (electrons are emitted nearly isotropically) in our $0\nu\beta\beta\phi$ scenarios with $\epsilon^\phi_{RL}$ and $\epsilon^\phi_{RR}$, respectively, for $^{82}\text{Se}$. For comparison, the angular correlation factor for Standard Model $2\nu\beta\beta$ decay is $k^{2\nu\beta\beta} = -0.66$ and $k^J = -0.80$ for ordinary Majoron emission; i.e. the electrons are dominantly emitted back-to-back.
\\

\begin{table}
    \centering
    \setlength{\tabcolsep}{6pt}
    \begin{tabular}{cccc}
        \hline
        Isotope & $T_{1/2}$~[y] & $|\epsilon^\phi_{RL}|$ & $|\epsilon^\phi_{RR}|$
        \\
        \hline
        {$^{82}\text{Se}$} & $3.7\times 10^{22}$ \cite{Arnold:2018tmo} & $ 4.1\times 10^{-4}$ & $4.6\times 10^{-2}$
        \\
        {$^{136}\text{Xe}$} & $2.6 \times 10^{24}$ \cite{KamLAND-Zen:2016pfg} & $1.1\times 10^{-4}$ & $1.1\times 10^{-2}$
        \\
        \hline
        {$^{82}\text{Se}$} & $1.0\times 10^{24}$ & $8.0\times 10^{-5}$ & $8.8\times 10^{-3}$
        \\
        {$^{136}\text{Xe}$} & $1.0\times 10^{25}$ & $5.7\times 10^{-5}$ & $5.8\times 10^{-3}$
        \\
        \hline
    \end{tabular}
\caption{Current limits and expected future sensitivity on the effective couplings $\epsilon^\phi_{RL}$ and $\epsilon^\phi_{RR}$ of $0\nu\beta\beta\phi$ decay for $^{82}\text{Se}$ and $^{136}\text{Xe}$. The limits are estimated based on the experimental half-life constraints for ordinary Majoron emission (spectral index $n = 1$) as given. Nuclear matrix elements from \tab{tab:0vbb:NMEs} \cite{Horoi:2015gdv, Caurier:1996bu, Caurier:2007qn} were used for this estimate.}
\label{tab:0vbb:limits}
\end{table}

Finally, we estimate the sensitivity of existing and planned future double-$\beta$ decay searches on the effective coupling strength $\epsilon^\phi_{RL}$ and $\epsilon^\phi_{RR}$ of $0\nu\beta\beta\phi$ decay. We would like to emphasise again that due to the different total electron energy distribution, a dedicated signal over background analysis is required to determine the constraints precisely. Experiments such as NEMO-3 and SuperNEMO can also improve their sensitivity due to the non-standard decay topology, especially for $\epsilon^\phi_{RR}$. A requirement that any one electron has a kinetic energy of $E_i - m_e > Q_{\beta\beta}/2$ can for example reduce the $2\nu\beta\beta$ background by an order of magnitude. Here, we simply estimate the sensitivity by comparing our predictions for the $0\nu\beta\beta\phi$ decay half-life $T_{1/2} = \ln 2/\Gamma$ \eq{eq:0vbb:TotalRate} with the experimental constraints on ordinary ($n=1$) Majoron emission. We use the most stringent limits for $^{82}\text{Se}$ by NEMO-3 \cite{Arnold:2018tmo} and for $^{136}\text{Xe}$ by KamLAND-Zen \cite{KamLAND-Zen:2016pfg}. For future prospects, we estimate that experimental Majoron search sensitivities may reach $T_{1/2}^{\rm Se} \approx 10^{24}$~y (e.g. with the help of angular and energy selection cuts at SuperNEMO) and $T_{1/2}^{\rm Xe} \approx 10^{25}$~y.\footnote{The corresponding $0\nu\beta\beta$ decay sensitivities of the planned SuperNEMO \cite{Macolino:2017vyd} and nEXO experiments \cite{gratta_giorgio_2018_1286892} may improve by $\mathcal{O}(100)$, but this requires an experimental approach that is essentially background-free. This is not possible for Majoron emission with a continuous total electron energy spectrum.} The corresponding limits on $\epsilon^\phi_{RL}$ and $\epsilon^\phi_{RR}$ are shown in \tab{tab:0vbb:limits}, where only one effective operator is assumed to be present at a time.

\section{Summary and discussion} \label{sec:0vbb:summary}

Searches for Majorons or Majoron-like particles are a staple in double-$\beta$ decay experiments. So far, they only cover the case where the neutrino involved couples via the Standard Model $(V-A)$ charged current interaction. This is clearly a well-motivated minimal choice but it is worthwhile to explore other scenarios. In this chapter, we have discussed one such alternative where a Majoron-like particle $\phi$ is emitted from effective operators with $(V+A)$ leptonic currents, cf. \fig{fig:0vbb:diagram}~(centre). The future sensitivities on the effective couplings $\epsilon^\phi_{RL}$ and $\epsilon^\phi_{RR}$ shown in \tab{tab:0vbb:limits} may be translated into effective operator scales $\Lambda_{\rm NP} \approx 1.3$~TeV and $270$~GeV, respectively, using $1/\Lambda_{\rm NP}^{3} = \epsilon^\phi_{RX} G_F\cos\theta_C/(\sqrt{2}m_p)$. As noted before, we assume a massless $\phi$ in deriving these limits; they remain essentially unchanged for masses small compared to $Q_{\beta\beta}$, $m_\phi \lesssim 0.2$~MeV and are of the same order for $m_\phi \lesssim 1$~MeV, but will deteriorate as $m_\phi \to Q_{\beta\beta}$ (for a recent analysis in ordinary Majoron emission see \cite{Brune:2018sab}). Constraints on our operators may also be set from other processes, such as exotic decay modes of the pion, $\pi^- \to e^- \bar\nu_e \phi$. As we consider only $V+A$ currents, helicity suppression will still apply and the limits are expected to be correspondingly weak, we roughly estimate $\Lambda_{\rm NP} \gtrsim 15$~GeV.
\\

An example of an ultraviolet scenario generating the effective operators in \eq{eq:0vbb:lagrangian} is suggested in Left-Right symmetric models \cite{Pati:1974yy, Mohapatra:1974hk, Mohapatra:1974gc, Senjanovic:1975rk, Bolton:2019bou} where the Standard Model $W$ and $\nu$ are replaced by their right-handed counterparts $W_R$ and $N$. The heavy neutrino $N$ then mixes with $\nu$ via a Yukawa coupling $y_\nu$ once the Standard Model Higgs boson acquires its vacuum expectation value $\langle H \rangle = 174$~GeV. A massless or light scalar $\phi$ is not part of the minimal Left-Right symmetric model which thus needs to be modified, e.g. by keeping the $\rm U(1)_{B-L}$ symmetry global or by extending its scalar sector. Charging $\phi$ under lepton number allows coupling to $N$ with a strength $y_N$. The corresponding diagram is shown in \fig{fig:0vbb:diagram}~(right). We can then identify,
\begin{align}
    \frac{G_F \cos\theta_C}{\sqrt{2} m_p} \epsilon^\phi_{RR}
    = \frac{g_R^2 y_N y_\nu \langle H \rangle\cos\theta_C^R}{8m_{W_R}^2 m_N^2} \, ,
\end{align} 
leading to the estimate,
\begin{align}
    \frac{T_{1/2}^{\rm Xe}}{10^{25}~\text{y}} \approx
    \left(\frac{3.5\times 10^{-4}}{g_R^2 y_N y_\nu\cos\theta_C^R}\right)^2
    \left(\frac{m_{W_R}}{4~\text{TeV}}\right)^4
    \left(\frac{m_N}{100~\text{MeV}}\right)^4 \, ,
\end{align} 
where $g_R$ is the gauge coupling constant and $\theta_C^R$ the equivalent of the Cabibbo angle, both associated with the $\rm SU(2)_R$ of the Left-Right symmetric model.

Alternatively, it is also possible to trigger the $\epsilon^\phi_{RL}$ mode through the $W_R$-$W$ mixing $\theta$. Its value is generically expected to be $\theta = \kappa g_R m_W^2/(g_L m_{W_R}^2)$ where $\kappa = \mathcal{O}(1)$. In this case one has,
\begin{align}
    \frac{G_F \cos\theta_C}{\sqrt{2} m_p} \epsilon^\phi_{RL}
    = \frac{g_R g_L\theta y_N y_\nu \langle H \rangle\cos\theta_C}{8 m_W^2 m_N^2} \, ,
\end{align} 
resulting in the estimate,
\begin{align}
    \frac{T_{1/2}^{\rm Xe}}{10^{25}~\text{y}} \approx
    \left(\frac{1.4\times 10^{-4}}{g_R^2 \kappa y_N y_\nu}\right)^2
    \left(\frac{m_{W_R}}{25~\text{TeV}}\right)^4
    \left(\frac{m_N}{100~\text{MeV}}\right)^4 \, .
\end{align} 
This is more stringent due to the better sensitivity on $\epsilon^\phi_{RL}$ in \tab{tab:0vbb:limits}. Choosing the right-handed neutrino mass $m_N$ to be as low as $100$~MeV is strictly speaking not allowed in the effective operator treatment which requires $m_N \gg p_F \approx 100$~MeV, but it may be more natural in a scenario where the mass of $N$ is generated through the VEV of $\phi$, $m_N = y_N\langle\phi\rangle$. In fact, choosing $m_N$ to be smaller and abandoning the effective operator treatment may be more natural; the qualitative arguments should hold as above though a dedicated calculation of $0\nu\beta\beta\phi$ would be required. In addition, the contribution to $0\nu\beta\beta\phi$ via a heavy neutrino is expected to peak at $m_N \approx p_F$ with the above estimates give a good approximation \cite{Simkovic:1999re}.\footnote{In addition to the operators discussed here, the Left-Right symmetric scenario will also induce a standard Majoron interaction $\phi\nu\nu$ (leading to standard Majoron emission with spectral index $n = 1$) after electroweak symmetry breaking from an operator of the form $\phi H H \nu\nu$. It is suppressed relative to our contributions by an additional power of $y_\nu$ but does not suffer from suppression by the heavy $W_R$ mass or the small $W_R-W$ mixing.}

Other ultraviolet completions do exist; to the lowest order, the effective operator $\epsilon^\phi_{RR}$ in \eq{eq:0vbb:lagrangian} can be matched to the Standard Model invariant operator $L e_R \bar d_R u_R H \phi$ ($= \mathcal{O}_8\phi$ in the counting of lepton number violating operators in \cite{Babu:2001ex}). All tree-level completions of the operator $\mathcal{O}_8$ were derived in \cite{Helo:2016vsi} which can be easily adapted to include the Standard Model singlet~$\phi$. These for example include heavy leptoquarks as well as heavy scalars and fermions as present in $R$-parity violating supersymmetry.
\\

The interactions in \eq{eq:0vbb:lagrangian} could also be extended in several directions. Most straightforwardly, one can generalise \eq{eq:0vbb:lagrangian} by including scalar and tensor fermion currents to incorporate all possible Lorentz-invariant combinations. The Majoron may also couple derivatively, if originating as a Goldstone boson; this would increase the number of possible Lorentz-invariant combinations. Alternatively, if the exotic particle is a vector boson $a^\mu$ \cite{Carone:1993jv}, such as a dark photon, the fermion currents can couple to it via the vector field itself as well as its field strength tensor $f^{\mu\nu}$. An even number of exotic neutral fermions $\chi$ may also be emitted but this would quickly increase the dimension of the corresponding effective operator. Instead, they may also originate from the internal neutrino via a dimension-6 operator of the form $\Lambda_{\rm NP}^{-2}\nu\nu\chi\chi$~\cite{Huang:2014bva}. Exploring such alternatives to the well-studied neutrinoless double-$\beta$ decay is imperative in order to be able to draw reliable conclusions on the nature of neutrino mass generation and motivate experimentalists to search for these exotic decays.

\pagebreak
\fancyhf{}

%% file: Conclusions/Conclusions.tex
\fancyhf{}
\fancyhead[LE,RO]{\thepage}
\fancyhead[RE]{\slshape{Conclusions}}
%\fancyhead[LO]{\slshape\nouppercase{\rightmark}}

\chapter*{Conclusions}
\markboth{CONCLUSIONS}{CONCLUSIONS}
\addcontentsline{toc}{chapter}{Conclusions}  

The observation of neutrino flavour oscillations and their explanation in terms of massive neutrinos provide the first and most clear evidence of physics beyond the Standard Model of particle physics. Nowadays, there is doubtless experimental certainty that neutrinos have a tiny, yet non-zero, mass, although we still do not know how these masses arise or even their scale.

As has been discussed in Chapters \ref{ch:nuphys} and \ref{ch:numass}, there are several mechanisms proposed to give masses to neutrinos. A very appealing option are radiative models, which explain the smallness of neutrino masses compared to the masses of the rest of the fermions: neutrinos are massless at tree-level and their masses are generated via loops. This allows to explain the lightness of neutrinos without introducing heavy scales, as neutrino masses are suppressed by the loop suppression factor $1/16\pi^2$ for each loop, as well as the possible presence of Standard Model masses and/or extra quartic and Yukawa couplings. The typical mass scale of radiative models lies roughly in the range of $\mathcal{O}(1 - 100)$ TeV. This is especially interesting for phenomenology, radiative models contain couplings which violate lepton flavour and, for Majorana mass models, lepton number, at a scale that may be low enough to be testable at colliders, like the LHC, and in low energy particle physics experiments, like searches for lepton flavour violating (LFV) and/or lepton number violating (LNV) processes, such as \mueg and $0\nu\beta\beta$ decay, respectively.
\\

In this thesis we classify and study several scenarios where neutrino masses are generated radiatively. In \ch{ch:Dim7_1loop} we discussed the generation of small neutrino masses from $d = 7$ one-loop diagrams. We systematically analysed all possible dimension $7$ one-loop topologies and diagrams, and organised them w.r.t. their particle content. From an initial set of $48$ topologies, only $8$ topologies can lead to genuine $d = 7$ neutrino masses. The remaining $40$ topologies are either non-renormalisable realisations, corrections to some other operator (infinite loop contributions), or they generate a lower order operator, for example $d=5$ at one-loop level, which would be the dominant contribution to neutrino masses. We then generated all possible diagrams and found there is only one diagram with no representation larger than an ${\rm SU(2)_L}$ triplet, while there are $22$ diagrams which require either one or two quadruplets. This classification is the first where neutrino masses are grouped into classes with similar particle content. This opens the possibility of obtaining general conclusions valid not only for a single model, but for a whole set of them, as models among the same class share a similar phenomenology (see \ch{ch:Dim7_pheno}).

In \ch{ch:3loop} we studied systematically the decomposition of the Weinberg operator at three-loop order. Due to the huge number of topologies we computationally implemented known algorithms from graph theory to generate a list of all three-loop connected topologies with 3- and 4-point vertices, and four external lines. We subsequently applied several cuts to obtain a list of $99$ genuine topologies, which can be divided into two sets according to their particle content, due to a loophole in the \textit{standard} procedure of previous classifications, first spotted in this thesis. Due to the antisymmetric nature of ${\rm SU(2)_L}$ couplings or the presence of massless fermions in the loops that forces derivatives to appear in the effective operator, there are certain diagrams, a priori non-genuine, which can be genuine for a very particular choice of fields. There are $55$ topologies that generate $271$ diagrams which are genuine due to the loophole just explained. We called them \textit{special} genuine, and they can be classified according to the particular particle content required to be genuine, which considerably restricts the possible models that can be generated. On the other hand, there are $44$ \textit{normal} genuine topologies with $228$ diagrams which do not required any specific fields. Finally, we estimated the typical parameter range for which dimension $5$ three-loop models can explain the measured neutrino oscillation data. We found that they can fit the data for a new physics mass scale of roughly $1 - 10^3$ TeV, which can be partially testable at current and future colliders, and experiments searching for lepton flavour violation.

In \ch{ch:dirac2l} we discussed the complete decomposition and classification of the Dirac neutrino mass operator $\bar L H^c \nu_R$ at two-loop order. We identified $70$ topologies with 3 external legs, two-loops and 3, 4-point vertices. Among them, only 5 satisfy the genuineness criteria explained in \ch{ch:numass}. These 5 topologies generate 18 renormalisable diagrams, which can be classified in three different categories depending on the requirements imposed on their possible particle content, similarly to the previous classification of the Weinberg operator at three-loop order. The first class contains diagrams which are genuine in general; the diagrams in the second class contains a one-loop realisation of a fermion-fermion-scalar vertex that can be genuine providing a symmetry transformation that forbids not only the tree-level but also the one-loop diagram and, then, breaking it softly allowing the two-loop mass diagram; and finally, the diagrams in the last class always contains an SM Higgs and a singlet $(1, 1, −1)$ running in the loop in order to be genuine. Examples of these classes are then analysed showing that they can fit the neutrino mass scale for couplings order one if the new physics scale is $\mathcal{O}(1)$ TeV.

In \ch{ch:dm_DiracMajo} we studied how the breaking pattern of lepton number determines whether neutrinos are Dirac or Majorana, depending on the residual $Z_n$ symmetry. We showed that only if neutrinos transform as the identity or under a subgroup $Z_2$ of $Z_n$, they can be Majorana fermions. Moreover, this remnant symmetry can be used to stabilise a dark matter candidate that participates in the radiative generation of the neutrino masses. We provided a generic framework to obtain stable dark matter along with naturally small Dirac or Majorana neutrino masses generated at the loop level. This is achieved through the spontaneous breaking of the global $U(1)_{B-L}$ symmetry to a residual even $Z_n$ ($n \geq 4$) subgroup. In this scenario, the residual $Z_n$ symmetry which guarantees dark matter stability and protects the radiative generation of neutrino masses is obtained dynamically from the breaking of lepton number.

In \ch{ch:loop_seesaw} we presented a class of falsifiable models where neutrino masses are generated via a type-I seesaw mechanism with radiative Dirac Yukawas. Neutrino mass is naturally suppressed due to the radiative generation of the Dirac Yukawas, rather than forcing a large Majorana mass scale for the right-handed neutrinos, as in the original type-I seesaw. Consequently, the new physics scale can be kept low, even below the TeV, depending on the specific realisation choosen to generate the effective Dirac Yukawa couplings. Parametrising, as general as possible, the effective Dirac Yukawa in terms of number of loops, mass insertions and number of dimensionful and dimensionless couplings, we found that the limits from Big Bang Nucleosynthesis and $\Delta N_{eff}$ rule out a big part of the possible models, while several of the remaining realisations are within the reach of future updates or experiments.

In \ch{ch:Dim7_pheno} we analysed in detail the rich phenomenology of dimension 7 one-loop neutrino mass models trying to derive general conclusions using LHC and low-energy particle physics experimental results. As already explained in \ch{ch:Dim7_1loop}, this class of models always contain large ${\rm SU(2)_L}$ representations and hypercharges in order to be genuine, i.e. particles with very high electrical charges. We studied the constraints coming from neutrino oscillation data, lepton flavour violating searches and colliders. Signals of highly charged fields with lepton number violation have a very low background from the Standard Model in colliders, such that most of the parameter space will be tested by future phases of LHC and the updates from charged lepton flavour violation experiments.

In \ch{ch:clfv} we studied the phenomenology of charged lepton flavour violation (CLFV) for the three most popular three-loop minimal Majorana neutrino mass models in the literature, i.e. the cocktail model, the KNT model and the AKS model. We call these models ``minimal'' since their particle content correspond to the minimal particle contents for which genuine three-loop models can be constructed. In all the three minimal models the neutrino mass matrix is proportional to some powers of Standard Model lepton masses, providing additional suppression factors on top of the expected loop suppression. Consequently, to correctly explain neutrino masses, large Yukawa couplings are needed. We have calculated charged lepton flavour violating observables and found that the models survive the current constraints only in very small regions of their parameter spaces for the present mass limits from LEP and the LHC. Only particular choices of the Dirac and Majorana phases survive for a narrow range of the lightest neutrino mass. We showed that this class of models, if not yet excluded, are severely constrained. Moreover, it opens the possibility to falsify some of the models in future CLFV experimental updates or double-$\beta$ decay experiments.

Finally, in \ch{ch:0vbb} we discussed a novel mode of neutrinoless double-$\beta$ decay with emission of a light Majoron-like scalar particle. We assumed an operator with a right vector-axial lepton current leading to a long-range contribution that is unsuppressed by the light neutrino mass. We have calculated the total double-$\beta$ decay rate and determined the fully differential shape for this mode. We found that future double-$\beta$ decay searches are sensitive to scales of the order $1$ TeV for a light scalar mass below $0.2$ MeV. The angular and energy distributions can deviate considerably from that of two-neutrino double-$\beta$ decay, which is the main background, encouraging future double-$\beta$ decay experiments to search for new physics in the analysis of the single electron energy spectrum and the angular distribution.
\\

Summing up, we have presented and discussed a great variety of radiative neutrino mass models from the model-building and phenomenological point of view. We have classified them and analysed specific scenarios or models we found particularly interesting due to their phenomenology and/or simplicity. In most of the cases, radiative models can explain the smallness of neutrino masses (and other drawbacks of the Standard Models) with a relative low energy scale, which can be tested by current and future experiments, offering a falsifiable window to new physics.

%% file: Appendices/Appendix_topos.tex
\fancyhf{}
\fancyhead[LE,RO]{\thepage}
\fancyhead[RE]{\slshape\nouppercase{\leftmark}}
\fancyhead[LO]{\slshape\nouppercase{\rightmark}}

\chapter{Complete lists of topologies and diagrams}
\label{app:topos}

In this appendix we list some of the relevant topologies and diagrams discussed in \ch{ch:Dim7_1loop} and \ch{ch:3loop}.

%%%%%%%%%%%%%%%%%%%%%%%%%%%%%%%%%%%%%%%%%%%%%%%%%%%%%%%%%%%%%%%
%%%%%%%%%%%%%%%%%%%%%%%%%%%%%%%%%%%%%%%%%%%%%%%%%%%%%%%%%%%%%%%
\section{$d=7$ one-loop}

Here, we present the list of all $d=7$ one-loop topologies, classified into genuine and non-genuine topologies, as discussed in \ch{ch:Dim7_1loop}. We also give the complete list of diagrams that can lead to ``genuine'' $d=7$ neutrino mass models with $SU(2)_L$ quadruplet representations.

%%%%%%%%%%%%%%%%%%%%%%%%%%%%%%%%%%%%%%%%%%%%%%%%%%%%%%%%%%%%%%%
\subsection*{Topologies}

\begin{figure}[t!]
    \centering
    \begin{tabular}{ c c c }
        \includegraphics{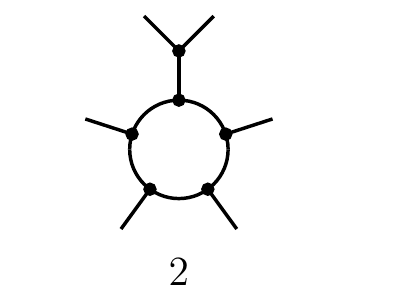}  &  \includegraphics{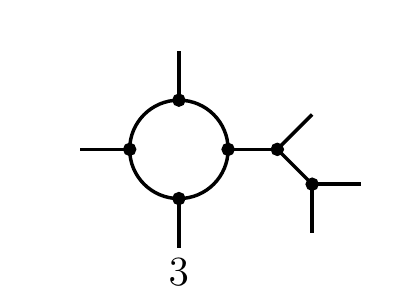}  &  
        \includegraphics{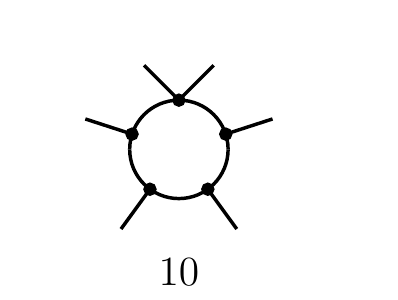}
        \\
        \includegraphics{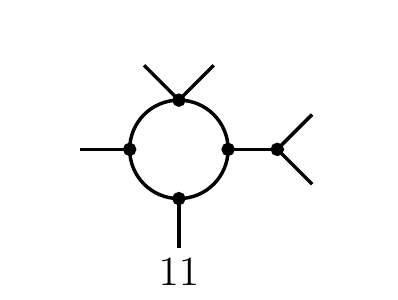} &  \includegraphics{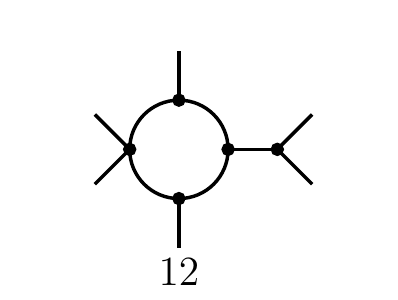}  &  
        \includegraphics{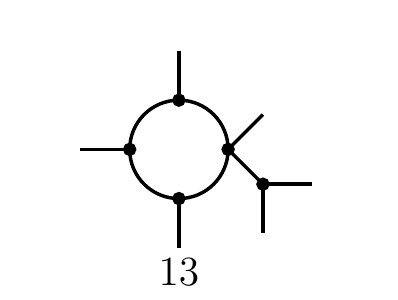}
    \end{tabular}
    \begin{tabular}{ c c}
        \includegraphics{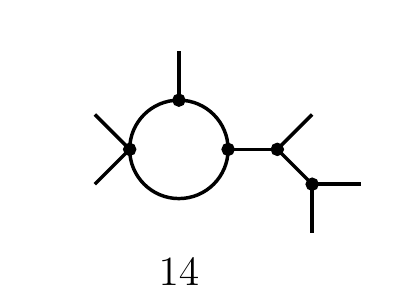} & \includegraphics{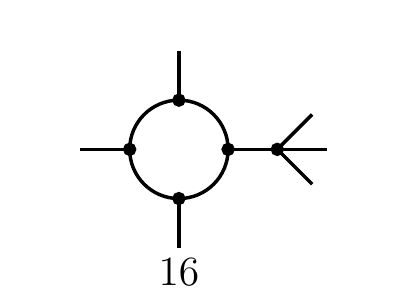}
    \end{tabular}
    \caption{Topologies that can lead to a genuine $d=7$ one-loop neutrino mass model. $T_{11}$ is the only topology for which the largest representation can be as small as a $SU(2)_L$ triplet. For all other topologies at least one quadruplet must appear in the diagram for the model to be genuine. The quadruplet diagrams based on those topologies are shown in \figs{fig:app:topos:Diags4pletsIn}{fig:app:topos:Diags4pletsInOut}. For the triplet model see \fig{fig:dim7loop:3plet_model} in \ch{ch:Dim7_1loop}.}
    \label{fig:app:topos:TopoGenuine_d7}
\end{figure}

In \fig{fig:app:topos:TopoGenuine_d7} shows the 8 topologies that can lead to genuine $d=7$ one-loop models. We stress again that not all diagrams, derived from these topologies, are necessarily genuine, as discussed in the main text. Note that only $T_{11}$ generate a model in which the largest representation can be as small as a $SU(2)_L$ triplet. All other 7 topologies require at least one quadruplet for genuine models.

\begin{figure}[t!]
    \centering
    \begin{tabular}{ c c c c }
        \includegraphics{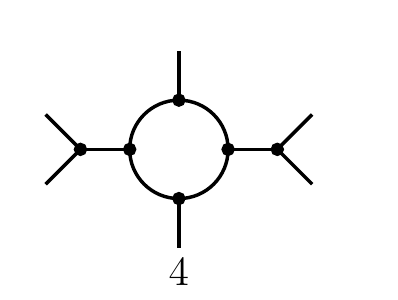}  &  \includegraphics{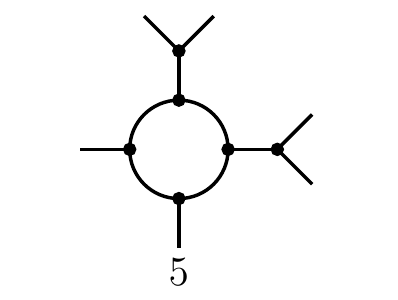}  &  
        \includegraphics{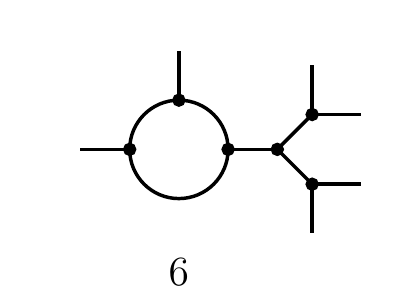}
        \\
        \includegraphics{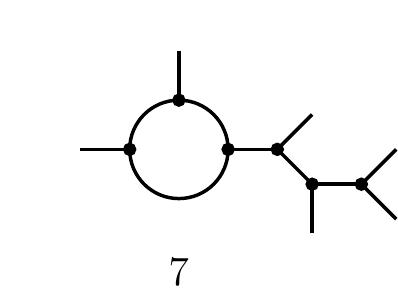}  &  \includegraphics{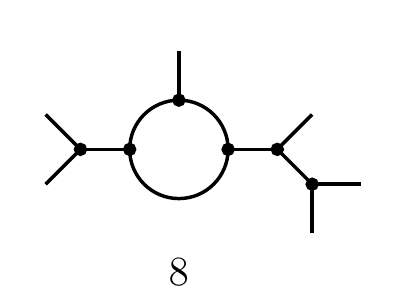}  &  
        \includegraphics{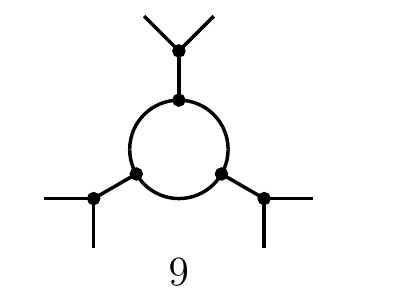}
        \\
        \includegraphics{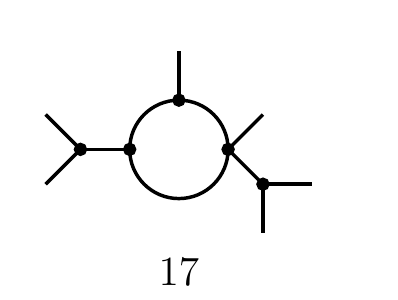}  &  \includegraphics{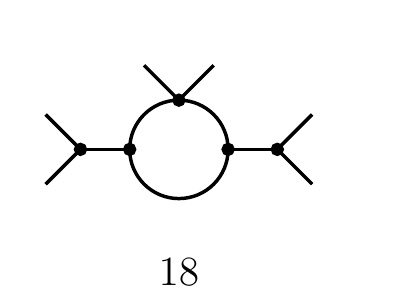}  &  
        \includegraphics{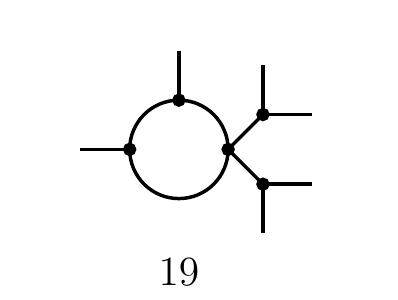}
        \\
        \includegraphics{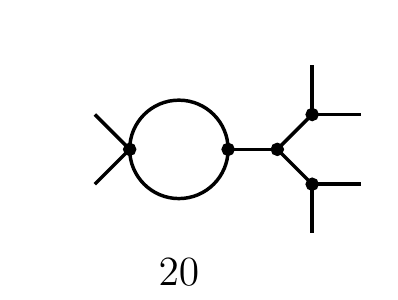}  &  \includegraphics{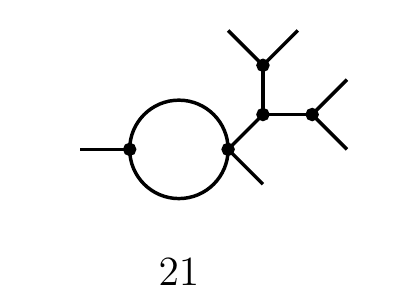}  &  
        \includegraphics{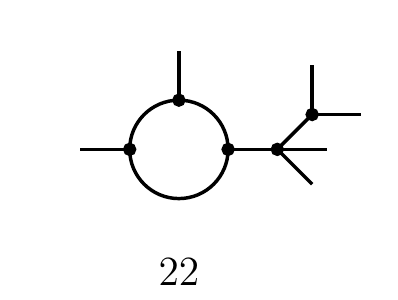}
        \\
        \includegraphics{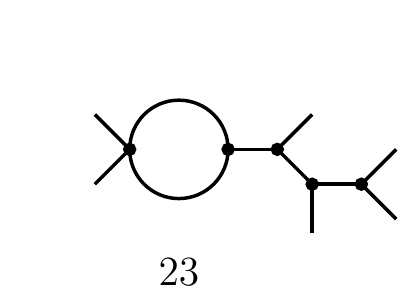}  &  \includegraphics{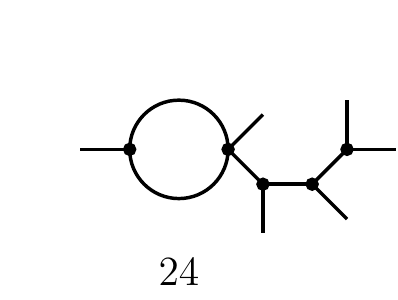}  &  
        \includegraphics{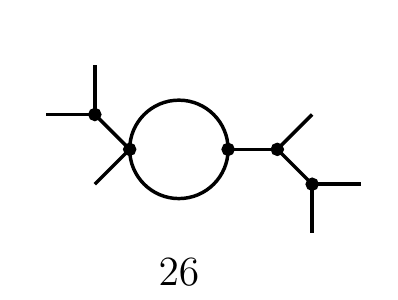}
\end{tabular}
    \caption{(Continues in \fig{fig:app:topos:TopologiesSeesaw}).}
    \label{fig:app:topos:TopologiesSeesaw1}
\end{figure}

\begin{figure}[t!]
    \centering
    \begin{tabular}{ c c c c }
        \includegraphics{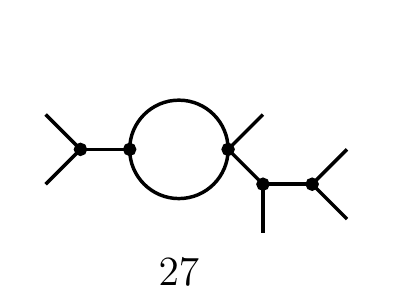}  &  \includegraphics{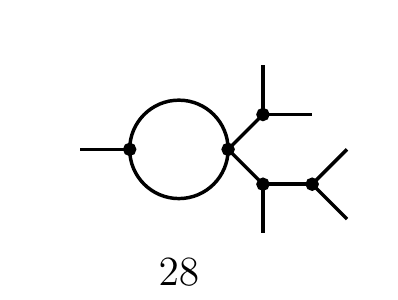}  &  
        \includegraphics{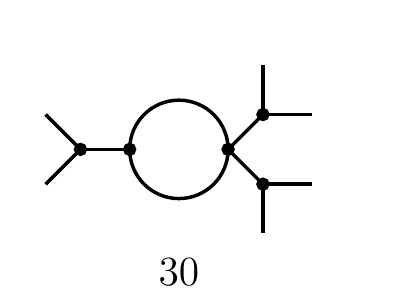}
        \\
        \includegraphics{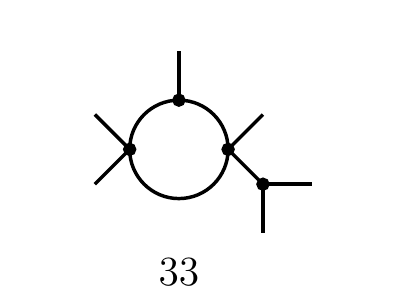}  &  \includegraphics{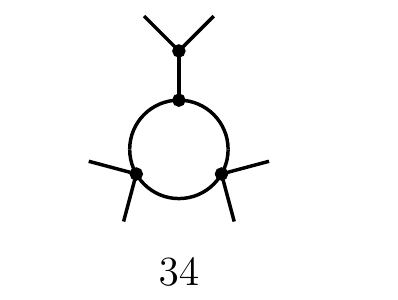}  &  
        \includegraphics{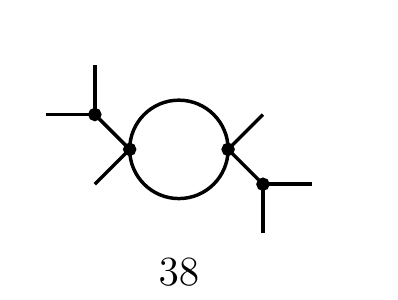}
        \\
        \includegraphics{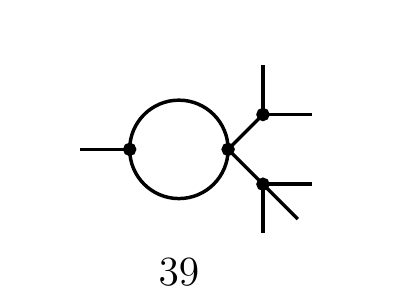} &  \includegraphics{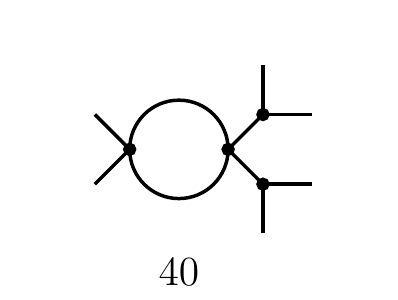}  &  
        \includegraphics{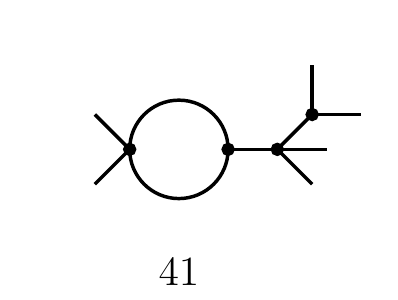}
        \\
        \includegraphics{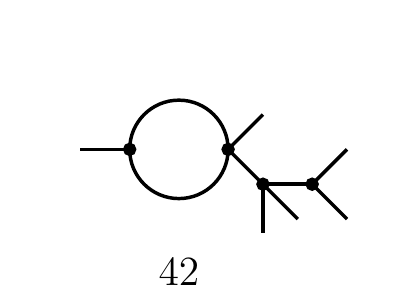}  &  \includegraphics{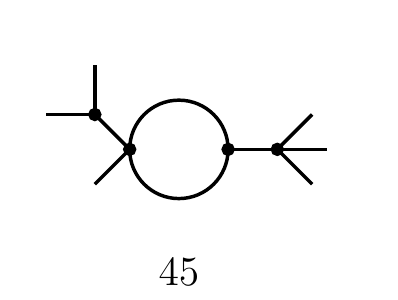} & 
        \includegraphics{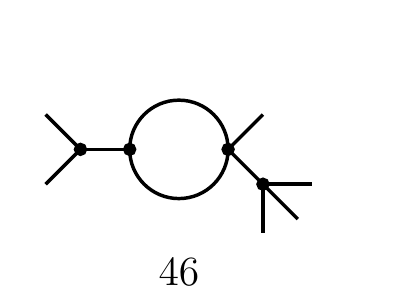}  &  
    \end{tabular}
    \caption{Topologies that necessarily lead to a $d=5$ tree-level neutrino mass. Continuation of \fig{fig:app:topos:TopologiesSeesaw1}.}
    \label{fig:app:topos:TopologiesSeesaw}
\end{figure}

In \fig{fig:app:topos:TopologiesSeesaw} we list all topologies for which every diagram that can be generated is excluded, since they contain either a singlet fermion $\nu_R\equiv (\textbf{1}, \textbf{1}, 0)$, a triplet scalar $\Delta \equiv (\textbf{1}, \textbf{3}, -1)$, or a triplet fermion $\Sigma \equiv (\textbf{1}, \textbf{3}, 0)$. All diagrams from these topologies, thus, will also generate a tree-level $d=5$ seesaw contribution to the neutrino mass matrix.

\begin{figure}[t!]
    \centering
    \begin{tabular}{ c c }
        \includegraphics{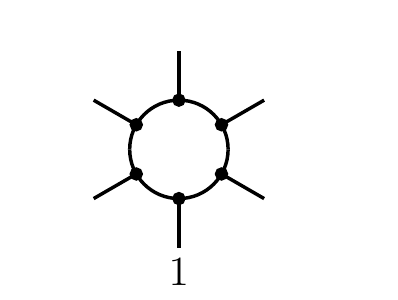}  &  \includegraphics{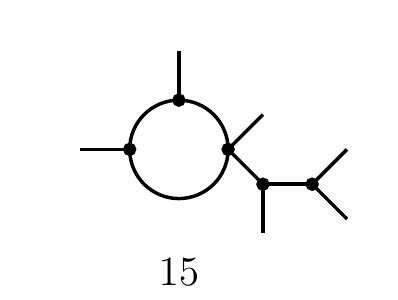}
        \\
        \includegraphics{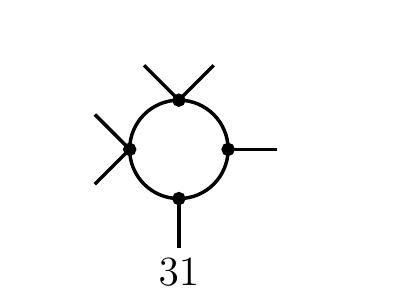}  &  \includegraphics{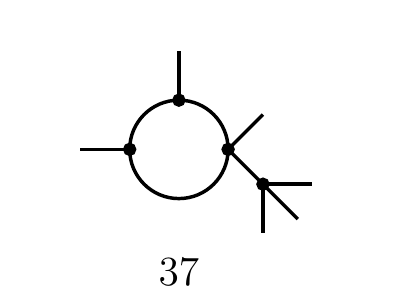}
    \end{tabular}
    \caption{Topologies that lead to a $d=5$ one-loop neutrino mass. All diagrams generated from the topologies in this class, not already excluded because they generate a $d=5$ tree-level mass, include the particle content necessary to generate neutrino mass through one of the four genuine $d=5$ one-loop diagrams, see \fig{fig:numass:loopd5}. Note that $T_{15}$ is an exceptional case, since it has diagrams for all three possibilities: tree-level $d=5$, one-loop $d=5$ and tree-level $d=7$. }
    \label{fig:app:topos:Topos_d=5_1-loop}
\end{figure}

In \fig{fig:app:topos:Topos_d=5_1-loop} we list the topologies for which many but not all diagrams are excluded by a $d=5$ tree-level seesaw. For these topologies, all remaining diagrams are excluded because a one-loop $d=5$ contribution to the neutrino mass necessarily exists.

\begin{figure}[t!]
    \centering
    \begin{tabular}{ c c c }
        \includegraphics{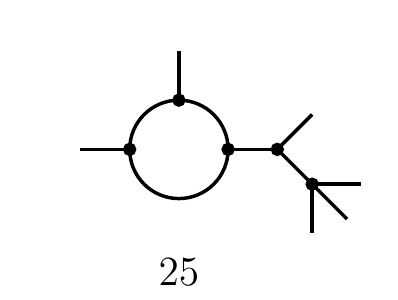}  &  \includegraphics{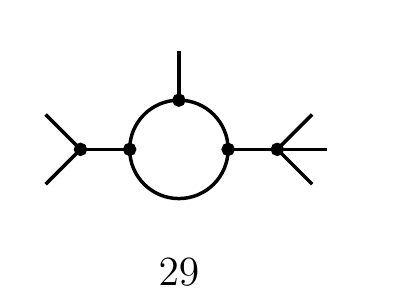}
        & \includegraphics{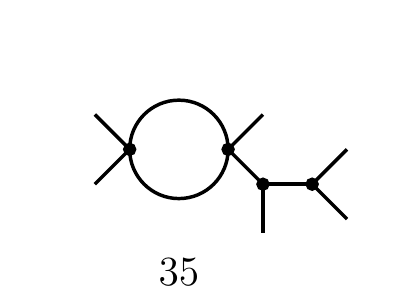} 
    \end{tabular}
    \caption{Topologies, which lead to a $d=7$ tree-level neutrino mass. For each of these topologies one can construct diagrams, which have a $d=5$ tree-level mass. All remaining diagrams contain the scalar $S\equiv(\textbf{1},\textbf{4},3/2)$ along with the fermion $\Psi\equiv(\textbf{1},\textbf{3},1)$ (\fig{fig:numass:bnt}) \cite{Babu:2009aq}.}
    \label{fig:app:topos:Topologiesd=7tree}
\end{figure}

In \fig{fig:app:topos:Topologiesd=7tree} we list topologies which lead to a $d=7$ tree-level neutrino mass. For each of these topologies, one can construct diagrams which have a $d=5$ tree-level mass. All remaining diagrams, contain the scalar $S\equiv(\textbf{1},\textbf{4},3/2)$ along with the fermion $\Psi\equiv(\textbf{1},\textbf{3},1)$ and, thus, generate the $d=7$ tree-level BNT model \cite{Babu:2009aq}. We note in passing that one can, in principle, use these diagrams to radiatively generate one of the vertices in the BNT model. This is very similar to the discussion for the radiative generation of a seesaw coupling given in \cite{Bonnet:2012kz} at $d=5$ level.

\begin{figure}[t!]
    \centering
    \begin{tabular}{ c c c}
        \includegraphics{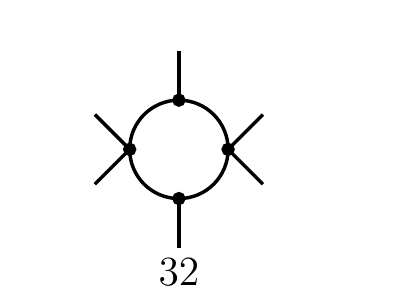}  &  \includegraphics{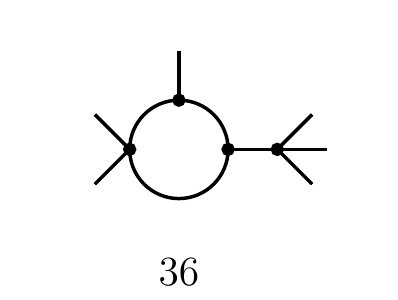} &
        \includegraphics{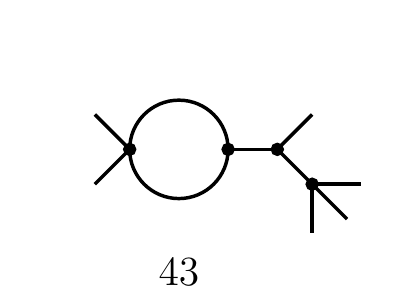}
        \\
         \includegraphics{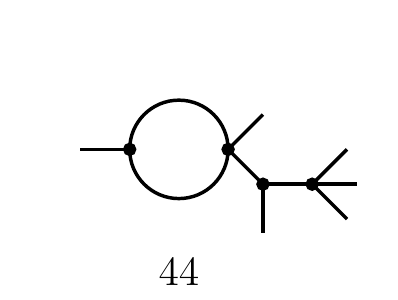}  &  \includegraphics{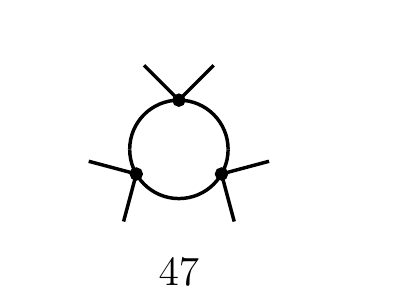}  &  
         \includegraphics{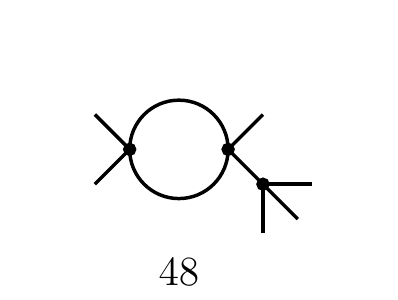}
    \end{tabular}
    \caption{Topologies discarded because they lead to non-renormalisable operators.}
    \label{fig:app:topos:TopologiesNRO}
\end{figure}

Finally, \fig{fig:app:topos:TopologiesNRO} contains, for completeness,  the topologies which are excluded since they can never lead to a renormalisable model.

%%%%%%%%%%%%%%%%%%%%%%%%%%%%%%%%%%%%%%%%%%%%%%%%%%%%%%%%%%%%%%%
\subsection*{Genuine diagrams}

\begin{figure}[t!]
    \centering
    \begin{tabular}{ c c }
        \includegraphics[width=0.33\textwidth]{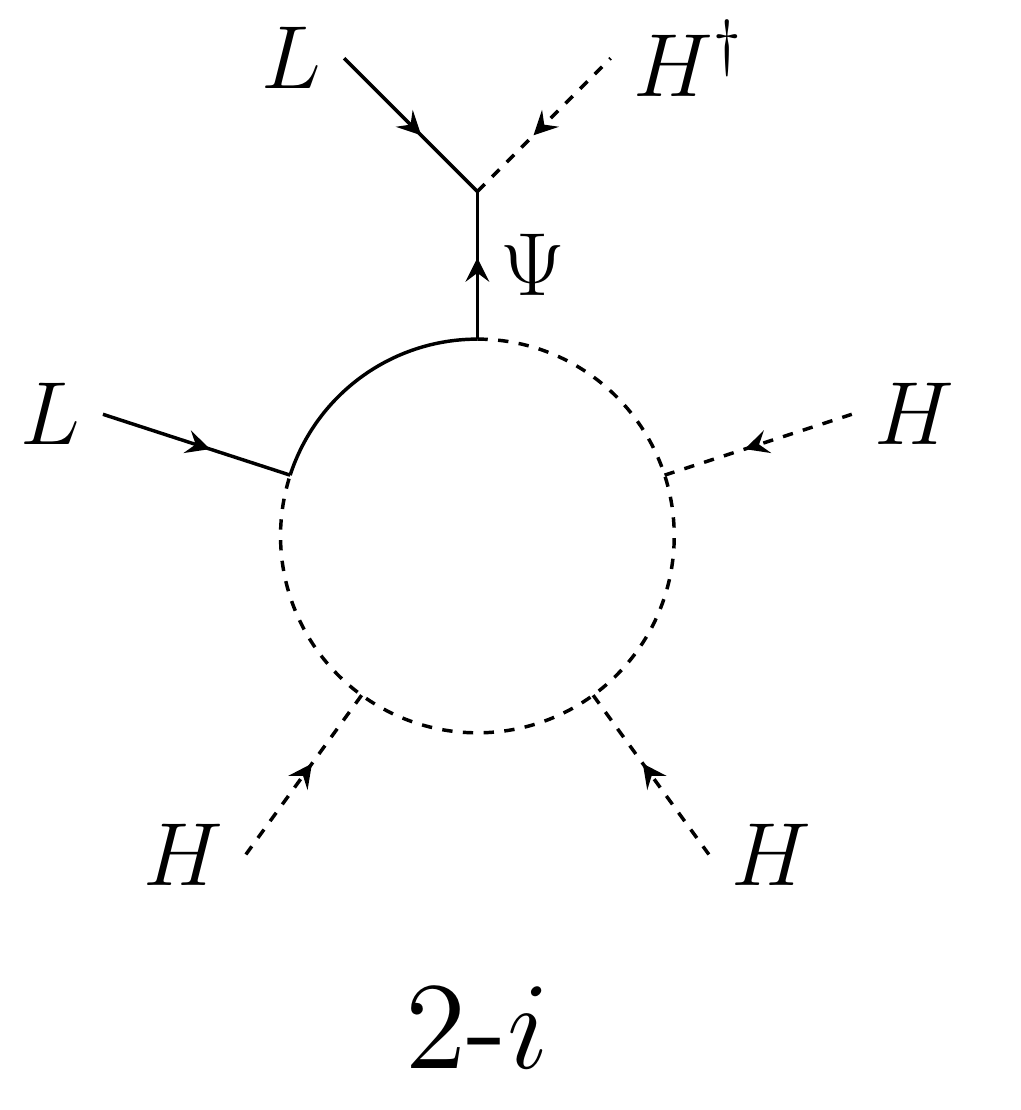} \quad & \quad 
        \includegraphics[width=0.33\textwidth]{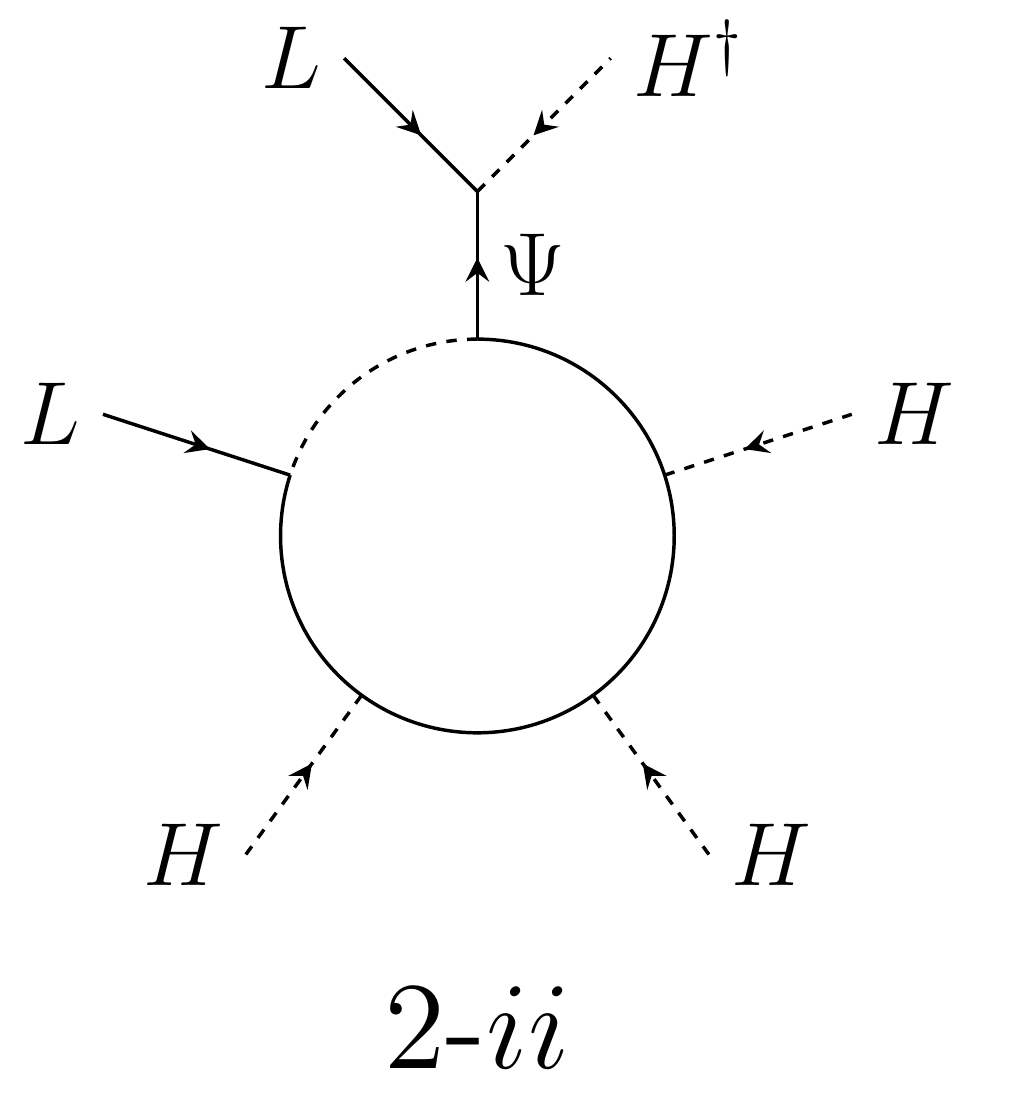}
    \end{tabular}
    \begin{tabular}{ c c c }
        \includegraphics[width=0.33\textwidth]{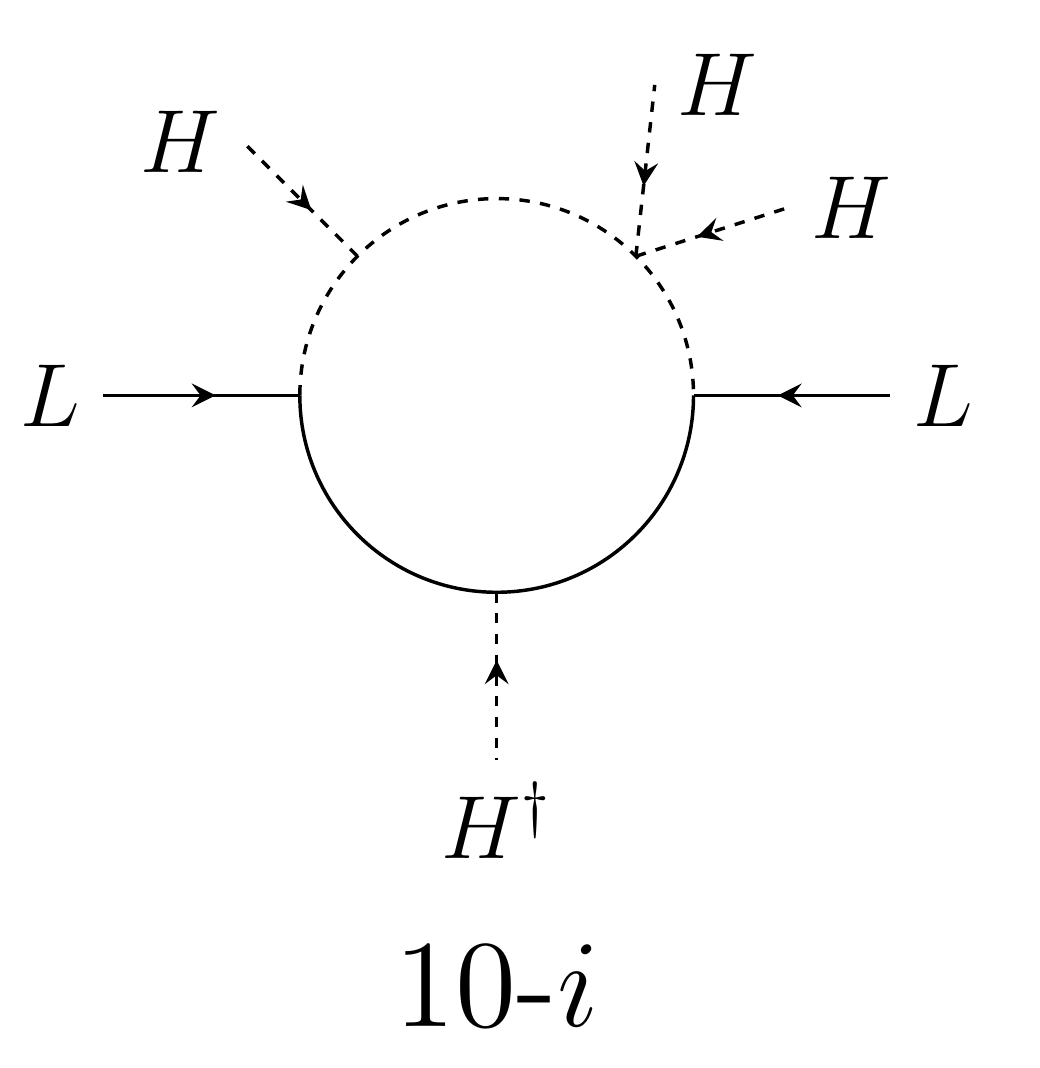} \hfill & \hfill 
        \includegraphics[width=0.33\textwidth]{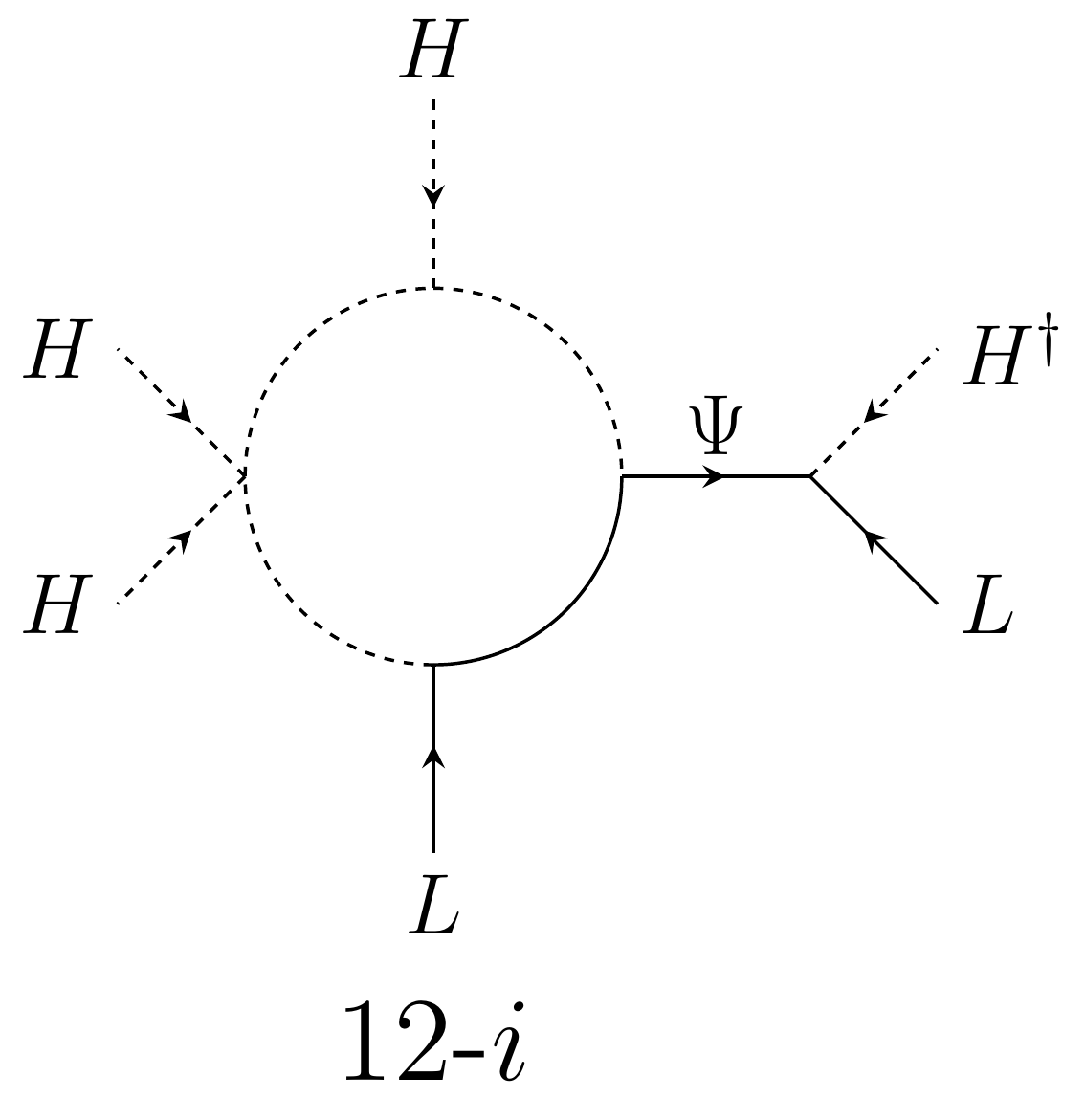} \hfill & \hfill
        \includegraphics[width=0.33\textwidth]{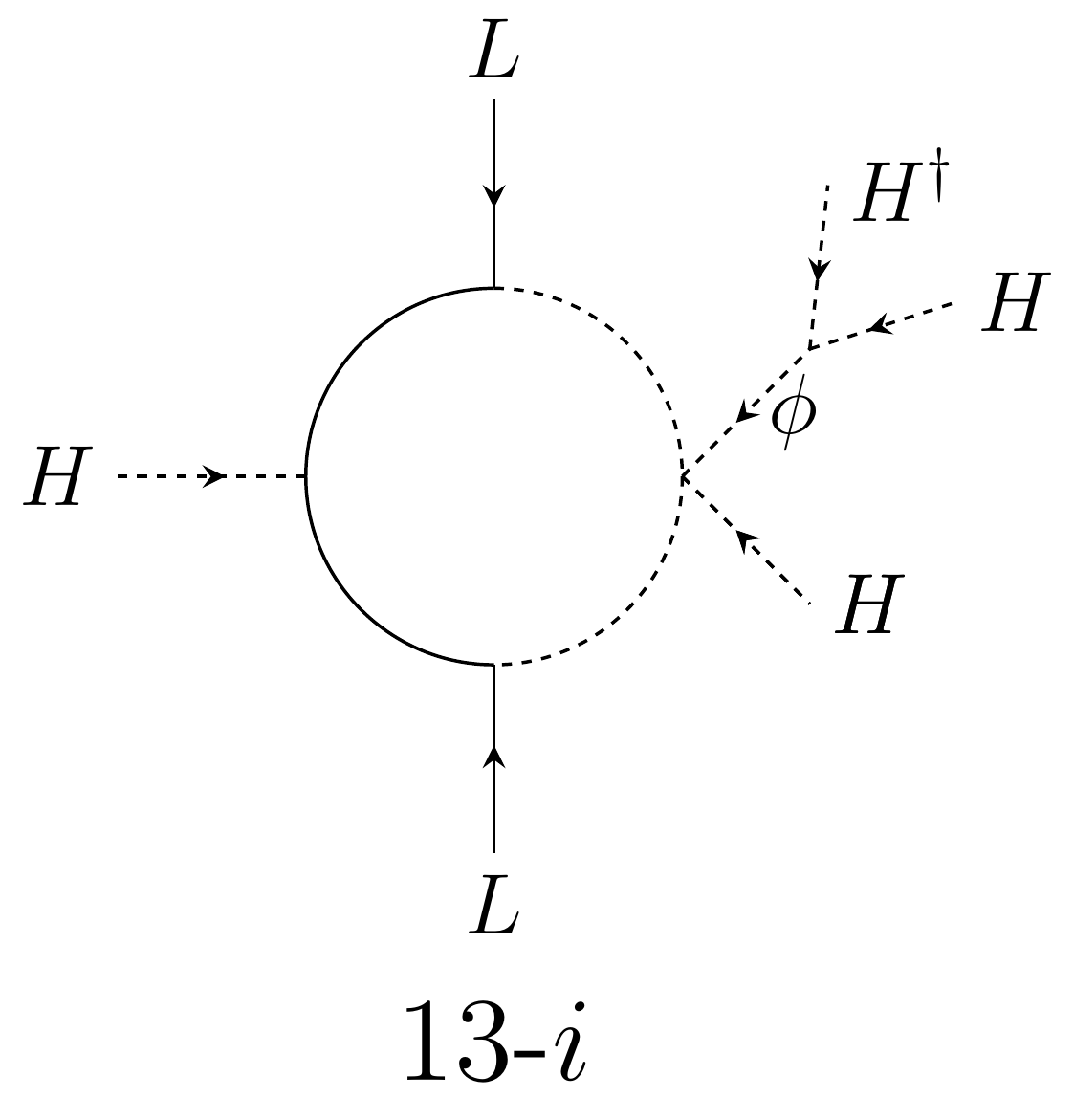}
    \end{tabular}
    \caption{Diagrams that can lead to a genuine $d=7$ one-loop neutrino mass for which the largest representations of $SU(2)_L$ is at least a quadruplet. This group of diagrams require the quadruplet to be one of the particles inside the loop to avoid lower order contributions.}
    \label{fig:app:topos:Diags4pletsIn}
\end{figure}

\begin{figure}[t!]
    \begin{tabular}{ c c c }
        \includegraphics[width=0.33\textwidth]{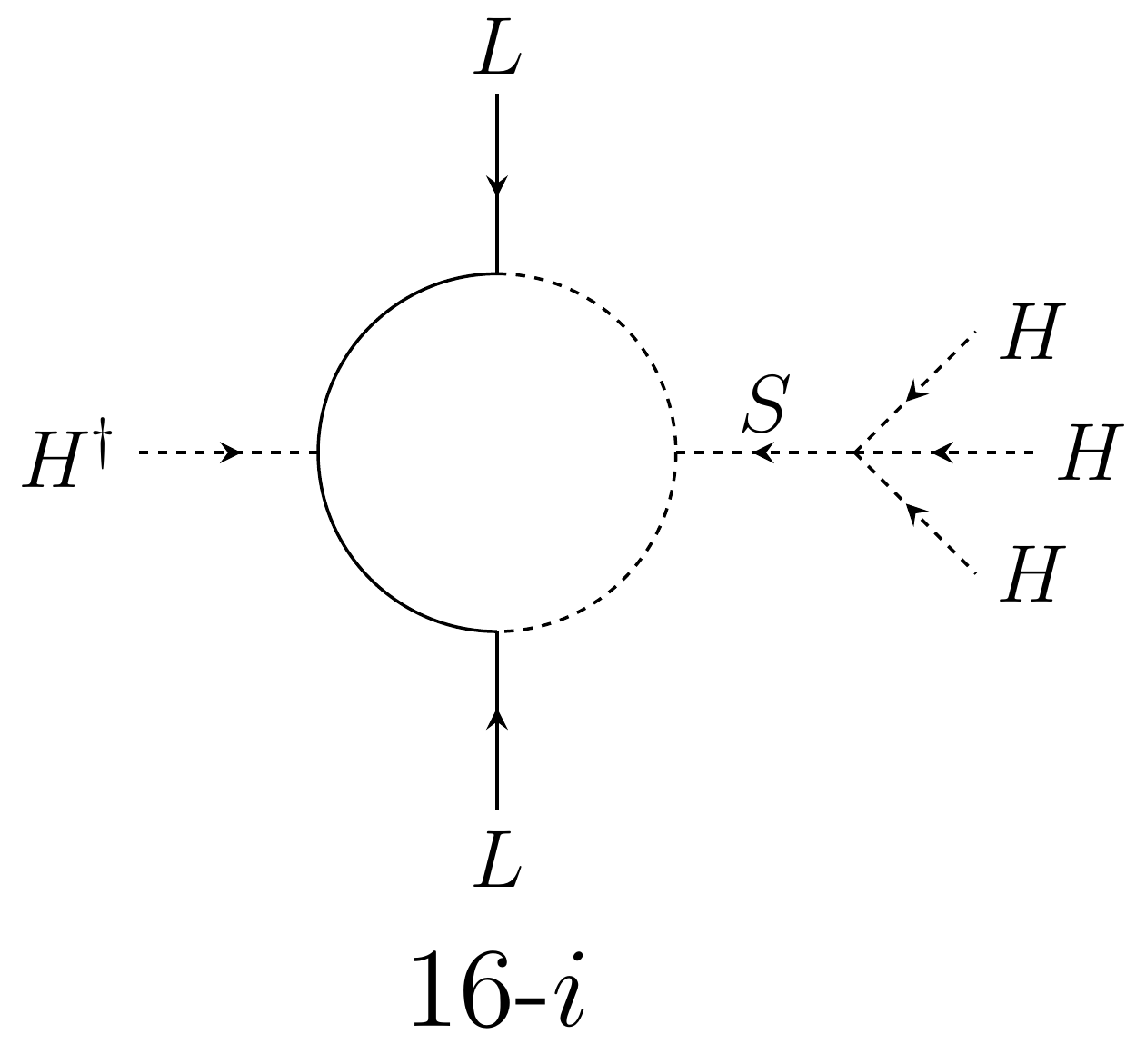} \hfill &  \hfill
        \includegraphics[width=0.33\textwidth]{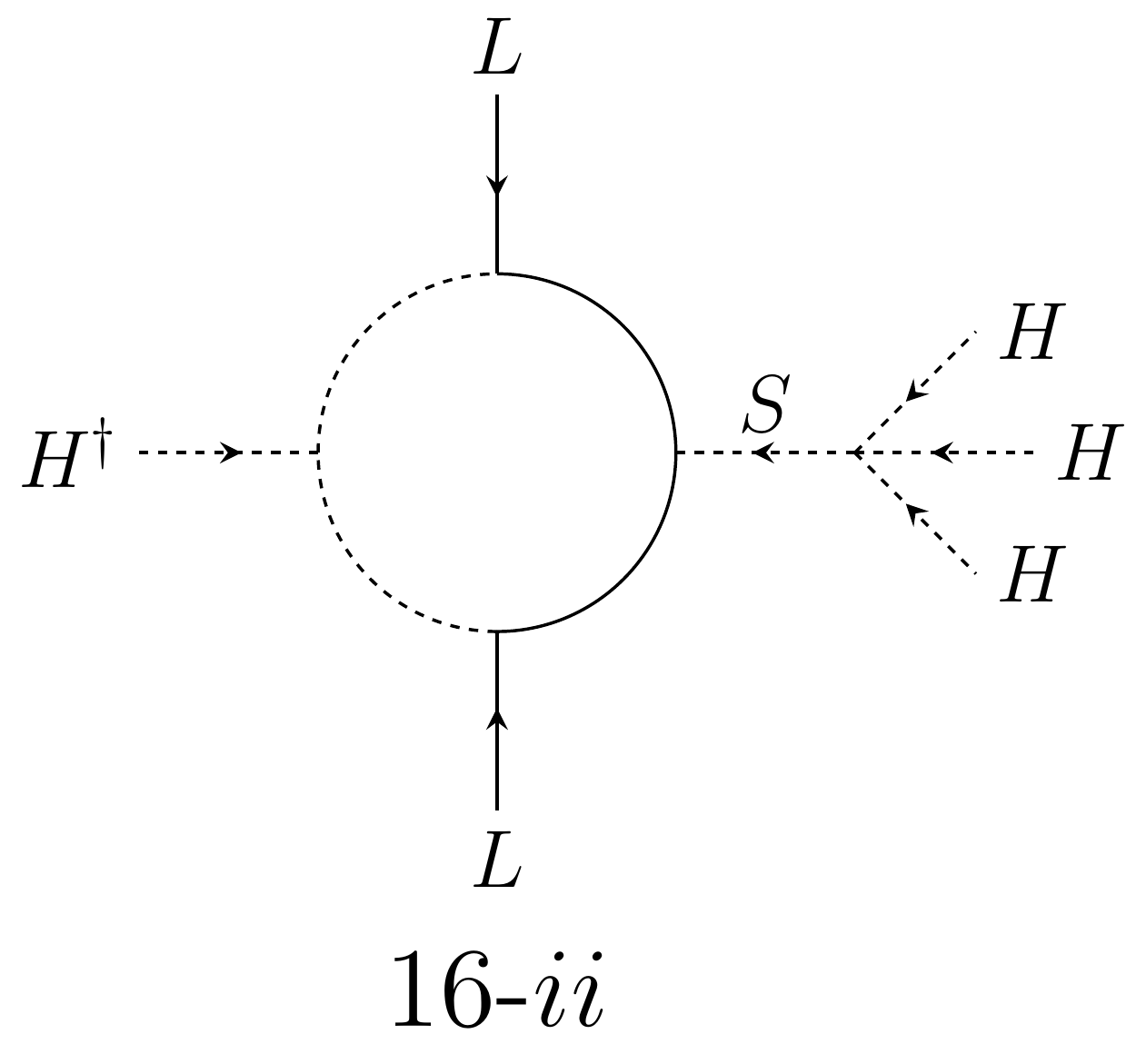} \hfill & \hfill
        \includegraphics[width=0.33\textwidth]{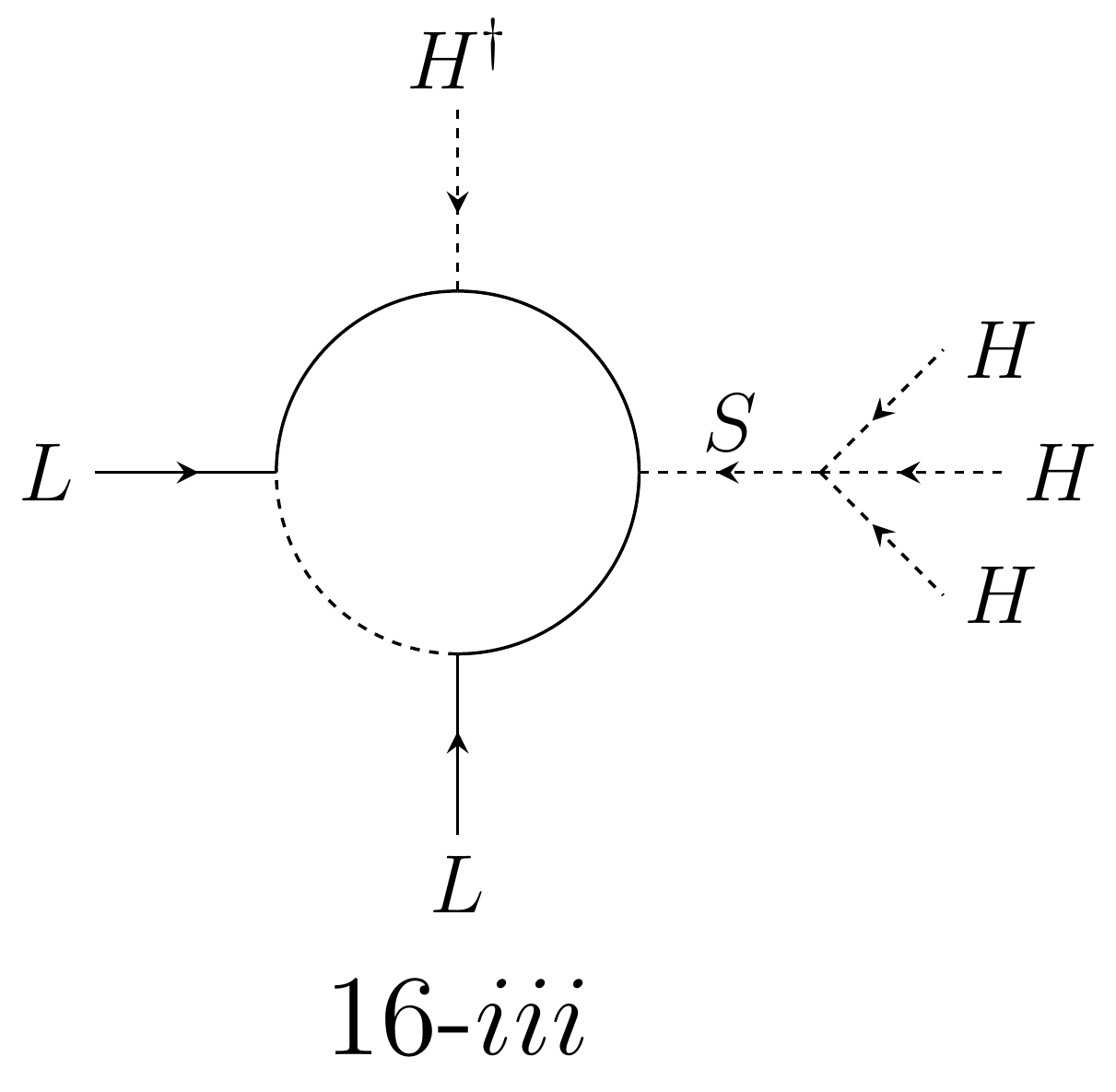}
    \end{tabular}
    \caption{Diagrams that lead to a genuine $d=7$ one-loop neutrino mass for which the largest representations of $SU(2)_L$ is at least a quadruplet. All these diagrams contain $S\equiv(\textbf{1},\textbf{4},3/2)$. The hypercharge of the scalar $S$ ensures the absence of a $d=5$ one-loop neutrino mass.}
    \label{fig:app:topos:Diags4pletsS}
\end{figure}

In this section we list diagrams with quadruplets. All diagrams are given in \figs{fig:app:topos:Diags4pletsIn}{fig:app:topos:Diags4pletsInOut}. We have divided these diagrams into three groups, depending on whether there is a quadruplet in the loop (\fig{fig:app:topos:Diags4pletsIn}), the scalar $S$ on the outside of the loop (\fig{fig:app:topos:Diags4pletsS}) or models with at least two different quadruplets (\fig{fig:app:topos:Diags4pletsInOut}). 

\begin{figure}[t!]
    \centering
    \begin{tabular}{ c c c }
        \includegraphics[width=0.33\textwidth]{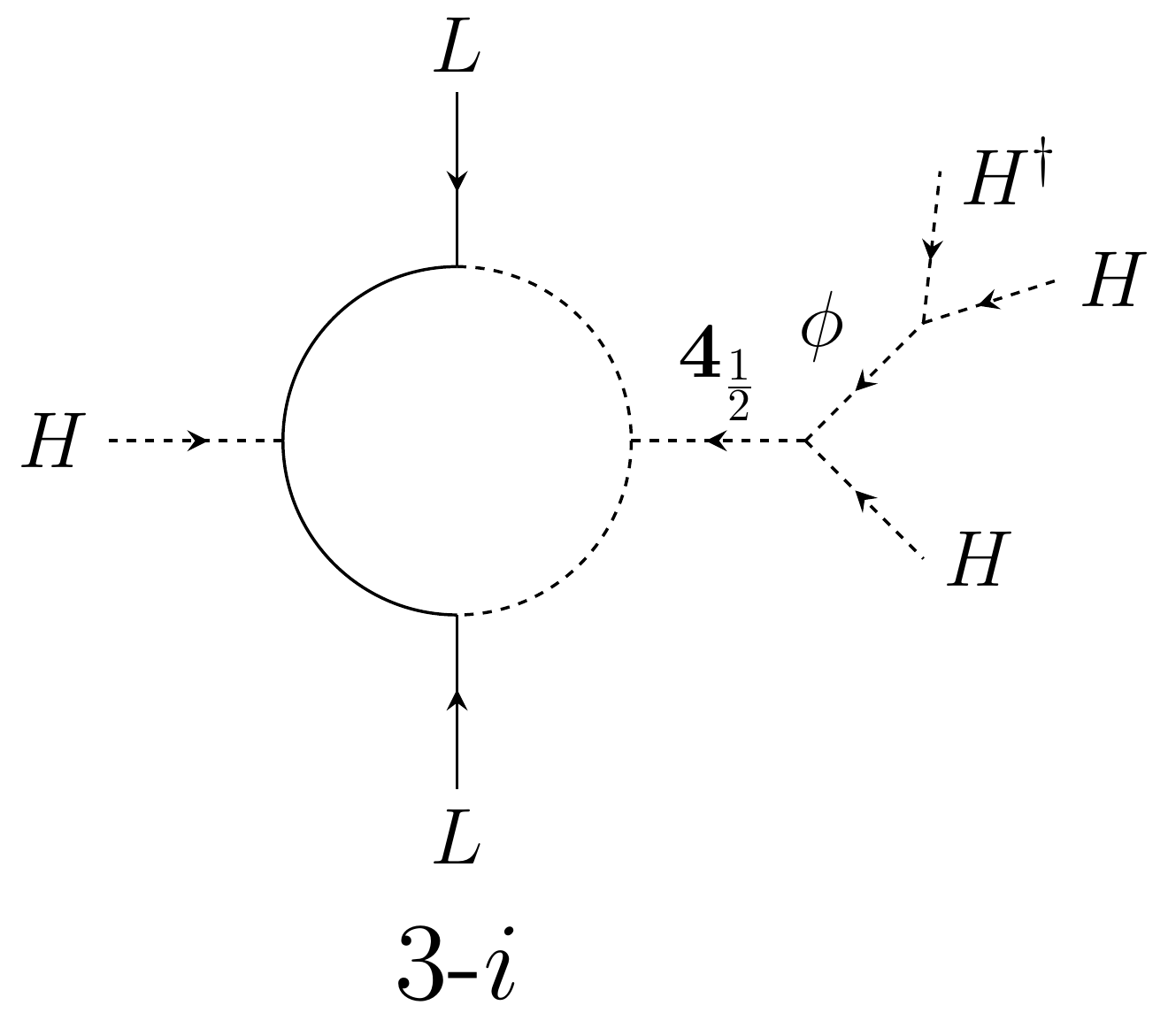} \hfill &  \hfill
        \includegraphics[width=0.33\textwidth]{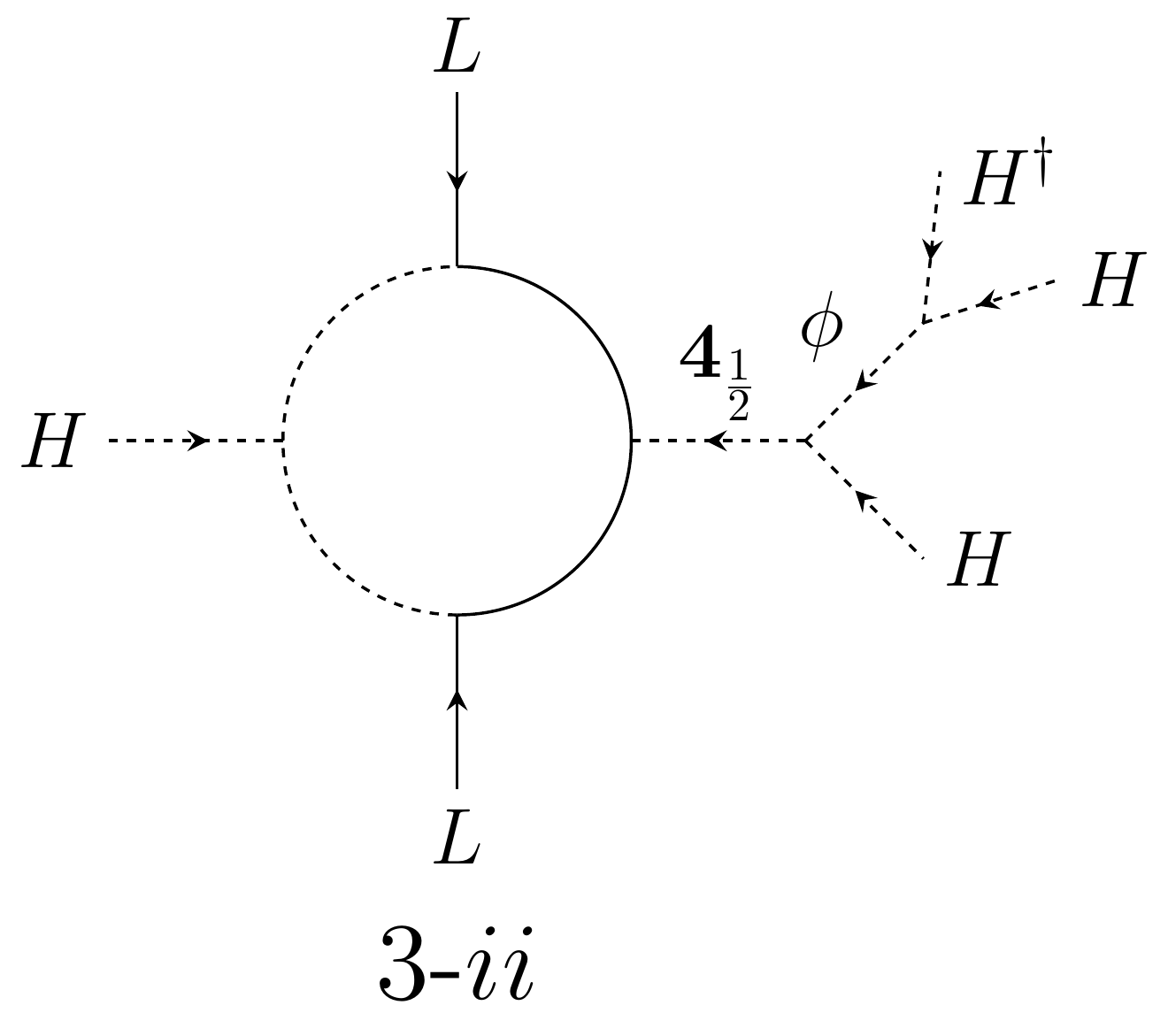} \hfill &  \hfill
        \includegraphics[width=0.33\textwidth]{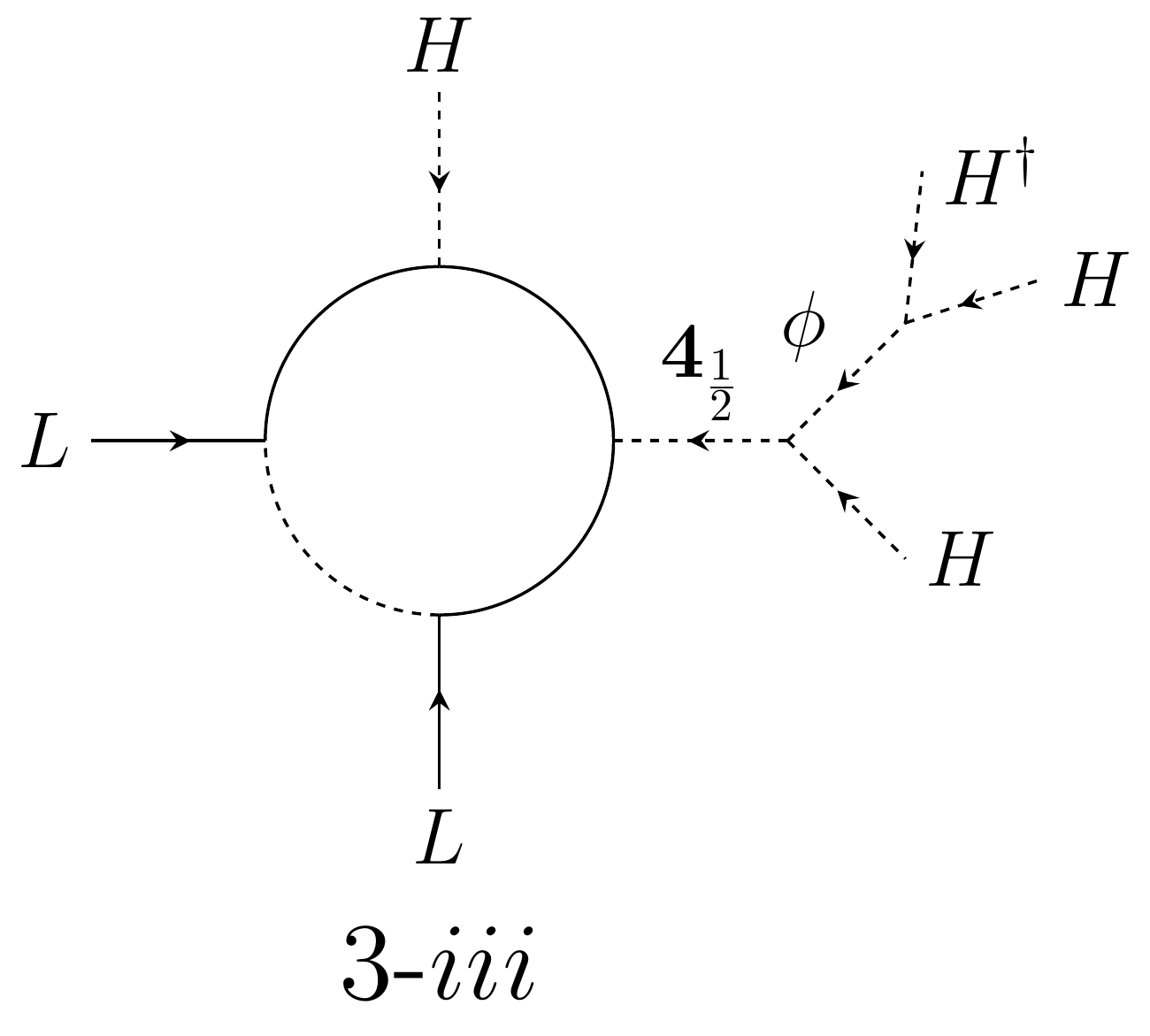}
        \\
        \includegraphics[width=0.33\textwidth]{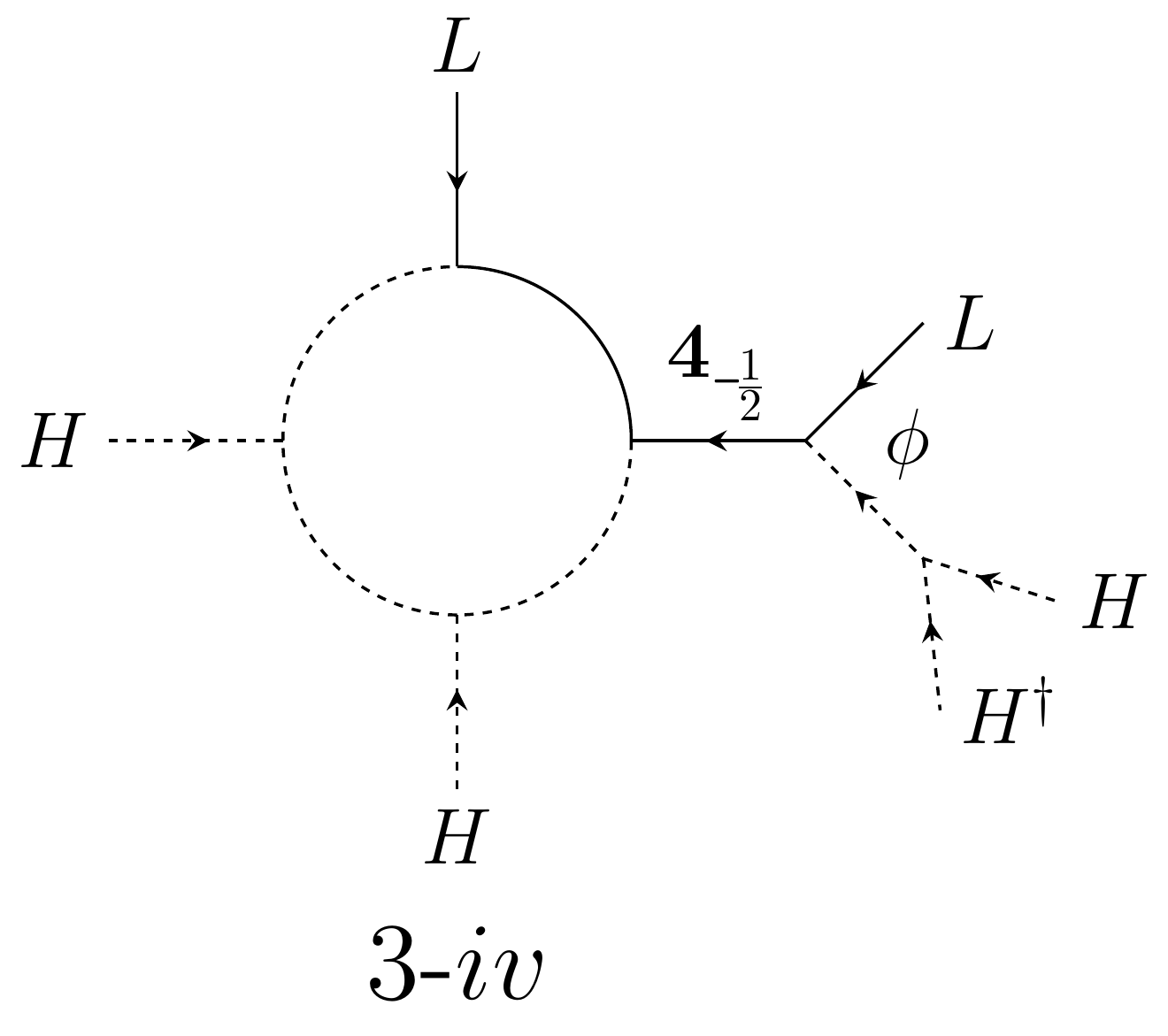} \hfill &  \hfill
        \includegraphics[width=0.33\textwidth]{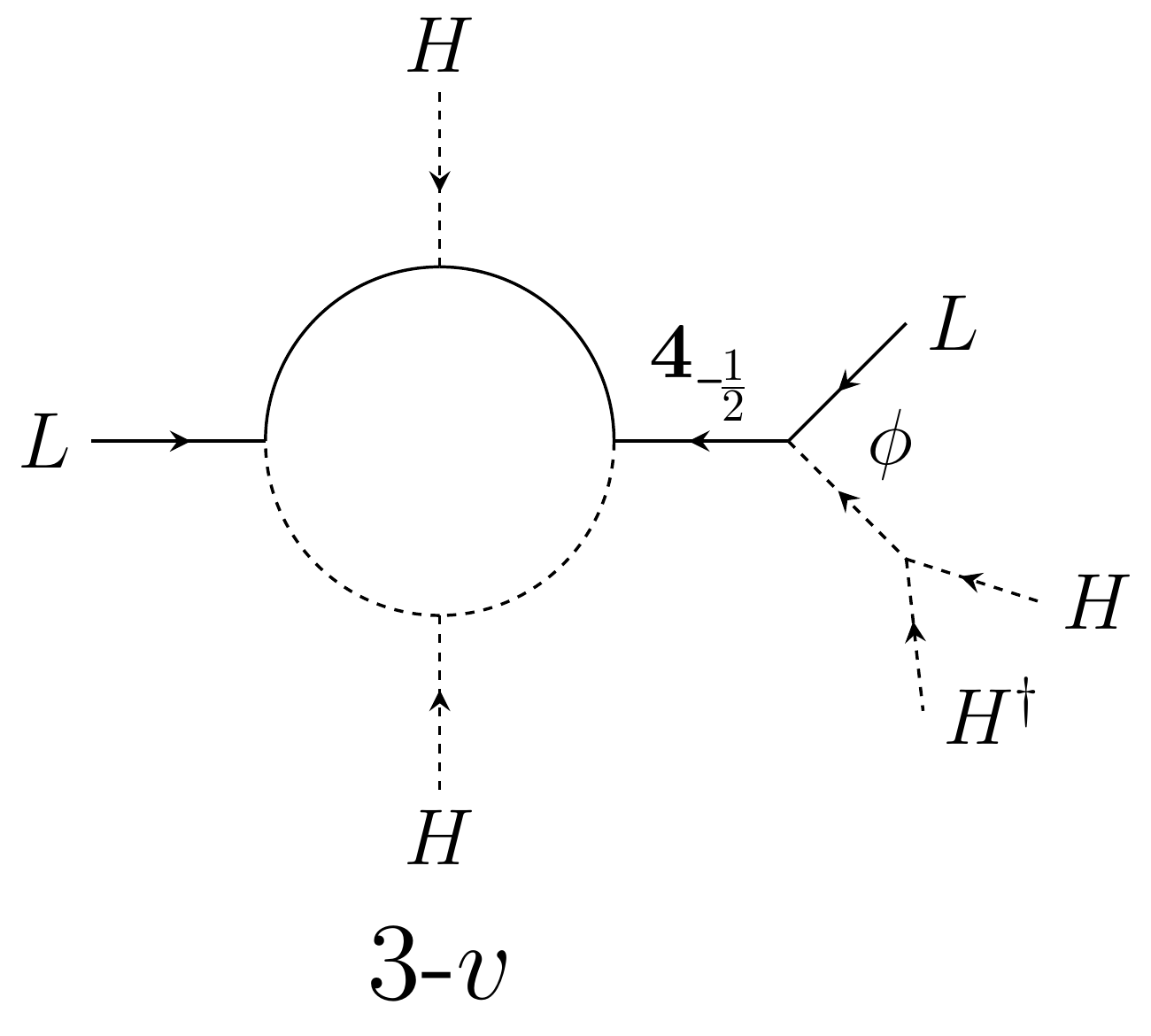} \hfill &  \hfill
        \includegraphics[width=0.33\textwidth]{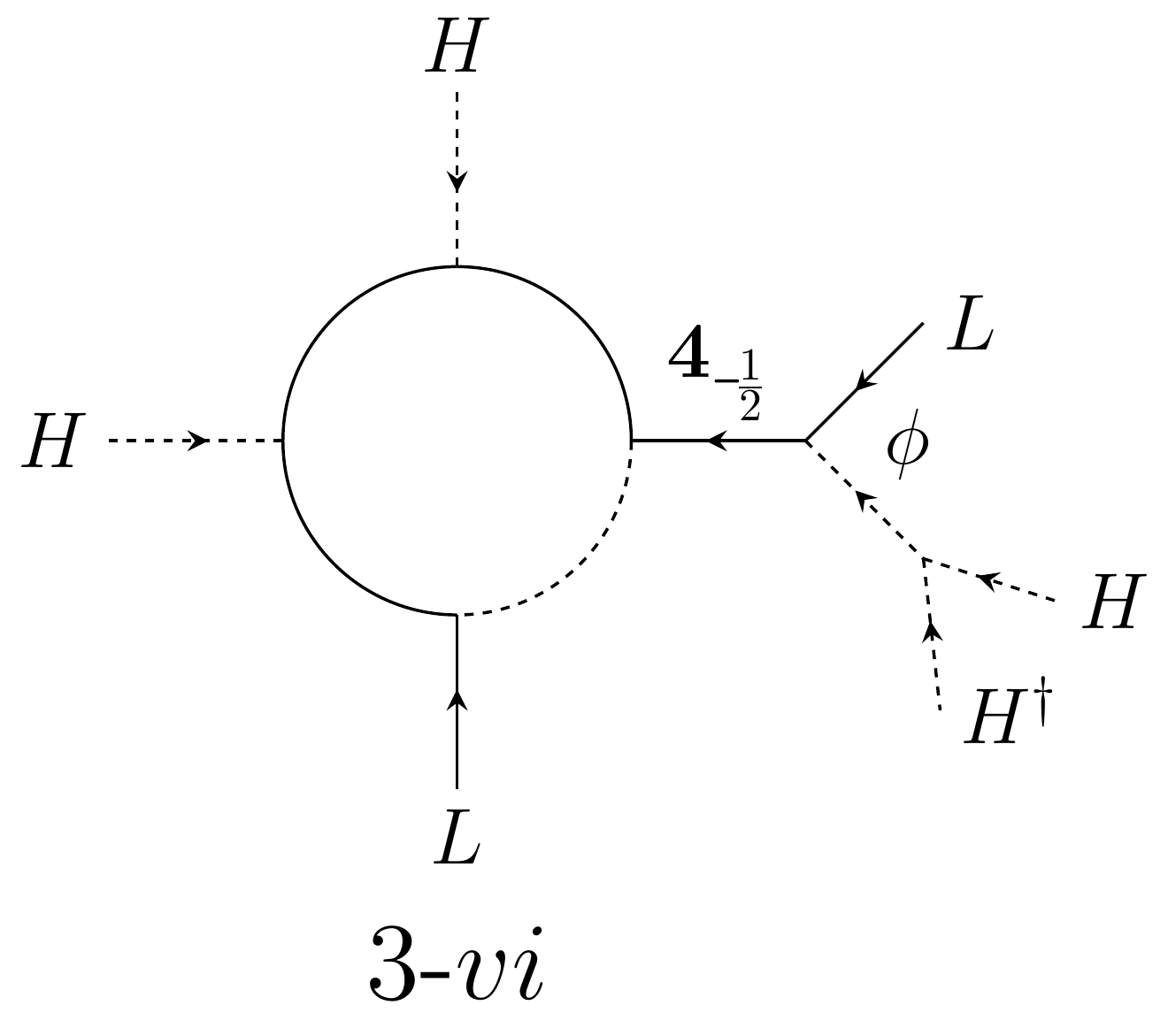}
        \\  
        \includegraphics[width=0.33\textwidth]{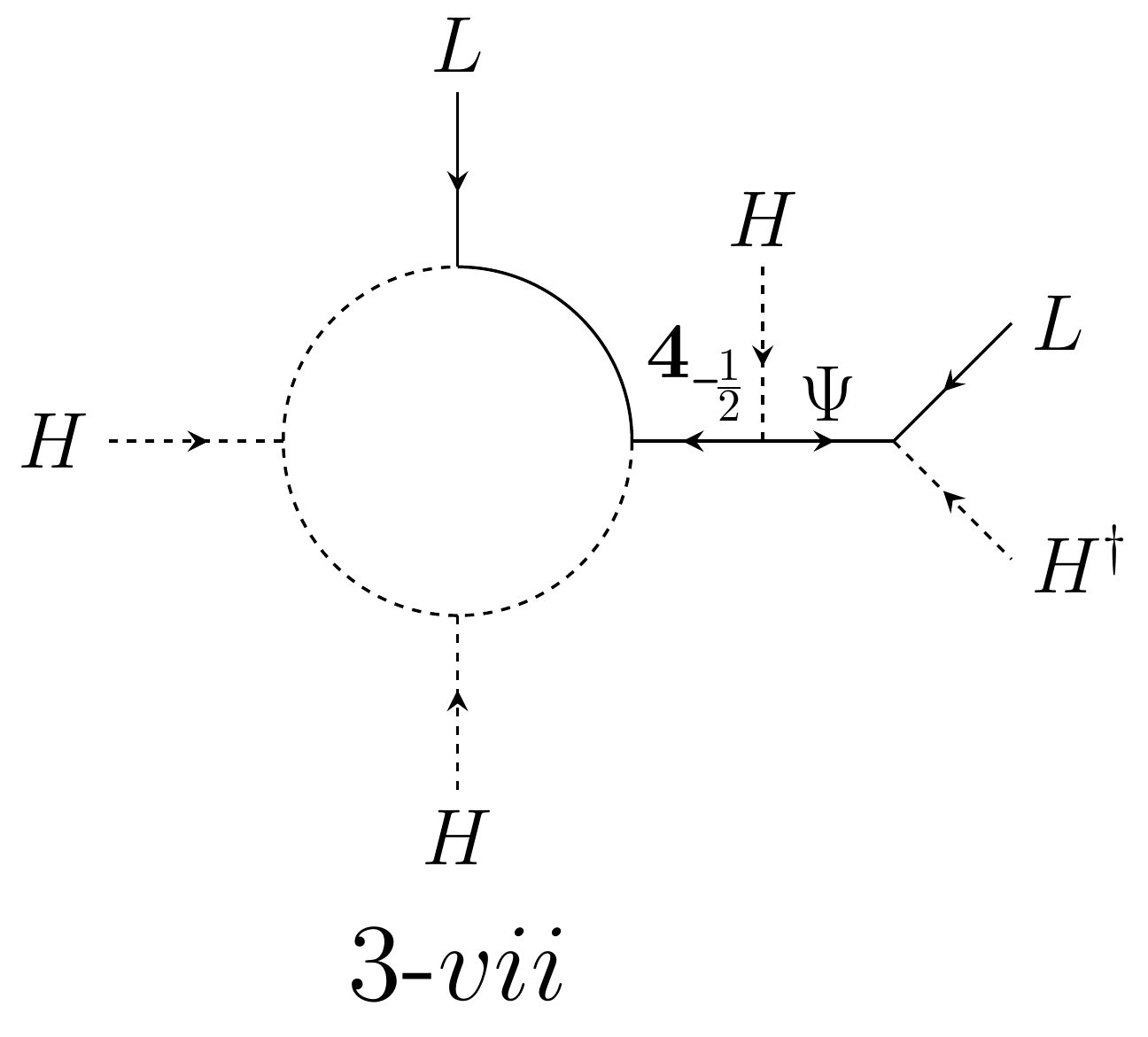} \hfill &  \hfill
        \includegraphics[width=0.33\textwidth]{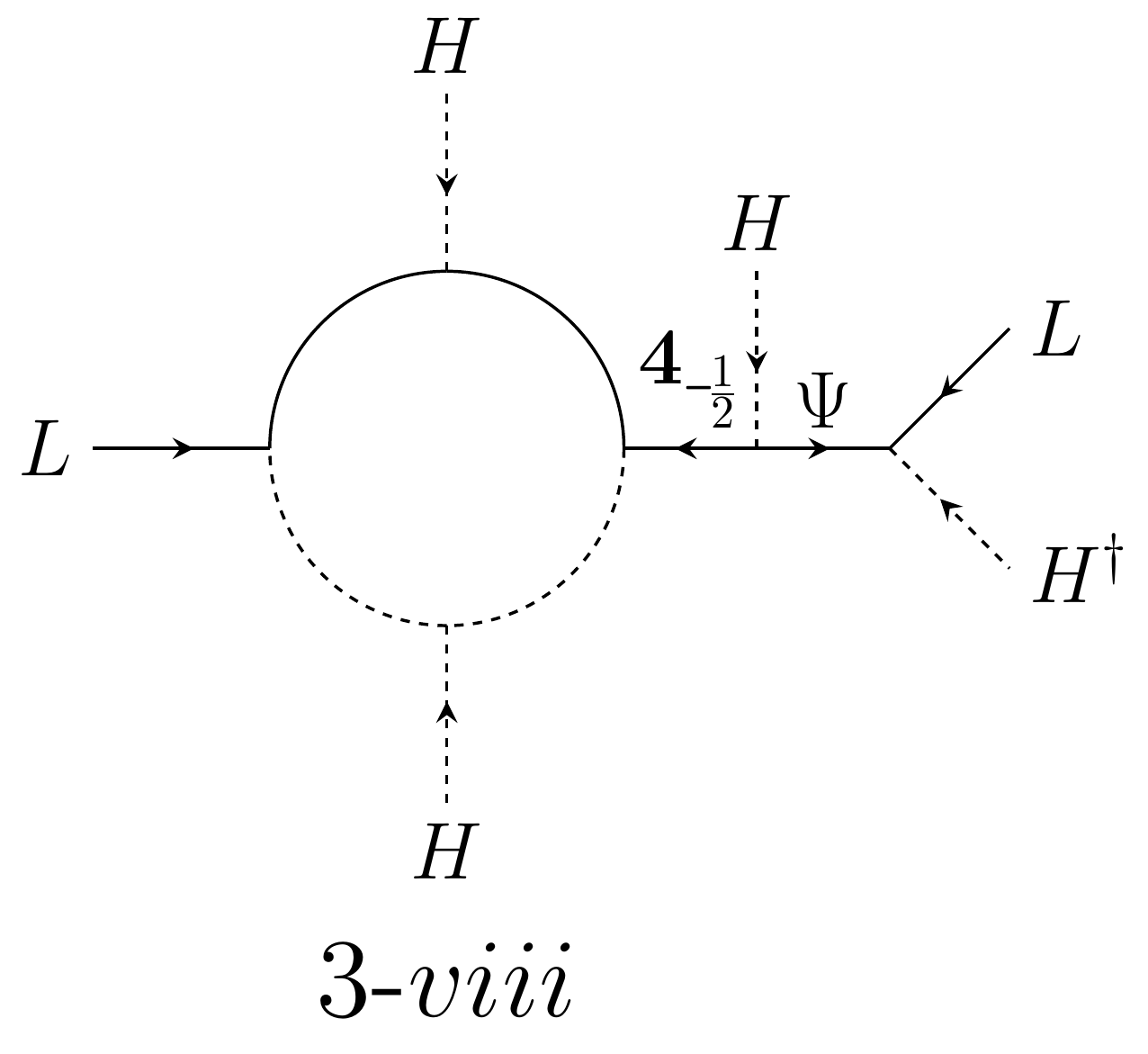} \hfill &  \hfill
        \includegraphics[width=0.33\textwidth]{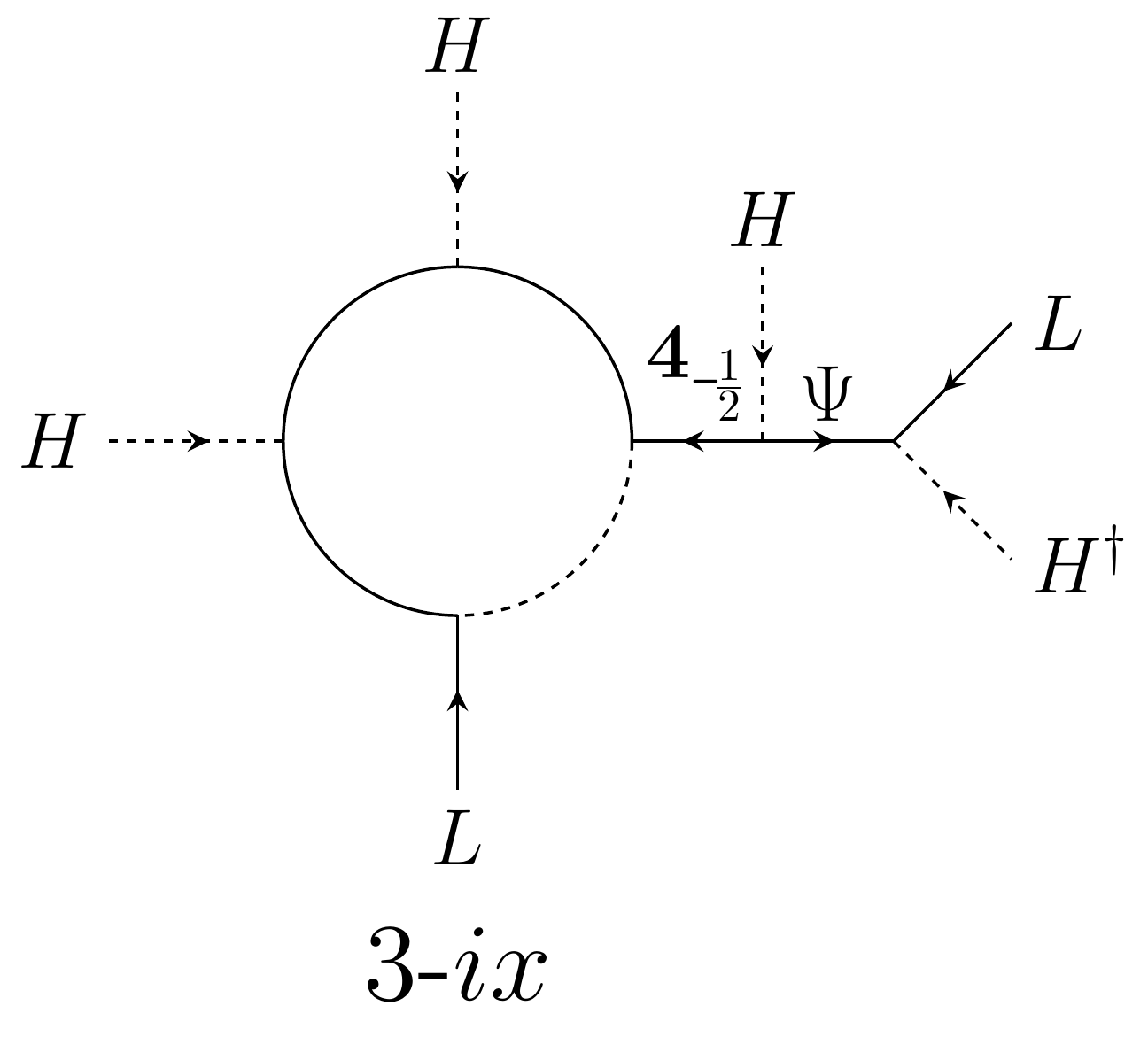}
    \end{tabular}
\caption{(Continues in \fig{fig:app:topos:Diags4pletsInOut}).}
    \label{fig:app:topos:Diags4pletsInOut1}
\end{figure}

\begin{figure}[t!]
    \centering
    \begin{tabular}{ c c }
        \includegraphics[width=0.33\textwidth]{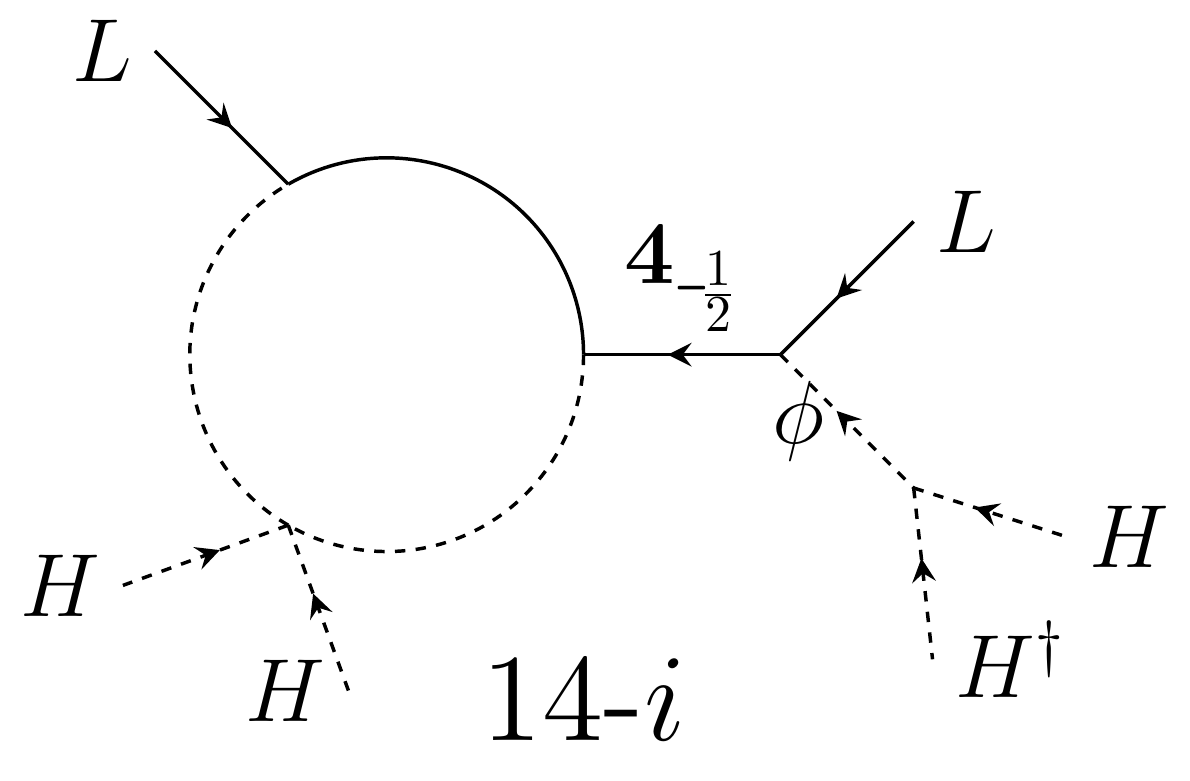} \quad &  \quad
        \includegraphics[width=0.33\textwidth]{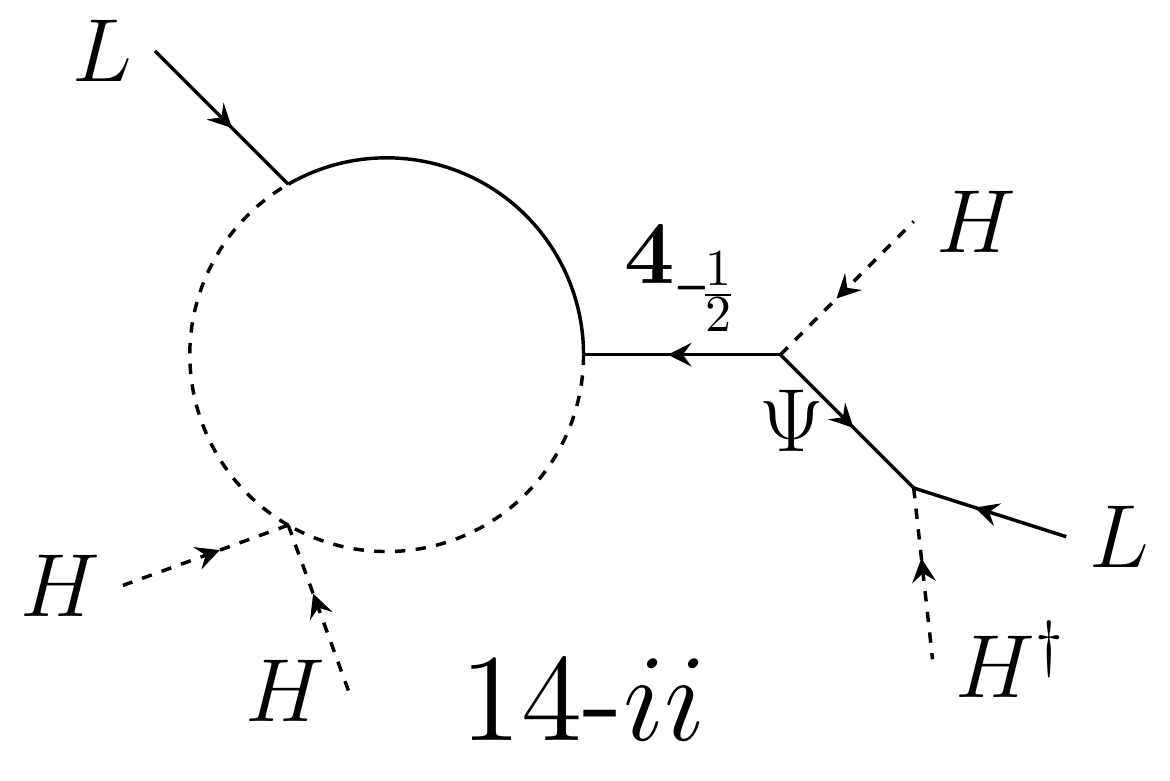}
    \end{tabular} 
    \begin{tabular}{ c c c }
        \includegraphics[width=0.33\textwidth]{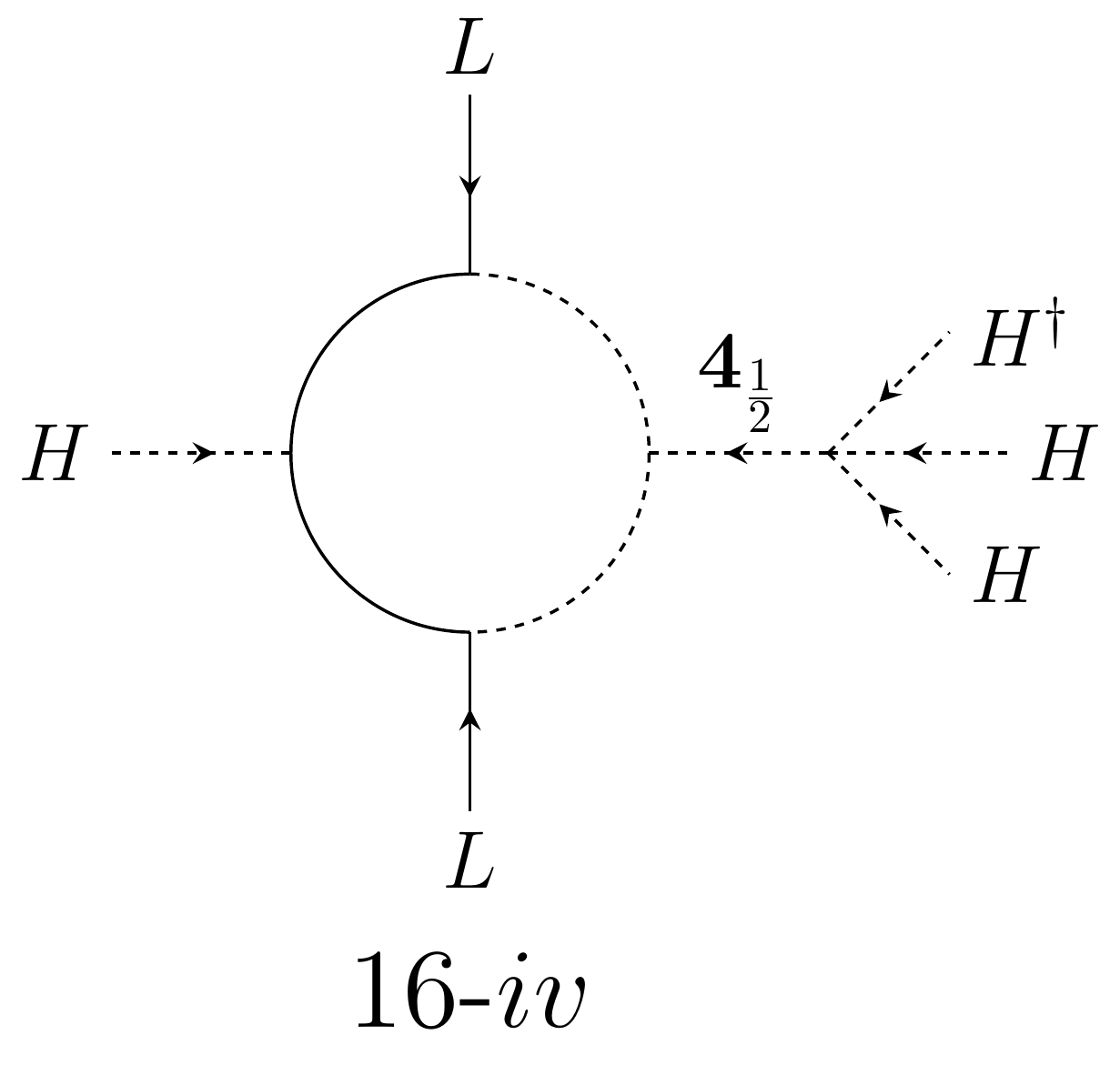} \hfill &  \hfill
        \includegraphics[width=0.33\textwidth]{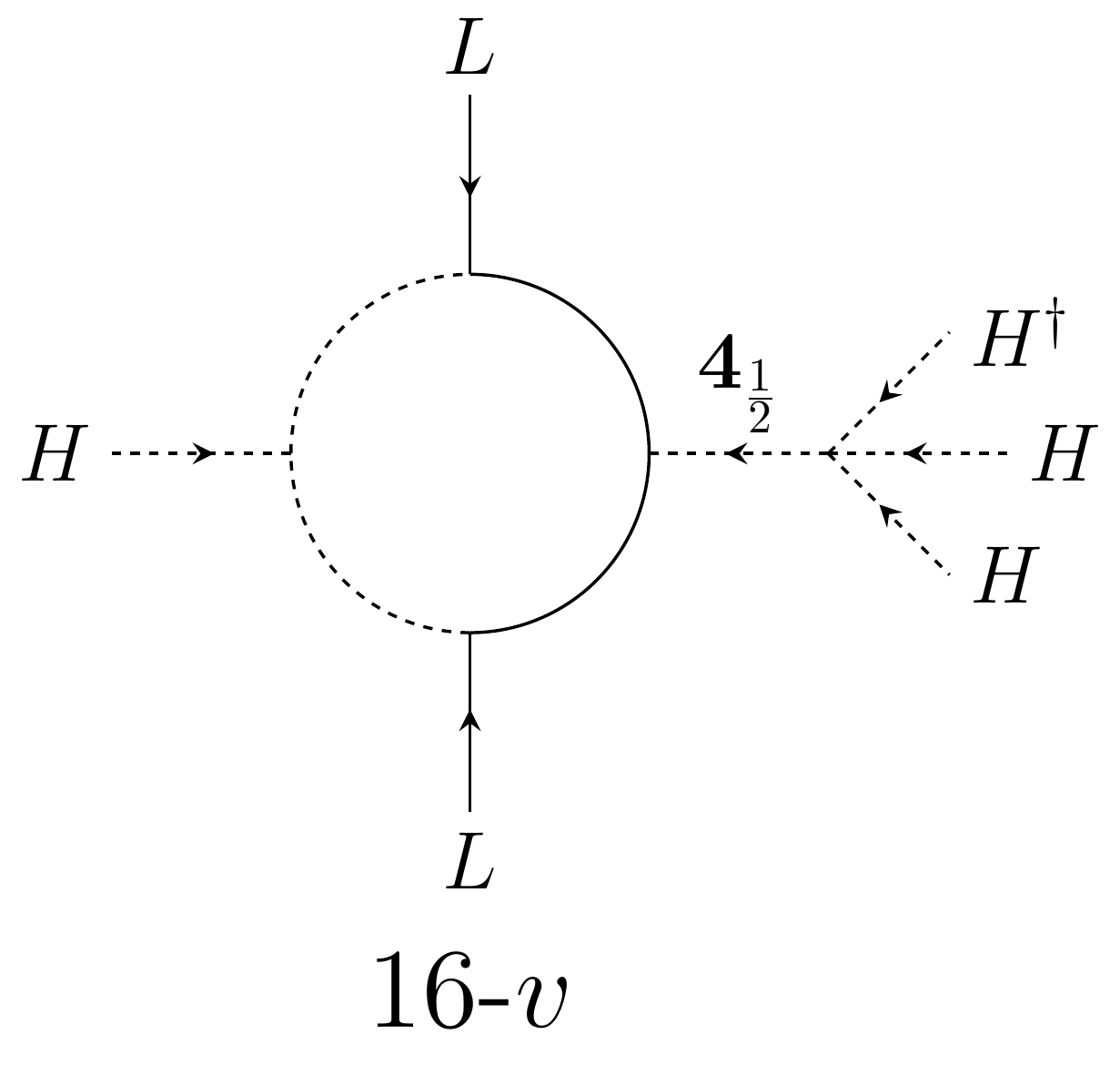} \hfill &  \hfill
        \includegraphics[width=0.33\textwidth]{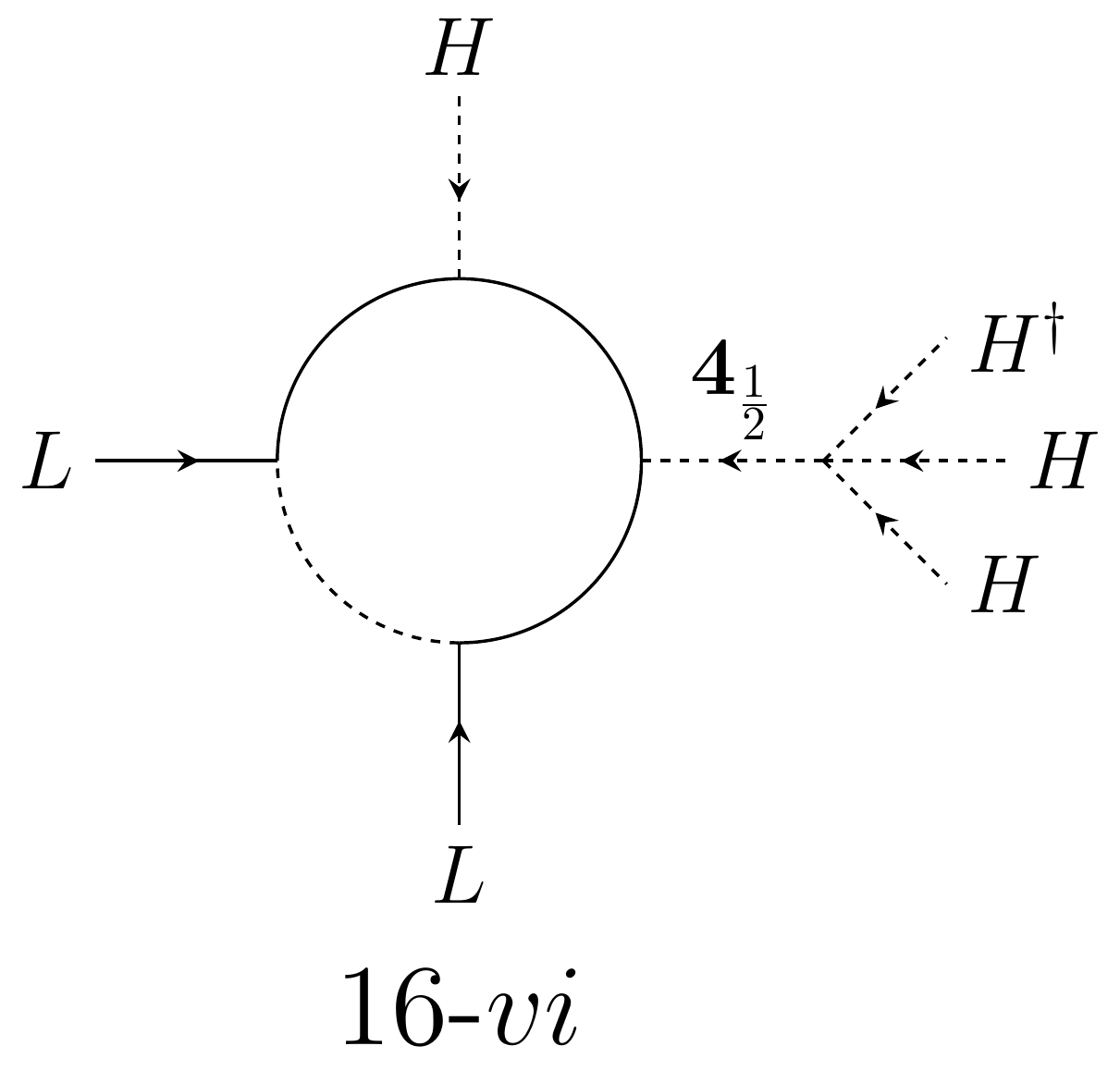}	
    \end{tabular}	 
    \caption{All remaining diagrams that lead to a genuine $d=7$ one-loop neutrino mass for which the maximum representations of $SU(2)_L$ is at least a quadruplet. In these diagrams, two quadruplets are needed. Along with the external fermion or scalar quadruplet, a genuine model always needs an internal quadruplet to distinguish between the scalar $(\textbf{1},\textbf{4},1/2)$ [fermion $(\textbf{1},\textbf{4},-1/2)$] and a Higgs [$L$]. Continuation of \fig{fig:app:topos:Diags4pletsInOut1}.}
    \label{fig:app:topos:Diags4pletsInOut}
\end{figure}

%%%%%%%%%%%%%%%%%%%%%%%%%%%%%%%%%%%%%%%%%%%%%%%%%%%%%%%%%%%%%%%
%%%%%%%%%%%%%%%%%%%%%%%%%%%%%%%%%%%%%%%%%%%%%%%%%%%%%%%%%%%%%%%
\section{$d=5$ three-loop}

\begin{figure}[t!]
    \centering
    \includegraphics[width=1\textwidth]{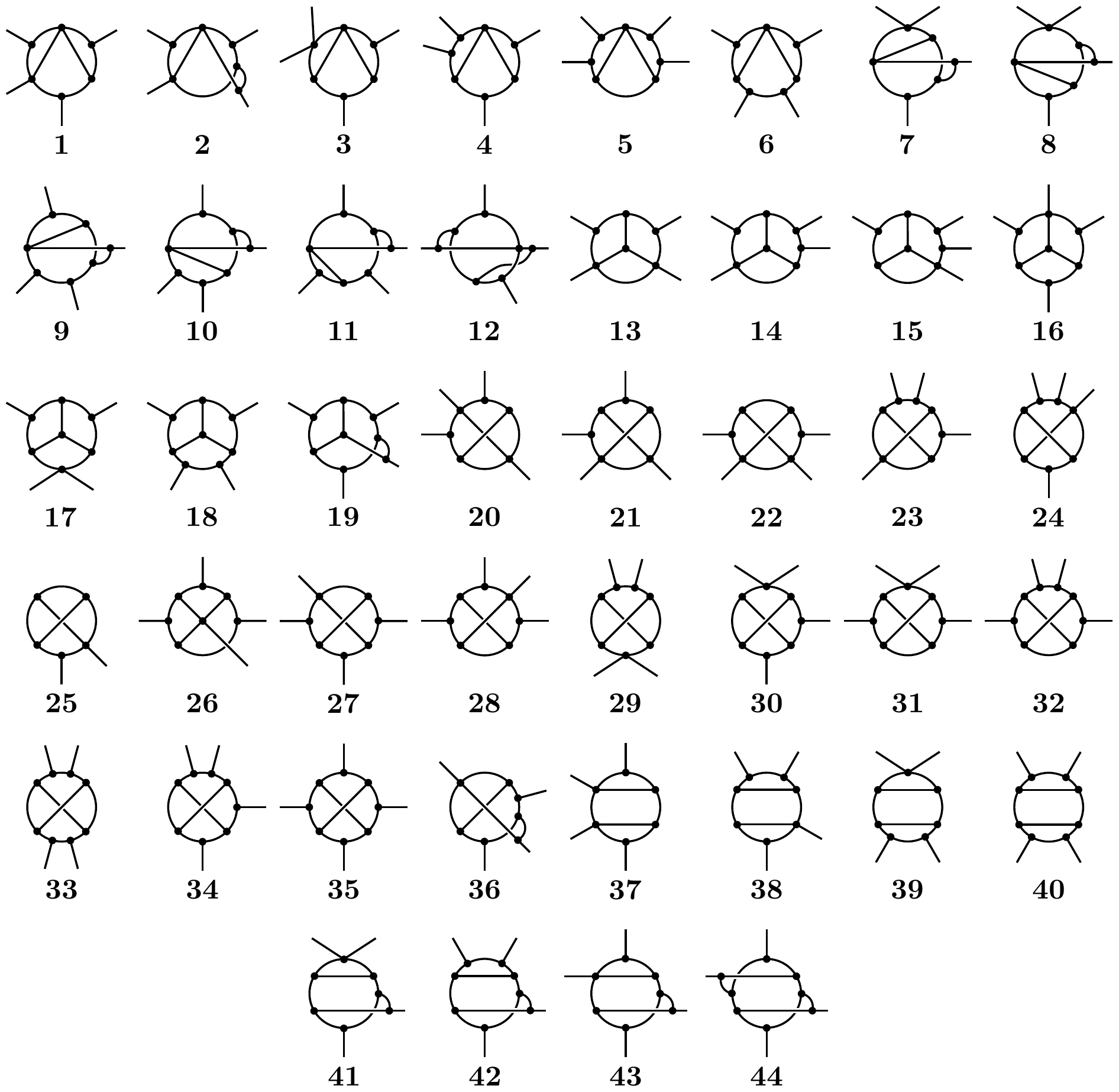}
    \caption{List of topologies associated to \textit{normal} genuine diagrams. We refer to them in the text as $T_i$ with $i=1,\cdots,44$.}
    \label{fig:app:topos:topologies_normal}
\end{figure}

In this section we are interested in those scenarios where the dominant contribution to neutrino masses arises from a three-loop realisation of the Weinberg operator. As explained in \ch{ch:3loop}, genuine neutrino mass diagrams must descend from one of the 44 topologies shown in \fig{fig:app:topos:topologies_normal}, otherwise it is not possible to forbid lower order contributions, independently of the assignment of fermions and scalars to the lines.

\begin{figure}[t!]
    \centering
    \includegraphics[width=1\textwidth]{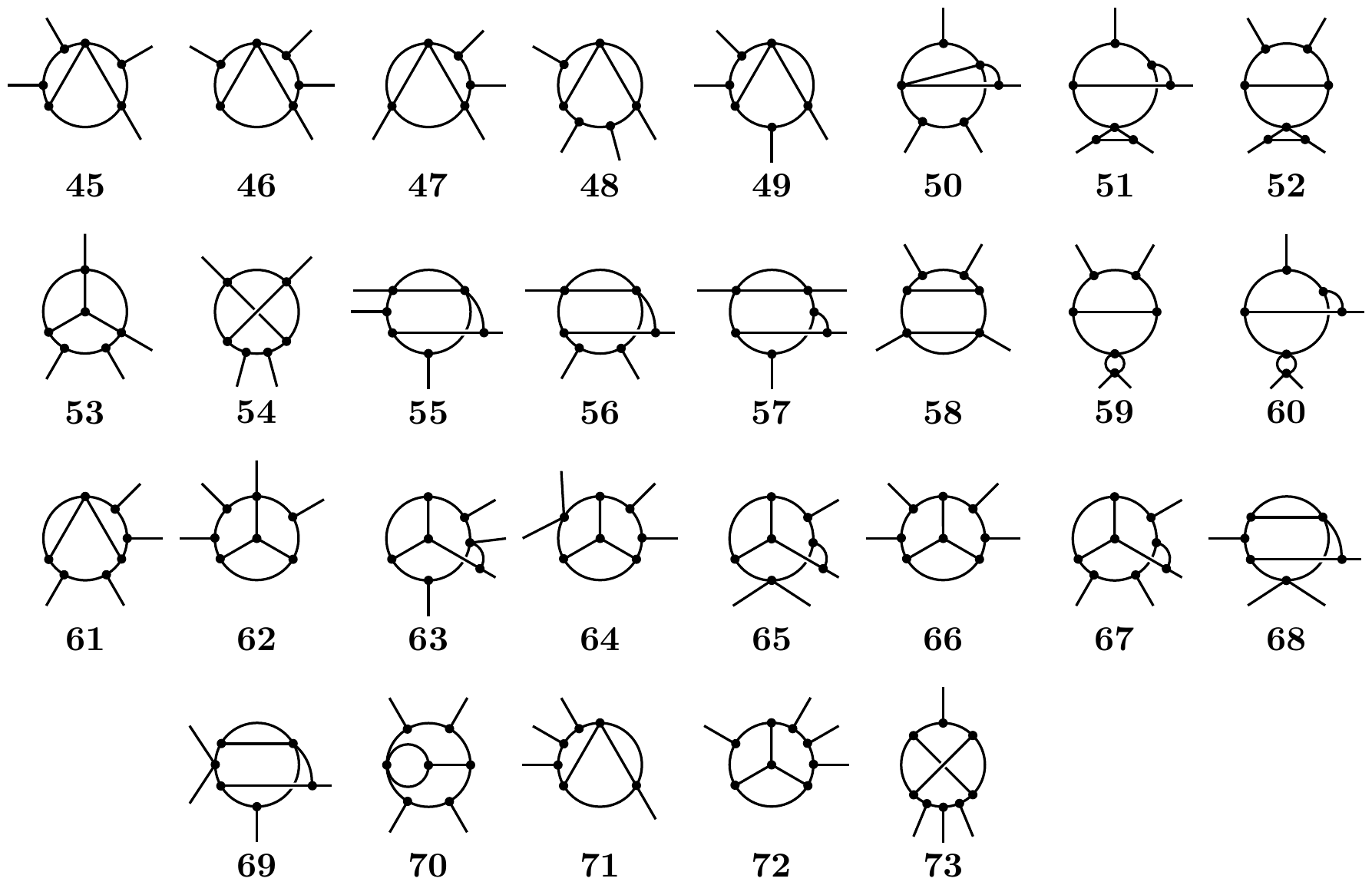}
    \caption{List of \textit{special} genuine topologies associated to the antisymmetric contractions of ${\rm SU(2)_L}$ ($T_i$ with $i=45,\cdots,73$). See \sect{sec:3loop:specgen} for details.}
    \label{fig:app:topos:topologies_special}
\end{figure}

However, the procedure used to identify these 44 \textit{normal} genuine topologies admits a loophole: in the presence of very special fields, it is possible to generate three-loop neutrino masses diagrams with other topologies, with no lower order contributions appearing (see \sect{sec:3loop:specgen}). In \figs{fig:app:topos:topologies_special}{fig:app:topos:derivative} we show these 55 \textit{special} genuine topologies. These topologies are separated in two groups: (i) at least one diagram exists which is genuine due to the presence of an antisymmetric $SU(2)_L$ contraction with two identical particles at loop level; or (ii) the diagram is genuine because it contains a one- or two-loop effective coupling of dimension five or above with at least one derivative.

\begin{figure}[t!]
    \centering
    \includegraphics[width=1\textwidth]{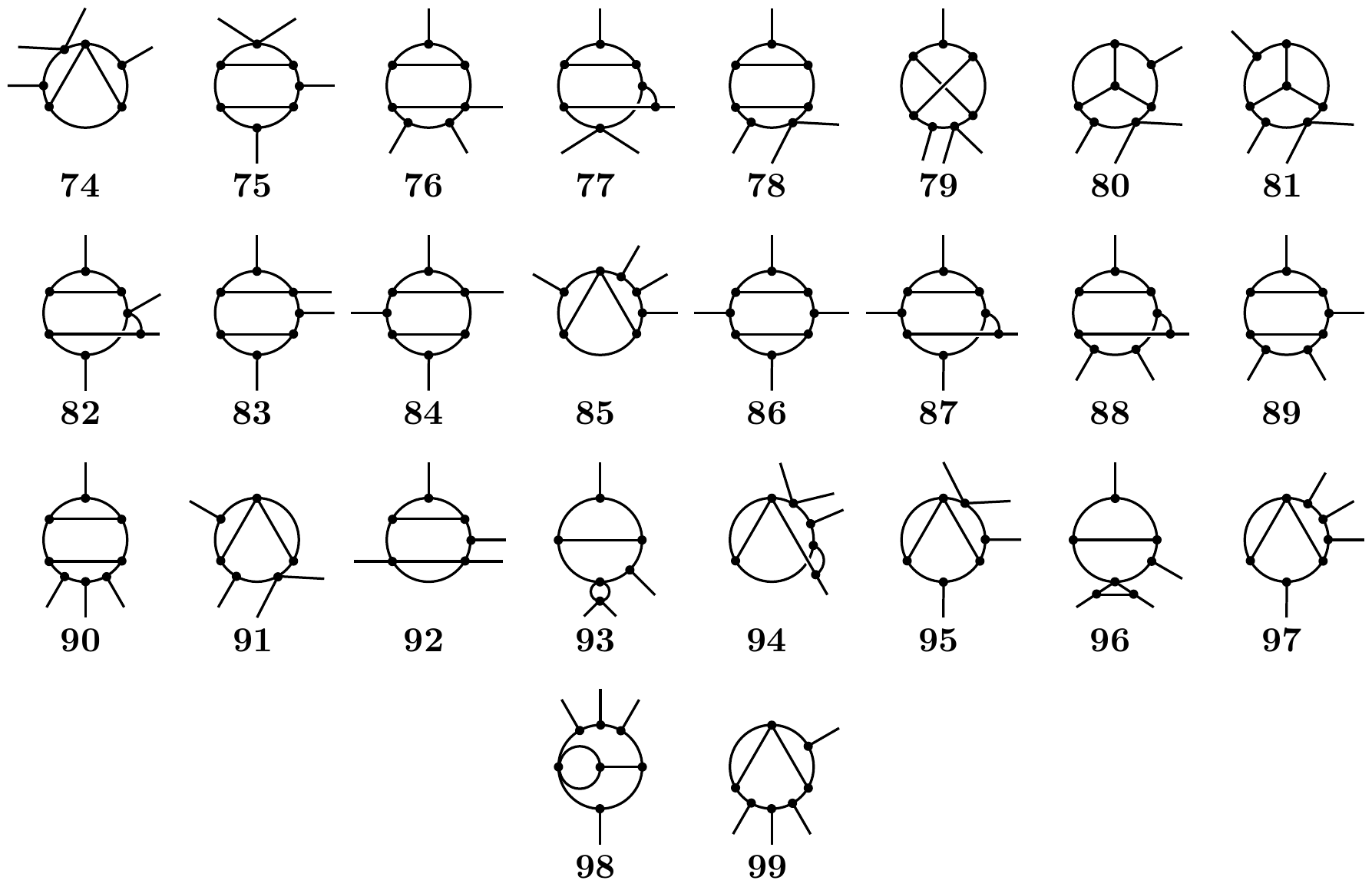}
    \caption{List of \textit{special} genuine topologies associated to internal massless fermion fields. See \sect{sec:3loop:specgen} for details.}
    \label{fig:app:topos:derivative}
\end{figure}

\pagebreak
\fancyhf{}

%% file: Appendices/Appendix_loops.tex
\fancyhf{}
\fancyhead[LE,RO]{\thepage}
\fancyhead[RE]{\slshape\nouppercase{\leftmark}}
\fancyhead[LO]{\slshape\nouppercase{\rightmark}}

\chapter{Loop integrals}
\label{app:loops}

In this appendix we give some useful results for the calculation of diagrams with loops. First, we briefly show how to compute two-loop integrals with one of the diagrams of the example models considered in \ch{ch:dirac2l}. Then, we have a short discussion about master integrals for three-loop neutrino masses, showing that every three-loop integral can be written as a combination of five master integrals. Finally, we compute the three-loop integrals in the cocktail, KNT and AKS models.

%%%%%%%%%%%%%%%%%%%%%%%%%%%%%%%%%%%%%%%%%%%%%%%%%%%%%%%%%%%%%%%
%%%%%%%%%%%%%%%%%%%%%%%%%%%%%%%%%%%%%%%%%%%%%%%%%%%%%%%%%%%%%%%
\section{Computation of two-loop integrals}

In this section, we summarise the main tools needed in order to write every two-loop integral in terms of two master integrals. Two-loop integrals have been evaluated before in the literature and here we will follow \cite{vanderBij:1983bw,Sierra:2014rxa,Martin:2016bgz}. 

To illustrate how the decomposition of two-loop integrals into an analytic expression works, we take the loop functions \eq{eq:dirac2l:Fintegrals1} from the first example in \sect{sec:dirac2l:models}. Rewriting them into explicitly dimensionless integrals we get,
\begin{subequations} \label{eq:app:loops:F}
    \begin{equation} %\label{eq:}
        F^{(1)} = \iint\limits_{(k,q)} \frac{ 1 }{ (k^2-x_1) (k^2-x_2) (q^2-x_3) (q^2-x_4) (q^2-x_5) ((k+q)^2-1) }
    \end{equation}
    \begin{equation} %\label{eq:}
        F^{(2)} = \iint\limits_{(k,q)} \frac{ k \cdot q }{ (k^2-x_1) (k^2-x_2) (q^2-x_3) (q^2-x_4) (q^2-x_5) ((k+q)^2-1) },
    \end{equation}
\end{subequations}
with the following definitions,
\begin{equation} %\label{eq:}
    x_1 = \frac{m_\eta^2}{m_{S_3}^2}, \quad x_2 = \frac{M_{Ei}^2}{m_{S_3}^2}, \quad x_3 = \frac{M_{Nj}^2}{m_{S_3}^2}, \quad x_4 = \frac{m_{S_1}^2}{m_{S_3}^2}, \quad x_5 = \frac{m_{S_2}^2}{m_{S_3}^2},
\end{equation}
and
\begin{equation} %\label{eq:}
    \int\limits_k \equiv (16\pi^2)\int \frac{d^4\!k}{(2\pi)^4}.
\end{equation}

By the use of partial fraction, when various propagators have the same momenta, the integral can be written as a sum over integrals with less denominators,
\begin{equation} %\label{eq:}
    \frac{1}{(k^2-x_1)(k^2-x_2)} = \frac{1}{x_1-x_2} \left( \frac{1}{k^2-x_1} - \frac{1}{k^2-x_2} \right).
\end{equation}
Moreover, integrals with momenta in the numerator which coincides with that of one of the propagators can be reduced as,
\begin{equation} %\label{eq:}
    \frac{q^2}{(k^2-x_1)(q^2-x_2)} = \frac{1}{k^2-x_1} + \frac{x_2}{(k^2-x_1)(q^2-x_2)}.
\end{equation} 
Making use of only these two expressions one can write every two-loop integral in terms of the basis,
\begin{subequations}
    \begin{equation} %\label{eq:}
        \mathbf{A}(x) = \int\limits_k \frac{1}{k^2-x},
    \end{equation}
\vspace*{-0.5cm}
    \begin{equation} %\label{eq:
        \mathbf{I}(x,y,z) = \iint\limits_{(k,q)} \frac{1}{(k^2-x)(q^2-y)((k+q)^2-z)},
    \end{equation}
\end{subequations}
for which analytical expression can be easily found in the literature, see for example \cite{Passarino:1978jh,Martin:2016bgz}.

Particularly, for the two-loop integrals given in \eq{eq:app:loops:F} and used for the numerical analysis of \fig{fig:dirac2l:NuMassplot1}, the decomposition in terms of the master integrals $\mathbf{A}$ and $\mathbf{I}$ is,
\begin{small}
\begin{eqnarray} %\label{eq:}
  \hspace*{-0.5cm}
  F^{(1)} &=& \frac{1}{(x_1-x_2)(x_3-x_4)} \, \times 
  \\ \nn
  && \left\lbrace \frac{1}{x_3-x_5} \left[ \mathbf{I}(x_1,x_3,1) - \mathbf{I}(x_1,x_5,1) - \mathbf{I}(x_2,x_3,1) + \mathbf{I}(x_2,x_5,1) \right] - \left( x_3 \leftrightarrow x_4 \right) \right\rbrace,
\end{eqnarray}
\end{small}
and
\begin{small} 
\begin{eqnarray}
        F^{(2)} &=& \frac 12 (1-x_2-x_5) F^{(1)} 
        \\ \nn
        &+& \frac{1}{(x_1-x_2)(x_3-x_4)} \left\lbrace \frac{1}{x_3-x_5} \left[ \mathbf{A}(x_1)\mathbf{A}(x_3) - \mathbf{A}(x_1)\mathbf{A}(x_5) - \mathbf{A}(x_2)\mathbf{A}(x_3) \right. \right.
        \\ \nn
        &&+ \left. \mathbf{A}(x_2)\mathbf{A}(x_5)
         - (x_1-x_2)( \mathbf{I}(x_1,x_3,1) - \mathbf{I}(x_1,x_5,1) ) \vphantom{\mathbf{A}} \right] 
         \\ \nn
         && - \left.  \mathbf{I}(x_1,x_3,1) + \mathbf{I}(x_2,x_3,1)
         - \left( x_3 \leftrightarrow x_4 \right) \right\rbrace,
\end{eqnarray}
\end{small}
\begin{small} 
\begin{eqnarray}
%\label{eq:}
  % \hspace*{-1cm}
        F^{(2)} &=& \frac 12 (1-x_2-x_5) F^{(1)} 
        \\ \nn
        &+& \frac{1}{(x_1-x_2)(x_3-x_4)} \left\lbrace \frac{1}{x_3-x_5} \left[ \mathbf{A}(x_1)\mathbf{A}(x_3) - \mathbf{A}(x_1)\mathbf{A}(x_5) - \mathbf{A}(x_2)\mathbf{A}(x_3)  \right. \right.
         \\ \nn
         &&+ \left. \mathbf{A}(x_2)\mathbf{A}(x_5) - (x_1-x_2)( \mathbf{I}(x_1,x_3,1) - \mathbf{I}(x_1,x_5,1) ) \vphantom{\mathbf{A}} \right] 
         \\ \nn
         && - \mathbf{I}(x_1,x_3,1) + \mathbf{I}(x_2,x_3,1) - \left. \left( x_3 \leftrightarrow x_4 \right) \right\rbrace,
\end{eqnarray}
\end{small}
where we have used that $k \cdot q = \frac 12 \left[ (k+q)^2-k^2-q^2 \right]$.

The decompositions for all the diagrams in \fig{fig:dirac2l:massdiagrams} in terms of the master integrals can be found in \cite{Sierra:2014rxa}.

%%%%%%%%%%%%%%%%%%%%%%%%%%%%%%%%%%%%%%%%%%%%%%%%%%%%%%%%%%%%%%%
%%%%%%%%%%%%%%%%%%%%%%%%%%%%%%%%%%%%%%%%%%%%%%%%%%%%%%%%%%%%%%%
\section{Master integrals for three-loop neutrino masses}

Here we give the minimal set of integrals that span the complete list of possible genuine models. In principle, starting from the list of 30 genuine diagrams in the mass eigenbasis (see \fig{fig:3loop:normalgenuinediagrams} and \fig{fig:3loop:specialgenuinediagrams}), one should obtain at least 30 integrals assigning momenta to the fields. This initial set, however, can be further reduced applying the results previously used in two-loop calculations in \cite{Sierra:2014rxa,McDonald:2003zj}, both based on \cite{vanderBij:1983bw}, and three-loop integrals in \cite{Freitas:2016zmy, Martin:2016bgz}. Here, we are going to summarise the results in these papers which we need.

Using the notation of \cite{Martin:2016bgz}, we can write
\begin{small}
\begin{equation} \label{eq:app:loops:Tdef}
    \mathbf{T}^{(n_1,n_2,n_3,n_4,n_5,n_6)} (x_1,x_2,x_3,x_4,x_5,x_6) =
\end{equation}
\begin{equation*}
\medmath{
     \iiint\limits_{(k_1,k_2,k_3)} \frac{1}{ [k_1^2-x_1]^{n_1} [k_2^2-x_2]^{n_2} [k_3^2-x_3]^{n_3} [(k_1-k_2)^2-x_4]^{n_4} [(k_2-k_3)^2-x_5]^{n_5} [(k_3-k_1)^2-x_6]^{n_6} } \, .
     }
\end{equation*}
\end{small}
Here we have used the abbreviation given in \eq{eq:3loop:intnotation} and the powers of the propagators $n_i$ can be any integer number. Note that the integral is invariant under the interchange of pairs $(n_i,x_i)$ and moreover, satisfies the nine identities obtained by integration by parts \cite{Chetyrkin:1981qh},
\begin{equation} \label{eq:app:loops:partialidentities}
    0 = \iiint\limits_{(k_1,k_2,k_3)} \frac{\partial}{\partial k_i^\mu} \left[ k_j^\mu \mathbf{X} \right] \, ,
\end{equation} 
for $i,j=1,2,3$, with $\mathbf{X}$ equal to any product of propagators of the form shown in \eq{eq:app:loops:Tdef}. One can find very useful identities for this kind of integrals, such as,
\begin{equation} %\label{eq:}
    \frac 3 2 d + \sum\limits_{j=1}^6 (x_j \mathbf{j}^+-1) n_j = 0 \, .
\end{equation}
Here $d$ is the dimension of the momentum integration in dimensional regularisation and $\mathbf{j}^\pm$ is short-hand notation for the following operator,
\begin{equation} %\label{eq:}
    \mathbf{j}^\pm \mathbf{T}^{(...,n_j,...)} = \mathbf{T}^{(...,n_j \pm 1,...)} \, .
\end{equation}

By repeated application of the identities \eq{eq:app:loops:partialidentities}, any of the three-loop integrals $\mathbf{T}$ can be reduced to a linear combination of five master integrals \cite{Martin:2016bgz},
\begin{align} %\label{eq:}
        \nonumber
    \mathbf{H}(x_1,x_2,x_3,x_4,x_5,x_6) &= \mathbf{T}^{(1,1,1,1,1,1)} (x_1,x_2,x_3,x_4,x_5,x_6) \, ,
    \\ \nonumber
    \mathbf{G}(x_3,x_1,x_6,x_2,x_5)     &= \mathbf{T}^{(1,1,1,0,1,1)} (x_1,x_2,x_3,x_4,x_5,x_6) \, ,
    \\ 
    \mathbf{F}(x_1,x_2,x_5,x_6)         &= \mathbf{T}^{(2,1,0,0,1,1)} (x_1,x_2,x_3,x_4,x_5,x_6) \, ,
    \\ \nonumber
    \mathbf{A}(x_1)\mathbf{I}(x_2,x_3,x_5) &= \mathbf{T}^{(1,1,1,0,1,0)} (x_1,x_2,x_3,x_4,x_5,x_6) \, ,
    \\ \nonumber
    \mathbf{A}(x_1) \mathbf{A}(x_2) \mathbf{A}(x_3) &= \mathbf{T}^{(1,1,1,0,0,0)} (x_1,x_2,x_3,x_4,x_5,x_6) \, .
\end{align}
where $\mathbf{A}$ is the standard one-loop Passarino-Veltman function \cite{Passarino:1978jh} and $\mathbf{I}$ is a two-loop integral described in \cite{vanderBij:1983bw}. It is worth mentioning that analytical expressions exist for the well-known integrals $\mathbf{A}$ and $\mathbf{I}$, while for the three-loop ones ($\mathbf{F}$, $\mathbf{G}$, and $\mathbf{H}$) results are known only for very particular cases (see \cite{Martin:2016bgz} for details).  \\

Particularising to our case, starting from the 30 diagrams in \fig{fig:3loop:normalgenuinediagrams} and \fig{fig:3loop:specialgenuinediagrams}, in the mass insertion approximation, and assigning momenta to the internal lines, one can find that the integrals have repeated propagators with equal momenta but different masses.\footnote{The momenta flowing into the diagrams is set to 0, given the smallness of neutrino masses.} One can prove that every three-loop integral in \fig{fig:3loop:normalgenuinediagrams} and \fig{fig:3loop:specialgenuinediagrams} can be written in terms of the integrals in \eq{eq:app:loops:Tdef}. We note here that partial fractions can be used to reduce the number of propagators with common momenta \cite{vanderBij:1983bw},
\begin{align} \label{eq:app:loops:propagatorreduce}
    \mathbf{T}^{(\left\lbrace n_{11},n_{12} \right\rbrace,n_2,n_3,n_4,n_5,n_6)}& ( \left\lbrace x_{11},x_{12} \right\rbrace, x_2,x_3,x_4,x_5,x_6) = 
    \\ \nonumber
    \frac{1}{x_{11}-x_{12}} \Big[ 
    &\mathbf{T}^{( \left\lbrace n_{11},n_{12}-1 \right\rbrace,n_2,n_3,n_4,n_5,n_6)} ( \left\lbrace x_{11},x_{12} \right\rbrace,x_2,x_3,x_4,x_5,x_6) 
    \\ \nonumber
    -&\mathbf{T}^{( \left\lbrace n_{11}-1,n_{12} \right\rbrace,n_2,n_3,n_4,n_5,n_6)} ( \left\lbrace x_{11},x_{12} \right\rbrace,x_2,x_3,x_4,x_5,x_6) \Big] \, ,
\end{align}
where $\mathbf{T}^{(\left\lbrace n_{11},n_{12}\right\rbrace,n_2,n_3,n_4,n_5,n_6)} ( \left\lbrace x_{11},x_{12} \right\rbrace,x_2,x_3,x_4,x_5,x_6)$ is the same as $\mathbf{T}$ without the braces, but with an extra propagator $[k_1^2-x_{12}]^{n_{12}}$.

On the other hand, some integrals with a non-trivial integrand numerator can be further simplified using the \textit{$p^2$-decomposition}, namely
\begin{equation} \label{eq:app:loops:p2decomposition}
    \frac{p^2}{(k^2-x_1)(p^2-x_2)} = \frac{1}{(k^2-x_1)} + \frac{x_2}{(k^2-x_1)(p^2-x_2)}.
\end{equation}

To demonstrate how this procedure works in practice, we can take for instance the loop integral of model 1, given in \sect{sec:3loop:examples}. Applying the identity \eq{eq:app:loops:propagatorreduce} twice to both propagators sharing $k_1$ and $k_2$ momenta, $F_{loop} \! \left( x_1, x_2 \right)$ can be directly decomposed in terms of a linear combination of $\mathbf{G}$'s,
\begin{eqnarray} %\label{eq:}
    F_{loop} \! \left( x_1, x_2 \right) = \frac{1}{x_1^2}  \bigg\{ &\mathbf{G}(1,x_1,x_2,x_1,x_1)& - \mathbf{G}(1,x_1,x_2,0,x_1) 
    \\
    \nn
    - &\mathbf{G}(1,0,x_2,x_1,x_1)& + \mathbf{G}(1,0,x_2,0,x_1)  \bigg\} \, .
\end{eqnarray}

For model 5 of \sect{sec:3loop:examples}, the decomposition of the loop integral $F_L( x_1,x_2 )$ in \eq{eq:3loop:Floop_model5} is straightforward given the previous example. One only has to apply \eq{eq:app:loops:propagatorreduce} three times to obtain a linear combination of eight $\mathbf{G}$ integrals. Here we focus on the decomposition of $F_R( x_1,x_2 )$, just to present an example of a integral with a non-trivial numerator. One should first notice that under the integral sign,
\begin{equation} %\label{eq:}
    \slashed{k_3}(\slashed{k_2}+\slashed{k_3}) \longrightarrow k_3 \cdot (k_2+k_3) = \frac 1 2 \left[ (k_2+k_3)^2-k_2^2+k_3^2 \right] \, .
\end{equation}
It is clear that one should apply the \textit{$p^2$-decomposition} in \eq{eq:app:loops:p2decomposition} along with the partial fractions decomposition \eq{eq:app:loops:propagatorreduce}, as in the previous case, to get rid of the numerator and the repeated propagators. The full process of the decomposition is rather lengthy and cumbersome, so here we give just the final result.
\begin{align} %\label{eq:}
    F_L( x_1,x_2,x_3,x_4 ) = \frac{\sqrt{x_1 x_3}}{x_1-x_2} 
    \bigg\{ 
        &\mathbf{G}(x_1,1,x_4,1,x_3) - \mathbf{G}(x_1,1,x_4,0,x_3) 
        \\ \nn
        - &\mathbf{G}(x_1,0,x_4,1,x_3) + \mathbf{G}(x_1,0,x_4,0,x_3) 
       \\ \nn
        -&\mathbf{G}(x_2,1,x_4,1,x_3) + \mathbf{G}(x_2,1,x_4,0,x_3) 
        \\ \nn
        + & \mathbf{G}(x_2,0,x_4,1,x_3) - \mathbf{G}(x_2,0,x_4,0,x_3)
    \bigg\},
\end{align}
\begin{align} %\label{eq:}
    F_R( x_1,x_2,x_3,x_4 ) = \frac 1 2 \frac{1}{x_1-x_2} 
    \bigg\{ 
        (x_1+x_3-1) &\Big[ \mathbf{G}(x_1,1,x_4,1,x_3) - \mathbf{G}(x_1,0,x_4,1,x_3) \Big] 
        \nonumber \\ \nonumber
        -(x_1+x_3) &\Big[ \mathbf{G}(x_1,1,x_4,0,x_3) - \mathbf{G}(x_1,0,x_4,0,x_3) \Big]
        \\ \nonumber
        -(x_2+x_3-1) &\Big[ \mathbf{G}(x_2,1,x_4,1,x_3) - \mathbf{G}(x_2,0,x_4,1,x_3) \Big]
        \\ \nonumber
        + (x_2+x_3) &\Big[ \mathbf{G}(x_2,1,x_4,0,x_3) - \mathbf{G}(x_2,0,x_4,0,x_3) \Big] 
        \\
        +\Big[ \mathbf{A}(1) - \mathbf{A}(0) \Big] \Big[ \mathbf{I}(x_1,1,x_4) - &\mathbf{I}(x_1,0,x_4) - \mathbf{I}(x_2,1,x_4) + \mathbf{I}(x_2,0,x_4) \Big]
    \bigg\}.
\end{align}
One can check that the loop integral decompositions are still symmetric under the interchange of $x_1$ and $x_2$, as it was the case with the original integral definitions.

%%%%%%%%%%%%%%%%%%%%%%%%%%%%%%%%%%%%%%%%%%%%%%%%%%%%%%%%%%%%%%%
%%%%%%%%%%%%%%%%%%%%%%%%%%%%%%%%%%%%%%%%%%%%%%%%%%%%%%%%%%%%%%%
\section{Minimal three-loop neutrino mass models: loop integrals}

Finally, we show the calculation of the loop integrals in the three-loop models discussed in \ch{ch:clfv}, the cocktail, KNT and AKS models. Here, we derive the loop functions used in the aforementioned chapter. For their computation, we did not rely on approximations, but implemented the full integral numerically using {\tt pySecDec}~\cite{Borowka:2017idc}.

Moreover, as explained in the previous section, all the integrals shown here, can be factorised in terms of five master integrals as normally done. Nevertheless, we decided not to do it, because there is still no analytical general solution to all the three-loop master integrals and their factorisation could lead to numerical precision issues. Note that these five master integrals have divergent parts, while the full integral is finite.

%%%%%%%%%%%%%%%%%%%%%%%%%%%%%%%%%%%%%%%%%%%%%%%%%%%%%%%%%%%%%%%
\subsection{Cocktail model}

To compute the dimensionless integral $F_{\rm Cocktail}$ in \eq{eq:clfv:mnucocktail}, we choose the Feynman-'t Hooft gauge $\xi=1$. In this gauge, the propagator of the $W_\mu$ boson has no momenta structure in the numerator, while the standard Goldstone $H^+$ contribution with a mass $m_W^2$ should be included. We decided to show the diagrams in the gauge basis to be able to identify the different contributions that enter in $F_{\rm Cocktail}$.

\begin{figure}
    \centering
    \includegraphics[width=0.4\textwidth]{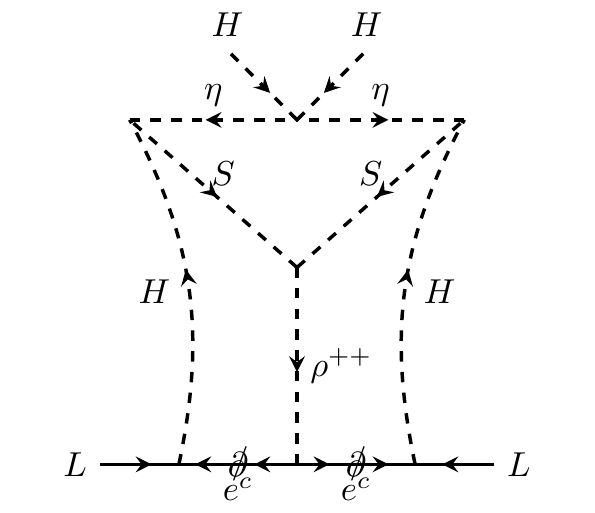}
    \includegraphics[width=0.4\textwidth]{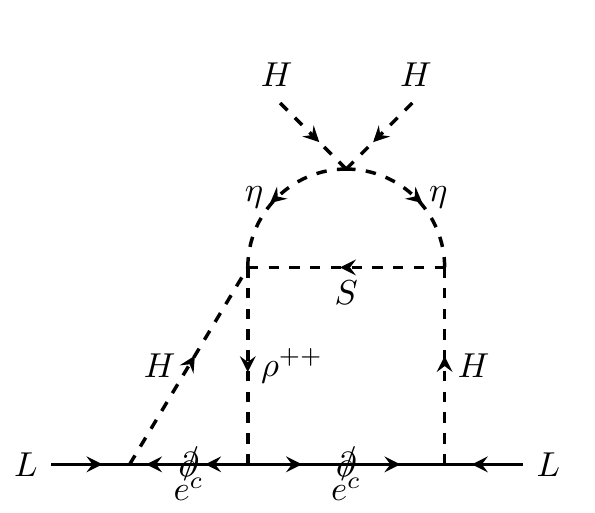}
    \caption{Dimension $5$ mass diagrams in the gauge basis. Note that given the chirality, the corresponding integral has two momenta in the numerator. This is denoted with $\slashed{\partial}$. When referring to these diagrams we will use the notation $\mathfrak{I}^{(5)}_i$ with $i=1,2$ following the order of the figures.}
    \label{fig:app:loops:CT_diag_d5}
\end{figure}

We identified $12$ different diagrams in the gauge basis with dimensions $5$, $7$ and $9$, see \figs{fig:app:loops:CT_diag_d5}{fig:app:loops:CT_diag_d9}. All of them are proportional to the mass of the charged leptons squared and with two derivatives. Naively, one could expect that the dominant contribution comes from the dimension $5$ diagrams. However, as we are considering $4\pi$ couplings and lowering the new physics scale as much as possible, all $12$ diagrams could be in principle relevant.

\begin{figure}[t]
    \centering
    \includegraphics[width=0.4\textwidth]{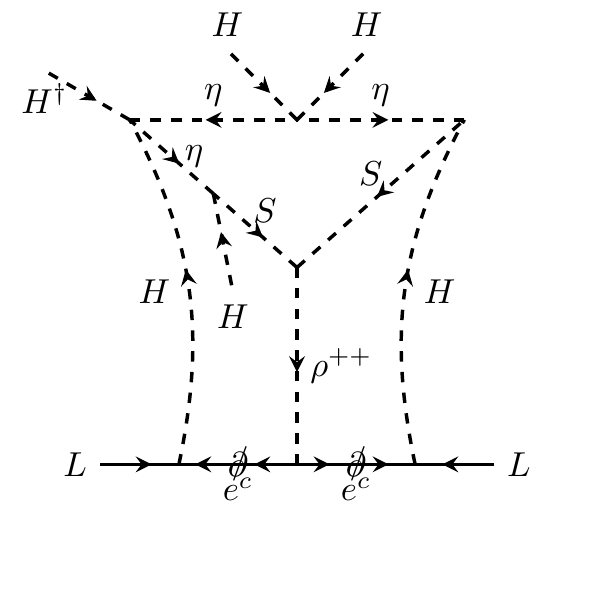}
    \includegraphics[width=0.4\textwidth]{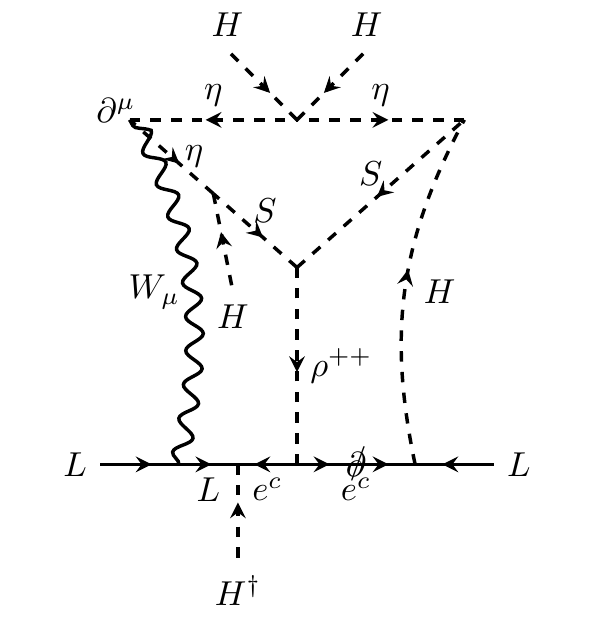}
    \\
    \includegraphics[width=0.4\textwidth]{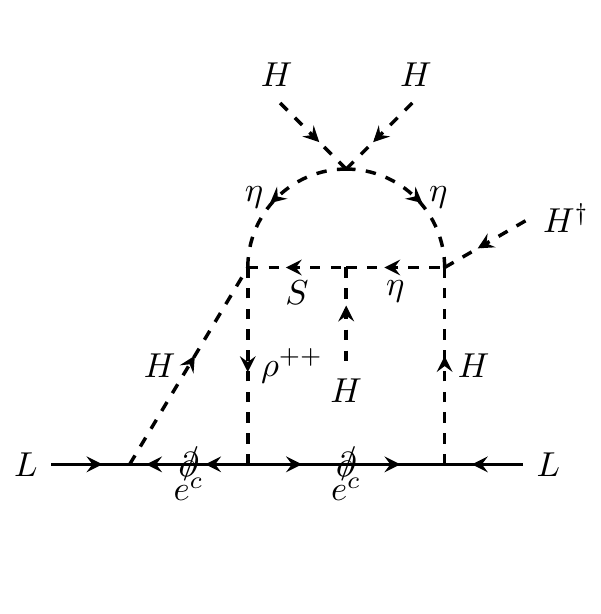}
    \includegraphics[width=0.4\textwidth]{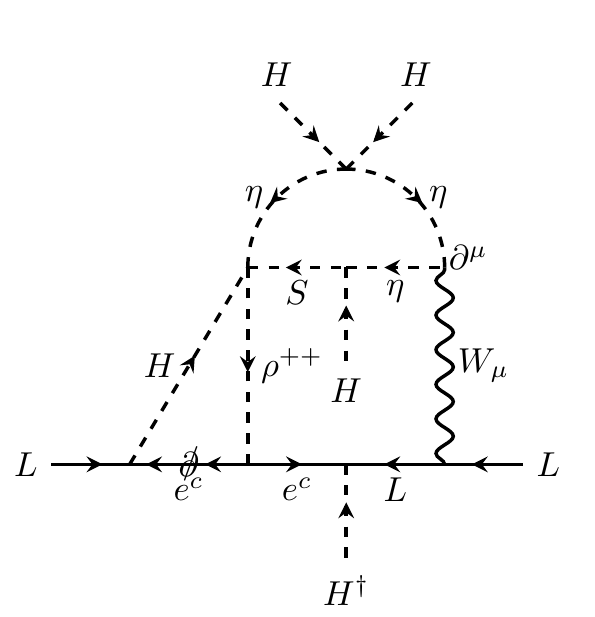}
    \caption{Dimension $7$ mass diagrams in the gauge basis. $W_\mu$ couples with a derivative to the scalars, so every diagram has the same number of derivatives. We will use the notation $\mathfrak{I}^{(7)}_i$ with $i=1,4$ following the usual order from left to right and top to bottom.}
    \label{fig:app:loops:CT_diag_d7}
\end{figure}

\begin{figure}[t]
    \centering
    \includegraphics[width=0.32\textwidth]{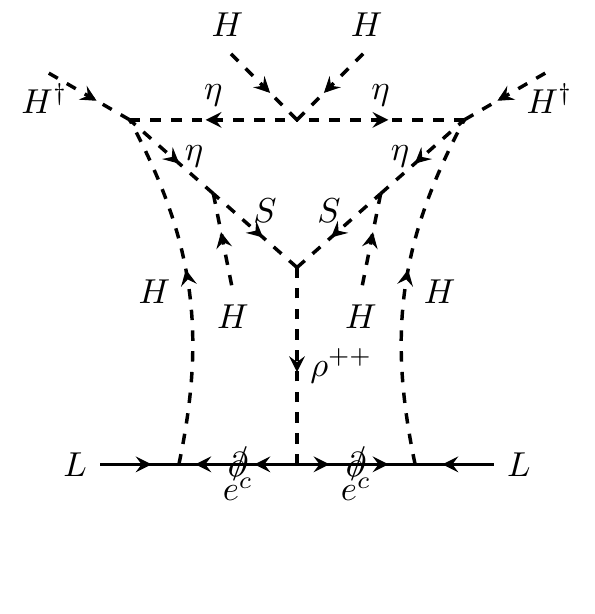}
    \includegraphics[width=0.32\textwidth]{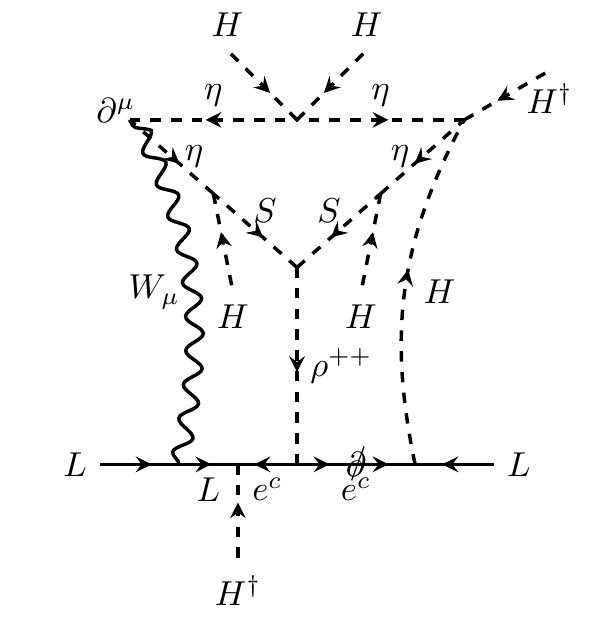}
    \includegraphics[width=0.32\textwidth]{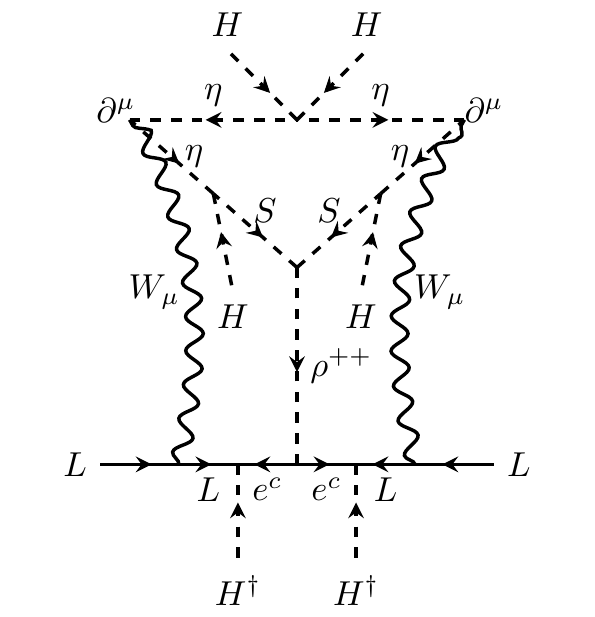}
    \\
    \includegraphics[width=0.32\textwidth]{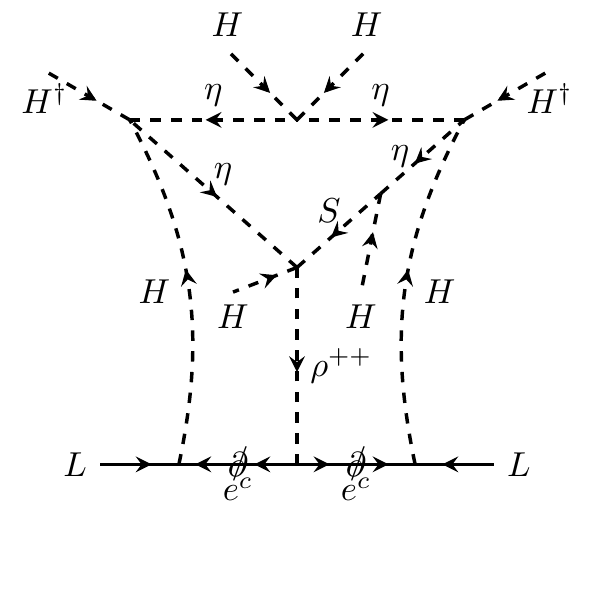}
    \includegraphics[width=0.32\textwidth]{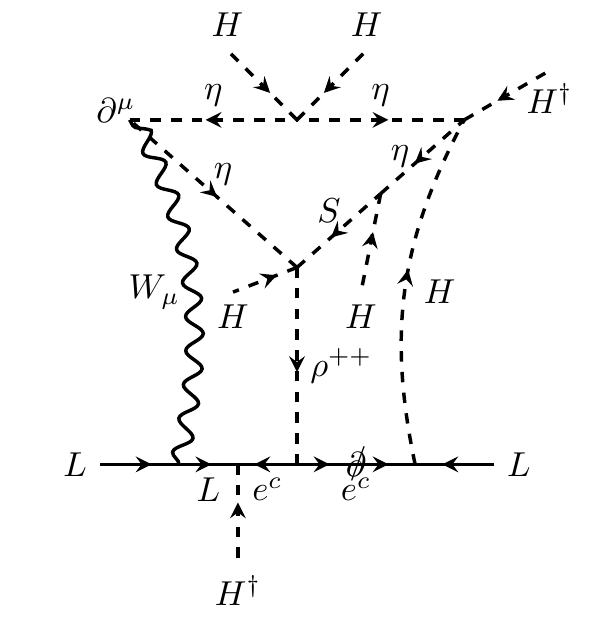}
    \includegraphics[width=0.32\textwidth]{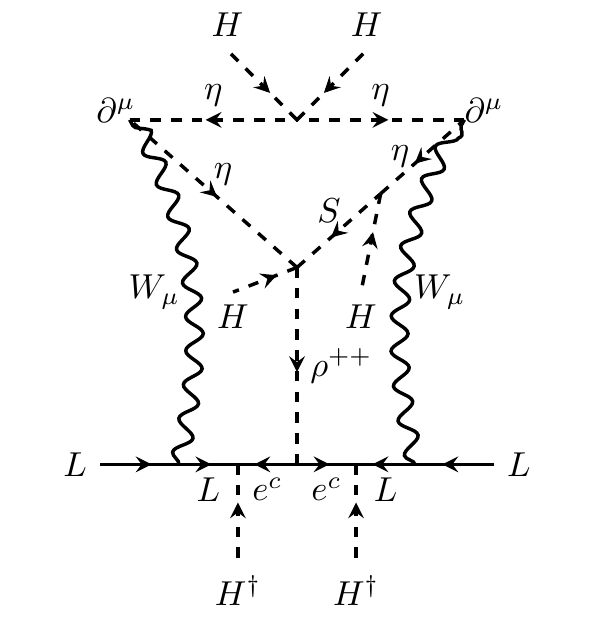}
    \caption{Dimension $9$ mass diagrams in the Feynman-'t Hooft gauge. Denoted as $\mathfrak{I}^{(9)}_i$ with $i=1,6$ when required, following the standard ordering (left to right and top to bottom).}
      \label{fig:app:loops:CT_diag_d9}
\end{figure}

\begin{figure}[t]
    \centering
    \includegraphics[width=0.4\textwidth]{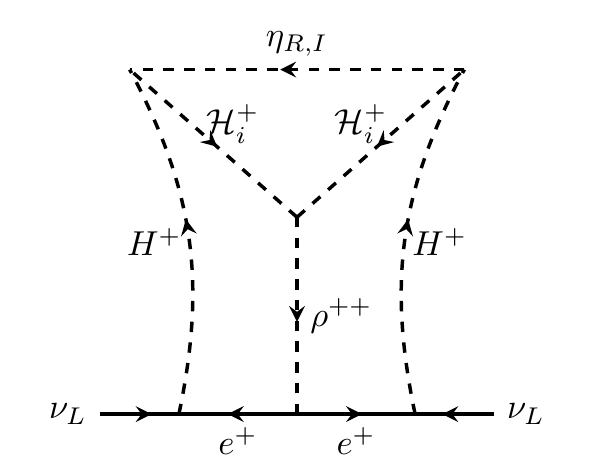}
    \caption{Mass diagram after EWSB in the Feynman-'t Hooft gauge. $H^+$ is the Goldstone boson associated to $W_\mu$ with mass $m_W$.}
    \label{fig:app:loops:CT_diag_mass}
\end{figure}

We shall show in detail how we derived the integral of the first diagram in \fig{fig:app:loops:CT_diag_d9} as an example, denoted as $\mathfrak{I}^{(9)}_1$, and give just the results for the rest. We chose this diagram as it gives a similar prefactor as in the original work \cite{Gustafsson:2012vj}. After electroweak symmetry breaking (EWSB), we rotate the diagram to the mass basis, see \fig{fig:app:loops:CT_diag_mass}. $H^+$ is the Goldstone boson associated to $W_\mu$, which appears explicitly with mass $m_W$ in the Feynman-'t Hooft gauge. $\mathcal{H}^+$ are the two mass eigenstates with eigenvalues $m_+^2$ coming from the mixing of $S^+$ and $\eta^+$,
\begin{equation} %\label{eq:}
    \mathcal{M}^2_{\mathcal{H}^+} = \left( \begin{array}{cc}
        M_S^2 + \frac 12 \lambda_{SH} v^2  &  \frac{1}{\sqrt{2}} \mu_1 v 
        \\
        \frac{1}{\sqrt{2}} \mu_1 v  &  M_\eta^2 + \frac 12 \left(\lambda_{\eta H}^{(1)} + \lambda_{\eta H}^{(3)} \right) v^2
\end{array} \right) \,,
\end{equation}
which can be trivially diagonalised by a $2 \times 2$ rotation matrix $R_{\mathcal{H}^+}$ with angle $\beta$. $\eta_{R,I}$ are the CP-even and CP-odd components of $\eta^0$, with masses $m_{R,I}^2 = M_\eta^2 \mp \frac 12 \lambda_5 v^2$.

Defining $\int_k \equiv (16\pi^2) \int d^4 k /(2\pi)^3$ and assigning momenta in the loop, the integral of the diagram in \fig{fig:app:loops:CT_diag_mass} in the mass insertion approximation is given by,
\begin{equation} \label{eq:app:loops:I9}
    \mathcal{I}^{(9)}_1 =  (R_{\mathcal{H}^+})_{1i} (R_{\mathcal{H}^+})_{2i} (R_{\mathcal{H}^+})_{1j} (R_{\mathcal{H}^+})_{2j} \iiint\limits_{k_1\, k_2 \, k_3} \, \frac{k_1 \cdot k_2}{ \mathcal{D}_1^{(9)} } \, ,
\end{equation}
with
\begin{eqnarray} %\label{eq:}
    \mathcal{D}_1^{(9)} &=& (k_1^2) (k_2^2) (k_1^2+m_W^2) (k_2^2+m_W^2)\, \times
    \\ \nn
    && ((k_1+k_2)^2+m_{\rho^{++}}^2)  (k_3^2 + m_{a}^2) ((k_1+k_3)^2 + m_{+i}^2) ((k_2-k_3)^2 + m_{+j}^2) \, ,
\end{eqnarray}
where $a=R,I$ and we have neglected the masses of the charged SM fermions. The sum over free indices can be explicitly done enlarging the denominator of the integral. For example,
\begin{equation} %\label{eq:}
    \sum_{a=1}^2 \frac{1}{k^2-m_a^2} = (m_1^2 - m_2^2) \frac{1}{(k^2-m_1^2)(k^2-m_2^2)} \, .
\end{equation}
Defining $\Delta m_0^2 = m_R^2- m_I^2$ and $\Delta m_+^2 = m_{+1}^2- m_{+2}^2$, \eq{eq:app:loops:I9} can be written as,
\begin{equation} %\label{eq:}
    \mathcal{I}^{(9)}_1 = \frac 14 \sin^2\! 2\beta \, \Delta m_0^2 \, (\Delta m_+^2)^2 \frac{1}{m_{\rho^{++}}^8} \, \widehat{\mathcal{I}}^{(1)}_1 \, ,
\end{equation}
with $\widehat{\mathcal{I}}^{(1)}_1$ a dimensionless integral defined in \eq{eq:app:loops:CT_integrals}, which depends only on mass ratios with $m_{\rho^{++}}$,
\begin{equation} \label{eq:app:loops:CT_massratios}
    x_W = \frac{m_W^2}{m_{\rho^{++}}^2}, \quad x_R = \frac{m_R^2}{m_{\rho^{++}}^2}, \quad x_I = \frac{m_I^2}{m_{\rho^{++}}^2}, \quad x_1 = \frac{m_{+1}^2}{m_{\rho^{++}}^2}, \quad x_2 = \frac{m_{+2}^2}{m_{\rho^{++}}^2} \, .
\end{equation}

Finally, including the corresponding couplings from the potential \eq{eq:clfv:PotCocktail}, the expression for the diagram in \fig{fig:app:loops:CT_diag_mass} reads,
\begin{equation} %\label{eq:}
    \mathfrak{I}^{(9)}_1 = \frac{1}{4} {\lambda_{\eta H}^{(3)}}^2 \, \sin^2\! 2\beta \, \frac{ \mu_2 \,  \Delta m_0^2 \, (\Delta m_+^2)^2}{m_{\rho^{++}}^8} \, \widehat{\mathcal{I}}^{(1)}_1 \, ,
\end{equation}
where the Yukawa $h$ and the SM charged fermion masses have been omitted.

The computation of the rest of the diagrams in \figs{fig:app:loops:CT_diag_d5}{fig:app:loops:CT_diag_d9} is very similar to the example shown. We only give the results here and omit their calculation. The function $F_{\rm Cocktail}$ in \eqref{eq:clfv:mnucocktail} is given by the sum of the different contributions from the diagrams, i.e.
\begin{equation} %\label{eq:}
    F_{\rm Cocktail} = \frac{m_{\rho^{++}}}{\lambda_5} \sum_{d,i} \, \mathfrak{I}^{(d)}_i \, ,
\end{equation}
with $d=5,7,9$ using the notation in \figs{fig:app:loops:CT_diag_d5}{fig:app:loops:CT_diag_d9}. The prefactor originates from the normalisation of \eq{eq:clfv:mnucocktail}. The corresponding $12$ contributions from each diagram are
\begingroup
\allowdisplaybreaks
\begin{eqnarray*} %\label{eq:}
    \mathfrak{I}^{(5)}_1 &=& 2 \, \frac{\mu_1^2 \, \mu_2 \, \Delta m_0^2}{v^2 m_{\rho^{++}}^4 } \left[ \cos^4\!\beta \, \frac{(\Delta m_+^2)^2}{m_{\rho^{++}}^4} \, \widehat{\mathcal{I}}^{(1)}_1 \, + 2 \cos 2\beta \, \frac{\Delta m_+^2}{m_{\rho^{++}}^2} \, \widehat{\mathcal{I}}^{(1)}_2 \, + \, \widehat{\mathcal{I}}^{(1)}_3 \right] ,
    \nn \\ \nn
    \mathfrak{I}^{(5)}_2 &=& 4 \, \kappa \, \frac{\mu_1 \, \Delta m_0^2}{v^2 m_{\rho^{++}}^2 } \left[ \cos^2\!\beta \, \frac{\Delta m_+^2}{m_{\rho^{++}}^2} \, \widehat{\mathcal{I}}^{(1)}_4 \, + \, \widehat{\mathcal{I}}^{(1)}_5 \right] ,
    \\\nn
    \mathfrak{I}^{(7)}_1 &=& \sqrt{2} \, \lambda_{\eta H}^{(3)} \, \frac{\mu_1 \, \mu_2 \, \Delta m_0^2 \, \Delta m_+^2}{v \, m_{\rho^{++}}^6 } \, \sin\!2\beta \, \left[ \cos\!\beta \, \sin\!2\beta \, \frac{\Delta m_+^2}{m_{\rho^{++}}^2} \, \widehat{\mathcal{I}}^{(1)}_1 \, + \, \widehat{\mathcal{I}}^{(1)}_2 \right] ,
    \\\nn
    \mathfrak{I}^{(7)}_2 &=& - \frac 12 \, g_2^2 \, \frac{\mu_1 \, \mu_2 \, \Delta m_0^2 \, \Delta m_+^2}{v \, m_{\rho^{++}}^6 } \, \sin\!2\beta \, \left[ \cos\!\beta \, \sin\!2\beta \, \frac{\Delta m_+^2}{m_{\rho^{++}}^2} \, \widehat{\mathcal{I}}^{(1)}_1 \, + \, \widehat{\mathcal{I}}^{(1)}_2 \right] ,
    \\\nn
    \mathfrak{I}^{(7)}_3 &=& \sqrt{2} \, \lambda_{\eta H}^{(3)} \, \kappa \, \frac{\Delta m_0^2 \, \Delta m_+^2}{v\, m_{\rho^{++}}^4} \, \sin 2\beta \, \widehat{\mathcal{I}}^{(1)}_4 \, ,
    \\\nn
    \mathfrak{I}^{(7)}_4 &=& - \frac 12 \, g_2^2 \, \kappa \, \frac{\Delta m_0^2 \, \Delta m_+^2}{v\, m_{\rho^{++}}^4} \, \sin 2\beta \, \widehat{\mathcal{I}}^{(2)}_4 \, ,
    \\\nn
    \mathfrak{I}^{(9)}_1 &=& \frac 14 {\lambda_{\eta H}^{(3)}}^2 \, \frac{ \mu_2 \,  \Delta m_0^2 \, (\Delta m_+^2)^2}{m_{\rho^{++}}^8} \, \sin^2\! 2\beta \, \widehat{\mathcal{I}}^{(1)}_1 \, ,
    \\\nn
    \mathfrak{I}^{(9)}_2 &=& - \frac{1}{4\sqrt{2}} \, g_2^2 \, \lambda_{\eta H}^{(3)} \, \frac{ \mu_2 \, \Delta m_0^2 \, (\Delta m_+^2)^2}{m_{\rho^{++}}^8} \, \sin^2\! 2\beta \, \widehat{\mathcal{I}}^{(2)}_1 \, ,
    \\\nn
    \mathfrak{I}^{(9)}_3 &=& \frac{1}{32} \, g_2^4 \, \frac{ \mu_2 \,  \Delta m_0^2 \, (\Delta m_+^2)^2}{m_{\rho^{++}}^8} \, \sin^2\! 2\beta \, \widehat{\mathcal{I}}^{(3)}_1 \, ,
    \\\nn
    \mathfrak{I}^{(9)}_4 &=& -\frac{1}{\sqrt{2}} \, {\lambda_{\eta H}^{(3)}}^2 \, \kappa \, \frac{v \, \Delta m_0^2 \, \Delta m_+^2}{m_{\rho^{++}}^6 } \, \sin\!2\beta \, \left[ \cos\!\beta \, \sin\!2\beta \, \frac{\Delta m_+^2}{m_{\rho^{++}}^2} \, \widehat{\mathcal{I}}^{(1)}_1 \, + \, \widehat{\mathcal{I}}^{(1)}_2 \right] ,
    \\\nn
    \mathfrak{I}^{(9)}_5 &=& \frac 14 \, g_2^2 \, \lambda_{\eta H}^{(3)} \, \kappa \, \frac{v \, \Delta m_0^2 \, \Delta m_+^2}{m_{\rho^{++}}^6 } \, \sin\!2\beta \, \left[ \cos\!\beta \, \sin\!2\beta \, \frac{\Delta m_+^2}{m_{\rho^{++}}^2} \, \widehat{\mathcal{I}}^{(1)}_1 \, + \, \widehat{\mathcal{I}}^{(1)}_2 \right] ,
    \\\nn
    \mathfrak{I}^{(9)}_6 &=& -\frac{1}{4\sqrt{2}} \, g_2^4 \, \kappa \, \frac{v \, \Delta m_0^2 \, \Delta m_+^2}{m_{\rho^{++}}^6 } \, \sin\!2\beta \, \left[ \cos\!\beta \, \sin\!2\beta \, \frac{\Delta m_+^2}{m_{\rho^{++}}^2} \, \widehat{\mathcal{I}}^{(1)}_1 \, + \, \widehat{\mathcal{I}}^{(1)}_2 \right] ,
\end{eqnarray*}
\endgroup
For simplicity, we introduced the notation,
\begin{equation} \label{eq:app:loops:CT_integrals}
    \widehat{\mathcal{I}}^{(a)}_i = \iiint\limits_{k_1\, k_2 \, k_3} \, \frac{ \mathcal{N}_a }{\mathcal{D}_i} \, ,
\end{equation}
where each numerator, associated to the derivatives depicted in the diagrams, is defined as
\begin{eqnarray} %\label{eq:}
    \mathcal{N}_1 &=& k_1 \cdot k_2 \, ,
    \nonumber
    \\
    \mathcal{N}_2 &=& k_2 \cdot (2 k_3 + k_1 ) \, ,
    \\ \nonumber
    \mathcal{N}_3 &=& (2 k_3 + k_1) \cdot (2 k_3 + k_2 ) \, ,
\end{eqnarray}
while for the denominators,
\begin{eqnarray} %\label{eq:}
    \mathcal{D}_0 &=& (k_1^2) (k_2^2) (k_1^2+x_W) (k_2^2+x_W) ((k_1+k_2)^2+1) (k_3^2 + x_R) (k_3^2 + x_I) \, ,
    \nonumber
    \\ \nonumber
    \mathcal{D}_1 &=& \mathcal{D}_0 \times ((k_1+k_3)^2 + x_1) ((k_2-k_3)^2 + x_1) ((k_1+k_3)^2 + x_2) ((k_2-k_3)^2 + x_2) \, ,
    \\ \nonumber
    \mathcal{D}_2 &=& \mathcal{D}_0 \times ((k_1+k_3)^2 + x_1) ((k_1+k_3)^2 + x_2) ((k_2-k_3)^2 + x_2) \, ,
    \\ 
    \mathcal{D}_3 &=& \mathcal{D}_0 \times ((k_1+k_3)^2 + x_1) ((k_2-k_3)^2 + x_2) \, ,
    \\ \nonumber
    \mathcal{D}_4 &=& \mathcal{D}_0 \times ((k_2-k_3)^2 + x_1) ((k_2-k_3)^2 + x_2) \, ,
    \\ \nonumber
    \mathcal{D}_5 &=& \mathcal{D}_0 \times ((k_2-k_3)^2 + x_2) \, ,
\end{eqnarray}
with the mass ratios $x$ defined in \eq{eq:app:loops:CT_massratios}. The integrals in \eq{eq:app:loops:CT_integrals} are evaluated numerically using {\tt pySecDec}.  \\

We now discuss the maximisation of $F_{\rm Cocktail}$. We are interested in this case, because the Yukawa $h$ in \eq{eq:clfv:hfit} is inversely proportional to $F_{\rm Cocktail}$, and we want to explore the parameter space where $h$ is small enough to be perturbative and avoid CLFV constraints. In general, every integral $\widehat{\mathcal{I}}^{(a)}_i$ gets larger for smaller masses or, equivalently, smaller ratios. We shall fix a limit of $100$ GeV on the scalar masses of $\eta_{R,I}$ and $\mathcal{H}^+$, and $800$ GeV for the doubly charged singlet $\rho^{++}$, see \sect{sec:clfv:cocktail} for details. We set the dimensionless couplings $\lambda_5$, $\lambda_{\eta H}^{(3)}$, and $\kappa$ to $4\pi$. For the dimensionful $\mu$ couplings we impose the limits $\mu_1 < 4 \, \text{max}[m_{+1}, m_{+2}]$ and $\mu_2 < 4 \, \text{max}[m_{+1}, m_{+2}, m_{\rho^{++}}]$, required to avoid the radiative generation of negative quartic scalar couplings \cite{Babu:2002uu}.

We found the maximum value of $F_{\rm Cocktail}^{\rm max} \simeq 192$ for $m_{R} = m_{+1} = 100$ GeV, $m_{\rho^{++}} = 800$ GeV, $m_{I} = 878$ GeV, and $m_{+2} = 1237$ GeV, with maximal mixing angle $\beta=\pi/4$. $\mu_1 = (\Delta m_+^2)/\sqrt{2} v = 4372$ GeV, while $\mu_2$ is simply $4\,m_{+2}$.

%%%%%%%%%%%%%%%%%%%%%%%%%%%%%%%%%%%%%%%%%%%%%%%%%%%%%%%%%%%%%%%
\subsection{KNT model}

\begin{figure}
    \centering
    \includegraphics[width=0.5\textwidth]{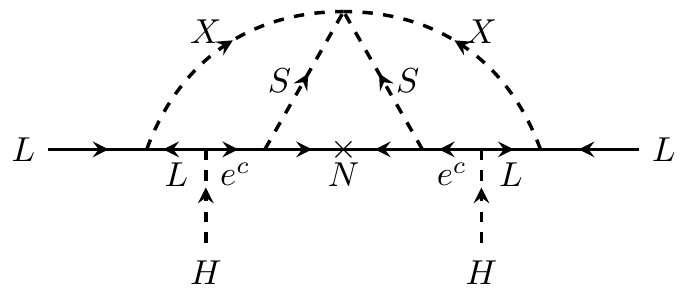}
    \caption{Neutrino mass diagram for the KNT model in the gauge basis.}
     \label{fig:app:loops:knt_diag}
\end{figure}

The computation of the mass diagram for the KNT model is much simpler than the one in the cocktail model. The main contribution to the neutrino mass comes only from the diagram in \fig{fig:app:loops:knt_diag} in the electroweak symmetric basis. Moreover, there is no mixing between the scalars participating in the loop, so $F_{\rm KNT}$ in \eq{eq:clfv:MnuKNT} is just the three-loop integral of the diagram in the mass basis shown in \fig{fig:clfv:KNT}. Neglecting the SM charged fermion masses, one finds 
\begin{equation} \label{eq:app:loops:FKNT}
\medmath{
    F_{\rm KNT} = \iiint\limits_{k_1\, k_2 \, k_3} \frac{1}{ (k_1^2) (k_2^2) (k_1^2+x_1) (k_2^2+x_1) (k_3^2 + 1) ((k_1-k_3)^2 + x_2) ((k_2-k_3)^2 + x_2) } \, , 
    }
\end{equation}
which is a dimensionless function of the ratios,
\begin{equation} %\label{eq:}
    x_1 = \frac{m_{s_1}^2}{M_{N_i}^2} \, , \quad x_2 = \frac{m_{s_2}^2}{M_{N_i}^2} \, .
\end{equation}
Note that \eq{eq:app:loops:FKNT} is simple enough to be easily decomposed in terms of three-loop master integrals \cite{Martin:2016bgz}. Due to the repetitions of the momenta in the denominator, using relation (22) from \cite{Cepedello:2018rfh}, one has
\begin{multline}
    F_{\rm KNT} = \iiint\limits_{k_1\, k_2\, k_3} \frac{1}{ (k_3^2-1) ((k_1-k_3)^2-x_2) ((k_2-k_3)^2-x_2) } \,\, \times
    \\
    \frac{1}{x_1} \left[ \frac{1}{ (k_1^2-x_1) (k_2^2-x_1) } - \frac{1}{ (k_1^2) (k_2^2-x_1) } - \frac{1}{ (k_1^2-x_1) (k_2^2) } + \frac{1}{ (k_1^2) (k_2^2) } \right] .
\end{multline}
As the second and third terms are identical under the exchange of $k_1$ and $k_2$, one can finally write $F_{\rm KNT}$ in terms of a combination of the master integral $\mathbf{G}$ integral given in \cite{Martin:2016bgz}. The resulting expression is
\begin{equation}
    F_{\rm KNT} = \frac{1}{x_1} \left[ \mathbf{G}(1,x_1,x_2,x_1,x_2) - 2 \, \mathbf{G}(1,x_1,x_2,0,x_2) + \mathbf{G}(1,0,x_2,0,x_2) \right].
\end{equation}
The integral has an analytical expression for $x_{1i}=x_{2i}=1$.
\\

About the maximum value of $F_{\rm KNT}$, we proceeded analogously to the cocktail model. In this case, we maximised $F_{\rm KNT} / M_{N_i}$, since the neutrino mass matrix in \eq{eq:clfv:MnuKNT} is proportional to this ratio. We set a lower limit on the mass of the singly charged scalars of $100$ GeV and let $M_{N_i} = M_N$ free. We found that the maximum is around $F_{\rm KNT} \simeq 60$ with $m_{S_1} = m_{S_2} = 100$ GeV and $M_{N_i} = 840$ GeV.

%%%%%%%%%%%%%%%%%%%%%%%%%%%%%%%%%%%%%%%%%%%%%%%%%%%%%%%%%%%%%%%
\subsection{AKS model}

\begin{figure}
    \centering
    \includegraphics[width=0.45\textwidth]{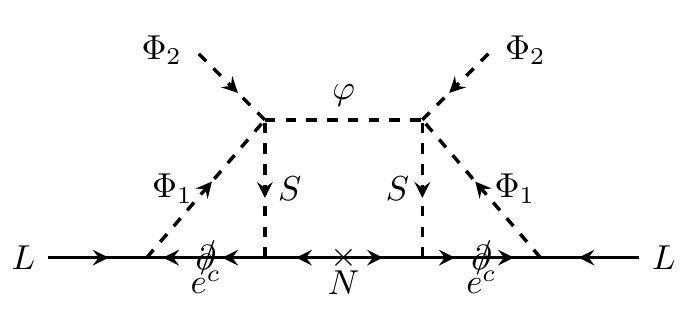}
    \quad
    \includegraphics[width=0.45\textwidth]{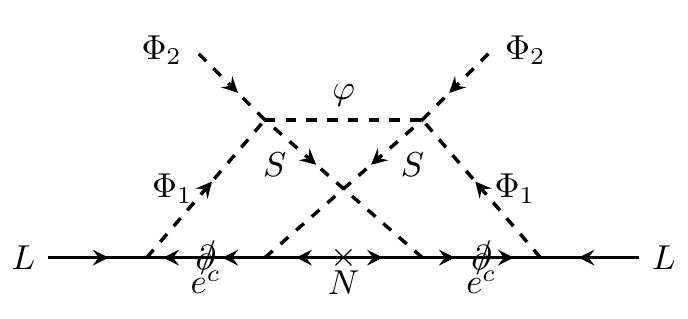}
    \caption{Neutrino mass diagrams for the AKS model in the gauge basis.}
    \label{fig:app:loops:aks_diag}
\end{figure}

In this case, there exists two non-equivalent diagrams shown in \fig{fig:app:loops:aks_diag}, which differ by the crossing of the internal $S$-lines. $F_{\rm AKS}$ is then the sum of the integrals from both diagrams with the correct normalisation, given in \eq{eq:clfv:MnuAKS},
\begin{equation} %\label{eq:}
    F_{\rm AKS} = \mathfrak{I}_1 \, + \, \mathfrak{I}_2 \, .
\end{equation}
By assigning momenta to the internal fields, the two dimensionless integrals can be compactly expressed as,
\begin{equation} %\label{eq:}
    \mathfrak{I}_i = \iiint\limits_{k_1\, k_2\, k_3} \frac{k_1 \cdot k_2}{\mathcal{D}_i} \, ,
\end{equation}
with the denominators,
\begin{eqnarray} %\label{eq:}
    \mathcal{D}_0 &=& (k_1^2) (k_1^2-x_1) (k_2^2) (k_2^2-x_1) (k_3^2-1) ((k_1-k_3)^2-x_S) ((k_2+k_3)^2-x_S) \, ,
    \nonumber\\
    \mathcal{D}_1 &=& \mathcal{D}_0 \, \times \, (k_3^2-x_\varphi) \, ,
    \\ \nonumber
    \mathcal{D}_2 &=& \mathcal{D}_0 \, \times \, ( (k1+k_2+k_3)^2-x_\varphi) \, ,
\end{eqnarray}
where we have neglected the SM charged fermion masses. The ratios of masses are then defined as
\begin{equation} %\label{eq:}
    x_1 = \frac{m_\mathcal{H}^2}{M_{N_i}^2} \, , \quad x_\varphi = \frac{m_\varphi^2}{M_{N_i}^2} \, , \quad x_S = \frac{m_{S^+}^2}{M_{N_i}^2} \,.
\end{equation}
\\

Similar to the previous models, we computed the maximum of the function $F_{\rm AKS}/M_{N_i}$ to minimise the absolute scale of the Yukawa $Y$. We considered $M_{N_i} = m_N$ and set a lower limit of $100$ GeV to the scalar masses. We found the maximum for $m_N = 272$ GeV and $m_\mathcal{H} = m_\varphi = m_S = 100$ GeV where $F_{\rm AKS} \simeq 0.45$.

\pagebreak
\fancyhf{}

%% file: Resumen_Tesis/Resumen_Tesis.tex
\fancyhf{}
\fancyhead[LE,RO]{\thepage}
\fancyhead[RE]{\slshape{Resumen de la Tesis}}
%\fancyhead[LO]{\slshape\nouppercase{\rightmark}}

\setcounter{figure}{0}
\setcounter{table}{0}
\renewcommand\figurename{Figura}
\renewcommand\tablename{Tabla}
\renewcommand\thefigure{\Roman{figure}}
\renewcommand\thetable{I}

\chapter*{Resumen de la Tesis}
\chaptermark{Resumen de la Tesis}
\addcontentsline{toc}{chapter}{Resumen de la Tesis}  

%To be done [... entre 4000 y 8000 palabras, aprox 12 páginas ...]

La observación de las oscilaciones de neutrinos ha supuesto la confirmación de que los neutrinos tienen masa, lo cual constituye la evidencia más clara de física más allá del Modelo Estándar. Las oscilaciones se observan debido a un desajuste entre los autoestados de masa y electrodébiles o de sabor, i.e. los autoestados electrodébiles son una superposición de autoestados de masa (o viceversa). Un neutrino de un sabor específico, al propagarse, oscila entre los distintos sabores con una probabilidad que depende de la energía del neutrino, la distancia recorrida, la diferencia entre los estados de masa y los elementos de la matriz que relaciona la base electrodébil con la base de masa, la matriz de mezcla $U$. Todos estos parámetros son medidos en experimentos (ver Tabla~\ref{tab:nuphys:globalfit}), algunos con una precisión notable, sin embargo sigue habiendo muchas incógnitas en la física de neutrinos. Desconocemos que valor tienen las masas de los neutrinos, solo tenemos cotas superiores a su masa o la suma de sus masas, no sabemos como están ordenadas estas masas e, incluso, ignoramos por completo como estas se generan o si los neutrinos son partículas de Dirac o Majorana.

El tema principal de esta tesis es el estudio de los modelos radiativos de masa de neutrinos. En estos modelos las masas de los neutrinos están prohibidas a nivel árbol, pero permitidas a un cierto orden en bucles (loops). Lo que se consigue con esto es explicar de una forma natural la pequeñez de las masas de los neutrinos comparadas con las masas de los otros fermiones del Modelo Estándar. Esto permite disminuir la escala de energía a la que aparece la nueva física llegando a rangos de energías observables para los experimentos actuales y futuros. Además estos modelos permiten incorporar fácilmente sectores con candidatos de materia oscura con una rica fenomenología, lo que ha atraído mucha atención sobre estos modelos.

Los modelos de masa de neutrinos de Majorana, ya sean a nivel árbol o radiativos, se basan mayormente en el operador de Weinberg $\mathcal{O}_W$, o en versiones extendidas del mismo de mayor dimensión, como $\mathcal{O}^d \, = \, \mathcal{O}_W \, \times \, (H^\dagger H)^{\frac{d-5}{2}}$. A nivel árbol el operador de Weinberg se puede generar en tres modelos mínimos, los conocidos \textit{seesaws} de tipo-I, II y III. Estos introducen, respectivamente, un singlete fermiónico, un triplete escalar y un triplete fermiónico. En cada modelo, la pequeñez de las masas de los neutrinos se debe, ora a la elevada escala de masa de los fermiones, ora al pequeño valor esperado en el vacío (VEV) del triplete escalar. Realizaciones más complejas del operador de Weinberg existen a nivel árbol con un interés ya sea fenomenológico o teórico. Por ejemplo, los \textit{seesaws} inverso y lineal son una extensión del \textit{seesaws} de tipo-I, donde se les añade un singlete fermiónico, de tal forma que se pueda explicar la pequeñez de las masas de los neutrinos sin recurrir a una escala muy elevada de nueva física. Así mismo existen ejemplos de realizaciones del operador $\mathcal{O}^d$ con $d=7,9,11,13$ en la literatura, donde se recurre a operadores dimensionalmente mayores para explicar el tamaño de las masas de los neutrinos. En analogía a los \textit{seesaws} explicados, recientemente se han propuesto varios modelos de masa para neutrinos de Dirac con mecanismos similares a los tres \textit{seesaws}. En estos modelos se genera una extensión del operador $\bar LH\nu_R$ añadiendo escalares cuyos VEVs rompen número leptónico en un número de unidades que no permite términos de Majorana ($\Delta L =2$) para los neutrinos.

Como se ha mencionado antes, otra posible forma de explicar la pequeñez de las masas de los neutrinos es mediante modelos radiativos. Cada loop añade una supresión de $1/16\pi^2$, lo que permite reducir la escala de nueva física. En la literatura existen ejemplos de modelos con varios loops, principalmente aprovechando la libertad de los mismos para explicar otros problemas del Modelo Estándar, como la materia oscura. A su vez, se han realizado clasificaciones de todos los diagramas/modelos que se pueden generar a partir de $\mathcal{O}_W$ y $\mathcal{O}^d$ para ciertas dimensiones $d$ y número de loops. Existen clasificaciones para el operador de Weinberg a uno y dos loops, mientras que para $d>5$ solo se han clasificado los operadores a nivel árbol. Igualmente, para neutrinos de Dirac algunas clasificaciones existen pero solo abarcan modelos hasta un loop. La idea principal de las clasificaciones es agrupar topologías o diagramas que generen modelos con características similares, ya sean debido a que dan una masa de neutrinos de un orden similar, a su fenomenología o a su contenido de partículas. Un concepto esencial en las clasificaciones es la de topologías o diagramas genuinos, entendido como la topología o diagrama de dimensión $d$ con $\ell$ loops que genera al menos un modelo que es la contribución dominante a la masa de los neutrinos con dimensión $d$ y $\ell$ loops, sin necesidad de una simetría adicional.

\section*{Clasificaciones de modelos radiativos}

\begin{figure}[h!]
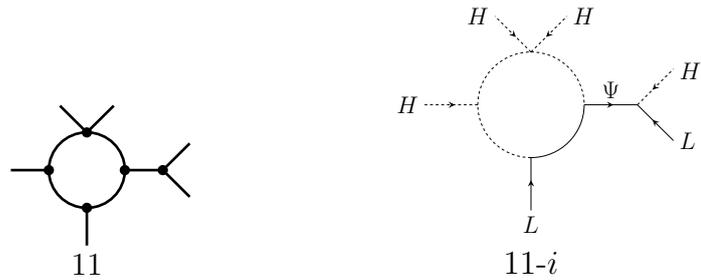

    \centering 
    \includegraphics{Chapters/Dim7_1loop/figures/T11.pdf}
    \qquad \qquad
    \includegraphics[scale=0.35]{Chapters/Dim7_1loop/figures/diagram_T11-i.pdf}
    \caption{Única topología (izq.) y diagrama (der.) para los cuales un modelo genuino existe sin representaciones mayores a tripletes. El resto de modelos generados de otras topologías y diagramas necesitan al menos un cuadruplete para ser genuinos.}
    \label{fig:resumen:Triplet}
\end{figure}

En el Capítulo \ref{ch:Dim7_1loop} hemos llevado a cabo la clasificación de las realizaciones del operador de Weinberg extendido de dimensión $7$ ($\mathcal{O}^{d=7}$) a un loop. Hemos construido todas las topologías y diagramas, organizándolas de acuerdo a su contenido en partículas. De un conjunto inicial de $48$ topologías, solo $8$ de ellas conducen a modelos genuinos de masas de neutrinos a nivel $d = 7$ un loop. Las $40$ topologías restantes son realizaciones no renormalizables, correcciones a algún otro operador (contribuciones con loops infinitos), o generan un operador de orden inferior, por ejemplo $d=5$ (Weinberg) a un loop o nivel árbol, que serían contribuciones dominantes a las masas de los neutrinos. Después hemos generado todos los diagramas posibles insertando fermiones y escalares, de tal forma que el diagrama resultante tuviese dos fermiones y dos escalares externos. Analizando el conjunto de diagramas, hemos encontrado que algunos de los diagramas, para ser genuinos, todos los modelos generados contienen ciertas representaciones de ${\rm SU(2)_L}$. Esto se debe a que el operador $\mathcal{O}^{d=7}$ contiene ($H^\dagger H$), término que hay que prevenir que se acople como un singlete, ya que entonces $\mathcal{O}^{d=7}$ se reduciría al operador Weinberg. Para ello hay que recurrir a representaciones por encima del doblete, de tal forma que se fuerce a ($H^\dagger H$) a acoplarse como un triplete. De esta forma, se obtienen $23$ diagramas genuinos, de los cuales uno de ellos no contiene ninguna representación mayor a un triplete de ${\rm SU(2)_L}$, mientras que los $22$ diagramas restantes requieren uno o dos cuadrupletes con hipercargas $3/2$ y/o $1/2$ para ser genuinos. A continuación, hemos construido tres modelos de ejemplo con la clasificación: el primero es el modelo más simple que hemos encontrado basado en el diagrama con triplete ($D_{11}^{(i)}$, figura \ref{fig:resumen:Triplet}); el segundo es un modelo que requiere un cuadruplete escalar para ser genuino; y el último un ejemplo de un modelo no mínimo que incluye un quintuplete de ${\rm SU(2)_L}$. Esta clasificación es la primera en la que los modelos de masas de neutrinos se agrupan en clases con un contenido similar de partículas. Esto abre la posibilidad de obtener conclusiones generales válidas, no solo aplicables a un modelo, sino generalizables a un conjunto completo de modelos. Resumiendo, para evitar masas de neutrinos de orden inferior, los modelos genuinos que hemos discutido siempre tienen que introducir al menos cinco nuevos multipletes de ${\rm SU(2)_L}$, generalmente con una hipercarga alta. Aunque estos modelos $d = 7$ son necesariamente construcciones más complicadas que el seesaw estándar, desde un punto de vista fenomenológico son particularmente atractivas. Una característica de los modelos basados en operadores con $d>5$, i.e. $\mathcal{O}^{d}$, es que debido a que contienen ($H^\dagger H$), cada una de estas combinaciones puede cerrarse y generar un loop. Es decir, $\mathcal{O}^{d}$ a $n$-loops, genera $\mathcal{O}^{d-2}$ a $(n-1)$-loops. Se puede calcular que aproximadamente la escala de nueve física tiene que ser menor a $2$ TeV si se quiere que $\mathcal{O}^{d}$ a $n$-loops sea la contribución dominante a la masa de los neutrinos sobre $\mathcal{O}^{d-2}$ a $(n-1)$-loops. Consecuentemente, estos modelos presentan de forma natural una cota superior a la masa de los campos más allá del Modelo Estándar. Si combinamos esto con el hecho de que estos modelos tienen siempre partículas con mucha carga eléctrica y la violación de número leptónico, se pueden esperar procesos interesantes en colisionadores. Esto implica que estos modelos se puede probar fácilmente en colisionadores dado el bajo fondo esperado para este tipo de señales y el límite superior a las masas de $2$ TeV.

\begin{figure}[h!]
    \centering
    \includegraphics[width=0.9\textwidth]{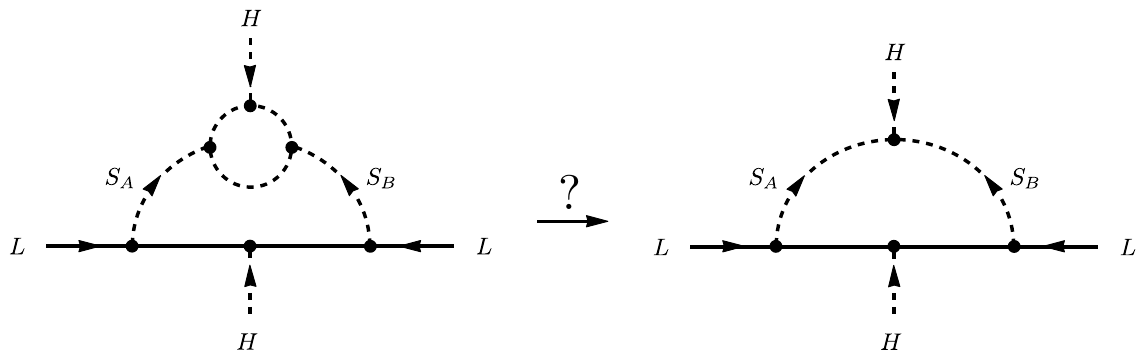}
    \caption{Realización de $\mathcal{O}_W$ a dos loops que ilustra una de las lagunas en el procedimiento de encontrar los diagramas y topologías genuinas. En particular, si uno de los escalares, $S_{A,B}$, es el Higgs, y el otro es un singlete de ${\rm SU(2)}$ con la hipercarga correcta, entonces la interacción puntual $H S_A S_B$ es cero. Por lo tanto, el diagrama de la izquierda no implica que se pueda construir el diagrama de la derecha con un loop menos. Esto es trivialmente generalizable para el caso de tres loops.}
    \label{fig:resumen:exception2}
\end{figure}

\begin{figure}[h!]
    \centering
    \includegraphics[width=1\textwidth]{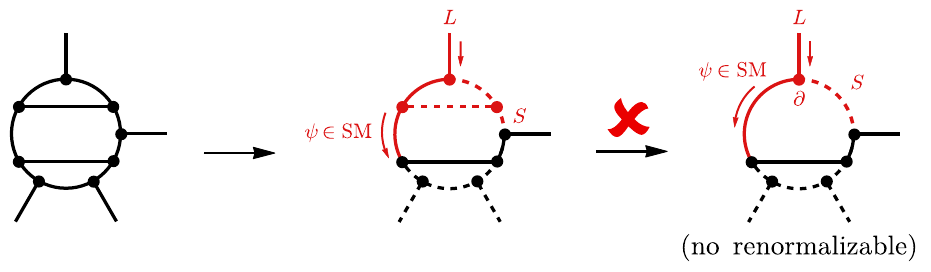}
    \caption{Ejemplo de un diagrama que contiene una interacción efectiva de fermión-fermión-escalar a un loop (señalado en rojo). Este loop es reducible salvo que $\psi$ sea un fermión del Modelo Estándar, en cuyo caso la presencia de una derivada en el propagador de $\psi$ hace que la interacción efectiva sea no renormalizable.}
    \label{fig:resumen:example}
\end{figure}

El Capítulo \ref{ch:3loop} continua con el estudio sistemático de la descomposición del operador de Weinberg a tres loops. Debido a la gran cantidad de topologías, hemos implementado computacionalmente algoritmos conocidos de teoría de grafos para generar la lista de todas las topologías conectadas con tres loops, vértices de 3 y 4 patas y cuatro líneas externas. Análogamente al capítulo anterior, aplicamos varios cortes para obtener una lista de $99$ topologías genuinas, que se pueden dividir en dos conjuntos de acuerdo con su contenido de partículas. Está clasificación se fundamenta en una laguna en el procedimiento \textit{estándar} de clasificaciones anteriores, que se vio por primera vez en esta tesis. A priori, se puede pensar que toda realización de un vértice a loops, si es renormalizable y está permitido por las simetrías, también lo debería estar el correspondiente vértice sin loops, es decir a nivel árbol. Esto implicaría que una topología o diagrama que contuviese una realización a loops de un vértice con tres patas externas o con cuatro patas escalares debería ser reducible y, por lo tanto, no genuino. Este fue el procedimiento adoptado para obtener la lista de topologías genuinas en clasificaciones anteriores. Sin embargo, lo mencionado anteriormente no es correcto para ciertas combinaciones de campos. Debido a la naturaleza antisimétrica de las contracciones de ${\rm SU(2)_L}$, ciertas contracciones se anulan cuando se tienen partículas idénticas. Es el caso de, por ejemplo, dos Higgses a singlete $S$, que resulta en la contracción $H_\alpha \, (i\sigma^2)^{\alpha \beta} \, H_\beta \, S = (H^+ H^0 - H^0 H^+)\, S = 0$, o tres Higgses acoplados a un doblete. Aunque esto es verdadero para el acoplamiento local $HHS$, la interacción no puntual, es decir mediante loops, no se anula (ver figura \ref{fig:resumen:exception2}). Así mismo, la presencia de fermiones sin masa dentro del loop puede evitar que ciertos diagramas sean reducibles. Como estamos trabajando con modelos antes de la rotura espontánea de la simetría, los únicos fermiones sin masa fenomenológicamente viables serían los del Modelo Estándar. Debido a la quiralidad de los fermiones estos contribuyen de forma efectiva con una derivada. Consecuentemente, el acoplamiento, a priori renormalizable y de dimensión $4$, pasa a ser no renormalizable al incluir la derivada y, por lo tanto, no reducible (ver figura \ref{fig:resumen:example}). Estos casos particulares afectan a $55$ topologías que generan $271$ diagramas que son genuinos debido a la laguna que se acaba de explicar. Los llamamos genuinos \textit{especiales} y se pueden clasificar de acuerdo con el contenido específico de partículas requerido para ser genuinos, lo que restringe considerablemente los posibles modelos que se pueden generar. Por otro lado, hay $44$ topologías genuinas \textit{normales} que generan $228$ diagramas que no requieren ninguna partícula específica para ser genuinos (ver resumen en la figura \ref{fig:3loop:summary}). Después de la rotura espontánea de la simetría, cuando el Higgs adquiere un VEV, estos diagramas se reducen a $18$ y $20$ en la base de masa. Todas las integrales de tres loops se pueden entonces escribir en función de $5$ integrales básicas. Finalmente, hemos estimado el rango de parámetros típicos para el cual los modelos de tres loops de dimensión $5$ pueden explicar los datos experimentales de oscilaciones de neutrinos. Descubrimos que pueden ajustar los datos para una escala de nueva física aproximadamente dentro del rango $1 - 10^3$ TeV, parcialmente comprobable en colisionadores actuales y futuros, así como experimentos que buscan violación del sabor leptónico. Por tanto, los modelos de tres loops son construcciones interesantes, ya que son comprobables experimentalmente.

Por último, en lo que respecta a las clasificaciones de modelos radiativos, en el Capítulo \ref{ch:dirac2l} hemos discutido la descomposición completa y la clasificación del operador de masa de neutrinos de Dirac de dimensión $4$, $\bar L H^c \nu_R$, a dos loops. Para generar una masa de Dirac y explicar los datos de oscilaciones de neutrinos es necesario añadir al menos dos copias de neutrinos dextrógiros, $\nu_R$, singletes bajo el grupo gauge del Modelo Estándar. Si no se supone ninguna simetría más allá del Modelo Estándar, esto generaría el seesaw de tipo-I, con una escala de nueva física del orden de $10^{14}$ TeV para acoplamientos de orden $1$, y los neutrinos serían partículas de Majorana. Para reducir esta escala de masas y mantener neutrinos de Dirac se añade una simetría al modelo, comúnmente abeliana discreta, $Z_n$, o continua, $U(1)$, que prohíbe el operador $\bar L H^c \nu_R$ a nivel árbol, pero una vez rota, permite su realización a nivel de loops y, a su vez, protege la naturaleza Dirac de los neutrinos. Siguiendo los procedimientos ya explicados en los casos anteriores, se han identificado $70$ topologías con 3 patas externas, dos loops y vértices con 3 o 4 inserciones. Entre estas topologías, solo 5 satisfacen los criterios de autenticidad explicados en el Capítulo \ref{ch:numass}. Una clara diferencia entre Dirac y Majorana es que los modelos de Dirac siempre necesitan una simetría más allá del Modelo Estándar, lo cual se traduce en que toda topología irreducible (1PI) es genuina ya que siempre existe una simetría que lo permite. Del total de las 5 topologías 1PI se generan 18 diagramas renormalizables, que se pueden clasificar en tres categorías diferentes en función de los requisitos impuestos sobre su posible contenido de partículas, de forma similar a la clasificación anterior del operador de Weinberg a tres loops. La primera clase contiene diagramas que son genuinos en general; los diagramas de la segunda clase contienen una realización a un loop de un vértice escalar-fermión-fermión; y finalmente, los diagramas en la última clase contienen siempre un loop con tres inserciones escalares. Las dos últimas clases contienen realizaciones a un loop que, en principio, son reducibles, sin embargo existen formas de evitarlo. La segunda clase puede ser genuina si se proporciona una simetría que prohíba no solo el diagrama a nivel árbol, sino también el acoplamiento efectivo de fermión-fermión-escalar a un loop, para luego romper suavemente esta simetría permitiendo el diagrama de masas a dos loops. Este procedimiento, aunque parezca general, falla para la tercera categoría, ya que el vértice a prohibir con tres escalares es un término \textit{suave} (dimensión $3$) y, por lo tanto, habría que incluirlo de todas formas una vez rota la simetría para que la teoría fuese consistente. Sin embargo, en estos diagramas se puede usar la antisimetría de las contracciones de ${\rm SU(2)_L}$ para evitar que sean reducibles. Por ello, estos diagramas deben contener siempre un Higgs $H$ y un singlete $(1, 1, −1)$ corriendo en el loop para ser genuinos. Esto limita considerablemente los posibles modelos, tal y como se observa al comprar las tablas dadas en la sección \ref{sec:dirac2l:genmodels} con los posibles campos que pueden ser empleados. Una vez caracterizadas las categorías, hemos construido un modelo de la primera y otro de la tercera clase como ejemplos, y analizado como funcionan las masas de los neutrinos en estos casos. Se muestra que se puede ajustar la escala de masa de neutrinos, suponiéndola en torno a $0.05$ eV, para acoplamientos de orden uno, si la nueva escala física es $ \mathcal{O}(1) $ TeV. A si mismo, se observa como el modelo de ejemplo de la tercera clase tiene un polo en la masa de los neutrinos. Esto se debe, otra vez, a la antisimetría de las contracciones de ${\rm SU(2)_L}$. Las topologías de esta clase contribuyen con dos diagramas a la masa de los neutrinos con signo opuesto, los cuales se anulan entre sí cuando todas las masas que participan en el loop interno son aproximadamente iguales.

\section*{Modelos específicos de generación de masa}

En los Capítulos \ref{ch:dm_DiracMajo} y \ref{ch:loop_seesaw} hemos estudiado un serie de modelos específicos inspirados en las clasificaciones anteriores. Se trata de dos capítulos donde se describen varios mecanismos generales para obtener masas de neutrinos radiativamente con una escala baja de nueva física, y relacionándolos con posibles candidatos de materia oscura.

Normalmente, la conservación o violación del número leptónico está asociada a neutrinos de Dirac o Majorana, respectivamente. Sin embargo, incluso si la simetría leptónica se rompe, los neutrinos pueden ser de Dirac. En el Capítulo \ref{ch:dm_DiracMajo} hemos estudiado cómo el patrón de ruptura del número leptónico determina si los neutrinos son Dirac o Majorana, dependiendo de la simetría residual $Z_n$. Se puede concluir que, si los neutrinos no transforman trivialmente bajo el grupo $Z_n$, estos son partículas de Dirac, salvo que el grupo $Z_n$ contenga un subgrupo $Z_2$ al cual pertenezcan los neutrinos. En este caso, los neutrinos serían partículas de Majorana. Con esta idea se pueden construir modelos radiativos  para Dirac y Majorana con materia oscura estable participando en el loop. Para ello se emplea la simetría de número leptónico, concretamente $U(1)_{B-L}$, prohibiendo la masa a nivel árbol, pero permitiéndola de forma radiativa al romper la simetría. Esta simetría, si no se rompe por completo, dejaría una simetría residual $Z_n$ (con $n\in \mathbb{N}$), que protegería la naturaleza Dirac o Majorana de los neutrinos y que, a su vez, puede usarse para estabilizar un candidato de materia oscura. Hemos construido un marco teórico del que se obtienen estos resultados dando un diagrama genérico de $n$-loops y $d$ dimensiones, y estableciendo ciertas condiciones que los campos y transformaciones deben de cumplir. De esta forma, se puede determinar cual es la simetría residual, y concretar si el modelo produce neutrinos de Dirac o Majorana, y cuales serían los campos que pertenecen al sector de materia oscura. La estabilidad de la materia oscura solo se logra si la simetría remanente $Z_n$ es tal que $n$ no es primo, es decir, que contiene al menos un subgrupo invariante. De esta forma, según como transforme el Modelo Estándar, si bajo el subgrupo o fuera de él, se puede saber que campos sería estables y, por lo tanto, posibles candidatos para materia oscura si son eléctricamente neutros. Finalmente, hemos mostrado dos ejemplos basados en los diagramas mínimos encontrados para cada caso. En el ejemplo de Dirac, el número leptónico se rompe a $Z_6$, de tal forma que el Modelo Estándar transforma bajo el subgrupo $Z_3$ y los campos dentro del loop fuera de dicho subgrupo. Siendo estos últimos posibles candidatos de materia oscura, ya que la simetría protege cualquier decaimiento de uno de los campos internos a solo partículas del Modelo Estándar. Para el ejemplo de Majorana, la simetría residual es $Z_4$, estando los neutrino cargados bajo el subgrupo $Z_2$. Con esto, cualquier campo que transforme como $\pm i$ bajo $Z_4$ no puede decaer solo al Modelo Estándar. El más ligero de estos, si es neutro, es por lo tanto un buen candidato de materia oscura estable a cualquier orden.

\begin{figure}[h!]
    \centering
    \includegraphics[width=0.6\textwidth]{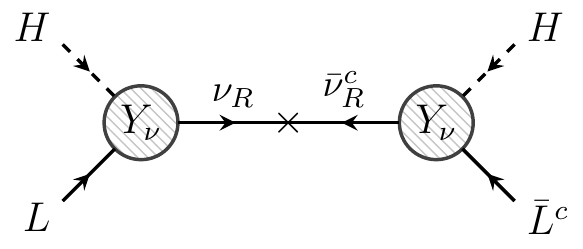}
    \caption{Seesaw efectivo de tipo-I. El Yukawa de Dirac $Y_\nu$ se genera radiativamente, suprimiendo de forma natural la masa de los neutrinos, no solo mediante la masa de Majorana de $\nu_R$, como en el seesaw clásico, sino también con la supresión debida a los factores de loop.}
    \label{fig:resumen:SeesawEff}
\end{figure}

\begin{figure}[h!]
    \centering
    \includegraphics[width=0.47\textwidth]{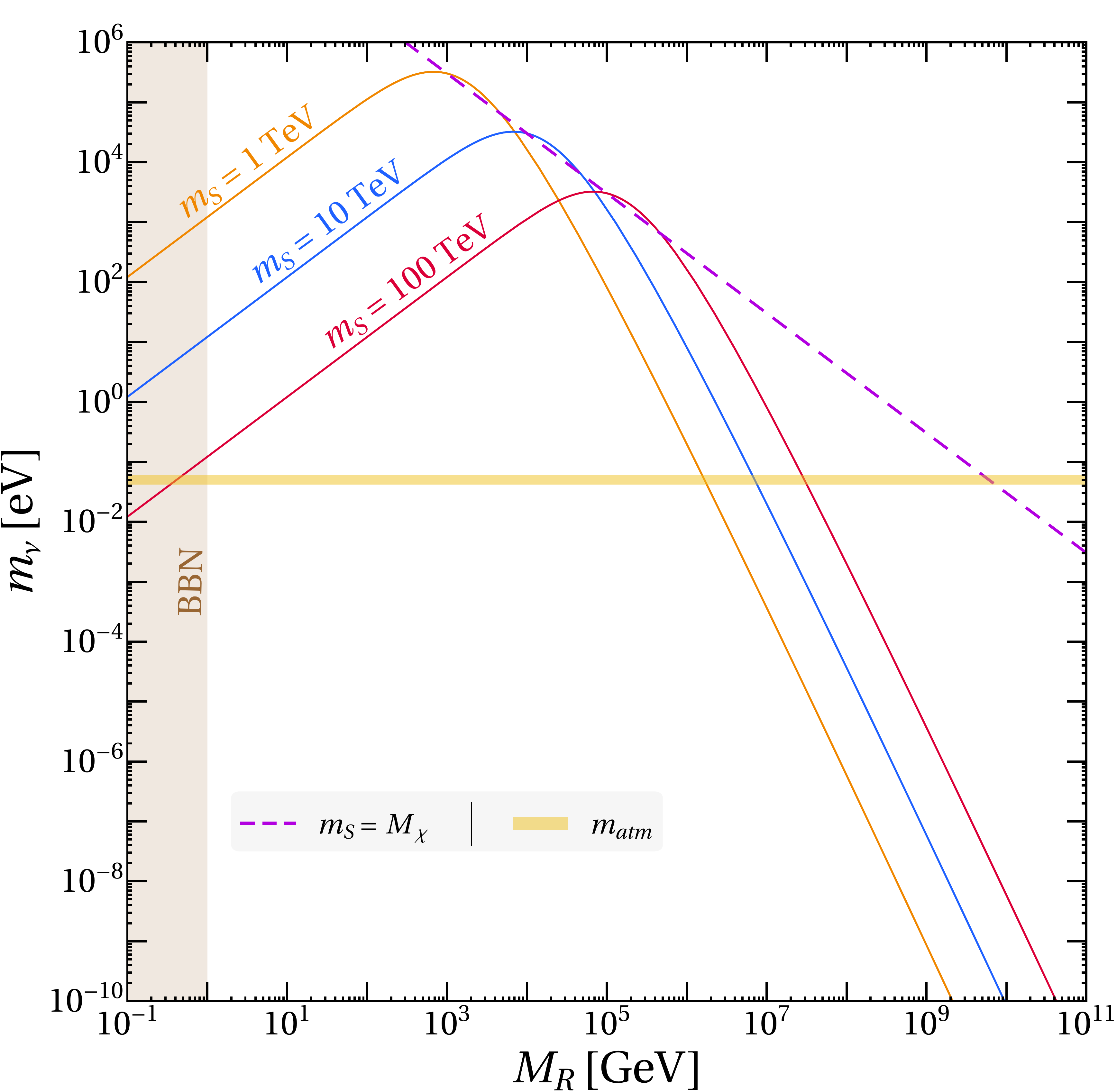}
    \hfill
    \includegraphics[width=0.47\textwidth]{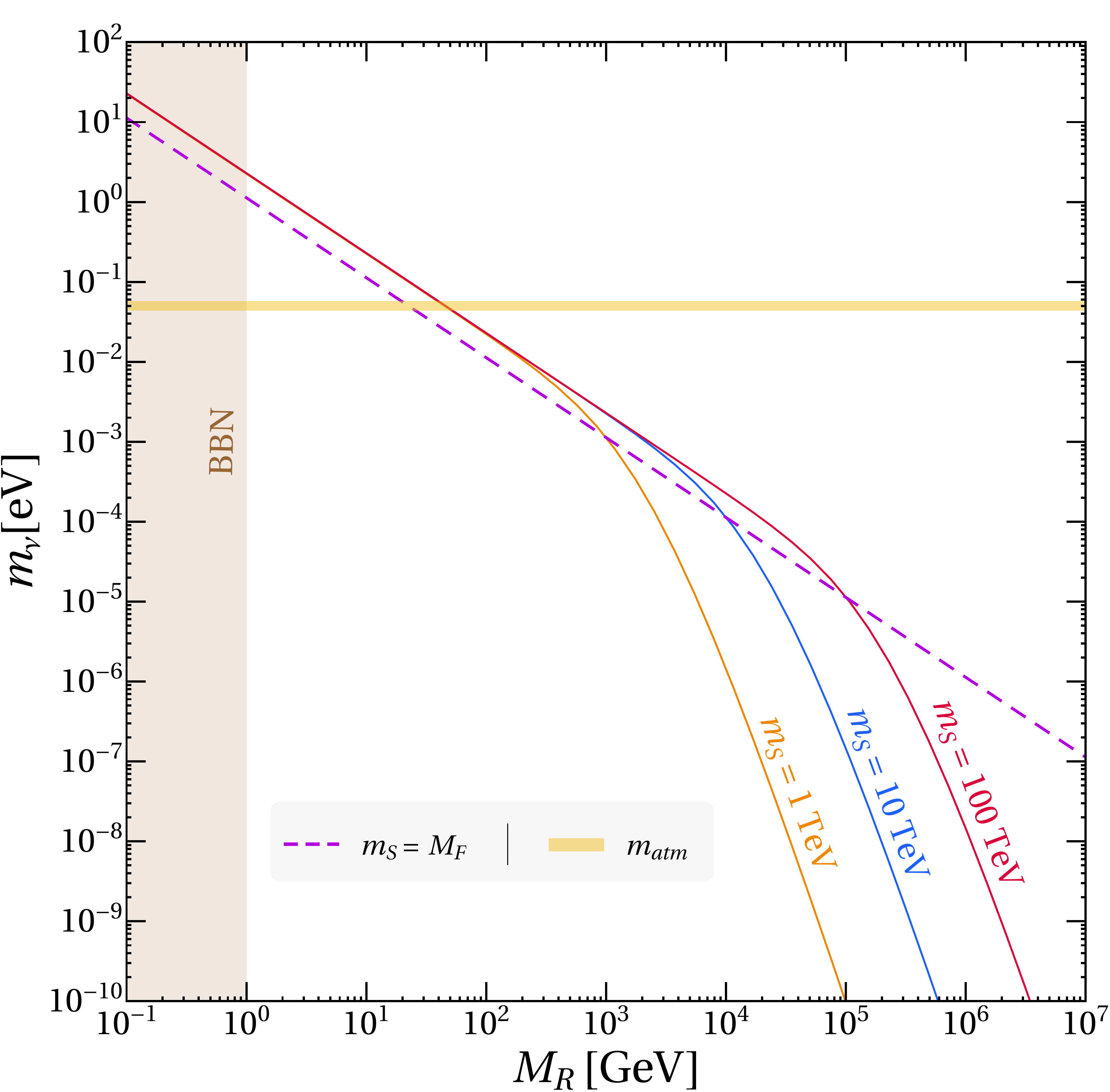}
    \caption{Escala de masa de los neutrinos para los ejemplos de modelos a un loop (izq.) y dos loops (der.) descritos en la Sección \ref{sec:loopss:examples}. La línea discontinua corresponde al caso en donde la masa de escalares y fermiones coincide, mientras que la línea continua describe el caso donde la masa de los fermiones es igual a $M_{R}$. BBN excluye las masas entre $M_R >(0.1-1)$ GeV para esta clase de modelos. La región de parámetros que se halla aproximadamente por debajo de la línea de $m_{atm}$, no sería capaz de explicar la escala de masa de neutrinos correctamente, ya que necesitaría acoplamientos no perturbativos.}
    \label{fig:resumen:massloopseesaw}
\end{figure}

En el Capítulo \ref{ch:loop_seesaw} hemos desarrollado y estudiado una extensión del seesaw de tipo-I original. Estos modelos de masa para neutrinos de Majorana, como en el seesaw, incluyen al menos dos copias del neutrino dextrógiro con una masa de Majorana. La diferencia radica en que los Yukawa de Dirac en lugar de ser a nivel árbol, se generan de forma radiativa. De esta forma, se pude explicar la masa de los neutrinos con acoplamientos de orden uno y una escala de nueva física entre $1-10^{10}$ GeV, según el número de loops (ver figura \ref{fig:resumen:SeesawEff}). Esto implica una mejora considerable respecto al seesaw original, que necesita una masa de $10^{14}$ GeV para los neutrinos dextrógiros. La generación radiativa de los Yukawas de Dirac se fundamenta en el mecanismo descrito en el capítulo anterior. Se empieza con una simetría que prohíbe el término $\bar L H^c \nu_R$ pero permite la masa de Majorana para $\nu_R$, esta se rompe, suavemente en nuestro caso, de tal forma que deja una simetría residual que permite la realización radiativa del operador $\bar L H^c \nu_R$ a $n$-loops, así como estabiliza a un posible candidato de materia oscura que participe en los loops. Hempos estudiado el mecanismo de la forma más general posible. Para ello, hemos parametrizado el Yukawa de Dirac efectivo en términos 5 exponentes relacionados con el número de loops, las inserciones de masa y el número de acoplamientos dimensionales y adimensionales. Variando estos parámetros, encontramos que los límites de nucleosíntesis primordial (BBN) y $\Delta N_{eff}$ descartan una gran parte de los posibles modelos, mientras que varias de las realizaciones restantes están al alcance de futuras búsquedas de nueva física y experimentos. Por último, hemos construido y estudiado dos ejemplos en detalle, una realización a un loop y otra a dos loops con un fermión interno del Modelo Estándar. Para ambos casos, se encuentra un valor máximo de la masa del neutrino dextrógiro $M_R$ considerando que el modelo tiene que generar una masa de neutrinos a la escala atmosférica, i.e. aproximadamente $0.05$ eV. A su vez, se tiene una cota mínima de $1$ GeV sobre $M_R$ procedente de nucleosíntesis primordial (ver figura \ref{fig:resumen:massloopseesaw}). También hemos estimado las cotas debidas a las búsquedas de violación de sabor leptónico en \mueg. En estos modelos existen dos escenarios extremos suponiendo que toda la mezcla de los neutrinos sea explicada por el Yukawa más externo que conecta con $L$, $Y_L$, o el más interno asociado a $\nu_R$, $Y_R$, siendo el resto de Yukawas diagonales. Se observa que ambas condiciones establecen un pequeño rango masas permitidas por el límite experimental de \mueg, capaz de reproducir los datos de oscilaciones de neutrinos y suponiendo todos los acoplamientos dentro del régimen perturbativo. Esto excluiría masas escalares menores a $4$ TeV, llegando aproximadamente a $6$ TeV si no se observa ninguna señal en la actualización de la colaboración MEG. 

\section*{Fenomenología}

Por último, los tres capítulos restantes se centran completamente en la fenomenología. En estos hemos estudiado distintos tipos de señales y límites que se pueden aplicar a los modelos radiativos de masas de neutrinos de dimensión 7 a un loop, a tres conocidos modelos mínimos a tres loops y, finalmente, hemos analizado un caso particular de decaimiento doble-$\beta$ sin emisión de neutrinos, pero acompañado de la emisión de un escalar ligero.

Primeramente, en el Capítulo \ref{ch:Dim7_pheno} hemos estudiado en detalle la rica feno- menología de los modelos de masa de neutrinos de dimensión 7 a un loop tratando de derivar conclusiones generales utilizando datos del LHC y resultados experimentales de física de partículas de baja energía. Para ello, partimos de la base del Capítulo \ref{ch:Dim7_1loop} y consideramos el modelo genuino mínimo que hemos encontrado dentro de esta clase, realizable con representaciones iguales o inferiores al triplete, así como un ejemplo de un modelo con un solo cuadruplete. Como ya se explicó en el Capítulo \ref{ch:Dim7_1loop}, esta clase de modelos siempre contiene grandes representaciones e hipercargas para ser genuinos, es decir, partículas con cargas eléctricas muy altas. A su vez, existe un límite superior a las masas de $2$ TeV, siempre y cuando se suponga que estos modelos son dominantes. Hemos estudiado las cotas provenientes de los datos de oscilación de neutrinos, las búsquedas de violación del sabor leptónico y del LHC. Hemos observado que ajustando los datos de neutrinos gran parte del espacio de parámetros queda excluido por la violación del sabor leptónico, mientras que la precisión esperada de los futuros experimentos debería ser capaz de testear casi todo el espacio de parámetros. Además, las cotas debidas al LHC excluyen una parte considerable del rango de masas capaz de explicar los datos experimentales, llegando en algunos casos a cotas por encima del límite fenomenológico de $2$ TeV considerando una futura luminosidad de $300$/fb en el LHC (ver tabla \ref{tab:resumen:Tablelnv}). Esto se debe a que estos modelos, al ser Majorana, violan número leptónico (LNV), en combinación con las partículas de alta carga eléctrica que contienen, proporcionan unas señales muy característica, con muy alta multiplicidad y leptones del mismo signo, con un fondo muy bajo del Modelo Estándar en colisionadores.

\begin{table}[t!]
    \begin{center}
        \small\addtolength{\tabcolsep}{-5pt}
        \begin{tabular}{|c|c|c|c|c|c||c|c|c|c|}
            \hline
            Multip.  & Señal de LNV & Partícula & Modelo & Masas  \\
            \hline
            4 (6) & $l^{\pm} l^{\pm}  +  W^{\mp} W^{\mp}$ & $S^{{\pm} {\pm}}$, $\phi_{1}^{{\pm}{\pm}}$, $\phi_{2}^{\pm\pm}$& Q & $m < 1.4$ TeV  \\
            \hline
            6 (8) & $l^{\pm} l^{\pm} l^{\pm}  +   W^{\mp} W^{\mp} l^{\mp}$ & $\chi_{2}^{3+}$& Q & $m < 2.6$  TeV \\
            \hline
             6 (10) & $l^{\pm} l^{\pm} W^{\pm} +  W^{\mp} W^{\mp} W^{\mp}$ & $S^{3+}$, $\phi_{2}^{3+}$& Q &   $m < 2.0$  TeV\\
            \hline
             8 (10) & $l^{\pm} l^{\pm} l^{\pm} l^{\pm}+  l^{\mp} l^{\mp} W^{\mp} W^{\mp}$ & $\eta_{3}^{4+}$& T & $m < 2.5$ TeV \\
            \hline
            8 (12) & $l^{\pm} W^{\pm} W^{\pm} W^{\pm}+   l^{\mp} l^{\mp} l^{\mp} W^{\mp}$ & $\chi_2^{4+}$ & Q & $m < 3.2$ TeV\\
            \hline
            8 (14) & $l^{\pm} l^{\pm} W^{\pm} W^{\pm}+   W^{\mp} W^{\mp} W^{\mp} W^{\mp}$ & -- & -- & -- \\
            \hline
            \hline
        \end{tabular}
    \end{center}
    \caption{Lista de estado finales simétricos con LNV. La primera y segunda columna muestran la multiplicidad y la señal correspondiente de LNV, respectivamente. Sin paréntesis se cuenta el número de campos en el estado final con $W$'s, y con paréntesis considerando que estos decaen a dos jets. Hemos separado el estado final en dos grupos de partículas, cada uno viene del campo dado en la tercera columna, producido siempre en pares en el LHC, y perteneciente al modelo del triplete o del cuadruplete (T o Q en la cuarta columna, ver Sección \ref{sec:pheno:dim7}). La última columna es una estimación aproximada del rango de masas que podría probarse en el LHC con una luminosidad de $300$/fb.}
    \label{tab:resumen:Tablelnv}
\end{table}

\begin{figure}[h!]
    \centering
    \includegraphics[width=0.495\textwidth]{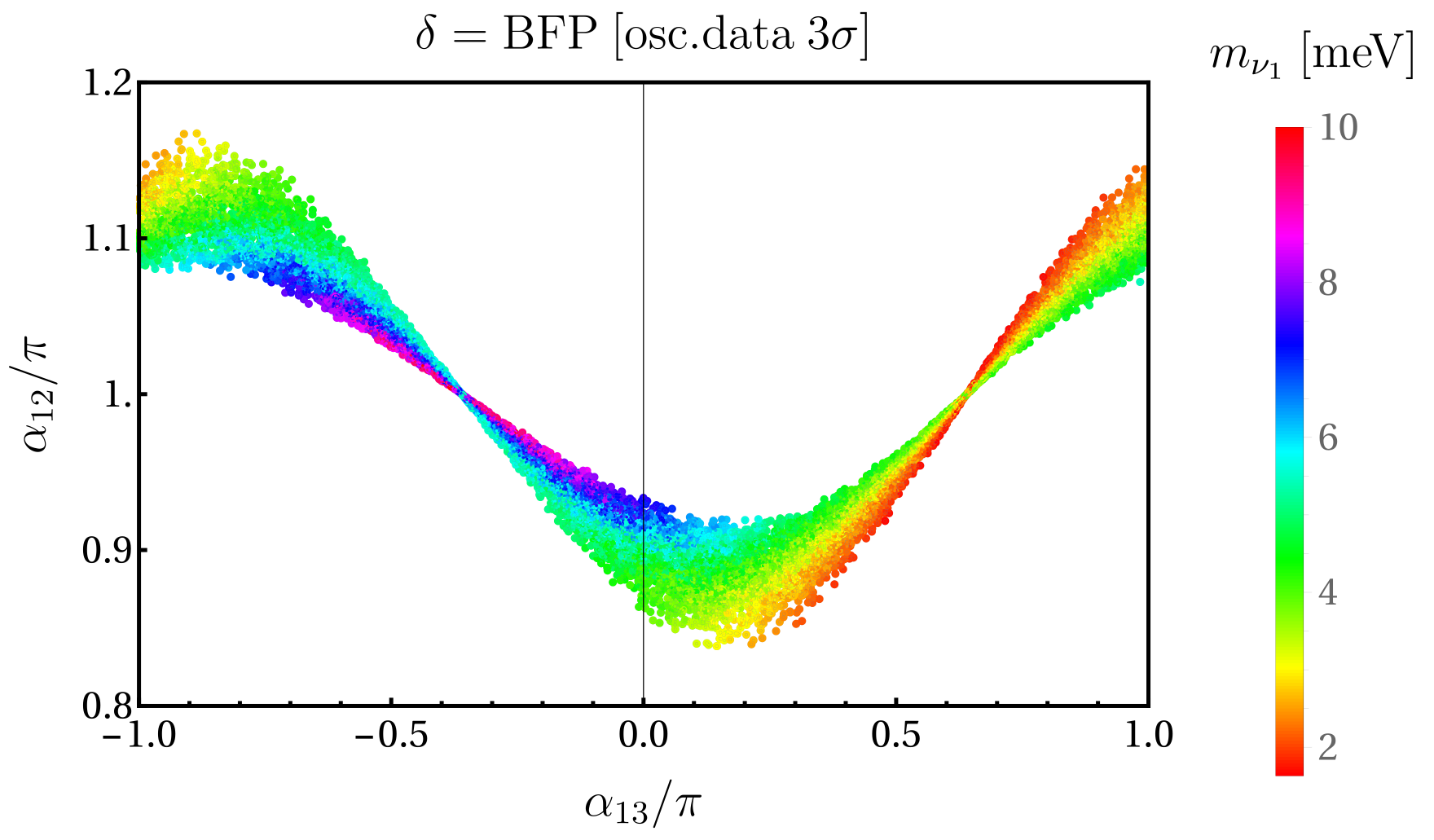}
    \hfill
    \includegraphics[width=0.47\textwidth]{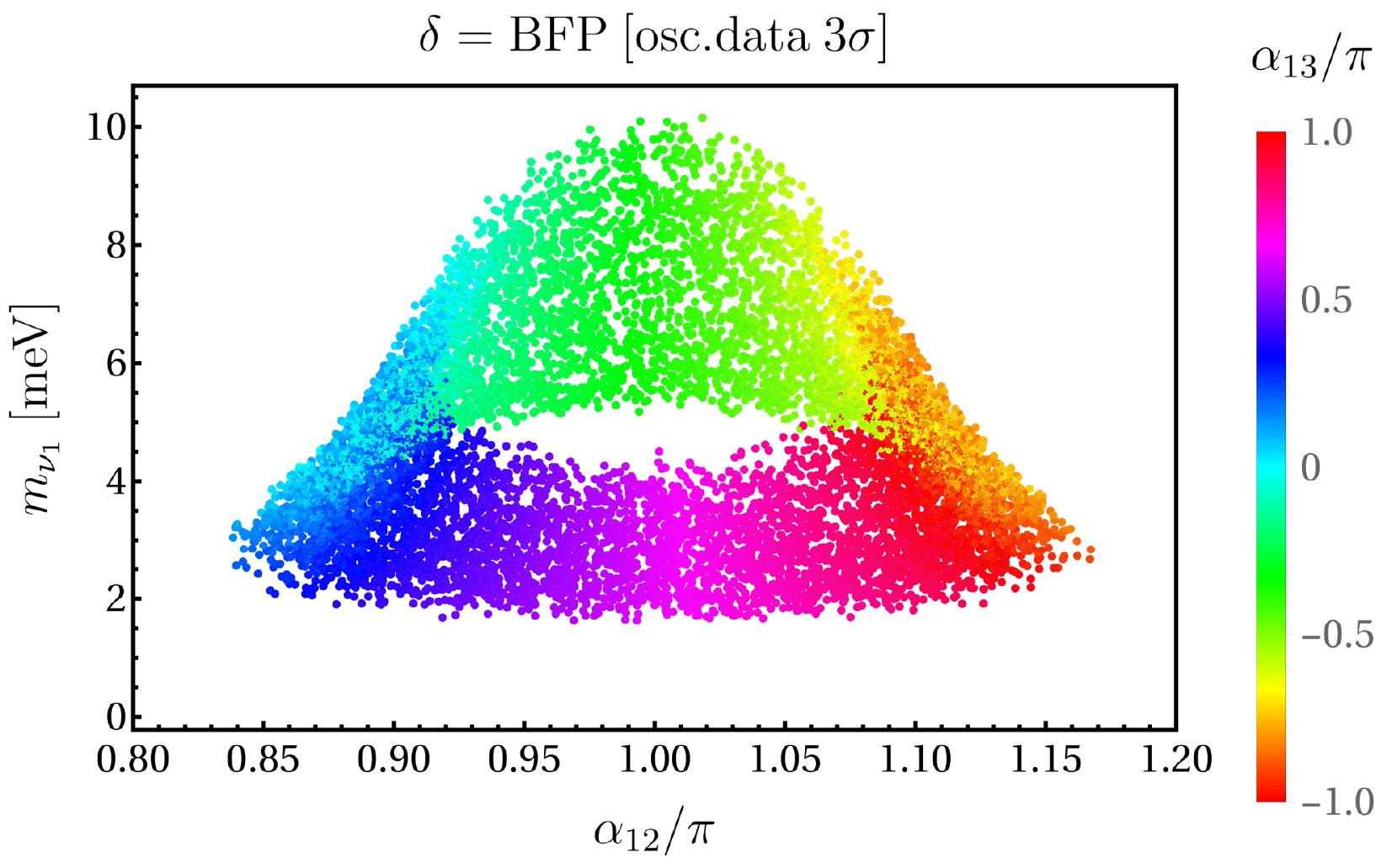}
    \caption{Espacio de parámetros permitido de $\alpha_{12}$, $\alpha_{13}$ y $m_{\nu_1}$ para el modelo cóctel. Los datos de oscilaciones de neutrinos se han escaneado en un rango de $3\sigma$, excepto $\delta$ que se ha tomado al valor esperado. Todos los puntos mostrados reproducen los datos de neutrinos, están por debajo de los límites experimentales de CLFV y no contienen acoplamientos no perturbativo.}
    \label{fig:resumen:ph}
\end{figure}

Seguidamente, en el Capítulo \ref{ch:clfv}, hemos estudiado tres modelos mínimos de masa de neutrinos de Majorana a tres loops. Estos modelos se conocen popularmente en al literatura como modelo cóctel, el modelo KNT y el modelo AKS. Llamamos a estos modelos ``mínimos'', ya que su contenido de partículas corresponde al contenido mínimo de partículas para el que se pueden construir modelos genuinos de tres loops. Nos hemos centrado en la fenomenología de la violación del sabor leptónico (CLFV), pues de esta provienen las cotas más restrictivas a estos modelos. En los tres modelos mínimos, la matriz de masa de neutrinos es proporcional a alguna potencia de las masas de los leptones del Modelo Estándar, lo que proporciona un factor de supresión adicional, además de la supresión esperada debida a los tres loops. En consecuencia, para explicar correctamente las masas de los neutrinos, se necesitan valores elevados de Yukawas, lo que dispara el valor que estos modelos predicen para la violación de sabor leptónico. Los modelos sobreviven solo en ciertas regiones del espacio de parámetros muy restringidas donde aparecen polos en alguna o varias de las señales de CLFV. Hemos observado que solamente sobreviven elecciones muy particulares de las fases de Dirac y Majorana para un rango estrecho de la masa del neutrino más ligero considerando los actuales límites provenientes de LEP y LHC (ver figura \ref{fig:resumen:ph} para el modelo cóctel). Finalmente, hemos concluido que esta clase de modelos, si aún no se han excluido, están severamente restringidos. Además, abre la posibilidad de testear algunos de los modelos en futuras actualizaciones experimentales de CLFV o experimentos de desintegración doble-$\beta$.

\begin{figure}[h!]
    \centering
    \includegraphics[width=0.29\textwidth]{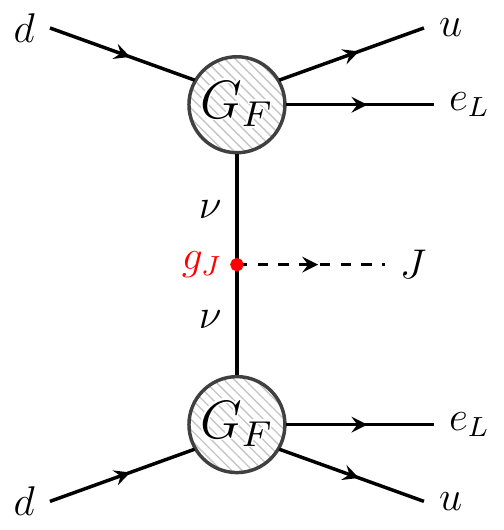}
    \hfill
    \includegraphics[width=0.29\textwidth]{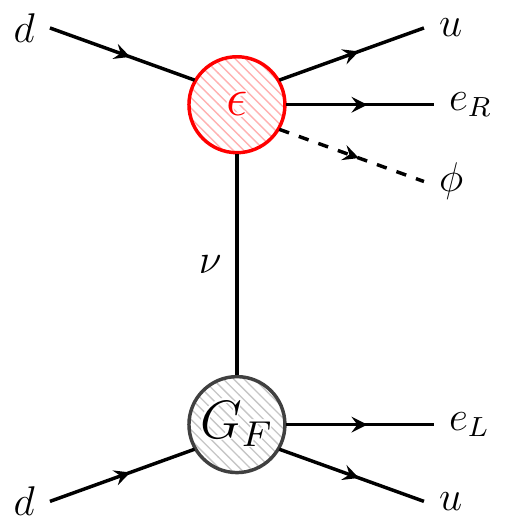}
    \hfill
    \includegraphics[width=0.35\textwidth]{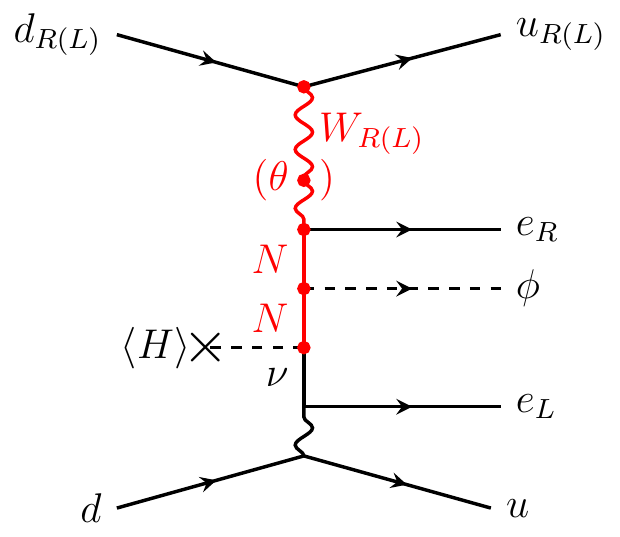}
    \caption{Diagrama del decaimiento con emisión de Majoron estándar $0\nu\beta\beta J$ (izq.), modo de decaimiento estudiado en esta tesis (centro) y posible realización ultravioleta en el modelo Left-Right (der.).}
\label{fig:resumen:0vbbdiagram} 
\end{figure}

\begin{figure}[h!]
    \centering
    \includegraphics[clip,trim={10 10 0 0},width=0.49\textwidth]{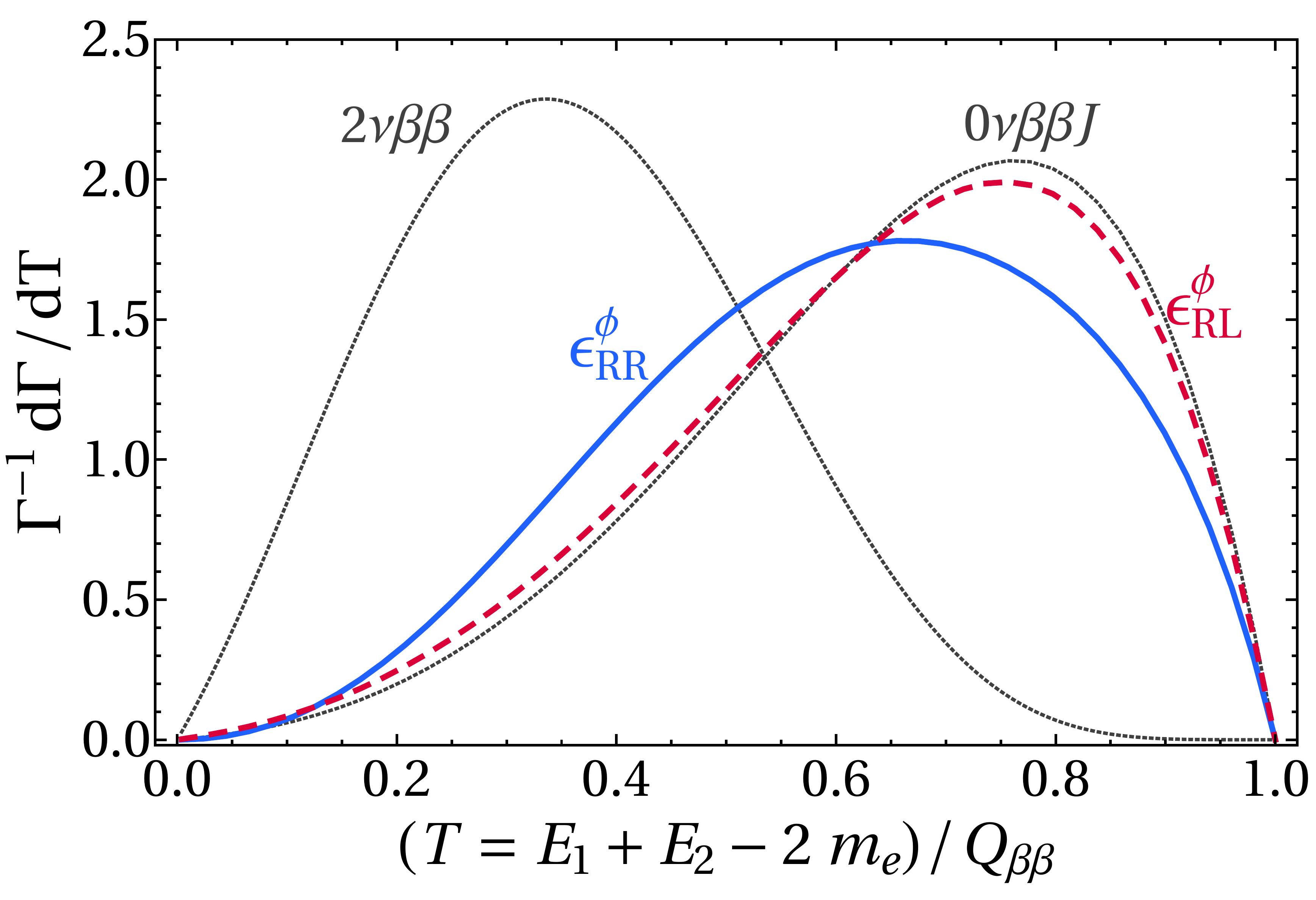}
    \hfill
    \includegraphics[clip,trim={10 10 0 0},width=0.49\textwidth]{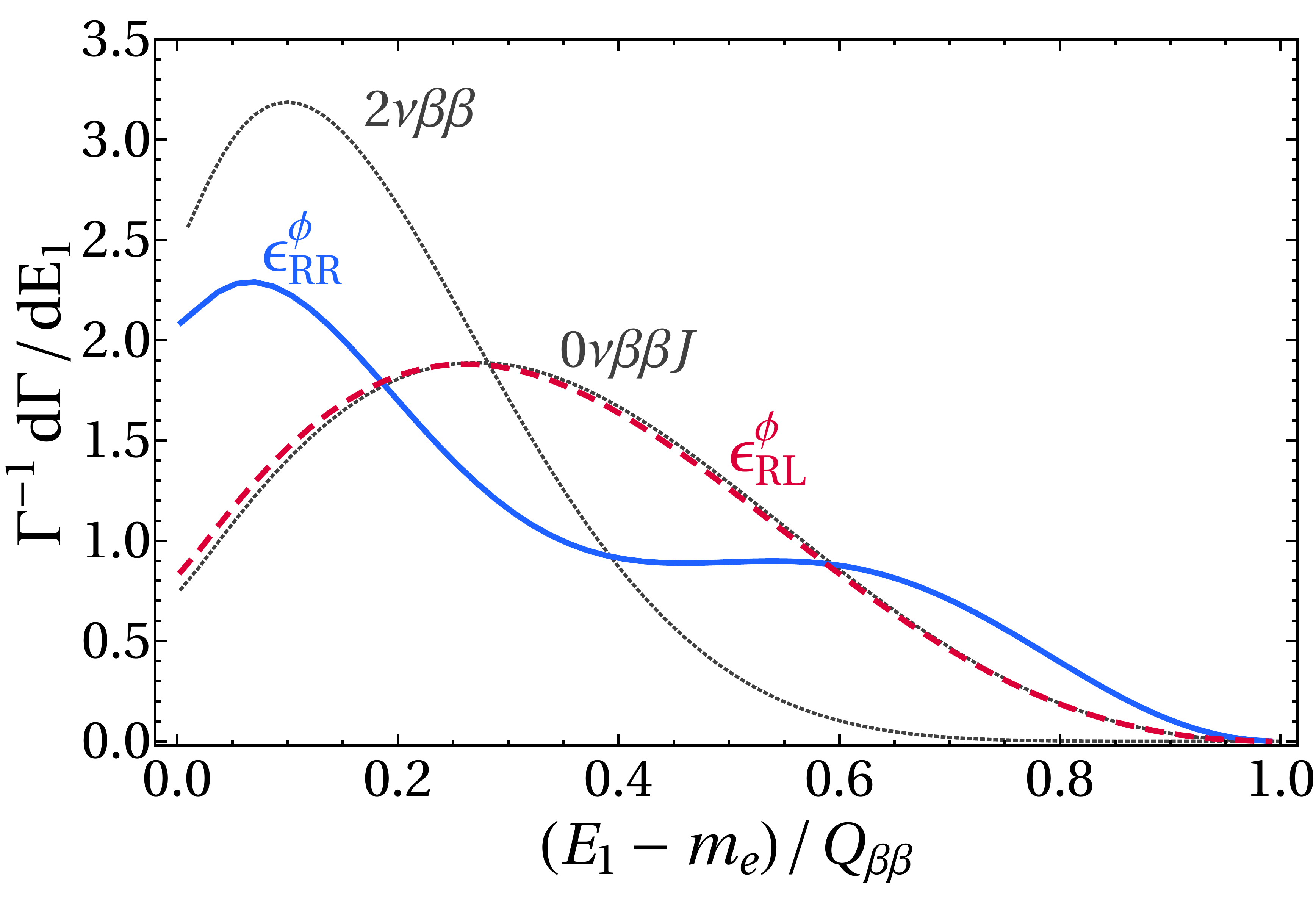}
    \caption{Distribución de decaimiento normalizada en función de la energía cinetica total de los electrones para el $^{136}\text{Xe}$ (izq.). Distribución de decaimiento normalizada respecto a la energía cinética de un solo electrón para el $^{82}\text{Se}$ (der.). Las líneas continuas azules y discontinuas rojas corresponden al mismo modo de emisión pero considerando corrientes hadrónicas dextrógiras o levógiras, respectivamente. Para comparar, se representa la distribución del Modelo Estándar para $2\nu\beta\beta$ y la distribución del Majoron ordinario $0\nu\beta\beta J$.}
    \label{fig:resumen:energydist}
\end{figure}

Por último, en el Capítulo \ref{ch:0vbb} discutimos un modo novedoso de desintegración doble-$\beta$ sin emisión de neutrinos, pero con la emisión de una partícula escalar ligera $\phi$, llamada comúnmente Majoron en la literatura, y que denotamos por $0 \nu \beta \beta \phi$. Para ello, hemos introducido un operador de dimensión $7$ con una corriente leptónica vector-axial dextrógira asociado al escalar $\phi$, $\bar u \gamma^\mu (1\pm\gamma_5) d \, \bar e \gamma_\mu (1+\gamma_5) \nu \, \phi$. De esta forma, el decaimiento se produce a través de una contribución de largo alcance que no está suprimida por la masa de los neutrinos ligeros, emitiendo en el proceso un escalar $\phi$ y dos neutrinos con quiralidades opuestas (ver figura \ref{fig:resumen:0vbbdiagram}). Hemos calculado las expresiones analíticas de las tasas de decaimiento total y diferencial asociadas a este proceso. Con ello, hemos obtenido límites a los operadores considerados, así como caracterizado el proceso mediante las distribuciones de decaimiento para la energía cinética total y de un solo electrón (ver figura \ref{fig:resumen:energydist}). Hemos concluido que las búsquedas futuras de decaimiento doble-$\beta$ son sensibles a escalas del orden de $1$ TeV para una masa escalar ligera por debajo de $0.2$ MeV. También, hemos observado que mientras que en la distribución en función de la energía cinética total la desviación respecto a la emisión clásica de un Majoron es pequeña, la de un solo electrón presenta un \textit{valle} muy característico en la distribución, debido a las quiralidades de los electrones, y cuyo máximo tiende al eje de abscisas, consecuencia de la emisión conjunta con el escalar. Una observación similar se obtiene si se representa la distribución angular del proceso. Esto podría utilizarse experimentalmente para mejorar la sensibilidad mediante criterios de selección cinemática, contrarrestando el efecto de la distribución de energía total menos pronunciada. Por ejemplo, sería posible requerir que cualquiera de los electrones en un evento de señal tenga una energía cinética mayor de la mitad de la máxima posible, esto reduciría la tasa de $0 \nu \beta \beta \phi$ solo en un factor de 2, pero suprimiría la tasa de $2\nu \beta \beta$ en un factor de 20. Este estudio supone un aliciente para futuros experimentos de decaimiento doble-$\beta$ dedicados a la búsqueda de nueva física en el análisis de la distribución de un solo electrón y la distribución angular.

\section*{Conclusiones}

En resumen, hemos presentado y discutido una gran variedad de modelos de masa de neutrinos radiativos desde un punto de vista fenomenológico y teórico de construcción de modelos. Los hemos clasificado prestando especial atención a su fenomenología, de esta forma se han obtenido clases de modelos con características y señales similares. Se han analizado en detalle escenarios o modelos específicos, especialmente interesantes por su fenomenología y/o sencillez. En la mayoría de casos, los modelos radiativos se presentan como una buena opción para explicar la pequeñez de las masas de los neutrinos de una forma natural, así como otros problemas del Modelo Estándar como la materia oscura, que puede ser fácilmente introducida en estos modelos. A su vez, al no requerir una escala de energía excesivamente alta para explicar los datos experimentales, los modelos radiativos ofrecen una ventana a la nueva física que puede ser probada por experimentos actuales y futuros.

\pagebreak
\fancyhf{}